\documentclass[a4paper,USenglish,cleveref,autoref,thm-restate,pdfa]{lipics-v2021}

\hideLIPIcs  

\usepackage{bm}
\usepackage{amsmath}
\usepackage{xcolor}
\usepackage{tikz}
  \usetikzlibrary{calc}
  \usetikzlibrary{arrows}
\usepackage{booktabs}
\usepackage{listings}
\usepackage{bussproofs}
\input{proglangtools/progbnf}
\input{proglangtools/proginfer}
\input{proglangtools/progindent}

\newcommand\defeq{\mathrel{:=}}
\newcommand\revdefeq{\mathrel{=:}}
\newcommand\dom{\mathop{\mathrm{dom}}}
\newcommand\fv{\mathop{\mathrm{fv}}}
\newcommand\arity[1]{\mathop{\mathrm{ar}}(#1)}
\newcommand\dfn[1]{\textit{#1}}
\newcommand\opVertConcat{%
  \mathbin{\text{\rlap{\raisebox{0.4em}{\footnotesize\(+\)}}}\raisebox{-0.1em}{\footnotesize\(+\)}}%
}
\newcommand\Aprime{A\({}^\prime\)}
\newcommand\Bprime{B\({}^\prime\)}
\newcommand\Cprime{C\({}^\prime\)}
\newcommand\TypeVec[1]{\mathtt{Vec}\ #1}
\newcommand\TypeMat[2]{\mathtt{Mat}\ #1\ #2}
\newcommand\TypeNat{\mathtt{Nat}}
\newcommand\LambdaBracketCast{\(\lambda^{\langle\rangle(\!\!\!\;||\!\!\!\;)}\)}
\newcommand\LambdaBracketCastImplicit{\(\lambda^{\langle\rangle(\!\!\!\;||\!\!\!\;)\{\}}\)}
\newcommand\ConstEnvZero{\mathcal{B}_{0}}
\newcommand\ConstEnvPers{\mathcal{B}_{\%}}
\newcommand\ConstEnvSurface{\mathcal{D}}
\newcommand\PositionZeroTurnstile{
  \mathrel{%
    \begin{tikzpicture}
      \useasboundingbox (-0.3em, 0) rectangle (0.9em, 0.75em);
      \draw
        (0em, -0.15em) -- (0em, 0.65em) -- (0.14em, 0.65em)
          -- (0.14em, 0.32em) -- (0.6em, 0.32em) -- (0.6em, 0.18em) -- (0.14em, 0.18em)
          -- (0.14em, -0.15em) -- cycle;
    \end{tikzpicture}%
    \mathstrut
  }%
}
\makeatletter
  \def\local@unique{\local@uniquesub}
  \def\local@ifempty#1#2#3{%
    \ifx\local@unique#1\local@unique#2\else#3\fi
  }
  \newcommand\recalltheorem[3][]{%
    {%
      \def\thetheorem{\ref{#2}}%
      \addtocounter{theorem}{-1}%
      \local@ifempty{#1}{%
        \begin{theorem}#3\end{theorem}%
      }{%
        \begin{theorem}[#1]#3\end{theorem}%
      }%
    }%
  }
  \newcommand\recalllemma[3][]{%
    {%
      \def\thetheorem{\ref{#2}}%
      \addtocounter{theorem}{-1}%
      \local@ifempty{#1}{%
        \begin{lemma}#3\end{lemma}%
      }{%
        \begin{lemma}[#1]#3\end{lemma}%
      }%
    }%
  }
\makeatother

\newcommand{\ottmv}[1]{\mathit{#1}}

\newcommand\token[1]{%
  \mathbf{#1}%
}
\newcommand\identifier[1]{%
  \mathit{#1}%
}
\newcommand\constant[1]{%
  #1%
}
\newcommand\stageOcolor[1]{%
  \textcolor[rgb]{0,0.4,0.8}{#1}%
}
\newcommand\stageOmetaColor[1]{%
  \textcolor[rgb]{0,0.4,0.8}{#1}%
}
\newcommand\ordO[1]{%
  \stageOcolor{#1}%
}
\newcommand\relO[1]{%
  \mathrel{%
    \stageOcolor{#1}%
  }%
}
\newcommand\binO[1]{%
  \mathbin{%
    \stageOcolor{#1}%
  }%
}
\newcommand\punctO[1]{%
  \mathpunct{%
    \stageOcolor{#1}%
  }%
}
\newcommand\openO[1]{%
  \mathopen{%
    \stageOcolor{#1}%
  }%
}
\newcommand\closeO[1]{%
  \mathclose{%
    \stageOcolor{#1}%
  }%
}
\newcommand\tokenO[1]{%
  \stageOcolor{%
    \token{#1}%
  }%
}
\newcommand\ttO[1]{%
  \stageOcolor{%
    \mathtt{#1}%
  }%
}
\newcommand\itO[1]{%
  \stageOcolor{%
    \mathit{#1}%
  }%
}
\newcommand\stageIcolor[1]{%
  \textcolor[rgb]{0.7,0.3,0}{#1}%
}
\newcommand\stageImetaColor[1]{%
  \textcolor[rgb]{0.7,0.3,0}{#1}%
}
\newcommand\ordI[1]{%
  \stageIcolor{#1}%
}
\newcommand\relI[1]{%
  \mathrel{%
    \stageIcolor{#1}%
  }%
}
\newcommand\binI[1]{%
  \mathbin{%
    \stageIcolor{#1}%
  }%
}
\newcommand\punctI[1]{%
  \mathpunct{%
    \stageIcolor{#1}%
  }%
}
\newcommand\openI[1]{%
  \mathopen{%
    \stageIcolor{#1}%
  }%
}
\newcommand\closeI[1]{%
  \mathclose{%
    \stageIcolor{#1}%
  }%
}
\newcommand\tokenI[1]{%
  \stageIcolor{%
    \token{#1}%
  }%
}
\newcommand\ttI[1]{%
  \stageIcolor{%
    \mathtt{#1}%
  }%
}
\newcommand\BlameSign{%
  \ordO{\Uparrow}%
}
\newcommand\LeftAssertParen{%
  \openO{(\!\!\!\;|}%
}
\newcommand\RightAssertParen{%
  \closeO{|\!)}%
}
\newcommand\CastArrow{%
  \triangleleft
}
\newcommand\ElabArrow{%
  \rightsquigarrow
}
\makeatletter
  \let\local@spaceLetter= \relax
  \def\local@chop#1#2\local@end{%
    \ifx\local@spaceLetter#1%
      local@chop#2\local@end
    \else
      \ifx V#1%
        \identifier{#2}%
      \else
        \ifx C#1%
          \constant{#2}%
        \else
          \PrefixOtherThanVOrC
        \fi
      \fi
    \fi
  }
  \newcommand\makeIdentOrConst[2]{%
    \mathtt{#1\local@chop#2\local@end}%
  }
  \def\local@space{ }
  \expandafter\def\expandafter\local@removeSpace\local@space{}
  \newcommand\possiblyWithSub[2]{%
    \def\local@cmd{#1}%
    \def\local@arg{#2}%
    \futurelet\local@nextToken\local@possiblyWithSub@trimSpace
  }
  \def\local@possiblyWithSub@trimSpace{%
    \expandafter\ifx\local@space\local@nextToken
      \expandafter\expandafter\expandafter\futurelet
        \expandafter\expandafter\expandafter\local@nextToken
        \expandafter\expandafter\expandafter\local@possiblyWithSub@trimSpace
        \expandafter\local@removeSpace
    \else
      \expandafter\local@possiblyWithSub
    \fi
  }
  \def\local@possiblyWithSub{%
    \ifx_\local@nextToken
      \expandafter\local@withSub
    \else
      \expandafter\local@withoutSub
    \fi
  }
  \def\local@withSub_#1{%
    \local@cmd{%
      \local@arg_{#1}%
    }%
  }
  \def\local@withoutSub{%
    \local@cmd{\local@arg}%
  }
\makeatother
\newcommand\superscriptO{\scriptscriptstyle(0)}
\newcommand\superscriptI{\scriptscriptstyle(1)}

\theoremstyle{acmplain}
\newtheorem{assumption}[theorem]{Assumption}

\title{Compile-Time Tensor Shape Checking via Staged Shape-Dependent Types}

\author{Takashi Suwa}{Kyoto University, Kyoto, Japan \and Imiron Co., Ltd., Tokyo, Japan}{tsuwa@fos.kuis.kyoto-u.ac.jp}{https://orcid.org/0009-0004-9845-7419}{}
\author{Atsushi Igarashi}{Kyoto University, Kyoto, Japan}{igarashi@kuis.kyoto-u.ac.jp}{0000-0002-5143-9764}{}
\authorrunning{T. Suwa and A. Igarashi}
\Copyright{Takashi Suwa and Atsushi Igarashi}



\funding{This work is partially supported by JSPS KAKENHI Grant Numbers 20H00582 and 26H02493, Japan.}

\acknowledgements{%
  The authors wish to thank the anonymous reviewers and (past) members of our laboratory
  for various fruitful comments and feedbacks.
  We are also grateful to Shivankur Gupta for implementing a basic code-generating backend for OCaml
  during his internship.
}

\nolinenumbers 

\EventEditors{Robbert Krebbers and Alexandra Silva}
\EventNoEds{2}
\EventLongTitle{40th European Conference on Object-Oriented Programming (ECOOP 2026)}
\EventShortTitle{ECOOP 2026}
\EventAcronym{ECOOP}
\EventYear{2026}
\EventDate{June 29--July 3, 2026}
\EventLocation{Brussels, Belgium}
\EventLogo{}
\SeriesVolume{372}
\ArticleNo{10}

\begin{document}

\maketitle

\begin{abstract}
  When writing programs involving matrices or \dfn{tensors} in general,
it is desirable to rule out the inconsistency of \dfn{tensor shapes}
(i.e.,~the generalization of matrix sizes)
before actual computation.
For this purpose, some languages provide dependent types such as \(\mathtt{Mat}\ m\ n\),
and others offer refinement types to track predicates for shapes.
Despite the theoretical maturity, however,
such methods are often unhandy for continuous software development
due to the requirement of proofs for judging type equality or subtyping;
even automated proving is often unsuitable due to its unforeseeable time consumption.
To remedy this, our study provides
an alternative formalization by using \dfn{staging}.
Based on the observation that conditions for the shape consistency
can be extracted before running the actual tensor computations in many typical cases,
we ensure such consistency by assertions evaluated as compile-time computations, not by proofs.
Under this formalization, we can verify the consistency \emph{virtually statically}
in the sense that inconsistencies
will be immediately detected as failures during compile-time computation.
Our work achieves a mathematical guarantee that
successfully generated code is always consistent with respect to tensor shapes.
Furthermore, to vastly lessen the burden of adding shape- or stage-related descriptions, we
(1)~allow shape-related arguments to be implicit and infer them in a best-effort manner,
and (2)~offer a non-staged surface language that seemingly resembles ordinary dependently-typed languages
and translate its programs into the staged core language.
By a prototype implementation, we confirm that our language is expressive enough to verify
a number of programs, including several examples offered by \texttt{ocaml-torch}.

\end{abstract}

\section{Introduction}
\subsection{Background: Tensor Computation and Shape Checking}
\indent
  Nowadays, \dfn{tensors} or \dfn{multi-dimensional arrays}
  (i.e.,~vectors, matrices, cuboids, and so on)
  are nearly everywhere;
  they have long been used in various methods for mathematical optimization,
  and recently, due to the growing demand for machine learning,
  tensors have been intensively used
  to represent various structures related to deep neural networks~(DNN).
\par
\indent
  When it comes to writing programs involving tensors,
  it is desirable to rule out the inconsistency of \dfn{tensor shapes}
  (i.e.,~the generalization of vector lengths or matrix sizes)
  before running actual computations.
  For example, consider the function~\(f\) that
  takes three matrices~\(A\), \(B\), and \(C\) and computes \(A (B \opVertConcat C \opVertConcat B)\),
  where \(\opVertConcat\) is a binary operator that vertically concatenates two matrices.
  For verifying consistency, the following constraints must be met:
  (1) for the use of \(\opVertConcat\),
  \(B\) and \(C\) must have the same number of columns; and
  (2) for the matrix multiplication,
  the number of \(A\)'s columns must be equal to
  the number of \((B \opVertConcat C \opVertConcat B)\)'s rows.
  Checking such constraints during program execution is problematic because it may cause
  a runtime error due to some trifling shape mismatch bugs only after heavy computations,
  consuming much time.
\par
\indent
  A popular approach to addressing this kind of problem is to use dependent type systems.
  Some languages, such as Idris~\cite{BradyJFP2013,BradyECOOP2021},
  provide dependent types like \(\TypeVec{n}\) or \(\TypeMat{m}{n}\),
  which are the type for vectors of length \(n\) and
  the one for matrices of size~\(m \times n\), respectively, and assign
  the matrix multiplication the following type:
  \(\forall p, q, r : \TypeNat.\ \TypeMat{p}{q} \to \TypeMat{q}{r} \to \TypeMat{p}{r}\).
  Other methods,
  such as
  GraTen~\cite{HattoriKobayashiSatoESOP2023},
  offer refinement types like
  \(\{\nu : \mathtt{Tensor}\mid \nu.\mathtt{shape} = [m, n]\}\)
  or
  \(\{\nu : \mathtt{Tensor}\mid \mathtt{len}(\nu.\mathtt{shape}) = 3\}\).
  As an example of this kind of approach,
  consider the function~\(f\) above implemented in an Idris-like hypothetical language:
  \begin{align*}
    &
      \token{lemma}\ \mathit{HeightsMatch}\ k\ m : m + k + m = k + 2 \ast m
    \ \token{proof}\ \cdots\ \token{end}
    \\&
      \token{let}\ f\ \{j : \TypeNat\}\ \{k : \TypeNat\}\ \{m : \TypeNat\}\ \{n : \TypeNat\}
    \\&\quad\quad
      \ (A : \TypeMat{j}{(k + 2 \ast m)})\ (B : \TypeMat{m}{n})\ (C : \TypeMat{k}{n}) =
    \\&\quad
        \token{let}\ D = \mathit{vertCat}\ (\mathit{vertCat}\ B\ C)\ B\ \token{in}
        \ \mathit{matMult}\ A\ (\token{rewrite}\ \mathit{HeightsMatch}\ k\ m\ \token{in}\ D)
  \end{align*}
  Here, \(\{j : \TypeNat\} \cdots \{n : \TypeNat\}\) are binders for implicit parameters, and
  \(\mathit{matMult}\) and \(\mathit{vertCat}\)
  stand for the matrix multiplication and \(\opVertConcat\),
  which are assigned
  types~\(\forall p, q, r : \TypeNat.\ \TypeMat{p}{q} \to \TypeMat{q}{r} \to \TypeMat{p}{r}\)
  and \(\forall p, q, r : \TypeNat.\ \TypeMat{p}{r} \to \TypeMat{q}{r} \to \TypeMat{(p + q)}{r}\),
  respectively.
  \(\mathit{HeightsMatch}\ k\ m\) is a manually proved lemma used for
  matching \(D\)'s actual type with the one required of \(D\);
  since \(D\) is of type~\(\TypeMat{(m + k + m)}{n}\)
  and the function~\((\mathit{matMult}\ A)\) requires its argument to be of \(\TypeMat{(k + 2 \ast m)}{r'}\)
  for some \(r'\), the type-checker must verify that \(m + k + m = k + 2 \ast m\) holds.
  Working as a proof, the lemma rewrites the former type to the latter
  through the \(\token{rewrite}\)-construct.
  In general, to type-check an application~\(M_1\ M_2\),
  where \(M_1\) and \(M_2\) are known to have
  types~\(\TypeMat{M'_{11}}{M'_{12}} \to T\) and \(\TypeMat{M'_{21}}{M'_{22}}\),
  respectively,
  we must prove the type equality~\(\TypeMat{M'_{11}}{M'_{12}} = \TypeMat{M'_{21}}{M'_{22}}\),
  i.e., that, for \(i \in \{1, 2\}\),
  \dfn{argument expressions}~\(M'_{1i}\) and \(M'_{2i}\) always describe
  the same value in the given context.
  Languages of such approaches provide some form of mechanism for proving this equality.
\par
\indent
  Despite their theoretic maturity and success in many safety-critical fields, however,
  such verification methods do not seem to be so eagerly applied to
  relatively typical, \emph{continuous} software development, especially cases in industry.
  Although various reasons can be considered for this,
  we suppose that the following situations would be major factors:
  \begin{enumerate}
    \item \textbf{The externality of requirements}:
      Requirements imposed on software continuously arise due to
      social situations and users' preferences,
      which are not predictable beforehand.
    \item \textbf{The sequential nature of software development}:
      Checking is performed repeatedly during development.
      Although we can restrict properties to check to some lightweight ones
      for frequently performed verification,
      it would still be better if we do not have to take much time for each run,
      in order not to harm productivity.
    \item \textbf{Time is literally money}:
      It is costly to let software engineers engaged in development.
      Moreover, each engineer is available for approximately only 40 hours per week.
  \end{enumerate}
  Under such circumstances,
  major issues that hamper adoption of existing methods for tensor shape checking
  would be some of the following:
  \begin{enumerate}
    \item[A.] \textbf{Cumbersomeness of future changes}:
      Methods that require manual proving easily make
      future changes of programs unwieldy;
      even slight modification of programs may demand nearly complete amendment of proofs.
      This does not go along with the condition~1 above,
      i.e.,~the unpredictable nature of requirements imposed on software.
    \item[B.] \textbf{Unpredictable time consumption by automated proving}:
      To reduce the burden of manual proving,
      some methods provide automated proving by using back-end solvers,
      possibly with some part of the syntax restricted to a subset suitable for automation.
      However, such approaches are often too time-consuming or at least take unpredictable time;
      slight change of properties to check may drastically increase the elapsed time.
    \item[C.] \textbf{Frequent false-positive errors}:
      As pointed out in some articles like Ascari~et~al.~\cite{AscariBruniGoriLogozzo2024}
      or a CACM article about the use of static analyses
      in Facebook (currently known as Meta)~\cite{DistefanoFahndrichLogozzoOHearn2019},
      methods for verifying detailed properties are likely to cause false positives too easily
      due to their nature of the overapproximation of program behaviors.
      Namely, even when the validity is clear for humans,
      analysis tools often warn, e.g.,~the existence of a type-level gap
      and require some form of annotations or proofs.
      Frequent false-positive errors also add additional cumbersomeness to future changes\footnote{%
        ``Theoreticians,'' including the authors,
        tend to take false-positive errors for granted
        (since we cannot achieve sound and complete verification
        and thereby overapproximations of some kind are necessarily introduced).
        However, for an affinity with development workflows,
        it would even be worth considering to strike a balance
        between soundness properties and the frequency of false positives.
      }.
    \item[D.] \textbf{Lack of concise support for flexible tensor-handling operations}:
      There are some implicit conversions of tensors
      frequently utilized in DNN-related programs,
      such as \dfn{broadcasting}~\cite{NumPyBroadcastingSemantics,PyTorchBroadcastingSemantics}.
      These conversions are flexible enough to make reasoning tensor shapes non-trivial,
      at least beyond decidable theories.
      For example,
      two tensors of shapes~\texttt{[5, 3, 1, 10]} and \texttt{[3, 4, 10]} are
      addable by broadcasting
      (specifically, by duplicating the former by \(4\) along with the third dimension
      and the latter by \(5\), respectively).
      Typical type systems have difficulty in supporting such conversions
      in a concise manner; they will require proofs for
      the feasibility of the conversion, and whether the proving is
      manual or automated, that will also lead to some of the issues~A--C above.
  \end{enumerate}
  In essence, for the adoption to continuous development,
  it would also be crucial to wipe off the concern about
  the burden arising from the mismatch between verification methods and development workflows,
  such as the one due to too many false positives or unpredictable time consumption,
  not only to establish a method to verify the correctness of programs.
\par
\subsection{Basics of Our Language Design}\label{subsec:our-work}
\indent
  To mitigate the issues~A--D above for tensor-manipulating programs,
  this work provides an alternative formalization of tensor shape checking
  by using \dfn{staging} (also called \dfn{staged computation} or \dfn{multi-stage programming}~\cite{%
    DaviesLICS1996,%
    DaviesJACM2017,%
    TahaSheardPEPM1997,%
    TahaSheard2000}).
  Our key idea is to split tensor computation into two stages: the stage~\(0\),
  which can be regarded as compile-time,
  is to verify
  the shape consistency by \emph{assertion checking} and generate a specialized program which is
  proven not to cause any run-time shape mismatch; and the stage~\(1\) is to do actual tensor computation by
  the specialized program.
  Our method is based on the observation\footnote{%
    Though our formalization may be general enough to be applied to other topics,
    we have not found other usages that fit this kind of phase separation
    and thereby focus on tensor shape checking for now.
  }
  that, in many tensor-manipulating programs, especially
  the ones that use sophisticated tensor libraries,
  the computation to check shape consistency can be independent of the actual tensor computations
  and thus is expected to be lightweight.
  Let us discuss the core of our idea using a concrete example.
  The previous example can be expressed as the following staged program\footnote{%
    Notes to those who are not very familiar to multi-stage programs:
    We use the notations from \dfn{MetaML}~\cite{TahaSheardPEPM1997,TahaSheard2000}.
    Here, expressions of the
    forms~\(\openO{\langle}\ordI{M}\closeO{\rangle}\) (called \dfn{bracket})
    and \(\ordI{\sim}\ordO{M}\) (called \dfn{escape})
    correspond to (hygienic) \dfn{quasiquotation} and \dfn{splicing} in Lisp, respectively.
    Intuitively, a bracket~\(\openO{\langle}\ordI{M}\closeO{\rangle}\)
    evaluates to a code value: for example,
    \(\openO{\langle}\ordI{1} \binI{+} \ordI{4}\closeO{\rangle}\) evaluates to itself,
    which stands for a piece of code that performs addition of 1 and 4.
    An escape~\(\ordI{\sim}\ordO{M}\) is supposed to appear inside a bracket;
    when \(\ordO{M}\) is evaluated to a code value~\(\openO{\langle}\ordI{M'}\closeO{\rangle}\),
    the code~\(\ordI{M'}\) is spliced into the surrounding code:
    For example, \(  \tokenO{let}\   \ordO{  \makeIdentOrConst{}{ Va }  }   \empty   \relO{=}   \openO{\langle}    \ordI{  \makeIdentOrConst{}{ C1 }  }   \binI{  +  }   \ordI{  \makeIdentOrConst{}{ C4 }  }    \closeO{\rangle}   \ \tokenO{in}\   \openO{\langle}    \ordI{  \makeIdentOrConst{}{ C2 }  }   \binI{  \ast  }  \ordI{\sim}   \ordO{  \makeIdentOrConst{}{ Va }  }     \closeO{\rangle}  \) evaluates to \( \openO{\langle}    \ordI{  \makeIdentOrConst{}{ C2 }  }   \binI{  \ast  }  \openI{(}    \ordI{  \makeIdentOrConst{}{ C1 }  }   \binI{  +  }   \ordI{  \makeIdentOrConst{}{ C4 }  }    \closeI{)}   \closeO{\rangle} \).
    The symbol~\(\ordI{\%}\) (called \dfn{cross-stage persistence}~\cite{%
      HanadaIgarashi2014,%
      KawataIgarashiAPLAS2019,%
      TahaSheardPEPM1997,%
      YuseIgarashiPPDP2006}
    in the literature) signifies that
    the argument comes from a lower stage.
    Unlike escapes, the argument can be of any type.
  }:
  \begin{align*}
     \progindent{  \tokenO{let}\   \ordO{  \makeIdentOrConst{}{ Vf' }  }   \empty   \relO{=}   \ordO{\lambda}  \ordO{  \makeIdentOrConst{}{ Vj }  }   \relO{:}    \ttO{Int}   \punctO{.}\   \ordO{\lambda}  \ordO{  \makeIdentOrConst{}{ Vk }  }   \relO{:}    \ttO{Int}   \punctO{.}\   \ordO{\lambda}  \ordO{  \makeIdentOrConst{}{ Vm }  }   \relO{:}    \ttO{Int}   \punctO{.}\   \ordO{\lambda}  \ordO{  \makeIdentOrConst{}{ Vn }  }   \relO{:}    \ttO{Int}   \punctO{.}\control\deepen{\control\br{}  \openO{\langle}  \ordI{\lambda}  \ordI{  \makeIdentOrConst{}{ VA }  }   \relI{:}   \ttI{Mat}\ \ordI{\%}   \ordO{  \makeIdentOrConst{}{ Vj }  }   \ \ordI{\%}  \openO{(}     \ordO{  \makeIdentOrConst{}{ Vk }  }   \binO{  +  }   \ordO{  \makeIdentOrConst{}{ C2 }  }    \binO{  \ast  }   \ordO{  \makeIdentOrConst{}{ Vm }  }    \closeO{)}   \punctI{.}\   \ordI{\lambda}  \ordI{  \makeIdentOrConst{}{ VB }  }   \relI{:}   \ttI{Mat}\ \ordI{\%}   \ordO{  \makeIdentOrConst{}{ Vm }  }   \ \ordI{\%}   \ordO{  \makeIdentOrConst{}{ Vn }  }    \punctI{.}\   \ordI{\lambda}  \ordI{  \makeIdentOrConst{}{ VC }  }   \relI{:}   \ttI{Mat}\ \ordI{\%}   \ordO{  \makeIdentOrConst{}{ Vk }  }   \ \ordI{\%}   \ordO{  \makeIdentOrConst{}{ Vn }  }    \punctI{.}\control\deepen{\control\br{}     \tokenI{let}\   \ordI{  \makeIdentOrConst{}{ VD }  }   \relI{=} \control\deepen{   \ordI{\sim}  \openO{(}      \ordO{  \makeIdentOrConst{}{ VgenVertCat }  }   \   \openO{(}    \ordO{  \makeIdentOrConst{}{ Vm }  }   \binO{  +  }   \ordO{  \makeIdentOrConst{}{ Vk }  }    \closeO{)}   \    \ordO{  \makeIdentOrConst{}{ Vm }  }    \    \ordO{  \makeIdentOrConst{}{ Vn }  }    \closeO{)}   \     \openI{(}    \ordI{\sim}  \openO{(}      \ordO{  \makeIdentOrConst{}{ VgenVertCat }  }   \    \ordO{  \makeIdentOrConst{}{ Vm }  }    \    \ordO{  \makeIdentOrConst{}{ Vk }  }    \    \ordO{  \makeIdentOrConst{}{ Vn }  }    \closeO{)}   \    \ordI{  \makeIdentOrConst{}{ VB }  }    \    \ordI{  \makeIdentOrConst{}{ VC }  }    \closeI{)}  \    \ordI{  \makeIdentOrConst{}{ VB }  }      }\ \tokenI{in}\control\br{}   \ordI{\sim}  \openO{(}      \ordO{  \makeIdentOrConst{}{ VgenMatMult }  }   \    \ordO{  \makeIdentOrConst{}{ Vj }  }    \   \openO{(}     \ordO{  \makeIdentOrConst{}{ Vk }  }   \binO{  +  }   \ordO{  \makeIdentOrConst{}{ C2 }  }    \binO{  \ast  }   \ordO{  \makeIdentOrConst{}{ Vm }  }    \closeO{)}   \    \ordO{  \makeIdentOrConst{}{ Vn }  }    \closeO{)}    \    \ordI{  \makeIdentOrConst{}{ VA }  }    \    \ordI{  \makeIdentOrConst{}{ VD }  }     }    \closeO{\rangle}  }      } 
  \end{align*}
  We use blue and orange to render stage-0 and stage-1 entities, respectively, throughout the paper\footnote{%
    Nonetheless, for accessibility reasons, we do not disambiguate stages just by colors.
  }.
  Numerous shape-related arguments are used in the program for now,
  but one can see later that many of them can actually be implicit.
\par
\indent
  Basically, \(  \ordO{  \makeIdentOrConst{}{ Vf' }  }  \) takes
  \(  \ordO{  \makeIdentOrConst{}{ Vj }  }  \), \(  \ordO{  \makeIdentOrConst{}{ Vk }  }  \), \(  \ordO{  \makeIdentOrConst{}{ Vm }  }  \), and \(  \ordO{  \makeIdentOrConst{}{ Vn }  }  \)
  as stage-\(0\) parameters,
  and produces code for \((A, B, C) \mapsto A (B \opVertConcat C \opVertConcat B)\)
  specialized for those parameters.
  Here, \( \ordO{  \makeIdentOrConst{}{ VgenVertCat }  } \) takes
  three parameters~\( \makeIdentOrConst{}{ Vp } \), \( \makeIdentOrConst{}{ Vq } \), and \( \makeIdentOrConst{}{ Vr } \)
  and returns \( \openO{\langle}   \ordI{  \makeIdentOrConst{}{ VvertCat }  _{   \makeIdentOrConst{}{ Vp }  ,    \makeIdentOrConst{}{ Vq }  ,    \makeIdentOrConst{}{ Vr }     } }   \closeO{\rangle} \), where
  \(  \ordI{  \makeIdentOrConst{}{ VvertCat }  _{   \makeIdentOrConst{}{ Vp }  ,    \makeIdentOrConst{}{ Vq }  ,    \makeIdentOrConst{}{ Vr }     } }  \) is
  the specialized operation that vertically concatenates
  \( \makeIdentOrConst{}{ Vp }  \times  \makeIdentOrConst{}{ Vr } \) and \( \makeIdentOrConst{}{ Vq }  \times  \makeIdentOrConst{}{ Vr } \)~matrices.
  The stage-\(0\) built-in function~\( \ordO{  \makeIdentOrConst{}{ VgenMatMult }  } \) also takes
  three parameters~\( \makeIdentOrConst{}{ Vp } \), \( \makeIdentOrConst{}{ Vq } \), and \( \makeIdentOrConst{}{ Vr } \)
  and returns \( \openO{\langle}   \ordI{  \makeIdentOrConst{}{ VmatMult }  _{   \makeIdentOrConst{}{ Vp }  ,    \makeIdentOrConst{}{ Vq }  ,    \makeIdentOrConst{}{ Vr }     } }   \closeO{\rangle} \), where
  \(  \ordI{  \makeIdentOrConst{}{ VmatMult }  _{   \makeIdentOrConst{}{ Vp }  ,    \makeIdentOrConst{}{ Vq }  ,    \makeIdentOrConst{}{ Vr }     } }  \) is the matrix multiplication operation
  specialized for \( \makeIdentOrConst{}{ Vp }  \times  \makeIdentOrConst{}{ Vq } \) and \( \makeIdentOrConst{}{ Vq }  \times  \makeIdentOrConst{}{ Vr } \)
  (if \( \makeIdentOrConst{}{ Vp } \), \( \makeIdentOrConst{}{ Vq } \), and \( \makeIdentOrConst{}{ Vr } \) are all non-negative;
  an assertion failure will be raised otherwise).
\par
\indent
  To reflect the operational behavior,
  code-generating functions are assigned types peculiar to staged computation.
  For example, \( \ordO{  \makeIdentOrConst{}{ VgenVertCat }  } \) is assigned
  type~\( \openO{(}  \ordO{  \makeIdentOrConst{}{ Vp }  }   \relO{:}    \ttO{Int}   \closeO{)} \relO{\to}   \openO{(}  \ordO{  \makeIdentOrConst{}{ Vq }  }   \relO{:}    \ttO{Int}   \closeO{)} \relO{\to}   \openO{(}  \ordO{  \makeIdentOrConst{}{ Vr }  }   \relO{:}    \ttO{Int}   \closeO{)} \relO{\to}   \openO{\langle}    \ttI{Mat}\ \ordI{\%}   \ordO{  \makeIdentOrConst{}{ Vp }  }   \ \ordI{\%}   \ordO{  \makeIdentOrConst{}{ Vr }  }     \relI{\to}   \ttI{Mat}\ \ordI{\%}   \ordO{  \makeIdentOrConst{}{ Vq }  }   \ \ordI{\%}   \ordO{  \makeIdentOrConst{}{ Vr }  }      \relI{\to}   \ttI{Mat}\ \ordI{\%}  \openO{(}    \ordO{  \makeIdentOrConst{}{ Vp }  }   \binO{  +  }   \ordO{  \makeIdentOrConst{}{ Vq }  }    \closeO{)}  \ \ordI{\%}   \ordO{  \makeIdentOrConst{}{ Vr }  }     \closeO{\rangle}    \)
  by combining dependent function types~\( \openO{(} \possiblyWithSub\stageOmetaColor{x}  \relO{:}  \possiblyWithSub\stageOmetaColor{T^{\superscriptO} }_{{\mathrm{1}}} \closeO{)} \relO{\to}  \possiblyWithSub\stageOmetaColor{T^{\superscriptO} }_{{\mathrm{2}}} \)
  and \dfn{code types}~\( \openO{\langle} \possiblyWithSub\stageImetaColor{T^{\superscriptI} } \closeO{\rangle} \),
  meaning that \(p\), \(q\), and \(r\) are available at stage~\(0\) and
  the resulting code for matrix computation is at stage~\(1\).
  By the same token, \( \ordO{  \makeIdentOrConst{}{ VgenMatMult }  } \) has
  type~\( \openO{(}  \ordO{  \makeIdentOrConst{}{ Vp }  }   \relO{:}    \ttO{Int}   \closeO{)} \relO{\to}   \openO{(}  \ordO{  \makeIdentOrConst{}{ Vq }  }   \relO{:}    \ttO{Int}   \closeO{)} \relO{\to}   \openO{(}  \ordO{  \makeIdentOrConst{}{ Vr }  }   \relO{:}    \ttO{Int}   \closeO{)} \relO{\to}   \openO{\langle}    \ttI{Mat}\ \ordI{\%}   \ordO{  \makeIdentOrConst{}{ Vp }  }   \ \ordI{\%}   \ordO{  \makeIdentOrConst{}{ Vq }  }     \relI{\to}   \ttI{Mat}\ \ordI{\%}   \ordO{  \makeIdentOrConst{}{ Vq }  }   \ \ordI{\%}   \ordO{  \makeIdentOrConst{}{ Vr }  }      \relI{\to}   \ttI{Mat}\ \ordI{\%}   \ordO{  \makeIdentOrConst{}{ Vp }  }   \ \ordI{\%}   \ordO{  \makeIdentOrConst{}{ Vr }  }     \closeO{\rangle}    \).
\par
\indent
  Although the type of matrices is indexed by size information as before,
  the type system is not equipped with non-trivial type equality to identify.
  Instead, the language offers stage-\(0\) \emph{casts} of the
  form~\(\LeftAssertParen \ordO{T_1} \relO{\triangleleft} \ordO{T_2} \RightAssertParen\)
  and inserts them through type-checking
  in order for the casts to be evaluated at compile time to assert the equality of two types.
  Basically, the above program will be elaborated to the following in a type-guided manner,
  for example:
  \begin{align*}
     \progindent{  \tokenO{let}\   \ordO{  \makeIdentOrConst{}{ Vf' }  }   \relO{=}   \ordO{\lambda}  \ordO{  \makeIdentOrConst{}{ Vj }  }   \relO{:}    \ttO{Int}   \punctO{.}\   \ordO{\lambda}  \ordO{  \makeIdentOrConst{}{ Vk }  }   \relO{:}    \ttO{Int}   \punctO{.}\   \ordO{\lambda}  \ordO{  \makeIdentOrConst{}{ Vm }  }   \relO{:}    \ttO{Int}   \punctO{.}\   \ordO{\lambda}  \ordO{  \makeIdentOrConst{}{ Vn }  }   \relO{:}    \ttO{Int}   \punctO{.}\control\deepen{\control\br{}  \openO{\langle}  \ordI{\lambda}  \ordI{  \makeIdentOrConst{}{ VA }  }   \relI{:}   \ttI{Mat}\ \ordI{\%}   \ordO{  \makeIdentOrConst{}{ Vj }  }   \ \ordI{\%}  \openO{(}     \ordO{  \makeIdentOrConst{}{ Vk }  }   \binO{  +  }   \ordO{  \makeIdentOrConst{}{ C2 }  }    \binO{  \ast  }   \ordO{  \makeIdentOrConst{}{ Vm }  }    \closeO{)}   \punctI{.}\   \ordI{\lambda}  \ordI{  \makeIdentOrConst{}{ VB }  }   \relI{:}   \ttI{Mat}\ \ordI{\%}   \ordO{  \makeIdentOrConst{}{ Vm }  }   \ \ordI{\%}   \ordO{  \makeIdentOrConst{}{ Vn }  }    \punctI{.}\   \ordI{\lambda}  \ordI{  \makeIdentOrConst{}{ VC }  }   \relI{:}   \ttI{Mat}\ \ordI{\%}   \ordO{  \makeIdentOrConst{}{ Vk }  }   \ \ordI{\%}   \ordO{  \makeIdentOrConst{}{ Vn }  }    \punctI{.}\ \control\deepen{\control\br{}     \tokenI{let}\   \ordI{  \makeIdentOrConst{}{ VD }  }   \relI{=} \control\deepen{   \ordI{\sim}  \openO{(}    \ordO{  \makeIdentOrConst{}{ VgenVertCat }  }   \     \openO{(}    \ordO{  \makeIdentOrConst{}{ Vm }  }   \binO{  +  }   \ordO{  \makeIdentOrConst{}{ Vk }  }    \closeO{)}  \    \ordO{  \makeIdentOrConst{}{ Vm }  }    \    \ordO{  \makeIdentOrConst{}{ Vn }  }     \closeO{)}   \     \openI{(}    \ordI{\sim}  \openO{(}    \ordO{  \makeIdentOrConst{}{ VgenVertCat }  }   \      \ordO{  \makeIdentOrConst{}{ Vm }  }   \    \ordO{  \makeIdentOrConst{}{ Vk }  }    \    \ordO{  \makeIdentOrConst{}{ Vn }  }     \closeO{)}   \    \ordI{  \makeIdentOrConst{}{ VB }  }    \    \ordI{  \makeIdentOrConst{}{ VC }  }    \closeI{)}  \    \ordI{  \makeIdentOrConst{}{ VB }  }      }\ \tokenI{in}\control\br{}   \ordI{\sim}  \openO{(}    \ordO{  \makeIdentOrConst{}{ VgenMatMult }  }   \      \ordO{  \makeIdentOrConst{}{ Vj }  }   \   \openO{(}     \ordO{  \makeIdentOrConst{}{ Vk }  }   \binO{  +  }   \ordO{  \makeIdentOrConst{}{ C2 }  }    \binO{  \ast  }   \ordO{  \makeIdentOrConst{}{ Vm }  }    \closeO{)}   \    \ordO{  \makeIdentOrConst{}{ Vn }  }     \closeO{)}    \    \ordI{  \makeIdentOrConst{}{ VA }  }    \control\deepen{\control\br{}  \ordI{\sim}  \openO{(}   \LeftAssertParen\openO{\langle}  \ttI{Mat}\ \ordI{\%}  \openO{(}   \openO{(}    \ordO{  \makeIdentOrConst{}{ Vm }  }   \binO{  +  }   \ordO{  \makeIdentOrConst{}{ Vk }  }    \closeO{)}  \binO{  +  }   \ordO{  \makeIdentOrConst{}{ Vm }  }    \closeO{)}  \ \ordI{\%}   \ordO{  \makeIdentOrConst{}{ Vn }  }    \closeO{\rangle} \relO{\CastArrow} \openO{\langle}  \ttI{Mat}\ \ordI{\%}  \openO{(}     \ordO{  \makeIdentOrConst{}{ Vk }  }   \binO{  +  }   \ordO{  \makeIdentOrConst{}{ C2 }  }    \binO{  \ast  }   \ordO{  \makeIdentOrConst{}{ Vm }  }    \closeO{)}  \ \ordI{\%}   \ordO{  \makeIdentOrConst{}{ Vn }  }    \closeO{\rangle}\RightAssertParen^{  \ell  }  \   \openO{\langle}   \ordI{  \makeIdentOrConst{}{ VD }  }   \closeO{\rangle}   \closeO{)}   }   }    \closeO{\rangle}  }      } 
  \end{align*}
  In this case, to identify \( \ttI{Mat}\ \ordI{\%}  \openO{(}   \openO{(}    \ordO{  \makeIdentOrConst{}{ Vm }  }   \binO{  +  }   \ordO{  \makeIdentOrConst{}{ Vk }  }    \closeO{)}  \binO{  +  }   \ordO{  \makeIdentOrConst{}{ Vm }  }    \closeO{)}  \ \ordI{\%}   \ordO{  \makeIdentOrConst{}{ Vn }  }   \) and \( \ttI{Mat}\ \ordI{\%}  \openO{(}     \ordO{  \makeIdentOrConst{}{ Vk }  }   \binO{  +  }   \ordO{  \makeIdentOrConst{}{ C2 }  }    \binO{  \ast  }   \ordO{  \makeIdentOrConst{}{ Vm }  }    \closeO{)}  \ \ordI{\%}   \ordO{  \makeIdentOrConst{}{ Vn }  }   \)
  to ensure the validity of the use of \( \ordO{  \makeIdentOrConst{}{ VgenMatMult }  } \),
  the cast~\( \LeftAssertParen\openO{\langle}  \ttI{Mat}\ \ordI{\%}  \openO{(}   \openO{(}    \ordO{  \makeIdentOrConst{}{ Vm }  }   \binO{  +  }   \ordO{  \makeIdentOrConst{}{ Vk }  }    \closeO{)}  \binO{  +  }   \ordO{  \makeIdentOrConst{}{ Vm }  }    \closeO{)}  \ \ordI{\%}   \ordO{  \makeIdentOrConst{}{ Vn }  }    \closeO{\rangle} \relO{\CastArrow} \openO{\langle}  \ttI{Mat}\ \ordI{\%}  \openO{(}     \ordO{  \makeIdentOrConst{}{ Vk }  }   \binO{  +  }   \ordO{  \makeIdentOrConst{}{ C2 }  }    \binO{  \ast  }   \ordO{  \makeIdentOrConst{}{ Vm }  }    \closeO{)}  \ \ordI{\%}   \ordO{  \makeIdentOrConst{}{ Vn }  }    \closeO{\rangle}\RightAssertParen^{  \ell  } \)
  was inserted by the type-checking procedure,
  where \(\ell\) is an attached label that points to the original application
  as the source of the failure if the assertion fails.
  This will check the corresponding arguments in the two types---specifically,
  \( \openO{(}   \openO{(}     \ordO{  \makeIdentOrConst{}{ Vm }  }    \binO{  +  }   \ordO{  \makeIdentOrConst{}{ Vk }  }    \closeO{)}  \binO{  +  }   \ordO{  \makeIdentOrConst{}{ Vm }  }    \closeO{)} \) and \(     \ordO{  \makeIdentOrConst{}{ Vk }  }    \binO{  +  }   \ordO{  \makeIdentOrConst{}{ C2 }  }    \binO{  \ast  }   \ordO{  \makeIdentOrConst{}{ Vm }  }   \)---are equal
  \emph{using the concrete values of \(  \ordO{  \makeIdentOrConst{}{ Vk }  }  \) and \(  \ordO{  \makeIdentOrConst{}{ Vm }  }  \)}.
  Thus, if \(  \ordO{  \makeIdentOrConst{}{ Vf' }  }  \) is applied to concrete integers, say,
  \( \ordO{  \makeIdentOrConst{}{ C4 }  } \), \( \ordO{  \makeIdentOrConst{}{ C1 }  } \), \( \ordO{  \makeIdentOrConst{}{ C2 }  } \), and \( \ordO{  \makeIdentOrConst{}{ C3 }  } \),
  the expression~\(       \ordO{  \makeIdentOrConst{}{ Vf' }  }    \    \ordO{  \makeIdentOrConst{}{ C4 }  }    \    \ordO{  \makeIdentOrConst{}{ C1 }  }    \    \ordO{  \makeIdentOrConst{}{ C2 }  }    \    \ordO{  \makeIdentOrConst{}{ C3 }  }   \)
  will evaluate to the following code without failure:
  \begin{align*}
     \progindent{  \openO{\langle}  \ordI{\lambda}  \ordI{  \makeIdentOrConst{}{ VA }  }   \relI{:}   \ttI{Mat}\ \ordI{\%}   \ordO{  \makeIdentOrConst{}{ C4 }  }   \ \ordI{\%}   \ordO{  \makeIdentOrConst{}{ C5 }  }    \punctI{.}\   \ordI{\lambda}  \ordI{  \makeIdentOrConst{}{ VB }  }   \relI{:}   \ttI{Mat}\ \ordI{\%}   \ordO{  \makeIdentOrConst{}{ C2 }  }   \ \ordI{\%}   \ordO{  \makeIdentOrConst{}{ C3 }  }    \punctI{.}\   \ordI{\lambda}  \ordI{  \makeIdentOrConst{}{ VC }  }   \relI{:}   \ttI{Mat}\ \ordI{\%}   \ordO{  \makeIdentOrConst{}{ C1 }  }   \ \ordI{\%}   \ordO{  \makeIdentOrConst{}{ C3 }  }    \punctI{.}\ \control\deepen{\control\br{}     \tokenI{let}\   \ordI{  \makeIdentOrConst{}{ VD }  }   \relI{=} \control\deepen{     \ordI{  \makeIdentOrConst{}{ VvertCat }  _{   \makeIdentOrConst{}{ C3 }  ,    \makeIdentOrConst{}{ C2 }  ,    \makeIdentOrConst{}{ C3 }     } }   \   \openI{(}     \ordI{  \makeIdentOrConst{}{ VvertCat }  _{   \makeIdentOrConst{}{ C2 }  ,    \makeIdentOrConst{}{ C1 }  ,    \makeIdentOrConst{}{ C3 }     } }   \    \ordI{  \makeIdentOrConst{}{ VB }  }    \    \ordI{  \makeIdentOrConst{}{ VC }  }    \closeI{)}   \    \ordI{  \makeIdentOrConst{}{ VB }  }    }\ \tokenI{in}\    \ordI{  \makeIdentOrConst{}{ VmatMult }  _{   \makeIdentOrConst{}{ C4 }  ,    \makeIdentOrConst{}{ C5 }  ,    \makeIdentOrConst{}{ C3 }     } }    \    \ordI{  \makeIdentOrConst{}{ VA }  }    \    \ordI{  \makeIdentOrConst{}{ VD }  }     }    \closeO{\rangle}  } 
  \end{align*}
  The body of the bracket is basically given type
  \(\TypeMat{4}{5} \to \TypeMat{2}{3} \to \TypeMat{1}{3} \to \TypeMat{4}{3}\)
  and it is guaranteed not to cause shape mismatch (as far as it is
  applied to matrices of the designated sizes).  Note that no
  sophisticated machinery is required if one wants to do some type-guided traversal on generated code,
  possibly for further optimization; a simple type system (with infinitely many
  base types~\(\TypeMat{1}{1}, \TypeMat{1}{2}, \TypeMat{2}{1}, \ldots\)) suffices\footnote{%
    The types of primitives are also simple.
    For example, \( \ordI{  \makeIdentOrConst{}{ VvertCat }  _{   \makeIdentOrConst{}{ C2 }  ,    \makeIdentOrConst{}{ C1 }  ,    \makeIdentOrConst{}{ C3 }     } } \) is given
    \(\TypeMat{2}{3} \to \TypeMat{1}{3} \to \TypeMat{3}{3}\).
  }.
\par
\indent
  In this way, we can verify the consistency
  during code-generating compile-time computation.
  We expect that, in many cases, compile-time computation does not take much time
  and that inconsistencies will be immediately detected as assertion failures.
  If the stage-\(0\) evaluation succeeds, i.e., produces a code fragment without causing any failure,
  the resulting code fragment is guaranteed to be consistent with respect to specific tensor shapes.
  In this sense, we can do tensor shape checking
  without relying on either manual or automated proving,
  and thus mitigate the aforementioned issues~A, B, and C
  (We defer to Section~\ref{subsec:implicit-conversion}
  how the remaining issue~D can be resolved in our method).
  One may see this approach as a compile-time version of
  \dfn{manifest contracts}~\cite{%
    FindlerFelleisenICFP2002,%
    GreenbergPierceWeirichPOPL2010,%
    Greenberg2013,%
    SekiyamaIgarashiGreenbergTOPLAS2017},
  \dfn{hybrid type checking}~\cite{FlanaganPOPL2006,KnowlesFlanaganTOPLAS2010},
  or some similar methods like Lemay~et~al.~\cite{LemayFuBlairZhangXiTYDE2023}.
  Our approach might also be able to be seen as
  some kind of staging-based foundation of
  the \dfn{template metaprogramming} in C++~\cite{Cpp}.
\par
\indent
  Nonetheless, now we have to write many ``annotation-like'' arguments
  in exchange for the absence of proofs
  (like the three arguments in \(      \ordO{  \makeIdentOrConst{}{ VgenVertCat }  }    \   \openO{(}    \ordO{  \makeIdentOrConst{}{ Vm }  }   \binO{  +  }   \ordO{  \makeIdentOrConst{}{ Vk }  }    \closeO{)}   \    \ordO{  \makeIdentOrConst{}{ Vm }  }    \    \ordO{  \makeIdentOrConst{}{ Vn }  }   \)).
  Manually adding staging constructs may also be cumbersome.
  In Section~\ref{sec:overview},
  we illustrate that most of these descriptions can actually be cleared away
  by introducing surface languages.
\par
\subsection{Our Contributions}
\indent
  Our contributions can be summarized as follows:
  \begin{enumerate}
    \item \textbf{Staged core language for compile-time tensor shape checking
    and the mathematical guarantee of its safety}:
      Based on the formalization of staged computation,
      we define \LambdaBracketCast, a two-stage language that enables us to ensure
      the consistency of tensor shapes at stage-\(0\) (i.e.,~compile-time) computation.
      This formalization can be considered handy for real-world use
      in continuous software development
      in that it requires neither manual nor automated proofs
      for type-checking programs.
      At the same time, our method achieves runtime safety in the sense that,
      \emph{%
        once a code fragment specialized for specific tensor shapes is generated after compile-time computation,
        it is guaranteed to contain no shape mismatches and thereby can be run safely}.
      Our method also accommodates
      complex tensor manipulations that are frequently used in DNN-related programs,
      such as \dfn{broadcasting}~\cite{NumPyBroadcastingSemantics,PyTorchBroadcastingSemantics}
      or \dfn{reshaping}.
      Furthermore, our formalization incorporates \dfn{refinement types}~\cite{%
        FlanaganPOPL2006,%
        RondonKawaguchiRanjitPLDI2008,%
        KnowlesFlanaganTOPLAS2010,%
        HattoriKobayashiSatoESOP2023}
      so that detailed preconditions expected of stage-\(0\) function parameters can be described by types
      for the sake of error localization.
    \item \textbf{Extension with implicit arguments and their reconstruction rules}:
      To alleviate the burden of specifying shape-related stage-\(0\) arguments,
      we define \LambdaBracketCastImplicit,
      an extended version of \LambdaBracketCast\ with implicit parameters/arguments,
      and provide algorithmic rules for the reconstruction of omitted arguments.
      Although this inference is not complete, it can reconstruct omitted arguments in many typical cases.
    \item \textbf{Horsea, an example non-staged surface language}:
      As an exemplification of adding a non-staged surface language on top of the staged core language,
      we design \dfn{Horsea}, which frees us from
      manually adding staging constructs and many shape-related arguments.
      By using \dfn{binding-time analysis} (\dfn{BTA})~\cite{JonesGomardSestoft1993,DaviesLICS1996,DaviesJACM2017},
      programs in this surface language are translated to
      \LambdaBracketCastImplicit\ (and finally to \LambdaBracketCast\ by reconstructing implicit arguments).
    \item \textbf{Prototype implementation}:
      We implemented a prototype type-checker of Horsea in Haskell based on our method
      and made it publicly available~\cite{HorseaRepository,HorseaArchive}.
      This type-checker exemplifies that our method is expressive enough to
      verify the shape consistency of
      10 example programs\footnote{%
        Although our method seems able to cover the other remaining examples as well,
        we could not port them simply due to the lack of time;
        assigning appropriate types to \texttt{ocaml-torch}'s API
        and understanding original programs as to unreconstructible implicit arguments
        requires a bit of time and effort.
      } offered by \texttt{ocaml-torch}~\cite{OCamlTorch},
      an OCaml binding of PyTorch~\cite{PaszkeGrossChintalaChananYangDevitoLinDesmaisonAntigaLerer2017},
      by porting them to Horsea manually and feed them to our type-checker.
  \end{enumerate}
\par
\indent
  We suppose that our method alleviates the issues~A--D from the following perspectives:
  \begin{enumerate}
    \item[A.]
      Since it only requires type annotations for binders and
      some implicit arguments that cannot be inferred,
      it would not hamper future changes to a large extent.
    \item[B.]
      Since compile-time computations can be
      described in a usual functional language,
      one can expect that users can easily estimate the elapsed time for shape checking.
      Furthermore, for typical cases, the elapsed time will be quite instant (say, less than 0.1~second).
    \item[C.]
      Our method does not cause false positive errors about tensor shapes
      in the sense that equations of tensor shapes are tested by concrete values after specializing them.
    \item[D.]
      As explained in Section~\ref{subsec:implicit-conversion},
      our method accommodates operations that perform implicit shape conversions such as broadcasting.
  \end{enumerate}
\par
\indent
  We here note that
  our formalization is not something that completely replaces theorem-proving approaches;
  our language ensures the shape consistency
  \emph{only after} concrete tensor shapes are given
  for compile-time code generation.
  By contrast, for verifying the consistency of a tensor-handling library,
  it is essentially necessary to prove that
  the library works for \emph{any possible combination} of tensor shapes,
  but our method cannot ensure properties that cannot be judged by evaluation,
  e.g., essentially universally quantified propositions.
  In this sense, the target population of our language design is
  end users who are implementing specific heavy tensor computations
  rather than authors of tensor-handling libraries.
  Nonetheless, it is possible to utilize our method for \emph{testing} libraries
  in a way like \dfn{property-based testing}~\cite{ClaessenHughesICFP2000},
  i.e., by running code generation with many randomly selected combinations of tensor shapes.
  Also, although in a somewhat awkward manner
  and with the aid of the so-called \(\mathbf{run}\)-primitive~\cite{%
    TahaSheardPEPM1997,%
    TsukadaIgarashi2009,%
    HanadaIgarashi2014,%
    KiselyovFLOPS2014},
  our method accommodates programs that handle
  tensors whose sizes are known only at runtime;
  Section~\ref{subsec:dynamic-shape} discusses how to achieve this.
\par
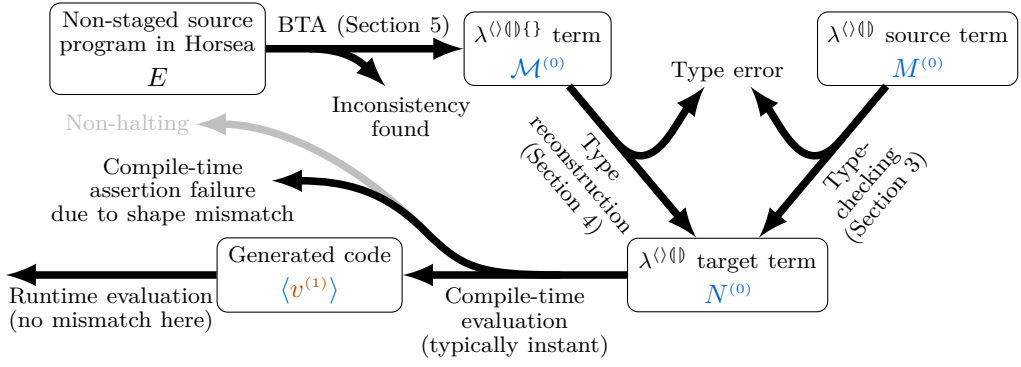
\begin{figure}[tb]
  \begin{tikzpicture}
    \node[draw,rounded corners,align=center,text width=2.6cm] (surface) at (-7cm, 3cm) {%
      {\footnotesize
        Non-staged source\\[-4pt]%
        program in Horsea}\\%
      \(E\)%
    };
    \node[draw,rounded corners,align=center,text width=1.7cm] (omissible) at (-2cm, 3cm) {%
      {\footnotesize \LambdaBracketCastImplicit\ term}\\%
      \(\possiblyWithSub\stageOmetaColor{\mathcal{M}^{\superscriptO} }\)%
    };
    \node[align=center,text width=3cm] at (-4.3cm, 3.3cm) {%
      {\footnotesize
        BTA (Section~\ref{sec:surface-language})}%
    };
    \draw[-latex,line width=1mm] (surface) -- (omissible);
    \draw[-latex,line width=1mm] (-5cm, 3cm) .. controls (-4.5cm, 3cm) and (-4.3cm, 2.8cm) .. (-4cm, 2.5cm);
    \node[align=center,text width=3cm] at (-3.8cm, 2.1cm) {%
      {\footnotesize
        Inconsistency\\[-4pt]%
        found}%
    };
    \node[draw,rounded corners,align=center,text width=2.4cm] (core) at (3cm, 3cm) {%
      {\footnotesize \LambdaBracketCast\ source term}\\%
      \(\possiblyWithSub\stageOmetaColor{M^{\scriptscriptstyle(0)} }\)%
    };
    \node[draw,rounded corners,align=center,text width=2.4cm] (target) at (0.5cm, 0cm) {%
      {\footnotesize \LambdaBracketCast\ target term}\\%
      \(\possiblyWithSub\stageOmetaColor{N^{\superscriptO} }\)%
    };
    \draw[-latex,line width=1mm] (core) -- node[sloped,below,align=center,text width=2cm] {%
      {\footnotesize
        \quad Type-\\%
        \quad checking\\[-4pt]%
        \quad (Section~\ref{sec:staged-language})%
      }} (target);
    \draw[-latex,line width=1mm] (omissible) -- node[sloped,below,align=center,text width=2.7cm] {%
      {\footnotesize
        Type\kern2em\relax\\%
        reconstruction\kern2em\relax\\[-4pt]%
        (Section~\ref{sec:implicit})\kern2em\relax
      }} (target);
    \draw[-latex,line width=1mm]
      ($(omissible)!0.4!(target)$) .. controls ($(omissible)!0.6!(target)$) and (-0.2cm, 2cm) .. (0.2cm, 2.5cm);
    \draw[-latex,line width=1mm]
      ($(core)!0.4!(target)$) .. controls ($(core)!0.6!(target)$) and (1.2cm, 2cm) .. (0.8cm, 2.5cm);
    \node[align=center,text width=3cm] at (0.5cm, 2.7cm) {%
      {\footnotesize
        Type error}%
    };
    \node[draw,rounded corners,align=center,text width=2.2cm] (code) at (-5cm, 0cm) {%
      {\footnotesize Generated code}\\%
      \( \openO{\langle} \possiblyWithSub\stageImetaColor{v^{\superscriptI} } \closeO{\rangle} \)%
    };
    \draw[-latex,line width=1mm] (target) -- node[below,align=center,text width=3cm] {%
        {\footnotesize
          Compile-time evaluation\\[-4pt]%
          (typically instant)}%
      } (code);
    \draw[-latex,line width=1mm,draw=lightgray]
       (-3.5cm, 0.6cm)
         .. controls (-4.5cm, 1.6cm) and (-5.5cm, 2cm) ..
         (-6.5cm, 2cm);
    \node[align=center,text width=2cm,text=lightgray] at (-7.4cm, 2cm) {%
      {\footnotesize
        Non-halting}%
    };
    \draw[-latex,line width=1mm]
      ($(target)!0.4!(code)$)
        .. controls ($(target)!0.6!(code)$) and (-3cm, 0.1cm) ..
        (-3.5cm, 0.6cm)
        .. controls (-4cm, 1.1cm) and (-4.5cm, 1.25cm) ..
        (-5.5cm, 1.25cm);
    \node[align=center,text width=3.2cm] at (-6.8cm, 1.1cm) {%
      {\footnotesize
        Compile-time\\[-1pt]%
        assertion failure\\[-4pt]%
        due to shape mismatch}%
    };
    \draw[-latex,line width=1mm] (code) -- node[below,align=center,text width=3cm] {%
        {\footnotesize
          Runtime evaluation\\[-4pt]%
          (no mismatch here)}%
      } (-9cm, 0cm);
  \end{tikzpicture}
  \caption{Overall picture of elaboration}
  \label{fig:overall-picture}
\end{figure}
\indent
  The rest of the paper is organized as follows:
  First, Section~\ref{sec:overview} gives an overview of our method by using running examples, and
  Section~\ref{sec:staged-language} describes
  the formalization of the staged core language~\LambdaBracketCast\ and
  proves its metatheoretic safety properties.
  Section~\ref{sec:implicit} extends \LambdaBracketCast\ with implicit parameters/arguments
  and gives how to infer omitted arguments.
  Section~\ref{sec:surface-language} provides \dfn{Horsea}
  as a proof-of-concept, non-staged surface language,
  and explains the basics of its translation to the staged language by using BTA.
  After that, Section~\ref{sec:further-discussions} discusses
  further extension of our language with some features necessary for real-world use.
  Finally, Section~\ref{sec:implementation} describes
  a prototype type-checker implemented based on our method
  and its example use cases involving \texttt{ocaml-torch},
  Section~\ref{sec:related-work} discusses the related work,
  and Section~\ref{sec:conclusion} concludes the paper.
  Figure~\ref{fig:overall-picture} depicts the overall procedure of
  our method explained in Sections~\ref{sec:staged-language}--\ref{sec:surface-language}.
\par

\section{Overview}\label{sec:overview}
\subsection{Implicit Arguments and Their Inference}\label{subsec:implicit-arguments}
\indent
  As we saw in Section~\ref{subsec:our-work}, in exchange for the absence of proofs,
  programs apparently require many ``annotation-like'' arguments for tensor shapes.
  Actually, we can infer many of such arguments even if they are omitted.
  With implicit arguments, we can write \( \makeIdentOrConst{}{ Vf' } \) as follows:
  \begin{align*}
     \progindent{  \tokenO{let}\   \ordO{  \makeIdentOrConst{}{ Vf' }  }   \empty   \relO{=}   \ordO{\lambda}\openO{\{}  \ordO{  \makeIdentOrConst{}{ Vj }  }   \relO{:}    \ttO{Int}   \closeO{\} }\punctO{.}\   \ordO{\lambda}\openO{\{}  \ordO{  \makeIdentOrConst{}{ Vk }  }   \relO{:}    \ttO{Int}   \closeO{\} }\punctO{.}\   \ordO{\lambda}\openO{\{}  \ordO{  \makeIdentOrConst{}{ Vm }  }   \relO{:}    \ttO{Int}   \closeO{\} }\punctO{.}\   \ordO{\lambda}\openO{\{}  \ordO{  \makeIdentOrConst{}{ Vn }  }   \relO{:}    \ttO{Int}   \closeO{\} }\punctO{.}\control\deepen{\control\br{}  \openO{\langle}  \ordI{\lambda}  \ordI{  \makeIdentOrConst{}{ VA }  }   \relI{:}   \ttI{Mat}\ \ordI{\%}   \ordO{  \makeIdentOrConst{}{ Vj }  }   \ \ordI{\%}  \openO{(}     \ordO{  \makeIdentOrConst{}{ Vk }  }   \binO{  +  }   \ordO{  \makeIdentOrConst{}{ C2 }  }    \binO{  \ast  }   \ordO{  \makeIdentOrConst{}{ Vm }  }    \closeO{)}   \punctI{.}\   \ordI{\lambda}  \ordI{  \makeIdentOrConst{}{ VB }  }   \relI{:}   \ttI{Mat}\ \ordI{\%}   \ordO{  \makeIdentOrConst{}{ Vm }  }   \ \ordI{\%}   \ordO{  \makeIdentOrConst{}{ Vn }  }    \punctI{.}\   \ordI{\lambda}  \ordI{  \makeIdentOrConst{}{ VC }  }   \relI{:}   \ttI{Mat}\ \ordI{\%}   \ordO{  \makeIdentOrConst{}{ Vk }  }   \ \ordI{\%}   \ordO{  \makeIdentOrConst{}{ Vn }  }    \punctI{.}\control\deepen{\control\br{}     \tokenI{let}\   \ordI{  \makeIdentOrConst{}{ VD }  }   \relI{=} \control\deepen{   \ordI{\sim}   \ordO{  \makeIdentOrConst{}{ VgenVertCat }  }    \     \openI{(}    \ordI{\sim}   \ordO{  \makeIdentOrConst{}{ VgenVertCat }  }    \    \ordI{  \makeIdentOrConst{}{ VB }  }    \    \ordI{  \makeIdentOrConst{}{ VC }  }    \closeI{)}  \    \ordI{  \makeIdentOrConst{}{ VB }  }      }\ \tokenI{in}\   \ordI{\sim}   \ordO{  \makeIdentOrConst{}{ VgenMatMult }  }     \    \ordI{  \makeIdentOrConst{}{ VA }  }    \    \ordI{  \makeIdentOrConst{}{ VD }  }     }    \closeO{\rangle}  }      } 
  \end{align*}
  All the shape-related arguments applied to \( \ordO{  \makeIdentOrConst{}{ VgenVertCat }  } \) and \( \ordO{  \makeIdentOrConst{}{ VgenMatMult }  } \)
  are now implicit; they will be inferred
  by using the type of already defined identifiers (such as built-in functions)
  and type annotations provided by the user.
\par
\indent
  For implicit parameters/arguments, we first introduce
  \(\openO{\{}\possiblyWithSub\stageOmetaColor{x} \relO{:} \possiblyWithSub\stageOmetaColor{T^{\superscriptO} }_{{\mathrm{1}}}\closeO{\}} \relO{\to} \possiblyWithSub\stageOmetaColor{T^{\superscriptO} }_{{\mathrm{2}}}\),
  a variant form of stage-\(0\) dependent function types.
  Functions of this type work exactly the same as
  those of type~\( \openO{(} \possiblyWithSub\stageOmetaColor{x}  \relO{:}  \possiblyWithSub\stageOmetaColor{T^{\superscriptO} }_{{\mathrm{1}}} \closeO{)} \relO{\to}  \possiblyWithSub\stageOmetaColor{T^{\superscriptO} }_{{\mathrm{2}}} \) from the operational perspective,
  but they allow users to omit arguments
  and require the type-checker to infer an expression that should substitute \(\possiblyWithSub\stageOmetaColor{x}\)
  for each context.
  For example,
  \( \ordO{  \makeIdentOrConst{}{ VgenVertCat }  } \) is now assigned
  type~\( \openO{\{}  \ordO{  \makeIdentOrConst{}{ Vp }  }   \relO{:}    \ttO{Nat}   \closeO{\} } \relO{\to}   \openO{\{}  \ordO{  \makeIdentOrConst{}{ Vq }  }   \relO{:}    \ttO{Nat}   \closeO{\} } \relO{\to}   \openO{\{}  \ordO{  \makeIdentOrConst{}{ Vr }  }   \relO{:}    \ttO{Nat}   \closeO{\} } \relO{\to}   \openO{\langle}    \ttI{Mat}\ \ordI{\%}   \ordO{  \makeIdentOrConst{}{ Vp }  }   \ \ordI{\%}   \ordO{  \makeIdentOrConst{}{ Vr }  }     \relI{\to}   \ttI{Mat}\ \ordI{\%}   \ordO{  \makeIdentOrConst{}{ Vq }  }   \ \ordI{\%}   \ordO{  \makeIdentOrConst{}{ Vr }  }      \relI{\to}   \ttI{Mat}\ \ordI{\%}  \openO{(}    \ordO{  \makeIdentOrConst{}{ Vp }  }   \binO{  +  }   \ordO{  \makeIdentOrConst{}{ Vq }  }    \closeO{)}  \ \ordI{\%}   \ordO{  \makeIdentOrConst{}{ Vr }  }     \closeO{\rangle}    \),
  meaning \(  \ordO{  \makeIdentOrConst{}{ Vp }  }  \), \(  \ordO{  \makeIdentOrConst{}{ Vq }  }  \), and \(  \ordO{  \makeIdentOrConst{}{ Vr }  }  \) can be implicit.
  Users can also define a function with implicit arguments by using
  variant \(\lambda\)-abstractions of the
  form~\(\openO{(}\ordO{\lambda}\openO{\{}\possiblyWithSub\stageOmetaColor{x} \relO{:} \possiblyWithSub\stageOmetaColor{T^{\superscriptO} }\closeO{\}}\punctO{.}\ \possiblyWithSub\stageOmetaColor{M^{\scriptscriptstyle(0)} }\closeO{)}\).
  In addition, when one wants to specify arguments explicitly for implicit parameters,
  applications of the form~\(\possiblyWithSub\stageOmetaColor{M^{\scriptscriptstyle(0)} }_{{\mathrm{1}}}\ \openO{\{}\possiblyWithSub\stageOmetaColor{M^{\scriptscriptstyle(0)} }_{{\mathrm{2}}}\closeO{\}}\) can be used.
\par
\indent
  The reconstruction of implicit arguments can be done in a type-guided manner.
  For example, consider the subexpression \(    \ordI{\sim}   \ordO{  \makeIdentOrConst{}{ VgenVertCat }  }     \    \ordI{  \makeIdentOrConst{}{ VB }  }    \    \ordI{  \makeIdentOrConst{}{ VC }  }   \),
  and let \(\possiblyWithSub\stageOmetaColor{N^{\superscriptO} }_{{\mathrm{1}}}\), \(\possiblyWithSub\stageOmetaColor{N^{\superscriptO} }_{{\mathrm{2}}}\), and \(\possiblyWithSub\stageOmetaColor{N^{\superscriptO} }_{{\mathrm{3}}}\) denote
  expressions that should be reconstructed for the three parameters.
  According to the type assigned to \( \ordO{  \makeIdentOrConst{}{ VgenVertCat }  } \),
  the reconstructed subexpression~\(\ordI{\sim}\openO{(} \ordO{  \makeIdentOrConst{}{ VgenVertCat }  } \ \openO{\{}\possiblyWithSub\stageOmetaColor{N^{\superscriptO} }_{{\mathrm{1}}}\closeO{\}}\ \openO{\{}\possiblyWithSub\stageOmetaColor{N^{\superscriptO} }_{{\mathrm{2}}}\closeO{\}}\ \openO{\{}\possiblyWithSub\stageOmetaColor{N^{\superscriptO} }_{{\mathrm{3}}}\closeO{\}}\closeO{)}\)
  will have type
  \(   \ttI{Mat}\ \ordI{\%} \possiblyWithSub\stageOmetaColor{N^{\superscriptO} }_{{\mathrm{1}}} \ \ordI{\%} \possiblyWithSub\stageOmetaColor{N^{\superscriptO} }_{{\mathrm{3}}}   \relI{\to}   \ttI{Mat}\ \ordI{\%} \possiblyWithSub\stageOmetaColor{N^{\superscriptO} }_{{\mathrm{2}}} \ \ordI{\%} \possiblyWithSub\stageOmetaColor{N^{\superscriptO} }_{{\mathrm{3}}}    \relI{\to}   \ttI{Mat}\ \ordI{\%}  \openO{(}  \possiblyWithSub\stageOmetaColor{N^{\superscriptO} }_{{\mathrm{1}}} \binO{  +  } \possiblyWithSub\stageOmetaColor{N^{\superscriptO} }_{{\mathrm{2}}}  \closeO{)}  \ \ordI{\%} \possiblyWithSub\stageOmetaColor{N^{\superscriptO} }_{{\mathrm{3}}}  \).
  Since this subexpression is applied to \(  \ordI{  \makeIdentOrConst{}{ VB }  }  \), which has type \( \ttI{Mat}\ \ordI{\%}   \ordO{  \makeIdentOrConst{}{ Vm }  }   \ \ordI{\%}   \ordO{  \makeIdentOrConst{}{ Vn }  }   \),
  the domain type \( \ttI{Mat}\ \ordI{\%} \possiblyWithSub\stageOmetaColor{N^{\superscriptO} }_{{\mathrm{1}}} \ \ordI{\%} \possiblyWithSub\stageOmetaColor{N^{\superscriptO} }_{{\mathrm{3}}} \) must be identical to \( \ttI{Mat}\ \ordI{\%}   \ordO{  \makeIdentOrConst{}{ Vm }  }   \ \ordI{\%}   \ordO{  \makeIdentOrConst{}{ Vn }  }   \) after evaluation.
  In this case, we can infer that
  it suffices to just substitute \(\possiblyWithSub\stageOmetaColor{N^{\superscriptO} }_{{\mathrm{1}}}\) and \(\possiblyWithSub\stageOmetaColor{N^{\superscriptO} }_{{\mathrm{3}}}\) with
  \(  \ordO{  \makeIdentOrConst{}{ Vm }  }  \) and \(  \ordO{  \makeIdentOrConst{}{ Vn }  }  \), respectively.
  Similarly, by comparing \( \ttI{Mat}\ \ordI{\%} \possiblyWithSub\stageOmetaColor{N^{\superscriptO} }_{{\mathrm{2}}} \ \ordI{\%} \possiblyWithSub\stageOmetaColor{N^{\superscriptO} }_{{\mathrm{3}}} \) with the type of \(  \ordI{  \makeIdentOrConst{}{ VC }  }  \) (i.e.,~\( \ttI{Mat}\ \ordI{\%}   \ordO{  \makeIdentOrConst{}{ Vk }  }   \ \ordI{\%}   \ordO{  \makeIdentOrConst{}{ Vn }  }   \)),
  we can judge that \(  \ordO{  \makeIdentOrConst{}{ Vk }  }  \) can be used as \(\possiblyWithSub\stageOmetaColor{N^{\superscriptO} }_{{\mathrm{2}}}\).
  We have found that this kind of reconstruction can be formalized by using
  a technique adapted from Xie and Oliveira's ``\dfn{let arguments go first}''~\cite{XieOliveiraESOP2018};
  other well-trodden approaches such as Hindley--Milner-like unification might also work,
  but our formalization seems more concise (and still effective enough) in that
  it suffices to track variables for substitution \emph{only locally}.
  Although the reconstruction is essentially incomplete
  (i.e.,~even when there exist appropriate expressions,
  it cannot always infer them),
  our reconstuction algorithm is fairly effective for typical use cases, as we will report later.
  Even when it cannot infer an argument, the algorithm can still report
  the position where the user should explicitly give arguments.
\par
\subsection{Support for Implicit Tensor Conversion}\label{subsec:implicit-conversion}
\indent
  As we have mentioned in the issue~D,
  it would be desirable to support
  some implicit conversions of tensors,
  such as \dfn{broadcasting}~\cite{NumPyBroadcastingSemantics,PyTorchBroadcastingSemantics},
  which are frequently used in DNN-related programs.
  Our language can safely support the tensor addition with broadcasting
  by providing a code-generating built-in function~\( \ordO{  \makeIdentOrConst{}{ VgenAdd }  } \) of the following type:
  {\fontsize{9.4pt}{0pt}\begin{align*}
     \openO{\{}  \ordO{  \makeIdentOrConst{}{ Vx }  }   \relO{:}   \ttO{List}\    \ttO{Int}    \closeO{\} } \relO{\to}   \openO{\{}  \ordO{  \makeIdentOrConst{}{ Vy }  }   \relO{:}   \ttO{List}\    \ttO{Int}    \closeO{\} } \relO{\to}   \openO{\langle}    \ttI{Tensor}\ \ordI{\%}   \ordO{  \makeIdentOrConst{}{ Vx }  }     \relI{\to}   \ttI{Tensor}\ \ordI{\%}   \ordO{  \makeIdentOrConst{}{ Vy }  }      \relI{\to}   \ttI{Tensor}\ \ordI{\%}  \openO{(}    \ordO{  \makeIdentOrConst{}{ Vbroadcast }  }   \     \ordO{  \makeIdentOrConst{}{ Vx }  }   \    \ordO{  \makeIdentOrConst{}{ Vy }  }     \closeO{)}    \closeO{\rangle}   ,
  \end{align*}}%
  where \( \ordO{  \makeIdentOrConst{}{ Vbroadcast }  }  :    \ttO{List}\    \ttO{Int}     \relO{\to}   \ttO{List}\    \ttO{Int}      \relO{\to}   \ttO{List}\    \ttO{Int}    \)
  is a partial function that returns the shape to which two given shapes
  can be commonly broadcastable.
  Receiving two lists~\( \openO{[}    \ordO{  \makeIdentOrConst{}{ C5 }  }   \punctO{,}     \ordO{  \makeIdentOrConst{}{ C3 }  }   \punctO{,}     \ordO{  \makeIdentOrConst{}{ C1 }  }   \punctO{,}     \ordO{  \makeIdentOrConst{}{ C10 }  }       \closeO{]} \) and \( \openO{[}    \ordO{  \makeIdentOrConst{}{ C3 }  }   \punctO{,}     \ordO{  \makeIdentOrConst{}{ C4 }  }   \punctO{,}     \ordO{  \makeIdentOrConst{}{ C10 }  }      \closeO{]} \), for example,
  \( \ordO{  \makeIdentOrConst{}{ VgenAdd }  } \) returns code~\( \openO{\langle}   \ordI{  \makeIdentOrConst{}{ Vadd }  _{  [    \makeIdentOrConst{}{ C5 }  ,    \makeIdentOrConst{}{ C3 }  ,    \makeIdentOrConst{}{ C1 }  ,    \makeIdentOrConst{}{ C10 }       ],   [    \makeIdentOrConst{}{ C3 }  ,    \makeIdentOrConst{}{ C4 }  ,    \makeIdentOrConst{}{ C10 }      ]   } }   \closeO{\rangle} \),
  where \( \ordI{  \makeIdentOrConst{}{ Vadd }  _{  [    \makeIdentOrConst{}{ C5 }  ,    \makeIdentOrConst{}{ C3 }  ,    \makeIdentOrConst{}{ C1 }  ,    \makeIdentOrConst{}{ C10 }       ],   [    \makeIdentOrConst{}{ C3 }  ,    \makeIdentOrConst{}{ C4 }  ,    \makeIdentOrConst{}{ C10 }      ]   } } \) is the specialized tensor addition
  of the following type:
  \begin{align*}
       \ttI{Tensor}\ \ordI{\%}  \openO{[}    \ordO{  \makeIdentOrConst{}{ C5 }  }   \punctO{,}     \ordO{  \makeIdentOrConst{}{ C3 }  }   \punctO{,}     \ordO{  \makeIdentOrConst{}{ C1 }  }   \punctO{,}     \ordO{  \makeIdentOrConst{}{ C10 }  }       \closeO{]}    \relI{\to}   \ttI{Tensor}\ \ordI{\%}  \openO{[}    \ordO{  \makeIdentOrConst{}{ C3 }  }   \punctO{,}     \ordO{  \makeIdentOrConst{}{ C4 }  }   \punctO{,}     \ordO{  \makeIdentOrConst{}{ C10 }  }      \closeO{]}     \relI{\to}   \ttI{Tensor}\ \ordI{\%}  \openO{[}    \ordO{  \makeIdentOrConst{}{ C5 }  }   \punctO{,}     \ordO{  \makeIdentOrConst{}{ C3 }  }   \punctO{,}     \ordO{  \makeIdentOrConst{}{ C4 }  }   \punctO{,}     \ordO{  \makeIdentOrConst{}{ C10 }  }       \closeO{]}   .
  \end{align*}
  On the other hand, if \( \ordO{  \makeIdentOrConst{}{ VgenAdd }  } \) is applied to
  two shapes that are not broadcastable to one common shape,
  it will emit a failure,
  just as \( \ordO{  \makeIdentOrConst{}{ VgenMatMult }  } \) will do for negative integers.
\par
\subsection{Error Localization by Stage-0 Refinement Types}
\indent
  One aspect that provides a source for improving the language design is
  how errors during compile-time computation are reported.
  Other than casts~\(\LeftAssertParen \ordO{T_1} \relO{\triangleleft} \ordO{T_2} \RightAssertParen\),
  stage-\(0\) functions may also emit a failure since some of them are partial
  (e.g.,~\( \ordO{  \makeIdentOrConst{}{ VgenMatMult }  } \) cannot take negative integers).
  Certainly, in both cases, failures can happen only at compile time
  and can be reported in a somewhat human-friendly manner
  since they point to a code position by \(\ell\),
  but the problem is that a reported position does not necessarily identify the direct source of the bug;
  it can be a position inside a function that does not contain erroneous descriptions,
  and the reported position should rather be one of the call sites.
  For example, consider \(       \ordO{  \makeIdentOrConst{}{ Vf' }  }    \   \openO{(}   \ordO{  \makeIdentOrConst{-}{ C1 }  }   \closeO{)}   \    \ordO{  \makeIdentOrConst{}{ C1 }  }    \    \ordO{  \makeIdentOrConst{}{ C1 }  }    \    \ordO{  \makeIdentOrConst{}{ C1 }  }   \).
  This will cause an error with the position in the definition of \(  \ordO{  \makeIdentOrConst{}{ Vf' }  }  \)
  at which \( \ordO{  \makeIdentOrConst{}{ VgenMatMult }  } \) is applied to \(  \ordO{  \makeIdentOrConst{}{ Vj }  }  \),
  i.e.,~the first parameter of \(  \ordO{  \makeIdentOrConst{}{ Vf' }  }  \).
  However, it is the call site of \(  \ordO{  \makeIdentOrConst{}{ Vf' }  }  \) passing \( \ordO{  \makeIdentOrConst{-}{ C1 }  } \),
  not the definition of \(  \ordO{  \makeIdentOrConst{}{ Vf' }  }  \),
  that this error should be attributed to.
  It will be better if we have a mechanism to emit a failure
  when \(  \ordO{  \makeIdentOrConst{}{ Vf' }  }  \) is applied to a negative value like \( \ordO{  \makeIdentOrConst{-}{ C1 }  } \).
\par
\indent
  To this end,
  we use \dfn{refinement types}~\cite{%
    FreemanPfenningPLDI1991,%
    FlanaganPOPL2006,%
    RondonKawaguchiRanjitPLDI2008,%
    KnowlesFlanaganTOPLAS2010}
  of the form~\( \openO{\{} \possiblyWithSub\stageOmetaColor{\nu}  \relO{:}  \possiblyWithSub\stageOmetaColor{B}  \relO{\mid}  \possiblyWithSub\stageOmetaColor{N^{\superscriptO} } \closeO{\} } \)
  for stage-0 types.
  The annotations
  can be modified from \(  \ttO{Int}  \) to \(   \ttO{Nat}   \),
  which abbreviates \( \openO{\{} \possiblyWithSub\stageOmetaColor{\nu}  \relO{:}   \ttO{Int}   \relO{\mid}    \possiblyWithSub\stageOmetaColor{\nu}  \binO{  \geq  }   \ordO{  \makeIdentOrConst{}{ C0 }  }    \closeO{\} } \):
  \begin{align*}
     \progindent{  \tokenO{let}\   \ordO{  \makeIdentOrConst{}{ Vf' }  }   \empty   \relO{=}   \ordO{\lambda}\openO{\{}  \ordO{  \makeIdentOrConst{}{ Vj }  }   \relO{:}    \ttO{Nat}   \closeO{\} }\punctO{.}\   \ordO{\lambda}\openO{\{}  \ordO{  \makeIdentOrConst{}{ Vk }  }   \relO{:}    \ttO{Nat}   \closeO{\} }\punctO{.}\   \ordO{\lambda}\openO{\{}  \ordO{  \makeIdentOrConst{}{ Vm }  }   \relO{:}    \ttO{Nat}   \closeO{\} }\punctO{.}\   \ordO{\lambda}\openO{\{}  \ordO{  \makeIdentOrConst{}{ Vn }  }   \relO{:}    \ttO{Nat}   \closeO{\} }\punctO{.}\control\deepen{\control\br{}  \openO{\langle}  \ordI{\lambda}  \ordI{  \makeIdentOrConst{}{ VA }  }   \relI{:}   \ttI{Mat}\ \ordI{\%}   \ordO{  \makeIdentOrConst{}{ Vj }  }   \ \ordI{\%}  \openO{(}     \ordO{  \makeIdentOrConst{}{ Vk }  }   \binO{  +  }   \ordO{  \makeIdentOrConst{}{ C2 }  }    \binO{  \ast  }   \ordO{  \makeIdentOrConst{}{ Vm }  }    \closeO{)}   \punctI{.}\   \ordI{\lambda}  \ordI{  \makeIdentOrConst{}{ VB }  }   \relI{:}   \ttI{Mat}\ \ordI{\%}   \ordO{  \makeIdentOrConst{}{ Vm }  }   \ \ordI{\%}   \ordO{  \makeIdentOrConst{}{ Vn }  }    \punctI{.}\   \ordI{\lambda}  \ordI{  \makeIdentOrConst{}{ VC }  }   \relI{:}   \ttI{Mat}\ \ordI{\%}   \ordO{  \makeIdentOrConst{}{ Vk }  }   \ \ordI{\%}   \ordO{  \makeIdentOrConst{}{ Vn }  }    \punctI{.}\   \ldots     \closeO{\rangle}  }      } 
  \end{align*}
  Then, consider an application \(   \ordO{  \makeIdentOrConst{}{ Vf' }  }   \  \possiblyWithSub\stageOmetaColor{M^{\scriptscriptstyle(0)} } \).
  Now that the type for \(  \ordO{  \makeIdentOrConst{}{ Vj }  }  \) tracks the precondition required of \(  \ordO{  \makeIdentOrConst{}{ Vj }  }  \),
  the above application elaborates to
  \(   \ordO{  \makeIdentOrConst{}{ Vf' }  }   \   \openO{(}   \LeftAssertParen \relO{\CastArrow}    \ttO{Nat}    \RightAssertParen^{  \ell  }  \  \possiblyWithSub\stageOmetaColor{N^{\superscriptO} }  \closeO{)}  \)
  (assuming \(\possiblyWithSub\stageOmetaColor{M^{\scriptscriptstyle(0)} }\) elaborates to \(\possiblyWithSub\stageOmetaColor{N^{\superscriptO} }\)),
  where \( \LeftAssertParen \relO{\CastArrow}  \stageOmetaColor{T^{\superscriptO}_{\mathrm{rfn} } }  \RightAssertParen^{  \ell  } \) is a cast function
  to assert that the argument satisfies the predicate of the refinement type~\(\stageOmetaColor{T^{\superscriptO}_{\mathrm{rfn} } }\),
  and \(\ell\) is a label that points to the original application.
  This will emit a failure if \(\possiblyWithSub\stageOmetaColor{N^{\superscriptO} }\) evaluates to a negative integer
  and report the position of the call site~\(\ell\).
  Compared to the original situation where failures are raised by
  \( \ordO{  \makeIdentOrConst{}{ VgenMatMult }  } \) used in \(  \ordO{  \makeIdentOrConst{}{ Vf' }  }  \),
  programmers can describe more detailed preconditions expected of arguments,
  and the type-checker can
  point to the location where some precondition was violated.
  We allow refinement types only for stage~\(0\), and thereby
  the evaluation of such casts happens only at compile time,
  i.e.,~runtime evaluation is still free from assertion failures.
\par
\indent
  We note that refinement types are also beneficial for
  error localization as to implicit conversion;
  with refinement types, we can assign \( \ordO{  \makeIdentOrConst{}{ VgenAdd }  } \) the following more natural type:
  \begin{align*}
     \progindent{  \openO{\{}  \ordO{  \makeIdentOrConst{}{ Vx }  }   \relO{:}    \ttO{NatList}   \closeO{\} } \relO{\to}   \openO{\{}  \ordO{  \makeIdentOrConst{}{ Vy }  }   \relO{:}    \openO{\{} \possiblyWithSub\stageOmetaColor{\nu}  \relO{:}   \ttO{NatList}   \relO{\mid}     \ordO{  \makeIdentOrConst{}{ Vbroadcastable }  }   \     \ordO{  \makeIdentOrConst{}{ Vx }  }   \   \possiblyWithSub\stageOmetaColor{\nu}    \closeO{\} }   \closeO{\} } \relO{\to} \control\deepen{\control\br{}  \openO{\langle}    \ttI{Tensor}\ \ordI{\%}   \ordO{  \makeIdentOrConst{}{ Vx }  }     \relI{\to}   \ttI{Tensor}\ \ordI{\%}   \ordO{  \makeIdentOrConst{}{ Vy }  }      \relI{\to}   \ttI{Tensor}\ \ordI{\%}  \openO{(}    \ordO{  \makeIdentOrConst{}{ Vbroadcast }  }   \     \ordO{  \makeIdentOrConst{}{ Vx }  }   \    \ordO{  \makeIdentOrConst{}{ Vy }  }     \closeO{)}    \closeO{\rangle}  }   } ,
  \end{align*}
  where \(  \ttO{NatList}  \) is a basetype for shapes that intuitively works as \( \ttO{List}\    \ttO{Nat}   \),
  and \( \ordO{  \makeIdentOrConst{}{ Vbroadcastable }  } \) judges whether a pair of two shapes
  is in the domain of \( \ordO{  \makeIdentOrConst{}{ Vbroadcast }  } \).
  As a sideline,
  \( \ordO{  \makeIdentOrConst{}{ Vbroadcast }  }  :
      \openO{(}  \ordO{  \makeIdentOrConst{}{ Vx }  }   \relO{:}    \ttO{NatList}   \closeO{)} \relO{\to}    \openO{\{} \possiblyWithSub\stageOmetaColor{\nu}  \relO{:}   \ttO{NatList}   \relO{\mid}     \ordO{  \makeIdentOrConst{}{ Vbroadcastable }  }   \     \ordO{  \makeIdentOrConst{}{ Vx }  }   \   \possiblyWithSub\stageOmetaColor{\nu}    \closeO{\} }     \relO{\to}    \ttO{NatList}   \)
  is now a total function.
\par
\subsection{Horsea: A Seemingly Dependently-Typed Surface Language}
\indent
  Although many shape-related arguments are now inferred,
  we still manually add staging constructs,
  i.e.,~brackets~\(\openO{\langle}\closeO{\rangle}\) and escapes~\(\ordI{\sim}\).
  This might be inconvenient for some users,
  especially those who are unfamiliar with staged computation.
  As a solution to this cumbersomeness,
  there are a number of ways to provide a surface language that is less explicit as to staging
  and to reconstruct where to insert gaps between stages.
  To provide a proof-of-concept language for the moment, we pick up
  \dfn{binding-time analysis}~(\dfn{BTA})~\cite{JonesGomardSestoft1993,DaviesLICS1996,DaviesJACM2017},
  a well-known classical technique in the literature of partial evaluation.
  On top of the staged language, we give a surface language named \dfn{Horsea}\footnote{%
    In Japanese, seahorses are called \dfn{tatsu-no-otoshi-go}
    (``dragon's lost children'') due to their resemblance to Asian-style dragons.
    We use the name of a seahorse character for our surface language,
    reflecting the fact that programs in the language are syntactically similar to
    but internally quite different from those in Idris~\cite{BradyJFP2013,BradyECOOP2021},
    which was named after a dragon character.
  } that is sheerly non-staged (i.e.,~does not require manual staging at all).
  In fact, users can describe \( \makeIdentOrConst{}{ Vf' } \) in Horsea as follows:
  \begin{align*}
     \progindent{  \token{let}\    \makeIdentOrConst{}{ Vf' }   \control\deepen{  \ \{   \makeIdentOrConst{}{ Vj }    :    \mathtt{Nat}   \}  \ \{   \makeIdentOrConst{}{ Vk }    :    \mathtt{Nat}   \}  \ \{   \makeIdentOrConst{}{ Vm }    :    \mathtt{Nat}   \}  \ \{   \makeIdentOrConst{}{ Vn }    :    \mathtt{Nat}   \}  \control\br{}\!\quad   \ (   \makeIdentOrConst{}{ VA }    :   \mathtt{Mat}\     \makeIdentOrConst{}{ Vj }    \   (      \makeIdentOrConst{}{ Vk }     +     \makeIdentOrConst{}{ C2 }      \ast     \makeIdentOrConst{}{ Vm }     )   )  \ (   \makeIdentOrConst{}{ VB }    :   \mathtt{Mat}\     \makeIdentOrConst{}{ Vm }    \     \makeIdentOrConst{}{ Vn }     )  \ (   \makeIdentOrConst{}{ VC }    :   \mathtt{Mat}\     \makeIdentOrConst{}{ Vk }    \     \makeIdentOrConst{}{ Vn }     )  \empty          } = \control\deepen{\control\br{}     \control\deepen{  \token{let}\    \makeIdentOrConst{}{ VD }   \control\deepen{  \empty  } =      \makeIdentOrConst{}{ VvertCat }    \     (      \makeIdentOrConst{}{ VvertCat }    \     \makeIdentOrConst{}{ VB }     \     \makeIdentOrConst{}{ VC }     )  \     \makeIdentOrConst{}{ VB }        }\ \token{in}\     \makeIdentOrConst{}{ VmatMult }     \     \makeIdentOrConst{}{ VA }     \     \makeIdentOrConst{}{ VD }      }  } 
  \end{align*}
  As one can see,
  thanks to the omission of some stage-\(0\) arguments and staging constructs,
  we can finally write a program that is syntactically quite similar to
  the first code in the Idris-like language.
  On the contrary, no \(\mathbf{rewrite}\)s are necessary for type equality here.
\par
\indent
  Given a program in Horsea, we perform BTA to find out
  which parts can be at compile time and
  insert brackets and escapes accordingly.
  For this conversion, built-in functions in Horsea (e.g.,~\( \makeIdentOrConst{}{ VvertCat } \))
  are associated with those in the staged language (e.g.,~\( \ordO{  \makeIdentOrConst{}{ VgenVertCat }  } \)).
\par

\section{Staged Language}\label{sec:staged-language}
\indent
  In this section, we explain how our staged core language \LambdaBracketCast\ is formalized.
  Since \LambdaBracketCast\ performs elaboration for cast insertion,
  it has a source syntax and a target syntax.
\par
\subsection{Syntax, Typing Rules, and Operational Semantics}
  The source syntax is defined by the following \(M^{(b)}\) and \(S^{(b)}\),
  which range over the set of stage-\(b\) expressions and that of stage-\(b\) type annotations, respectively
  (for \(b \in \{0, 1\}\)):
  \begin{center}
    \vspace{-2.5em}
    \begin{minipage}{0.75\textwidth}
      \begin{align*}
        \bnf{\possiblyWithSub\stageOmetaColor{M^{\scriptscriptstyle(0)} }}{%
          \possiblyWithSub\stageOmetaColor{p} | \possiblyWithSub\stageOmetaColor{c} | \possiblyWithSub\stageOmetaColor{x} |  \ordO{\lambda} \possiblyWithSub\stageOmetaColor{x}  \relO{:}  \possiblyWithSub\stageOmetaColor{S^{\superscriptO} } \punctO{.}\  \possiblyWithSub\stageOmetaColor{M^{\scriptscriptstyle(0)} }  |  \openO{(} \possiblyWithSub\stageOmetaColor{M^{\scriptscriptstyle(0)} } \  \possiblyWithSub\stageOmetaColor{M^{\scriptscriptstyle(0)} } \closeO{)}_{ \ell }  |  \openO{\langle} \possiblyWithSub\stageImetaColor{M^{\superscriptI} } \closeO{\rangle} 
        }
      \\
        \bnf{\possiblyWithSub\stageImetaColor{M^{\superscriptI} }}{%
          \possiblyWithSub\stageImetaColor{c} | \possiblyWithSub\stageImetaColor{x} |  \ordI{\lambda} \possiblyWithSub\stageImetaColor{x}  \relI{:}  \possiblyWithSub\stageImetaColor{S^{\superscriptI} } \punctI{.}\  \possiblyWithSub\stageImetaColor{M^{\superscriptI} }  |  \openI{(} \possiblyWithSub\stageImetaColor{M^{\superscriptI} } \  \possiblyWithSub\stageImetaColor{M^{\superscriptI} } \closeI{)}_{ \ell }  |  \ordI{\sim} \possiblyWithSub\stageOmetaColor{M^{\scriptscriptstyle(0)} } 
        }
      \\
        \bnf{\possiblyWithSub\stageOmetaColor{S^{\superscriptO} }}{%
           \openO{\{} \possiblyWithSub\stageOmetaColor{x}  \relO{:}  \possiblyWithSub\stageOmetaColor{B}  \relO{\mid}  \possiblyWithSub\stageOmetaColor{M^{\scriptscriptstyle(0)} } \closeO{\} }  |  \ttO{Tensor}\  \possiblyWithSub\stageOmetaColor{s}  |  \openO{(} \possiblyWithSub\stageOmetaColor{x}  \relO{:}  \possiblyWithSub\stageOmetaColor{S^{\superscriptO} } \closeO{)} \relO{\to}  \possiblyWithSub\stageOmetaColor{S^{\superscriptO} }  |  \openO{\langle} \possiblyWithSub\stageImetaColor{S^{\superscriptI} } \closeO{\rangle} 
        }
      \\
        \bnf{\possiblyWithSub\stageImetaColor{S^{\superscriptI} }}{%
          \possiblyWithSub\stageImetaColor{B} |  \ttI{Tensor}\ \ordI{\%} \possiblyWithSub\stageOmetaColor{M^{\scriptscriptstyle(0)} }  |  \possiblyWithSub\stageImetaColor{S^{\superscriptI} }  \relI{\to}  \possiblyWithSub\stageImetaColor{S^{\superscriptI} } 
        }
      \end{align*}
    \end{minipage}%
    \begin{minipage}{0.23\textwidth}
      \begin{align*}
        \bnf{B}{%
            \mathtt{Bool}  
        |*  \mathtt{Int}  
        |*  \mathtt{NatList}  
        |*  \mathtt{Float}   | \cdots
        }
      \end{align*}
    \end{minipage}
  \end{center}
  Here, \(\possiblyWithSub\stageOmetaColor{p}\) and \(c\) respectively range over
  the set of built-in functions available only at stage~\(0\) (e.g., \( \ordO{  \makeIdentOrConst{}{ VgenMatMult }  } \))
  and the set of constants usable at both stages,
  which includes base constants (e.g., \(   \makeIdentOrConst{}{ C42 }   \) or \(  \mathtt{true}  \)) and
  stage-agnostic built-in functions (e.g., \(+\) or \(   \makeIdentOrConst{}{ VmatMult }  _{   \makeIdentOrConst{}{ C3 }  ,    \makeIdentOrConst{}{ C4 }  ,    \makeIdentOrConst{}{ C5 }     }  \)).
  While the application of \(\possiblyWithSub\stageOmetaColor{p}\) may be restricted by refinement predicates
  (e.g.,~one cannot pass \( \ordO{  \makeIdentOrConst{-}{ C1 }  } \) to \( \ordO{  \makeIdentOrConst{}{ VgenMatMult }  } \)),
  \(c\) must be simply-typed.
  Each constant ranged over by \(\possiblyWithSub\stageOmetaColor{p}\) or \(c\) has its own arity,
  and in particular, base constants are stage-agnostic constants of arity~\(0\).
  We denote these arities by \(\arity{\possiblyWithSub\stageOmetaColor{p}}\) and \(\arity{c}\).
  To report the cause of compile-time assertion failures,
  each occurrence of function application is equipped with a unique label~\(\ell\)
  that points to its code position.
  Users do not have to write these labels; they are simply attached by a pre-processor.
\par
\indent
  The most essential part of the syntax is that
  stage-\(1\) tensor types~\( \ttI{Tensor}\ \ordI{\%} \possiblyWithSub\stageOmetaColor{M^{\scriptscriptstyle(0)} } \)
  have a stage-\(0\) expression~\(\possiblyWithSub\stageOmetaColor{M^{\scriptscriptstyle(0)} }\) of type~\(  \ttO{NatList}  \)
  to represent a tensor shape\footnote{%
    While \(\ttO{Nat}\) is just a shorthand for \( \openO{\{} \possiblyWithSub\stageOmetaColor{\nu}  \relO{:}   \ttO{Int}   \relO{\mid}    \possiblyWithSub\stageOmetaColor{\nu}  \binO{  \geq  }   \ordO{  \makeIdentOrConst{}{ C0 }  }    \closeO{\} } \),
    \(  \ttO{NatList}  \) is a base type.
    This will simplify metatheory.
  }.
  This gap between tensor types and their argument expressions as to stages
  ensures, for example, that
  all the stage-\(1\) binders of the form \( \openI{(}  \ordI{\lambda} \possiblyWithSub\stageImetaColor{x}  \relI{:}   \ttI{Tensor}\ \ordI{\%}  \openO{[}  \possiblyWithSub\stageOmetaColor{M^{\scriptscriptstyle(0)} }_{{\mathrm{1}}} \punctO{,}   \possiblyWithSub\stageOmetaColor{M^{\scriptscriptstyle(0)} }_{{\mathrm{2}}}   \closeO{]}   \punctI{.}\   \ldots   \closeI{)} \)
  will be \( \openI{(}  \ordI{\lambda} \possiblyWithSub\stageImetaColor{x}  \relI{:}   \ttI{Tensor}\ \ordI{\%}  \openO{[}    \possiblyWithSub\stageOmetaColor{n}_{{\mathrm{1}}}   \punctO{,}     \possiblyWithSub\stageOmetaColor{n}_{{\mathrm{2}}}     \closeO{]}   \punctI{.}\   \ldots   \closeI{)} \) after compile-time computation
  and thereby that every tensor in generated code has a specialized shape.
  Stage-\(1\) types~\( \ttI{Vec}\ \ordI{\%} \possiblyWithSub\stageOmetaColor{M^{\scriptscriptstyle(0)} }_{{\mathrm{1}}} \) and \( \ttI{Mat}\ \ordI{\%} \possiblyWithSub\stageOmetaColor{M^{\scriptscriptstyle(0)} }_{{\mathrm{1}}} \ \ordI{\%} \possiblyWithSub\stageOmetaColor{M^{\scriptscriptstyle(0)} }_{{\mathrm{2}}} \)
  for vectors and matrices can be provided as
  syntax sugar of
  \( \ttI{Tensor}\ \ordI{\%}  \openO{[}  \possiblyWithSub\stageOmetaColor{M^{\scriptscriptstyle(0)} }_{{\mathrm{1}}}  \closeO{]}  \) and \( \ttI{Tensor}\ \ordI{\%}  \openO{[}  \possiblyWithSub\stageOmetaColor{M^{\scriptscriptstyle(0)} }_{{\mathrm{1}}} \punctO{,}   \possiblyWithSub\stageOmetaColor{M^{\scriptscriptstyle(0)} }_{{\mathrm{2}}}   \closeO{]}  \),
  respectively.
  The symbol~\(\%\) in the notation indicates this gap
  by an analogy to the notion of cross-stage persistence~\cite{%
    HanadaIgarashi2014,%
    KawataIgarashiAPLAS2019,%
    TahaSheardPEPM1997,%
    YuseIgarashiPPDP2006}.
  By contrast, stage-\(0\) tensor types~\( \ttO{Tensor}\  \possiblyWithSub\stageOmetaColor{s} \) have fixed shapes;
  the metavariable~\(\possiblyWithSub\stageOmetaColor{s}\) ranges over the set of finite sequences of natural numbers,
  and these types can be thought of as a family of countably infinite base types.
  Stage-\(0\) tensor types occur mostly in stage-\(0\) terms obtained by unlifting generated code
  and are seldom written by users.
\par
\indent
  In response to the setting of stage-\(1\) tensor types,
  stage-\(0\) function types can be dependent ones.
  This allows, for instance, \( \ordO{  \makeIdentOrConst{}{ VgenMatMult }  } \) to be assigned
  \( \openO{(}  \ordO{  \makeIdentOrConst{}{ Vp }  }   \relO{:}    \ttO{Nat}   \closeO{)} \relO{\to}   \openO{(}  \ordO{  \makeIdentOrConst{}{ Vq }  }   \relO{:}    \ttO{Nat}   \closeO{)} \relO{\to}   \openO{(}  \ordO{  \makeIdentOrConst{}{ Vr }  }   \relO{:}    \ttO{Nat}   \closeO{)} \relO{\to}   \openO{\langle}    \ttI{Mat}\ \ordI{\%}   \ordO{  \makeIdentOrConst{}{ Vp }  }   \ \ordI{\%}   \ordO{  \makeIdentOrConst{}{ Vq }  }     \relI{\to}   \ttI{Mat}\ \ordI{\%}   \ordO{  \makeIdentOrConst{}{ Vq }  }   \ \ordI{\%}   \ordO{  \makeIdentOrConst{}{ Vr }  }      \relI{\to}   \ttI{Mat}\ \ordI{\%}   \ordO{  \makeIdentOrConst{}{ Vp }  }   \ \ordI{\%}   \ordO{  \makeIdentOrConst{}{ Vr }  }     \closeO{\rangle}    \).
  However, this does not apply to stage~\(1\);
  by restricting stage-\(1\) function types to non-dependent ones,
  compile-time assertions are considerably simplified, and
  many realistic programs can still be supported,
  as exemplified in Section~\ref{sec:implementation}.
  We also use the notation~\( \possiblyWithSub\stageOmetaColor{T^{\superscriptO} }_{{\mathrm{1}}}  \relO{\to}  \possiblyWithSub\stageOmetaColor{T^{\superscriptO} }_{{\mathrm{2}}} \) for \( \openO{(} \possiblyWithSub\stageOmetaColor{x}  \relO{:}  \possiblyWithSub\stageOmetaColor{T^{\superscriptO} }_{{\mathrm{1}}} \closeO{)} \relO{\to}  \possiblyWithSub\stageOmetaColor{T^{\superscriptO} }_{{\mathrm{2}}} \),
  where \(\possiblyWithSub\stageOmetaColor{x} \not\in \fv(\possiblyWithSub\stageOmetaColor{T^{\superscriptO} }_{{\mathrm{2}}})\).
\par
\indent
  The target syntax is basically an ``assertion-included'' variant of the source syntax.
  It consists of
  stage-\(b\) \dfn{assertive terms} \(N^{(b)}\) and \dfn{assertive types} \(T^{(b)}\)
  defined by the following:
  \begin{align*}
    \bnf{\possiblyWithSub\stageOmetaColor{N^{\superscriptO} }}{%
      \possiblyWithSub\stageOmetaColor{p} | \possiblyWithSub\stageOmetaColor{c} | \possiblyWithSub\stageOmetaColor{x} |  \ordO{\lambda} \possiblyWithSub\stageOmetaColor{x}  \relO{:}  \possiblyWithSub\stageOmetaColor{T^{\superscriptO} } \punctO{.}\  \possiblyWithSub\stageOmetaColor{N^{\superscriptO} }  |  \possiblyWithSub\stageOmetaColor{N^{\superscriptO} } \  \possiblyWithSub\stageOmetaColor{N^{\superscriptO} }  |  \openO{\langle} \possiblyWithSub\stageImetaColor{N^{\superscriptI} } \closeO{\rangle} 
    |* \LeftAssertParen\openO{\langle} \possiblyWithSub\stageImetaColor{T^{\superscriptI} } \closeO{\rangle} \relO{\CastArrow} \openO{\langle} \possiblyWithSub\stageImetaColor{T^{\superscriptI} } \closeO{\rangle}\RightAssertParen^{ L }  |  \LeftAssertParen \relO{\CastArrow}   \openO{\{} \possiblyWithSub\stageOmetaColor{x}  \relO{:}  \possiblyWithSub\stageOmetaColor{B}  \relO{\mid}  \possiblyWithSub\stageOmetaColor{N^{\superscriptO} } \closeO{\} }   \RightAssertParen^{ L }  |  \LeftAssertParen   \openO{\{} \possiblyWithSub\stageOmetaColor{x}  \relO{:}  \possiblyWithSub\stageOmetaColor{B}  \relO{\mid}  \possiblyWithSub\stageOmetaColor{N^{\superscriptO} } \closeO{\} }  \punctO{,}  \possiblyWithSub\stageOmetaColor{N^{\superscriptO} } \punctO{,}  \possiblyWithSub\stageOmetaColor{c}  \RightAssertParen^{ L } 
    }
  \\
    \bnf{\possiblyWithSub\stageImetaColor{N^{\superscriptI} }}{%
      \possiblyWithSub\stageImetaColor{c} | \possiblyWithSub\stageImetaColor{x} |  \ordI{\lambda} \possiblyWithSub\stageImetaColor{x}  \relI{:}  \possiblyWithSub\stageImetaColor{T^{\superscriptI} } \punctI{.}\  \possiblyWithSub\stageImetaColor{N^{\superscriptI} }  |  \possiblyWithSub\stageImetaColor{N^{\superscriptI} } \  \possiblyWithSub\stageImetaColor{N^{\superscriptI} }  |  \ordI{\sim} \possiblyWithSub\stageOmetaColor{N^{\superscriptO} } 
    }
  \\
    \bnf{\possiblyWithSub\stageOmetaColor{T^{\superscriptO} }}{%
       \openO{\{} \possiblyWithSub\stageOmetaColor{x}  \relO{:}  \possiblyWithSub\stageOmetaColor{B}  \relO{\mid}  \possiblyWithSub\stageOmetaColor{N^{\superscriptO} } \closeO{\} }  |  \ttO{Tensor}\  \possiblyWithSub\stageOmetaColor{s}  |  \openO{(} \possiblyWithSub\stageOmetaColor{x}  \relO{:}  \possiblyWithSub\stageOmetaColor{T^{\superscriptO} } \closeO{)} \relO{\to}  \possiblyWithSub\stageOmetaColor{T^{\superscriptO} }  |  \openO{\langle} \possiblyWithSub\stageImetaColor{T^{\superscriptI} } \closeO{\rangle} 
    }
  \\
    \bnf{\possiblyWithSub\stageImetaColor{T^{\superscriptI} }}{%
      \possiblyWithSub\stageImetaColor{B} |  \ttI{Tensor}\ \ordI{\%} \possiblyWithSub\stageOmetaColor{N^{\superscriptO} }  |  \possiblyWithSub\stageImetaColor{T^{\superscriptI} }  \relI{\to}  \possiblyWithSub\stageImetaColor{T^{\superscriptI} } 
    }
  \qquad
    \bnfnotab{L}{%
      \ell |  L .\mathbf{dom}  |  L .\mathbf{cod} 
    }
  \end{align*}
  Stage-\(0\) assertions have two forms.
  The first one is \( \LeftAssertParen\openO{\langle} \possiblyWithSub\stageImetaColor{T^{\superscriptI} }_{{\mathrm{1}}} \closeO{\rangle} \relO{\CastArrow} \openO{\langle} \possiblyWithSub\stageImetaColor{T^{\superscriptI} }_{{\mathrm{2}}} \closeO{\rangle}\RightAssertParen^{ L } \),
  which judges that the stage-\(1\) types \(\possiblyWithSub\stageImetaColor{T^{\superscriptI} }_{{\mathrm{1}}}\) and \(\possiblyWithSub\stageImetaColor{T^{\superscriptI} }_{{\mathrm{2}}}\)
  syntactically coincide after evaluation.
  Interestingly, adding assertions of this form covers
  all the necessary cases for checking type equality itself.
  The attachment~\(L\) stands for the source of errors when the assertion fails.
  The other one~\( \LeftAssertParen \relO{\CastArrow}   \openO{\{} \possiblyWithSub\stageOmetaColor{\nu}  \relO{:}  \possiblyWithSub\stageOmetaColor{B}  \relO{\mid}  \possiblyWithSub\stageOmetaColor{N^{\superscriptO} }_{{\mathrm{1}}} \closeO{\} }   \RightAssertParen^{ L } \) is used for checking the validity of downcasts.
  In operational terms,
  it basically works as an identity function, but before returning the argument (say \(\possiblyWithSub\stageOmetaColor{c}\)) as is,
  it checks whether the argument satisfies the predicate~\(\possiblyWithSub\stageOmetaColor{N^{\superscriptO} }_{{\mathrm{1}}}\),
  i.e., whether \(  [    \possiblyWithSub\stageOmetaColor{c}    /  \possiblyWithSub\stageOmetaColor{\nu}  ]    \possiblyWithSub\stageOmetaColor{N^{\superscriptO} }_{{\mathrm{1}}} \) evaluates to \(  \ttO{true}  \).
  The variant~\( \LeftAssertParen   \openO{\{} \possiblyWithSub\stageOmetaColor{\nu}  \relO{:}  \possiblyWithSub\stageOmetaColor{B}  \relO{\mid}  \possiblyWithSub\stageOmetaColor{N^{\superscriptO} }_{{\mathrm{1}}} \closeO{\} }  \punctO{,}  \possiblyWithSub\stageOmetaColor{N^{\superscriptO} }_{{\mathrm{2}}} \punctO{,}  \possiblyWithSub\stageOmetaColor{c}  \RightAssertParen^{ L } \) called an \dfn{active check} is
  the intermediate form of this assertion process,
  where \(\possiblyWithSub\stageOmetaColor{N^{\superscriptO} }_{{\mathrm{2}}}\) is a ``refinement proposition''
  (i.e., the application of the refinement predicate~\(\possiblyWithSub\stageOmetaColor{N^{\superscriptO} }_{{\mathrm{1}}}\) to the tested value~\(\possiblyWithSub\stageOmetaColor{c}\))
  under evaluation.
  It also keeps \(\possiblyWithSub\stageOmetaColor{c}\) separately so that it will evaluate to \(\possiblyWithSub\stageOmetaColor{c}\) when the assertion passes.
  Only the first one of these two assertion forms is new;
  the latter device is ported from the context of
  manifest contracts~\cite{%
    GreenbergPierceWeirichPOPL2010,%
    SekiyamaIgarashiGreenbergTOPLAS2017}.
\par
\begin{figure}[tb]
\small
  \begin{flushleft}
    \fbox{\(\mathit{\Gamma} \vdash^{b} M^{(b)} : T^{(b)} \ElabArrow N^{(b)}\)}
  \end{flushleft}
  \vspace{-4em}
  \begin{center}
  \hspace{12.5em}%
    \derive[S0-Brkt]{%
       \mathit{\Gamma}  \vdash^{1}  \possiblyWithSub\stageImetaColor{M^{\superscriptI} }  :  \possiblyWithSub\stageImetaColor{T^{\superscriptI} }  \ElabArrow  \possiblyWithSub\stageImetaColor{N^{\superscriptI} } 
    }{%
       \mathit{\Gamma}  \vdash^{0}   \openO{\langle} \possiblyWithSub\stageImetaColor{M^{\superscriptI} } \closeO{\rangle}   :   \openO{\langle} \possiblyWithSub\stageImetaColor{T^{\superscriptI} } \closeO{\rangle}   \ElabArrow   \openO{\langle} \possiblyWithSub\stageImetaColor{N^{\superscriptI} } \closeO{\rangle}  
    }
  \quad
    \derive[S0-Cst0]{%
      \ConstEnvZero(\possiblyWithSub\stageOmetaColor{p}) = \possiblyWithSub\stageOmetaColor{T^{\superscriptO} }
    }{%
       \mathit{\Gamma}  \vdash^{0}   \possiblyWithSub\stageOmetaColor{p}   :  \possiblyWithSub\stageOmetaColor{T^{\superscriptO} }  \ElabArrow    \possiblyWithSub\stageOmetaColor{p}   
    }
  \\[0.7em]
    \derive[S0-CstP]{%
      \ConstEnvPers(c) = \possiblyWithSub\stageImetaColor{\tau^{\superscriptI} }
    }{%
       \mathit{\Gamma}  \vdash^{0}   \possiblyWithSub\stageOmetaColor{c}   :   \mathop{\downarrow}( \possiblyWithSub\stageImetaColor{\tau^{\superscriptI} } )   \ElabArrow    \possiblyWithSub\stageOmetaColor{c}   
    }
  \quad
    \derive[S0-Abs]{%
       \mathit{\Gamma}  \vdash^{0}  \possiblyWithSub\stageOmetaColor{S^{\superscriptO} }_{{\mathrm{1}}}  \ElabArrow  \possiblyWithSub\stageOmetaColor{T^{\superscriptO} }_{{\mathrm{1}}} 
    \andalso
        \mathit{\Gamma} ,  \possiblyWithSub\stageOmetaColor{x}  : ( \possiblyWithSub\stageOmetaColor{T^{\superscriptO} }_{{\mathrm{1}}} )^{0}   \vdash^{0}  \possiblyWithSub\stageOmetaColor{M^{\scriptscriptstyle(0)} }_{{\mathrm{2}}}  :  \possiblyWithSub\stageOmetaColor{T^{\superscriptO} }_{{\mathrm{2}}}  \ElabArrow  \possiblyWithSub\stageOmetaColor{N^{\superscriptO} }_{{\mathrm{2}}} 
    }{%
       \mathit{\Gamma}  \vdash^{0}   \openO{(}  \ordO{\lambda} \possiblyWithSub\stageOmetaColor{x}  \relO{:}  \possiblyWithSub\stageOmetaColor{S^{\superscriptO} }_{{\mathrm{1}}} \punctO{.}\  \possiblyWithSub\stageOmetaColor{M^{\scriptscriptstyle(0)} }_{{\mathrm{2}}}  \closeO{)}   :   \openO{(} \possiblyWithSub\stageOmetaColor{x}  \relO{:}  \possiblyWithSub\stageOmetaColor{T^{\superscriptO} }_{{\mathrm{1}}} \closeO{)} \relO{\to}  \possiblyWithSub\stageOmetaColor{T^{\superscriptO} }_{{\mathrm{2}}}   \ElabArrow   \openO{(}  \ordO{\lambda} \possiblyWithSub\stageOmetaColor{x}  \relO{:}  \possiblyWithSub\stageOmetaColor{T^{\superscriptO} }_{{\mathrm{1}}} \punctO{.}\  \possiblyWithSub\stageOmetaColor{N^{\superscriptO} }_{{\mathrm{2}}}  \closeO{)}  
    }
  \\[0.7em]
    \derive[S0-App]{%
       \mathit{\Gamma}  \vdash^{0}  \possiblyWithSub\stageOmetaColor{M^{\scriptscriptstyle(0)} }_{{\mathrm{1}}}  :   \openO{(} \possiblyWithSub\stageOmetaColor{x}  \relO{:}  \possiblyWithSub\stageOmetaColor{T^{\superscriptO} }_{{\mathrm{11}}} \closeO{)} \relO{\to}  \possiblyWithSub\stageOmetaColor{T^{\superscriptO} }_{{\mathrm{12}}}   \ElabArrow  \possiblyWithSub\stageOmetaColor{N^{\superscriptO} }_{{\mathrm{1}}} 
    \\
       \mathit{\Gamma}  \vdash^{0}  \possiblyWithSub\stageOmetaColor{M^{\scriptscriptstyle(0)} }_{{\mathrm{2}}}  :  \possiblyWithSub\stageOmetaColor{T^{\superscriptO} }_{{\mathrm{2}}}  \ElabArrow  \possiblyWithSub\stageOmetaColor{N^{\superscriptO} }_{{\mathrm{2}}} 
    \andalso
       \mathit{\Gamma}  \vdash_{  \ell  }  \possiblyWithSub\stageOmetaColor{T^{\superscriptO} }_{{\mathrm{2}}}  \CastArrow  \possiblyWithSub\stageOmetaColor{T^{\superscriptO} }_{{\mathrm{11}}}  \ElabArrow  \possiblyWithSub\stageOmetaColor{N^{\superscriptO} }_{{\mathrm{0}}} 
    }{%
       \mathit{\Gamma}  \vdash^{0}   \openO{(} \possiblyWithSub\stageOmetaColor{M^{\scriptscriptstyle(0)} }_{{\mathrm{1}}} \  \possiblyWithSub\stageOmetaColor{M^{\scriptscriptstyle(0)} }_{{\mathrm{2}}} \closeO{)}_{ \ell }   :    [   \possiblyWithSub\stageOmetaColor{N^{\superscriptO} }_{{\mathrm{0}}} \  \possiblyWithSub\stageOmetaColor{N^{\superscriptO} }_{{\mathrm{2}}}   /  \possiblyWithSub\stageOmetaColor{x}  ]    \possiblyWithSub\stageOmetaColor{T^{\superscriptO} }_{{\mathrm{12}}}   \ElabArrow   \possiblyWithSub\stageOmetaColor{N^{\superscriptO} }_{{\mathrm{1}}} \   \openO{(}  \possiblyWithSub\stageOmetaColor{N^{\superscriptO} }_{{\mathrm{0}}} \  \possiblyWithSub\stageOmetaColor{N^{\superscriptO} }_{{\mathrm{2}}}  \closeO{)}   
    }
  \qquad
    \derive[S0-Var]{%
      \mathit{\Gamma}(\possiblyWithSub\stageOmetaColor{x}) = (\possiblyWithSub\stageOmetaColor{T^{\superscriptO} })^{0}
    }{%
       \mathit{\Gamma}  \vdash^{0}   \possiblyWithSub\stageOmetaColor{x}   :  \possiblyWithSub\stageOmetaColor{T^{\superscriptO} }  \ElabArrow   \possiblyWithSub\stageOmetaColor{x}  
    }
  \\[0.7em]
    \derive[S1-Esc]{%
       \mathit{\Gamma}  \vdash^{0}  \possiblyWithSub\stageOmetaColor{M^{\scriptscriptstyle(0)} }  :   \openO{\langle} \possiblyWithSub\stageImetaColor{T^{\superscriptI} } \closeO{\rangle}   \ElabArrow  \possiblyWithSub\stageOmetaColor{N^{\superscriptO} } 
    }{%
       \mathit{\Gamma}  \vdash^{1}   \ordI{\sim} \possiblyWithSub\stageOmetaColor{M^{\scriptscriptstyle(0)} }   :  \possiblyWithSub\stageImetaColor{T^{\superscriptI} }  \ElabArrow   \ordI{\sim} \possiblyWithSub\stageOmetaColor{N^{\superscriptO} }  
    }
  \qquad
    \derive[S1-CstP]{%
      \ConstEnvPers(c) = \possiblyWithSub\stageImetaColor{\tau^{\superscriptI} }
    }{%
       \mathit{\Gamma}  \vdash^{1}   \possiblyWithSub\stageImetaColor{c}   :   \possiblyWithSub\stageImetaColor{\tau^{\superscriptI} }   \ElabArrow   \possiblyWithSub\stageImetaColor{c}  
    }
  \qquad
    \derive[S1-Var]{%
      \mathit{\Gamma}(\possiblyWithSub\stageImetaColor{x}) = (\possiblyWithSub\stageImetaColor{T^{\superscriptI} })^{1}
    }{%
       \mathit{\Gamma}  \vdash^{1}   \possiblyWithSub\stageImetaColor{x}   :  \possiblyWithSub\stageImetaColor{T^{\superscriptI} }  \ElabArrow   \possiblyWithSub\stageImetaColor{x}  
    }
  \\[0.7em]
    \derive[S1-Abs]{%
       \mathit{\Gamma}  \vdash^{1}  \possiblyWithSub\stageImetaColor{S^{\superscriptI} }_{{\mathrm{1}}}  \ElabArrow  \possiblyWithSub\stageImetaColor{T^{\superscriptI} }_{{\mathrm{1}}} 
    \andalso
        \mathit{\Gamma} ,  \possiblyWithSub\stageImetaColor{x}  : ( \possiblyWithSub\stageImetaColor{T^{\superscriptI} }_{{\mathrm{1}}} )^{1}   \vdash^{1}  \possiblyWithSub\stageImetaColor{M^{\superscriptI} }_{{\mathrm{2}}}  :  \possiblyWithSub\stageImetaColor{T^{\superscriptI} }_{{\mathrm{2}}}  \ElabArrow  \possiblyWithSub\stageImetaColor{N^{\superscriptI} }_{{\mathrm{2}}} 
    \andalso
      \possiblyWithSub\stageImetaColor{x} \not\in \fv(\possiblyWithSub\stageImetaColor{T^{\superscriptI} }_{{\mathrm{2}}})
    }{%
       \mathit{\Gamma}  \vdash^{1}   \openI{(}  \ordI{\lambda} \possiblyWithSub\stageImetaColor{x}  \relI{:}  \possiblyWithSub\stageImetaColor{S^{\superscriptI} }_{{\mathrm{1}}} \punctI{.}\  \possiblyWithSub\stageImetaColor{M^{\superscriptI} }_{{\mathrm{2}}}  \closeI{)}   :   \possiblyWithSub\stageImetaColor{T^{\superscriptI} }_{{\mathrm{1}}}  \relI{\to}  \possiblyWithSub\stageImetaColor{T^{\superscriptI} }_{{\mathrm{2}}}   \ElabArrow   \openI{(}  \ordI{\lambda} \possiblyWithSub\stageImetaColor{x}  \relI{:}  \possiblyWithSub\stageImetaColor{T^{\superscriptI} }_{{\mathrm{1}}} \punctI{.}\  \possiblyWithSub\stageImetaColor{N^{\superscriptI} }_{{\mathrm{2}}}  \closeI{)}  
    }
  \\[0.7em]
    \derive[S1-App]{%
       \mathit{\Gamma}  \vdash^{1}  \possiblyWithSub\stageImetaColor{M^{\superscriptI} }_{{\mathrm{1}}}  :   \possiblyWithSub\stageImetaColor{T^{\superscriptI} }_{{\mathrm{11}}}  \relI{\to}  \possiblyWithSub\stageImetaColor{T^{\superscriptI} }_{{\mathrm{12}}}   \ElabArrow  \possiblyWithSub\stageImetaColor{N^{\superscriptI} }_{{\mathrm{1}}} 
    \andalso
       \mathit{\Gamma}  \vdash^{1}  \possiblyWithSub\stageImetaColor{M^{\superscriptI} }_{{\mathrm{2}}}  :  \possiblyWithSub\stageImetaColor{T^{\superscriptI} }_{{\mathrm{2}}}  \ElabArrow  \possiblyWithSub\stageImetaColor{N^{\superscriptI} }_{{\mathrm{2}}} 
    \andalso
       \possiblyWithSub\stageImetaColor{T^{\superscriptI} }_{{\mathrm{2}}}  \mathrel{||}^{1}  \possiblyWithSub\stageImetaColor{T^{\superscriptI} }_{{\mathrm{11}}} 
    }{%
       \mathit{\Gamma}  \vdash^{1}   \openI{(} \possiblyWithSub\stageImetaColor{M^{\superscriptI} }_{{\mathrm{1}}} \  \possiblyWithSub\stageImetaColor{M^{\superscriptI} }_{{\mathrm{2}}} \closeI{)}_{ \ell }   :  \possiblyWithSub\stageImetaColor{T^{\superscriptI} }_{{\mathrm{12}}}  \ElabArrow   \possiblyWithSub\stageImetaColor{N^{\superscriptI} }_{{\mathrm{1}}} \   \ordI{\sim}  \openO{(}   \LeftAssertParen\openO{\langle} \possiblyWithSub\stageImetaColor{T^{\superscriptI} }_{{\mathrm{2}}} \closeO{\rangle} \relO{\CastArrow} \openO{\langle} \possiblyWithSub\stageImetaColor{T^{\superscriptI} }_{{\mathrm{11}}} \closeO{\rangle}\RightAssertParen^{  \ell  }  \   \openO{\langle} \possiblyWithSub\stageImetaColor{N^{\superscriptI} }_{{\mathrm{2}}} \closeO{\rangle}   \closeO{)}    
    }
  \end{center}
  \vspace{-1em}
  \begin{flushleft}
    \fbox{\(\mathit{\Gamma} \vdash^{b} S^{(b)} \ElabArrow T^{(b)}\)}
  \end{flushleft}
  \vspace{-4em}
  \begin{center}
  \hspace{8em}%
    \derive[ST0-Base]{%
        \mathit{\Gamma} ,  \possiblyWithSub\stageOmetaColor{\nu}  : (   \openO{\{} \possiblyWithSub\stageOmetaColor{\nu}_{{\mathrm{1}}}  \relO{:}  \possiblyWithSub\stageOmetaColor{B}  \relO{\mid}     \ttO{true}    \closeO{\} }   )^{0}   \vdash^{0}  \possiblyWithSub\stageOmetaColor{M^{\scriptscriptstyle(0)} }  :    \openO{\{} \possiblyWithSub\stageOmetaColor{\nu}_{{\mathrm{2}}}  \relO{:}   \ttO{Bool}   \relO{\mid}  \possiblyWithSub\stageOmetaColor{N'^{\superscriptO} } \closeO{\} }    \ElabArrow  \possiblyWithSub\stageOmetaColor{N^{\superscriptO} } 
    }{%
       \mathit{\Gamma}  \vdash^{0}    \openO{\{} \possiblyWithSub\stageOmetaColor{\nu}  \relO{:}  \possiblyWithSub\stageOmetaColor{B}  \relO{\mid}  \possiblyWithSub\stageOmetaColor{M^{\scriptscriptstyle(0)} } \closeO{\} }    \ElabArrow    \openO{\{} \possiblyWithSub\stageOmetaColor{\nu}  \relO{:}  \possiblyWithSub\stageOmetaColor{B}  \relO{\mid}  \possiblyWithSub\stageOmetaColor{N^{\superscriptO} } \closeO{\} }   
    }
  \\[0.7em]
    \derive[ST0-Tensor]{}{%
       \mathit{\Gamma}  \vdash^{0}   \ttO{Tensor}\  \possiblyWithSub\stageOmetaColor{s}   \ElabArrow   \ttO{Tensor}\  \possiblyWithSub\stageOmetaColor{s}  
    }
  \qquad
    \derive[ST1-Base]{}{%
       \mathit{\Gamma}  \vdash^{1}   \possiblyWithSub\stageImetaColor{B}   \ElabArrow   \possiblyWithSub\stageImetaColor{B}  
    }
  \\[0.7em]
    \derive[ST0-Arr]{%
       \mathit{\Gamma}  \vdash^{0}  \possiblyWithSub\stageOmetaColor{S^{\superscriptO} }_{{\mathrm{1}}}  \ElabArrow  \possiblyWithSub\stageOmetaColor{T^{\superscriptO} }_{{\mathrm{1}}} 
    \andalso
        \mathit{\Gamma} ,  \possiblyWithSub\stageOmetaColor{x}  : ( \possiblyWithSub\stageOmetaColor{T^{\superscriptO} }_{{\mathrm{1}}} )^{0}   \vdash^{0}  \possiblyWithSub\stageOmetaColor{S^{\superscriptO} }_{{\mathrm{2}}}  \ElabArrow  \possiblyWithSub\stageOmetaColor{T^{\superscriptO} }_{{\mathrm{2}}} 
    }{%
       \mathit{\Gamma}  \vdash^{0}   \openO{(} \possiblyWithSub\stageOmetaColor{x}  \relO{:}  \possiblyWithSub\stageOmetaColor{S^{\superscriptO} }_{{\mathrm{1}}} \closeO{)} \relO{\to}  \possiblyWithSub\stageOmetaColor{S^{\superscriptO} }_{{\mathrm{2}}}   \ElabArrow   \openO{(} \possiblyWithSub\stageOmetaColor{x}  \relO{:}  \possiblyWithSub\stageOmetaColor{T^{\superscriptO} }_{{\mathrm{1}}} \closeO{)} \relO{\to}  \possiblyWithSub\stageOmetaColor{T^{\superscriptO} }_{{\mathrm{2}}}  
    }
  \qquad
    \derive[ST0-Code]{%
       \mathit{\Gamma}  \vdash^{1}  \possiblyWithSub\stageImetaColor{S^{\superscriptI} }  \ElabArrow  \possiblyWithSub\stageImetaColor{T^{\superscriptI} } 
    }{%
       \mathit{\Gamma}  \vdash^{0}   \openO{\langle} \possiblyWithSub\stageImetaColor{S^{\superscriptI} } \closeO{\rangle}   \ElabArrow   \openO{\langle} \possiblyWithSub\stageImetaColor{T^{\superscriptI} } \closeO{\rangle}  
    }
  \\[0.7em]
    \derive[ST1-Arr]{%
       \mathit{\Gamma}  \vdash^{1}  \possiblyWithSub\stageImetaColor{S^{\superscriptI} }_{{\mathrm{1}}}  \ElabArrow  \possiblyWithSub\stageImetaColor{T^{\superscriptI} }_{{\mathrm{1}}} 
      \quad\text{(for \(i \in \{1, 2\}\))}
    }{%
       \mathit{\Gamma}  \vdash^{1}   \possiblyWithSub\stageImetaColor{S^{\superscriptI} }_{{\mathrm{1}}}  \relI{\to}  \possiblyWithSub\stageImetaColor{S^{\superscriptI} }_{{\mathrm{2}}}   \ElabArrow   \possiblyWithSub\stageImetaColor{T^{\superscriptI} }_{{\mathrm{1}}}  \relI{\to}  \possiblyWithSub\stageImetaColor{T^{\superscriptI} }_{{\mathrm{2}}}  
    }
  \quad
    \derive[ST1-Tensor]{%
       \mathit{\Gamma}  \vdash^{0}  \possiblyWithSub\stageOmetaColor{M^{\scriptscriptstyle(0)} }  :    \openO{\{} \possiblyWithSub\stageOmetaColor{\nu}  \relO{:}   \ttO{NatList}   \relO{\mid}  \possiblyWithSub\stageOmetaColor{N'^{\superscriptO} } \closeO{\} }    \ElabArrow  \possiblyWithSub\stageOmetaColor{N^{\superscriptO} } 
    }{%
       \mathit{\Gamma}  \vdash^{1}   \ttI{Tensor}\ \ordI{\%} \possiblyWithSub\stageOmetaColor{M^{\scriptscriptstyle(0)} }   \ElabArrow   \ttI{Tensor}\ \ordI{\%} \possiblyWithSub\stageOmetaColor{N^{\superscriptO} }  
    }
  \end{center}
  \vspace{-0.75em}%
  \caption{Source typing rules with assertion insertion}
  \label{fig:assertion-insertion}
\end{figure}
\indent
  Typing judgments are defined as 
  \(\mathit{\Gamma} \vdash^{b} M^{(b)} : T^{(b)} \ElabArrow N^{(b)}\) (for \(b \in \{0, 1\}\)),
  which can be read as ``under the type environment~\(\mathit{\Gamma}\),
  the source expression~\(M^{(b)}\) has type~\(T^{(b)}\)
  and elaborates to \(N^{(b)}\) by assertion insertion.''
  The structure of type environments are defined by:
  \(\bnfnotab{\mathit{\Gamma}}{  \bullet  |  \mathit{\Gamma} ,  \possiblyWithSub\stageOmetaColor{x}  : ( \possiblyWithSub\stageOmetaColor{T^{\superscriptO} } )^{0}  |  \mathit{\Gamma} ,  \possiblyWithSub\stageImetaColor{x}  : ( \possiblyWithSub\stageImetaColor{T^{\superscriptI} } )^{1}  }\),
  i.e, \(\mathit{\Gamma}\) tracks the stage at which each variable was bound.
  Figure~\ref{fig:assertion-insertion} shows the typing rules for these judgments.
  The rules \rulename{S0-Brkt} and \rulename{S1-Esc} are peculiar to staged computation
  and are natural extensions from the literature.
  \rulename{S0-Abs} and \rulename{S1-Abs} elaborate the type annotations
  by using the judgments \(\mathit{\Gamma} \vdash^{b} S^{(b)} \ElabArrow T^{(b)}\).
  Also, for constants, we have \rulename{S0-Cst0}, \rulename{S0-CstP}, and \rulename{S1-CstP}.
  Here, we use two environments~\(\ConstEnvZero\) and \(\ConstEnvPers\);
  the former maps stage-\(0\)-specific built-in functions~\(\possiblyWithSub\stageOmetaColor{p}\)
  to stage-\(0\) types~\(\possiblyWithSub\stageOmetaColor{T^{\superscriptO} }\),
  and the latter works similarly for stage-agnostic constants~\(c\).
  The metavariable~\(\possiblyWithSub\stageImetaColor{\tau^{\superscriptI} }\) and its \dfn{unlifting}~\( \mathop{\downarrow}( \possiblyWithSub\stageImetaColor{\tau^{\superscriptI} } )  = \possiblyWithSub\stageOmetaColor{T^{\superscriptO} }\)
  will be introduced shortly.
  Entries are like the following:
  \begin{gather*}
    \ConstEnvZero( \ordO{  \makeIdentOrConst{}{ VgenMatMult }  } ) =
      \begin{aligned}[t]
         \progindent{  \openO{(}  \ordO{  \makeIdentOrConst{}{ Vp }  }   \relO{:}    \ttO{Nat}   \closeO{)} \relO{\to}   \openO{(}  \ordO{  \makeIdentOrConst{}{ Vq }  }   \relO{:}    \ttO{Nat}   \closeO{)} \relO{\to}   \openO{(}  \ordO{  \makeIdentOrConst{}{ Vr }  }   \relO{:}    \ttO{Nat}   \closeO{)} \relO{\to} \control\deepen{\control\br{}  \openO{\langle}    \ttI{Mat}\ \ordI{\%}   \ordO{  \makeIdentOrConst{}{ Vp }  }   \ \ordI{\%}   \ordO{  \makeIdentOrConst{}{ Vq }  }     \relI{\to}   \ttI{Mat}\ \ordI{\%}   \ordO{  \makeIdentOrConst{}{ Vq }  }   \ \ordI{\%}   \ordO{  \makeIdentOrConst{}{ Vr }  }      \relI{\to}   \ttI{Mat}\ \ordI{\%}   \ordO{  \makeIdentOrConst{}{ Vp }  }   \ \ordI{\%}   \ordO{  \makeIdentOrConst{}{ Vr }  }     \closeO{\rangle}  }    } ,
      \end{aligned}
  \\
    \ConstEnvPers(   \makeIdentOrConst{}{ VmatMult }  _{   \makeIdentOrConst{}{ C3 }  ,    \makeIdentOrConst{}{ C4 }  ,    \makeIdentOrConst{}{ C5 }     }  ) =
         \ttI{Mat}\ \ordI{\%}   \ordO{  \makeIdentOrConst{}{ C3 }  }   \ \ordI{\%}   \ordO{  \makeIdentOrConst{}{ C4 }  }     \relI{\to}   \ttI{Mat}\ \ordI{\%}   \ordO{  \makeIdentOrConst{}{ C4 }  }   \ \ordI{\%}   \ordO{  \makeIdentOrConst{}{ C5 }  }      \relI{\to}   \ttI{Mat}\ \ordI{\%}   \ordO{  \makeIdentOrConst{}{ C3 }  }   \ \ordI{\%}   \ordO{  \makeIdentOrConst{}{ C5 }  }    ,
  \qquad
    \ConstEnvPers(   \makeIdentOrConst{}{ C42 }   ) =   \ttI{Int}  .
  \end{gather*}
\par
\begin{figure}[tb]
\small
  \begin{flushleft}
    \fbox{\( \possiblyWithSub\stageImetaColor{T^{\superscriptI} }_{{\mathrm{1}}}  \mathrel{||}^{1}  \possiblyWithSub\stageImetaColor{T^{\superscriptI} }_{{\mathrm{2}}} \)}
    \fbox{\( \mathit{\Gamma}  \vdash_{ L }  \possiblyWithSub\stageOmetaColor{T^{\superscriptO} }_{{\mathrm{1}}}  \CastArrow  \possiblyWithSub\stageOmetaColor{T^{\superscriptO} }_{{\mathrm{2}}}  \ElabArrow  \possiblyWithSub\stageOmetaColor{N^{\superscriptO} } \)}
  \end{flushleft}
  \vspace{-4.5em}%
  \begin{center}
  \hspace{17em}%
    \derive[I-Tensor]{}{%
       \mathit{\Gamma}  \vdash_{ L }   \ttO{Tensor}\  \possiblyWithSub\stageOmetaColor{s}   \CastArrow   \ttO{Tensor}\  \possiblyWithSub\stageOmetaColor{s}   \ElabArrow   \ordO{\lambda} \possiblyWithSub\stageOmetaColor{x}  \relO{:}   \ttO{Tensor}\  \possiblyWithSub\stageOmetaColor{s}  \punctO{.}\   \possiblyWithSub\stageOmetaColor{x}   
    }
  \\[0.7em]
    \derive{}{%
        \ttI{Tensor}\ \ordI{\%} \possiblyWithSub\stageOmetaColor{N^{\superscriptO} }_{{\mathrm{1}}}   \mathrel{||}^{1}   \ttI{Tensor}\ \ordI{\%} \possiblyWithSub\stageOmetaColor{N^{\superscriptO} }_{{\mathrm{2}}}  
    }
  \quad\!%
    \derive[I-Rfn]{}{%
       \mathit{\Gamma}  \vdash_{ L }    \openO{\{} \possiblyWithSub\stageOmetaColor{\nu}  \relO{:}  \possiblyWithSub\stageOmetaColor{B}  \relO{\mid}  \possiblyWithSub\stageOmetaColor{N^{\superscriptO} }_{{\mathrm{1}}} \closeO{\} }    \CastArrow    \openO{\{} \possiblyWithSub\stageOmetaColor{\nu}  \relO{:}  \possiblyWithSub\stageOmetaColor{B}  \relO{\mid}  \possiblyWithSub\stageOmetaColor{N^{\superscriptO} }_{{\mathrm{2}}} \closeO{\} }    \ElabArrow   \LeftAssertParen \relO{\CastArrow}   \openO{\{} \possiblyWithSub\stageOmetaColor{\nu}  \relO{:}  \possiblyWithSub\stageOmetaColor{B}  \relO{\mid}  \possiblyWithSub\stageOmetaColor{N^{\superscriptO} }_{{\mathrm{2}}} \closeO{\} }   \RightAssertParen^{ L }  
    }
  \\[0.7em]
    \derive{}{%
        \possiblyWithSub\stageImetaColor{B}   \mathrel{||}^{1}   \possiblyWithSub\stageImetaColor{B}  
    }
  \qquad
    \derive{%
       \possiblyWithSub\stageImetaColor{T^{\superscriptI} }_{{\mathrm{11}}}  \mathrel{||}^{1}  \possiblyWithSub\stageImetaColor{T^{\superscriptI} }_{{\mathrm{21}}} 
    \andalso
       \possiblyWithSub\stageImetaColor{T^{\superscriptI} }_{{\mathrm{12}}}  \mathrel{||}^{1}  \possiblyWithSub\stageImetaColor{T^{\superscriptI} }_{{\mathrm{22}}} 
    }{%
        \possiblyWithSub\stageImetaColor{T^{\superscriptI} }_{{\mathrm{11}}}  \relI{\to}  \possiblyWithSub\stageImetaColor{T^{\superscriptI} }_{{\mathrm{12}}}   \mathrel{||}^{1}   \possiblyWithSub\stageImetaColor{T^{\superscriptI} }_{{\mathrm{21}}}  \relI{\to}  \possiblyWithSub\stageImetaColor{T^{\superscriptI} }_{{\mathrm{22}}}  
    }
  \qquad
    \derive[I-Code]{%
       \possiblyWithSub\stageImetaColor{T^{\superscriptI} }_{{\mathrm{1}}}  \mathrel{||}^{1}  \possiblyWithSub\stageImetaColor{T^{\superscriptI} }_{{\mathrm{2}}} 
    }{%
       \mathit{\Gamma}  \vdash_{ L }   \openO{\langle} \possiblyWithSub\stageImetaColor{T^{\superscriptI} }_{{\mathrm{1}}} \closeO{\rangle}   \CastArrow   \openO{\langle} \possiblyWithSub\stageImetaColor{T^{\superscriptI} }_{{\mathrm{2}}} \closeO{\rangle}   \ElabArrow   \LeftAssertParen\openO{\langle} \possiblyWithSub\stageImetaColor{T^{\superscriptI} }_{{\mathrm{1}}} \closeO{\rangle} \relO{\CastArrow} \openO{\langle} \possiblyWithSub\stageImetaColor{T^{\superscriptI} }_{{\mathrm{2}}} \closeO{\rangle}\RightAssertParen^{ L }  
    }
  \\[0.7em]
    \derive[I-Arr]{%
       \mathit{\Gamma}  \vdash_{  L .\mathbf{dom}  }  \possiblyWithSub\stageOmetaColor{T^{\superscriptO} }_{{\mathrm{21}}}  \CastArrow  \possiblyWithSub\stageOmetaColor{T^{\superscriptO} }_{{\mathrm{11}}}  \ElabArrow  \possiblyWithSub\stageOmetaColor{N^{\superscriptO} }_{{\mathrm{1}}} 
    \andalso
      \possiblyWithSub\stageOmetaColor{f} \not\in \dom(\mathit{\Gamma}) \uplus \{\possiblyWithSub\stageOmetaColor{x}\}
    \\
      \possiblyWithSub\stageOmetaColor{x'} \not\in \dom(\mathit{\Gamma}) \uplus \{\possiblyWithSub\stageOmetaColor{x}, \possiblyWithSub\stageOmetaColor{f}\}
    \andalso
        \mathit{\Gamma} ,  \possiblyWithSub\stageOmetaColor{x'}  : ( \possiblyWithSub\stageOmetaColor{T^{\superscriptO} }_{{\mathrm{11}}} )^{0}   \vdash_{  L .\mathbf{cod}  }    [   \possiblyWithSub\stageOmetaColor{x'}   /  \possiblyWithSub\stageOmetaColor{x}  ]    \possiblyWithSub\stageOmetaColor{T^{\superscriptO} }_{{\mathrm{12}}}   \CastArrow  \possiblyWithSub\stageOmetaColor{T^{\superscriptO} }_{{\mathrm{22}}}  \ElabArrow  \possiblyWithSub\stageOmetaColor{N^{\superscriptO} }_{{\mathrm{2}}} 
    }{%
      \begin{aligned}
        &
        \mathit{\Gamma} \vdash_{L}  \openO{(} \possiblyWithSub\stageOmetaColor{x}  \relO{:}  \possiblyWithSub\stageOmetaColor{T^{\superscriptO} }_{{\mathrm{11}}} \closeO{)} \relO{\to}  \possiblyWithSub\stageOmetaColor{T^{\superscriptO} }_{{\mathrm{12}}}  \CastArrow  \openO{(} \possiblyWithSub\stageOmetaColor{x}  \relO{:}  \possiblyWithSub\stageOmetaColor{T^{\superscriptO} }_{{\mathrm{21}}} \closeO{)} \relO{\to}  \possiblyWithSub\stageOmetaColor{T^{\superscriptO} }_{{\mathrm{22}}} 
        \ElabArrow
      \\[-0.25em]
        &\qquad\qquad
         \ordO{\lambda} \possiblyWithSub\stageOmetaColor{f}  \relO{:}   \openO{(} \possiblyWithSub\stageOmetaColor{x}  \relO{:}  \possiblyWithSub\stageOmetaColor{T^{\superscriptO} }_{{\mathrm{11}}} \closeO{)} \relO{\to}  \possiblyWithSub\stageOmetaColor{T^{\superscriptO} }_{{\mathrm{12}}}  \punctO{.}\    \ordO{\lambda} \possiblyWithSub\stageOmetaColor{x}  \relO{:}  \possiblyWithSub\stageOmetaColor{T^{\superscriptO} }_{{\mathrm{21}}} \punctO{.}\    \tokenO{let}\  \possiblyWithSub\stageOmetaColor{x'}  \relO{:}  \possiblyWithSub\stageOmetaColor{T^{\superscriptO} }_{{\mathrm{11}}}  \relO{=}   \possiblyWithSub\stageOmetaColor{N^{\superscriptO} }_{{\mathrm{1}}} \   \possiblyWithSub\stageOmetaColor{x}    \ \tokenO{in}\  \possiblyWithSub\stageOmetaColor{N^{\superscriptO} }_{{\mathrm{2}}}   \   \openO{(}   \possiblyWithSub\stageOmetaColor{f}  \   \possiblyWithSub\stageOmetaColor{x'}   \closeO{)}   
      \end{aligned}
    }
  \end{center}
  \vspace{-0.75em}%
  \caption{The rules for type compatibility and cast term generation}
  \label{fig:compatibility-and-cast-insertion}
\end{figure}
\indent
  The most distinctive typing rules are those for applications,
  i.e., \rulename{S0-App} and \rulename{S1-App}.
  Here, \( \possiblyWithSub\stageImetaColor{T^{\superscriptI} }_{{\mathrm{1}}}  \mathrel{||}^{1}  \possiblyWithSub\stageImetaColor{T^{\superscriptI} }_{{\mathrm{2}}} \) judges \dfn{type compatibility}, i.e.,
  type equivalence ignoring the difference of argument expressions,
  and \( \mathit{\Gamma}  \vdash_{ L }  \possiblyWithSub\stageOmetaColor{T^{\superscriptO} }_{{\mathrm{1}}}  \CastArrow  \possiblyWithSub\stageOmetaColor{T^{\superscriptO} }_{{\mathrm{2}}}  \ElabArrow  \possiblyWithSub\stageOmetaColor{N^{\superscriptO} }_{{\mathrm{0}}} \) generates \dfn{assertive cast terms} \(\possiblyWithSub\stageOmetaColor{N^{\superscriptO} }_{{\mathrm{0}}}\).
  Figure~\ref{fig:compatibility-and-cast-insertion} displays the rules for these judgments.
  Among these rules, \rulename{I-Code} is the core of the cast term generation;
  it checks the compatibility of given two types
  and simply produces an ``atomic'' cast term \( \LeftAssertParen\openO{\langle} \possiblyWithSub\stageImetaColor{T^{\superscriptI} }_{{\mathrm{1}}} \closeO{\rangle} \relO{\CastArrow} \openO{\langle} \possiblyWithSub\stageImetaColor{T^{\superscriptI} }_{{\mathrm{2}}} \closeO{\rangle}\RightAssertParen^{ L } \).
  \rulename{I-Arr}, which works for functions, is another characteristic rule.
  It produces two cast terms for domain types and codomain types, respectively,
  and combines them.
  Thanks to this rule, we can handle higher-order programs without any hindrance.
  This mechanism is partially inspired by
  GraTen~\cite{HattoriKobayashiSatoESOP2023}
  and \(\mathrm{F}_{\mathrm{H}}^{\sigma}\)~\cite{SekiyamaIgarashiGreenbergTOPLAS2017}.
\par
\begin{figure}[tb]
\small
  \begin{flushleft}
    \fbox{\(\possiblyWithSub\stageOmetaColor{N^{\superscriptO} } \longrightarrow^{0} (\possiblyWithSub\stageOmetaColor{N'^{\superscriptO} } \mid  \BlameSign^{ L } )\)}
  \end{flushleft}
  \vspace{-4.25em}
  \begin{center}
    \hspace{11em}%
    \derive[E0-App1]{%
       \possiblyWithSub\stageOmetaColor{N^{\superscriptO} }_{{\mathrm{1}}}  \longrightarrow^{0}   \possiblyWithSub\stageOmetaColor{N'^{\superscriptO} }_{{\mathrm{1}}}  
    }{%
        \possiblyWithSub\stageOmetaColor{N^{\superscriptO} }_{{\mathrm{1}}} \  \possiblyWithSub\stageOmetaColor{N^{\superscriptO} }_{{\mathrm{2}}}   \longrightarrow^{0}    \possiblyWithSub\stageOmetaColor{N'^{\superscriptO} }_{{\mathrm{1}}} \  \possiblyWithSub\stageOmetaColor{N^{\superscriptO} }_{{\mathrm{2}}}   
    }
  \qquad
    \derive[E0-App1F]{%
       \possiblyWithSub\stageOmetaColor{N^{\superscriptO} }_{{\mathrm{1}}}  \longrightarrow^{0}   \BlameSign^{ L }  
    }{%
        \possiblyWithSub\stageOmetaColor{N^{\superscriptO} }_{{\mathrm{1}}} \  \possiblyWithSub\stageOmetaColor{N^{\superscriptO} }_{{\mathrm{2}}}   \longrightarrow^{0}   \BlameSign^{ L }  
    }
  \\[0.7em]
    \derive[E0-Ass1]{%
       \possiblyWithSub\stageImetaColor{T^{\superscriptI} }_{{\mathrm{1}}}  \longrightarrow^{1}   \possiblyWithSub\stageImetaColor{T'^{\superscriptI} }_{{\mathrm{1}}}  
    }{%
        \LeftAssertParen\openO{\langle} \possiblyWithSub\stageImetaColor{T^{\superscriptI} }_{{\mathrm{1}}} \closeO{\rangle} \relO{\CastArrow} \openO{\langle} \possiblyWithSub\stageImetaColor{T^{\superscriptI} }_{{\mathrm{2}}} \closeO{\rangle}\RightAssertParen^{ L }   \longrightarrow^{0}    \LeftAssertParen\openO{\langle} \possiblyWithSub\stageImetaColor{T'^{\superscriptI} }_{{\mathrm{1}}} \closeO{\rangle} \relO{\CastArrow} \openO{\langle} \possiblyWithSub\stageImetaColor{T^{\superscriptI} }_{{\mathrm{2}}} \closeO{\rangle}\RightAssertParen^{ L }   
    }
  \qquad
    \derive[E0-AssFail]{%
      \possiblyWithSub\stageImetaColor{\tau^{\superscriptI} }_{{\mathrm{1}}} \neq \possiblyWithSub\stageImetaColor{\tau^{\superscriptI} }_{{\mathrm{2}}}
    }{%
        \LeftAssertParen\openO{\langle}  \possiblyWithSub\stageImetaColor{\tau^{\superscriptI} }_{{\mathrm{1}}}  \closeO{\rangle} \relO{\CastArrow} \openO{\langle}  \possiblyWithSub\stageImetaColor{\tau^{\superscriptI} }_{{\mathrm{2}}}  \closeO{\rangle}\RightAssertParen^{ L }   \longrightarrow^{0}   \BlameSign^{ L }  
    }
  \\[0.7em]
    \derive[E0-AssPass]{}{%
        \LeftAssertParen\openO{\langle}  \possiblyWithSub\stageImetaColor{\tau^{\superscriptI} }  \closeO{\rangle} \relO{\CastArrow} \openO{\langle}  \possiblyWithSub\stageImetaColor{\tau^{\superscriptI} }  \closeO{\rangle}\RightAssertParen^{ L }   \longrightarrow^{0}    \ordO{\lambda} \possiblyWithSub\stageOmetaColor{x}  \relO{:}   \openO{\langle}  \possiblyWithSub\stageImetaColor{\tau^{\superscriptI} }  \closeO{\rangle}  \punctO{.}\   \possiblyWithSub\stageOmetaColor{x}    
    }
  \qquad
    \derive[E0-Beta]{}{%
         \openO{(}  \ordO{\lambda} \possiblyWithSub\stageOmetaColor{x}  \relO{:}  \possiblyWithSub\stageOmetaColor{T^{\superscriptO} }_{{\mathrm{11}}} \punctO{.}\  \possiblyWithSub\stageOmetaColor{N^{\superscriptO} }_{{\mathrm{12}}}  \closeO{)}  \   \possiblyWithSub\stageOmetaColor{v^{\superscriptO} }_{{\mathrm{2}}}    \longrightarrow^{0}     [   \possiblyWithSub\stageOmetaColor{v^{\superscriptO} }_{{\mathrm{2}}}   /  \possiblyWithSub\stageOmetaColor{x}  ]    \possiblyWithSub\stageOmetaColor{N^{\superscriptO} }_{{\mathrm{12}}}   
    }
  \\[0.7em]
    \derive[E0-Delta]{%
      \delta(\possiblyWithSub\stageOmetaColor{a}_{{\mathrm{1}}}, \possiblyWithSub\stageOmetaColor{c}_{{\mathrm{2}}}) = \possiblyWithSub\stageOmetaColor{q}
    }{%
         \possiblyWithSub\stageOmetaColor{a}_{{\mathrm{1}}}  \    \possiblyWithSub\stageOmetaColor{c}_{{\mathrm{2}}}     \longrightarrow^{0}    \possiblyWithSub\stageOmetaColor{q}   
    }
  \qquad
    \derive[E0-RfnStart]{}{%
         \LeftAssertParen \relO{\CastArrow}   \openO{\{} \possiblyWithSub\stageOmetaColor{\nu}  \relO{:}  \possiblyWithSub\stageOmetaColor{B}  \relO{\mid}  \possiblyWithSub\stageOmetaColor{N^{\superscriptO} }_{{\mathrm{1}}} \closeO{\} }   \RightAssertParen^{ L }  \    \possiblyWithSub\stageOmetaColor{c}_{{\mathrm{2}}}     \longrightarrow^{0}    \LeftAssertParen   \openO{\{} \possiblyWithSub\stageOmetaColor{\nu}  \relO{:}  \possiblyWithSub\stageOmetaColor{B}  \relO{\mid}  \possiblyWithSub\stageOmetaColor{N^{\superscriptO} }_{{\mathrm{1}}} \closeO{\} }  \punctO{,}    [    \possiblyWithSub\stageOmetaColor{c}_{{\mathrm{2}}}    /  \possiblyWithSub\stageOmetaColor{\nu}  ]    \possiblyWithSub\stageOmetaColor{N^{\superscriptO} }_{{\mathrm{1}}}  \punctO{,}  \possiblyWithSub\stageOmetaColor{c}_{{\mathrm{2}}}  \RightAssertParen^{ L }   
    }
  \\[0.7em]
    \derive[E0-Brkt]{%
       \possiblyWithSub\stageImetaColor{N^{\superscriptI} }  \longrightarrow^{1}   \possiblyWithSub\stageImetaColor{N'^{\superscriptI} }  
    }{%
        \openO{\langle} \possiblyWithSub\stageImetaColor{N^{\superscriptI} } \closeO{\rangle}   \longrightarrow^{0}    \openO{\langle} \possiblyWithSub\stageImetaColor{N'^{\superscriptI} } \closeO{\rangle}   
    }
  \quad
    \derive[E0-RfnAct]{%
       \possiblyWithSub\stageOmetaColor{N^{\superscriptO} }  \longrightarrow^{0}   \possiblyWithSub\stageOmetaColor{N'^{\superscriptO} }  
    }{%
        \LeftAssertParen   \openO{\{} \possiblyWithSub\stageOmetaColor{\nu}  \relO{:}  \possiblyWithSub\stageOmetaColor{B}  \relO{\mid}  \possiblyWithSub\stageOmetaColor{N^{\superscriptO} }_{{\mathrm{1}}} \closeO{\} }  \punctO{,}  \possiblyWithSub\stageOmetaColor{N^{\superscriptO} } \punctO{,}  \possiblyWithSub\stageOmetaColor{c}_{{\mathrm{2}}}  \RightAssertParen^{ L }   \longrightarrow^{0}    \LeftAssertParen   \openO{\{} \possiblyWithSub\stageOmetaColor{\nu}  \relO{:}  \possiblyWithSub\stageOmetaColor{B}  \relO{\mid}  \possiblyWithSub\stageOmetaColor{N^{\superscriptO} }_{{\mathrm{1}}} \closeO{\} }  \punctO{,}  \possiblyWithSub\stageOmetaColor{N'^{\superscriptO} } \punctO{,}  \possiblyWithSub\stageOmetaColor{c}_{{\mathrm{2}}}  \RightAssertParen^{ L }   
    }
  \\[0.7em]
    \derive[E0-RfnPass]{}{%
        \LeftAssertParen   \openO{\{} \possiblyWithSub\stageOmetaColor{\nu}  \relO{:}  \possiblyWithSub\stageOmetaColor{B}  \relO{\mid}  \possiblyWithSub\stageOmetaColor{N^{\superscriptO} }_{{\mathrm{1}}} \closeO{\} }  \punctO{,}     \ttO{true}    \punctO{,}  \possiblyWithSub\stageOmetaColor{c}_{{\mathrm{2}}}  \RightAssertParen^{ L }   \longrightarrow^{0}     \possiblyWithSub\stageOmetaColor{c}_{{\mathrm{2}}}    
    }
  \qquad
    \derive[E0-RfnFail]{}{%
        \LeftAssertParen   \openO{\{} \possiblyWithSub\stageOmetaColor{\nu}  \relO{:}  \possiblyWithSub\stageOmetaColor{B}  \relO{\mid}  \possiblyWithSub\stageOmetaColor{N^{\superscriptO} }_{{\mathrm{1}}} \closeO{\} }  \punctO{,}     \ttO{false}    \punctO{,}  \possiblyWithSub\stageOmetaColor{c}_{{\mathrm{2}}}  \RightAssertParen^{ L }   \longrightarrow^{0}   \BlameSign^{ L }  
    }
  \end{center}
  \vspace{-1em}
  \begin{flushleft}
    \fbox{\(\possiblyWithSub\stageImetaColor{N^{\superscriptI} } \longrightarrow^{1} (\possiblyWithSub\stageImetaColor{N'^{\superscriptI} } \mid  \BlameSign^{ L } )\)}
  \end{flushleft}
  \vspace{-4.25em}
  \begin{center}
    \hspace{9em}%
    \derive[E1-App1]{%
       \possiblyWithSub\stageImetaColor{N^{\superscriptI} }_{{\mathrm{1}}}  \longrightarrow^{1}   \possiblyWithSub\stageImetaColor{N'^{\superscriptI} }_{{\mathrm{1}}}  
    }{%
        \possiblyWithSub\stageImetaColor{N^{\superscriptI} }_{{\mathrm{1}}} \  \possiblyWithSub\stageImetaColor{N^{\superscriptI} }_{{\mathrm{2}}}   \longrightarrow^{1}    \possiblyWithSub\stageImetaColor{N'^{\superscriptI} }_{{\mathrm{1}}} \  \possiblyWithSub\stageImetaColor{N^{\superscriptI} }_{{\mathrm{2}}}   
    }
  \qquad
    \derive[E1-App1F]{%
       \possiblyWithSub\stageImetaColor{N^{\superscriptI} }_{{\mathrm{1}}}  \longrightarrow^{1}   \BlameSign^{ L }  
    }{%
        \possiblyWithSub\stageImetaColor{N^{\superscriptI} }_{{\mathrm{1}}} \  \possiblyWithSub\stageImetaColor{N^{\superscriptI} }_{{\mathrm{2}}}   \longrightarrow^{1}   \BlameSign^{ L }  
    }
  \\[0.7em]
    \derive[E1-Abs1]{%
       \possiblyWithSub\stageImetaColor{T^{\superscriptI} }  \longrightarrow^{1}   \possiblyWithSub\stageImetaColor{T'^{\superscriptI} }  
    }{%
        \ordI{\lambda} \possiblyWithSub\stageImetaColor{x}  \relI{:}  \possiblyWithSub\stageImetaColor{T^{\superscriptI} } \punctI{.}\  \possiblyWithSub\stageImetaColor{N^{\superscriptI} }   \longrightarrow^{1}    \ordI{\lambda} \possiblyWithSub\stageImetaColor{x}  \relI{:}  \possiblyWithSub\stageImetaColor{T'^{\superscriptI} } \punctI{.}\  \possiblyWithSub\stageImetaColor{N^{\superscriptI} }   
    }
  \qquad
    \derive[E1-Abs2]{%
       \possiblyWithSub\stageImetaColor{N^{\superscriptI} }  \longrightarrow^{1}   \possiblyWithSub\stageImetaColor{N'^{\superscriptI} }  
    }{%
        \ordI{\lambda} \possiblyWithSub\stageImetaColor{x}  \relI{:}   \possiblyWithSub\stageImetaColor{\tau^{\superscriptI} }  \punctI{.}\  \possiblyWithSub\stageImetaColor{N^{\superscriptI} }   \longrightarrow^{1}    \ordI{\lambda} \possiblyWithSub\stageImetaColor{x}  \relI{:}   \possiblyWithSub\stageImetaColor{\tau^{\superscriptI} }  \punctI{.}\  \possiblyWithSub\stageImetaColor{N'^{\superscriptI} }   
    }
  \\[0.7em]
    \derive[E1-Esc]{%
       \possiblyWithSub\stageOmetaColor{N^{\superscriptO} }  \longrightarrow^{0}   \possiblyWithSub\stageOmetaColor{N'^{\superscriptO} }  
    }{%
        \ordI{\sim} \possiblyWithSub\stageOmetaColor{N^{\superscriptO} }   \longrightarrow^{1}    \ordI{\sim} \possiblyWithSub\stageOmetaColor{N'^{\superscriptO} }   
    }
  \qquad
    \derive[E1-EscF]{%
       \possiblyWithSub\stageOmetaColor{N^{\superscriptO} }  \longrightarrow^{0}   \BlameSign^{ L }  
    }{%
        \ordI{\sim} \possiblyWithSub\stageOmetaColor{N^{\superscriptO} }   \longrightarrow^{1}   \BlameSign^{ L }  
    }
  \qquad
    \derive[E1-Cancel]{}{%
        \ordI{\sim}  \openO{\langle}  \possiblyWithSub\stageImetaColor{v^{\superscriptI} }  \closeO{\rangle}    \longrightarrow^{1}    \possiblyWithSub\stageImetaColor{v^{\superscriptI} }   
    }
  \end{center}
  \vspace{-1em}
  \begin{flushleft}
    \fbox{\(\possiblyWithSub\stageImetaColor{T^{\superscriptI} } \longrightarrow^{1} (\possiblyWithSub\stageImetaColor{T'^{\superscriptI} } \mid  \BlameSign^{ L } )\)}
  \end{flushleft}
  \vspace{-4.25em}
  \begin{center}
    \hspace{10em}%
    \derive[ET1-TensorF]{%
       \possiblyWithSub\stageOmetaColor{N^{\superscriptO} }  \longrightarrow^{0}   \BlameSign^{ L }  
    }{%
        \ttI{Tensor}\ \ordI{\%} \possiblyWithSub\stageOmetaColor{N^{\superscriptO} }   \longrightarrow^{1}   \BlameSign^{ L }  
    }
  \quad
    \derive[ET1-Arr1F]{%
       \possiblyWithSub\stageImetaColor{T^{\superscriptI} }_{{\mathrm{1}}}  \longrightarrow^{1}   \BlameSign^{ L }  
    }{%
        \openI{(}  \possiblyWithSub\stageImetaColor{T^{\superscriptI} }_{{\mathrm{1}}}  \relI{\to}  \possiblyWithSub\stageImetaColor{T^{\superscriptI} }_{{\mathrm{2}}}  \closeI{)}   \longrightarrow^{1}   \BlameSign^{ L }  
    }
  \\[0.7em]
    \derive[ET1-Tensor]{%
       \possiblyWithSub\stageOmetaColor{N^{\superscriptO} }  \longrightarrow^{0}   \possiblyWithSub\stageOmetaColor{N'^{\superscriptO} }  
    }{%
        \ttI{Tensor}\ \ordI{\%} \possiblyWithSub\stageOmetaColor{N^{\superscriptO} }   \longrightarrow^{1}    \ttI{Tensor}\ \ordI{\%} \possiblyWithSub\stageOmetaColor{N'^{\superscriptO} }   
    }
  \qquad
    \derive[ET1-Arr1]{%
       \possiblyWithSub\stageImetaColor{T^{\superscriptI} }_{{\mathrm{1}}}  \longrightarrow^{1}   \possiblyWithSub\stageImetaColor{T'^{\superscriptI} }_{{\mathrm{1}}}  
    }{%
        \openI{(}  \possiblyWithSub\stageImetaColor{T^{\superscriptI} }_{{\mathrm{1}}}  \relI{\to}  \possiblyWithSub\stageImetaColor{T^{\superscriptI} }_{{\mathrm{2}}}  \closeI{)}   \longrightarrow^{1}    \openI{(}  \possiblyWithSub\stageImetaColor{T'^{\superscriptI} }_{{\mathrm{1}}}  \relI{\to}  \possiblyWithSub\stageImetaColor{T^{\superscriptI} }_{{\mathrm{2}}}  \closeI{)}   
    }
  \end{center}
  \vspace{-0.75em}%
  \caption{%
    Reduction relations
    (selective; see Figures~\ref{fig:term-reduction-full}
    and \ref{fig:type-reduction-full} for full definition)%
  }
  \label{fig:reduction}
\end{figure}
\indent
  Assertive terms are equipped with
  staged call-by-value small-step reduction relations
  \(N^{(b)} \longrightarrow^{b} (N'^{(b)} \mid  \BlameSign^{ L } )\)
  defined in a straightforward manner except that
  (1)~%
    argument expressions in assertions are evaluated,
    and for this purpose, type expressions in the programs are evaluated as well as terms;
  (2)~%
    when an assertion for type equality passes, it evaluates to an identity function; and that
  (3)~%
    when assertion fails, it evaluates to a special symbol \( \BlameSign^{ L } \) standing for failures,
    and then the result \( \BlameSign^{ L } \) is propagated to the whole program.
  Figure~\ref{fig:reduction} displays the rules for this operational semantics,
  where stage-\(b\) values~\(v^{(b)}\) (for \(b \in \{0, 1\}\))
  and \dfn{stage-\(1\) type values}~\(\possiblyWithSub\stageImetaColor{\tau^{\superscriptI} }\) are defined by the following:
  \begin{gather*}
    \bnfnotab{\possiblyWithSub\stageOmetaColor{v^{\superscriptO} }}{\possiblyWithSub\stageOmetaColor{a} |  \ordO{\lambda} \possiblyWithSub\stageOmetaColor{x}  \relO{:}  \possiblyWithSub\stageOmetaColor{T^{\superscriptO} } \punctO{.}\  \possiblyWithSub\stageOmetaColor{N^{\superscriptO} }  |  \LeftAssertParen \relO{\CastArrow}   \openO{\{} \possiblyWithSub\stageOmetaColor{x}  \relO{:}  \possiblyWithSub\stageOmetaColor{B}  \relO{\mid}  \possiblyWithSub\stageOmetaColor{N^{\superscriptO} } \closeO{\} }   \RightAssertParen^{ L }  |  \openO{\langle} \possiblyWithSub\stageImetaColor{v^{\superscriptI} } \closeO{\rangle} }
  \\
    \bnfnotab{\possiblyWithSub\stageImetaColor{v^{\superscriptI} }}{\possiblyWithSub\stageImetaColor{c} | \possiblyWithSub\stageImetaColor{x} |  \ordI{\lambda} \possiblyWithSub\stageImetaColor{x}  \relI{:}  \possiblyWithSub\stageImetaColor{\tau^{\superscriptI} } \punctI{.}\  \possiblyWithSub\stageImetaColor{v^{\superscriptI} }  |  \possiblyWithSub\stageImetaColor{v^{\superscriptI} } \  \possiblyWithSub\stageImetaColor{v^{\superscriptI} } }
  \qquad
    \bnfnotab{\possiblyWithSub\stageImetaColor{\tau^{\superscriptI} }}{\possiblyWithSub\stageImetaColor{B} |  \ttI{Tensor}\ \ordI{\%} \possiblyWithSub\stageOmetaColor{s}  |  \possiblyWithSub\stageImetaColor{\tau^{\superscriptI} }  \relI{\to}  \possiblyWithSub\stageImetaColor{\tau^{\superscriptI} } }
  \end{gather*}
  Intuitively, \(\possiblyWithSub\stageImetaColor{v^{\superscriptI} }\) and \(\possiblyWithSub\stageImetaColor{\tau^{\superscriptI} }\) correspond to
  ``completed'' code fragments and type annotations, respectively.
  Here, \(\possiblyWithSub\stageOmetaColor{a}\) ranges over the set of \dfn{runtime constants}, which consists of
  base constants and possibly partially applied built-in functions
  (e.g., \( \ordO{  \makeIdentOrConst{}{ VgenMatMult }  } \) or \(    \ordO{  \makeIdentOrConst{}{ VgenMatMult }  }    \     \ordO{  \makeIdentOrConst{}{ C3 }  }    \)).
  More formally, \(\possiblyWithSub\stageOmetaColor{a}\) ranges over the following:
  (i)~\(\possiblyWithSub\stageOmetaColor{c}\) with \(\arity{\possiblyWithSub\stageOmetaColor{c}} = 0\) (i.e.,~base constants),
  (ii)~\( \openO{(}    \possiblyWithSub\stageOmetaColor{p}   \    \possiblyWithSub\stageOmetaColor{c}_{{\mathrm{1}}}   \ \cdots\    \possiblyWithSub\stageOmetaColor{c}_{\ottmv{k}}    \closeO{)} \) such that \(k < \arity{\possiblyWithSub\stageOmetaColor{p}}\),
  and (iii)~\( \openO{(}    \possiblyWithSub\stageOmetaColor{c}   \    \possiblyWithSub\stageOmetaColor{c}_{{\mathrm{1}}}   \ \cdots\    \possiblyWithSub\stageOmetaColor{c}_{\ottmv{k}}    \closeO{)} \) such that \(k < \arity{\possiblyWithSub\stageOmetaColor{c}}\).
  Note that, unlike stage-\(1\) type expressions,
  we do not have to evaluate stage-\(0\) type expressions;
  while stage-\(1\) types should remain in produced code
  (possibly for further post-process optimization),
  stage-\(0\) types are all thrown away by \(\beta\)-reduction
  through stage-\(0\) computation.
\par
\indent
  To deal with applications of built-in functions,
  \rulename{E0-Delta} uses
  the so-called delta reduction~\(\delta\),
  which maps a pair consisting of an operation and a complete array of operands to
  a value of the form~\(\bnfnotab{\possiblyWithSub\stageOmetaColor{q}}{ \possiblyWithSub\stageOmetaColor{c} |  \openO{\langle}  \possiblyWithSub\stageImetaColor{c}  \closeO{\rangle}  }\).
  Entries of \(\delta\) are like the following:
  \begin{gather*}
    \delta(+, ( \ordO{  \makeIdentOrConst{}{ C42 }  } ,  \ordO{  \makeIdentOrConst{}{ C57 }  } )) =  \ordO{  \makeIdentOrConst{}{ C99 }  } ,
  \qquad
    \delta( \ordO{  \makeIdentOrConst{}{ VgenMatMult }  } , ( \ordO{  \makeIdentOrConst{}{ C3 }  } ,  \ordO{  \makeIdentOrConst{}{ C4 }  } ,  \ordO{  \makeIdentOrConst{}{ C5 }  } )) =  \openO{\langle}   \ordI{  \makeIdentOrConst{}{ VmatMult }  _{   \makeIdentOrConst{}{ C3 }  ,    \makeIdentOrConst{}{ C4 }  ,    \makeIdentOrConst{}{ C5 }     } }   \closeO{\rangle} .
  \end{gather*}
  As an abuse of notations, we often write
  \(\delta(  \possiblyWithSub\stageOmetaColor{a}  \    \possiblyWithSub\stageOmetaColor{c}   )\) for \(\delta(\possiblyWithSub\stageOmetaColor{\Hat{c} }, (\possiblyWithSub\stageOmetaColor{c}_{{\mathrm{1}}}, \ldots, \possiblyWithSub\stageOmetaColor{c}_{\ottmv{k}}, \possiblyWithSub\stageOmetaColor{c}))\)
  (resp.~\(\delta(\possiblyWithSub\stageOmetaColor{p}, (\possiblyWithSub\stageOmetaColor{c}_{{\mathrm{1}}}, \ldots, \possiblyWithSub\stageOmetaColor{c}_{\ottmv{k}}, \possiblyWithSub\stageOmetaColor{c}))\))
  when \(\possiblyWithSub\stageOmetaColor{a} =  \openO{(}    \possiblyWithSub\stageOmetaColor{\Hat{c} }   \    \possiblyWithSub\stageOmetaColor{c}_{{\mathrm{1}}}   \ \cdots\    \possiblyWithSub\stageOmetaColor{c}_{\ottmv{k}}    \closeO{)} \) (resp.~\(\possiblyWithSub\stageOmetaColor{a} =  \openO{(}    \possiblyWithSub\stageOmetaColor{p}   \    \possiblyWithSub\stageOmetaColor{c}_{{\mathrm{1}}}   \ \cdots\    \possiblyWithSub\stageOmetaColor{c}_{\ottmv{k}}    \closeO{)} \)).
\par
\indent
  When code generation successfully terminates with a code value \( \openO{\langle} \possiblyWithSub\stageImetaColor{v^{\superscriptI} } \closeO{\rangle} \),
  such \(\possiblyWithSub\stageImetaColor{v^{\superscriptI} }\) can be regarded as a stage-\(0\) term by
  \dfn{unlifting} operations~\( \mathop{\downarrow}( \possiblyWithSub\stageImetaColor{v^{\superscriptI} } )  = \possiblyWithSub\stageOmetaColor{N^{\superscriptO} }\) and \( \mathop{\downarrow}( \possiblyWithSub\stageImetaColor{\tau^{\superscriptI} } )  = \possiblyWithSub\stageOmetaColor{T^{\superscriptO} }\):
  \begin{gather*}
     \mathop{\downarrow}(  \possiblyWithSub\stageImetaColor{x}  )  := \possiblyWithSub\stageOmetaColor{x}
  \qquad
     \mathop{\downarrow}(  \ordI{\lambda} \possiblyWithSub\stageImetaColor{x}  \relI{:}  \possiblyWithSub\stageImetaColor{\tau^{\superscriptI} } \punctI{.}\  \possiblyWithSub\stageImetaColor{v^{\superscriptI} }  )  :=  \ordO{\lambda} \possiblyWithSub\stageOmetaColor{x}  \relO{:}   \mathop{\downarrow}( \possiblyWithSub\stageImetaColor{\tau^{\superscriptI} } )  \punctO{.}\   \mathop{\downarrow}( \possiblyWithSub\stageImetaColor{v^{\superscriptI} } )  
  \qquad
     \mathop{\downarrow}(  \possiblyWithSub\stageImetaColor{v^{\superscriptI} }_{{\mathrm{1}}} \  \possiblyWithSub\stageImetaColor{v^{\superscriptI} }_{{\mathrm{2}}}  )  :=   \mathop{\downarrow}( \possiblyWithSub\stageImetaColor{v^{\superscriptI} }_{{\mathrm{1}}} )  \   \mathop{\downarrow}( \possiblyWithSub\stageImetaColor{v^{\superscriptI} }_{{\mathrm{2}}} )  
  \\
     \mathop{\downarrow}(  \possiblyWithSub\stageImetaColor{c}  )  := \possiblyWithSub\stageOmetaColor{c}
  \qquad
     \mathop{\downarrow}(  \ttI{Tensor}\ \ordI{\%} \possiblyWithSub\stageOmetaColor{s}  )  :=  \ttO{Tensor}\  \possiblyWithSub\stageOmetaColor{s} 
  \qquad
     \mathop{\downarrow}(  \possiblyWithSub\stageImetaColor{B}  )  :=  \openO{\{} \possiblyWithSub\stageOmetaColor{\nu}  \relO{:}  \possiblyWithSub\stageOmetaColor{B}  \relO{\mid}    \ttO{true}   \closeO{\} } 
  \\
     \mathop{\downarrow}(  \possiblyWithSub\stageImetaColor{\tau^{\superscriptI} }_{{\mathrm{1}}}  \relI{\to}  \possiblyWithSub\stageImetaColor{\tau^{\superscriptI} }_{{\mathrm{2}}}  )  :=  \openO{(} \possiblyWithSub\stageOmetaColor{x}  \relO{:}   \mathop{\downarrow}( \possiblyWithSub\stageImetaColor{\tau^{\superscriptI} }_{{\mathrm{1}}} )  \closeO{)} \relO{\to}   \mathop{\downarrow}( \possiblyWithSub\stageImetaColor{\tau^{\superscriptI} }_{{\mathrm{2}}} )  
    \quad\text{(where \(\possiblyWithSub\stageOmetaColor{x} \not\in \fv( \mathop{\downarrow}( \possiblyWithSub\stageImetaColor{\tau^{\superscriptI} }_{{\mathrm{2}}} ) )\))}
  \end{gather*}
  One can then evaluate \(\possiblyWithSub\stageOmetaColor{N^{\superscriptO} } :=  \mathop{\downarrow}( \possiblyWithSub\stageImetaColor{v^{\superscriptI} } ) \) as ordinary runtime execution,
  which will not cause any failure since \(\possiblyWithSub\stageOmetaColor{N^{\superscriptO} }\) contains no assertions.
\par
\indent
  Lastly, we note that, while reduction rules make sense to open terms as well,
  we suppose that only closed terms can step,
  i.e.,~all the reduction rules implicitly require that the reduced term be closed.
  Here, by closed terms, we mean those in which no \emph{stage-0} variables freely occur;
  stage-1 variables are not considered,
  and hence \( \openO{\langle}    \ordI{  \makeIdentOrConst{}{ Vx }  }   \binI{  +  }   \ordI{  \makeIdentOrConst{}{ C1 }  }    \closeO{\rangle} \) is a closed term, for example.
  This restriction is crucial for our metatheory,
  specifically for proving \dfn{cotermination}~\cite{SekiyamaIgarashiGreenbergTOPLAS2017}.
\par
\subsection{Metatheory}\label{subsec:staged-language-metatheory}
\begin{figure}[tb]
\small
  \begin{flushleft}
    \fbox{\(\mathit{\Gamma} \vdash^{b} N^{(b)} : T^{(b)}\)}
  \end{flushleft}
  \vspace{-4em}
  \begin{center}
  \hspace{8.5em}%
    \derive[T0-RfnPred]{%
       \mathit{\Gamma}  \vdash^{0}    \openO{\{} \possiblyWithSub\stageOmetaColor{\nu}  \relO{:}  \possiblyWithSub\stageOmetaColor{B}  \relO{\mid}  \possiblyWithSub\stageOmetaColor{N^{\superscriptO} } \closeO{\} }   
    \andalso
      \ConstEnvPers(c) = \possiblyWithSub\stageImetaColor{B}
    \andalso
         [    \possiblyWithSub\stageOmetaColor{c}    /  \possiblyWithSub\stageOmetaColor{\nu}  ]    \possiblyWithSub\stageOmetaColor{N^{\superscriptO} }   \longrightarrow^{0\,\ast}      \ttO{true}     
    }{%
       \mathit{\Gamma}  \vdash^{0}    \possiblyWithSub\stageOmetaColor{c}    :    \openO{\{} \possiblyWithSub\stageOmetaColor{\nu}  \relO{:}  \possiblyWithSub\stageOmetaColor{B}  \relO{\mid}  \possiblyWithSub\stageOmetaColor{N^{\superscriptO} } \closeO{\} }   
    }
  \\[0.7em]
    \derive[T0-Var]{%
       \vdash  \mathit{\Gamma} 
    \andalso
      \mathit{\Gamma}(\possiblyWithSub\stageOmetaColor{x}) = (\possiblyWithSub\stageOmetaColor{T^{\superscriptO} })^{0}
    }{%
       \mathit{\Gamma}  \vdash^{0}   \possiblyWithSub\stageOmetaColor{x}   :  \possiblyWithSub\stageOmetaColor{T^{\superscriptO} } 
    }
  \qquad
    \derive[T0-App]{%
       \mathit{\Gamma}  \vdash^{0}  \possiblyWithSub\stageOmetaColor{N^{\superscriptO} }_{{\mathrm{1}}}  :   \openO{(} \possiblyWithSub\stageOmetaColor{x}  \relO{:}  \possiblyWithSub\stageOmetaColor{T^{\superscriptO} }_{{\mathrm{11}}} \closeO{)} \relO{\to}  \possiblyWithSub\stageOmetaColor{T^{\superscriptO} }_{{\mathrm{12}}}  
    \andalso
       \mathit{\Gamma}  \vdash^{0}  \possiblyWithSub\stageOmetaColor{N^{\superscriptO} }_{{\mathrm{2}}}  :  \possiblyWithSub\stageOmetaColor{T^{\superscriptO} }_{{\mathrm{11}}} 
    }{%
       \mathit{\Gamma}  \vdash^{0}   \possiblyWithSub\stageOmetaColor{N^{\superscriptO} }_{{\mathrm{1}}} \  \possiblyWithSub\stageOmetaColor{N^{\superscriptO} }_{{\mathrm{2}}}   :    [  \possiblyWithSub\stageOmetaColor{N^{\superscriptO} }_{{\mathrm{2}}}  /  \possiblyWithSub\stageOmetaColor{x}  ]    \possiblyWithSub\stageOmetaColor{T^{\superscriptO} }_{{\mathrm{12}}}  
    }
  \\[0.7em]
    \derive[T0-Ass]{%
       \mathit{\Gamma}  \vdash^{1}  \possiblyWithSub\stageImetaColor{T^{\superscriptI} }_{{\mathrm{1}}} 
    \andalso
       \mathit{\Gamma}  \vdash^{1}  \possiblyWithSub\stageImetaColor{T^{\superscriptI} }_{{\mathrm{2}}} 
    \andalso
       \possiblyWithSub\stageImetaColor{T^{\superscriptI} }_{{\mathrm{1}}}  \mathrel{||}^{1}  \possiblyWithSub\stageImetaColor{T^{\superscriptI} }_{{\mathrm{2}}} 
    \andalso
      \possiblyWithSub\stageOmetaColor{x}\ \not\in \dom \mathit{\Gamma}
    }{%
       \mathit{\Gamma}  \vdash^{0}   \LeftAssertParen\openO{\langle} \possiblyWithSub\stageImetaColor{T^{\superscriptI} }_{{\mathrm{1}}} \closeO{\rangle} \relO{\CastArrow} \openO{\langle} \possiblyWithSub\stageImetaColor{T^{\superscriptI} }_{{\mathrm{2}}} \closeO{\rangle}\RightAssertParen^{ L }   :   \openO{(} \possiblyWithSub\stageOmetaColor{x}  \relO{:}   \openO{\langle} \possiblyWithSub\stageImetaColor{T^{\superscriptI} }_{{\mathrm{1}}} \closeO{\rangle}  \closeO{)} \relO{\to}   \openO{\langle} \possiblyWithSub\stageImetaColor{T^{\superscriptI} }_{{\mathrm{2}}} \closeO{\rangle}   
    }
  \qquad
    \derive[T0-CstP]{%
       \vdash  \mathit{\Gamma} 
    \andalso
      \ConstEnvPers(c) = \possiblyWithSub\stageImetaColor{\tau^{\superscriptI} }
    }{%
       \mathit{\Gamma}  \vdash^{0}    \possiblyWithSub\stageOmetaColor{c}    :   \mathop{\downarrow}( \possiblyWithSub\stageImetaColor{\tau^{\superscriptI} } )  
    }
  \\[0.7em]
    \derive[T0-TyEquiv]{%
       \mathit{\Gamma}  \vdash^{0}  \possiblyWithSub\stageOmetaColor{N^{\superscriptO} }  :  \possiblyWithSub\stageOmetaColor{T'^{\superscriptO} } 
    \andalso
       \possiblyWithSub\stageOmetaColor{T'^{\superscriptO} }  \equiv^{0}  \possiblyWithSub\stageOmetaColor{T^{\superscriptO} } 
    \andalso
       \mathit{\Gamma}  \vdash^{0}  \possiblyWithSub\stageOmetaColor{T^{\superscriptO} } 
    }{%
       \mathit{\Gamma}  \vdash^{0}  \possiblyWithSub\stageOmetaColor{N^{\superscriptO} }  :  \possiblyWithSub\stageOmetaColor{T^{\superscriptO} } 
    }
  \qquad
    \derive[T0-Cst0]{%
       \vdash  \mathit{\Gamma} 
    \andalso
      \ConstEnvZero(\possiblyWithSub\stageOmetaColor{p}) = \possiblyWithSub\stageOmetaColor{T^{\superscriptO} }
    }{%
       \mathit{\Gamma}  \vdash^{0}    \possiblyWithSub\stageOmetaColor{p}    :  \possiblyWithSub\stageOmetaColor{T^{\superscriptO} } 
    }
  \\[0.7em]
    \derive[T0-Abs]{%
       \mathit{\Gamma}  \vdash^{0}  \possiblyWithSub\stageOmetaColor{T^{\superscriptO} }_{{\mathrm{1}}} 
    \andalso
        \mathit{\Gamma} ,  \possiblyWithSub\stageOmetaColor{x}  : ( \possiblyWithSub\stageOmetaColor{T^{\superscriptO} }_{{\mathrm{1}}} )^{0}   \vdash^{0}  \possiblyWithSub\stageOmetaColor{N^{\superscriptO} }_{{\mathrm{2}}}  :  \possiblyWithSub\stageOmetaColor{T^{\superscriptO} }_{{\mathrm{2}}} 
    }{%
       \mathit{\Gamma}  \vdash^{0}   \openO{(}  \ordO{\lambda} \possiblyWithSub\stageOmetaColor{x}  \relO{:}  \possiblyWithSub\stageOmetaColor{T^{\superscriptO} }_{{\mathrm{1}}} \punctO{.}\  \possiblyWithSub\stageOmetaColor{N^{\superscriptO} }_{{\mathrm{2}}}  \closeO{)}   :   \openO{(} \possiblyWithSub\stageOmetaColor{x}  \relO{:}  \possiblyWithSub\stageOmetaColor{T^{\superscriptO} }_{{\mathrm{1}}} \closeO{)} \relO{\to}  \possiblyWithSub\stageOmetaColor{T^{\superscriptO} }_{{\mathrm{2}}}  
    }
  \qquad
    \derive[T0-Brkt]{%
       \mathit{\Gamma}  \vdash^{1}  \possiblyWithSub\stageImetaColor{N^{\superscriptI} }  :  \possiblyWithSub\stageImetaColor{T^{\superscriptI} } 
    }{%
       \mathit{\Gamma}  \vdash^{0}   \openO{\langle} \possiblyWithSub\stageImetaColor{N^{\superscriptI} } \closeO{\rangle}   :   \openO{\langle} \possiblyWithSub\stageImetaColor{T^{\superscriptI} } \closeO{\rangle}  
    }
  \\[0.7em]
    \derive[T0-Rfn]{%
       \mathit{\Gamma}  \vdash^{0}    \openO{\{} \possiblyWithSub\stageOmetaColor{\nu}  \relO{:}  \possiblyWithSub\stageOmetaColor{B}  \relO{\mid}  \possiblyWithSub\stageOmetaColor{N^{\superscriptO} } \closeO{\} }   
    \andalso
       \mathit{\Gamma}  \vdash^{0}    \openO{\{} \possiblyWithSub\stageOmetaColor{\nu}  \relO{:}  \possiblyWithSub\stageOmetaColor{B}  \relO{\mid}  \possiblyWithSub\stageOmetaColor{N'^{\superscriptO} } \closeO{\} }   
    \andalso
      \possiblyWithSub\stageOmetaColor{x} \not\in \dom \mathit{\Gamma}
    }{%
       \mathit{\Gamma}  \vdash^{0}   \LeftAssertParen \relO{\CastArrow}   \openO{\{} \possiblyWithSub\stageOmetaColor{\nu}  \relO{:}  \possiblyWithSub\stageOmetaColor{B}  \relO{\mid}  \possiblyWithSub\stageOmetaColor{N^{\superscriptO} } \closeO{\} }   \RightAssertParen^{ L }   :   \openO{(} \possiblyWithSub\stageOmetaColor{x}  \relO{:}    \openO{\{} \possiblyWithSub\stageOmetaColor{\nu}  \relO{:}  \possiblyWithSub\stageOmetaColor{B}  \relO{\mid}  \possiblyWithSub\stageOmetaColor{N'^{\superscriptO} } \closeO{\} }   \closeO{)} \relO{\to}    \openO{\{} \possiblyWithSub\stageOmetaColor{\nu}  \relO{:}  \possiblyWithSub\stageOmetaColor{B}  \relO{\mid}  \possiblyWithSub\stageOmetaColor{N^{\superscriptO} } \closeO{\} }    
    }
  \quad
    \derive[T1-Esc]{%
       \mathit{\Gamma}  \vdash^{0}  \possiblyWithSub\stageOmetaColor{N^{\superscriptO} }  :   \openO{\langle} \possiblyWithSub\stageImetaColor{T^{\superscriptI} } \closeO{\rangle}  
    }{%
       \mathit{\Gamma}  \vdash^{1}   \ordI{\sim} \possiblyWithSub\stageOmetaColor{N^{\superscriptO} }   :  \possiblyWithSub\stageImetaColor{T^{\superscriptI} } 
    }
  \\[0.7em]
    \derive[T0-RfnAct]{%
       \mathit{\Gamma}  \vdash^{0}    \openO{\{} \possiblyWithSub\stageOmetaColor{\nu}  \relO{:}  \possiblyWithSub\stageOmetaColor{B}  \relO{\mid}  \possiblyWithSub\stageOmetaColor{N^{\superscriptO} }_{{\mathrm{1}}} \closeO{\} }   
    \andalso
       \mathit{\Gamma}  \vdash^{0}  \possiblyWithSub\stageOmetaColor{N^{\superscriptO} }_{{\mathrm{2}}}  :    \openO{\{} \possiblyWithSub\stageOmetaColor{\nu}_{{\mathrm{0}}}  \relO{:}   \ttO{Bool}   \relO{\mid}     \ttO{true}    \closeO{\} }   
    \\
      \ConstEnvPers(c) = \possiblyWithSub\stageImetaColor{B}
    \andalso
         [    \possiblyWithSub\stageOmetaColor{c}    /  \possiblyWithSub\stageOmetaColor{\nu}  ]    \possiblyWithSub\stageOmetaColor{N^{\superscriptO} }_{{\mathrm{1}}}   \longrightarrow^{0\,\ast}   \possiblyWithSub\stageOmetaColor{N^{\superscriptO} }_{{\mathrm{2}}}  
    }{%
       \mathit{\Gamma}  \vdash^{0}   \LeftAssertParen   \openO{\{} \possiblyWithSub\stageOmetaColor{\nu}  \relO{:}  \possiblyWithSub\stageOmetaColor{B}  \relO{\mid}  \possiblyWithSub\stageOmetaColor{N^{\superscriptO} }_{{\mathrm{1}}} \closeO{\} }  \punctO{,}  \possiblyWithSub\stageOmetaColor{N^{\superscriptO} }_{{\mathrm{2}}} \punctO{,}  \possiblyWithSub\stageOmetaColor{c}  \RightAssertParen^{ L }   :    \openO{\{} \possiblyWithSub\stageOmetaColor{\nu}  \relO{:}  \possiblyWithSub\stageOmetaColor{B}  \relO{\mid}  \possiblyWithSub\stageOmetaColor{N^{\superscriptO} }_{{\mathrm{1}}} \closeO{\} }   
    }
  \quad
    \derive[T1-App]{%
       \mathit{\Gamma}  \vdash^{1}  \possiblyWithSub\stageImetaColor{N^{\superscriptI} }_{{\mathrm{1}}}  :   \possiblyWithSub\stageImetaColor{T^{\superscriptI} }_{{\mathrm{11}}}  \relI{\to}  \possiblyWithSub\stageImetaColor{T^{\superscriptI} }_{{\mathrm{12}}}  
    \\
       \mathit{\Gamma}  \vdash^{1}  \possiblyWithSub\stageImetaColor{N^{\superscriptI} }_{{\mathrm{2}}}  :  \possiblyWithSub\stageImetaColor{T^{\superscriptI} }_{{\mathrm{11}}} 
    }{%
       \mathit{\Gamma}  \vdash^{1}   \possiblyWithSub\stageImetaColor{N^{\superscriptI} }_{{\mathrm{1}}} \  \possiblyWithSub\stageImetaColor{N^{\superscriptI} }_{{\mathrm{2}}}   :  \possiblyWithSub\stageImetaColor{T^{\superscriptI} }_{{\mathrm{12}}} 
    }
  \\[0.7em]
    \derive[T1-Abs]{%
       \mathit{\Gamma}  \vdash^{1}  \possiblyWithSub\stageImetaColor{T^{\superscriptI} }_{{\mathrm{1}}} 
    \andalso
        \mathit{\Gamma} ,  \possiblyWithSub\stageImetaColor{x}  : ( \possiblyWithSub\stageImetaColor{T^{\superscriptI} }_{{\mathrm{1}}} )^{1}   \vdash^{1}  \possiblyWithSub\stageImetaColor{N^{\superscriptI} }_{{\mathrm{2}}}  :  \possiblyWithSub\stageImetaColor{T^{\superscriptI} }_{{\mathrm{2}}} 
    \andalso
      \possiblyWithSub\stageImetaColor{x} \not\in \fv(\possiblyWithSub\stageImetaColor{T^{\superscriptI} }_{{\mathrm{2}}})
    }{%
       \mathit{\Gamma}  \vdash^{1}   \openI{(}  \ordI{\lambda} \possiblyWithSub\stageImetaColor{x}  \relI{:}  \possiblyWithSub\stageImetaColor{T^{\superscriptI} }_{{\mathrm{1}}} \punctI{.}\  \possiblyWithSub\stageImetaColor{N^{\superscriptI} }_{{\mathrm{2}}}  \closeI{)}   :   \possiblyWithSub\stageImetaColor{T^{\superscriptI} }_{{\mathrm{1}}}  \relI{\to}  \possiblyWithSub\stageImetaColor{T^{\superscriptI} }_{{\mathrm{2}}}  
    }
  \quad
    \derive[T1-CstP]{%
       \vdash  \mathit{\Gamma} 
    \andalso
      \ConstEnvPers(c) = \possiblyWithSub\stageImetaColor{\tau^{\superscriptI} }
    }{%
       \mathit{\Gamma}  \vdash^{1}   \possiblyWithSub\stageImetaColor{c}   :   \possiblyWithSub\stageImetaColor{\tau^{\superscriptI} }  
    }
  \\[0.7em]
    \derive[T1-TyEquiv]{%
       \mathit{\Gamma}  \vdash^{1}  \possiblyWithSub\stageImetaColor{N^{\superscriptI} }  :  \possiblyWithSub\stageImetaColor{T'^{\superscriptI} } 
    \andalso
       \possiblyWithSub\stageImetaColor{T'^{\superscriptI} }  \equiv^{1}  \possiblyWithSub\stageImetaColor{T^{\superscriptI} } 
    \andalso
       \mathit{\Gamma}  \vdash^{1}  \possiblyWithSub\stageImetaColor{T^{\superscriptI} } 
    }{%
       \mathit{\Gamma}  \vdash^{1}  \possiblyWithSub\stageImetaColor{N^{\superscriptI} }  :  \possiblyWithSub\stageImetaColor{T^{\superscriptI} } 
    }
  \quad
    \derive[T1-Var]{%
       \vdash  \mathit{\Gamma} 
    \andalso
      \mathit{\Gamma}(\possiblyWithSub\stageImetaColor{x}) = (\possiblyWithSub\stageImetaColor{T^{\superscriptI} })^{1}
    }{%
       \mathit{\Gamma}  \vdash^{1}   \possiblyWithSub\stageImetaColor{x}   :  \possiblyWithSub\stageImetaColor{T^{\superscriptI} } 
    }
  \end{center}
  \vspace{-1em}
  \caption{Target typing for metatheory}
  \label{fig:target-typing}
\end{figure}
\indent
  For the purpose of proving type safety, assertive terms are also assigned types
  by declarative target typing of the form~\(\mathit{\Gamma} \vdash^{b} N^{(b)} : T^{(b)}\).
  Figure~\ref{fig:target-typing} displays the rules for these judgments.
  Some of the rules depend on type equivalences~\(T^{(b)}_1 \equiv^b T^{(b)}_2\)
  and well-formedness judgments~\(\mathit{\Gamma} \vdash^{b} T^{(b)}\) and \( \vdash  \mathit{\Gamma} \)
  defined respectively in Figures~\ref{fig:type-equivalence} and \ref{fig:well-formedness} in Appendix.
  The equivalences~\(\equiv^b\) are necessarily introduced
  to prove Preservation as to function applications.
\par
\indent
  We first prove that well-typed source terms are always elaborated to
  assertive terms well-typed under target typing,
  by relatively straightforward induction:
\par
\begin{theorem}[Soundness of Assertion Insertion]\label{thm:soundness-of-elaboration}
  If \( \vdash  \mathit{\Gamma} \) and \(\mathit{\Gamma} \vdash^b M^{(b)} : T^{(b)} \ElabArrow N^{(b)}\),
  then \(\mathit{\Gamma} \vdash^b N^{(b)} : T^{(b)}\).
\end{theorem}
\indent
  Proving Preservation and Progress~\cite{Pierce2002} is much more challenging
  due to the combination of the \(\delta\)-reduction, the type equivalences~\(\equiv^b\),
  and the dependent nature of our typing.
  To prove Preservation as to built-in functions,
  we must assume some natural properties on \(\ConstEnvZero\), \(\ConstEnvPers\), and \(\delta\).
  For example, since
  \(\delta( \ordO{  \makeIdentOrConst{}{ VgenVertCat }  } , ( \ordO{  \makeIdentOrConst{}{ C3 }  } ,  \ordO{  \makeIdentOrConst{}{ C4 }  } ,  \ordO{  \makeIdentOrConst{}{ C5 }  } ))
    =  \openO{\langle}   \ordI{  \makeIdentOrConst{}{ VvertCat }  _{   \makeIdentOrConst{}{ C3 }  ,    \makeIdentOrConst{}{ C4 }  ,    \makeIdentOrConst{}{ C5 }     } }   \closeO{\rangle} \)
  holds,
  \(     \ordO{  \makeIdentOrConst{}{ VgenVertCat }  }   \    \ordO{  \makeIdentOrConst{}{ C3 }  }    \    \ordO{  \makeIdentOrConst{}{ C4 }  }    \    \ordO{  \makeIdentOrConst{}{ C5 }  }   \) and \( \openO{\langle}   \ordI{  \makeIdentOrConst{}{ VvertCat }  _{   \makeIdentOrConst{}{ C3 }  ,    \makeIdentOrConst{}{ C4 }  ,    \makeIdentOrConst{}{ C5 }     } }   \closeO{\rangle} \)
  must have equivalent types.
  However, we cannot use the type equivalences to describe such assumptions;
  because the definition of equivalences will depend on \(\delta\),
  we have to avoid the dependency of the reverse direction
  in order not to make the validity circular.
  To this end, we have to use reduction relations, rather than equivalences.
  We put the resulting descriptions as Assumption~\ref{assump:type-of-constants} in Appendix
  because they are one-page long.
\par
\indent
  Another challenging point arising after identifying the above assumptions is
  how to define the type equivalences~\(\equiv^b\) precisely.
  They must be at least compatible with the \(\beta\)-equivalence,
  but the \(\beta\)-equivalence itself is actually too loose;
  in order to prove Preservation as to \(\delta\),
  we must define \(\equiv^b\) carefully so that
  they preserve the ``reducibility'' of refinement predicates to \(  \ttO{true}  \).
  For this purpose, following Sekiyama~et~al.~\cite{SekiyamaIgarashiGreenbergTOPLAS2017},
  we can extend and use the \dfn{common subexpression reduction} (\dfn{CSR}) equivalence~\cite{%
    Greenberg2013,%
    SekiyamaIgarashiGreenbergTOPLAS2017},
  which is the equivalence spanned by the relation that allows
  the usual call-by-value reduction for arbitrary \emph{closed} subexpressions.
\par
\indent
  Under these settings,
  we have proved the following safety properties,
  where \( \vdash^{1}  \mathit{\Gamma} \) means that all the entries in \(\mathit{\Gamma}\) are of the form
  \((\possiblyWithSub\stageImetaColor{x} : (\possiblyWithSub\stageImetaColor{T^{\superscriptI} })^{1})\):
\par
\begin{theorem}[Preservation]\label{thm:preservation}
  If \(\mathit{\Gamma} \vdash^b N^{(b)} : T^{(b)}\) and \(N^{(b)} \longrightarrow^b N'^{(b)}\),
  then \(\mathit{\Gamma} \vdash^b N'^{(b)} : T^{(b)}\).
\end{theorem}
\begin{theorem}[Progress]\label{thm:progress}
  If \( \vdash^{1}  \mathit{\Gamma} \) and \(\mathit{\Gamma} \vdash^{b} N^{(b)} : T^{(b)}\), then we have one of the following:
  (1)~\(N^{(b)} \longrightarrow^b  \BlameSign^{ L } \);
  (2)~there exists \(N'^{(b)}\) such that \(N^{(b)} \longrightarrow^b N'^{(b)}\); or
  (3)~\(N^{(b)}\) is a value.
\end{theorem}
\begin{figure}[tb]
\small
  \begin{flushleft}
    \fbox{\( \gamma  \PositionZeroTurnstile  \possiblyWithSub\stageImetaColor{v^{\superscriptI} }  :  \possiblyWithSub\stageImetaColor{\tau^{\superscriptI} } \)}
  \end{flushleft}
  \vspace{-4.25em}
  \begin{center}
  \hspace{5em}%
    \derive[G-Var]{%
      \gamma(\possiblyWithSub\stageImetaColor{x}) = \possiblyWithSub\stageImetaColor{\tau^{\superscriptI} }
    }{%
       \gamma  \PositionZeroTurnstile   \possiblyWithSub\stageImetaColor{x}   :  \possiblyWithSub\stageImetaColor{\tau^{\superscriptI} } 
    }
  \qquad
    \derive[G-Abs]{%
        \gamma ,  \possiblyWithSub\stageImetaColor{x}  :  \possiblyWithSub\stageImetaColor{\tau^{\superscriptI} }_{{\mathrm{1}}}   \PositionZeroTurnstile  \possiblyWithSub\stageImetaColor{v^{\superscriptI} }_{{\mathrm{2}}}  :  \possiblyWithSub\stageImetaColor{\tau^{\superscriptI} }_{{\mathrm{2}}} 
    }{%
       \gamma  \PositionZeroTurnstile   \openI{(}  \ordI{\lambda} \possiblyWithSub\stageImetaColor{x}  \relI{:}  \possiblyWithSub\stageImetaColor{\tau^{\superscriptI} }_{{\mathrm{1}}} \punctI{.}\  \possiblyWithSub\stageImetaColor{v^{\superscriptI} }_{{\mathrm{2}}}  \closeI{)}   :   \possiblyWithSub\stageImetaColor{\tau^{\superscriptI} }_{{\mathrm{1}}}  \relI{\to}  \possiblyWithSub\stageImetaColor{\tau^{\superscriptI} }_{{\mathrm{2}}}  
    }
  \\[0.7em]
    \derive[G-Cst]{%
      \ConstEnvPers(\possiblyWithSub\stageImetaColor{c}) = \possiblyWithSub\stageImetaColor{\tau^{\superscriptI} }
    }{%
       \gamma  \PositionZeroTurnstile   \possiblyWithSub\stageImetaColor{c}   :  \possiblyWithSub\stageImetaColor{\tau^{\superscriptI} } 
    }
  \qquad
    \derive[G-App]{%
       \gamma  \PositionZeroTurnstile  \possiblyWithSub\stageImetaColor{v^{\superscriptI} }_{{\mathrm{1}}}  :   \possiblyWithSub\stageImetaColor{\tau^{\superscriptI} }_{{\mathrm{2}}}  \relI{\to}  \possiblyWithSub\stageImetaColor{\tau^{\superscriptI} }  
    \andalso
       \gamma  \PositionZeroTurnstile  \possiblyWithSub\stageImetaColor{v^{\superscriptI} }_{{\mathrm{2}}}  :  \possiblyWithSub\stageImetaColor{\tau^{\superscriptI} }_{{\mathrm{2}}} 
    }{%
       \gamma  \PositionZeroTurnstile   \possiblyWithSub\stageImetaColor{v^{\superscriptI} }_{{\mathrm{1}}} \  \possiblyWithSub\stageImetaColor{v^{\superscriptI} }_{{\mathrm{2}}}   :  \possiblyWithSub\stageImetaColor{\tau^{\superscriptI} } 
    }
  \end{center}
  \vspace{-0.75em}
  \caption{The typing rules for generated code}
  \label{fig:code-typing}
\end{figure}
\indent
  For proving the safety of generated code, we use
  an additional type judgment~\( \gamma  \PositionZeroTurnstile  \possiblyWithSub\stageImetaColor{v^{\superscriptI} }  :  \possiblyWithSub\stageImetaColor{\tau^{\superscriptI} } \)
  defined by the rules in Figure~\ref{fig:code-typing},
  where \(\gamma\) is defined by: \(\bnfnotab{\gamma}{  \bullet  |  \gamma ,  \possiblyWithSub\stageImetaColor{x}  :  \possiblyWithSub\stageImetaColor{\tau^{\superscriptI} }  }\).
  Unlike the target typing, this type system uses only value types~\(\possiblyWithSub\stageImetaColor{\tau^{\superscriptI} }\)
  and does not depend on the type equivalences;
  as mentioned in Section~\ref{subsec:our-work},
  this is essentially a simply-typed setting with countably infinite number of base types
  (\(\TypeMat{1}{1}\), \(\TypeMat{1}{2}\), \(\TypeMat{2}{1}\), and so on).
  The following states that generated code is always well-typed under
  the \(\PositionZeroTurnstile\)-rules,
  where \( \mathit{\Gamma}  \equiv  \gamma \) is the evident pointwise equivalence based on \(\equiv^1\):
\par
\begin{lemma}\label{lem:final-target-typing}
  If \( \vdash^{1}  \mathit{\Gamma} \), \( \mathit{\Gamma}  \vdash^{1}   \possiblyWithSub\stageImetaColor{v^{\superscriptI} }   :  \possiblyWithSub\stageImetaColor{T^{\superscriptI} } \), and \( \mathit{\Gamma}  \equiv  \gamma \),
  then there exists \(\possiblyWithSub\stageImetaColor{\tau^{\superscriptI} }\) such that \( \gamma  \PositionZeroTurnstile  \possiblyWithSub\stageImetaColor{v^{\superscriptI} }  :  \possiblyWithSub\stageImetaColor{\tau^{\superscriptI} } \) and \( \possiblyWithSub\stageImetaColor{T^{\superscriptI} }  \equiv^{1}   \possiblyWithSub\stageImetaColor{\tau^{\superscriptI} }  \).
\end{lemma}
\indent
  However, there is one remaining challenge as to proving this lemma;
  the following property is crucial
  (since the \(\PositionZeroTurnstile\)-rules are independent of \(\equiv^b\)):
\par
\begin{lemma}\label{lem:value-type-csr-equiv-implies-equal}
  \(  \possiblyWithSub\stageImetaColor{\tau^{\superscriptI} }_{{\mathrm{1}}}   \equiv^{1}   \possiblyWithSub\stageImetaColor{\tau^{\superscriptI} }_{{\mathrm{2}}}  \) implies \(\possiblyWithSub\stageImetaColor{\tau^{\superscriptI} }_{{\mathrm{1}}} = \possiblyWithSub\stageImetaColor{\tau^{\superscriptI} }_{{\mathrm{2}}}\).
\end{lemma}
\noindent
  We prove this by separating it into two:
  (1)~\(  \possiblyWithSub\stageImetaColor{\tau^{\superscriptI} }_{{\mathrm{1}}}   \equiv^{1}   \possiblyWithSub\stageImetaColor{\tau^{\superscriptI} }_{{\mathrm{2}}}  \) implies \(  \possiblyWithSub\stageImetaColor{\tau^{\superscriptI} }_{{\mathrm{1}}}   \cong^{1}   \possiblyWithSub\stageImetaColor{\tau^{\superscriptI} }_{{\mathrm{2}}}  \),
  where \(\cong^b\) is the standard \(\beta\)-equivalence; and
  (2)~\(  \possiblyWithSub\stageImetaColor{\tau^{\superscriptI} }_{{\mathrm{1}}}   \cong^{1}   \possiblyWithSub\stageImetaColor{\tau^{\superscriptI} }_{{\mathrm{2}}}  \) implies \(\possiblyWithSub\stageImetaColor{\tau^{\superscriptI} }_{{\mathrm{1}}} = \possiblyWithSub\stageImetaColor{\tau^{\superscriptI} }_{{\mathrm{2}}}\).
  While the former is straightforward, the latter requires slight attention:
  the reduction includes the rule~%
  \(  \LeftAssertParen\openO{\langle}  \possiblyWithSub\stageImetaColor{\tau^{\superscriptI} }  \closeO{\rangle} \relO{\CastArrow} \openO{\langle}  \possiblyWithSub\stageImetaColor{\tau^{\superscriptI} }  \closeO{\rangle}\RightAssertParen^{ L }   \longrightarrow^{0}    \openO{(}  \ordO{\lambda} \possiblyWithSub\stageOmetaColor{x}  \relO{:}   \openO{\langle}  \possiblyWithSub\stageImetaColor{\tau^{\superscriptI} }  \closeO{\rangle}  \punctO{.}\   \possiblyWithSub\stageOmetaColor{x}   \closeO{)}   \),
  which is not left-linear in the sense that
  the metavariable~\(\possiblyWithSub\stageImetaColor{\tau^{\superscriptI} }\) appears more than once on the left-hand side.
  In such a reduction system, confluence is often hard to show or even broken.
  Nonetheless, in our cases,
  it turns out to be sufficient to use a standard syntactic approach to consistency,
  which defines parallel reduction corresponding to the equivalence and prove
  the uniqueness of normal forms via confluence.
  Thanks to Lemma~\ref{lem:value-type-csr-equiv-implies-equal},
  we can finally prove Lemma~\ref{lem:final-target-typing}.
\par
\indent
  The following lemma can easily be shown,
  where \( \mathop{\downarrow}( \gamma )  = \mathit{\Gamma}'\) is the evident unlifting:
\par
\begin{lemma}
  \( \gamma  \PositionZeroTurnstile  \possiblyWithSub\stageImetaColor{v^{\superscriptI} }  :  \possiblyWithSub\stageImetaColor{\tau^{\superscriptI} } \) implies \(  \mathop{\downarrow}( \gamma )   \vdash^{0}   \mathop{\downarrow}( \possiblyWithSub\stageImetaColor{v^{\superscriptI} } )   :   \mathop{\downarrow}( \possiblyWithSub\stageImetaColor{\tau^{\superscriptI} } )  \).
\end{lemma}
\indent
  Since unlifted code does not contain
  \(\possiblyWithSub\stageOmetaColor{p}\) or assertions and is typeable without the type equivalences,
  the combination of the theorems and lemmata above ensures that,
  if the source program elaborates to a term and
  the term evaluates to a code value without failures,
  we will not have runtime shape mismatches when running the produced code:
\par
  \begin{corollary}
    If \(  \bullet   \vdash^{0}  \possiblyWithSub\stageOmetaColor{M^{\scriptscriptstyle(0)} }  :   \openO{\langle} \possiblyWithSub\stageImetaColor{T^{\superscriptI} } \closeO{\rangle}   \ElabArrow  \possiblyWithSub\stageOmetaColor{N^{\superscriptO} } \) and \(\possiblyWithSub\stageOmetaColor{N^{\superscriptO} } \longrightarrow^{0\,\ast} \possiblyWithSub\stageOmetaColor{v^{\superscriptO} }_{{\mathrm{0}}}\),
    then \(\possiblyWithSub\stageOmetaColor{v^{\superscriptO} }_{{\mathrm{0}}}\) is of the form \( \openO{\langle} \possiblyWithSub\stageImetaColor{v^{\superscriptI} } \closeO{\rangle} \), and
    this \(\possiblyWithSub\stageImetaColor{v^{\superscriptI} }\) satisfies one of the following:
    (1)~there exists \(\possiblyWithSub\stageOmetaColor{v^{\superscriptO} }_{{\mathrm{1}}}\) such that
    \( \mathop{\downarrow}( \possiblyWithSub\stageImetaColor{v^{\superscriptI} } )  \longrightarrow^{0\,\ast} \possiblyWithSub\stageOmetaColor{v^{\superscriptO} }_{{\mathrm{1}}}\), or
    (2)~the evaluation of \( \mathop{\downarrow}( \possiblyWithSub\stageImetaColor{v^{\superscriptI} } ) \) does not halt\footnote{%
      More precisely, unless extended with recursive functions,
      \LambdaBracketCast\ is strongly normalizing,
      and thus (2) does not happen.
    }.
  \end{corollary}
\par

\section{Implicit Arguments and Their Inference}\label{sec:implicit}
\indent
  To alleviate the burden of manually adding
  shape-related stage-\(0\) arguments in \LambdaBracketCast,
  we offer \LambdaBracketCastImplicit,
  an extension of \LambdaBracketCast\ with implicit parameters/arguments,
  as mentioned in Section~\ref{subsec:implicit-arguments}.
  This section formalizes \LambdaBracketCastImplicit\ and explains how to infer implicit arguments.
\par
\indent
  First, the source syntax is
  extended
  with a variant of
  \(\lambda\)-abstractions and applications\footnote{%
    Some contexts will newly allow \(\lambda\)-abstractions without a type annotation
    (i.e.,~\( \ordO{\lambda} \possiblyWithSub\stageOmetaColor{x} \punctO{.}\  \possiblyWithSub\stageOmetaColor{\mathcal{M}^{\superscriptO} } \)).
    This is closely related to how the reconstruction algorithm is designed;
    \rulename{B0-AbsNoAnnot} and \rulename{B1-AbsNoAnnot}
    in Figures~\ref{fig:option-inference-main-full-1} and \ref{fig:option-inference-main-full-2}
    deal with such \(\lambda\)-abstractions.
  }:
  \begin{align*}
    \bnf{\possiblyWithSub\stageOmetaColor{\mathcal{M}^{\superscriptO} }}{%
      \possiblyWithSub\stageOmetaColor{p} | \possiblyWithSub\stageOmetaColor{c} | \possiblyWithSub\stageOmetaColor{x} |  \ordO{\lambda} \possiblyWithSub\stageOmetaColor{x}  \relO{:}  \possiblyWithSub\stageOmetaColor{\mathcal{S}^{\superscriptO} } \punctO{.}\  \possiblyWithSub\stageOmetaColor{\mathcal{M}^{\superscriptO} }  |  \ordO{\lambda} \possiblyWithSub\stageOmetaColor{x} \punctO{.}\  \possiblyWithSub\stageOmetaColor{\mathcal{M}^{\superscriptO} }  |  \openO{(} \possiblyWithSub\stageOmetaColor{\mathcal{M}^{\superscriptO} } \  \possiblyWithSub\stageOmetaColor{\mathcal{M}^{\superscriptO} } \closeO{)}_{ \ell } 
    |* \ordO{\lambda}\openO{\{} \possiblyWithSub\stageOmetaColor{x}  \relO{:}  \possiblyWithSub\stageOmetaColor{\mathcal{S}^{\superscriptO} } \closeO{\} }\punctO{.}\  \possiblyWithSub\stageOmetaColor{\mathcal{M}^{\superscriptO} }  |  \openO{(} \possiblyWithSub\stageOmetaColor{\mathcal{M}^{\superscriptO} } \ \openO{\{} \possiblyWithSub\stageOmetaColor{\mathcal{M}^{\superscriptO} } \closeO{\} }\closeO{)}_{ \ell }  |  \possiblyWithSub\stageOmetaColor{\mathcal{M}^{\superscriptO} } \ \ordO{\_}  |  \openO{\langle} \possiblyWithSub\stageImetaColor{\mathcal{M}^{\superscriptI} } \closeO{\rangle} 
    }
  \\
    \bnf{\possiblyWithSub\stageImetaColor{\mathcal{M}^{\superscriptI} }}{%
      \possiblyWithSub\stageImetaColor{c} | \possiblyWithSub\stageImetaColor{x} |  \ordI{\lambda} \possiblyWithSub\stageImetaColor{x}  \relI{:}  \possiblyWithSub\stageImetaColor{\mathcal{S}^{\superscriptI} } \punctI{.}\  \possiblyWithSub\stageImetaColor{\mathcal{M}^{\superscriptI} }  |  \ordI{\lambda} \possiblyWithSub\stageImetaColor{x} \punctI{.}\  \possiblyWithSub\stageImetaColor{\mathcal{M}^{\superscriptI} }  |  \openI{(} \possiblyWithSub\stageImetaColor{\mathcal{M}^{\superscriptI} } \  \possiblyWithSub\stageImetaColor{\mathcal{M}^{\superscriptI} } \closeI{)}_{ \ell }  |  \ordI{\sim} \possiblyWithSub\stageOmetaColor{\mathcal{M}^{\superscriptO} } 
    }
  \\
    \bnf{\possiblyWithSub\stageOmetaColor{\mathcal{S}^{\superscriptO} }}{%
      \possiblyWithSub\stageOmetaColor{B} |  \ttO{Tensor}\  \possiblyWithSub\stageOmetaColor{s}  |  \openO{(} \possiblyWithSub\stageOmetaColor{x}  \relO{:}  \possiblyWithSub\stageOmetaColor{\mathcal{S}^{\superscriptO} } \closeO{)} \relO{\to}  \possiblyWithSub\stageOmetaColor{\mathcal{S}^{\superscriptO} }  |  \openO{\{} \possiblyWithSub\stageOmetaColor{x}  \relO{:}  \possiblyWithSub\stageOmetaColor{\mathcal{S}^{\superscriptO} } \closeO{\} } \relO{\to}  \possiblyWithSub\stageOmetaColor{\mathcal{S}^{\superscriptO} }  |  \openO{\langle} \possiblyWithSub\stageImetaColor{\mathcal{S}^{\superscriptI} } \closeO{\rangle} 
    }
  \\
    \bnf{\possiblyWithSub\stageImetaColor{\mathcal{S}^{\superscriptI} }}{%
      \possiblyWithSub\stageImetaColor{B} |  \openI{(}\ttI{Tensor}\ \ordI{\%} \possiblyWithSub\stageOmetaColor{\mathcal{M}^{\superscriptO} } \closeI{)}_{ \ell }  |  \possiblyWithSub\stageImetaColor{\mathcal{S}^{\superscriptI} }  \relI{\to}  \possiblyWithSub\stageImetaColor{\mathcal{S}^{\superscriptI} } 
    }
  \end{align*}
  The construct~\( \openO{(}  \ordO{\lambda}\openO{\{} \possiblyWithSub\stageOmetaColor{x}  \relO{:}  \possiblyWithSub\stageOmetaColor{\mathcal{S}^{\superscriptO} } \closeO{\} }\punctO{.}\  \possiblyWithSub\stageOmetaColor{\mathcal{M}^{\superscriptO} }  \closeO{)} \) works as
  \(\lambda\)-abstractions whose parameter can be implicit,
  and \( \openO{(} \possiblyWithSub\stageOmetaColor{\mathcal{M}^{\superscriptO} } \ \openO{\{} \possiblyWithSub\stageOmetaColor{\mathcal{M}^{\superscriptO} } \closeO{\} }\closeO{)}_{ \ell } \) can be used to specify an argument for such parameters explicitly.
  The form~\( \possiblyWithSub\stageOmetaColor{\mathcal{M}^{\superscriptO} } \ \ordO{\_} \) lies between complete omission and explicit designation;
  it indicates the existence of \(\possiblyWithSub\stageOmetaColor{\mathcal{M}^{\superscriptO} }\)'s implicit parameter,
  but does not specify a concrete expression for it.
  This is useful, e.g.,~for specifying an argument for only the second implicit parameter
  by~\( \openO{(}  \possiblyWithSub\stageOmetaColor{\mathcal{M}^{\superscriptO} } \ \ordO{\_}  \ \openO{\{} \possiblyWithSub\stageOmetaColor{\mathcal{M}^{\superscriptO} }_{{\mathrm{2}}} \closeO{\} }\closeO{)}_{ \ell } \).
\par
\indent
  As mentioned earlier in Section~\ref{subsec:implicit-arguments},
  we reconstruct implicit arguments through type-checking
  as well as inserting compile-time assertions.
  This process can be formalized by utilizing
  ``\dfn{let arguments go first}''~\cite{XieOliveiraESOP2018},
  which is a variant of \dfn{bidirectional type-checking}~\cite{PierceTurner2000,DunfieldNeel2021};
  when checking applications, we first traverse the argument to obtain its type
  and then inspect the function, not the other way around as usual.
  To handle dependent function types and cast insertion,
  our formalization extends Xie and Oliveira's original work
  in several aspects, such as the form of \dfn{application contexts} and return types,
  which will be explained below.
\par
\begin{figure}[tbp]
\small
  \begin{flushleft}
    \fbox{\( \mathcal{G}  \mid  \mathit{\Psi}  \vdash^{0}  \possiblyWithSub\stageOmetaColor{\mathcal{M}^{\superscriptO} }  \Rightarrow  \possiblyWithSub\stageOmetaColor{R^{\superscriptO} }  \ElabArrow  \possiblyWithSub\stageOmetaColor{N^{\superscriptO} } \)}
  \end{flushleft}
  \vspace{-4.25em}
  \begin{center}
  \hspace{10em}%
    \derive[B0-Var]{%
      \mathcal{G}(\possiblyWithSub\stageOmetaColor{x}) = (\possiblyWithSub\stageOmetaColor{\mathcal{T}^{\superscriptO} })^0
    \andalso
       \mathcal{G}  \mid  \mathit{\Psi}  \vdash^{0}  \possiblyWithSub\stageOmetaColor{\mathcal{T}^{\superscriptO} }  \mathrel{<:}  \possiblyWithSub\stageOmetaColor{R^{\superscriptO} } 
    }{%
       \mathcal{G}  \mid  \mathit{\Psi}  \vdash^{0}   \possiblyWithSub\stageOmetaColor{x}   \Rightarrow  \possiblyWithSub\stageOmetaColor{R^{\superscriptO} }  \ElabArrow   \possiblyWithSub\stageOmetaColor{x}  
    }
  \\[0.7em]
    \derive[B0-App]{%
       \mathcal{G}  \mid   \bullet   \vdash^{0}  \possiblyWithSub\stageOmetaColor{\mathcal{M}^{\superscriptO} }_{{\mathrm{2}}}  \Rightarrow   \possiblyWithSub\stageOmetaColor{\mathcal{T}^{\superscriptO} }_{{\mathrm{2}}}   \ElabArrow  \possiblyWithSub\stageOmetaColor{N^{\superscriptO} }_{{\mathrm{2}}} 
    \\
       \mathcal{G}  \mid   \mathit{\Psi} , ( \possiblyWithSub\stageOmetaColor{N^{\superscriptO} }_{{\mathrm{2}}}  :  \possiblyWithSub\stageOmetaColor{\mathcal{T}^{\superscriptO} }_{{\mathrm{2}}} )^{0}_{ \ell }   \vdash^{0}  \possiblyWithSub\stageOmetaColor{\mathcal{M}^{\superscriptO} }_{{\mathrm{1}}}  \Rightarrow    ( \possiblyWithSub\stageOmetaColor{N^{\superscriptO} }_{{\mathrm{0}}} \ /\  \possiblyWithSub\stageOmetaColor{\mathcal{T}^{\superscriptO} }_{{\mathrm{11}}} )^{0}   \relO{\to}  \possiblyWithSub\stageOmetaColor{R^{\superscriptO} }_{{\mathrm{12}}}   \ElabArrow  \possiblyWithSub\stageOmetaColor{N^{\superscriptO} }_{{\mathrm{1}}} 
    }{%
       \mathcal{G}  \mid  \mathit{\Psi}  \vdash^{0}   \openO{(} \possiblyWithSub\stageOmetaColor{\mathcal{M}^{\superscriptO} }_{{\mathrm{1}}} \  \possiblyWithSub\stageOmetaColor{\mathcal{M}^{\superscriptO} }_{{\mathrm{2}}} \closeO{)}_{ \ell }   \Rightarrow  \possiblyWithSub\stageOmetaColor{R^{\superscriptO} }_{{\mathrm{12}}}  \ElabArrow   \possiblyWithSub\stageOmetaColor{N^{\superscriptO} }_{{\mathrm{1}}} \   \openO{(}  \possiblyWithSub\stageOmetaColor{N^{\superscriptO} }_{{\mathrm{0}}} \  \possiblyWithSub\stageOmetaColor{N^{\superscriptO} }_{{\mathrm{2}}}  \closeO{)}   
    }
  \\[0.7em]
    \derive[B0-AbsAnnot1]{%
       \mathcal{G}  \vdash^{0}  \possiblyWithSub\stageOmetaColor{\mathcal{S}^{\superscriptO} }_{{\mathrm{1}}}  \ElabArrow  \possiblyWithSub\stageOmetaColor{\mathcal{T}^{\superscriptO} }_{{\mathrm{1}}} 
    \andalso
        \mathcal{G} ,  \possiblyWithSub\stageOmetaColor{x}  : ( \possiblyWithSub\stageOmetaColor{\mathcal{T}^{\superscriptO} }_{{\mathrm{1}}} )^{0}   \mid   \bullet   \vdash^{0}  \possiblyWithSub\stageOmetaColor{\mathcal{M}^{\superscriptO} }_{{\mathrm{2}}}  \Rightarrow   \possiblyWithSub\stageOmetaColor{\mathcal{T}^{\superscriptO} }_{{\mathrm{2}}}   \ElabArrow  \possiblyWithSub\stageOmetaColor{N^{\superscriptO} }_{{\mathrm{2}}} 
    }{%
       \mathcal{G}  \mid   \bullet   \vdash^{0}   \ordO{\lambda} \possiblyWithSub\stageOmetaColor{x}  \relO{:}  \possiblyWithSub\stageOmetaColor{\mathcal{S}^{\superscriptO} }_{{\mathrm{1}}} \punctO{.}\  \possiblyWithSub\stageOmetaColor{\mathcal{M}^{\superscriptO} }_{{\mathrm{2}}}   \Rightarrow    \openO{(} \possiblyWithSub\stageOmetaColor{x}  \relO{:}  \possiblyWithSub\stageOmetaColor{\mathcal{T}^{\superscriptO} }_{{\mathrm{1}}} \closeO{)} \relO{\to}  \possiblyWithSub\stageOmetaColor{\mathcal{T}^{\superscriptO} }_{{\mathrm{2}}}    \ElabArrow   \ordO{\lambda} \possiblyWithSub\stageOmetaColor{x}  \relO{:}   \lfloor  \possiblyWithSub\stageOmetaColor{\mathcal{T}^{\superscriptO} }_{{\mathrm{1}}} \rfloor  \punctO{.}\  \possiblyWithSub\stageOmetaColor{N^{\superscriptO} }_{{\mathrm{2}}}  
    }
  \\[0.7em]
    \derive[B0-Brkt]{%
      \mathcal{G} \mid \mathit{\Psi} \vdash^{1} \possiblyWithSub\stageImetaColor{\mathcal{M}^{\superscriptI} } \Rightarrow (D_{\ottmv{i}} \to)_{i = 1}^{m}  \possiblyWithSub\stageImetaColor{T^{\superscriptI} }  \ElabArrow \possiblyWithSub\stageImetaColor{N^{\superscriptI} }
    }{%
      \mathcal{G} \mid \mathit{\Psi} \vdash^{0}  \openO{\langle} \possiblyWithSub\stageImetaColor{\mathcal{M}^{\superscriptI} } \closeO{\rangle}  \Rightarrow (D_{\ottmv{i}} \to)_{i = 1}^{m}   \openO{\langle} \possiblyWithSub\stageImetaColor{T^{\superscriptI} } \closeO{\rangle}   \ElabArrow  \openO{\langle} \possiblyWithSub\stageImetaColor{N^{\superscriptI} } \closeO{\rangle} 
    }
  \\[0.7em]
    \derive[B0-AppImp]{%
       \mathcal{G}  \mid   \bullet   \vdash^{0}  \possiblyWithSub\stageOmetaColor{\mathcal{M}^{\superscriptO} }_{{\mathrm{2}}}  \Rightarrow   \possiblyWithSub\stageOmetaColor{\mathcal{T}^{\superscriptO} }_{{\mathrm{2}}}   \ElabArrow  \possiblyWithSub\stageOmetaColor{N^{\superscriptO} }_{{\mathrm{2}}} 
    \\
       \mathcal{G}  \mid   \mathit{\Psi} , ( \possiblyWithSub\stageOmetaColor{N^{\superscriptO} }_{{\mathrm{2}}}  :  \possiblyWithSub\stageOmetaColor{\mathcal{T}^{\superscriptO} }_{{\mathrm{2}}} )^{0}_{ \ell }   \vdash^{0}  \possiblyWithSub\stageOmetaColor{\mathcal{M}^{\superscriptO} }_{{\mathrm{1}}}  \Rightarrow    \{ \possiblyWithSub\stageOmetaColor{N^{\superscriptO} }_{{\mathrm{0}}} \ /\  \possiblyWithSub\stageOmetaColor{\mathcal{T}^{\superscriptO} }_{{\mathrm{11}}} \}   \relO{\to}  \possiblyWithSub\stageOmetaColor{R^{\superscriptO} }_{{\mathrm{12}}}   \ElabArrow  \possiblyWithSub\stageOmetaColor{N^{\superscriptO} }_{{\mathrm{2}}} 
    }{%
       \mathcal{G}  \mid  \mathit{\Psi}  \vdash^{0}   \openO{(} \possiblyWithSub\stageOmetaColor{\mathcal{M}^{\superscriptO} }_{{\mathrm{1}}} \ \openO{\{} \possiblyWithSub\stageOmetaColor{\mathcal{M}^{\superscriptO} }_{{\mathrm{2}}} \closeO{\} }\closeO{)}_{ \ell }   \Rightarrow  \possiblyWithSub\stageOmetaColor{R^{\superscriptO} }_{{\mathrm{12}}}  \ElabArrow   \possiblyWithSub\stageOmetaColor{N^{\superscriptO} }_{{\mathrm{1}}} \   \openO{(}  \possiblyWithSub\stageOmetaColor{N^{\superscriptO} }_{{\mathrm{0}}} \  \possiblyWithSub\stageOmetaColor{N^{\superscriptO} }_{{\mathrm{2}}}  \closeO{)}   
    }
  \\[0.7em]
    \derive[B0-FillImp]{%
       \mathcal{G}  \mid   \mathit{\Psi} , \__{ \ell }   \vdash^{0}  \possiblyWithSub\stageOmetaColor{\mathcal{M}^{\superscriptO} }_{{\mathrm{1}}}  \Rightarrow    \mathbf{fill}\ \{ \possiblyWithSub\stageOmetaColor{N^{\superscriptO} }_{{\mathrm{2}}}  :  \possiblyWithSub\stageOmetaColor{\mathcal{T}^{\superscriptO} }_{{\mathrm{2}}} \}   \relO{\to}  \possiblyWithSub\stageOmetaColor{R^{\superscriptO} }   \ElabArrow  \possiblyWithSub\stageOmetaColor{N^{\superscriptO} }_{{\mathrm{1}}} 
    }{%
       \mathcal{G}  \mid  \mathit{\Psi}  \vdash^{0}   \possiblyWithSub\stageOmetaColor{\mathcal{M}^{\superscriptO} }_{{\mathrm{1}}} \ \ordO{\_}   \Rightarrow  \possiblyWithSub\stageOmetaColor{R^{\superscriptO} }  \ElabArrow   \possiblyWithSub\stageOmetaColor{N^{\superscriptO} }_{{\mathrm{1}}} \  \possiblyWithSub\stageOmetaColor{N^{\superscriptO} }_{{\mathrm{2}}}  
    }
  \\[0.7em]
    \derive[B0-InsertImp]{%
       \mathcal{G}  \mid  \mathit{\Psi}  \vdash^{0}  \possiblyWithSub\stageOmetaColor{\mathcal{M}^{\superscriptO} }_{{\mathrm{1}}}  \Rightarrow    \mathbf{insert}\ \{ \possiblyWithSub\stageOmetaColor{N^{\superscriptO} }_{{\mathrm{2}}}  :  \possiblyWithSub\stageOmetaColor{\mathcal{T}^{\superscriptO} }_{{\mathrm{2}}} \}   \relO{\to}  \possiblyWithSub\stageOmetaColor{R^{\superscriptO} }   \ElabArrow  \possiblyWithSub\stageOmetaColor{N^{\superscriptO} }_{{\mathrm{1}}} 
    }{%
       \mathcal{G}  \mid  \mathit{\Psi}  \vdash^{0}  \possiblyWithSub\stageOmetaColor{\mathcal{M}^{\superscriptO} }_{{\mathrm{1}}}  \Rightarrow  \possiblyWithSub\stageOmetaColor{R^{\superscriptO} }  \ElabArrow   \possiblyWithSub\stageOmetaColor{N^{\superscriptO} }_{{\mathrm{1}}} \  \possiblyWithSub\stageOmetaColor{N^{\superscriptO} }_{{\mathrm{2}}}  
    }
  \end{center}
  \begin{flushleft}
    \fbox{\( \mathcal{G}  \mid  \mathit{\Psi}  \vdash^{1}  \possiblyWithSub\stageImetaColor{\mathcal{M}^{\superscriptI} }  \Rightarrow  \possiblyWithSub\stageImetaColor{R^{\superscriptI} }  \ElabArrow  \possiblyWithSub\stageImetaColor{N^{\superscriptI} } \)}
  \end{flushleft}
  \vspace{-4.25em}
  \begin{center}
  \hspace{12em}%
    \derive[B1-Esc]{%
      \mathcal{G} \mid \mathit{\Psi} \vdash^{0} \possiblyWithSub\stageOmetaColor{\mathcal{M}^{\superscriptO} } \Rightarrow (D_{\ottmv{i}} \to)_{i = 1}^{m}   \openO{\langle} \possiblyWithSub\stageImetaColor{T^{\superscriptI} } \closeO{\rangle}   \ElabArrow \possiblyWithSub\stageOmetaColor{N^{\superscriptO} }
    }{%
      \mathcal{G} \mid \mathit{\Psi} \vdash^{1}  \ordI{\sim} \possiblyWithSub\stageOmetaColor{\mathcal{M}^{\superscriptO} }  \Rightarrow (D_{\ottmv{i}} \to)_{i = 1}^{m}  \possiblyWithSub\stageImetaColor{T^{\superscriptI} }  \ElabArrow  \ordI{\sim} \possiblyWithSub\stageOmetaColor{N^{\superscriptO} } 
    }
  \\[0.7em]
    \derive[B1-Var]{%
      \mathcal{G}(\possiblyWithSub\stageImetaColor{x}) = (\possiblyWithSub\stageImetaColor{T^{\superscriptI} })^1
    \andalso
       \mathcal{G}  \mid  \mathit{\Psi}  \vdash^{1}  \possiblyWithSub\stageImetaColor{T^{\superscriptI} }  \mathrel{<:}  \possiblyWithSub\stageImetaColor{R^{\superscriptI} } 
    }{%
       \mathcal{G}  \mid  \mathit{\Psi}  \vdash^{1}   \possiblyWithSub\stageImetaColor{x}   \Rightarrow  \possiblyWithSub\stageImetaColor{R^{\superscriptI} }  \ElabArrow   \possiblyWithSub\stageImetaColor{x}  
    }
  \qquad
    \derive[B1-CstP]{%
      \ConstEnvZero(c) = \possiblyWithSub\stageImetaColor{\tau^{\superscriptI} }
    \andalso
       \mathcal{G}  \mid  \mathit{\Psi}  \vdash^{1}   \possiblyWithSub\stageImetaColor{\tau^{\superscriptI} }   \mathrel{<:}  \possiblyWithSub\stageImetaColor{R^{\superscriptI} } 
    }{%
       \mathcal{G}  \mid  \mathit{\Psi}  \vdash^{1}   \possiblyWithSub\stageImetaColor{c}   \Rightarrow  \possiblyWithSub\stageImetaColor{R^{\superscriptI} }  \ElabArrow   \possiblyWithSub\stageImetaColor{c}  
    }
  \\[0.7em]
    \derive[B1-App]{%
       \mathcal{G}  \mid   \bullet   \vdash^{1}  \possiblyWithSub\stageImetaColor{\mathcal{M}^{\superscriptI} }_{{\mathrm{2}}}  \Rightarrow   \possiblyWithSub\stageImetaColor{T^{\superscriptI} }_{{\mathrm{2}}}   \ElabArrow  \possiblyWithSub\stageImetaColor{N^{\superscriptI} }_{{\mathrm{2}}} 
    \andalso
       \mathcal{G}  \mid   \mathit{\Psi} , ( \possiblyWithSub\stageImetaColor{T^{\superscriptI} }_{{\mathrm{2}}} )^{1}_{ \ell }   \vdash^{1}  \possiblyWithSub\stageImetaColor{\mathcal{M}^{\superscriptI} }_{{\mathrm{1}}}  \Rightarrow    ( \possiblyWithSub\stageOmetaColor{N^{\superscriptO} }_{{\mathrm{0}}} \ /\  \possiblyWithSub\stageImetaColor{T^{\superscriptI} }_{{\mathrm{11}}} )^{1}   \relI{\to}  \possiblyWithSub\stageImetaColor{R^{\superscriptI} }_{{\mathrm{12}}}   \ElabArrow  \possiblyWithSub\stageImetaColor{N^{\superscriptI} }_{{\mathrm{1}}} 
    }{%
       \mathcal{G}  \mid  \mathit{\Psi}  \vdash^{1}   \openI{(} \possiblyWithSub\stageImetaColor{\mathcal{M}^{\superscriptI} }_{{\mathrm{1}}} \  \possiblyWithSub\stageImetaColor{\mathcal{M}^{\superscriptI} }_{{\mathrm{2}}} \closeI{)}_{ \ell }   \Rightarrow  \possiblyWithSub\stageImetaColor{R^{\superscriptI} }_{{\mathrm{12}}}  \ElabArrow   \possiblyWithSub\stageImetaColor{N^{\superscriptI} }_{{\mathrm{1}}} \   \ordI{\sim}  \openO{(}  \possiblyWithSub\stageOmetaColor{N^{\superscriptO} }_{{\mathrm{0}}} \   \openO{\langle} \possiblyWithSub\stageImetaColor{N^{\superscriptI} }_{{\mathrm{2}}} \closeO{\rangle}   \closeO{)}    
    }
  \end{center}
  \begin{flushleft}
    \fbox{\( \mathcal{G}  \vdash^{0}  \possiblyWithSub\stageOmetaColor{\mathcal{S}^{\superscriptO} }  \ElabArrow  \possiblyWithSub\stageOmetaColor{\mathcal{T}^{\superscriptO} } \)}
    \fbox{\( \mathcal{G}  \vdash^{1}  \possiblyWithSub\stageImetaColor{\mathcal{S}^{\superscriptI} }  \ElabArrow  \possiblyWithSub\stageImetaColor{T^{\superscriptI} } \)}
  \end{flushleft}
  \begin{center}
    \derive[BT1-Tensor]{%
       \mathcal{G}  \mid   \bullet   \vdash^{0}  \possiblyWithSub\stageOmetaColor{\mathcal{M}^{\superscriptO} }  \Rightarrow   \possiblyWithSub\stageOmetaColor{\mathcal{T}^{\superscriptO} }   \ElabArrow  \possiblyWithSub\stageOmetaColor{N^{\superscriptO} } 
    \andalso
        \lfloor  \mathcal{G} \rfloor   \vdash_{  \ell  }   \lfloor  \possiblyWithSub\stageOmetaColor{\mathcal{T}^{\superscriptO} } \rfloor   \CastArrow    \openO{\{} \possiblyWithSub\stageOmetaColor{\nu}  \relO{:}   \ttO{NatList}   \relO{\mid}     \ttO{true}    \closeO{\} }    \ElabArrow  \possiblyWithSub\stageOmetaColor{N^{\superscriptO} }_{{\mathrm{0}}} 
    }{%
       \mathcal{G}  \vdash^{1}   \openI{(}\ttI{Tensor}\ \ordI{\%} \possiblyWithSub\stageOmetaColor{\mathcal{M}^{\superscriptO} } \closeI{)}_{ \ell }   \ElabArrow   \ttI{Tensor}\ \ordI{\%}  \openO{(}  \possiblyWithSub\stageOmetaColor{N^{\superscriptO} }_{{\mathrm{0}}} \  \possiblyWithSub\stageOmetaColor{N^{\superscriptO} }  \closeO{)}   
    }
  \\[0.7em]
    \derive[BT0-Code]{%
       \mathcal{G}  \vdash^{1}  \possiblyWithSub\stageImetaColor{\mathcal{S}^{\superscriptI} }  \ElabArrow  \possiblyWithSub\stageImetaColor{T^{\superscriptI} } 
    }{%
       \mathcal{G}  \vdash^{0}   \openO{\langle} \possiblyWithSub\stageImetaColor{\mathcal{S}^{\superscriptI} } \closeO{\rangle}   \ElabArrow   \openO{\langle} \possiblyWithSub\stageImetaColor{T^{\superscriptI} } \closeO{\rangle}  
    }
  \qquad
    \derive[BT0-Imp]{%
       \mathcal{G}  \vdash^{0}  \possiblyWithSub\stageOmetaColor{\mathcal{S}^{\superscriptO} }_{{\mathrm{1}}}  \ElabArrow  \possiblyWithSub\stageOmetaColor{\mathcal{T}^{\superscriptO} }_{{\mathrm{1}}} 
    \andalso
        \mathcal{G} ,  \possiblyWithSub\stageOmetaColor{x}  : ( \possiblyWithSub\stageOmetaColor{\mathcal{T}^{\superscriptO} }_{{\mathrm{1}}} )^{0}   \vdash^{0}  \possiblyWithSub\stageOmetaColor{\mathcal{S}^{\superscriptO} }_{{\mathrm{2}}}  \ElabArrow  \possiblyWithSub\stageOmetaColor{\mathcal{T}^{\superscriptO} }_{{\mathrm{2}}} 
    }{%
       \mathcal{G}  \vdash^{0}   \openO{\{} \possiblyWithSub\stageOmetaColor{x}  \relO{:}  \possiblyWithSub\stageOmetaColor{\mathcal{S}^{\superscriptO} }_{{\mathrm{1}}} \closeO{\} } \relO{\to}  \possiblyWithSub\stageOmetaColor{\mathcal{S}^{\superscriptO} }_{{\mathrm{2}}}   \ElabArrow   \openO{\{} \possiblyWithSub\stageOmetaColor{x}  \relO{:}  \possiblyWithSub\stageOmetaColor{\mathcal{T}^{\superscriptO} }_{{\mathrm{1}}} \closeO{\} } \relO{\to}  \possiblyWithSub\stageOmetaColor{\mathcal{T}^{\superscriptO} }_{{\mathrm{2}}}  
    }
  \end{center}
  \vspace{-0.75em}%
  \caption{%
    Rules for inferring implicit arguments
    (selective; see Figures~\ref{fig:option-inference-main-full-1}
    and \ref{fig:option-inference-main-full-2} for full definition)%
  }
  \label{fig:option-inference-main}
\end{figure}
\indent
  Figure~\ref{fig:option-inference-main} displays the rules for
  the new elaboration that simultaneously performs type-checking,
  assertion insertion, and the reconstruction of implicit arguments.
  The main judgments for stage-\(b\) expressions are
  \(\mathcal{G} \mid \mathit{\Psi} \vdash^{b} \mathcal{M}^{(b)} \Rightarrow R^{(b)} \rightsquigarrow N^{(b)}\)
  (for \(b \in \{0, 1\}\)), which can be understood as
  ``under the type environment \(\mathcal{G}\) and the \dfn{application context} \(\mathit{\Psi}\),
  the expression \(\mathcal{M}^{(b)}\) is well-typed and can be elaborated to the target term \(N^{(b)}\),
  and its type is \emph{instantiated} to \(R^{(b)}\) at this context.''
  A number of new devices are introduced for this elaboration,
  so we will explain them one by one.
  First, as can be seen from the judgments,
  the syntax of target terms remains the same as \LambdaBracketCast; we keep using \(N^{(b)}\).
  Type environments~\(\mathcal{G}\) used here are normal entities
  that just associate stage-\(b\) variables with stage-\(b\) types for \(b \in \{0, 1\}\), respectively:
  \(\bnfnotab{\mathcal{G}}{  \bullet  |  \mathcal{G} ,  \possiblyWithSub\stageOmetaColor{x}  : ( \possiblyWithSub\stageOmetaColor{\mathcal{T}^{\superscriptO} } )^{0}  |  \mathcal{G} ,  \possiblyWithSub\stageImetaColor{x}  : ( \possiblyWithSub\stageImetaColor{T^{\superscriptI} } )^{1}  }\).
  The sole notable thing is that, while we continue using \(\possiblyWithSub\stageImetaColor{T^{\superscriptI} }\) for stage-\(1\) types,
  the syntax of stage-\(0\) types is extended with those for functions with implicit parameters as follows:
  \begin{align*}
    \bnf{\possiblyWithSub\stageOmetaColor{\mathcal{T}^{\superscriptO} }}{%
      \possiblyWithSub\stageOmetaColor{B} |  \ttO{Tensor}\  \possiblyWithSub\stageOmetaColor{s}  |  \openO{(} \possiblyWithSub\stageOmetaColor{x}  \relO{:}  \possiblyWithSub\stageOmetaColor{\mathcal{T}^{\superscriptO} } \closeO{)} \relO{\to}  \possiblyWithSub\stageOmetaColor{\mathcal{T}^{\superscriptO} }  |  \openO{\{} \possiblyWithSub\stageOmetaColor{x}  \relO{:}  \possiblyWithSub\stageOmetaColor{\mathcal{T}^{\superscriptO} } \closeO{\} } \relO{\to}  \possiblyWithSub\stageOmetaColor{\mathcal{T}^{\superscriptO} }  |  \openO{\langle} \possiblyWithSub\stageImetaColor{T^{\superscriptI} } \closeO{\rangle} 
    }
  \end{align*}
  Some rules in Figure~\ref{fig:option-inference-main} uses the evident injection~\( \lfloor  \possiblyWithSub\stageOmetaColor{\mathcal{T}^{\superscriptO} } \rfloor \)
  of \(\possiblyWithSub\stageOmetaColor{\mathcal{T}^{\superscriptO} }\)-types to \(\possiblyWithSub\stageOmetaColor{T^{\superscriptO} }\)-types, which simply forgets the difference between
  \( \openO{(} \possiblyWithSub\stageOmetaColor{x}  \relO{:}  \possiblyWithSub\stageOmetaColor{\mathcal{T}^{\superscriptO} }_{{\mathrm{1}}} \closeO{)} \relO{\to}  \possiblyWithSub\stageOmetaColor{\mathcal{T}^{\superscriptO} }_{{\mathrm{2}}} \) and \( \openO{\{} \possiblyWithSub\stageOmetaColor{x}  \relO{:}  \possiblyWithSub\stageOmetaColor{\mathcal{T}^{\superscriptO} }_{{\mathrm{1}}} \closeO{\} } \relO{\to}  \possiblyWithSub\stageOmetaColor{\mathcal{T}^{\superscriptO} }_{{\mathrm{2}}} \).
\par
\indent
  One of the essential devices to do ``let arguments go first'' is
  \dfn{application contexts}~\(\mathit{\Psi}\).
  When checking an expression \(\possiblyWithSub\stageOmetaColor{\mathcal{M}^{\superscriptO} }\) that is expected to be a function,
  the stack \(\mathit{\Psi}\) provides information about the outside,
  i.e., what kind of sequence is passed as arguments of \(\possiblyWithSub\stageOmetaColor{\mathcal{M}^{\superscriptO} }\):
  \begin{align*}
    \bnf{\mathit{\Psi}}{%
       \bullet  |  \mathit{\Psi} , ( \possiblyWithSub\stageOmetaColor{N^{\superscriptO} }  :  \possiblyWithSub\stageOmetaColor{\mathcal{T}^{\superscriptO} } )^{0}_{ \ell }  |  \mathit{\Psi} , \{ \possiblyWithSub\stageOmetaColor{N^{\superscriptO} }  :  \possiblyWithSub\stageOmetaColor{\mathcal{T}^{\superscriptO} } \}_{ \ell }  |  \mathit{\Psi} , \__{ \ell }  |  \mathit{\Psi} , ( \possiblyWithSub\stageImetaColor{T^{\superscriptI} } )^{1}_{ \ell } 
    }
  \end{align*}
  An entry~\((\possiblyWithSub\stageOmetaColor{N^{\superscriptO} } : \possiblyWithSub\stageOmetaColor{\mathcal{T}^{\superscriptO} })^{0}_{\ell}\) (resp.~\(\{\possiblyWithSub\stageOmetaColor{N^{\superscriptO} } : \possiblyWithSub\stageOmetaColor{\mathcal{T}^{\superscriptO} }\}^{0}_{\ell}\))
  stands for the existence of a stage-\(0\) mandatory argument
  (resp.~an explicitly specified argument for an implicit parameter)
  of type~\(\possiblyWithSub\stageOmetaColor{\mathcal{T}^{\superscriptO} }\) whose elaboration result is \(\possiblyWithSub\stageOmetaColor{N^{\superscriptO} }\).
  Unlike the original work~\cite{XieOliveiraESOP2018},
  we push actual arguments to \(\mathit{\Psi}\) as well as the types of the arguments
  to instantiate the type of the function applied to them, as explained later.
  The other two forms can be understood in the same way;
  \(\__{\ell}\) and \((\possiblyWithSub\stageImetaColor{T^{\superscriptI} })^{1}_{\ell}\)
  represents the existence of \(\ordO{\_}\) and a stage-\(1\) argument of type \(\possiblyWithSub\stageImetaColor{T^{\superscriptI} }\), respectively.
  The rules \rulename{B0-App}, \rulename{B0-AppImp}, \rulename{B0-FillImp}, and \rulename{B1-App}
  appropriately push these entries to the stack~\(\mathit{\Psi}\)
  to check the function after traversing the argument first.
  The stacked entries will be popped by some rules, such as
  \rulename{B0-Var} or \rulename{B1-Var}.
  \rulename{B0-Var} instantiates the type of the variable
  guided by~\(\mathit{\Psi}\)
  by using an auxiliary judgment~\( \mathcal{G}  \mid  \mathit{\Psi}  \vdash^{0}  \possiblyWithSub\stageOmetaColor{\mathcal{T}^{\superscriptO} }  \mathrel{<:}  \possiblyWithSub\stageOmetaColor{R^{\superscriptO} } \).
  This instantiates \(\possiblyWithSub\stageOmetaColor{\mathcal{T}^{\superscriptO} }\) to an \dfn{extended return type}~\(\possiblyWithSub\stageOmetaColor{R^{\superscriptO} }\)
  so that the domain types can match the types of the arguments passed to the variable.
  \rulename{B1-Var} and \( \mathcal{G}  \mid  \mathit{\Psi}  \vdash^{1}  \possiblyWithSub\stageImetaColor{T^{\superscriptI} }  \mathrel{<:}  \possiblyWithSub\stageImetaColor{R^{\superscriptI} } \) do basically the same thing
  for stage-\(1\) variables.
  The syntax of \(\possiblyWithSub\stageOmetaColor{R^{\superscriptO} }\) and \(\possiblyWithSub\stageImetaColor{R^{\superscriptI} }\) is defined by the following:
  \begin{center}
    \vspace{-1.5em}
    \begin{minipage}{0.33\textwidth}
      \begin{align*}
        \bnf{\possiblyWithSub\stageOmetaColor{R^{\superscriptO} }}{%
           D  \relO{\to}  \possiblyWithSub\stageOmetaColor{R^{\superscriptO} }  |  \possiblyWithSub\stageOmetaColor{\mathcal{T}^{\superscriptO} } 
        }
      \\
        \bnf{\possiblyWithSub\stageImetaColor{R^{\superscriptI} }}{%
           D  \relI{\to}  \possiblyWithSub\stageImetaColor{R^{\superscriptI} }  |  \possiblyWithSub\stageImetaColor{T^{\superscriptI} } 
        }
      \end{align*}
    \end{minipage}%
    \begin{minipage}{0.65\textwidth}
      \begin{align*}
        \bnf{D}{%
           ( \possiblyWithSub\stageOmetaColor{N^{\superscriptO} } \ /\  \possiblyWithSub\stageOmetaColor{\mathcal{T}^{\superscriptO} } )^{0}  |  \{ \possiblyWithSub\stageOmetaColor{N^{\superscriptO} } \ /\  \possiblyWithSub\stageOmetaColor{\mathcal{T}^{\superscriptO} } \}  |  ( \possiblyWithSub\stageOmetaColor{N^{\superscriptO} } \ /\  \possiblyWithSub\stageImetaColor{T^{\superscriptI} } )^{1} 
        |* \mathbf{fill}\ \{ \possiblyWithSub\stageOmetaColor{N^{\superscriptO} }  :  \possiblyWithSub\stageOmetaColor{\mathcal{T}^{\superscriptO} } \}  |  \mathbf{insert}\ \{ \possiblyWithSub\stageOmetaColor{N^{\superscriptO} }  :  \possiblyWithSub\stageOmetaColor{\mathcal{T}^{\superscriptO} } \} 
        }
      \end{align*}
    \end{minipage}
  \end{center}
  As the name suggests, extended return types basically work as types;
  \( D  \relO{\to}  \possiblyWithSub\stageOmetaColor{R^{\superscriptO} } \) can be seen as a function type
  just by regarding the \dfn{extended domain}~\(D\) as \(\possiblyWithSub\stageOmetaColor{\mathcal{T}^{\superscriptO} }\) or \( \openO{\langle} \possiblyWithSub\stageImetaColor{T^{\superscriptI} } \closeO{\rangle} \).
  The difference from usual types is that extended domains are
  equipped with a cast term or a reconstructed term that works as
  a ``feedback'' for the corresponding application site.
  \rulename{B0-App} inserts into the elaborated application
  the cast term~\(\possiblyWithSub\stageOmetaColor{N^{\superscriptO} }_{{\mathrm{0}}}\) returned by the traversal of \(\possiblyWithSub\stageOmetaColor{\mathcal{M}^{\superscriptO} }_{{\mathrm{1}}}\),
  and \rulename{B0-FillImp} compensates for the hole~\(\ordO{\_}\) with the inferred term~\(\possiblyWithSub\stageOmetaColor{N^{\superscriptO} }_{{\mathrm{2}}}\).
  \rulename{B0-InsertImp} works basically the same as \rulename{B0-FillImp},
  but it inserts the inferred term~\(\possiblyWithSub\stageOmetaColor{N^{\superscriptO} }_{{\mathrm{2}}}\) rather than filling \(\ordO{\_}\).
  Some other rules also use terms conveyed by extended domains for elaboration.
  One can easily see that the shape of \(R^{(b)}\) is determined
  basically in response to \(\mathit{\Psi}\) in each rule.
\par
\indent
  Due to the existence of \rulename{B0-InsertImp},
  one may think that
  typing derivation is not syntax-directed and thereby uninterpretable as an algorithm.
  However,
  \( \mathcal{G}  \mid  \mathit{\Psi}  \vdash^{0}  \possiblyWithSub\stageOmetaColor{\mathcal{M}^{\superscriptO} }  \Rightarrow  \possiblyWithSub\stageOmetaColor{R^{\superscriptO} }  \ElabArrow  \possiblyWithSub\stageOmetaColor{N^{\superscriptO} } \)~can actually be read as an algorithm in the following ways,
  where \((\mathcal{G}, \mathit{\Psi}, \possiblyWithSub\stageOmetaColor{\mathcal{M}^{\superscriptO} })\) and \((\possiblyWithSub\stageOmetaColor{R^{\superscriptO} }, \possiblyWithSub\stageOmetaColor{N^{\superscriptO} })\) are regarded as inputs and outputs, respectively:
  Suppose that, during the check of an expression~\(\possiblyWithSub\stageOmetaColor{\mathcal{M}^{\superscriptO} }\),
  the traversal of a subexpression~\(\possiblyWithSub\stageOmetaColor{\mathcal{M}^{\superscriptO} }_{\ottmv{i}}\) returned \((\possiblyWithSub\stageOmetaColor{R^{\superscriptO} }_{\ottmv{i}}, \possiblyWithSub\stageOmetaColor{N^{\superscriptO} }_{\ottmv{i}})\).
  \begin{enumerate}
    \item[(i)]
      If \(\possiblyWithSub\stageOmetaColor{R^{\superscriptO} }_{\ottmv{i}}\) is of the form~\(  \mathbf{insert}\ \{ \possiblyWithSub\stageOmetaColor{N'^{\superscriptO} }  :  \possiblyWithSub\stageOmetaColor{\mathcal{T}'^{\superscriptO} } \}   \relO{\to}  \possiblyWithSub\stageOmetaColor{R'^{\superscriptO} } \),
      then, the sole possible rule used directly below is \rulename{B0-InsertImp},
      and one can thereby replace the output with \((\possiblyWithSub\stageOmetaColor{R'^{\superscriptO} },  \openO{(}  \possiblyWithSub\stageOmetaColor{N^{\superscriptO} } \  \possiblyWithSub\stageOmetaColor{N'^{\superscriptO} }  \closeO{)} )\).
    \item[(ii)]
      Repeat the process~(i) until \(\possiblyWithSub\stageOmetaColor{R^{\superscriptO} }_{\ottmv{i}}\) is no longer of the form~\(  \mathbf{insert}\ \{ \possiblyWithSub\stageOmetaColor{N'^{\superscriptO} }  :  \possiblyWithSub\stageOmetaColor{\mathcal{T}'^{\superscriptO} } \}   \relO{\to}  \possiblyWithSub\stageOmetaColor{R'^{\superscriptO} } \).
      Then, one can use the rule corresponding to the form of \(\possiblyWithSub\stageOmetaColor{\mathcal{M}^{\superscriptO} }\).
  \end{enumerate}
  That is, in a broad sense, typing rules are syntax-directed
  with respect to \(\possiblyWithSub\stageOmetaColor{\mathcal{M}^{\superscriptO} }\) \emph{and all the \(\possiblyWithSub\stageOmetaColor{R^{\superscriptO} }\)'s produced by its subexpressions},
  not just to \(\possiblyWithSub\stageOmetaColor{\mathcal{M}^{\superscriptO} }\).
\par
\begin{figure}[tbp]
\small
  \begin{flushleft}
    \fbox{\( \mathcal{G}  \mid  \mathit{\Psi}  \vdash^{0}  \possiblyWithSub\stageOmetaColor{\mathcal{T}^{\superscriptO} }  \mathrel{<:}  \possiblyWithSub\stageOmetaColor{R^{\superscriptO} } \)}
  \end{flushleft}
  \vspace{-4.25em}
  \begin{center}
  \hspace{11em}%
    \derive[BI0-ImpGuess1]{%
        \lfloor  \mathcal{G} \rfloor   \vdash^{0}  \possiblyWithSub\stageOmetaColor{N^{\superscriptO} }_{{\mathrm{1}}}  :   \lfloor  \possiblyWithSub\stageOmetaColor{\mathcal{T}^{\superscriptO} }_{{\mathrm{1}}} \rfloor  
    \andalso
       \mathcal{G}  \mid  \mathit{\Psi}  \vdash^{0}    [  \possiblyWithSub\stageOmetaColor{N^{\superscriptO} }_{{\mathrm{1}}}  /  \possiblyWithSub\stageOmetaColor{x}  ]    \possiblyWithSub\stageOmetaColor{\mathcal{T}^{\superscriptO} }_{{\mathrm{2}}}   \mathrel{<:}  \possiblyWithSub\stageOmetaColor{R^{\superscriptO} }_{{\mathrm{2}}} 
    }{%
       \mathcal{G}  \mid   \mathit{\Psi} , \__{ \ell }   \vdash^{0}   \openO{\{} \possiblyWithSub\stageOmetaColor{x}  \relO{:}  \possiblyWithSub\stageOmetaColor{\mathcal{T}^{\superscriptO} }_{{\mathrm{1}}} \closeO{\} } \relO{\to}  \possiblyWithSub\stageOmetaColor{\mathcal{T}^{\superscriptO} }_{{\mathrm{2}}}   \mathrel{<:}    \mathbf{fill}\ \{ \possiblyWithSub\stageOmetaColor{N^{\superscriptO} }_{{\mathrm{1}}}  :  \possiblyWithSub\stageOmetaColor{\mathcal{T}^{\superscriptO} }_{{\mathrm{1}}} \}   \relO{\to}  \possiblyWithSub\stageOmetaColor{R^{\superscriptO} }_{{\mathrm{2}}}  
    }
  \\[0.7em]
    \derive[BI0-ImpGuess2]{%
      \text{\(\mathit{\Psi}\) is neither of the form~\(( \mathit{\Psi}' , \__{ \ell } )\) nor \(( \mathit{\Psi}' , \{ \possiblyWithSub\stageOmetaColor{N'^{\superscriptO} }  :  \possiblyWithSub\stageOmetaColor{\mathcal{T}'^{\superscriptO} } \}_{ \ell } )\)}
    \\
        \lfloor  \mathcal{G} \rfloor   \vdash^{0}  \possiblyWithSub\stageOmetaColor{N^{\superscriptO} }_{{\mathrm{1}}}  :   \lfloor  \possiblyWithSub\stageOmetaColor{\mathcal{T}^{\superscriptO} }_{{\mathrm{1}}} \rfloor  
    \andalso
       \mathcal{G}  \mid  \mathit{\Psi}  \vdash^{0}    [  \possiblyWithSub\stageOmetaColor{N^{\superscriptO} }_{{\mathrm{1}}}  /  \possiblyWithSub\stageOmetaColor{x}  ]    \possiblyWithSub\stageOmetaColor{\mathcal{T}^{\superscriptO} }_{{\mathrm{2}}}   \mathrel{<:}  \possiblyWithSub\stageOmetaColor{R^{\superscriptO} }_{{\mathrm{2}}} 
    }{%
       \mathcal{G}  \mid  \mathit{\Psi}  \vdash^{0}   \openO{\{} \possiblyWithSub\stageOmetaColor{x}  \relO{:}  \possiblyWithSub\stageOmetaColor{\mathcal{T}^{\superscriptO} }_{{\mathrm{1}}} \closeO{\} } \relO{\to}  \possiblyWithSub\stageOmetaColor{\mathcal{T}^{\superscriptO} }_{{\mathrm{2}}}   \mathrel{<:}    \mathbf{insert}\ \{ \possiblyWithSub\stageOmetaColor{N^{\superscriptO} }_{{\mathrm{1}}}  :  \possiblyWithSub\stageOmetaColor{\mathcal{T}^{\superscriptO} }_{{\mathrm{1}}} \}   \relO{\to}  \possiblyWithSub\stageOmetaColor{R^{\superscriptO} }_{{\mathrm{2}}}  
    }
  \\[0.7em]
    \derive[BI0-ImpGiven]{%
        \lfloor  \mathcal{G} \rfloor   \vdash_{  \ell  }   \lfloor  \possiblyWithSub\stageOmetaColor{\mathcal{T}'^{\superscriptO} }_{{\mathrm{1}}} \rfloor   \CastArrow   \lfloor  \possiblyWithSub\stageOmetaColor{\mathcal{T}^{\superscriptO} }_{{\mathrm{1}}} \rfloor   \ElabArrow  \possiblyWithSub\stageOmetaColor{N^{\superscriptO} }_{{\mathrm{0}}} 
    \andalso
       \mathcal{G}  \mid  \mathit{\Psi}  \vdash^{0}    [   \possiblyWithSub\stageOmetaColor{N^{\superscriptO} }_{{\mathrm{0}}} \  \possiblyWithSub\stageOmetaColor{N^{\superscriptO} }_{{\mathrm{1}}}   /  \possiblyWithSub\stageOmetaColor{x}  ]    \possiblyWithSub\stageOmetaColor{\mathcal{T}^{\superscriptO} }_{{\mathrm{2}}}   \mathrel{<:}  \possiblyWithSub\stageOmetaColor{R^{\superscriptO} }_{{\mathrm{2}}} 
    }{%
       \mathcal{G}  \mid   \mathit{\Psi} , \{ \possiblyWithSub\stageOmetaColor{N^{\superscriptO} }_{{\mathrm{1}}}  :  \possiblyWithSub\stageOmetaColor{\mathcal{T}'^{\superscriptO} }_{{\mathrm{1}}} \}_{ \ell }   \vdash^{0}   \openO{\{} \possiblyWithSub\stageOmetaColor{x}  \relO{:}  \possiblyWithSub\stageOmetaColor{\mathcal{T}^{\superscriptO} }_{{\mathrm{1}}} \closeO{\} } \relO{\to}  \possiblyWithSub\stageOmetaColor{\mathcal{T}^{\superscriptO} }_{{\mathrm{2}}}   \mathrel{<:}    \{ \possiblyWithSub\stageOmetaColor{N^{\superscriptO} }_{{\mathrm{0}}} \ /\  \possiblyWithSub\stageOmetaColor{\mathcal{T}^{\superscriptO} }_{{\mathrm{1}}} \}   \relO{\to}  \possiblyWithSub\stageOmetaColor{R^{\superscriptO} }_{{\mathrm{2}}}  
    }
  \\[0.7em]
    \derive[BI0-Arr]{%
        \lfloor  \mathcal{G} \rfloor   \vdash_{  \ell  }   \lfloor  \possiblyWithSub\stageOmetaColor{\mathcal{T}'^{\superscriptO} }_{{\mathrm{1}}} \rfloor   \CastArrow   \lfloor  \possiblyWithSub\stageOmetaColor{\mathcal{T}^{\superscriptO} }_{{\mathrm{1}}} \rfloor   \ElabArrow  \possiblyWithSub\stageOmetaColor{N^{\superscriptO} }_{{\mathrm{0}}} 
    \andalso
       \mathcal{G}  \mid  \mathit{\Psi}  \vdash^{0}    [   \possiblyWithSub\stageOmetaColor{N^{\superscriptO} }_{{\mathrm{0}}} \  \possiblyWithSub\stageOmetaColor{N^{\superscriptO} }_{{\mathrm{1}}}   /  \possiblyWithSub\stageOmetaColor{x}  ]    \possiblyWithSub\stageOmetaColor{\mathcal{T}^{\superscriptO} }_{{\mathrm{2}}}   \mathrel{<:}  \possiblyWithSub\stageOmetaColor{R^{\superscriptO} }_{{\mathrm{2}}} 
    }{%
       \mathcal{G}  \mid   \mathit{\Psi} , ( \possiblyWithSub\stageOmetaColor{N^{\superscriptO} }_{{\mathrm{1}}}  :  \possiblyWithSub\stageOmetaColor{\mathcal{T}'^{\superscriptO} }_{{\mathrm{1}}} )^{0}_{ \ell }   \vdash^{0}   \openO{(} \possiblyWithSub\stageOmetaColor{x}  \relO{:}  \possiblyWithSub\stageOmetaColor{\mathcal{T}^{\superscriptO} }_{{\mathrm{1}}} \closeO{)} \relO{\to}  \possiblyWithSub\stageOmetaColor{\mathcal{T}^{\superscriptO} }_{{\mathrm{2}}}   \mathrel{<:}    ( \possiblyWithSub\stageOmetaColor{N^{\superscriptO} }_{{\mathrm{0}}} \ /\  \possiblyWithSub\stageOmetaColor{\mathcal{T}^{\superscriptO} }_{{\mathrm{1}}} )^{0}   \relO{\to}  \possiblyWithSub\stageOmetaColor{R^{\superscriptO} }_{{\mathrm{2}}}  
    }
  \\[0.7em]
    \derive[BI0-Empty]{}{%
       \mathcal{G}  \mid   \bullet   \vdash^{0}  \possiblyWithSub\stageOmetaColor{\mathcal{T}^{\superscriptO} }  \mathrel{<:}   \possiblyWithSub\stageOmetaColor{\mathcal{T}^{\superscriptO} }  
    }
  \qquad
    \derive[BI0-Code]{%
      \mathcal{G} \mid \mathit{\Psi} \vdash^{1} \possiblyWithSub\stageImetaColor{T^{\superscriptI} } \mathrel{<:} (D_{\ottmv{i}} \to)_{i = 1}^{m}  \possiblyWithSub\stageImetaColor{T'^{\superscriptI} } 
    }{%
      \mathcal{G} \mid \mathit{\Psi} \vdash^{0}  \openO{\langle} \possiblyWithSub\stageImetaColor{T^{\superscriptI} } \closeO{\rangle}  \mathrel{<:} (D_{\ottmv{i}} \to)_{i = 1}^{m}   \openO{\langle} \possiblyWithSub\stageImetaColor{T'^{\superscriptI} } \closeO{\rangle}  
    }
  \end{center}
  \begin{flushleft}
    \fbox{\( \mathcal{G}  \mid  \mathit{\Psi}  \vdash^{1}  \possiblyWithSub\stageImetaColor{T^{\superscriptI} }  \mathrel{<:}  \possiblyWithSub\stageImetaColor{R^{\superscriptI} } \)}
  \end{flushleft}
  \vspace{-4em}
  \begin{center}
    \derive[BI1-Empty]{}{%
       \mathcal{G}  \mid   \bullet   \vdash^{1}  \possiblyWithSub\stageImetaColor{T^{\superscriptI} }  \mathrel{<:}   \possiblyWithSub\stageImetaColor{T^{\superscriptI} }  
    }
  \qquad
    \derive[BI1-Arr]{%
       \possiblyWithSub\stageImetaColor{T'^{\superscriptI} }_{{\mathrm{1}}}  \mathrel{||}^{1}  \possiblyWithSub\stageImetaColor{T^{\superscriptI} }_{{\mathrm{1}}} 
    \andalso
       \mathcal{G}  \mid  \mathit{\Psi}  \vdash^{1}  \possiblyWithSub\stageImetaColor{T^{\superscriptI} }_{{\mathrm{2}}}  \mathrel{<:}  \possiblyWithSub\stageImetaColor{R^{\superscriptI} }_{{\mathrm{2}}} 
    \\
      \possiblyWithSub\stageOmetaColor{N^{\superscriptO} } :=  \LeftAssertParen\openO{\langle} \possiblyWithSub\stageImetaColor{T'^{\superscriptI} }_{{\mathrm{1}}} \closeO{\rangle} \relO{\CastArrow} \openO{\langle} \possiblyWithSub\stageImetaColor{T^{\superscriptI} }_{{\mathrm{1}}} \closeO{\rangle}\RightAssertParen^{  \ell  } 
    }{%
       \mathcal{G}  \mid   \mathit{\Psi} , ( \possiblyWithSub\stageImetaColor{T'^{\superscriptI} }_{{\mathrm{1}}} )^{1}_{ \ell }   \vdash^{1}   \possiblyWithSub\stageImetaColor{T^{\superscriptI} }_{{\mathrm{1}}}  \relI{\to}  \possiblyWithSub\stageImetaColor{T^{\superscriptI} }_{{\mathrm{2}}}   \mathrel{<:}    ( \possiblyWithSub\stageOmetaColor{N^{\superscriptO} } \ /\  \possiblyWithSub\stageImetaColor{T^{\superscriptI} }_{{\mathrm{1}}} )^{1}   \relI{\to}  \possiblyWithSub\stageImetaColor{R^{\superscriptI} }_{{\mathrm{2}}}  
    }
  \end{center}
  \vspace{-0.75em}%
  \caption{The declarative rules for the inference of implicit arguments}
  \label{fig:option-inference-instantiation}
\end{figure}
\indent
  The core of our inference lies in
  the judgments~\( \mathcal{G}  \mid  \mathit{\Psi}  \vdash^{0}  \possiblyWithSub\stageOmetaColor{\mathcal{T}^{\superscriptO} }  \mathrel{<:}  \possiblyWithSub\stageOmetaColor{R^{\superscriptO} } \) and \( \mathcal{G}  \mid  \mathit{\Psi}  \vdash^{1}  \possiblyWithSub\stageImetaColor{T^{\superscriptI} }  \mathrel{<:}  \possiblyWithSub\stageImetaColor{R^{\superscriptI} } \).
  Figure~\ref{fig:option-inference-instantiation} first shows
  the declarative rules for these judgments.
  Guided by the application context, \rulename{BI0-Arr} and \rulename{BI0-ImpGiven}
  instantiate the codomain type~\(\possiblyWithSub\stageOmetaColor{\mathcal{T}^{\superscriptO} }_{{\mathrm{2}}}\)
  by substituting \(\possiblyWithSub\stageOmetaColor{x}\) with \( \openO{(}  \possiblyWithSub\stageOmetaColor{N^{\superscriptO} }_{{\mathrm{0}}} \  \possiblyWithSub\stageOmetaColor{N^{\superscriptO} }_{{\mathrm{1}}}  \closeO{)} \),
  i.e., the given argument wrapped by an appropriate cast function.
\par
\indent
  \rulename{BI0-ImpGuess} guesses an implicit argument~\(\possiblyWithSub\stageOmetaColor{N^{\superscriptO} }_{{\mathrm{1}}}\);
  for actual implementation, we have to infer it algorithmically.
  The intuition for defining an algorithmic version is quite simple:
  we can track a finite set~\(I\) of variables
  that should be resolved into a term through the traversal, and
  if about to insert casts for a type
  that contains unresolved variables and has some suitable structure,
  we can simply interpret that equation as a solution.
  For space reasons, the algorithmic rules are described
  in Figure~\ref{fig:algorithmic-option-inference} in Appendix.
  We mention only the most important one: \rulename{ABI0-ImpGuess}.
  It adds \(\possiblyWithSub\stageOmetaColor{x}\) to \(I\) as unresolved
  and extracts the solution for \(\possiblyWithSub\stageOmetaColor{x}\) from the substitution~\(\theta_{{\mathrm{2}}}\)
  after traversing \(\possiblyWithSub\stageOmetaColor{\mathcal{T}^{\superscriptO} }_{{\mathrm{2}}}\).
  If \(\theta_{{\mathrm{2}}}\) does not contain \(\possiblyWithSub\stageOmetaColor{x}\),
  it means that the inference failed;
  in this case, the type-checker will emit an error with the label~\(\ell\).
\par

\section{Basics of Horsea}\label{sec:surface-language}
\indent
  To lessen the cumbersomeness of manually inserting staging constructs,
  we provide a proof-of-concept surface language named \dfn{Horsea}.
  Programs in this language are translated into
  \LambdaBracketCastImplicit\ by reconstructing staging constructs.
  This section explains the basics of
  how to properly complement \(\openO{\langle}\closeO{\rangle}\) and \(\ordI{\sim}\).
  The reconstruction process is formalized as an elaboration through type-checking-like traversal.
  The target language of the elaboration is \LambdaBracketCastImplicit,
  which we have introduced in Section~\ref{sec:implicit}.
  The formalization is fairly standard;
  we can reconstruct \(\openO{\langle}\closeO{\rangle}\) and \(\ordI{\sim}\)
  by a well-known technique in the literature of partial evaluation
  called \dfn{binding-time analysis} (\dfn{BTA})~\cite{JonesGomardSestoft1993,DaviesLICS1996,DaviesJACM2017}.
  Because most part of the formalization is simply a repetition of classical results,
  precise descriptions are deferred to Appendix~\ref{sec:formalization-of-surface-language}.
  Nonetheless, care must be taken to handle dependent function types and implicit arguments in our case.
\par
\indent
  The syntax of source expressions \(E\) and type annotations \(T\)
  is quite concise:
  \begin{align*}
    \bnf{E}{%
      d | x |  \lambda  x  :  T .\  E  |  ( E_{{\mathrm{1}}} \  E_{{\mathrm{2}}} )_{ \ell } 
    |  \lambda \{ x  :  T \}.\  E  |  ( E_{{\mathrm{1}}} \ \{ E_{{\mathrm{2}}} \})_{ \ell } 
    |  E \ \_ 
    }
  \\
    \bnf{T}{%
       B  |  \{ x  :  B  \mid  E \}  |  \mathtt{Tensor}\  E  |  ( x  :  T ) \to  T 
    }
  \end{align*}
  As one can see, this is basically ``\LambdaBracketCastImplicit\ without staging constructs''.
  Here, \(d\) ranges over the set of constants (e.g.,~\( \makeIdentOrConst{}{ VmatMult } \))
  associated with corresponding ones
  in \LambdaBracketCastImplicit\ (e.g.,~\( \ordO{  \makeIdentOrConst{}{ VgenMatMult }  } \)).
\par
\indent
  BTA assigns a \dfn{binding time}~\(b\)
  to every subexpression to produce~\(\Hat{e}\) defined below:
  \begin{gather*}
    \bnfnotab{\Hat{e}}{%
       e ^{ b } 
    }
  \qquad\quad
    \bnfnotab{e}{%
      d | x |  \lambda  x  :  \Hat{\tau} .\  \Hat{e}  |  ( \Hat{e}_{{\mathrm{1}}} \  \Hat{e}_{{\mathrm{2}}} )_{ \ell } 
    |  \lambda \{ x  :  \Hat{\tau} \}.\  \Hat{e}  |  ( \Hat{e}_{{\mathrm{1}}} \ \{ \Hat{e}_{{\mathrm{2}}} \})_{ \ell }  |  \Hat{e} \ \_ 
    }
  \\
    \bnfnotab{\Hat{\tau}}{%
       \tau ^{ b } 
    }
  \qquad\quad
    \bnfnotab{\tau}{%
      B |  \mathtt{Tensor}\  \Hat{e}  |  ( x  :  \Hat{\tau} ) \to  \Hat{\tau} 
    }
  \qquad\quad
    \bnfnotab{b}{%
       0  |  1 
    }
  \end{gather*}
  By the binding times (\(\approx\) stages) assigned to subexpressions,
  we can evidently reconstruct
  brackets~\(\openO{\langle}\closeO{\rangle}\) and escapes~\(\ordI{\sim}\)
  by inserting them at each gap of the two binding times on the syntax tree,
  as described in Figure~\ref{fig:reconstruction-of-staging-construct} in Appendix.
  The assignment can be done by extracting binding-time constraints from programs and solving them.
  For example, if the given non-staged program contains a subexpression
  \( (  \lambda    \makeIdentOrConst{}{ Vx }    :   \mathtt{Mat}\     \makeIdentOrConst{}{ Vm }    \   (     \makeIdentOrConst{}{ Vn }     +     \makeIdentOrConst{}{ C1 }     )   .\   \ldots   ) \), then we can extract the information that
  variables~\( \makeIdentOrConst{}{ Vm } \) and \( \makeIdentOrConst{}{ Vn } \) must be bound at stage~\(0\) and
  that the type annotation~\( \mathtt{Mat}\     \makeIdentOrConst{}{ Vm }    \   (     \makeIdentOrConst{}{ Vn }     +     \makeIdentOrConst{}{ C1 }     )  \) must live in stage~\(1\),
  in order to appropriately transform the program into the staged core language.
  Such constraints can be extracted in a type-checking-like manner,
  as explained in Appendix~\ref{sec:formalization-of-surface-language}.
\par

\section{Further Discussions}\label{sec:further-discussions}
\indent
  Although the formalization we have given so far basically satisfies our goal,
  we still have room for improvement for real-world use.
  This section briefly touch on some of such aspects.
\par
\subsection{Adding Conditionals Is Unexpectedly Non-Trivial But Viable}
\indent
  In addition to the constructs in our formalization,
  recursive functions and conditionals are desiderata for real-world use.
  While adding the former does not incur much difficulty as to typing,
  designing a rule for stage-\(0\) \(\tokenO{if}\)-expressions is
  actually not as straightforward as expected.
  One may imagine that something like the following \(\tokenO{then}\)-biased rule would work:
  \begin{center}
  \small
    \derive{%
       \mathit{\Gamma}  \vdash^{0}  \possiblyWithSub\stageOmetaColor{M^{\scriptscriptstyle(0)} }_{{\mathrm{0}}}  :    \ttO{Bool}    \ElabArrow  \possiblyWithSub\stageOmetaColor{N^{\superscriptO} }_{{\mathrm{0}}} 
    \andalso
       \mathit{\Gamma}  \vdash^{0}  \possiblyWithSub\stageOmetaColor{M^{\scriptscriptstyle(0)} }_{\ottmv{i}}  :  \possiblyWithSub\stageOmetaColor{T^{\superscriptO} }_{\ottmv{i}}  \ElabArrow  \possiblyWithSub\stageOmetaColor{N^{\superscriptO} }_{\ottmv{i}} 
      \ \text{(for \(i \in \{1, 2\}\))}
    \andalso
       \mathit{\Gamma}  \vdash_{  \ell  }  \possiblyWithSub\stageOmetaColor{T^{\superscriptO} }_{{\mathrm{2}}}  \CastArrow  \possiblyWithSub\stageOmetaColor{T^{\superscriptO} }_{{\mathrm{1}}}  \ElabArrow  \possiblyWithSub\stageOmetaColor{N'^{\superscriptO} } 
    }{%
       \mathit{\Gamma}  \vdash^{0}   \openO{(}\tokenO{if}\  \possiblyWithSub\stageOmetaColor{M^{\scriptscriptstyle(0)} }_{{\mathrm{0}}} \ \tokenO{then}\  \possiblyWithSub\stageOmetaColor{M^{\scriptscriptstyle(0)} }_{{\mathrm{1}}} \ \tokenO{else}\  \possiblyWithSub\stageOmetaColor{M^{\scriptscriptstyle(0)} }_{{\mathrm{2}}} \closeO{)}^{ \ell }   :  \possiblyWithSub\stageOmetaColor{T^{\superscriptO} }_{{\mathrm{1}}}  \ElabArrow   \tokenO{if}\  \possiblyWithSub\stageOmetaColor{N^{\superscriptO} }_{{\mathrm{0}}} \ \tokenO{then}\  \possiblyWithSub\stageOmetaColor{N^{\superscriptO} }_{{\mathrm{1}}} \ \tokenO{else}\    \possiblyWithSub\stageOmetaColor{N'^{\superscriptO} } \  \possiblyWithSub\stageOmetaColor{N^{\superscriptO} }_{{\mathrm{2}}}    
    }
  \end{center}
  However, this is not the actual way we want to check conditionals;
  for example, it cannot handle programs like the following:
  {\small\begin{align*}
     \progindent{  \tokenO{let}\ \tokenO{rec}\   \ordO{  \makeIdentOrConst{}{ VgenRep }  }   \ \openO{(}  \ordO{  \makeIdentOrConst{}{ Vm }  }   \relO{:}    \ttO{Nat}   \closeO{)}  \ \openO{(}  \ordO{  \makeIdentOrConst{}{ Vn }  }   \relO{:}    \ttO{Nat}   \closeO{)}  \relO{:}   \openO{\langle}   \ttI{Vec}\ \ordI{\%}   \ordO{  \makeIdentOrConst{}{ Vm }  }     \relI{\to}   \ttI{Vec}\ \ordI{\%}  \openO{(}    \ordO{  \makeIdentOrConst{}{ Vm }  }   \binO{  \ast  }   \ordO{  \makeIdentOrConst{}{ Vn }  }    \closeO{)}    \closeO{\rangle}      \relO{=}   \openO{\langle}  \ordI{\lambda}  \ordI{  \makeIdentOrConst{}{ Vv }  }   \relI{:}   \ttI{Vec}\ \ordI{\%}   \ordO{  \makeIdentOrConst{}{ Vm }  }    \punctI{.}\control\deepen{\control\br{}   \ordI{\sim}  \openO{(}  \tokenO{if}\     \ordO{  \makeIdentOrConst{}{ Vn }  }   \binO{  \leq  }   \ordO{  \makeIdentOrConst{}{ C0 }  }    \ \tokenO{then}\   \openO{\langle}   \ordI{  \makeIdentOrConst{}{ VvecNil }  }   \closeO{\rangle}  \ \tokenO{else}\   \openO{\langle}    \ordI{\sim}  \openO{(}     \ordO{  \makeIdentOrConst{}{ VgenVecCat }  }   \ \openO{\{}   \ordO{  \makeIdentOrConst{}{ Vm }  }   \closeO{\} }  \ \openO{\{}   \openO{(}    \ordO{  \makeIdentOrConst{}{ Vn }  }   \binO{  -  }   \ordO{  \makeIdentOrConst{}{ C1 }  }    \closeO{)}  \binO{  \ast  }   \ordO{  \makeIdentOrConst{}{ Vm }  }    \closeO{\} }  \closeO{)}   \    \ordI{  \makeIdentOrConst{}{ Vv }  }    \   \ordI{\sim}  \openO{(}      \ordO{  \makeIdentOrConst{}{ VgenRep }  }   \    \ordO{  \makeIdentOrConst{}{ Vm }  }    \   \openO{(}    \ordO{  \makeIdentOrConst{}{ Vn }  }   \binO{  -  }   \ordO{  \makeIdentOrConst{}{ C1 }  }    \closeO{)}   \   \openO{\langle}   \ordI{  \makeIdentOrConst{}{ Vv }  }   \closeO{\rangle}   \closeO{)}    \closeO{\rangle}   \closeO{)}    }  \closeO{\rangle}   } 
  \end{align*}}\noindent
  Here, \( \ordI{  \makeIdentOrConst{}{ VvecNil }  } \) is a built-in constant for the vector of length~\(0\),
  and \( \ordO{  \makeIdentOrConst{}{ VgenVecCat }  } \) is the one that produces code of vector concatenation functions.
  The application \(     \ordO{  \makeIdentOrConst{}{ VgenRep }  }    \    \ordO{  \makeIdentOrConst{}{ Vm }  }    \    \ordO{  \makeIdentOrConst{}{ Vn }  }   \) produces code of a function
  that takes a vector of length~\(  \ordO{  \makeIdentOrConst{}{ Vm }  }  \) and duplicate it \(  \ordO{  \makeIdentOrConst{}{ Vn }  }  \)~times
  to make the vector of length~\( \openO{(}     \ordO{  \makeIdentOrConst{}{ Vm }  }    \binO{  \ast  }   \ordO{  \makeIdentOrConst{}{ Vn }  }    \closeO{)} \).
  Because the \(\tokenO{then}\)-branch and the \(\tokenO{else}\)-branch
  have type \(\possiblyWithSub\stageOmetaColor{T^{\superscriptO} }_{{\mathrm{1}}} :=  \openO{\langle}  \ttI{Vec}\ \ordI{\%}   \ordO{  \makeIdentOrConst{}{ C0 }  }    \closeO{\rangle} \)
  and \(\possiblyWithSub\stageOmetaColor{T^{\superscriptO} }_{{\mathrm{2}}} :=  \openO{\langle}  \ttI{Vec}\ \ordI{\%}  \openO{(}    \ordO{  \makeIdentOrConst{}{ Vm }  }   \binO{  +  }  \openO{(}   \openO{(}    \ordO{  \makeIdentOrConst{}{ Vn }  }   \binO{  -  }   \ordO{  \makeIdentOrConst{}{ C1 }  }    \closeO{)}  \binO{  \ast  }   \ordO{  \makeIdentOrConst{}{ Vm }  }    \closeO{)}   \closeO{)}   \closeO{\rangle} \), respectively,
  the inserted assertion~\(\LeftAssertParen \possiblyWithSub\stageOmetaColor{T^{\superscriptO} }_{{\mathrm{2}}} \relO{\CastArrow} \possiblyWithSub\stageOmetaColor{T^{\superscriptO} }_{{\mathrm{1}}}\RightAssertParen^{L}\)
  will pass only when either \(  \ordO{  \makeIdentOrConst{}{ Vm }  }  \) or \(  \ordO{  \makeIdentOrConst{}{ Vn }  }  \) equals \( \ordO{  \makeIdentOrConst{}{ C0 }  } \).
  This is clearly different from the intention;
  what we wanted to assert here is that both branches have type \( \openO{\langle}  \ttI{Vec}\ \ordI{\%}  \openO{(}    \ordO{  \makeIdentOrConst{}{ Vm }  }   \binO{  \ast  }   \ordO{  \makeIdentOrConst{}{ Vn }  }    \closeO{)}   \closeO{\rangle} \).
\par
\indent
  This can be resolved by some rule that ``merges'' two types by conditionals as follows:
  \begin{center}
  \small
    \derive{
       \mathit{\Gamma}  \vdash^{0}  \possiblyWithSub\stageOmetaColor{M^{\scriptscriptstyle(0)} }_{{\mathrm{0}}}  :    \ttO{Bool}    \ElabArrow  \possiblyWithSub\stageOmetaColor{N^{\superscriptO} }_{{\mathrm{0}}} 
    \andalso
       \mathit{\Gamma}  \vdash^{0}  \possiblyWithSub\stageOmetaColor{M^{\scriptscriptstyle(0)} }_{\ottmv{i}}  :   \openO{\langle}  \ttI{Vec}\ \ordI{\%} \possiblyWithSub\stageOmetaColor{N'^{\superscriptO} }_{\ottmv{i}}  \closeO{\rangle}   \ElabArrow  \possiblyWithSub\stageOmetaColor{N^{\superscriptO} }_{\ottmv{i}} 
      \ \text{(for \(i \in \{1, 2\}\))}
    }{
      \begin{aligned}
        \mathit{\Gamma} \vdash^{0}  \openO{(}\tokenO{if}\  \possiblyWithSub\stageOmetaColor{M^{\scriptscriptstyle(0)} }_{{\mathrm{0}}} \ \tokenO{then}\  \possiblyWithSub\stageOmetaColor{M^{\scriptscriptstyle(0)} }_{{\mathrm{1}}} \ \tokenO{else}\  \possiblyWithSub\stageOmetaColor{M^{\scriptscriptstyle(0)} }_{{\mathrm{2}}} \closeO{)}^{ \ell }  :  \openO{\langle}  \ttI{Vec}\ \ordI{\%}  \openO{(}  \tokenO{if}\  \possiblyWithSub\stageOmetaColor{N^{\superscriptO} }_{{\mathrm{0}}} \ \tokenO{then}\  \possiblyWithSub\stageOmetaColor{N'^{\superscriptO} }_{{\mathrm{1}}} \ \tokenO{else}\  \possiblyWithSub\stageOmetaColor{N'^{\superscriptO} }_{{\mathrm{2}}}  \closeO{)}   \closeO{\rangle} 
        \qquad&
      \\
        \ElabArrow  \tokenO{if}\  \possiblyWithSub\stageOmetaColor{N^{\superscriptO} }_{{\mathrm{0}}} \ \tokenO{then}\  \possiblyWithSub\stageOmetaColor{N^{\superscriptO} }_{{\mathrm{1}}} \ \tokenO{else}\  \possiblyWithSub\stageOmetaColor{N^{\superscriptO} }_{{\mathrm{2}}} 
        &
      \end{aligned}
    }
  \end{center}
  Such merging can be generalized to arbitrary compatible pairs of types.
\par
\subsection{Handling Tensors with ``Essentially Dynamic'' Shapes}\label{subsec:dynamic-shape}
\indent
  As we have seen so far, our language is basically designed so that all the computations
  will be specialized to certain tensor shapes at compile time.
  However, sometimes one wants programs to deal with tensors
  whose shapes are essentially unknown at compile time and available only at runtime.
  Indeed, it would be quite common, for example, to set up a server-side application
  that can receive image files (\(\approx\) matrices) of arbitrary sizes by request from users
  to perform some computation on them using tensors.
  To this end, with the aid of the \( \tokenO{run} \)-primitive~\cite{%
    TahaSheardPEPM1997,%
    TsukadaIgarashi2009,%
    HanadaIgarashi2014,%
    KiselyovFLOPS2014},
  our language can also handle tensors with essentially dynamic shapes to some extent
  within its design principle.
  This is not completely free from runtime size mismatch,
  but even if a failure happens, it will be emitted immediately,
  not during actual tensor computation,
  and thus the user can still avoid wasting time.
\par
\indent
  The \( \tokenO{run} \)-primitive is a special construct
  that can unlift code fragments like the following:
  {\small\begin{align*}
  &
        \tokenO{let}\   \ordO{  \makeIdentOrConst{}{ Vg }  }   \ \openO{(}  \ordO{  \makeIdentOrConst{}{ Vc }  }   \relO{:}   \openO{\langle}   \ttI{Int}   \closeO{\rangle}  \closeO{)}  \empty    \relO{=}    \tokenO{run}  \   \openO{\langle}   \ordI{\lambda}  \ordI{  \makeIdentOrConst{}{ Vx }  }   \relI{:}    \ttI{Int}   \punctI{.}\    \ordI{  \makeIdentOrConst{}{ Vx }  }    \binI{  +  }  \ordI{\sim}   \ordO{  \makeIdentOrConst{}{ Vc }  }     \closeO{\rangle}    \ \tokenO{in}\    \ordO{  \makeIdentOrConst{}{ Vg }  }    \   \openO{\langle}   \ordI{  \makeIdentOrConst{}{ C42 }  }   \closeO{\rangle}   \    \ordO{  \makeIdentOrConst{}{ C57 }  }   
  \\&\longrightarrow^{0\,\ast}
      \openO{(}   \tokenO{run}  \   \openO{\langle}   \ordI{\lambda}  \ordI{  \makeIdentOrConst{}{ Vx }  }   \relI{:}    \ttI{Int}   \punctI{.}\    \ordI{  \makeIdentOrConst{}{ Vx }  }    \binI{  +  }   \ordI{  \makeIdentOrConst{}{ C42 }  }    \closeO{\rangle}   \closeO{)}  \    \ordO{  \makeIdentOrConst{}{ C57 }  }   
  \longrightarrow^{0}
      \openO{(}   \ordO{\lambda}  \ordO{  \makeIdentOrConst{}{ Vx }  }   \relO{:}    \ttO{Int}   \punctO{.}\     \ordO{  \makeIdentOrConst{}{ Vx }  }     \binO{  +  }   \ordO{  \makeIdentOrConst{}{ C42 }  }    \closeO{)}  \    \ordO{  \makeIdentOrConst{}{ C57 }  }   
  \longrightarrow^{0\,\ast}
      \ordO{  \makeIdentOrConst{}{ C99 }  } 
  \end{align*}}%
  Note that, in general, adding \( \tokenO{run} \) could require ingenious modification to the type system.
  This is because one cannot always unlift code fragments;
  even if the program is well-typed under na\"ive typing for staging,
  variables occurring in code fragments passed to \( \tokenO{run} \) might be locally unbound\footnote{%
    For example, the following term gets stuck:
    \(
        \tokenO{let}\   \ordO{  \makeIdentOrConst{}{ Vh }  }   \ \openO{(}  \ordO{  \makeIdentOrConst{}{ Vt }  }   \relO{:}    \ttO{Bool}   \closeO{)}  \empty    \relO{=}   \openO{\langle}   \ordI{  \makeIdentOrConst{}{ C0 }  }   \closeO{\rangle}   \ \tokenO{in}\   \openO{\langle}  \ordI{\lambda}  \ordI{  \makeIdentOrConst{}{ Vb }  }   \relI{:}    \ttI{Bool}   \punctI{.}\   \ordI{\sim}  \openO{(}    \ordO{  \makeIdentOrConst{}{ Vh }  }   \   \openO{(}   \tokenO{run}  \   \openO{\langle}   \ordI{  \makeIdentOrConst{}{ Vb }  }   \closeO{\rangle}   \closeO{)}   \closeO{)}    \closeO{\rangle}  
    \).
  }.
  However, this can basically be solved by a method orthogonal to ours,
  and also, as explained later, we do not have to care too much about this in our setting.
\par
\indent
  The basic idea to achieve the dynamic feature by using \( \tokenO{run} \) is quite easy:
  when receiving some form of a tensor whose shape has not been fixed at compile time,
  we can (1)~generate code of the necessary function specialized for its shape,
  (2)~lift the tensor to the upper stage and embed it as an argument of the produced function,
  and (3)~run the code to compute the final result.
  An implementation for doing this would be something like the following:
  {\small\begin{align*}
  &
    \tokenO{let}\ \itO{handle}\ \openO{(}\itO{out} \relO{:} \ttO{FilePath}\closeO{)}\ \openO{(}\itO{dynMat} \relO{:} \ttO{List}\ \openO{(}\ttO{List}\ \ttO{Int}\closeO{)}\closeO{)} \relO{=}
      \tokenO{match}\ \itO{rectSize}\ \itO{dynMat}\ \tokenO{with}
  \\&\quad
      \ordO{|}\ \ttO{None} \relO{\to} \itO{printError}\ \ordO{\text{\texttt{``The given matrix is not rectangular''}}}
  \\&\quad
      \ordO{|}\ \ttO{Some}\ \openO{(}\ordO{m}\punctO{,} \ordO{n}\closeO{)} \relO{\to}
        \tokenO{match}\ \tokenO{run}
          \ \openI{\langle}
            \ordI{\sim}\openO{(}\itO{forgetSize}\ \ordO{n}\ \ordO{m}\closeO{)}
            \ \openI{(}\ordI{\sim}\openO{(}\itO{genF}\ \ordO{m}\ \ordO{n}\closeO{)}
              \ \ordI{\sim}\ordO{(}\itO{liftMat}\ \ordO{m}\ \ordO{n}\ \itO{dynMat}\closeO{)}\closeI{)}\closeI{\rangle}\ \tokenO{with}
      \\&\quad\quad
        \ordO{|}\ \ttO{Error}\ \itO{msg} \relO{\to} \itO{printError}\ \openO{(}\ordO{\text{\texttt{``Bug: ''}}} \binO{+\!+} \itO{msg}\closeO{)}
      \quad
        \ordO{|}\ \ttO{Ok}\ \itO{result} \relO{\to} \itO{writeMatrixToFile}\ \itO{out}\ \itO{result}
  \end{align*}}\noindent
  The function~\(  \ordO{  \makeIdentOrConst{}{ VrectSize }  }   :   \ttO{List}\   \openO{(}  \ttO{List}\    \ttO{Int}    \closeO{)}    \relO{\to}   \ttO{Option}\   \openO{(}    \ttO{Nat}    \binO{\ast}    \ttO{Nat}    \closeO{)}   \)
  first takes a dynamic matrix
  and returns its size~\(\openO{(}  \ordO{  \makeIdentOrConst{}{ Vm }  }  \punctO{,}   \ordO{  \makeIdentOrConst{}{ Vn }  }  \closeO{)}\)
  wrapped by \(\ttO{Some}\) if it is rectangular
  (or returns \(\ttO{None}\) otherwise).
  Then, by using this size,
  \(  \ordO{  \makeIdentOrConst{}{ VgenF }  }  \) produces code of the function necessary for the user's purpose,
  and \(  \ordO{  \makeIdentOrConst{}{ VliftMat }  }   :   \openO{(}  \ordO{  \makeIdentOrConst{}{ Vp }  }   \relO{:}    \ttO{Nat}   \closeO{)} \relO{\to}   \openO{(}  \ordO{  \makeIdentOrConst{}{ Vq }  }   \relO{:}    \ttO{Nat}   \closeO{)} \relO{\to}   \ttO{List}\   \openO{(}  \ttO{List}\    \ttO{Int}    \closeO{)}      \relO{\to}   \openO{\langle}  \ttI{Mat}\ \ordI{\%}   \ordO{  \makeIdentOrConst{}{ Vp }  }   \ \ordI{\%}   \ordO{  \makeIdentOrConst{}{ Vq }  }    \closeO{\rangle}  \)
  lifts the matrix to code.
  As an example, we here suppose the case where the function produced by \(  \ordO{  \makeIdentOrConst{}{ VgenSomeFun }  }  \) returns
  a matrix of the transposed size, so the type of \(  \ordO{  \makeIdentOrConst{}{ VgenSomeFun }  }  \) is
  \( \openO{(}  \ordO{  \makeIdentOrConst{}{ Vp }  }   \relO{:}    \ttO{Nat}   \closeO{)} \relO{\to}   \openO{(}  \ordO{  \makeIdentOrConst{}{ Vq }  }   \relO{:}    \ttO{Nat}   \closeO{)} \relO{\to}   \openO{\langle}   \ttI{Mat}\ \ordI{\%}   \ordO{  \makeIdentOrConst{}{ Vp }  }   \ \ordI{\%}   \ordO{  \makeIdentOrConst{}{ Vq }  }     \relI{\to}   \ttI{Mat}\ \ordI{\%}   \ordO{  \makeIdentOrConst{}{ Vq }  }   \ \ordI{\%}   \ordO{  \makeIdentOrConst{}{ Vp }  }     \closeO{\rangle}   \).
  After the main computation, we use
  \(  \ordO{  \makeIdentOrConst{}{ VforgetSize }  }   :  \openO{(}  \ordO{  \makeIdentOrConst{}{ Vj }  }   \relO{:}    \ttO{Nat}   \closeO{)} \relO{\to}   \openO{(}  \ordO{  \makeIdentOrConst{}{ Vk }  }   \relO{:}    \ttO{Nat}   \closeO{)} \relO{\to}   \openO{\langle}   \ttI{Mat}\ \ordI{\%}   \ordO{  \makeIdentOrConst{}{ Vj }  }   \ \ordI{\%}   \ordO{  \makeIdentOrConst{}{ Vk }  }     \relI{\to}   \ttI{List}\   \openI{(}  \ttI{List}\    \ttI{Int}    \closeI{)}    \closeO{\rangle}   \)
  so that the size of the matrix will be discarded from the type.
  Finally, by using \( \tokenO{run} \), we run the code constructed so far.
  Here, \( \tokenO{run} \) wraps resulting values with a sum type equivalent to OCaml's \(\mathtt{result}\),
  and it will return an error message if some shape mismatch or unlifting failure has happened.
  Namely, we cannot perfectly eliminate the possibility of runtime shape mismatch after all,
  but the important point here is that,
  even if the implementation of \(  \ordO{  \makeIdentOrConst{}{ VgenSomeFun }  }  \) causes a shape mismatch for the given size
  (or the occurrence of locally unbound variables),
  \( \tokenO{run} \) emit errors \emph{immediately} for typical cases;
  such errors happen during code generation or unlifting, not during heavy computation involving tensors.
  Thus, owing to staging,
  one can still successfully avoid wasting time even if the program contains some shape mismatch.
\par

\section{Implementation Report}\label{sec:implementation}
\indent
  Based on the formalization we have given so far,
  we implemented a prototype type-checker in Haskell
  and made it publicly available~\cite{HorseaRepository,HorseaArchive}.
  This type-checker accepts
  both programs in \LambdaBracketCastImplicit\ and those in Horsea;
  when receiving a program in Horsea,
  the type-checker internally converts it to the one
  in \LambdaBracketCastImplicit\ in a manner
  explained in Section~\ref{sec:surface-language}.
\par
\indent
  To support realistic examples,
  the type-checker extends various aspects of the languages,
  such as some of the features mentioned in Section~\ref{sec:further-discussions}.
  Specifically, to handle example programs of \texttt{ocaml-torch}~\cite{OCamlTorch},
  an OCaml binding of PyTorch~\cite{PaszkeGrossChintalaChananYangDevitoLinDesmaisonAntigaLerer2017},
  we utilize stage-\(0\) refinement types
  to express \dfn{broadcasting}~\cite{%
    NumPyBroadcastingSemantics,%
    PyTorchBroadcastingSemantics}
  of tensors.
  Broadcasting is a kind of implicit conversion of tensors,
  and the use of refinement types for this purpose is
  largely inspired by GraTen~\cite{HattoriKobayashiSatoESOP2023}.
\par
\indent
  Our aims for implementing this are the following:
  (1)~%
    Because our method relies on the observation that
    tensor shapes are ``not very dynamic''
    (i.e.,~that typical programs contain a limited number of operations
    where the resulting shape can be determined only at runtime),
    it is unclear whether our method sufficiently accommodates realistic programs.
    By porting examples offered by \texttt{ocaml-torch}~\cite{OCamlTorch}
    into our language, we substantiate the applicability of our method.
  (2)~%
    We demonstrate that our inference algorithm can reconstruct
    sufficiently many implicit arguments for the ported example programs.
  (3)~%
    While the core language is proven to be type-safe,
    its extension with implicit arguments and the surface language
    have yet to establish a mathematical guarantee
    (although they are heavily inspired by existing methods that fulfill type safety).
    By feeding various programs to the type-checker,
    we exemplify that the extensions indeed work fine.
\par
\indent
  Figure~\ref{fig:built-in-functions-in-implementation}
  displays the declaration of some built-in values,
  where \(\texttt{val}\ \texttt{\textasciitilde}x\ \ldots\)
  and \(\texttt{val}\ x\ \ldots\) bind \(x\) at stage~\(0\) and \(1\), respectively,
  and \(\texttt{\&(}\cdots\texttt{)}\) denotes a bracket~\(\openO{\langle} \cdots \closeO{\rangle}\).
  Here, \((M\texttt{ as }T)\) is a construct for manually inserting assertions,
  i.e.,~works like the following\footnote{%
    The actual implementation as to \(\tokenO{as}\)-expressions
    is, however, somewhat different from the rule shown here.
    It is rather something that enforces
    the so-called \dfn{checking mode} of bidirectional type-checking~\cite{DunfieldNeel2021}.
  }:
  {\small\begin{center}
    \derive{%
       \mathit{\Gamma}  \vdash^{0}  \possiblyWithSub\stageOmetaColor{M^{\scriptscriptstyle(0)} }_{{\mathrm{1}}}  :  \possiblyWithSub\stageOmetaColor{T^{\superscriptO} }_{{\mathrm{1}}}  \ElabArrow  \possiblyWithSub\stageOmetaColor{N^{\superscriptO} }_{{\mathrm{1}}} 
    \andalso
       \mathit{\Gamma}  \vdash^{0}  \possiblyWithSub\stageOmetaColor{S^{\superscriptO} }_{{\mathrm{2}}}  \ElabArrow  \possiblyWithSub\stageOmetaColor{T^{\superscriptO} }_{{\mathrm{2}}} 
    \andalso
       \mathit{\Gamma}  \vdash_{  \ell  }  \possiblyWithSub\stageOmetaColor{T^{\superscriptO} }_{{\mathrm{1}}}  \CastArrow  \possiblyWithSub\stageOmetaColor{T^{\superscriptO} }_{{\mathrm{2}}}  \ElabArrow  \possiblyWithSub\stageOmetaColor{N^{\superscriptO} }_{{\mathrm{0}}} 
    }{%
       \mathit{\Gamma}  \vdash^{0}   ( \possiblyWithSub\stageOmetaColor{M^{\scriptscriptstyle(0)} }_{{\mathrm{1}}} \ \tokenO{as}\  \possiblyWithSub\stageOmetaColor{S^{\superscriptO} }_{{\mathrm{2}}} )^{ \ell }   :  \possiblyWithSub\stageOmetaColor{T^{\superscriptO} }_{{\mathrm{2}}}  \ElabArrow   \possiblyWithSub\stageOmetaColor{N^{\superscriptO} }_{{\mathrm{0}}} \  \possiblyWithSub\stageOmetaColor{N^{\superscriptO} }_{{\mathrm{1}}}  
    }
  \end{center}}\noindent
  The function~\texttt{broadcast} takes two tensor shapes,
  and, if the two shapes are compatible
  (i.e.,~if tensors of the two shapes can be injected to one common shape),
  it returns that common shape, or raises a failure otherwise.
  Although having \texttt{broadcast} suffices for expressing broadcasting,
  expecting better error localization, we also provide \texttt{broadcastable},
  a function that judges whether given two shapes are compatible,
  and use it in refinement predicates.
\par
\begin{figure}[tb]
\begin{lstlisting}
val ~broadcast : List Nat -> List Nat -> List Nat external ...
val ~broadcastable : List Nat -> List Nat -> Bool external ...
...
module Tensor = struct
  val f : Float -> Tensor %[] external ...
  val ~gen_zeros : (s : List Nat) -> &(Tensor %s) external ...
  val ~gen_add : {x : List Nat} -> {y : {s : List Nat | broadcastable x s}} ->
    &(Tensor %x -> Tensor %y -> Tensor %(broadcast x y)) external ...
  val ~gen_grad : {s : List Nat} -> &(Tensor %s -> Tensor %s) external ...
  val ~gen_zero_grad : {s : List Nat} -> &(Tensor %s -> Unit) external ...
  val ~gen_mm : {a : Nat} -> {b : Nat} -> {c : Nat} ->
    &(Mat %a %b -> Mat %b %c -> Mat %a %c) external ...
  val ~gen_cross_entropy_for_logits :
    {s : {v : List Nat | List.length v == 2}} ->
    &(Tensor %s -> Tensor %[List.nth 0 s] -> Tensor %[]) external ...
  ...
end
module MnistHelper = struct
  val ~image_dim = (28 * 28) as Nat
  val ~num_train_images = 60000 as Nat
  ...
  val train_images : Tensor %[num_train_images, image_dim] external ...
  val train_labels : Tensor %[num_train_images] external ...
  ...
end
\end{lstlisting}
  \vspace{-1em}
  \caption{An excerpt of declarations about built-in functions}
  \label{fig:built-in-functions-in-implementation}
\end{figure}
\indent
  Figure~\ref{fig:code-example} shows
  an example program~\texttt{mnist/linear.hrs} in Horsea,
  which is ported from an example program~\texttt{mnist/linear.ml}
  offered by \texttt{ocaml-torch}~\cite{OCamlTorch}.
  This program tries linear regression for the well-known MNIST dataset~\cite{Deng2012}.
  Built-in functions used in this example are mapped to those of the core language like the following:
  \begin{gather*}
    {\small\texttt{Tensor.(+)}} \mapsto {\small\texttt{Tensor.gen\_add}},
  \qquad
    {\small\texttt{Tensor.mm}} \mapsto {\small\texttt{Tensor.gen\_mm}},
  \\
    {\small\texttt{Tensor.cross\_entropy\_for\_logits}} \mapsto
      {\small\texttt{Tensor.gen\_cross\_entropy\_for\_logits}},
  \qquad
    \ldots
  \end{gather*}
  Interestingly, the type-checker successfully reconstructs all the 20 implicit arguments in this program;
  several functions, such as \texttt{+} and \texttt{mm}, have implicit parameters,
  and our inference algorithm can compensate all of them,
  in combination with the interface of modules such as \texttt{MnistHelper}.
  Compared to the original program in OCaml,
  the essential differences are only three annotations highlighted by
  a \texttt{\textbf{\textcolor{blue}{bold blue}}} typeface;
  simply adding these three suffices for tensor shape checking for this case.
  The first two are direct annotations for tensor shapes,
  and the last one, \texttt{\textbf{\textcolor{blue}{lift\_int}}}, is
  an annotation for BTA that turns compile-time integers into ones available at runtime as well.
\par
\begin{figure}[tb]
\begin{lstlisting}
let learning_rate = Tensor.f 1.0 in
let ws = Tensor.zeros (let open MnistHelper in [image_dim, label_count]) in
let bs = Tensor.zeros (let open MnistHelper in [label_count]) in
let model @\textbf{\textcolor{blue}{\{n :\ Nat\}}}@ (xs : @\textbf{\textcolor{blue}{Tensor [n, MnistHelper.image\_dim]}}@) =
  let open Tensor in mm xs ws + bs
in
range 1 200 |> List.iter (fun(i : Int) ->
  let loss =
    Tensor.cross_entropy_for_logits
      (model MnistHelper.train_images) MnistHelper.train_labels in
  Tensor.backward loss;
  Tensor.no_grad (fun(u : Unit) -> let open Tensor in
    ws -= grad ws * learning_rate; bs -= grad bs * learning_rate);
  Tensor.zero_grad ws; Tensor.zero_grad bs;
  let got = model MnistHelper.test_images in
  let estimated = Tensor.argmax 1 got in
  let sum = Tensor.count_equal estimated MnistHelper.test_labels in
  let test_accuracy = float sum / float (@\textbf{\textcolor{blue}{lift\_int}}@ MnistHelper.num_test_images) in
  print_float test_accuracy)
\end{lstlisting}
  \vspace{-1em}
  \caption{An example program in Horsea that can be checked by the prototype type-checker}
  \label{fig:code-example}
\end{figure}
\indent
  Table~\ref{table:stats} shows similar results for 10 example programs (including \texttt{mnist/linear.hrs}).
  The columns~``total'' and ``inferred'' display
  the total number of implicit arguments in each program and
  the number of successfully inferred ones among those arguments, respectively.
  Those that cannot be inferred are manually specified in the programs
  (e.g., \texttt{char\_rnn/char\_rnn.hrs} contains \(39 - 37 = 2\)~manually specified implicit arguments,
  and the other 37 are appropriately reconstructed by the type-checker).
  We also show the number of shape-related descriptions
  contained in type annotations in each program on the column~``\#annot''
  because adding these descriptions often helps the inference.
\par
\indent
  As shown by these results, the inference algorithm works unexpectedly effective,
  despite its (intentionally) plain strategy;
  it can reconstruct approximately 90\% of the implicit arguments.
  The major source of the inference failure is, on the other hand, the use of higher-order functions.
  It would thus be even better if we use Hindley--Milner-like unification-based algorithm.
  It might also be beneficial for the inference to compute function applications in types
  that are known to be pure and halting, such as those of {\small\texttt{broadcast}}, during type-checking.
\par
\begin{table}[tb]
\centering
  {\small
    \begin{tabular}{lrrrr}
    \toprule
      program & total & inferred & \#annot & \#lines
    \\\midrule
      \texttt{char\_rnn/char\_rnn.hrs} & 39 & 37 & 12 & 118
    \\\hline
      \texttt{gan/mnist\_cgan.hrs} & 59 & 55 & 5 & 154
    \\\hline
      \texttt{gan/mnist\_dcgan.hrs} & 112 & 106 & 4 & 195
    \\\hline
      \texttt{gan/mnist\_gan.hrs} & 51 & 47 & 4 & 142
    \\\hline
      \texttt{jit/load\_and\_run.hrs} & 5 & 3 & 0 & 17
    \\\hline
      \texttt{min-gpt/mingpt.hrs} & 108 & 96 & 17 & 321
    \\\hline
      \texttt{mnist/conv.hrs} & 28 & 25 & 3 & 72
    \\\hline
      \texttt{mnist/linear.hrs} & 20 & 20 & 1 & 29
    \\\hline
      \texttt{pretrained/finetuning.hrs} & 45 & 35 & 6 & 89
    \\\hline
      \texttt{pretrained/predict.hrs} & 8 & 5 & 0 & 77
    \\\bottomrule
    \end{tabular}%
  }%
  \vspace{1em}
  \caption{Properties of example programs and the inference results of implicit arguments}
  \label{table:stats}
\end{table}

\section{Related Work}\label{sec:related-work}
\subsection{Staged Computation}
\indent
  Typed languages with staging constructs,
  such as MetaML~\cite{TahaSheardPEPM1997,Taha1999,TahaSheard2000}
  or \(\lambda^{\bigcirc}\)~\cite{DaviesLICS1996,DaviesJACM2017},
  arose from the context of partial evaluation~\cite{%
    NielsonNielsonPOPL1988,%
    GluckJorgensen1995,%
    GluckJorgensen1997},
  and a considerable amount of studies
  have been done subsequently~\cite{%
    DaviesPfenningJACM2001,%
    GanzSabryTahaICFP2001,%
    TahaNielsenPOPL2003,%
    CalcagnoMoggiSheardJFP2003,%
    YuseIgarashiPPDP2006,%
    TsukadaIgarashi2009,%
    HanadaIgarashi2014,%
    KawataIgarashiAPLAS2019}.
  Although they differ from one another in
  the design choice of constructs,
  many of them aim at ensuring the validity of produced code
  statically by checking code-generating programs.
  As to implementation,
  BER~MetaOCaml~\cite{KiselyovFLOPS2014,KiselyovFLOPS2024}
  extends OCaml with MetaML-style staging constructs,
  and Scala~3~\cite{StuckiBiboudisOderskyGPCE2018}
  recently adopted a staging-based formalization for the core of its macro system.
\par
\indent
  Combining staging with types dependent on values in some sense
  seems to be investigated by somewhat limited number of studies.
  Concoqtion~\cite{FogartyPasalicSiekTahaPEPM2007} has
  both indexed types and staging constructs,
  but their use is for establishing a tagless staged interpreter~\cite{PasalicTahaSheardICFP2002},
  and the language for computation and that for indices are separated.
  Kawata and Igarashi~\cite{KawataIgarashiAPLAS2019} propose
  \(\lambda^{\mathrm{MD}}\),
  a dependently-typed multi-stage language that can be regarded as a theoretical foundation.
  It also supports a kind of cross-stage persistence
  based on \(\lambda^{\triangleright\%}\)~\cite{HanadaIgarashi2014}.
  Unlike our staged core language,
  \(\lambda^{\mathrm{MD}}\) does not have some mechanism like
  stage-\(0\) assertions~\( \LeftAssertParen\openO{\langle} \possiblyWithSub\stageImetaColor{T^{\superscriptI} }_{{\mathrm{1}}} \closeO{\rangle} \relO{\CastArrow} \openO{\langle} \possiblyWithSub\stageImetaColor{T^{\superscriptI} }_{{\mathrm{2}}} \closeO{\rangle}\RightAssertParen^{ L } \),
  and its type equality for checking function applications is based on the \(\beta\)-equivalence.
\par
\subsection{Tensor Shape Checking}
\indent
  Checking the consistency of programs as to tensor shapes is also a classical topic.
  The incorporation of length-indexed vector types into realistic languages
  dates back at least to Dependent~ML~\cite{XiPfenningPLDI1998,XiJFP2007},
  and handling data-independent sizes of structures as type-level information
  has already been targeted by \dfn{shapely types}~\cite{JayCockettESOP1994}.
  Since then, a variety of studies have been done to design a mechanism
  that is less general than full dependent types but handier in some sense
  for major use cases of tensor computation.
\par
\indent
  Repa~\cite{KellerChakravartyLeshchinskiyPeytonJonesLippmeierICFP2010}
  uses some kind of shape information at type level
  and supports \dfn{shape polymorphism}~\cite{ScholzJFP2003},
  i.e., the ability to reuse functions for some class of the shapes,
  by exploiting usual type classes,
  \dfn{associated data types}~\cite{ChakravartyKellerPeytonJonesMarlowPOPL2005},
  and \dfn{type families}~\cite{SchrijversPeytonJonesChakravartySulzmannICFP2008}
  offered by GHC~\cite{GHC}.
  Its main purpose of the use is, however, to achieve high-performance tensor computation
  while keeping high-level description in source programs at the same time,
  and seems not to pursue the complete elimination of shape mismatches.
  Accelerate~\cite{ChakravartyKellerLeeMcDonellGroverDAMP2011}
  follows a Repa-based interface for its frontend of GPGPU programs.
  It also performs dynamic code generation for producing GPU kernel functions,
  so the utilization of our staging-based formalization
  for systems like Accelerate might be worth investigating.
\par
\indent
  To achieve strict safety, Gibbons~\cite{GibbonsESOP2017} proposes
  an elegant embedded-DSL approach to
  checking shape consistency as to operations similar to those of APL~\cite{Iverson1962,APL}
  by using \dfn{Naperian} functors, type classes,
  and the extension of GHC~\cite{GHC} for dependent types.
  This also supports \dfn{rank polymorphism},
  a mechanism to allow conversions close to
  broadcasting~\cite{NumPyBroadcastingSemantics,PyTorchBroadcastingSemantics}.
  To support realistic programs, however,
  it would be good if tensor shapes are more ``first-class,''
  as pointed out in Hattori~et~al.~\cite{HattoriKobayashiSatoESOP2023}.
  Henriksen and Elsman~\cite{HenriksenElsmanARRAY2021}
  and Bailly~et~al.~\cite{BaillyHenriksenElsmanFHPNC2023}
  give another interesting formalization
  mainly for the use in Futhark~\cite{HenriksenSerupElsmanHengleinOanceaPLDI2017}:
  a system with \dfn{size-dependent types}.
  While it may cause runtime errors due to array indexing or
  runtime coercion,
  it allows term-level variables for type-level indices and
  accommodates dynamically determined shapes by existential quantification on indices.
  Somewhat similar mechanisms are also proposed by
  Abe and Sumii~\cite{AbeSumii2014} and Xi~\cite{XiJFP2007},
  the former of which exploits \dfn{phantom types} in OCaml.
  It might be worth considering to combine our work with such existential quantification
  to achieve better handling of tensors with dynamic shapes.
\par
\indent
  As another line of work, those based on
  refinement types~\cite{%
    FreemanPfenningPLDI1991,%
    FlanaganPOPL2006,%
    RondonKawaguchiRanjitPLDI2008,%
    KnowlesFlanaganTOPLAS2010}
  or manifest contracts~\cite{%
    FindlerFelleisenICFP2002,%
    WadlerFindler2009,%
    GreenbergPierceWeirichPOPL2010,%
    Greenberg2013,%
    SekiyamaIgarashiGreenbergTOPLAS2017}
  are also prominent.
  Some early researches, such as
  hybrid type checking~(\(\lambda^{\mathrm{H}}\))~\cite{FlanaganPOPL2006,KnowlesFlanaganTOPLAS2010}
  or liquid types~\cite{RondonKawaguchiRanjitPLDI2008},
  already mention array bounds checking as one of their applications.
  GraTen~\cite{HattoriKobayashiSatoESOP2023} propels this approach forward
  to support many tensor-related operations used in realistic DNN-related programs,
  such as examples of \texttt{ocaml-torch}~\cite{OCamlTorch},
  with a flavor of gradual typing~\cite{SiekTaha2006}.
  Migeed, Reed, Ansel, and Palsberg~\cite{MigeedReedAnselPalsbergECOOP2024} propose
  a less expressive gradually-typed system
  in order to strike a balance between the coverage of the verification
  and the affinity with the existing tool support.
  Our use of refinement types for stage-\(0\) types is highly inspired by that of GraTen,
  although we provide shape-related functions like \texttt{broadcast} just as literally usual functions
  while GraTen handles them carefully so that the back-end SMT solver can ensure subtyping relations.
\par
\subsection{Bidirectional Type-Checking for Implicit Arguments}
\indent
  The idea of utilizing bidirectional type-checking~\cite{DunfieldNeel2021}
  for reconstructing implicit arguments is not very new;
  Odersky~et~al.~\cite{OderskyBlanvillainLiuBiboudisMillerStuckiPOPL2017}
  give such a formalization as a foundation of Scala~3's \(\token{implicit}\)-parameters.
  Due to the dependent nature of our formalization, however,
  we have found that
  extending Xie and Oliveira's ``let arguments go first''~\cite{XieOliveiraESOP2018}
  better fits our purpose.
\par
\subsection{Distinction between Compile Time and Runtime}
\indent
  Although ours seems to give a staging-based foundation for tensor shape checking
  with a mathematical safety guarantee for the first time,
  the distinction between data available only at compile time (e.g.,~parameters for matrix sizes)
  and those at runtime (e.g., matrices themselves) by some other forms
  appears to have long been done by a number of existing studies.
  In particular, Idris~2~\cite{BradyECOOP2021},
  which has a quite different core language from the previous version of Idris~\cite{BradyJFP2013},
  adopts \dfn{quantitative type theory}~(\dfn{QTT})~\cite{McBride2016,AtkeyLICS2018}
  for distinguishing the two.
  Investigating the relationship
  between QTT-based type systems and staging-based ones,
  such as interoperability,
  might be another interesting topic.
  Other than that, LMS-Verify~\cite{AminRompfPOPL2017}
  proposes a method somewhat similar to ours from the operational perspective
  and provides an example of how to check size consistency of matrix operations,
  based on \dfn{lightweight modular staging} (\dfn{LMS})~\cite{RompfOderskyGPCE2010}
  with a flavor of higher-order contracts~\cite{FindlerFelleisenICFP2002}.
  It seems that, however, the consistency of generated code as to shapes
  is not guaranteed by language-level metatheory in this method;
  it is not the language (in particular, not the type system)
  but programmers who write contracts as to shapes
  that are responsible for ensuring such consistency.
\par

\section{Conclusion and Future Work}\label{sec:conclusion}
\indent
  By utilizing staged computation,
  we have proposed a method that ensures the consistency as to tensor shapes
  through compile-time computation,
  aiming at an affinity with continuous development.
  We have also offered features
  for further reducing the burden of writing tensor-involving programs,
  such as implicit parameters or a non-staged surface language.
  Various future directions can be considered furthermore,
  as some of them are mentioned in Section~\ref{sec:related-work},
  and others are as follows:
  (1)~%
    support basic type-related devices such as polymorphism or algebraic datatypes;
  (2)~%
    generalize \LambdaBracketCast\ to a multi-stage version
    and inspect the relationship between that language and
    \(\lambda^{\mathrm{MD}}\)~\cite{KawataIgarashiAPLAS2019};
  (3)~%
    improve elaboration rules so that
    fewer terms will be duplicated through assertion insertion; and
  (4)~%
    investigate the interoperability of our method with
    programs written in Idris or some dependently-typed languages.
\par



\bibliography{bibliography}

\appendix
\section{Full definitions}
  \begin{itemize}
    \item Figure~\ref{fig:term-reduction-full}:
      all rules for reduction relations on terms
    \item Figure~\ref{fig:type-reduction-full}
      all rules for reduction relations on types
    \item Figure~\ref{fig:type-equivalence}:
      the CSR equivalence on types
    \item Figure~\ref{fig:well-formedness}:
      the well-formedness of types and type environments
    \item Figures~\ref{fig:option-inference-main-full-1} and \ref{fig:option-inference-main-full-2}:
      the declarative rules for the inference of implicit arguments
    \item Figure~\ref{fig:algorithmic-option-inference}:
      the algorithmic rules for the inference of implicit arguments
  \end{itemize}
\begin{figure}[p]
\small
  \begin{flushleft}
    \fbox{\(\possiblyWithSub\stageOmetaColor{N^{\superscriptO} } \longrightarrow^{0} (\possiblyWithSub\stageOmetaColor{N'^{\superscriptO} } \mid  \BlameSign^{ L } )\)}
  \end{flushleft}
  \vspace{-3.25em}
  \begin{center}
    \hspace{9em}%
    \derive[E0-App1]{%
       \possiblyWithSub\stageOmetaColor{N^{\superscriptO} }_{{\mathrm{1}}}  \longrightarrow^{0}   \possiblyWithSub\stageOmetaColor{N'^{\superscriptO} }_{{\mathrm{1}}}  
    }{%
        \possiblyWithSub\stageOmetaColor{N^{\superscriptO} }_{{\mathrm{1}}} \  \possiblyWithSub\stageOmetaColor{N^{\superscriptO} }_{{\mathrm{2}}}   \longrightarrow^{0}    \possiblyWithSub\stageOmetaColor{N'^{\superscriptO} }_{{\mathrm{1}}} \  \possiblyWithSub\stageOmetaColor{N^{\superscriptO} }_{{\mathrm{2}}}   
    }
  \qquad
    \derive[E0-App1F]{%
       \possiblyWithSub\stageOmetaColor{N^{\superscriptO} }_{{\mathrm{1}}}  \longrightarrow^{0}   \BlameSign^{ L }  
    }{%
        \possiblyWithSub\stageOmetaColor{N^{\superscriptO} }_{{\mathrm{1}}} \  \possiblyWithSub\stageOmetaColor{N^{\superscriptO} }_{{\mathrm{2}}}   \longrightarrow^{0}   \BlameSign^{ L }  
    }
  \\[0.7em]
    \derive[E0-App2]{%
       \possiblyWithSub\stageOmetaColor{N^{\superscriptO} }_{{\mathrm{2}}}  \longrightarrow^{0}   \possiblyWithSub\stageOmetaColor{N'^{\superscriptO} }_{{\mathrm{2}}}  
    }{%
         \possiblyWithSub\stageOmetaColor{v^{\superscriptO} }_{{\mathrm{1}}}  \  \possiblyWithSub\stageOmetaColor{N^{\superscriptO} }_{{\mathrm{2}}}   \longrightarrow^{0}     \possiblyWithSub\stageOmetaColor{v^{\superscriptO} }_{{\mathrm{1}}}  \  \possiblyWithSub\stageOmetaColor{N'^{\superscriptO} }_{{\mathrm{2}}}   
    }
  \qquad
    \derive[E0-App2F]{%
       \possiblyWithSub\stageOmetaColor{N^{\superscriptO} }_{{\mathrm{2}}}  \longrightarrow^{0}   \BlameSign^{ L }  
    }{%
         \possiblyWithSub\stageOmetaColor{v^{\superscriptO} }_{{\mathrm{1}}}  \  \possiblyWithSub\stageOmetaColor{N^{\superscriptO} }_{{\mathrm{2}}}   \longrightarrow^{0}   \BlameSign^{ L }  
    }
  \qquad
    \derive[E0-Brkt]{%
       \possiblyWithSub\stageImetaColor{N^{\superscriptI} }  \longrightarrow^{1}   \possiblyWithSub\stageImetaColor{N'^{\superscriptI} }  
    }{%
        \openO{\langle} \possiblyWithSub\stageImetaColor{N^{\superscriptI} } \closeO{\rangle}   \longrightarrow^{0}    \openO{\langle} \possiblyWithSub\stageImetaColor{N'^{\superscriptI} } \closeO{\rangle}   
    }
  \\[0.7em]
    \derive[E0-BrktF]{%
       \possiblyWithSub\stageImetaColor{N^{\superscriptI} }  \longrightarrow^{1}   \BlameSign^{ L }  
    }{%
        \openO{\langle} \possiblyWithSub\stageImetaColor{N^{\superscriptI} } \closeO{\rangle}   \longrightarrow^{0}   \BlameSign^{ L }  
    }
  \quad
    \derive[E0-Beta]{}{%
         \openO{(}  \ordO{\lambda} \possiblyWithSub\stageOmetaColor{x}  \relO{:}  \possiblyWithSub\stageOmetaColor{T^{\superscriptO} }_{{\mathrm{11}}} \punctO{.}\  \possiblyWithSub\stageOmetaColor{N^{\superscriptO} }_{{\mathrm{12}}}  \closeO{)}  \   \possiblyWithSub\stageOmetaColor{v^{\superscriptO} }_{{\mathrm{2}}}    \longrightarrow^{0}     [   \possiblyWithSub\stageOmetaColor{v^{\superscriptO} }_{{\mathrm{2}}}   /  \possiblyWithSub\stageOmetaColor{x}  ]    \possiblyWithSub\stageOmetaColor{N^{\superscriptO} }_{{\mathrm{12}}}   
    }
  \quad
    \derive[E0-Delta]{%
      \delta(  \possiblyWithSub\stageOmetaColor{a}_{{\mathrm{1}}}  \    \possiblyWithSub\stageOmetaColor{c}_{{\mathrm{2}}}   ) = \possiblyWithSub\stageOmetaColor{q}
    }{%
         \possiblyWithSub\stageOmetaColor{a}_{{\mathrm{1}}}  \    \possiblyWithSub\stageOmetaColor{c}_{{\mathrm{2}}}     \longrightarrow^{0}    \possiblyWithSub\stageOmetaColor{q}   
    }
  \\[0.7em]
    \derive[E0-Ass1]{%
       \possiblyWithSub\stageImetaColor{T^{\superscriptI} }_{{\mathrm{1}}}  \longrightarrow^{1}   \possiblyWithSub\stageImetaColor{T'^{\superscriptI} }_{{\mathrm{1}}}  
    }{%
        \LeftAssertParen\openO{\langle} \possiblyWithSub\stageImetaColor{T^{\superscriptI} }_{{\mathrm{1}}} \closeO{\rangle} \relO{\CastArrow} \openO{\langle} \possiblyWithSub\stageImetaColor{T^{\superscriptI} }_{{\mathrm{2}}} \closeO{\rangle}\RightAssertParen^{ L }   \longrightarrow^{0}    \LeftAssertParen\openO{\langle} \possiblyWithSub\stageImetaColor{T'^{\superscriptI} }_{{\mathrm{1}}} \closeO{\rangle} \relO{\CastArrow} \openO{\langle} \possiblyWithSub\stageImetaColor{T^{\superscriptI} }_{{\mathrm{2}}} \closeO{\rangle}\RightAssertParen^{ L }   
    }
  \qquad
    \derive[E0-Ass1F]{%
       \possiblyWithSub\stageImetaColor{T^{\superscriptI} }_{{\mathrm{1}}}  \longrightarrow^{1}   \BlameSign^{ L }  
    }{%
        \LeftAssertParen\openO{\langle} \possiblyWithSub\stageImetaColor{T^{\superscriptI} }_{{\mathrm{1}}} \closeO{\rangle} \relO{\CastArrow} \openO{\langle} \possiblyWithSub\stageImetaColor{T^{\superscriptI} }_{{\mathrm{2}}} \closeO{\rangle}\RightAssertParen^{ L' }   \longrightarrow^{0}   \BlameSign^{ L }  
    }
  \\[0.7em]
    \derive[E0-Ass2]{%
       \possiblyWithSub\stageImetaColor{T^{\superscriptI} }_{{\mathrm{2}}}  \longrightarrow^{1}   \possiblyWithSub\stageImetaColor{T'^{\superscriptI} }_{{\mathrm{2}}}  
    }{%
        \LeftAssertParen\openO{\langle}  \possiblyWithSub\stageImetaColor{\tau^{\superscriptI} }_{{\mathrm{1}}}  \closeO{\rangle} \relO{\CastArrow} \openO{\langle} \possiblyWithSub\stageImetaColor{T^{\superscriptI} }_{{\mathrm{2}}} \closeO{\rangle}\RightAssertParen^{ L }   \longrightarrow^{0}    \LeftAssertParen\openO{\langle}  \possiblyWithSub\stageImetaColor{\tau^{\superscriptI} }_{{\mathrm{1}}}  \closeO{\rangle} \relO{\CastArrow} \openO{\langle} \possiblyWithSub\stageImetaColor{T'^{\superscriptI} }_{{\mathrm{2}}} \closeO{\rangle}\RightAssertParen^{ L }   
    }
  \qquad
    \derive[E0-Ass2F]{%
       \possiblyWithSub\stageImetaColor{T^{\superscriptI} }_{{\mathrm{2}}}  \longrightarrow^{1}   \BlameSign^{ L }  
    }{%
        \LeftAssertParen\openO{\langle}  \possiblyWithSub\stageImetaColor{\tau^{\superscriptI} }_{{\mathrm{1}}}  \closeO{\rangle} \relO{\CastArrow} \openO{\langle} \possiblyWithSub\stageImetaColor{T^{\superscriptI} }_{{\mathrm{2}}} \closeO{\rangle}\RightAssertParen^{ L' }   \longrightarrow^{0}   \BlameSign^{ L }  
    }
  \\[0.7em]
    \derive[E0-AssPass]{}{%
        \LeftAssertParen\openO{\langle}  \possiblyWithSub\stageImetaColor{\tau^{\superscriptI} }  \closeO{\rangle} \relO{\CastArrow} \openO{\langle}  \possiblyWithSub\stageImetaColor{\tau^{\superscriptI} }  \closeO{\rangle}\RightAssertParen^{ L }   \longrightarrow^{0}    \ordO{\lambda} \possiblyWithSub\stageOmetaColor{x}  \relO{:}   \openO{\langle}  \possiblyWithSub\stageImetaColor{\tau^{\superscriptI} }  \closeO{\rangle}  \punctO{.}\   \possiblyWithSub\stageOmetaColor{x}    
    }
  \qquad
    \derive[E0-AssFail]{%
      \possiblyWithSub\stageImetaColor{\tau^{\superscriptI} }_{{\mathrm{1}}} \neq \possiblyWithSub\stageImetaColor{\tau^{\superscriptI} }_{{\mathrm{2}}}
    }{%
        \LeftAssertParen\openO{\langle}  \possiblyWithSub\stageImetaColor{\tau^{\superscriptI} }_{{\mathrm{1}}}  \closeO{\rangle} \relO{\CastArrow} \openO{\langle}  \possiblyWithSub\stageImetaColor{\tau^{\superscriptI} }_{{\mathrm{2}}}  \closeO{\rangle}\RightAssertParen^{ L }   \longrightarrow^{0}   \BlameSign^{ L }  
    }
  \\[0.7em]
    \derive[E0-RfnStart]{}{%
         \LeftAssertParen \relO{\CastArrow}   \openO{\{} \possiblyWithSub\stageOmetaColor{\nu}  \relO{:}  \possiblyWithSub\stageOmetaColor{B}  \relO{\mid}  \possiblyWithSub\stageOmetaColor{N^{\superscriptO} }_{{\mathrm{1}}} \closeO{\} }   \RightAssertParen^{ L }  \    \possiblyWithSub\stageOmetaColor{c}_{{\mathrm{2}}}     \longrightarrow^{0}    \LeftAssertParen   \openO{\{} \possiblyWithSub\stageOmetaColor{\nu}  \relO{:}  \possiblyWithSub\stageOmetaColor{B}  \relO{\mid}  \possiblyWithSub\stageOmetaColor{N^{\superscriptO} }_{{\mathrm{1}}} \closeO{\} }  \punctO{,}    [    \possiblyWithSub\stageOmetaColor{c}_{{\mathrm{2}}}    /  \possiblyWithSub\stageOmetaColor{\nu}  ]    \possiblyWithSub\stageOmetaColor{N^{\superscriptO} }_{{\mathrm{1}}}  \punctO{,}  \possiblyWithSub\stageOmetaColor{c}_{{\mathrm{2}}}  \RightAssertParen^{ L }   
    }
  \\[0.7em]
    \derive[E0-RfnAct]{%
       \possiblyWithSub\stageOmetaColor{N^{\superscriptO} }  \longrightarrow^{0}   \possiblyWithSub\stageOmetaColor{N'^{\superscriptO} }  
    }{%
        \LeftAssertParen   \openO{\{} \possiblyWithSub\stageOmetaColor{\nu}  \relO{:}  \possiblyWithSub\stageOmetaColor{B}  \relO{\mid}  \possiblyWithSub\stageOmetaColor{N^{\superscriptO} }_{{\mathrm{1}}} \closeO{\} }  \punctO{,}  \possiblyWithSub\stageOmetaColor{N^{\superscriptO} } \punctO{,}  \possiblyWithSub\stageOmetaColor{c}_{{\mathrm{2}}}  \RightAssertParen^{ L }   \longrightarrow^{0}    \LeftAssertParen   \openO{\{} \possiblyWithSub\stageOmetaColor{\nu}  \relO{:}  \possiblyWithSub\stageOmetaColor{B}  \relO{\mid}  \possiblyWithSub\stageOmetaColor{N^{\superscriptO} }_{{\mathrm{1}}} \closeO{\} }  \punctO{,}  \possiblyWithSub\stageOmetaColor{N'^{\superscriptO} } \punctO{,}  \possiblyWithSub\stageOmetaColor{c}_{{\mathrm{2}}}  \RightAssertParen^{ L }   
    }
  \\[0.7em]
    \derive[E0-RfnActF]{%
       \possiblyWithSub\stageOmetaColor{N^{\superscriptO} }  \longrightarrow^{0}   \BlameSign^{ L }  
    }{%
        \LeftAssertParen   \openO{\{} \possiblyWithSub\stageOmetaColor{\nu}  \relO{:}  \possiblyWithSub\stageOmetaColor{B}  \relO{\mid}  \possiblyWithSub\stageOmetaColor{N^{\superscriptO} }_{{\mathrm{1}}} \closeO{\} }  \punctO{,}  \possiblyWithSub\stageOmetaColor{N^{\superscriptO} } \punctO{,}  \possiblyWithSub\stageOmetaColor{c}_{{\mathrm{2}}}  \RightAssertParen^{ L' }   \longrightarrow^{0}   \BlameSign^{ L }  
    }
  \\[0.7em]
    \derive[E0-RfnPass]{}{%
        \LeftAssertParen   \openO{\{} \possiblyWithSub\stageOmetaColor{\nu}  \relO{:}  \possiblyWithSub\stageOmetaColor{B}  \relO{\mid}  \possiblyWithSub\stageOmetaColor{N^{\superscriptO} }_{{\mathrm{1}}} \closeO{\} }  \punctO{,}     \ttO{true}    \punctO{,}  \possiblyWithSub\stageOmetaColor{c}_{{\mathrm{2}}}  \RightAssertParen^{ L }   \longrightarrow^{0}     \possiblyWithSub\stageOmetaColor{c}_{{\mathrm{2}}}    
    }
  \qquad
    \derive[E0-RfnFail]{}{%
        \LeftAssertParen   \openO{\{} \possiblyWithSub\stageOmetaColor{\nu}  \relO{:}  \possiblyWithSub\stageOmetaColor{B}  \relO{\mid}  \possiblyWithSub\stageOmetaColor{N^{\superscriptO} }_{{\mathrm{1}}} \closeO{\} }  \punctO{,}     \ttO{false}    \punctO{,}  \possiblyWithSub\stageOmetaColor{c}_{{\mathrm{2}}}  \RightAssertParen^{ L }   \longrightarrow^{0}   \BlameSign^{ L }  
    }
  \end{center}
  \begin{flushleft}
    \fbox{\(\possiblyWithSub\stageImetaColor{N^{\superscriptI} } \longrightarrow^{1} (\possiblyWithSub\stageImetaColor{N'^{\superscriptI} } \mid  \BlameSign^{ L } )\)}
  \end{flushleft}
  \vspace{-3.25em}
  \begin{center}
    \hspace{7em}%
    \derive[E1-App1F]{%
       \possiblyWithSub\stageImetaColor{N^{\superscriptI} }_{{\mathrm{1}}}  \longrightarrow^{1}   \BlameSign^{ L }  
    }{%
        \possiblyWithSub\stageImetaColor{N^{\superscriptI} }_{{\mathrm{1}}} \  \possiblyWithSub\stageImetaColor{N^{\superscriptI} }_{{\mathrm{2}}}   \longrightarrow^{1}   \BlameSign^{ L }  
    }
  \qquad
    \derive[E1-App2F]{%
       \possiblyWithSub\stageImetaColor{N^{\superscriptI} }_{{\mathrm{2}}}  \longrightarrow^{1}   \BlameSign^{ L }  
    }{%
         \possiblyWithSub\stageImetaColor{v^{\superscriptI} }_{{\mathrm{1}}}  \  \possiblyWithSub\stageImetaColor{N^{\superscriptI} }_{{\mathrm{2}}}   \longrightarrow^{1}   \BlameSign^{ L }  
    }
  \\[0.7em]
    \derive[E1-App1]{%
       \possiblyWithSub\stageImetaColor{N^{\superscriptI} }_{{\mathrm{1}}}  \longrightarrow^{1}   \possiblyWithSub\stageImetaColor{N'^{\superscriptI} }_{{\mathrm{1}}}  
    }{%
        \possiblyWithSub\stageImetaColor{N^{\superscriptI} }_{{\mathrm{1}}} \  \possiblyWithSub\stageImetaColor{N^{\superscriptI} }_{{\mathrm{2}}}   \longrightarrow^{1}    \possiblyWithSub\stageImetaColor{N'^{\superscriptI} }_{{\mathrm{1}}} \  \possiblyWithSub\stageImetaColor{N^{\superscriptI} }_{{\mathrm{2}}}   
    }
  \qquad
    \derive[E1-App2]{%
       \possiblyWithSub\stageImetaColor{N^{\superscriptI} }_{{\mathrm{2}}}  \longrightarrow^{1}   \possiblyWithSub\stageImetaColor{N'^{\superscriptI} }_{{\mathrm{2}}}  
    }{%
         \possiblyWithSub\stageImetaColor{v^{\superscriptI} }_{{\mathrm{1}}}  \  \possiblyWithSub\stageImetaColor{N^{\superscriptI} }_{{\mathrm{2}}}   \longrightarrow^{1}     \possiblyWithSub\stageImetaColor{v^{\superscriptI} }_{{\mathrm{1}}}  \  \possiblyWithSub\stageImetaColor{N'^{\superscriptI} }_{{\mathrm{2}}}   
    }
  \\[0.7em]
    \derive[E1-Abs1]{%
       \possiblyWithSub\stageImetaColor{T^{\superscriptI} }  \longrightarrow^{1}   \possiblyWithSub\stageImetaColor{T'^{\superscriptI} }  
    }{%
        \ordI{\lambda} \possiblyWithSub\stageImetaColor{x}  \relI{:}  \possiblyWithSub\stageImetaColor{T^{\superscriptI} } \punctI{.}\  \possiblyWithSub\stageImetaColor{N^{\superscriptI} }   \longrightarrow^{1}    \ordI{\lambda} \possiblyWithSub\stageImetaColor{x}  \relI{:}  \possiblyWithSub\stageImetaColor{T'^{\superscriptI} } \punctI{.}\  \possiblyWithSub\stageImetaColor{N^{\superscriptI} }   
    }
  \qquad
    \derive[E1-Abs1F]{%
       \possiblyWithSub\stageImetaColor{T^{\superscriptI} }  \longrightarrow^{1}   \BlameSign^{ L }  
    }{%
        \ordI{\lambda} \possiblyWithSub\stageImetaColor{x}  \relI{:}  \possiblyWithSub\stageImetaColor{T^{\superscriptI} } \punctI{.}\  \possiblyWithSub\stageImetaColor{N^{\superscriptI} }   \longrightarrow^{1}   \BlameSign^{ L }  
    }
  \\[0.7em]
    \derive[E1-Abs2]{%
       \possiblyWithSub\stageImetaColor{N^{\superscriptI} }  \longrightarrow^{1}   \possiblyWithSub\stageImetaColor{N'^{\superscriptI} }  
    }{%
        \ordI{\lambda} \possiblyWithSub\stageImetaColor{x}  \relI{:}   \possiblyWithSub\stageImetaColor{\tau^{\superscriptI} }  \punctI{.}\  \possiblyWithSub\stageImetaColor{N^{\superscriptI} }   \longrightarrow^{1}    \ordI{\lambda} \possiblyWithSub\stageImetaColor{x}  \relI{:}   \possiblyWithSub\stageImetaColor{\tau^{\superscriptI} }  \punctI{.}\  \possiblyWithSub\stageImetaColor{N'^{\superscriptI} }   
    }
  \qquad
    \derive[E1-Abs2F]{%
       \possiblyWithSub\stageImetaColor{N^{\superscriptI} }  \longrightarrow^{1}   \BlameSign^{ L }  
    }{%
        \ordI{\lambda} \possiblyWithSub\stageImetaColor{x}  \relI{:}   \possiblyWithSub\stageImetaColor{\tau^{\superscriptI} }  \punctI{.}\  \possiblyWithSub\stageImetaColor{N^{\superscriptI} }   \longrightarrow^{1}   \BlameSign^{ L }  
    }
  \\[0.7em]
    \derive[E1-Esc]{%
       \possiblyWithSub\stageOmetaColor{N^{\superscriptO} }  \longrightarrow^{0}   \possiblyWithSub\stageOmetaColor{N'^{\superscriptO} }  
    }{%
        \ordI{\sim} \possiblyWithSub\stageOmetaColor{N^{\superscriptO} }   \longrightarrow^{1}    \ordI{\sim} \possiblyWithSub\stageOmetaColor{N'^{\superscriptO} }   
    }
  \qquad
    \derive[E1-EscF]{%
       \possiblyWithSub\stageOmetaColor{N^{\superscriptO} }  \longrightarrow^{0}   \BlameSign^{ L }  
    }{%
        \ordI{\sim} \possiblyWithSub\stageOmetaColor{N^{\superscriptO} }   \longrightarrow^{1}   \BlameSign^{ L }  
    }
  \qquad
    \derive[E1-Cancel]{}{%
        \ordI{\sim}  \openO{\langle} \possiblyWithSub\stageImetaColor{N^{\superscriptI} } \closeO{\rangle}    \longrightarrow^{1}   \possiblyWithSub\stageImetaColor{N^{\superscriptI} }  
    }
  \end{center}
  \caption{Reduction relation on terms}
  \label{fig:term-reduction-full}
\end{figure}
\begin{figure}[p]
\small
  \begin{flushleft}
    \fbox{\(\possiblyWithSub\stageImetaColor{T^{\superscriptI} } \longrightarrow^{1} (\possiblyWithSub\stageImetaColor{T'^{\superscriptI} } \mid  \BlameSign^{ L } )\)}
  \end{flushleft}
  \vspace{-4.25em}
  \begin{center}
    \hspace{10em}%
    \derive[ET1-Arr1F]{%
       \possiblyWithSub\stageImetaColor{T^{\superscriptI} }_{{\mathrm{1}}}  \longrightarrow^{1}   \BlameSign^{ L }  
    }{%
        \openI{(}  \possiblyWithSub\stageImetaColor{T^{\superscriptI} }_{{\mathrm{1}}}  \relI{\to}  \possiblyWithSub\stageImetaColor{T^{\superscriptI} }_{{\mathrm{2}}}  \closeI{)}   \longrightarrow^{1}   \BlameSign^{ L }  
    }
  \quad
    \derive[ET1-Arr2F]{%
       \possiblyWithSub\stageImetaColor{T^{\superscriptI} }_{{\mathrm{2}}}  \longrightarrow^{1}   \BlameSign^{ L }  
    }{%
        \openI{(}   \possiblyWithSub\stageImetaColor{\tau^{\superscriptI} }_{{\mathrm{1}}}   \relI{\to}  \possiblyWithSub\stageImetaColor{T^{\superscriptI} }_{{\mathrm{2}}}  \closeI{)}   \longrightarrow^{1}   \BlameSign^{ L }  
    }
  \\[0.7em]
    \derive[ET1-Arr1]{%
       \possiblyWithSub\stageImetaColor{T^{\superscriptI} }_{{\mathrm{1}}}  \longrightarrow^{1}   \possiblyWithSub\stageImetaColor{T'^{\superscriptI} }_{{\mathrm{1}}}  
    }{%
        \openI{(}  \possiblyWithSub\stageImetaColor{T^{\superscriptI} }_{{\mathrm{1}}}  \relI{\to}  \possiblyWithSub\stageImetaColor{T^{\superscriptI} }_{{\mathrm{2}}}  \closeI{)}   \longrightarrow^{1}    \openI{(}  \possiblyWithSub\stageImetaColor{T'^{\superscriptI} }_{{\mathrm{1}}}  \relI{\to}  \possiblyWithSub\stageImetaColor{T^{\superscriptI} }_{{\mathrm{2}}}  \closeI{)}   
    }
  \qquad
    \derive[ET1-Arr2]{%
       \possiblyWithSub\stageImetaColor{T^{\superscriptI} }_{{\mathrm{2}}}  \longrightarrow^{1}   \possiblyWithSub\stageImetaColor{T'^{\superscriptI} }_{{\mathrm{2}}}  
    }{%
        \openI{(}   \possiblyWithSub\stageImetaColor{\tau^{\superscriptI} }_{{\mathrm{1}}}   \relI{\to}  \possiblyWithSub\stageImetaColor{T^{\superscriptI} }_{{\mathrm{2}}}  \closeI{)}   \longrightarrow^{1}    \openI{(}   \possiblyWithSub\stageImetaColor{\tau^{\superscriptI} }_{{\mathrm{1}}}   \relI{\to}  \possiblyWithSub\stageImetaColor{T'^{\superscriptI} }_{{\mathrm{2}}}  \closeI{)}   
    }
  \\[0.7em]
    \derive[ET1-Tensor]{%
       \possiblyWithSub\stageOmetaColor{N^{\superscriptO} }  \longrightarrow^{0}   \possiblyWithSub\stageOmetaColor{N'^{\superscriptO} }  
    }{%
        \ttI{Tensor}\ \ordI{\%} \possiblyWithSub\stageOmetaColor{N^{\superscriptO} }   \longrightarrow^{1}    \ttI{Tensor}\ \ordI{\%} \possiblyWithSub\stageOmetaColor{N'^{\superscriptO} }   
    }
  \qquad
    \derive[ET1-TensorF]{%
       \possiblyWithSub\stageOmetaColor{N^{\superscriptO} }  \longrightarrow^{0}   \BlameSign^{ L }  
    }{%
        \ttI{Tensor}\ \ordI{\%} \possiblyWithSub\stageOmetaColor{N^{\superscriptO} }   \longrightarrow^{1}   \BlameSign^{ L }  
    }
  \end{center}
  \caption{Reduction relation on stage-1 types}
  \label{fig:type-reduction-full}
\end{figure}
\begin{figure}[p]
\small
  \begin{flushleft}
    \fbox{\( \possiblyWithSub\stageOmetaColor{T^{\superscriptO} }_{{\mathrm{1}}}  \equiv^{0}  \possiblyWithSub\stageOmetaColor{T^{\superscriptO} }_{{\mathrm{2}}} \)}
  \end{flushleft}
  \vspace{-4.25em}
  \begin{center}
    \derive[CqT0-Rfn]{%
       \sigma_{{\mathrm{1}}}  \longrightarrow  \sigma_{{\mathrm{2}}} 
    }{%
         \openO{\{} \possiblyWithSub\stageOmetaColor{\nu}  \relO{:}  \possiblyWithSub\stageOmetaColor{B}  \relO{\mid}   \sigma_{{\mathrm{1}}}   \possiblyWithSub\stageOmetaColor{N^{\superscriptO} }  \closeO{\} }    \equiv^{0}    \openO{\{} \possiblyWithSub\stageOmetaColor{\nu}  \relO{:}  \possiblyWithSub\stageOmetaColor{B}  \relO{\mid}   \sigma_{{\mathrm{2}}}   \possiblyWithSub\stageOmetaColor{N^{\superscriptO} }  \closeO{\} }   
    }
  \\[0.7em]
    \derive[CqT0-Refl]{}{%
       \possiblyWithSub\stageOmetaColor{T^{\superscriptO} }  \equiv^{0}  \possiblyWithSub\stageOmetaColor{T^{\superscriptO} } 
    }
  \qquad
    \derive[CqT0-Sym]{%
       \possiblyWithSub\stageOmetaColor{T^{\superscriptO} }_{{\mathrm{1}}}  \equiv^{0}  \possiblyWithSub\stageOmetaColor{T^{\superscriptO} }_{{\mathrm{2}}} 
    }{%
       \possiblyWithSub\stageOmetaColor{T^{\superscriptO} }_{{\mathrm{2}}}  \equiv^{0}  \possiblyWithSub\stageOmetaColor{T^{\superscriptO} }_{{\mathrm{1}}} 
    }
  \qquad
    \derive[CqT0-Trans]{%
       \possiblyWithSub\stageOmetaColor{T^{\superscriptO} }_{{\mathrm{1}}}  \equiv^{0}  \possiblyWithSub\stageOmetaColor{T^{\superscriptO} }_{{\mathrm{2}}} 
    \andalso
       \possiblyWithSub\stageOmetaColor{T^{\superscriptO} }_{{\mathrm{2}}}  \equiv^{0}  \possiblyWithSub\stageOmetaColor{T^{\superscriptO} }_{{\mathrm{3}}} 
    }{%
       \possiblyWithSub\stageOmetaColor{T^{\superscriptO} }_{{\mathrm{1}}}  \equiv^{0}  \possiblyWithSub\stageOmetaColor{T^{\superscriptO} }_{{\mathrm{3}}} 
    }
  \\[0.7em]
    \derive[CqT0-Code]{%
       \possiblyWithSub\stageImetaColor{T^{\superscriptI} }_{{\mathrm{1}}}  \equiv^{1}  \possiblyWithSub\stageImetaColor{T^{\superscriptI} }_{{\mathrm{2}}} 
    }{%
        \openO{\langle} \possiblyWithSub\stageImetaColor{T^{\superscriptI} }_{{\mathrm{1}}} \closeO{\rangle}   \equiv^{0}   \openO{\langle} \possiblyWithSub\stageImetaColor{T^{\superscriptI} }_{{\mathrm{2}}} \closeO{\rangle}  
    }
  \qquad
    \derive[CqT0-Arr]{%
       \possiblyWithSub\stageOmetaColor{T^{\superscriptO} }_{{\mathrm{11}}}  \equiv^{0}  \possiblyWithSub\stageOmetaColor{T^{\superscriptO} }_{{\mathrm{21}}} 
    \andalso
       \possiblyWithSub\stageOmetaColor{T^{\superscriptO} }_{{\mathrm{12}}}  \equiv^{0}  \possiblyWithSub\stageOmetaColor{T^{\superscriptO} }_{{\mathrm{22}}} 
    }{%
        \openO{(} \possiblyWithSub\stageOmetaColor{x}  \relO{:}  \possiblyWithSub\stageOmetaColor{T^{\superscriptO} }_{{\mathrm{11}}} \closeO{)} \relO{\to}  \possiblyWithSub\stageOmetaColor{T^{\superscriptO} }_{{\mathrm{12}}}   \equiv^{0}   \openO{(} \possiblyWithSub\stageOmetaColor{x}  \relO{:}  \possiblyWithSub\stageOmetaColor{T^{\superscriptO} }_{{\mathrm{21}}} \closeO{)} \relO{\to}  \possiblyWithSub\stageOmetaColor{T^{\superscriptO} }_{{\mathrm{22}}}  
    }
  \end{center}
  \vspace{1em}
  \begin{flushleft}
    \fbox{\( \possiblyWithSub\stageImetaColor{T^{\superscriptI} }_{{\mathrm{1}}}  \equiv^{1}  \possiblyWithSub\stageImetaColor{T^{\superscriptI} }_{{\mathrm{2}}} \)}
  \end{flushleft}
  \vspace{-4.25em}
  \begin{center}
    \derive[CqT1-Tensor]{%
       \sigma_{{\mathrm{1}}}  \longrightarrow  \sigma_{{\mathrm{2}}} 
    }{%
        \ttI{Tensor}\ \ordI{\%}  \openO{(}  \sigma_{{\mathrm{1}}}   \possiblyWithSub\stageOmetaColor{N^{\superscriptO} }  \closeO{)}    \equiv^{1}   \ttI{Tensor}\ \ordI{\%}  \openO{(}  \sigma_{{\mathrm{2}}}   \possiblyWithSub\stageOmetaColor{N^{\superscriptO} }  \closeO{)}   
    }
  \\[0.7em]
    \derive[CqT1-Refl]{}{%
       \possiblyWithSub\stageImetaColor{T^{\superscriptI} }  \equiv^{1}  \possiblyWithSub\stageImetaColor{T^{\superscriptI} } 
    }
  \qquad
    \derive[CqT1-Trans]{%
       \possiblyWithSub\stageImetaColor{T^{\superscriptI} }_{{\mathrm{1}}}  \equiv^{1}  \possiblyWithSub\stageImetaColor{T^{\superscriptI} }_{{\mathrm{2}}} 
    \andalso
       \possiblyWithSub\stageImetaColor{T^{\superscriptI} }_{{\mathrm{2}}}  \equiv^{1}  \possiblyWithSub\stageImetaColor{T^{\superscriptI} }_{{\mathrm{3}}} 
    }{%
       \possiblyWithSub\stageImetaColor{T^{\superscriptI} }_{{\mathrm{1}}}  \equiv^{1}  \possiblyWithSub\stageImetaColor{T^{\superscriptI} }_{{\mathrm{3}}} 
    }
  \\[0.7em]
    \derive[CqT1-Sym]{%
       \possiblyWithSub\stageImetaColor{T^{\superscriptI} }_{{\mathrm{1}}}  \equiv^{1}  \possiblyWithSub\stageImetaColor{T^{\superscriptI} }_{{\mathrm{2}}} 
    }{%
       \possiblyWithSub\stageImetaColor{T^{\superscriptI} }_{{\mathrm{2}}}  \equiv^{1}  \possiblyWithSub\stageImetaColor{T^{\superscriptI} }_{{\mathrm{1}}} 
    }
  \qquad
    \derive[CqT1-Arr]{%
       \possiblyWithSub\stageImetaColor{T^{\superscriptI} }_{{\mathrm{11}}}  \equiv^{1}  \possiblyWithSub\stageImetaColor{T^{\superscriptI} }_{{\mathrm{21}}} 
    \andalso
       \possiblyWithSub\stageImetaColor{T^{\superscriptI} }_{{\mathrm{12}}}  \equiv^{1}  \possiblyWithSub\stageImetaColor{T^{\superscriptI} }_{{\mathrm{22}}} 
    }{%
        \possiblyWithSub\stageImetaColor{T^{\superscriptI} }_{{\mathrm{11}}}  \relI{\to}  \possiblyWithSub\stageImetaColor{T^{\superscriptI} }_{{\mathrm{12}}}   \equiv^{1}   \possiblyWithSub\stageImetaColor{T^{\superscriptI} }_{{\mathrm{21}}}  \relI{\to}  \possiblyWithSub\stageImetaColor{T^{\superscriptI} }_{{\mathrm{22}}}  
    }
  \end{center}
  \caption{CSR equivalence relations on types}
  \label{fig:type-equivalence}
\end{figure}
\begin{figure}[p]
\small
  \begin{flushleft}
    \fbox{\( \mathit{\Gamma}  \vdash^{0}  \possiblyWithSub\stageOmetaColor{T^{\superscriptO} } \)}
  \end{flushleft}
  \vspace{-4em}
  \begin{center}
  \hspace{5em}%
    \derive[WfT0-Tensor]{%
       \vdash  \mathit{\Gamma} 
    }{%
       \mathit{\Gamma}  \vdash^{0}   \ttO{Tensor}\  \possiblyWithSub\stageOmetaColor{s}  
    }
  \qquad
    \derive[WfT0-Arr]{%
       \mathit{\Gamma}  \vdash^{0}  \possiblyWithSub\stageOmetaColor{T^{\superscriptO} }_{{\mathrm{1}}} 
    \andalso
        \mathit{\Gamma} ,  \possiblyWithSub\stageOmetaColor{x}  : ( \possiblyWithSub\stageOmetaColor{T^{\superscriptO} }_{{\mathrm{1}}} )^{0}   \vdash^{0}  \possiblyWithSub\stageOmetaColor{T^{\superscriptO} }_{{\mathrm{2}}} 
    }{%
       \mathit{\Gamma}  \vdash^{0}   \openO{(} \possiblyWithSub\stageOmetaColor{x}  \relO{:}  \possiblyWithSub\stageOmetaColor{T^{\superscriptO} }_{{\mathrm{1}}} \closeO{)} \relO{\to}  \possiblyWithSub\stageOmetaColor{T^{\superscriptO} }_{{\mathrm{2}}}  
    }
  \\[0.7em]
    \derive[WfT0-Rfn]{%
        \mathit{\Gamma} ,  \possiblyWithSub\stageOmetaColor{\nu}  : (   \openO{\{} \possiblyWithSub\stageOmetaColor{\nu}_{{\mathrm{1}}}  \relO{:}  \possiblyWithSub\stageOmetaColor{B}  \relO{\mid}     \ttO{true}    \closeO{\} }   )^{0}   \vdash^{0}  \possiblyWithSub\stageOmetaColor{N^{\superscriptO} }  :    \openO{\{} \possiblyWithSub\stageOmetaColor{\nu}_{{\mathrm{2}}}  \relO{:}   \ttO{Bool}   \relO{\mid}  \possiblyWithSub\stageOmetaColor{N'^{\superscriptO} } \closeO{\} }   
    }{%
       \mathit{\Gamma}  \vdash^{0}    \openO{\{} \possiblyWithSub\stageOmetaColor{\nu}  \relO{:}  \possiblyWithSub\stageOmetaColor{B}  \relO{\mid}  \possiblyWithSub\stageOmetaColor{N^{\superscriptO} } \closeO{\} }   
    }
  \qquad
    \derive[WfT0-Code]{%
       \mathit{\Gamma}  \vdash^{1}  \possiblyWithSub\stageImetaColor{T^{\superscriptI} } 
    }{%
       \mathit{\Gamma}  \vdash^{0}   \openO{\langle} \possiblyWithSub\stageImetaColor{T^{\superscriptI} } \closeO{\rangle}  
    }
  \end{center}
  \begin{flushleft}
    \fbox{\( \mathit{\Gamma}  \vdash^{1}  \possiblyWithSub\stageImetaColor{T^{\superscriptI} } \)}
  \end{flushleft}
  \vspace{-4em}
  \begin{center}
    \derive[WfT1-Tensor]{%
       \mathit{\Gamma}  \vdash^{0}  \possiblyWithSub\stageOmetaColor{N^{\superscriptO} }  :    \openO{\{} \possiblyWithSub\stageOmetaColor{\nu}  \relO{:}   \ttO{NatList}   \relO{\mid}  \possiblyWithSub\stageOmetaColor{N'^{\superscriptO} } \closeO{\} }   
    }{%
       \mathit{\Gamma}  \vdash^{1}   \ttI{Tensor}\ \ordI{\%} \possiblyWithSub\stageOmetaColor{N^{\superscriptO} }  
    }
  \\[0.7em]
    \derive[WfT1-Base]{%
       \vdash  \mathit{\Gamma} 
    }{%
       \mathit{\Gamma}  \vdash^{1}   \possiblyWithSub\stageImetaColor{B}  
    }
  \qquad
    \derive[WfT1-Arr]{%
       \mathit{\Gamma}  \vdash^{1}  \possiblyWithSub\stageImetaColor{T^{\superscriptI} }_{{\mathrm{1}}} 
    \andalso
       \mathit{\Gamma}  \vdash^{1}  \possiblyWithSub\stageImetaColor{T^{\superscriptI} }_{{\mathrm{2}}} 
    }{%
       \mathit{\Gamma}  \vdash^{1}   \possiblyWithSub\stageImetaColor{T^{\superscriptI} }_{{\mathrm{1}}}  \relI{\to}  \possiblyWithSub\stageImetaColor{T^{\superscriptI} }_{{\mathrm{2}}}  
    }
  \end{center}
  \begin{flushleft}
    \fbox{\( \vdash  \mathit{\Gamma} \)}
  \end{flushleft}
  \vspace{-4.25em}
  \begin{center}
  \hspace{4em}%
    \derive[WfEnv-Nil]{}{%
       \vdash   \bullet  
    }
  \quad
    \derive[WfEnv-Cons0]{%
       \vdash  \mathit{\Gamma} 
    \andalso
       \mathit{\Gamma}  \vdash^{0}  \possiblyWithSub\stageOmetaColor{T^{\superscriptO} } 
    }{%
       \vdash   \mathit{\Gamma} ,  \possiblyWithSub\stageOmetaColor{x}  : ( \possiblyWithSub\stageOmetaColor{T^{\superscriptO} } )^{0}  
    }
  \quad
    \derive[WfEnv-Cons1]{%
       \vdash  \mathit{\Gamma} 
    \andalso
       \mathit{\Gamma}  \vdash^{1}  \possiblyWithSub\stageImetaColor{T^{\superscriptI} } 
    }{%
       \vdash   \mathit{\Gamma} ,  \possiblyWithSub\stageImetaColor{x}  : ( \possiblyWithSub\stageImetaColor{T^{\superscriptI} } )^{1}  
    }
  \end{center}
  \caption{Well-formedness of types and type environments}
  \label{fig:well-formedness}
\end{figure}
\begin{figure}[p]
\small
  \begin{flushleft}
    \fbox{\( \mathcal{G}  \mid  \mathit{\Psi}  \vdash^{0}  \possiblyWithSub\stageOmetaColor{\mathcal{M}^{\superscriptO} }  \Rightarrow  \possiblyWithSub\stageOmetaColor{R^{\superscriptO} }  \ElabArrow  \possiblyWithSub\stageOmetaColor{N^{\superscriptO} } \)}
  \end{flushleft}
  \vspace{-4.25em}
  \begin{center}
  \hspace{10em}%
    \derive[B0-Var]{%
      \mathcal{G}(\possiblyWithSub\stageOmetaColor{x}) = (\possiblyWithSub\stageOmetaColor{\mathcal{T}^{\superscriptO} })^0
    \andalso
       \mathcal{G}  \mid  \mathit{\Psi}  \vdash^{0}  \possiblyWithSub\stageOmetaColor{\mathcal{T}^{\superscriptO} }  \mathrel{<:}  \possiblyWithSub\stageOmetaColor{R^{\superscriptO} } 
    }{%
       \mathcal{G}  \mid  \mathit{\Psi}  \vdash^{0}   \possiblyWithSub\stageOmetaColor{x}   \Rightarrow  \possiblyWithSub\stageOmetaColor{R^{\superscriptO} }  \ElabArrow   \possiblyWithSub\stageOmetaColor{x}  
    }
  \\[1em]
    \derive[B0-Cst0]{%
      \ConstEnvZero(\possiblyWithSub\stageOmetaColor{p}) = \possiblyWithSub\stageOmetaColor{\mathcal{T}^{\superscriptO} }
    \andalso
       \mathcal{G}  \mid  \mathit{\Psi}  \vdash^{0}  \possiblyWithSub\stageOmetaColor{\mathcal{T}^{\superscriptO} }  \mathrel{<:}  \possiblyWithSub\stageOmetaColor{R^{\superscriptO} } 
    }{%
       \mathcal{G}  \mid  \mathit{\Psi}  \vdash^{0}   \possiblyWithSub\stageOmetaColor{p}   \Rightarrow  \possiblyWithSub\stageOmetaColor{R^{\superscriptO} }  \ElabArrow    \possiblyWithSub\stageOmetaColor{p}   
    }
  \quad
    \derive[B0-CstP]{%
      \ConstEnvZero(c) = \possiblyWithSub\stageImetaColor{\tau^{\superscriptI} }
    \andalso
       \mathcal{G}  \mid  \mathit{\Psi}  \vdash^{0}   \mathop{\downarrow}( \possiblyWithSub\stageImetaColor{\tau^{\superscriptI} } )   \mathrel{<:}  \possiblyWithSub\stageOmetaColor{R^{\superscriptO} } 
    }{%
       \mathcal{G}  \mid  \mathit{\Psi}  \vdash^{0}   \possiblyWithSub\stageOmetaColor{c}   \Rightarrow  \possiblyWithSub\stageOmetaColor{R^{\superscriptO} }  \ElabArrow    \possiblyWithSub\stageOmetaColor{c}   
    }
  \\[1em]
    \derive[B0-App]{%
       \mathcal{G}  \mid   \bullet   \vdash^{0}  \possiblyWithSub\stageOmetaColor{\mathcal{M}^{\superscriptO} }_{{\mathrm{2}}}  \Rightarrow   \possiblyWithSub\stageOmetaColor{\mathcal{T}^{\superscriptO} }_{{\mathrm{2}}}   \ElabArrow  \possiblyWithSub\stageOmetaColor{N^{\superscriptO} }_{{\mathrm{2}}} 
    \\
       \mathcal{G}  \mid   \mathit{\Psi} , ( \possiblyWithSub\stageOmetaColor{N^{\superscriptO} }_{{\mathrm{2}}}  :  \possiblyWithSub\stageOmetaColor{\mathcal{T}^{\superscriptO} }_{{\mathrm{2}}} )^{0}_{ \ell }   \vdash^{0}  \possiblyWithSub\stageOmetaColor{\mathcal{M}^{\superscriptO} }_{{\mathrm{1}}}  \Rightarrow    ( \possiblyWithSub\stageOmetaColor{N^{\superscriptO} }_{{\mathrm{0}}} \ /\  \possiblyWithSub\stageOmetaColor{\mathcal{T}^{\superscriptO} }_{{\mathrm{11}}} )^{0}   \relO{\to}  \possiblyWithSub\stageOmetaColor{R^{\superscriptO} }_{{\mathrm{12}}}   \ElabArrow  \possiblyWithSub\stageOmetaColor{N^{\superscriptO} }_{{\mathrm{1}}} 
    }{%
       \mathcal{G}  \mid  \mathit{\Psi}  \vdash^{0}   \openO{(} \possiblyWithSub\stageOmetaColor{\mathcal{M}^{\superscriptO} }_{{\mathrm{1}}} \  \possiblyWithSub\stageOmetaColor{\mathcal{M}^{\superscriptO} }_{{\mathrm{2}}} \closeO{)}_{ \ell }   \Rightarrow  \possiblyWithSub\stageOmetaColor{R^{\superscriptO} }_{{\mathrm{12}}}  \ElabArrow   \possiblyWithSub\stageOmetaColor{N^{\superscriptO} }_{{\mathrm{1}}} \   \openO{(}  \possiblyWithSub\stageOmetaColor{N^{\superscriptO} }_{{\mathrm{0}}} \  \possiblyWithSub\stageOmetaColor{N^{\superscriptO} }_{{\mathrm{2}}}  \closeO{)}   
    }
  \\[1em]
    \derive[B0-AbsNoAnnot]{%
        \mathcal{G} ,  \possiblyWithSub\stageOmetaColor{x}  : ( \possiblyWithSub\stageOmetaColor{\mathcal{T}^{\superscriptO} }_{{\mathrm{1}}} )^{0}   \mid  \mathit{\Psi}  \vdash^{0}  \possiblyWithSub\stageOmetaColor{\mathcal{M}^{\superscriptO} }_{{\mathrm{2}}}  \Rightarrow  \possiblyWithSub\stageOmetaColor{R^{\superscriptO} }_{{\mathrm{2}}}  \ElabArrow  \possiblyWithSub\stageOmetaColor{N^{\superscriptO} }_{{\mathrm{2}}} 
    \andalso
      \possiblyWithSub\stageOmetaColor{N^{\superscriptO} }_{{\mathrm{0}}} :=  \ordO{\lambda} \possiblyWithSub\stageOmetaColor{x'}  \relO{:}   \lfloor  \possiblyWithSub\stageOmetaColor{\mathcal{T}^{\superscriptO} }_{{\mathrm{1}}} \rfloor  \punctO{.}\   \possiblyWithSub\stageOmetaColor{x'}  
    }{%
       \mathcal{G}  \mid   \mathit{\Psi} , ( \possiblyWithSub\stageOmetaColor{N'^{\superscriptO} }_{{\mathrm{1}}}  :  \possiblyWithSub\stageOmetaColor{\mathcal{T}^{\superscriptO} }_{{\mathrm{1}}} )^{0}_{ \ell }   \vdash^{0}   \ordO{\lambda} \possiblyWithSub\stageOmetaColor{x} \punctO{.}\  \possiblyWithSub\stageOmetaColor{\mathcal{M}^{\superscriptO} }_{{\mathrm{2}}}   \Rightarrow    ( \possiblyWithSub\stageOmetaColor{N^{\superscriptO} }_{{\mathrm{0}}} \ /\  \possiblyWithSub\stageOmetaColor{\mathcal{T}^{\superscriptO} }_{{\mathrm{1}}} )^{0}   \relO{\to}  \possiblyWithSub\stageOmetaColor{R^{\superscriptO} }_{{\mathrm{2}}}   \ElabArrow   \ordO{\lambda} \possiblyWithSub\stageOmetaColor{x}  \relO{:}   \lfloor  \possiblyWithSub\stageOmetaColor{\mathcal{T}^{\superscriptO} }_{{\mathrm{1}}} \rfloor  \punctO{.}\  \possiblyWithSub\stageOmetaColor{N^{\superscriptO} }_{{\mathrm{2}}}  
    }
  \\[1em]
    \derive[B0-AbsAnnot1]{%
       \mathcal{G}  \vdash^{0}  \possiblyWithSub\stageOmetaColor{\mathcal{S}^{\superscriptO} }_{{\mathrm{1}}}  \ElabArrow  \possiblyWithSub\stageOmetaColor{\mathcal{T}^{\superscriptO} }_{{\mathrm{1}}} 
    \andalso
        \mathcal{G} ,  \possiblyWithSub\stageOmetaColor{x}  : ( \possiblyWithSub\stageOmetaColor{\mathcal{T}^{\superscriptO} }_{{\mathrm{1}}} )^{0}   \mid   \bullet   \vdash^{0}  \possiblyWithSub\stageOmetaColor{\mathcal{M}^{\superscriptO} }_{{\mathrm{2}}}  \Rightarrow   \possiblyWithSub\stageOmetaColor{\mathcal{T}^{\superscriptO} }_{{\mathrm{2}}}   \ElabArrow  \possiblyWithSub\stageOmetaColor{N^{\superscriptO} }_{{\mathrm{2}}} 
    }{%
       \mathcal{G}  \mid   \bullet   \vdash^{0}   \ordO{\lambda} \possiblyWithSub\stageOmetaColor{x}  \relO{:}  \possiblyWithSub\stageOmetaColor{\mathcal{S}^{\superscriptO} }_{{\mathrm{1}}} \punctO{.}\  \possiblyWithSub\stageOmetaColor{\mathcal{M}^{\superscriptO} }_{{\mathrm{2}}}   \Rightarrow    \openO{(} \possiblyWithSub\stageOmetaColor{x}  \relO{:}  \possiblyWithSub\stageOmetaColor{\mathcal{T}^{\superscriptO} }_{{\mathrm{1}}} \closeO{)} \relO{\to}  \possiblyWithSub\stageOmetaColor{\mathcal{T}^{\superscriptO} }_{{\mathrm{2}}}    \ElabArrow   \ordO{\lambda} \possiblyWithSub\stageOmetaColor{x}  \relO{:}   \lfloor  \possiblyWithSub\stageOmetaColor{\mathcal{T}^{\superscriptO} }_{{\mathrm{1}}} \rfloor  \punctO{.}\  \possiblyWithSub\stageOmetaColor{N^{\superscriptO} }_{{\mathrm{2}}}  
    }
  \\[1em]
    \derive[B0-AbsAnnot2]{%
       \mathcal{G}  \vdash^{0}  \possiblyWithSub\stageOmetaColor{\mathcal{S}^{\superscriptO} }_{{\mathrm{1}}}  \ElabArrow  \possiblyWithSub\stageOmetaColor{\mathcal{T}^{\superscriptO} }_{{\mathrm{1}}} 
    \andalso
        \lfloor  \mathcal{G} \rfloor   \vdash_{  \ell  }   \lfloor  \possiblyWithSub\stageOmetaColor{\mathcal{T}'^{\superscriptO} }_{{\mathrm{1}}} \rfloor   \CastArrow   \lfloor  \possiblyWithSub\stageOmetaColor{\mathcal{T}^{\superscriptO} } \rfloor   \ElabArrow  \possiblyWithSub\stageOmetaColor{N^{\superscriptO} }_{{\mathrm{0}}} 
    \\
        \mathcal{G} ,  \possiblyWithSub\stageOmetaColor{x}  : ( \possiblyWithSub\stageOmetaColor{\mathcal{T}^{\superscriptO} }_{{\mathrm{1}}} )^{0}   \mid  \mathit{\Psi}  \vdash^{0}  \possiblyWithSub\stageOmetaColor{\mathcal{M}^{\superscriptO} }_{{\mathrm{2}}}  \Rightarrow  \possiblyWithSub\stageOmetaColor{R^{\superscriptO} }_{{\mathrm{2}}}  \ElabArrow  \possiblyWithSub\stageOmetaColor{N^{\superscriptO} }_{{\mathrm{2}}} 
    }{%
       \mathcal{G}  \mid   \mathit{\Psi} , ( \possiblyWithSub\stageOmetaColor{N'^{\superscriptO} }_{{\mathrm{1}}}  :  \possiblyWithSub\stageOmetaColor{\mathcal{T}'^{\superscriptO} }_{{\mathrm{1}}} )^{0}_{ \ell }   \vdash^{0}   \ordO{\lambda} \possiblyWithSub\stageOmetaColor{x}  \relO{:}  \possiblyWithSub\stageOmetaColor{\mathcal{S}^{\superscriptO} }_{{\mathrm{1}}} \punctO{.}\  \possiblyWithSub\stageOmetaColor{\mathcal{M}^{\superscriptO} }_{{\mathrm{2}}}   \Rightarrow    ( \possiblyWithSub\stageOmetaColor{N^{\superscriptO} }_{{\mathrm{0}}} \ /\  \possiblyWithSub\stageOmetaColor{\mathcal{T}^{\superscriptO} }_{{\mathrm{1}}} )^{0}   \relO{\to}  \possiblyWithSub\stageOmetaColor{R^{\superscriptO} }_{{\mathrm{2}}}   \ElabArrow   \ordO{\lambda} \possiblyWithSub\stageOmetaColor{x}  \relO{:}   \lfloor  \possiblyWithSub\stageOmetaColor{\mathcal{T}^{\superscriptO} }_{{\mathrm{1}}} \rfloor  \punctO{.}\  \possiblyWithSub\stageOmetaColor{N^{\superscriptO} }_{{\mathrm{2}}}  
    }
  \\[1em]
    \derive[B0-Brkt]{%
      \mathcal{G} \mid \mathit{\Psi} \vdash^{1} \possiblyWithSub\stageImetaColor{\mathcal{M}^{\superscriptI} } \Rightarrow (D_{\ottmv{i}} \to)_{i = 1}^{m}  \possiblyWithSub\stageImetaColor{T^{\superscriptI} }  \ElabArrow \possiblyWithSub\stageImetaColor{N^{\superscriptI} }
    }{%
      \mathcal{G} \mid \mathit{\Psi} \vdash^{0}  \openO{\langle} \possiblyWithSub\stageImetaColor{\mathcal{M}^{\superscriptI} } \closeO{\rangle}  \Rightarrow (D_{\ottmv{i}} \to)_{i = 1}^{m}   \openO{\langle} \possiblyWithSub\stageImetaColor{T^{\superscriptI} } \closeO{\rangle}   \ElabArrow  \openO{\langle} \possiblyWithSub\stageImetaColor{N^{\superscriptI} } \closeO{\rangle} 
    }
  \\[1em]
    \derive[B0-AbsImp]{%
       \mathcal{G}  \vdash^{0}  \possiblyWithSub\stageOmetaColor{\mathcal{S}^{\superscriptO} }_{{\mathrm{1}}}  \ElabArrow  \possiblyWithSub\stageOmetaColor{\mathcal{T}^{\superscriptO} }_{{\mathrm{1}}} 
    \andalso
        \mathcal{G} ,  \possiblyWithSub\stageOmetaColor{x}  : ( \possiblyWithSub\stageOmetaColor{\mathcal{T}^{\superscriptO} }_{{\mathrm{1}}} )^{0}   \mid   \bullet   \vdash^{0}  \possiblyWithSub\stageOmetaColor{\mathcal{M}^{\superscriptO} }_{{\mathrm{2}}}  \Rightarrow   \possiblyWithSub\stageOmetaColor{\mathcal{T}^{\superscriptO} }_{{\mathrm{2}}}   \ElabArrow  \possiblyWithSub\stageOmetaColor{N^{\superscriptO} }_{{\mathrm{2}}} 
    }{%
       \mathcal{G}  \mid   \bullet   \vdash^{0}   \ordO{\lambda}\openO{\{} \possiblyWithSub\stageOmetaColor{x}  \relO{:}  \possiblyWithSub\stageOmetaColor{\mathcal{S}^{\superscriptO} }_{{\mathrm{1}}} \closeO{\} }\punctO{.}\  \possiblyWithSub\stageOmetaColor{\mathcal{M}^{\superscriptO} }_{{\mathrm{2}}}   \Rightarrow    \openO{\{} \possiblyWithSub\stageOmetaColor{x}  \relO{:}  \possiblyWithSub\stageOmetaColor{\mathcal{T}^{\superscriptO} }_{{\mathrm{1}}} \closeO{\} } \relO{\to}  \possiblyWithSub\stageOmetaColor{\mathcal{T}^{\superscriptO} }_{{\mathrm{2}}}    \ElabArrow   \ordO{\lambda} \possiblyWithSub\stageOmetaColor{x}  \relO{:}   \lfloor  \possiblyWithSub\stageOmetaColor{\mathcal{T}^{\superscriptO} }_{{\mathrm{1}}} \rfloor  \punctO{.}\  \possiblyWithSub\stageOmetaColor{N^{\superscriptO} }_{{\mathrm{2}}}  
    }
  \\[1em]
    \derive[B0-AppImp]{%
       \mathcal{G}  \mid   \bullet   \vdash^{0}  \possiblyWithSub\stageOmetaColor{\mathcal{M}^{\superscriptO} }_{{\mathrm{2}}}  \Rightarrow   \possiblyWithSub\stageOmetaColor{\mathcal{T}^{\superscriptO} }_{{\mathrm{2}}}   \ElabArrow  \possiblyWithSub\stageOmetaColor{N^{\superscriptO} }_{{\mathrm{2}}} 
    \\
       \mathcal{G}  \mid   \mathit{\Psi} , ( \possiblyWithSub\stageOmetaColor{N^{\superscriptO} }_{{\mathrm{2}}}  :  \possiblyWithSub\stageOmetaColor{\mathcal{T}^{\superscriptO} }_{{\mathrm{2}}} )^{0}_{ \ell }   \vdash^{0}  \possiblyWithSub\stageOmetaColor{\mathcal{M}^{\superscriptO} }_{{\mathrm{1}}}  \Rightarrow    \{ \possiblyWithSub\stageOmetaColor{N^{\superscriptO} }_{{\mathrm{0}}} \ /\  \possiblyWithSub\stageOmetaColor{\mathcal{T}^{\superscriptO} }_{{\mathrm{11}}} \}   \relO{\to}  \possiblyWithSub\stageOmetaColor{R^{\superscriptO} }_{{\mathrm{12}}}   \ElabArrow  \possiblyWithSub\stageOmetaColor{N^{\superscriptO} }_{{\mathrm{2}}} 
    }{%
       \mathcal{G}  \mid  \mathit{\Psi}  \vdash^{0}   \openO{(} \possiblyWithSub\stageOmetaColor{\mathcal{M}^{\superscriptO} }_{{\mathrm{1}}} \ \openO{\{} \possiblyWithSub\stageOmetaColor{\mathcal{M}^{\superscriptO} }_{{\mathrm{2}}} \closeO{\} }\closeO{)}_{ \ell }   \Rightarrow  \possiblyWithSub\stageOmetaColor{R^{\superscriptO} }_{{\mathrm{12}}}  \ElabArrow   \possiblyWithSub\stageOmetaColor{N^{\superscriptO} }_{{\mathrm{1}}} \   \openO{(}  \possiblyWithSub\stageOmetaColor{N^{\superscriptO} }_{{\mathrm{0}}} \  \possiblyWithSub\stageOmetaColor{N^{\superscriptO} }_{{\mathrm{2}}}  \closeO{)}   
    }
  \\[1em]
    \derive[B0-FillImp]{%
       \mathcal{G}  \mid   \mathit{\Psi} , \__{ \ell }   \vdash^{0}  \possiblyWithSub\stageOmetaColor{\mathcal{M}^{\superscriptO} }_{{\mathrm{1}}}  \Rightarrow    \mathbf{fill}\ \{ \possiblyWithSub\stageOmetaColor{N^{\superscriptO} }_{{\mathrm{2}}}  :  \possiblyWithSub\stageOmetaColor{\mathcal{T}^{\superscriptO} }_{{\mathrm{2}}} \}   \relO{\to}  \possiblyWithSub\stageOmetaColor{R^{\superscriptO} }   \ElabArrow  \possiblyWithSub\stageOmetaColor{N^{\superscriptO} }_{{\mathrm{1}}} 
    }{%
       \mathcal{G}  \mid  \mathit{\Psi}  \vdash^{0}   \possiblyWithSub\stageOmetaColor{\mathcal{M}^{\superscriptO} }_{{\mathrm{1}}} \ \ordO{\_}   \Rightarrow  \possiblyWithSub\stageOmetaColor{R^{\superscriptO} }  \ElabArrow   \possiblyWithSub\stageOmetaColor{N^{\superscriptO} }_{{\mathrm{1}}} \  \possiblyWithSub\stageOmetaColor{N^{\superscriptO} }_{{\mathrm{2}}}  
    }
  \\[1em]
    \derive[B0-InsertImp]{%
       \mathcal{G}  \mid  \mathit{\Psi}  \vdash^{0}  \possiblyWithSub\stageOmetaColor{\mathcal{M}^{\superscriptO} }_{{\mathrm{1}}}  \Rightarrow    \mathbf{insert}\ \{ \possiblyWithSub\stageOmetaColor{N^{\superscriptO} }_{{\mathrm{2}}}  :  \possiblyWithSub\stageOmetaColor{\mathcal{T}^{\superscriptO} }_{{\mathrm{2}}} \}   \relO{\to}  \possiblyWithSub\stageOmetaColor{R^{\superscriptO} }   \ElabArrow  \possiblyWithSub\stageOmetaColor{N^{\superscriptO} }_{{\mathrm{1}}} 
    }{%
       \mathcal{G}  \mid  \mathit{\Psi}  \vdash^{0}  \possiblyWithSub\stageOmetaColor{\mathcal{M}^{\superscriptO} }_{{\mathrm{1}}}  \Rightarrow  \possiblyWithSub\stageOmetaColor{R^{\superscriptO} }  \ElabArrow   \possiblyWithSub\stageOmetaColor{N^{\superscriptO} }_{{\mathrm{1}}} \  \possiblyWithSub\stageOmetaColor{N^{\superscriptO} }_{{\mathrm{2}}}  
    }
  \end{center}
  \caption{The rules for the inference of implicit arguments (1/2)}
  \label{fig:option-inference-main-full-1}
\end{figure}
\begin{figure}[p]
\small
  \begin{flushleft}
    \fbox{\( \mathcal{G}  \mid  \mathit{\Psi}  \vdash^{1}  \possiblyWithSub\stageImetaColor{\mathcal{M}^{\superscriptI} }  \Rightarrow  \possiblyWithSub\stageImetaColor{R^{\superscriptI} }  \ElabArrow  \possiblyWithSub\stageImetaColor{N^{\superscriptI} } \)}
  \end{flushleft}
  \vspace{-4.25em}
  \begin{center}
  \hspace{12em}%
    \derive[B1-Esc]{%
      \mathcal{G} \mid \mathit{\Psi} \vdash^{0} \possiblyWithSub\stageOmetaColor{\mathcal{M}^{\superscriptO} } \Rightarrow (D_{\ottmv{i}} \to)_{i = 1}^{m}   \openO{\langle} \possiblyWithSub\stageImetaColor{T^{\superscriptI} } \closeO{\rangle}   \ElabArrow \possiblyWithSub\stageOmetaColor{N^{\superscriptO} }
    }{%
      \mathcal{G} \mid \mathit{\Psi} \vdash^{1}  \ordI{\sim} \possiblyWithSub\stageOmetaColor{\mathcal{M}^{\superscriptO} }  \Rightarrow (D_{\ottmv{i}} \to)_{i = 1}^{m}  \possiblyWithSub\stageImetaColor{T^{\superscriptI} }  \ElabArrow  \ordI{\sim} \possiblyWithSub\stageOmetaColor{N^{\superscriptO} } 
    }
  \\[1em]
    \derive[B1-Var]{%
      \mathcal{G}(\possiblyWithSub\stageImetaColor{x}) = (\possiblyWithSub\stageImetaColor{T^{\superscriptI} })^1
    \andalso
       \mathcal{G}  \mid  \mathit{\Psi}  \vdash^{1}  \possiblyWithSub\stageImetaColor{T^{\superscriptI} }  \mathrel{<:}  \possiblyWithSub\stageImetaColor{R^{\superscriptI} } 
    }{%
       \mathcal{G}  \mid  \mathit{\Psi}  \vdash^{1}   \possiblyWithSub\stageImetaColor{x}   \Rightarrow  \possiblyWithSub\stageImetaColor{R^{\superscriptI} }  \ElabArrow   \possiblyWithSub\stageImetaColor{x}  
    }
  \qquad
    \derive[B1-CstP]{%
      \ConstEnvZero(c) = \possiblyWithSub\stageImetaColor{\tau^{\superscriptI} }
    \andalso
       \mathcal{G}  \mid  \mathit{\Psi}  \vdash^{1}   \possiblyWithSub\stageImetaColor{\tau^{\superscriptI} }   \mathrel{<:}  \possiblyWithSub\stageImetaColor{R^{\superscriptI} } 
    }{%
       \mathcal{G}  \mid  \mathit{\Psi}  \vdash^{1}   \possiblyWithSub\stageImetaColor{c}   \Rightarrow  \possiblyWithSub\stageImetaColor{R^{\superscriptI} }  \ElabArrow   \possiblyWithSub\stageImetaColor{c}  
    }
  \\[1em]
    \derive[B1-App]{%
       \mathcal{G}  \mid   \bullet   \vdash^{1}  \possiblyWithSub\stageImetaColor{\mathcal{M}^{\superscriptI} }_{{\mathrm{2}}}  \Rightarrow   \possiblyWithSub\stageImetaColor{T^{\superscriptI} }_{{\mathrm{2}}}   \ElabArrow  \possiblyWithSub\stageImetaColor{N^{\superscriptI} }_{{\mathrm{2}}} 
    \\
       \mathcal{G}  \mid   \mathit{\Psi} , ( \possiblyWithSub\stageImetaColor{T^{\superscriptI} }_{{\mathrm{2}}} )^{1}_{ \ell }   \vdash^{1}  \possiblyWithSub\stageImetaColor{\mathcal{M}^{\superscriptI} }_{{\mathrm{1}}}  \Rightarrow    ( \possiblyWithSub\stageOmetaColor{N^{\superscriptO} }_{{\mathrm{0}}} \ /\  \possiblyWithSub\stageImetaColor{T^{\superscriptI} }_{{\mathrm{11}}} )^{1}   \relI{\to}  \possiblyWithSub\stageImetaColor{R^{\superscriptI} }_{{\mathrm{12}}}   \ElabArrow  \possiblyWithSub\stageImetaColor{N^{\superscriptI} }_{{\mathrm{1}}} 
    }{%
       \mathcal{G}  \mid  \mathit{\Psi}  \vdash^{1}   \openI{(} \possiblyWithSub\stageImetaColor{\mathcal{M}^{\superscriptI} }_{{\mathrm{1}}} \  \possiblyWithSub\stageImetaColor{\mathcal{M}^{\superscriptI} }_{{\mathrm{2}}} \closeI{)}_{ \ell }   \Rightarrow  \possiblyWithSub\stageImetaColor{R^{\superscriptI} }_{{\mathrm{12}}}  \ElabArrow   \possiblyWithSub\stageImetaColor{N^{\superscriptI} }_{{\mathrm{1}}} \   \ordI{\sim}  \openO{(}  \possiblyWithSub\stageOmetaColor{N^{\superscriptO} }_{{\mathrm{0}}} \   \openO{\langle} \possiblyWithSub\stageImetaColor{N^{\superscriptI} }_{{\mathrm{2}}} \closeO{\rangle}   \closeO{)}    
    }
  \\[1em]
    \derive[B1-AbsAnnot1]{%
       \mathcal{G}  \vdash^{1}  \possiblyWithSub\stageImetaColor{\mathcal{S}^{\superscriptI} }_{{\mathrm{1}}}  \ElabArrow  \possiblyWithSub\stageImetaColor{T^{\superscriptI} }_{{\mathrm{1}}} 
    \andalso
        \mathcal{G} ,  \possiblyWithSub\stageImetaColor{x}  : ( \possiblyWithSub\stageImetaColor{T^{\superscriptI} }_{{\mathrm{1}}} )^{1}   \mid   \bullet   \vdash^{1}  \possiblyWithSub\stageImetaColor{\mathcal{M}^{\superscriptI} }_{{\mathrm{2}}}  \Rightarrow   \possiblyWithSub\stageImetaColor{T^{\superscriptI} }_{{\mathrm{2}}}   \ElabArrow  \possiblyWithSub\stageImetaColor{N^{\superscriptI} }_{{\mathrm{2}}} 
    }{%
       \mathcal{G}  \mid   \bullet   \vdash^{1}   \ordI{\lambda} \possiblyWithSub\stageImetaColor{x}  \relI{:}  \possiblyWithSub\stageImetaColor{\mathcal{S}^{\superscriptI} }_{{\mathrm{1}}} \punctI{.}\  \possiblyWithSub\stageImetaColor{\mathcal{M}^{\superscriptI} }_{{\mathrm{2}}}   \Rightarrow    \possiblyWithSub\stageImetaColor{T^{\superscriptI} }_{{\mathrm{1}}}  \relI{\to}  \possiblyWithSub\stageImetaColor{T^{\superscriptI} }_{{\mathrm{2}}}    \ElabArrow   \ordI{\lambda} \possiblyWithSub\stageImetaColor{x}  \relI{:}  \possiblyWithSub\stageImetaColor{T^{\superscriptI} }_{{\mathrm{1}}} \punctI{.}\  \possiblyWithSub\stageImetaColor{N^{\superscriptI} }_{{\mathrm{2}}}  
    }
  \\[1em]
    \derive[B1-AbsAnnot2]{%
       \mathcal{G}  \vdash^{1}  \possiblyWithSub\stageImetaColor{\mathcal{S}^{\superscriptI} }_{{\mathrm{1}}}  \ElabArrow  \possiblyWithSub\stageImetaColor{T^{\superscriptI} }_{{\mathrm{1}}} 
    \andalso
        \mathcal{G} ,  \possiblyWithSub\stageImetaColor{x}  : ( \possiblyWithSub\stageImetaColor{T^{\superscriptI} }_{{\mathrm{1}}} )^{1}   \mid  \mathit{\Psi}  \vdash^{1}  \possiblyWithSub\stageImetaColor{\mathcal{M}^{\superscriptI} }_{{\mathrm{2}}}  \Rightarrow  \possiblyWithSub\stageImetaColor{R^{\superscriptI} }_{{\mathrm{2}}}  \ElabArrow  \possiblyWithSub\stageImetaColor{N^{\superscriptI} }_{{\mathrm{2}}} 
    \\
      \possiblyWithSub\stageOmetaColor{N^{\superscriptO} }_{{\mathrm{0}}} :=  \LeftAssertParen\openO{\langle} \possiblyWithSub\stageImetaColor{T'^{\superscriptI} }_{{\mathrm{1}}} \closeO{\rangle} \relO{\CastArrow} \openO{\langle} \possiblyWithSub\stageImetaColor{T^{\superscriptI} }_{{\mathrm{1}}} \closeO{\rangle}\RightAssertParen^{  \ell  } 
    }{%
       \mathcal{G}  \mid   \mathit{\Psi} , ( \possiblyWithSub\stageImetaColor{T'^{\superscriptI} }_{{\mathrm{1}}} )^{1}_{ \ell }   \vdash^{1}   \ordI{\lambda} \possiblyWithSub\stageImetaColor{x}  \relI{:}  \possiblyWithSub\stageImetaColor{\mathcal{S}^{\superscriptI} }_{{\mathrm{1}}} \punctI{.}\  \possiblyWithSub\stageImetaColor{\mathcal{M}^{\superscriptI} }_{{\mathrm{2}}}   \Rightarrow    ( \possiblyWithSub\stageOmetaColor{N^{\superscriptO} }_{{\mathrm{0}}} \ /\  \possiblyWithSub\stageImetaColor{T^{\superscriptI} }_{{\mathrm{1}}} )^{1}   \relI{\to}  \possiblyWithSub\stageImetaColor{R^{\superscriptI} }_{{\mathrm{2}}}   \ElabArrow   \ordI{\lambda} \possiblyWithSub\stageImetaColor{x}  \relI{:}  \possiblyWithSub\stageImetaColor{T^{\superscriptI} }_{{\mathrm{1}}} \punctI{.}\  \possiblyWithSub\stageImetaColor{N^{\superscriptI} }_{{\mathrm{2}}}  
    }
  \\[1em]
    \derive[B1-AbsNoAnnot]{%
        \mathcal{G} ,  \possiblyWithSub\stageImetaColor{x}  : ( \possiblyWithSub\stageImetaColor{T^{\superscriptI} }_{{\mathrm{1}}} )^{1}   \mid  \mathit{\Psi}  \vdash^{1}  \possiblyWithSub\stageImetaColor{\mathcal{M}^{\superscriptI} }_{{\mathrm{2}}}  \Rightarrow  \possiblyWithSub\stageImetaColor{R^{\superscriptI} }_{{\mathrm{2}}}  \ElabArrow  \possiblyWithSub\stageImetaColor{N^{\superscriptI} }_{{\mathrm{2}}} 
    \andalso
      \possiblyWithSub\stageOmetaColor{N^{\superscriptO} }_{{\mathrm{0}}} :=  \LeftAssertParen\openO{\langle} \possiblyWithSub\stageImetaColor{T'^{\superscriptI} }_{{\mathrm{1}}} \closeO{\rangle} \relO{\CastArrow} \openO{\langle} \possiblyWithSub\stageImetaColor{T^{\superscriptI} }_{{\mathrm{1}}} \closeO{\rangle}\RightAssertParen^{  \ell  } 
    }{%
       \mathcal{G}  \mid   \mathit{\Psi} , ( \possiblyWithSub\stageImetaColor{T'^{\superscriptI} }_{{\mathrm{1}}} )^{1}_{ \ell }   \vdash^{1}   \ordI{\lambda} \possiblyWithSub\stageImetaColor{x} \punctI{.}\  \possiblyWithSub\stageImetaColor{\mathcal{M}^{\superscriptI} }_{{\mathrm{2}}}   \Rightarrow    ( \possiblyWithSub\stageOmetaColor{N^{\superscriptO} }_{{\mathrm{0}}} \ /\  \possiblyWithSub\stageImetaColor{T^{\superscriptI} }_{{\mathrm{1}}} )^{1}   \relI{\to}  \possiblyWithSub\stageImetaColor{R^{\superscriptI} }_{{\mathrm{2}}}   \ElabArrow   \ordI{\lambda} \possiblyWithSub\stageImetaColor{x}  \relI{:}  \possiblyWithSub\stageImetaColor{T^{\superscriptI} }_{{\mathrm{1}}} \punctI{.}\  \possiblyWithSub\stageImetaColor{N^{\superscriptI} }_{{\mathrm{2}}}  
    }
  \end{center}
  \vspace{0.5em}
  \begin{flushleft}
    \fbox{\( \mathcal{G}  \vdash^{0}  \possiblyWithSub\stageOmetaColor{\mathcal{S}^{\superscriptO} }  \ElabArrow  \possiblyWithSub\stageOmetaColor{\mathcal{T}^{\superscriptO} } \)}
    \fbox{\( \mathcal{G}  \vdash^{1}  \possiblyWithSub\stageImetaColor{\mathcal{S}^{\superscriptI} }  \ElabArrow  \possiblyWithSub\stageImetaColor{T^{\superscriptI} } \)}
  \end{flushleft}
  \begin{center}
    \derive[BT0-Code]{%
       \mathcal{G}  \vdash^{1}  \possiblyWithSub\stageImetaColor{\mathcal{S}^{\superscriptI} }  \ElabArrow  \possiblyWithSub\stageImetaColor{T^{\superscriptI} } 
    }{%
       \mathcal{G}  \vdash^{0}   \openO{\langle} \possiblyWithSub\stageImetaColor{\mathcal{S}^{\superscriptI} } \closeO{\rangle}   \ElabArrow   \openO{\langle} \possiblyWithSub\stageImetaColor{T^{\superscriptI} } \closeO{\rangle}  
    }
  \qquad
    \derive[BT0-Imp]{%
       \mathcal{G}  \vdash^{0}  \possiblyWithSub\stageOmetaColor{\mathcal{S}^{\superscriptO} }_{{\mathrm{1}}}  \ElabArrow  \possiblyWithSub\stageOmetaColor{\mathcal{T}^{\superscriptO} }_{{\mathrm{1}}} 
    \andalso
        \mathcal{G} ,  \possiblyWithSub\stageOmetaColor{x}  : ( \possiblyWithSub\stageOmetaColor{\mathcal{T}^{\superscriptO} }_{{\mathrm{1}}} )^{0}   \vdash^{0}  \possiblyWithSub\stageOmetaColor{\mathcal{S}^{\superscriptO} }_{{\mathrm{2}}}  \ElabArrow  \possiblyWithSub\stageOmetaColor{\mathcal{T}^{\superscriptO} }_{{\mathrm{2}}} 
    }{%
       \mathcal{G}  \vdash^{0}   \openO{\{} \possiblyWithSub\stageOmetaColor{x}  \relO{:}  \possiblyWithSub\stageOmetaColor{\mathcal{S}^{\superscriptO} }_{{\mathrm{1}}} \closeO{\} } \relO{\to}  \possiblyWithSub\stageOmetaColor{\mathcal{S}^{\superscriptO} }_{{\mathrm{2}}}   \ElabArrow   \openO{\{} \possiblyWithSub\stageOmetaColor{x}  \relO{:}  \possiblyWithSub\stageOmetaColor{\mathcal{T}^{\superscriptO} }_{{\mathrm{1}}} \closeO{\} } \relO{\to}  \possiblyWithSub\stageOmetaColor{\mathcal{T}^{\superscriptO} }_{{\mathrm{2}}}  
    }
  \\[1em]
    \derive[BT1-Tensor]{%
       \mathcal{G}  \mid   \bullet   \vdash^{0}  \possiblyWithSub\stageOmetaColor{\mathcal{M}^{\superscriptO} }  \Rightarrow   \possiblyWithSub\stageOmetaColor{\mathcal{T}^{\superscriptO} }   \ElabArrow  \possiblyWithSub\stageOmetaColor{N^{\superscriptO} } 
    \andalso
        \lfloor  \mathcal{G} \rfloor   \vdash_{  \ell  }   \lfloor  \possiblyWithSub\stageOmetaColor{\mathcal{T}^{\superscriptO} } \rfloor   \CastArrow    \openO{\{} \possiblyWithSub\stageOmetaColor{\nu}  \relO{:}   \ttO{NatList}   \relO{\mid}     \ttO{true}    \closeO{\} }    \ElabArrow  \possiblyWithSub\stageOmetaColor{N^{\superscriptO} }_{{\mathrm{0}}} 
    }{%
       \mathcal{G}  \vdash^{1}   \openI{(}\ttI{Tensor}\ \ordI{\%} \possiblyWithSub\stageOmetaColor{\mathcal{M}^{\superscriptO} } \closeI{)}_{ \ell }   \ElabArrow   \ttI{Tensor}\ \ordI{\%}  \openO{(}  \possiblyWithSub\stageOmetaColor{N^{\superscriptO} }_{{\mathrm{0}}} \  \possiblyWithSub\stageOmetaColor{N^{\superscriptO} }  \closeO{)}   
    }
  \end{center}
  \caption{The rules for the inference of implicit arguments (2/2)}
  \label{fig:option-inference-main-full-2}
\end{figure}
\begin{figure}[p]
\small
  \begin{flushleft}
    \fbox{\( \mathcal{G}  \mid  I  \mid  \mathit{\Psi}  \vdash^{0}  \possiblyWithSub\stageOmetaColor{\mathcal{T}^{\superscriptO} }  \mathrel{<:}  \possiblyWithSub\stageOmetaColor{R^{\superscriptO} }  \mid  \theta \)}
  \end{flushleft}
  \vspace{-4.25em}
  \begin{center}
    \hspace{10em}%
    \derive[ABI0-Empty]{}{%
       \mathcal{G}  \mid  I  \mid   \bullet   \vdash^{0}  \possiblyWithSub\stageOmetaColor{\mathcal{T}^{\superscriptO} }  \mathrel{<:}   \possiblyWithSub\stageOmetaColor{\mathcal{T}^{\superscriptO} }   \mid   \varnothing  
    }
  \\[1em]
    \derive[ABI0-Arr]{%
       \mathcal{G}  \mid  I  \vdash_{  \ell  }  \possiblyWithSub\stageOmetaColor{\mathcal{T}'^{\superscriptO} }_{{\mathrm{1}}}  \CastArrow  \possiblyWithSub\stageOmetaColor{\mathcal{T}^{\superscriptO} }_{{\mathrm{1}}}  \ElabArrow  \possiblyWithSub\stageOmetaColor{N^{\superscriptO} }_{{\mathrm{0}}}  \mid  \theta_{{\mathrm{1}}} 
    \\
       \mathcal{G}  \mid   I  \setminus   \mathop{\mathrm{dom} }  \theta_{{\mathrm{1}}}    \mid  \mathit{\Psi}  \vdash^{0}    [   \possiblyWithSub\stageOmetaColor{N^{\superscriptO} }_{{\mathrm{0}}} \  \possiblyWithSub\stageOmetaColor{N^{\superscriptO} }_{{\mathrm{1}}}   /  \possiblyWithSub\stageOmetaColor{x}  ]     \theta_{{\mathrm{1}}}   \possiblyWithSub\stageOmetaColor{\mathcal{T}^{\superscriptO} }_{{\mathrm{2}}}    \mathrel{<:}  \possiblyWithSub\stageOmetaColor{R^{\superscriptO} }_{{\mathrm{2}}}  \mid  \theta_{{\mathrm{2}}} 
    }{%
       \mathcal{G}  \mid  I  \mid   \mathit{\Psi} , ( \possiblyWithSub\stageOmetaColor{N^{\superscriptO} }_{{\mathrm{1}}}  :  \possiblyWithSub\stageOmetaColor{\mathcal{T}'^{\superscriptO} }_{{\mathrm{1}}} )^{0}_{ \ell }   \vdash^{0}   \openO{(} \possiblyWithSub\stageOmetaColor{x}  \relO{:}  \possiblyWithSub\stageOmetaColor{\mathcal{T}^{\superscriptO} }_{{\mathrm{1}}} \closeO{)} \relO{\to}  \possiblyWithSub\stageOmetaColor{\mathcal{T}^{\superscriptO} }_{{\mathrm{2}}}   \mathrel{<:}    ( \possiblyWithSub\stageOmetaColor{N^{\superscriptO} } \ /\    \theta_{{\mathrm{1}}}   \possiblyWithSub\stageOmetaColor{\mathcal{T}^{\superscriptO} }_{{\mathrm{1}}}   )^{0}   \relO{\to}  \possiblyWithSub\stageOmetaColor{R^{\superscriptO} }_{{\mathrm{2}}}   \mid  \theta_{{\mathrm{2}}} 
    }
  \\[1em]
    \derive[ABI0-ImpGiven]{%
       \mathcal{G}  \mid  I  \vdash_{  \ell  }  \possiblyWithSub\stageOmetaColor{\mathcal{T}'^{\superscriptO} }_{{\mathrm{1}}}  \CastArrow  \possiblyWithSub\stageOmetaColor{\mathcal{T}^{\superscriptO} }_{{\mathrm{1}}}  \ElabArrow  \possiblyWithSub\stageOmetaColor{N^{\superscriptO} }_{{\mathrm{0}}}  \mid  \theta_{{\mathrm{1}}} 
    \\
       \mathcal{G}  \mid   I  \setminus   \mathop{\mathrm{dom} }  \theta_{{\mathrm{1}}}    \mid  \mathit{\Psi}  \vdash^{0}    [   \possiblyWithSub\stageOmetaColor{N^{\superscriptO} }_{{\mathrm{0}}} \  \possiblyWithSub\stageOmetaColor{N^{\superscriptO} }_{{\mathrm{1}}}   /  \possiblyWithSub\stageOmetaColor{x}  ]     \theta_{{\mathrm{1}}}   \possiblyWithSub\stageOmetaColor{\mathcal{T}^{\superscriptO} }_{{\mathrm{2}}}    \mathrel{<:}  \possiblyWithSub\stageOmetaColor{R^{\superscriptO} }_{{\mathrm{2}}}  \mid  \theta_{{\mathrm{2}}} 
    }{%
       \mathcal{G}  \mid  I  \mid   \mathit{\Psi} , \{ \possiblyWithSub\stageOmetaColor{N^{\superscriptO} }_{{\mathrm{1}}}  :  \possiblyWithSub\stageOmetaColor{\mathcal{T}'^{\superscriptO} }_{{\mathrm{1}}} \}_{ \ell }   \vdash^{0}   \openO{\{} \possiblyWithSub\stageOmetaColor{x}  \relO{:}  \possiblyWithSub\stageOmetaColor{\mathcal{T}^{\superscriptO} }_{{\mathrm{1}}} \closeO{\} } \relO{\to}  \possiblyWithSub\stageOmetaColor{\mathcal{T}^{\superscriptO} }_{{\mathrm{2}}}   \mathrel{<:}    \{ \possiblyWithSub\stageOmetaColor{N^{\superscriptO} }_{{\mathrm{0}}} \ /\    \theta_{{\mathrm{1}}}   \possiblyWithSub\stageOmetaColor{\mathcal{T}^{\superscriptO} }_{{\mathrm{1}}}   \}   \relO{\to}  \possiblyWithSub\stageOmetaColor{R^{\superscriptO} }_{{\mathrm{2}}}   \mid  \theta_{{\mathrm{2}}} 
    }
  \\[1em]
    \derive[ABI0-ImpGuess1]{%
       \mathcal{G}  \mid   I  \uplus   \{ \possiblyWithSub\stageOmetaColor{x} \}    \mid  \mathit{\Psi}  \vdash^{0}  \possiblyWithSub\stageOmetaColor{\mathcal{T}^{\superscriptO} }_{{\mathrm{2}}}  \mathrel{<:}  \possiblyWithSub\stageOmetaColor{R^{\superscriptO} }_{{\mathrm{2}}}  \mid  \theta_{{\mathrm{2}}} 
    \andalso
      \possiblyWithSub\stageOmetaColor{N^{\superscriptO} }_{{\mathrm{1}}} := \theta_{{\mathrm{2}}}(\possiblyWithSub\stageOmetaColor{x})
    }{%
       \mathcal{G}  \mid  I  \mid   \mathit{\Psi} , \__{ \ell }   \vdash^{0}   \openO{\{} \possiblyWithSub\stageOmetaColor{x}  \relO{:}  \possiblyWithSub\stageOmetaColor{\mathcal{T}^{\superscriptO} }_{{\mathrm{1}}} \closeO{\} } \relO{\to}  \possiblyWithSub\stageOmetaColor{\mathcal{T}^{\superscriptO} }_{{\mathrm{2}}}   \mathrel{<:}    \mathbf{fill}\ \{ \possiblyWithSub\stageOmetaColor{N^{\superscriptO} }_{{\mathrm{1}}}  :  \possiblyWithSub\stageOmetaColor{\mathcal{T}^{\superscriptO} }_{{\mathrm{1}}} \}   \relO{\to}  \possiblyWithSub\stageOmetaColor{R^{\superscriptO} }_{{\mathrm{2}}}   \mid   \theta_{{\mathrm{2}}}  \setminus   \{ \possiblyWithSub\stageOmetaColor{x} \}   
    }
  \\[1em]
    \derive[ABI0-ImpGuess2]{%
      \text{\(\mathit{\Psi}\) is neither of the form~\(( \mathit{\Psi}' , \__{ \ell } )\) nor \(( \mathit{\Psi}' , \{ \possiblyWithSub\stageOmetaColor{N'^{\superscriptO} }  :  \possiblyWithSub\stageOmetaColor{\mathcal{T}'^{\superscriptO} } \}_{ \ell } )\)}
    \\
       \mathcal{G}  \mid   I  \uplus   \{ \possiblyWithSub\stageOmetaColor{x} \}    \mid  \mathit{\Psi}  \vdash^{0}  \possiblyWithSub\stageOmetaColor{\mathcal{T}^{\superscriptO} }_{{\mathrm{2}}}  \mathrel{<:}  \possiblyWithSub\stageOmetaColor{R^{\superscriptO} }_{{\mathrm{2}}}  \mid  \theta_{{\mathrm{2}}} 
    \andalso
      \possiblyWithSub\stageOmetaColor{N^{\superscriptO} }_{{\mathrm{1}}} := \theta_{{\mathrm{2}}}(\possiblyWithSub\stageOmetaColor{x})
    }{%
       \mathcal{G}  \mid  I  \mid  \mathit{\Psi}  \vdash^{0}   \openO{\{} \possiblyWithSub\stageOmetaColor{x}  \relO{:}  \possiblyWithSub\stageOmetaColor{\mathcal{T}^{\superscriptO} }_{{\mathrm{1}}} \closeO{\} } \relO{\to}  \possiblyWithSub\stageOmetaColor{\mathcal{T}^{\superscriptO} }_{{\mathrm{2}}}   \mathrel{<:}    \mathbf{insert}\ \{ \possiblyWithSub\stageOmetaColor{N^{\superscriptO} }_{{\mathrm{1}}}  :  \possiblyWithSub\stageOmetaColor{\mathcal{T}^{\superscriptO} }_{{\mathrm{1}}} \}   \relO{\to}  \possiblyWithSub\stageOmetaColor{R^{\superscriptO} }_{{\mathrm{2}}}   \mid   \theta_{{\mathrm{2}}}  \setminus   \{ \possiblyWithSub\stageOmetaColor{x} \}   
    }
  \\[1em]
    \derive[ABI0-Code]{%
      \mathcal{G} \mid I \mid \mathit{\Psi} \vdash^{1} \possiblyWithSub\stageImetaColor{T^{\superscriptI} } \mathrel{<:} (D_{\ottmv{i}} \to)_{i = 1}^{m}  \possiblyWithSub\stageImetaColor{T'^{\superscriptI} }  \mid \theta
    }{%
      \mathcal{G} \mid I \mid \mathit{\Psi} \vdash^{0}  \openO{\langle} \possiblyWithSub\stageImetaColor{T^{\superscriptI} } \closeO{\rangle}  \mathrel{<:} (D_{\ottmv{i}} \to)_{i = 1}^{m}   \openO{\langle} \possiblyWithSub\stageImetaColor{T'^{\superscriptI} } \closeO{\rangle}   \mid \theta
    }
  \end{center}
  \vspace{0.5em}
  \begin{flushleft}
    \fbox{\( \mathcal{G}  \mid  I  \mid  \mathit{\Psi}  \vdash^{1}  \possiblyWithSub\stageImetaColor{T^{\superscriptI} }  \mathrel{<:}  \possiblyWithSub\stageImetaColor{R^{\superscriptI} }  \mid  \theta \)}
  \end{flushleft}
  \vspace{-4.25em}
  \begin{center}
  \hspace{10em}%
    \derive[ABI1-Empty]{}{%
       \mathcal{G}  \mid  I  \mid   \bullet   \vdash^{1}  \possiblyWithSub\stageImetaColor{T^{\superscriptI} }  \mathrel{<:}   \possiblyWithSub\stageImetaColor{T^{\superscriptI} }   \mid   \varnothing  
    }
  \\[1em]
    \derive[ABI1-Arr]{%
       I  \vdash  \possiblyWithSub\stageImetaColor{T'^{\superscriptI} }_{{\mathrm{1}}}  \mathrel{||}^{1}  \possiblyWithSub\stageImetaColor{T^{\superscriptI} }_{{\mathrm{1}}}  \mid  \theta_{{\mathrm{1}}} 
    \andalso
       \mathcal{G}  \mid   I  \setminus   \mathop{\mathrm{dom} }  \theta_{{\mathrm{1}}}    \mid  \mathit{\Psi}  \vdash^{1}   \theta_{{\mathrm{1}}}   \possiblyWithSub\stageImetaColor{T^{\superscriptI} }_{{\mathrm{2}}}   \mathrel{<:}  \possiblyWithSub\stageImetaColor{R^{\superscriptI} }_{{\mathrm{2}}}  \mid  \theta_{{\mathrm{2}}} 
    }{%
       \mathcal{G}  \mid   \mathit{\Psi} , ( \possiblyWithSub\stageImetaColor{T'^{\superscriptI} }_{{\mathrm{1}}} )^{1}_{ \ell }   \vdash^{1}   \possiblyWithSub\stageImetaColor{T^{\superscriptI} }_{{\mathrm{1}}}  \relI{\to}  \possiblyWithSub\stageImetaColor{T^{\superscriptI} }_{{\mathrm{2}}}   \mathrel{<:}    (  \LeftAssertParen\openO{\langle}  \theta_{{\mathrm{1}}}   \possiblyWithSub\stageImetaColor{T'^{\superscriptI} }_{{\mathrm{1}}}  \closeO{\rangle} \relO{\CastArrow} \openO{\langle}  \theta_{{\mathrm{1}}}   \possiblyWithSub\stageImetaColor{T^{\superscriptI} }_{{\mathrm{1}}}  \closeO{\rangle}\RightAssertParen^{  \ell  }  \ /\   \theta_{{\mathrm{1}}}   \possiblyWithSub\stageImetaColor{T^{\superscriptI} }_{{\mathrm{1}}}  )^{1}   \relI{\to}  \possiblyWithSub\stageImetaColor{R^{\superscriptI} }_{{\mathrm{2}}}  
    }
  \end{center}
  \vspace{0.5em}
  \begin{flushleft}
    \fbox{\( I  \vdash  \possiblyWithSub\stageImetaColor{T'^{\superscriptI} }_{{\mathrm{1}}}  \mathrel{||}^{1}  \possiblyWithSub\stageImetaColor{T^{\superscriptI} }_{{\mathrm{1}}}  \mid  \theta \)}
  \end{flushleft}
  \vspace{-4.25em}
  \begin{center}
  \hspace{8em}%
    \derive{}{%
       I  \vdash   \possiblyWithSub\stageImetaColor{B}   \mathrel{||}^{1}   \possiblyWithSub\stageImetaColor{B}   \mid   \varnothing  
    }
  \qquad
    \derive{%
       I  \vdash  \possiblyWithSub\stageOmetaColor{N^{\superscriptO} }_{{\mathrm{1}}}  \mathrel{||}^{0}  \possiblyWithSub\stageOmetaColor{N^{\superscriptO} }_{{\mathrm{2}}}  \mid  \theta 
    }{%
       I  \vdash   \ttI{Tensor}\ \ordI{\%} \possiblyWithSub\stageOmetaColor{N^{\superscriptO} }_{{\mathrm{1}}}   \mathrel{||}^{1}   \ttI{Tensor}\ \ordI{\%} \possiblyWithSub\stageOmetaColor{N^{\superscriptO} }_{{\mathrm{2}}}   \mid  \theta 
    }
  \\[1em]
    \derive{%
       I  \vdash  \possiblyWithSub\stageImetaColor{T^{\superscriptI} }_{{\mathrm{11}}}  \mathrel{||}^{1}  \possiblyWithSub\stageImetaColor{T^{\superscriptI} }_{{\mathrm{21}}}  \mid  \theta_{{\mathrm{1}}} 
    \andalso
        I  \setminus   \mathop{\mathrm{dom} }  \theta_{{\mathrm{1}}}    \vdash  \possiblyWithSub\stageImetaColor{T^{\superscriptI} }_{{\mathrm{12}}}  \mathrel{||}^{1}  \possiblyWithSub\stageImetaColor{T^{\superscriptI} }_{{\mathrm{22}}}  \mid  \theta_{{\mathrm{2}}} 
    }{%
       I  \vdash   \possiblyWithSub\stageImetaColor{T^{\superscriptI} }_{{\mathrm{11}}}  \relI{\to}  \possiblyWithSub\stageImetaColor{T^{\superscriptI} }_{{\mathrm{12}}}   \mathrel{||}^{1}   \possiblyWithSub\stageImetaColor{T^{\superscriptI} }_{{\mathrm{21}}}  \relI{\to}  \possiblyWithSub\stageImetaColor{T^{\superscriptI} }_{{\mathrm{22}}}   \mid   \theta_{{\mathrm{1}}}  \uplus  \theta_{{\mathrm{2}}}  
    }
  \end{center}
  \vspace{0.5em}
  \begin{flushleft}
    \fbox{\( I  \vdash  \possiblyWithSub\stageOmetaColor{N'^{\superscriptO} }_{{\mathrm{1}}}  \mathrel{||}^{0}  \possiblyWithSub\stageOmetaColor{N^{\superscriptO} }_{{\mathrm{1}}}  \mid  \theta \)}
  \end{flushleft}
  \vspace{-4.25em}
  \begin{center}
  \hspace{7em}%
    \derive{%
      \possiblyWithSub\stageOmetaColor{x} \in I
    }{%
       I  \vdash  \possiblyWithSub\stageOmetaColor{N'^{\superscriptO} }_{{\mathrm{1}}}  \mathrel{||}^{0}   \possiblyWithSub\stageOmetaColor{x}   \mid   \{ \possiblyWithSub\stageOmetaColor{x}  \mapsto  \possiblyWithSub\stageOmetaColor{N'^{\superscriptO} }_{{\mathrm{1}}} \}  
    }
  \qquad
    \derive{%
      \possiblyWithSub\stageOmetaColor{x} \in I
    }{%
       I  \vdash   \possiblyWithSub\stageOmetaColor{x}   \mathrel{||}^{0}  \possiblyWithSub\stageOmetaColor{N^{\superscriptO} }_{{\mathrm{1}}}  \mid   \{ \possiblyWithSub\stageOmetaColor{x}  \mapsto  \possiblyWithSub\stageOmetaColor{N^{\superscriptO} }_{{\mathrm{1}}} \}  
    }
  \\[1em]
    \derive{%
      \text{(If no other rules apply)}
    }{%
       I  \vdash  \possiblyWithSub\stageOmetaColor{N'^{\superscriptO} }_{{\mathrm{1}}}  \mathrel{||}^{0}  \possiblyWithSub\stageOmetaColor{N^{\superscriptO} }_{{\mathrm{1}}}  \mid   \varnothing  
    }
  \end{center}
  \caption{%
    The algorithmic rules for inferring implicit arguments
    (traversal starts with \(I =  \varnothing \))%
  }
  \label{fig:algorithmic-option-inference}
\end{figure}

\clearpage
\section{Notes on the Inference of Implicit Arguments}
\indent
  This section briefly discusses the sensitiveness of the inference algorithm
  to the description of types.
  As we have seen in Section~\ref{sec:implementation},
  the function~\small\texttt{Tensor.cross\_entropy\_for\_logits}
  has the following interface:
\begin{lstlisting}
  val ~gen_cross_entropy_for_logits :
    {s : {v : List Nat | List.length v == 2}} ->
    &(Tensor %s -> Tensor %[List.nth 0 s] -> Tensor %[]) external ...
\end{lstlisting}
  Apparently, this looks equivalent to the following:
\begin{lstlisting}
  val ~gen_cross_entropy_for_logits :
    {m : Nat} -> {n : Nat} ->
    &(Tensor %[m, n] -> Tensor %[m] -> Tensor %[]) external ...
\end{lstlisting}
  However, the former is actually easier to handle than the latter
  in terms of the inference of implicit arguments.
  For example, consider checking
  the application~\(\ordI{\sim}\ordO{\text{{\small\texttt{gen\_cross\_entropy\_for\_logits}}}}\ \possiblyWithSub\stageImetaColor{M^{\superscriptI} }_{{\mathrm{1}}}\).
  If we use the former interface, we can infer that
  the argument corresponding to the parameter~\(\ordI{\text{{\small\texttt{s}}}}\)
  can be \(\possiblyWithSub\stageOmetaColor{N^{\superscriptO} }\), simply looking at the type~\( \ttI{Tensor}\ \ordI{\%} \possiblyWithSub\stageOmetaColor{N^{\superscriptO} } \)
  of the first runtime argument~\(\possiblyWithSub\stageImetaColor{M^{\superscriptI} }_{{\mathrm{1}}}\).
  Precisely speaking, the reconstructed argument should be \( \possiblyWithSub\stageOmetaColor{N^{\superscriptO} }_{{\mathrm{0}}} \  \possiblyWithSub\stageOmetaColor{N^{\superscriptO} } \),
  where \(\possiblyWithSub\stageOmetaColor{N^{\superscriptO} }_{{\mathrm{0}}}\) is a cast used to confirm that
  \(\possiblyWithSub\stageOmetaColor{N^{\superscriptO} }\) is certainly a list of exactly two natural numbers.
  Things become a bit more complicated when using the latter;
  matching \(\possiblyWithSub\stageOmetaColor{N^{\superscriptO} }\) against the ``pattern''~\(\ordO{\text{{\small\texttt{[m, n]}}}}\),
  we can deduce that the arguments corresponding to
  \(\ordO{\text{{\small\texttt{m}}}}\) and
  \(\ordO{\text{{\small\texttt{n}}}}\) can be
  \(\ordO{\text{{\small\texttt{List.nth}}}}\  \ordO{  \makeIdentOrConst{}{ C0 }  } \  \openO{(}  \possiblyWithSub\stageOmetaColor{N^{\superscriptO} }_{{\mathrm{0}}} \  \possiblyWithSub\stageOmetaColor{N^{\superscriptO} }  \closeO{)} \) and
  \(\ordO{\text{{\small\texttt{List.nth}}}}\  \ordO{  \makeIdentOrConst{}{ C1 }  } \  \openO{(}  \possiblyWithSub\stageOmetaColor{N^{\superscriptO} }_{{\mathrm{0}}} \  \possiblyWithSub\stageOmetaColor{N^{\superscriptO} }  \closeO{)} \), respectively.
\par
\indent
  The current implementation of our prototype type-checker is, however,
  not wise enough to handle the latter in the manner explained above.
  This is simply because it did not need to be so ingenious;
  we can basically avoid interfaces like the latter.
  Using the latter interface of \small\texttt{Tensor.cross\_entropy\_for\_logits}
  to type-check the example program~\texttt{mnist/linear.hrs}
  (which is displayed on Figure~\ref{fig:code-example})
  indeed results in the following error:
\begin{lstlisting}
 Cannot infer an implicit argument for n (line 9, columns 4-35)
   Tensor.cross_entropy_for_logits
   ^^^^^^^^^^^^^^^^^^^^^^^^^^^^^^^
 application context:
   _,
   Tensor %(BROADCAST [num_train_images, label_count] [label_count]),
   Vec %num_train_images
 type:
   {n : {Int | <= 0}} -> &(Mat %m %n -> Vec %m -> Tensor %[])
\end{lstlisting}
  That is, matching the expression~%
    \(\ordO{\text{{\small\texttt{broadcast [num\_train\_images, label\_count] [label\_count]}}}}\)
  against \(\ordO{\text{{\small\texttt{[m, n]}}}}\) failed
  (Recall: {\small\texttt{Mat \%m \%n}} is a shorthand for {\small\texttt{Tensor \%[m, n]}}).
\par

\clearpage
\section{Formalization of Horsea}\label{sec:formalization-of-surface-language}
\begin{figure}[tbp]
\small
  \begin{flushleft}
    \fbox{\( \mathit{\Delta}  \vdash  E  :  \Hat{A}  \ElabArrow  \Hat{e} \)}
  \end{flushleft}
  \vspace{-4.25em}
  \begin{center}
    \hspace{5em}%
    \derive[AE-AbsOpt]{%
       \mathit{\Delta}  \vdash  T_{{\mathrm{1}}}  \ElabArrow   A_{{\mathrm{1}}} ^{ b_{{\mathrm{1}}} }   \mid  \Hat{\tau}_{{\mathrm{1}}} 
    \andalso
        \mathit{\Delta} ,  x  :   A_{{\mathrm{1}}} ^{ b_{{\mathrm{1}}} }   \mathbin{@}   0    \vdash  E_{{\mathrm{2}}}  :   A_{{\mathrm{2}}} ^{ b_{{\mathrm{2}}} }   \ElabArrow  \Hat{e}_{{\mathrm{2}}} 
    }{%
       \mathit{\Delta}  \vdash   (  \lambda \{ x  :  T_{{\mathrm{1}}} \}.\  E_{{\mathrm{2}}}  )   :    (  \{  A_{{\mathrm{1}}} ^{ b_{{\mathrm{1}}} }  \} \to   A_{{\mathrm{2}}} ^{ b_{{\mathrm{2}}} }   )  ^{  0  }   \ElabArrow    (  \lambda \{ x  :  \Hat{\tau}_{{\mathrm{1}}} \}.\  \Hat{e}_{{\mathrm{2}}}  )  ^{  0  }  
    }
  \\[0.7em]
    \derive[AE-AppOpt]{%
       \mathit{\Delta}  \vdash  E_{{\mathrm{1}}}  :    (  \{ \Hat{A}_{{\mathrm{2}}} \} \to  \Hat{A}  )  ^{  0  }   \ElabArrow  \Hat{e}_{{\mathrm{1}}} 
    \\
       \mathit{\Delta}  \vdash  E_{{\mathrm{2}}}  :  \Hat{A}_{{\mathrm{2}}}  \ElabArrow  \Hat{e}_{{\mathrm{2}}} 
    }{%
       \mathit{\Delta}  \vdash   ( E_{{\mathrm{1}}} \ \{ E_{{\mathrm{2}}} \})_{ \ell }   :  \Hat{A}  \ElabArrow    ( \Hat{e}_{{\mathrm{1}}} \ \{ \Hat{e}_{{\mathrm{2}}} \})_{ \ell }  ^{  0  }  
    }
  \qquad
    \derive[AE-Omit]{%
       \mathit{\Delta}  \vdash  E_{{\mathrm{1}}}  :    (  \{ \Hat{A}_{{\mathrm{11}}} \} \to  \Hat{A}_{{\mathrm{12}}}  )  ^{  0  }   \ElabArrow  \Hat{e}_{{\mathrm{1}}} 
    }{%
       \mathit{\Delta}  \vdash   E_{{\mathrm{1}}} \ \_   :  \Hat{A}_{{\mathrm{12}}}  \ElabArrow    (  \Hat{e}_{{\mathrm{1}}} \ \_  )  ^{  0  }  
    }
  \\[0.7em]
    \derive[AE-Cst0]{%
      \ConstEnvSurface(d) = \possiblyWithSub\stageOmetaColor{p}
    \andalso
      \ConstEnvZero(\possiblyWithSub\stageOmetaColor{p}) = \possiblyWithSub\stageOmetaColor{\mathcal{T}^{\superscriptO} }
    }{%
       \mathit{\Delta}  \vdash   d   :   \lceil  \possiblyWithSub\stageOmetaColor{\mathcal{T}^{\superscriptO} } \rceil_{0}   \ElabArrow    d  ^{  0  }  
    }
  \qquad
    \derive[AE-CstP]{%
      \ConstEnvSurface(d) = c
    \andalso
      \ConstEnvPers(c) = \possiblyWithSub\stageImetaColor{\tau^{\superscriptI} }
    }{%
       \mathit{\Delta}  \vdash   d   :   \lceil  b ,  \possiblyWithSub\stageImetaColor{\tau^{\superscriptI} } \rceil   \ElabArrow    d  ^{ b }  
    }
  \\[0.7em]
    \derive[AE-Var]{%
      \mathit{\Delta}(x) = \Hat{A} \mathbin{@} b
    }{%
       \mathit{\Delta}  \vdash   x   :  \Hat{A}  \ElabArrow    x  ^{ b }  
    }
  \qquad
    \derive[AE-Abs]{%
       \mathit{\Delta}  \vdash  T_{{\mathrm{1}}}  \ElabArrow   A_{{\mathrm{1}}} ^{ b_{{\mathrm{1}}} }   \mid  \Hat{\tau}_{{\mathrm{1}}} 
    \andalso
        \mathit{\Delta} ,  x  :   A_{{\mathrm{1}}} ^{ b_{{\mathrm{1}}} }   \mathbin{@}  b   \vdash  E_{{\mathrm{2}}}  :   A_{{\mathrm{2}}} ^{ b_{{\mathrm{2}}} }   \ElabArrow  \Hat{e}_{{\mathrm{2}}} 
    }{%
       \mathit{\Delta}  \vdash   (  \lambda  x  :  T_{{\mathrm{1}}} .\  E_{{\mathrm{2}}}  )   :    (   A_{{\mathrm{1}}} ^{ b_{{\mathrm{1}}} }   \to   A_{{\mathrm{2}}} ^{ b_{{\mathrm{2}}} }   )  ^{ b }   \ElabArrow    (  \lambda  x  :  \Hat{\tau}_{{\mathrm{1}}} .\  \Hat{e}_{{\mathrm{2}}}  )  ^{ b }  
    }
  \\[0.7em]
    \derive[AE-App]{%
       \mathit{\Delta}  \vdash  E_{{\mathrm{1}}}  :    (  \{ \Hat{A}'_{{\mathrm{1}}} \} \to   \cdots \to    (  \{ \Hat{A}'_{\ottmv{m}} \} \to    (  \Hat{A}_{{\mathrm{2}}}  \to  \Hat{A}  )  ^{ b }   )  ^{  0  }   \cdots   )  ^{  0  }   \ElabArrow  \Hat{e}_{{\mathrm{1}}} 
    \andalso
       \mathit{\Delta}  \vdash  E_{{\mathrm{2}}}  :  \Hat{A}_{{\mathrm{2}}}  \ElabArrow  \Hat{e}_{{\mathrm{2}}} 
    }{%
      \mathit{\Delta} \vdash  ( E_{{\mathrm{1}}} \  E_{{\mathrm{2}}} )_{ \ell }  : \Hat{A} \ElabArrow (\Hat{e}_{{\mathrm{1}}}\ \underbrace{\_\ \cdots\ \_}_{m}\ \Hat{e}_{{\mathrm{2}}})_{\ell}^{b}
    }
  \end{center}
  \begin{flushleft}
    \fbox{\( \mathit{\Delta}  \vdash  T  \ElabArrow  \Hat{A}  \mid  \Hat{\tau} \)}
  \end{flushleft}
  \vspace{-4.25em}
  \begin{center}
  \hspace{6em}%
    \derive[AT-Rfn]{%
       \mathit{\Delta}  \vdash  E  :    \mathbf{base}  ^{  0  }   \ElabArrow  \Hat{e} 
    }{%
       \mathit{\Delta}  \vdash    \{ x  :  B  \mid  E \}    \ElabArrow    \mathbf{base}  ^{  0  }   \mid     \{ x  :  B  \mid  \Hat{e} \}   ^{  0  }  
    }
  \\[0.7em]
    \derive[AT-Base]{}{%
       \mathit{\Delta}  \vdash   B   \ElabArrow    \mathbf{base}  ^{ b }   \mid    B  ^{ b }  
    }
  \qquad
    \derive[AT-Tensor1]{}{%
       \mathit{\Delta}  \vdash   \mathtt{Tensor}\    s     \ElabArrow    \mathbf{base}  ^{ b }   \mid    (  \mathtt{Tensor}\     s   ^{  0  }   )  ^{ b }  
    }
  \\[0.7em]
    \derive[AT-Tensor2]{%
      \text{\(E\) is not of the form~\(s\)}
    \andalso
       \mathit{\Delta}  \vdash  E  :    \mathbf{base}  ^{  0  }   \ElabArrow  \Hat{e} 
    }{%
       \mathit{\Delta}  \vdash   \mathtt{Tensor}\  E   \ElabArrow    \mathbf{base}  ^{  1  }   \mid    (  \mathtt{Tensor}\  \Hat{e}  )  ^{  1  }  
    }
  \\[0.7em]
    \derive[AT-ArrNonDep]{%
      x \not\in \fv(T_{{\mathrm{2}}})
    \andalso
       \mathit{\Delta}  \vdash  T_{{\mathrm{1}}}  \ElabArrow   A_{{\mathrm{1}}} ^{ b_{{\mathrm{1}}} }   \mid  \Hat{\tau}_{{\mathrm{1}}} 
    \andalso
       \mathit{\Delta}  \vdash  T_{{\mathrm{2}}}  \ElabArrow   A_{{\mathrm{2}}} ^{ b_{{\mathrm{2}}} }   \mid  \Hat{\tau}_{{\mathrm{2}}} 
    \andalso
      b \leq b_{{\mathrm{1}}}
    \andalso
      b \leq b_{{\mathrm{2}}}
    }{%
       \mathit{\Delta}  \vdash   ( x  :  T_{{\mathrm{1}}} ) \to  T_{{\mathrm{2}}}   \ElabArrow    (   A_{{\mathrm{1}}} ^{ b_{{\mathrm{1}}} }   \to   A_{{\mathrm{2}}} ^{ b_{{\mathrm{2}}} }   )  ^{ b }   \mid    (  ( x  :  \Hat{\tau}_{{\mathrm{1}}} ) \to  \Hat{\tau}_{{\mathrm{2}}}  )  ^{ b }  
    }
  \\[0.7em]
    \derive[AT-ArrDep]{%
      x \in \fv(T_{{\mathrm{2}}})
    \andalso
       \mathit{\Delta}  \vdash  T_{{\mathrm{1}}}  \ElabArrow   A_{{\mathrm{1}}} ^{ b_{{\mathrm{1}}} }   \mid  \Hat{\tau}_{{\mathrm{1}}} 
    \andalso
        \mathit{\Delta} ,  x  :   A_{{\mathrm{1}}} ^{ b_{{\mathrm{1}}} }   \mathbin{@}   0    \vdash  T_{{\mathrm{2}}}  \ElabArrow   A_{{\mathrm{2}}} ^{ b_{{\mathrm{2}}} }   \mid  \Hat{\tau}_{{\mathrm{2}}} 
    }{%
       \mathit{\Delta}  \vdash   ( x  :  T_{{\mathrm{1}}} ) \to  T_{{\mathrm{2}}}   \ElabArrow    (   A_{{\mathrm{1}}} ^{ b_{{\mathrm{1}}} }   \to   A_{{\mathrm{2}}} ^{ b_{{\mathrm{2}}} }   )  ^{  0  }   \mid    (  ( x  :  \Hat{\tau}_{{\mathrm{1}}} ) \to  \Hat{\tau}_{{\mathrm{2}}}  )  ^{  0  }  
    }
  \end{center}
  \begin{gather*}
     \lceil    \openO{\{} \possiblyWithSub\stageOmetaColor{\nu}  \relO{:}  \possiblyWithSub\stageOmetaColor{B}  \relO{\mid}  \possiblyWithSub\stageOmetaColor{N^{\superscriptO} } \closeO{\} }   \rceil_{0}  =
     \lceil   \ttO{Tensor}\  \possiblyWithSub\stageOmetaColor{s}  \rceil_{0}  :=   \mathbf{base}  ^{  0  } 
  \qquad
     \lceil   \openO{(} \possiblyWithSub\stageOmetaColor{x}  \relO{:}  \possiblyWithSub\stageOmetaColor{\mathcal{T}^{\superscriptO} }_{{\mathrm{1}}} \closeO{)} \relO{\to}  \possiblyWithSub\stageOmetaColor{\mathcal{T}^{\superscriptO} }_{{\mathrm{2}}}  \rceil_{0}  :=   (   \lceil  \possiblyWithSub\stageOmetaColor{\mathcal{T}^{\superscriptO} }_{{\mathrm{1}}} \rceil_{0}   \to   \lceil  \possiblyWithSub\stageOmetaColor{\mathcal{T}^{\superscriptO} }_{{\mathrm{2}}} \rceil_{0}   )  ^{  0  } 
  \\
     \lceil   \openO{\{} \possiblyWithSub\stageOmetaColor{x}  \relO{:}  \possiblyWithSub\stageOmetaColor{\mathcal{T}^{\superscriptO} }_{{\mathrm{1}}} \closeO{\} } \relO{\to}  \possiblyWithSub\stageOmetaColor{\mathcal{T}^{\superscriptO} }_{{\mathrm{2}}}  \rceil_{0}  :=   (  \{  \lceil  \possiblyWithSub\stageOmetaColor{\mathcal{T}^{\superscriptO} }_{{\mathrm{1}}} \rceil_{0}  \} \to   \lceil  \possiblyWithSub\stageOmetaColor{\mathcal{T}^{\superscriptO} }_{{\mathrm{2}}} \rceil_{0}   )  ^{  0  } 
  \qquad
     \lceil   \openO{\langle} \possiblyWithSub\stageImetaColor{T^{\superscriptI} } \closeO{\rangle}  \rceil_{0}  :=  \lceil  \possiblyWithSub\stageImetaColor{T^{\superscriptI} } \rceil_{1} 
  \\
     \lceil   \possiblyWithSub\stageImetaColor{B}  \rceil_{1}  =
     \lceil   \ttI{Tensor}\ \ordI{\%} \possiblyWithSub\stageOmetaColor{N^{\superscriptO} }  \rceil_{1}  :=   \mathbf{base}  ^{  1  } 
  \qquad
     \lceil   \possiblyWithSub\stageImetaColor{T^{\superscriptI} }_{{\mathrm{1}}}  \relI{\to}  \possiblyWithSub\stageImetaColor{T^{\superscriptI} }_{{\mathrm{2}}}  \rceil_{1}  :=   (   \lceil  \possiblyWithSub\stageImetaColor{T^{\superscriptI} }_{{\mathrm{1}}} \rceil_{1}   \to   \lceil  \possiblyWithSub\stageImetaColor{T^{\superscriptI} }_{{\mathrm{2}}} \rceil_{1}   )  ^{  1  } 
  \\
     \lceil  b ,   \possiblyWithSub\stageImetaColor{B}  \rceil  =
     \lceil  b ,   \ttI{Tensor}\ \ordI{\%} \possiblyWithSub\stageOmetaColor{s}  \rceil  :=   \mathbf{base}  ^{ b } 
  \qquad
     \lceil  b ,   \possiblyWithSub\stageImetaColor{\tau^{\superscriptI} }_{{\mathrm{1}}}  \relI{\to}  \possiblyWithSub\stageImetaColor{\tau^{\superscriptI} }_{{\mathrm{2}}}  \rceil  :=   (   \lceil  b ,  \possiblyWithSub\stageImetaColor{\tau^{\superscriptI} }_{{\mathrm{1}}} \rceil   \to   \lceil  b ,  \possiblyWithSub\stageImetaColor{\tau^{\superscriptI} }_{{\mathrm{2}}} \rceil   )  ^{ b } 
  \end{gather*}
  \caption{The (declarative) rules for binding-time analysis}
  \label{fig:declarative-bta}
\end{figure}
\indent
  Figure~\ref{fig:declarative-bta} describes the declarative BTA rules
  by using \(\Hat{A}\) and \(\mathit{\Delta}\) defined as follows:
  \begin{align*}
    \bnf{\Hat{A}}{%
       A ^{ b } 
    }
  &
    \bnf{A}{%
       \mathbf{base}  
    |  \Hat{A}  \to  \Hat{A}  |  \{ \Hat{A} \} \to  \Hat{A} 
    }
  &
    \bnf{\mathit{\Delta}}{%
       \bullet  |  \mathit{\Delta} ,  x  :  \Hat{A}  \mathbin{@}  b 
    }
  \end{align*}
  Type-like entities \(\Hat{A}\) are basically ``skeletons'' of usual types
  where each subpart is equipped with a binding time;
  they represent only the structure of functions.
  The judgment~\( \mathit{\Delta}  \vdash  E  :  \Hat{A}  \ElabArrow  \Hat{e} \) can be read
  ``under \(\mathit{\Delta}\), the expression~\(E\) is assigned \(\Hat{A}\)
  and elaborates to \(\Hat{e}\).''
\par
\indent
  Let us look through some rules in Figure~\ref{fig:declarative-bta}.
  First, \rulename{AE-AbsOpt} and \rulename{AE-AppOpt} translate
  abstractions and applications for implicit parameters/arguments, respectively.
  Because implicit parameters/arguments must be at stage~\(0\)
  in the target language~\LambdaBracketCastImplicit,
  these rules require the binding time of elaborated expressions to be \(0\).
  Guided by the binding time of the function, \rulename{AE-App} inserts holes~\(\_\)
  for the positions where omitted arguments should be completed.
  Since arguments can be elided at other positions than the function part of applications,
  this rule does not support all the possible cases of omission,
  but this is sufficient to cover most of the typical use cases.
  \rulename{AE-Abs} translates type annotations for binders
  by the judgment~\( \mathit{\Delta}  \vdash  T_{{\mathrm{1}}}  \ElabArrow   A_{{\mathrm{1}}} ^{ b_{{\mathrm{1}}} }   \mid  \Hat{\tau}_{{\mathrm{1}}} \).
  \rulename{AE-Cst0} and \rulename{AE-CstP} check constants
  by using the map~\(\ConstEnvSurface \colon d \mapsto (\possiblyWithSub\stageOmetaColor{p} \mid c)\),
  which associates each constant in Horsea
  with the corresponding one in \LambdaBracketCastImplicit, and
  the conversions~\( \lceil  \possiblyWithSub\stageOmetaColor{\mathcal{T}^{\superscriptO} } \rceil_{0}  = \Hat{A}\), \( \lceil  \possiblyWithSub\stageImetaColor{T^{\superscriptI} } \rceil_{1}  = \Hat{A}\),
  and \( \lceil  b ,  \possiblyWithSub\stageImetaColor{\tau^{\superscriptI} } \rceil  = \Hat{A}\),
  which extract binding-time types from staged types.
\par
\indent
  \rulename{AT-Tensor2} is the most essential rule for this conversion.
  It requires argument expressions of tensor types to be at stage~\(0\)
  and tensor types themselves to be at stage~\(1\).
  Another characteristic case is about function types:
  \rulename{AT-Arr} and \rulename{AT-ArrNonDep}.
  The restrictions \( b  \leq  b_{{\mathrm{1}}} \) and \( b  \leq  b_{{\mathrm{2}}} \) imposed by \rulename{AT-ArrNonDep}
  come from a line of previous studies about partial evaluation~\cite{%
    NielsonNielsonPOPL1988,%
    GluckJorgensen1997,%
    AsaiPEPM2016};
  these intuitively state that
  values passed to or returned by a function must be available at runtime
  if the function is available at runtime.
  These two also apply to \rulename{AT-Arr}, but
  since dependent function types can only be used at stage-\(0\) in the target language,
  they are assigned \(b := 0\), and thus the restrictions trivially hold.
\par
\indent
  The BTA rules introduced in this section are
  basically a natural extension of those in the literature,
  but there is one interesting point:
  while we must be careful about
  whether each \(\lambda\)-abstraction will have a dependent function type
  so that we can properly judge binding times,
  we do not have to track dependent type-like entities through the analysis;
  using \(\Hat{A}\) suffices.
  This can be explained intuitively like the following:
  Suppose \( (  \lambda  x  :  T_{{\mathrm{1}}} .\  E_{{\mathrm{2}}}  ) \) will have
  an essentially dependent function type~\( \openO{(} \possiblyWithSub\stageOmetaColor{x}  \relO{:}  \possiblyWithSub\stageOmetaColor{S^{\superscriptO} }_{{\mathrm{1}}} \closeO{)} \relO{\to}  \possiblyWithSub\stageOmetaColor{S^{\superscriptO} }_{{\mathrm{2}}} \)
  after BTA and thereby
  we must judge beforehand that this \(\lambda\)-abstraction be at stage~\(0\).
  Since \(x\) occurs in \(\possiblyWithSub\stageOmetaColor{S^{\superscriptO} }_{{\mathrm{2}}}\), \(x\) must have occurred in \(E_{{\mathrm{2}}}\) as well.
  This occurrence will generate some constraint that forces the binding time of \(x\) to be \(0\).
  Therefore, we do not have to track the occurrence of \(x\) in types
  to judge whether the \(\lambda\)-abstraction must reside at stage~\(0\) or not.
\par
\begin{figure}[tbp]
\small
  \begin{flushleft}
    \fbox{\( \mathit{\Delta}  \vdash  E  :  \Hat{A}  \ElabArrow  \Hat{e}  \mid  C \)}
  \end{flushleft}
  \vspace{-4.25em}
  \begin{center}
    \hspace{9em}%
    \derive[AAE-App]{%
       \mathit{\Delta}  \vdash  E_{{\mathrm{1}}}  :    (  \{ \Hat{A}'_{{\mathrm{1}}} \} \to   \cdots \to    (  \{ \Hat{A}'_{\ottmv{m}} \} \to    (  \Hat{A}_{{\mathrm{11}}}  \to  \Hat{A}_{{\mathrm{12}}}  )  ^{ b }   )  ^{  0  }   \cdots   )  ^{  0  }   \ElabArrow  \Hat{e}_{{\mathrm{1}}}  \mid  C_{{\mathrm{1}}} 
    \\
       \mathit{\Delta}  \vdash  E_{{\mathrm{2}}}  :  \Hat{A}_{{\mathrm{2}}}  \ElabArrow  \Hat{e}_{{\mathrm{2}}}  \mid  C_{{\mathrm{2}}} 
    \andalso
       \Hat{A}_{{\mathrm{11}}}  =  \Hat{A}_{{\mathrm{2}}}  \ElabArrow  C 
    }{%
      \mathit{\Delta} \vdash  ( E_{{\mathrm{1}}} \  E_{{\mathrm{2}}} )_{ \ell }  : \Hat{A}_{{\mathrm{12}}} \ElabArrow (\Hat{e}_{{\mathrm{1}}}\ \underbrace{\_\ \cdots\ \_}_{m}\ \Hat{e}_{{\mathrm{2}}})_{\ell}^{b}
        \mid   C_{{\mathrm{1}}}  \cup  C_{{\mathrm{2}}}   \cup  C 
    }
  \\[0.5em]
    \derive[AAE-Var]{%
      \mathit{\Delta}(x) = \Hat{A} \mathbin{@} b
    }{%
       \mathit{\Delta}  \vdash   x   :  \Hat{A}  \ElabArrow    x  ^{ b }   \mid   \varnothing  
    }
  \qquad
    \derive[AAE-Omit]{%
       \mathit{\Delta}  \vdash  E_{{\mathrm{1}}}  :    (  \{ \Hat{A}_{{\mathrm{11}}} \} \to  \Hat{A}_{{\mathrm{12}}}  )  ^{  0  }   \ElabArrow  \Hat{e}_{{\mathrm{1}}}  \mid  C_{{\mathrm{1}}} 
    }{%
       \mathit{\Delta}  \vdash   E_{{\mathrm{1}}} \ \_   :  \Hat{A}_{{\mathrm{12}}}  \ElabArrow    (  \Hat{e}_{{\mathrm{1}}} \ \_  )  ^{  0  }   \mid  C_{{\mathrm{1}}} 
    }
  \\[0.7em]
    \derive[AAE-Abs]{%
       \mathit{\Delta}  \vdash  T_{{\mathrm{1}}}  \ElabArrow   A_{{\mathrm{1}}} ^{ b_{{\mathrm{1}}} }   \mid  \Hat{\tau}_{{\mathrm{1}}}  \mid  C_{{\mathrm{1}}} 
    \andalso
      \beta\ \text{fresh}
    \andalso
        \mathit{\Delta} ,  x  :   A_{{\mathrm{1}}} ^{ b_{{\mathrm{1}}} }   \mathbin{@}   \beta    \vdash  E_{{\mathrm{2}}}  :   A_{{\mathrm{2}}} ^{ b_{{\mathrm{2}}} }   \ElabArrow  \Hat{e}_{{\mathrm{2}}}  \mid  C_{{\mathrm{2}}} 
    }{%
       \mathit{\Delta}  \vdash   (  \lambda  x  :  T_{{\mathrm{1}}} .\  E_{{\mathrm{2}}}  )   :    (   A_{{\mathrm{1}}} ^{ b_{{\mathrm{1}}} }   \to   A_{{\mathrm{2}}} ^{ b_{{\mathrm{2}}} }   )  ^{  \beta  }   \ElabArrow    (  \lambda  x  :  \Hat{\tau}_{{\mathrm{1}}} .\  \Hat{e}_{{\mathrm{2}}}  )  ^{  \beta  }   \mid    C_{{\mathrm{1}}}  \cup  C_{{\mathrm{2}}}   \cup   \{    \beta   \leq  b_{{\mathrm{1}}}  ,     \beta   \leq  b_{{\mathrm{2}}}    \}   
    }
  \\[1em]
    \derive[AAE-AbsOpt]{%
       \mathit{\Delta}  \vdash  T_{{\mathrm{1}}}  \ElabArrow   A_{{\mathrm{1}}} ^{ b_{{\mathrm{1}}} }   \mid  \Hat{\tau}_{{\mathrm{1}}}  \mid  C_{{\mathrm{1}}} 
    \andalso
        \mathit{\Delta} ,  x  :   A_{{\mathrm{1}}} ^{ b_{{\mathrm{1}}} }   \mathbin{@}   0    \vdash  E_{{\mathrm{2}}}  :   A_{{\mathrm{2}}} ^{ b_{{\mathrm{2}}} }   \ElabArrow  \Hat{e}_{{\mathrm{2}}}  \mid  C_{{\mathrm{2}}} 
    }{%
       \mathit{\Delta}  \vdash   (  \lambda \{ x  :  T_{{\mathrm{1}}} \}.\  E_{{\mathrm{2}}}  )   :    (  \{  A_{{\mathrm{1}}} ^{ b_{{\mathrm{1}}} }  \} \to   A_{{\mathrm{2}}} ^{ b_{{\mathrm{2}}} }   )  ^{  0  }   \ElabArrow    (  \lambda \{ x  :  \Hat{\tau}_{{\mathrm{1}}} \}.\  \Hat{e}_{{\mathrm{2}}}  )  ^{  0  }   \mid   C_{{\mathrm{1}}}  \cup  C_{{\mathrm{2}}}  
    }
  \\[1em]
    \derive[AAE-AppOpt]{%
       \mathit{\Delta}  \vdash  E_{{\mathrm{1}}}  :    (  \{ \Hat{A}_{{\mathrm{11}}} \} \to  \Hat{A}_{{\mathrm{12}}}  )  ^{  0  }   \ElabArrow  \Hat{e}_{{\mathrm{1}}}  \mid  C_{{\mathrm{1}}} 
    \andalso
       \mathit{\Delta}  \vdash  E_{{\mathrm{2}}}  :  \Hat{A}_{{\mathrm{2}}}  \ElabArrow  \Hat{e}_{{\mathrm{2}}}  \mid  C_{{\mathrm{2}}} 
    \andalso
       \Hat{A}_{{\mathrm{11}}}  =  \Hat{A}_{{\mathrm{2}}}  \ElabArrow  C 
    }{%
       \mathit{\Delta}  \vdash   ( E_{{\mathrm{1}}} \ \{ E_{{\mathrm{2}}} \})_{ \ell }   :  \Hat{A}_{{\mathrm{12}}}  \ElabArrow    ( \Hat{e}_{{\mathrm{1}}} \ \{ \Hat{e}_{{\mathrm{2}}} \})_{ \ell }  ^{  0  }   \mid    C_{{\mathrm{1}}}  \cup  C_{{\mathrm{2}}}   \cup  C  
    }
  \end{center}
  \vspace{1em}
  \begin{flushleft}
    \fbox{\( \mathit{\Delta}  \vdash  T  \ElabArrow  \Hat{A}  \mid  \Hat{\tau}  \mid  C \)}
  \end{flushleft}
  \vspace{-4.25em}
  \begin{center}
  \hspace{7em}%
    \derive[AAT-Rfn]{%
       \mathit{\Delta}  \vdash  E  :    \mathbf{base}  ^{ b }   \ElabArrow  \Hat{e}  \mid  C 
    }{%
       \mathit{\Delta}  \vdash    \{ x  :  B  \mid  E \}    \ElabArrow    \mathbf{base}  ^{  0  }   \mid     \{ x  :  B  \mid  \Hat{e} \}   ^{  0  }   \mid   C  \cup   \{   b  =   0    \}   
    }
  \\[0.7em]
    \derive[AAT-Base]{%
      \beta\ \text{fresh}
    }{%
       \mathit{\Delta}  \vdash   B   \ElabArrow    \mathbf{base}  ^{  \beta  }   \mid    B  ^{  \beta  }   \mid   \varnothing  
    }
  \quad
    \derive[AAT-Tensor1]{%
      \beta\ \text{fresh}
    }{%
       \mathit{\Delta}  \vdash   \mathtt{Tensor}\    s     \ElabArrow    \mathbf{base}  ^{  \beta  }   \mid    (  \mathtt{Tensor}\     s   ^{  0  }   )  ^{  \beta  }   \mid   \varnothing  
    }
  \\[0.7em]
    \derive[AAT-Tensor2]{%
      \text{\(E\) is not of the form~\(s\)}
    \andalso
       \mathit{\Delta}  \vdash  E  :    \mathbf{base}  ^{ b }   \ElabArrow  \Hat{e}  \mid  C 
    }{%
       \mathit{\Delta}  \vdash   \mathtt{Tensor}\  E   \ElabArrow    \mathbf{base}  ^{  1  }   \mid    (  \mathtt{Tensor}\  \Hat{e}  )  ^{  1  }   \mid   C  \cup   \{   b  =   0    \}   
    }
  \\[1em]
    \derive[AAT-ArrNonDep]{%
      x \not\in \fv(T_{{\mathrm{2}}})
    \andalso
       \mathit{\Delta}  \vdash  T_{{\mathrm{1}}}  \ElabArrow   A_{{\mathrm{1}}} ^{ b_{{\mathrm{1}}} }   \mid  \Hat{\tau}_{{\mathrm{1}}}  \mid  C_{{\mathrm{1}}} 
    \\
      \beta\ \text{fresh}
    \andalso
        \mathit{\Delta} ,  x  :   A_{{\mathrm{1}}} ^{ b_{{\mathrm{1}}} }   \mathbin{@}   \beta    \vdash  T_{{\mathrm{2}}}  \ElabArrow   A_{{\mathrm{2}}} ^{ b_{{\mathrm{2}}} }   \mid  \Hat{\tau}_{{\mathrm{2}}}  \mid  C_{{\mathrm{2}}} 
    }{%
       \mathit{\Delta}  \vdash   ( x  :  T_{{\mathrm{1}}} ) \to  T_{{\mathrm{2}}}   \ElabArrow    (   A_{{\mathrm{1}}} ^{ b_{{\mathrm{1}}} }   \to   A_{{\mathrm{2}}} ^{ b_{{\mathrm{2}}} }   )  ^{  \beta  }   \mid    (  ( x  :  \Hat{\tau}_{{\mathrm{1}}} ) \to  \Hat{\tau}_{{\mathrm{2}}}  )  ^{  \beta  }   \mid    C_{{\mathrm{1}}}  \cup  C_{{\mathrm{2}}}   \cup   \{    \beta   \leq  b_{{\mathrm{1}}}  ,     \beta   \leq  b_{{\mathrm{2}}}    \}   
    }
  \\[1em]
    \derive[AAT-ArrDep]{%
      x \in \fv(T_{{\mathrm{2}}})
    \andalso
       \mathit{\Delta}  \vdash  T_{{\mathrm{1}}}  \ElabArrow   A_{{\mathrm{1}}} ^{ b_{{\mathrm{1}}} }   \mid  \Hat{\tau}_{{\mathrm{1}}}  \mid  C_{{\mathrm{1}}} 
    \andalso
        \mathit{\Delta} ,  x  :   A_{{\mathrm{1}}} ^{ b_{{\mathrm{1}}} }   \mathbin{@}   0    \vdash  T_{{\mathrm{2}}}  \ElabArrow   A_{{\mathrm{2}}} ^{ b_{{\mathrm{2}}} }   \mid  \Hat{\tau}_{{\mathrm{2}}}  \mid  C_{{\mathrm{2}}} 
    }{%
       \mathit{\Delta}  \vdash   ( x  :  T_{{\mathrm{1}}} ) \to  T_{{\mathrm{2}}}   \ElabArrow    (   A_{{\mathrm{1}}} ^{ b_{{\mathrm{1}}} }   \to   A_{{\mathrm{2}}} ^{ b_{{\mathrm{2}}} }   )  ^{  0  }   \mid    (  ( x  :  \Hat{\tau}_{{\mathrm{1}}} ) \to  \Hat{\tau}_{{\mathrm{2}}}  )  ^{  0  }   \mid   C_{{\mathrm{1}}}  \cup  C_{{\mathrm{2}}}  
    }
  \end{center}
  \vspace{1em}
  \begin{flushleft}
    \fbox{\( \Hat{A}_{{\mathrm{1}}}  =  \Hat{A}_{{\mathrm{2}}}  \ElabArrow  C \)}
  \end{flushleft}
  \vspace{-4.25em}
  \begin{center}
  \hspace{4em}%
    \derive{%
       \Hat{A}_{{\mathrm{11}}}  =  \Hat{A}_{{\mathrm{21}}}  \ElabArrow  C_{{\mathrm{1}}} 
    \andalso
       \Hat{A}_{{\mathrm{12}}}  =  \Hat{A}_{{\mathrm{22}}}  \ElabArrow  C_{{\mathrm{2}}} 
    }{%
         (  \Hat{A}_{{\mathrm{11}}}  \to  \Hat{A}_{{\mathrm{12}}}  )  ^{ b_{{\mathrm{1}}} }   =    (  \Hat{A}_{{\mathrm{21}}}  \to  \Hat{A}_{{\mathrm{22}}}  )  ^{ b_{{\mathrm{2}}} }   \ElabArrow    C_{{\mathrm{1}}}  \cup  C_{{\mathrm{2}}}   \cup   \{   b_{{\mathrm{1}}}  =  b_{{\mathrm{2}}}   \}   
    }
  \\[1em]
    \derive{%
       \Hat{A}_{{\mathrm{11}}}  =  \Hat{A}_{{\mathrm{21}}}  \ElabArrow  C_{{\mathrm{1}}} 
    \andalso
       \Hat{A}_{{\mathrm{12}}}  =  \Hat{A}_{{\mathrm{22}}}  \ElabArrow  C_{{\mathrm{2}}} 
    }{%
         (  \{ \Hat{A}_{{\mathrm{11}}} \} \to  \Hat{A}_{{\mathrm{12}}}  )  ^{  0  }   =    (  \{ \Hat{A}_{{\mathrm{21}}} \} \to  \Hat{A}_{{\mathrm{22}}}  )  ^{  0  }   \ElabArrow   C_{{\mathrm{1}}}  \cup  C_{{\mathrm{2}}}  
    }
  \qquad
    \derive{}{%
         \mathbf{base}  ^{ b_{{\mathrm{1}}} }   =    \mathbf{base}  ^{ b_{{\mathrm{2}}} }   \ElabArrow   \{   b_{{\mathrm{1}}}  =  b_{{\mathrm{2}}}   \}  
    }
  \end{center}
  \caption{The algorithmic constraint generation rules for binding-time analysis}
  \label{fig:algorithmic-bta}
\end{figure}
\begin{figure}[tb]
  \begin{align*}
     ( \varphi ,   C  \uplus   \{    \beta   \leq   \beta    \}   ) &\longrightarrow ( \varphi ,  C ) 
  \\
     ( \varphi ,   C  \uplus   \{    \beta   \leq   0    \}   ) &\longrightarrow (    [   0   /  \beta  ]  \varphi    \uplus   \{ \beta  \mapsto   0  \}   ,   [   0   /  \beta  ]  C  ) 
  \\
     ( \varphi ,   C  \uplus   \{    1   \leq   \beta    \}   ) &\longrightarrow (    [   1   /  \beta  ]  \varphi    \uplus   \{ \beta  \mapsto   1  \}   ,   [   1   /  \beta  ]  C  ) 
  \\
     ( \varphi ,   C  \uplus   \{    0   \leq  b   \}   ) &\longrightarrow ( \varphi ,  C ) 
  \\
     ( \varphi ,   C  \uplus   \{   b  \leq   1    \}   ) &\longrightarrow ( \varphi ,  C ) 
  \\
     ( \varphi ,   C  \uplus   \{    1   \leq   0    \}   ) &\longrightarrow \token{fail} 
  \\
     ( \varphi ,   C  \uplus   \{   b  =  b   \}   ) &\longrightarrow ( \varphi ,  C ) 
  \\
     ( \varphi ,   C  \uplus   \{    \beta   =  b   \}   ) &\longrightarrow (    [  b  /  \beta  ]  \varphi    \uplus   \{ \beta  \mapsto  b \}   ,   [  b  /  \beta  ]  C  ) 
  \\
     ( \varphi ,   C  \uplus   \{   b  =   \beta    \}   ) &\longrightarrow (    [  b  /  \beta  ]  \varphi    \uplus   \{ \beta  \mapsto  b \}   ,   [  b  /  \beta  ]  C  ) 
  \\
     ( \varphi ,   C  \uplus   \{    1   =   0    \}   ) &\longrightarrow \token{fail} 
  \\
     ( \varphi ,   C  \uplus   \{    0   =   1    \}   ) &\longrightarrow \token{fail} 
  \\
     ( \varphi ,  C ) &\longrightarrow \token{solved}\  \varphi 
    \tag{if all the constraints in \(C\) are of the form \(  \beta_{{\mathrm{1}}}   \leq   \beta_{{\mathrm{2}}}  \)}
  \end{align*}
  \caption{The reduction rules for the constraint solving after binding-time analysis (should start with \(( \varnothing , C)\))}
  \label{fig:bta-constraint-solving}
\end{figure}
\begin{figure}[tbp]
\small
  \begin{flushleft}
    \fbox{\(\vdash^{b} \Hat{e} \ElabArrow \mathcal{M}^{(b)}\)}
  \end{flushleft}
  \vspace{-4em}
  \begin{center}
  \hspace{7em}%
    \derive{%
       \vdash^{1}  e  \ElabArrow  \possiblyWithSub\stageImetaColor{\mathcal{M}^{\superscriptI} } 
    }{%
       \vdash^{0}   e ^{  1  }   \ElabArrow   \openO{\langle} \possiblyWithSub\stageImetaColor{\mathcal{M}^{\superscriptI} } \closeO{\rangle}  
    }
  \quad
    \derive{%
       \vdash^{0}  e  \ElabArrow  \possiblyWithSub\stageOmetaColor{\mathcal{M}^{\superscriptO} } 
    }{%
       \vdash^{0}   e ^{  0  }   \ElabArrow  \possiblyWithSub\stageOmetaColor{\mathcal{M}^{\superscriptO} } 
    }
  \quad
    \derive{%
       \vdash^{1}  e  \ElabArrow  \possiblyWithSub\stageImetaColor{\mathcal{M}^{\superscriptI} } 
    }{%
       \vdash^{1}   e ^{  1  }   \ElabArrow  \possiblyWithSub\stageImetaColor{\mathcal{M}^{\superscriptI} } 
    }
  \quad
    \derive{%
       \vdash^{0}  e  \ElabArrow  \possiblyWithSub\stageOmetaColor{\mathcal{M}^{\superscriptO} } 
    }{%
       \vdash^{1}   e ^{  0  }   \ElabArrow   \ordI{\sim} \possiblyWithSub\stageOmetaColor{\mathcal{M}^{\superscriptO} }  
    }
  \end{center}
  \begin{flushleft}
    \fbox{\(\vdash^{b} e \ElabArrow \mathcal{M}^{(b)}\)}
  \end{flushleft}
  \vspace{-4.25em}
  \begin{center}
    \hspace{6em}%
    \derive{%
      \ConstEnvSurface(d) = \possiblyWithSub\stageOmetaColor{p}
    }{%
      \vdash^{ 0 } d \ElabArrow \possiblyWithSub\stageOmetaColor{p}
    }
  \qquad
    \derive{%
      \ConstEnvSurface(d) = c
    }{%
      \vdash^{b} d \ElabArrow c
    }
  \qquad
    \derive{%
      \vdash^{b} \Hat{\tau} \ElabArrow \mathcal{S}^{(b)}_1
    \andalso
      \vdash^{b} \Hat{e}_{{\mathrm{2}}} \ElabArrow \mathcal{M}^{(b)}_2
    }{%
      \vdash^{b}  \lambda  x  :  \Hat{\tau}_{{\mathrm{1}}} .\  \Hat{e}_{{\mathrm{2}}}  \ElabArrow \lambda x : \mathcal{S}^{(b)}_1.\ \mathcal{M}^{(b)}_2
    }
  \\[1em]
    \derive{%
      \vdash^{b} \Hat{e}_{{\mathrm{1}}} \ElabArrow \mathcal{M}^{(b)}_1
    \andalso
      \vdash^{b} \Hat{e}_{{\mathrm{2}}} \ElabArrow \mathcal{M}^{(b)}_2
    }{%
      \vdash^{b}  ( \Hat{e}_{{\mathrm{1}}} \  \Hat{e}_{{\mathrm{2}}} )_{ \ell }  \ElabArrow (\mathcal{M}^{(b)}_1\ \mathcal{M}^{(b)}_2)_{\ell}
    }
  \qquad
    \derive{%
       \vdash^{0}  \Hat{\tau}_{{\mathrm{1}}}  \ElabArrow  \possiblyWithSub\stageOmetaColor{\mathcal{S}^{\superscriptO} }_{{\mathrm{1}}} 
    \andalso
       \vdash^{0}  \Hat{e}_{{\mathrm{2}}}  \ElabArrow  \possiblyWithSub\stageOmetaColor{\mathcal{M}^{\superscriptO} }_{{\mathrm{2}}} 
    }{%
       \vdash^{0}   \lambda \{ x  :  \Hat{\tau}_{{\mathrm{1}}} \}.\  \Hat{e}_{{\mathrm{2}}}   \ElabArrow   \ordO{\lambda}\openO{\{} \possiblyWithSub\stageOmetaColor{x}  \relO{:}  \possiblyWithSub\stageOmetaColor{\mathcal{S}^{\superscriptO} }_{{\mathrm{1}}} \closeO{\} }\punctO{.}\  \possiblyWithSub\stageOmetaColor{\mathcal{M}^{\superscriptO} }_{{\mathrm{2}}}  
    }
  \\[1em]
    \derive{}{%
      \vdash^{b} x \ElabArrow x
    }
  \qquad
    \derive{%
       \vdash^{0}  \Hat{e}_{{\mathrm{1}}}  \ElabArrow  \possiblyWithSub\stageOmetaColor{\mathcal{M}^{\superscriptO} }_{{\mathrm{1}}} 
    \andalso
       \vdash^{0}  \Hat{e}_{{\mathrm{2}}}  \ElabArrow  \possiblyWithSub\stageOmetaColor{\mathcal{M}^{\superscriptO} }_{{\mathrm{2}}} 
    }{%
       \vdash^{0}   ( \Hat{e}_{{\mathrm{1}}} \ \{ \Hat{e}_{{\mathrm{2}}} \})_{ \ell }   \ElabArrow   \openO{(} \possiblyWithSub\stageOmetaColor{\mathcal{M}^{\superscriptO} }_{{\mathrm{1}}} \ \openO{\{} \possiblyWithSub\stageOmetaColor{\mathcal{M}^{\superscriptO} }_{{\mathrm{2}}} \closeO{\} }\closeO{)}_{ \ell }  
    }
  \qquad
    \derive{%
       \vdash^{0}  \Hat{e}_{{\mathrm{1}}}  \ElabArrow  \possiblyWithSub\stageOmetaColor{\mathcal{M}^{\superscriptO} }_{{\mathrm{1}}} 
    }{%
       \vdash^{0}   \Hat{e}_{{\mathrm{1}}} \ \_   \ElabArrow   \possiblyWithSub\stageOmetaColor{\mathcal{M}^{\superscriptO} }_{{\mathrm{1}}} \ \ordO{\_}  
    }
  \end{center}
  \begin{flushleft}
    \fbox{\(\vdash^{b} \Hat{\tau} \ElabArrow \mathcal{S}^{(b)}\)}
  \end{flushleft}
  \vspace{-4.25em}
  \begin{center}
    \hspace{6em}%
    \derive{%
       \vdash^{1}  \tau  \ElabArrow  \possiblyWithSub\stageImetaColor{\mathcal{S}^{\superscriptI} } 
    }{%
       \vdash^{0}   \tau ^{  1  }   \ElabArrow   \openO{\langle} \possiblyWithSub\stageImetaColor{\mathcal{S}^{\superscriptI} } \closeO{\rangle}  
    }
  \qquad
    \derive{%
       \vdash^{0}  \tau  \ElabArrow  \possiblyWithSub\stageOmetaColor{\mathcal{S}^{\superscriptO} } 
    }{%
       \vdash^{0}   \tau ^{  0  }   \ElabArrow  \possiblyWithSub\stageOmetaColor{\mathcal{S}^{\superscriptO} } 
    }
  \qquad
    \derive{%
       \vdash^{1}  \tau  \ElabArrow  \possiblyWithSub\stageImetaColor{\mathcal{S}^{\superscriptI} } 
    }{%
       \vdash^{1}   \tau ^{  1  }   \ElabArrow  \possiblyWithSub\stageImetaColor{\mathcal{S}^{\superscriptI} } 
    }
  \end{center}
  \begin{flushleft}
    \fbox{\(\vdash^{b} \tau \ElabArrow \mathcal{S}^{(b)}\)}
  \end{flushleft}
  \vspace{-4.25em}
  \begin{center}
    \hspace{6em}%
    \derive{}{%
      \vdash^{b} B \ElabArrow B
    }
  \qquad
    \derive{}{%
       \vdash^{0}   \mathtt{Tensor}\     s   ^{  0  }    \ElabArrow   \ttO{Tensor}\  \possiblyWithSub\stageOmetaColor{s}  
    }
  \\[0.7em]
    \derive{%
       \vdash^{0}  \Hat{e}  \ElabArrow  \possiblyWithSub\stageOmetaColor{\mathcal{M}^{\superscriptO} } 
    }{%
       \vdash^{0}    \{ x  :  B  \mid  \Hat{e} \}    \ElabArrow    \openO{\{} \possiblyWithSub\stageOmetaColor{x}  \relO{:}  \possiblyWithSub\stageOmetaColor{B}  \relO{\mid}  \possiblyWithSub\stageOmetaColor{\mathcal{M}^{\superscriptO} } \closeO{\} }   
    }
  \qquad
    \derive{%
       \vdash^{0}  \Hat{e}  \ElabArrow  \possiblyWithSub\stageOmetaColor{\mathcal{M}^{\superscriptO} } 
    }{%
       \vdash^{1}   \mathtt{Tensor}\  \Hat{e}   \ElabArrow   \ttI{Tensor}\ \ordI{\%} \possiblyWithSub\stageOmetaColor{\mathcal{M}^{\superscriptO} }  
    }
  \\[0.7em]
    \derive{%
       \vdash^{0}  \Hat{\tau}_{{\mathrm{1}}}  \ElabArrow  \possiblyWithSub\stageOmetaColor{\mathcal{S}^{\superscriptO} }_{{\mathrm{1}}} 
    \andalso
       \vdash^{0}  \Hat{\tau}_{{\mathrm{2}}}  \ElabArrow  \possiblyWithSub\stageOmetaColor{\mathcal{S}^{\superscriptO} }_{{\mathrm{2}}} 
    }{%
       \vdash^{0}   ( x  :  \Hat{\tau}_{{\mathrm{1}}} ) \to  \Hat{\tau}_{{\mathrm{2}}}   \ElabArrow   \openO{(} \possiblyWithSub\stageOmetaColor{x}  \relO{:}  \possiblyWithSub\stageOmetaColor{\mathcal{S}^{\superscriptO} }_{{\mathrm{1}}} \closeO{)} \relO{\to}  \possiblyWithSub\stageOmetaColor{\mathcal{S}^{\superscriptO} }_{{\mathrm{2}}}  
    }
  \qquad
    \derive{%
       \vdash^{1}  \Hat{\tau}_{{\mathrm{1}}}  \ElabArrow  \possiblyWithSub\stageImetaColor{\mathcal{S}^{\superscriptI} }_{{\mathrm{1}}} 
    \andalso
       \vdash^{1}  \Hat{\tau}_{{\mathrm{2}}}  \ElabArrow  \possiblyWithSub\stageImetaColor{\mathcal{S}^{\superscriptI} }_{{\mathrm{2}}} 
    }{%
       \vdash^{1}   ( x  :  \Hat{\tau}_{{\mathrm{1}}} ) \to  \Hat{\tau}_{{\mathrm{2}}}   \ElabArrow   \possiblyWithSub\stageImetaColor{\mathcal{S}^{\superscriptI} }_{{\mathrm{1}}}  \relI{\to}  \possiblyWithSub\stageImetaColor{\mathcal{S}^{\superscriptI} }_{{\mathrm{2}}}  
    }
  \end{center}
  \caption{Insertion of staging constructs by using binding times}
  \label{fig:reconstruction-of-staging-construct}
\end{figure}
\indent
  Lastly, we touch on how to ``implement'' the binding-time analysis.
  The rules for the judgment \( \mathit{\Delta}  \vdash  E  :  \Hat{A}  \ElabArrow  \Hat{e} \) are declarative
  in the sense that each binding times cannot be obtained in a syntax-directed traversal.
  For actual implementation,
  we have to use an algorithmic version \( \mathit{\Delta}  \vdash  E  :  \Hat{A}  \ElabArrow  \Hat{e}  \mid  C \),
  where \(C\) ranges over the set of finite sets of constraints
  of the form \( b_{{\mathrm{1}}}  =  b_{{\mathrm{2}}} \) or \( b_{{\mathrm{1}}}  \leq  b_{{\mathrm{2}}} \),
  and binding times \(b\) are extended with \dfn{binding-time variables} \(\beta\)
  (i.e., \(\bnfnotab{b}{  0  |  1  | \beta }\)).
  The derivation rules generate
  constraints \(C\) on binding-time variables through traversal
  to solve them afterwards, just as the standard Hindley---Milner type inference~\cite{%
    Hindley1969,%
    Milner1978}
  does by tracking types containing type variables.
  Figures~\ref{fig:algorithmic-bta} and \ref{fig:bta-constraint-solving}
  describe these rules and the subsequent constraint-solving algorithm, respectively.
  Note that solutions are not necessarily unique;
  some variables can remain unresolved.
  One can substitute such variables with either \(0\) or \(1\).
\par

\clearpage
\section{Correctness Proofs}
\begin{figure}[tbp]
\small
  \begin{flushleft}
    \fbox{\( \vdash^{0}_{\mathrm{dom} }  \possiblyWithSub\stageOmetaColor{T^{\superscriptO} } \)}
    \fbox{\( \vdash^{0}_{\mathrm{cod} }  \possiblyWithSub\stageOmetaColor{T^{\superscriptO} } \)}
    \fbox{\( \vdash^{0}_{\mathrm{fo} }  \possiblyWithSub\stageOmetaColor{T^{\superscriptO} } \)}
    \fbox{\( \vdash^{1}_{\mathrm{oz} }  \possiblyWithSub\stageImetaColor{\tau^{\superscriptI} } \)}
    \fbox{\( \possiblyWithSub\stageOmetaColor{q}  \vDash  \possiblyWithSub\stageOmetaColor{T^{\superscriptO} } \)}
    \fbox{\( \possiblyWithSub\stageImetaColor{\tau^{\superscriptI} }  \gg  \possiblyWithSub\stageOmetaColor{T^{\superscriptO} } \)}
  \end{flushleft}
  \begin{center}
    \derive{}{
       \vdash^{0}_{\mathrm{dom} }   \ttO{Tensor}\  \possiblyWithSub\stageOmetaColor{s}  
    }
  \qquad
    \derive{}{%
       \vdash^{0}_{\mathrm{dom} }    \openO{\{} \possiblyWithSub\stageOmetaColor{x}  \relO{:}  \possiblyWithSub\stageOmetaColor{B}  \relO{\mid}  \possiblyWithSub\stageOmetaColor{N^{\superscriptO} } \closeO{\} }   
    }
  \qquad
    \derive{%
       \vdash^{0}_{\mathrm{dom} }  \possiblyWithSub\stageOmetaColor{T^{\superscriptO} }_{{\mathrm{1}}} 
    \andalso
       \vdash^{0}_{\mathrm{fo} }  \possiblyWithSub\stageOmetaColor{T^{\superscriptO} }_{{\mathrm{2}}} 
    }{%
       \vdash^{0}_{\mathrm{fo} }   \openO{(} \possiblyWithSub\stageOmetaColor{x}  \relO{:}  \possiblyWithSub\stageOmetaColor{T^{\superscriptO} }_{{\mathrm{1}}} \closeO{)} \relO{\to}  \possiblyWithSub\stageOmetaColor{T^{\superscriptO} }_{{\mathrm{2}}}  
    }
  \qquad
    \derive{%
       \vdash^{0}_{\mathrm{cod} }  \possiblyWithSub\stageOmetaColor{T^{\superscriptO} } 
    }{%
       \vdash^{0}_{\mathrm{fo} }  \possiblyWithSub\stageOmetaColor{T^{\superscriptO} } 
    }
  \\[0.7em]
    \derive{}{%
       \vdash^{0}_{\mathrm{cod} }   \ttO{Tensor}\  \possiblyWithSub\stageOmetaColor{s}  
    }
  \qquad
    \derive{}{%
       \vdash^{0}_{\mathrm{cod} }    \openO{\{} \possiblyWithSub\stageOmetaColor{x}  \relO{:}  \possiblyWithSub\stageOmetaColor{B}  \relO{\mid}  \possiblyWithSub\stageOmetaColor{N^{\superscriptO} } \closeO{\} }   
    }
  \qquad
    \derive{}{%
       \vdash^{0}_{\mathrm{cod} }   \openO{\langle} \possiblyWithSub\stageImetaColor{T^{\superscriptI} } \closeO{\rangle}  
    }
  \qquad
    \derive{}{
       \vdash^{1}_{\mathrm{oz} }   \ttI{Tensor}\ \ordI{\%} \possiblyWithSub\stageOmetaColor{s}  
    }
  \qquad
    \derive{}{%
       \vdash^{1}_{\mathrm{oz} }   \possiblyWithSub\stageImetaColor{B}  
    }
  \\[0.7em]
    \derive{%
      \ConstEnvPers(c) = \possiblyWithSub\stageImetaColor{B}
    \andalso
         [    \possiblyWithSub\stageOmetaColor{c}    /  \possiblyWithSub\stageOmetaColor{\nu}  ]    \possiblyWithSub\stageOmetaColor{N^{\superscriptO} }   \longrightarrow^{0\,\ast}      \ttO{true}     
    }{%
        \possiblyWithSub\stageOmetaColor{c}   \vDash    \openO{\{} \possiblyWithSub\stageOmetaColor{\nu}  \relO{:}  \possiblyWithSub\stageOmetaColor{B}  \relO{\mid}  \possiblyWithSub\stageOmetaColor{N^{\superscriptO} } \closeO{\} }   
    }
  \quad
    \derive{%
      \ConstEnvPers(c) =  \ttI{Tensor}\ \ordI{\%} \possiblyWithSub\stageOmetaColor{s} 
    }{%
        \possiblyWithSub\stageOmetaColor{c}   \vDash   \ttO{Tensor}\  \possiblyWithSub\stageOmetaColor{s}  
    }
  \quad
    \derive{%
      \ConstEnvPers(c) = \possiblyWithSub\stageImetaColor{\tau^{\superscriptI} }
    \andalso
       \possiblyWithSub\stageImetaColor{T^{\superscriptI} }  \longrightarrow^{1\,\ast}    \possiblyWithSub\stageImetaColor{\tau^{\superscriptI} }   
    }{%
        \openO{\langle} \possiblyWithSub\stageImetaColor{c} \closeO{\rangle}   \vDash   \openO{\langle} \possiblyWithSub\stageImetaColor{T^{\superscriptI} } \closeO{\rangle}  
    }
  \\[0.7em]
    \derive{}{%
        \possiblyWithSub\stageImetaColor{B}   \gg    \openO{\{} \possiblyWithSub\stageOmetaColor{\nu}  \relO{:}  \possiblyWithSub\stageOmetaColor{B}  \relO{\mid}  \possiblyWithSub\stageOmetaColor{N^{\superscriptO} } \closeO{\} }   
    }
  \qquad
    \derive{}{%
        \ttI{Tensor}\ \ordI{\%} \possiblyWithSub\stageOmetaColor{s}   \gg   \ttO{Tensor}\  \possiblyWithSub\stageOmetaColor{s}  
    }
  \qquad
    \derive{%
       \possiblyWithSub\stageImetaColor{\tau^{\superscriptI} }_{{\mathrm{1}}}  \gg  \possiblyWithSub\stageOmetaColor{T^{\superscriptO} }_{{\mathrm{1}}} 
    \andalso
       \possiblyWithSub\stageImetaColor{\tau^{\superscriptI} }_{{\mathrm{2}}}  \gg  \possiblyWithSub\stageOmetaColor{T^{\superscriptO} }_{{\mathrm{2}}} 
    }{%
        \possiblyWithSub\stageImetaColor{\tau^{\superscriptI} }_{{\mathrm{1}}}  \relI{\to}  \possiblyWithSub\stageImetaColor{\tau^{\superscriptI} }_{{\mathrm{2}}}   \gg   \openO{(} \possiblyWithSub\stageOmetaColor{x}  \relO{:}  \possiblyWithSub\stageOmetaColor{T^{\superscriptO} }_{{\mathrm{1}}} \closeO{)} \relO{\to}  \possiblyWithSub\stageOmetaColor{T^{\superscriptO} }_{{\mathrm{2}}}  
    }
  \end{center}
  \caption{Miscellaneous rules for correctness proof}
\end{figure}
  \indent
    For simplicity, we assume the so-called \textit{Barendregt convention} henceforth,
    i.e., by appropriate \(\alpha\)-renaming, we can assume w.l.o.g.
    that no variable names are bound more than once in a scope.
    As mentioned in Section~\ref{subsec:staged-language-metatheory},
    we also assume the following:
  \par
  \begin{assumption}[Consistency of Built-in Constants]\label{assump:type-of-constants}
    \noindent
    \begin{enumerate}
      \item
        For each \(c\), we have a natural number~\(\arity{c}\) called
        the \dfn{arity} of \(c\).
      \item
        For each \(c\), its type~\(\possiblyWithSub\stageImetaColor{\tau^{\superscriptI} } := \ConstEnvPers(c)\) is a first-order type,
        i.e., there exist \(\possiblyWithSub\stageImetaColor{\tau'^{\superscriptI} }\) and a sequence~\((\possiblyWithSub\stageImetaColor{\tau^{\superscriptI} }_{\ottmv{i}})_{i = 1}^{m}\),
        where \(m := \arity{c}\), such that:
        \begin{itemize}
          \item
            \(\possiblyWithSub\stageImetaColor{\tau^{\superscriptI} } =   \possiblyWithSub\stageImetaColor{\tau^{\superscriptI} }_{{\mathrm{1}}}  \relI{\to}   \cdots \relI{\to}  \possiblyWithSub\stageImetaColor{\tau^{\superscriptI} }_{\ottmv{m}}    \relI{\to}  \possiblyWithSub\stageImetaColor{\tau'^{\superscriptI} } \),
          \item
            \( \vdash^{1}_{\mathrm{oz} }  \possiblyWithSub\stageImetaColor{\tau^{\superscriptI} }_{\ottmv{i}} \) for each \(i \in \{1, \ldots, m\}\), and
          \item
            \( \vdash^{1}_{\mathrm{oz} }  \possiblyWithSub\stageImetaColor{\tau'^{\superscriptI} } \).
        \end{itemize}
      \item
        For a function~\(c\) of arity~\(m := \arity{c} \geq 1\)
        and a sequence~\((c_{\ottmv{i}})_{i = 1}^{m}\)
        such that \(\ConstEnvPers(c_{\ottmv{i}}) = \possiblyWithSub\stageImetaColor{\tau^{\superscriptI} }_{\ottmv{i}}\) for each \(i \in \{1, \ldots, m\}\)
        where \(\ConstEnvPers(c) =:   \possiblyWithSub\stageImetaColor{\tau^{\superscriptI} }_{{\mathrm{1}}}  \relI{\to}   \cdots \relI{\to}  \possiblyWithSub\stageImetaColor{\tau^{\superscriptI} }_{\ottmv{m}}    \relI{\to}  \possiblyWithSub\stageImetaColor{\tau'^{\superscriptI} } \),
        there exists \(c'\) such that
        \(\delta(\possiblyWithSub\stageOmetaColor{c}, (\possiblyWithSub\stageOmetaColor{c}_{\ottmv{i}})_{i = 1}^{m}) = \possiblyWithSub\stageOmetaColor{c'}\)
        and \(\ConstEnvPers(c') = \possiblyWithSub\stageImetaColor{\tau'^{\superscriptI} }\).
      \item
        If \(\delta(\possiblyWithSub\stageOmetaColor{c}, (\possiblyWithSub\stageOmetaColor{c}_{\ottmv{i}})_{i = 1}^{k}) = \possiblyWithSub\stageOmetaColor{q}\),
        then:
        \begin{itemize}
          \item
            \(k = \arity{c}\),
          \item
            \(\possiblyWithSub\stageOmetaColor{q}\) is of the form~\(\possiblyWithSub\stageOmetaColor{c'}\), and
          \item
            there exist a sequence~\((\possiblyWithSub\stageImetaColor{\tau^{\superscriptI} }_{\ottmv{i}})_{i = 1}^{k}\) and \(\possiblyWithSub\stageImetaColor{\tau'^{\superscriptI} }\) such that:
            \begin{itemize}
              \item \(\ConstEnvPers(c) =   \possiblyWithSub\stageImetaColor{\tau^{\superscriptI} }_{{\mathrm{1}}}  \relI{\to}   \cdots \relI{\to}  \possiblyWithSub\stageImetaColor{\tau^{\superscriptI} }_{\ottmv{k}}    \relI{\to}  \possiblyWithSub\stageImetaColor{\tau'^{\superscriptI} } \), and
              \item \(\ConstEnvPers(c_{\ottmv{i}}) = \possiblyWithSub\stageImetaColor{\tau^{\superscriptI} }_{\ottmv{i}}\) for each \(i \in \{1, \ldots, k\}\).
            \end{itemize}
        \end{itemize}
      \item
        For each \(\possiblyWithSub\stageOmetaColor{p}\), we have a natural number~\(\arity{\possiblyWithSub\stageOmetaColor{p}}\) called
        the \dfn{arity} of \(\possiblyWithSub\stageOmetaColor{p}\).
      \item
        For each \(\possiblyWithSub\stageOmetaColor{p}\), its type~\(\possiblyWithSub\stageOmetaColor{T^{\superscriptO} } := \ConstEnvZero(\possiblyWithSub\stageOmetaColor{p})\) is
        a well-formed first-order function type, i.e.,
        we have \(  \bullet   \vdash^{0}  \possiblyWithSub\stageOmetaColor{T^{\superscriptO} } \), and
        there exist \(\possiblyWithSub\stageOmetaColor{T'^{\superscriptO} }\) and a sequence \((\possiblyWithSub\stageOmetaColor{T^{\superscriptO} }_{\ottmv{i}})_{i = 1}^{m}\),
        where \(m := \arity{\possiblyWithSub\stageOmetaColor{p}}\), such that:
        \begin{itemize}
          \item
            \(\possiblyWithSub\stageOmetaColor{T^{\superscriptO} } =  \openO{(} \possiblyWithSub\stageOmetaColor{x}_{{\mathrm{1}}}  \relO{:}  \possiblyWithSub\stageOmetaColor{T^{\superscriptO} }_{{\mathrm{1}}} \closeO{)} \relO{\to}   \cdots \relO{\to}   \openO{(} \possiblyWithSub\stageOmetaColor{x}_{\ottmv{m}}  \relO{:}  \possiblyWithSub\stageOmetaColor{T^{\superscriptO} }_{\ottmv{m}} \closeO{)} \relO{\to}  \possiblyWithSub\stageOmetaColor{T'^{\superscriptO} }   \),
          \item
            \( \vdash^{0}_{\mathrm{dom} }  \possiblyWithSub\stageOmetaColor{T^{\superscriptO} }_{\ottmv{i}} \) for each \(i \in \{1, \ldots, m\}\), and
          \item
            \( \vdash^{0}_{\mathrm{cod} }  \possiblyWithSub\stageOmetaColor{T'^{\superscriptO} } \).
        \end{itemize}
      \item
        For a function~\(\possiblyWithSub\stageOmetaColor{p}\) of arity~\(m := \arity{\possiblyWithSub\stageOmetaColor{p}} \geq 1\)
        and a sequence~\((c_{\ottmv{i}})_{i = 1}^{m}\) such that
        \(\possiblyWithSub\stageOmetaColor{c}_{\ottmv{i}} \vDash [ \possiblyWithSub\stageOmetaColor{c}_{i - 1}/\possiblyWithSub\stageOmetaColor{x}_{i - 1} ] \cdots [ \possiblyWithSub\stageOmetaColor{c}_{{\mathrm{1}}}/\possiblyWithSub\stageOmetaColor{x}_{{\mathrm{1}}} ] \possiblyWithSub\stageOmetaColor{T^{\superscriptO} }_{\ottmv{i}}\)
        for each \(i \in \{1, \ldots, m\}\)
        where \(\ConstEnvZero(\possiblyWithSub\stageOmetaColor{p}) =:  \openO{(} \possiblyWithSub\stageOmetaColor{x}_{{\mathrm{1}}}  \relO{:}  \possiblyWithSub\stageOmetaColor{T^{\superscriptO} }_{{\mathrm{1}}} \closeO{)} \relO{\to}   \cdots \relO{\to}   \openO{(} \possiblyWithSub\stageOmetaColor{x}_{\ottmv{m}}  \relO{:}  \possiblyWithSub\stageOmetaColor{T^{\superscriptO} }_{\ottmv{m}} \closeO{)} \relO{\to}  \possiblyWithSub\stageOmetaColor{T'^{\superscriptO} }   \),
        there exists \(\possiblyWithSub\stageOmetaColor{q}\) such that
        \(\delta(\possiblyWithSub\stageOmetaColor{p}, (\possiblyWithSub\stageOmetaColor{c}_{\ottmv{i}})_{i = 1}^{m}) = \possiblyWithSub\stageOmetaColor{q}\) and
        \(\possiblyWithSub\stageOmetaColor{q} \vDash [ \possiblyWithSub\stageOmetaColor{c}_{\ottmv{m}}/\possiblyWithSub\stageOmetaColor{x}_{\ottmv{m}} ] \cdots [ \possiblyWithSub\stageOmetaColor{c}_{{\mathrm{1}}}/\possiblyWithSub\stageOmetaColor{x}_{{\mathrm{1}}} ] \possiblyWithSub\stageOmetaColor{T'^{\superscriptO} }\).
      \item
        If \(\delta(\possiblyWithSub\stageOmetaColor{p}, (\possiblyWithSub\stageOmetaColor{c}_{\ottmv{i}})_{i = 1}^{k}) = \possiblyWithSub\stageOmetaColor{q}\),
        then:
        \begin{itemize}
          \item \(k = \arity{\possiblyWithSub\stageOmetaColor{p}}\), and
          \item there exist a sequence~\((\possiblyWithSub\stageOmetaColor{T^{\superscriptO} }_{\ottmv{i}})_{i = 1}^{k}\) and \(\possiblyWithSub\stageOmetaColor{T'^{\superscriptO} }\) such that:
            \begin{itemize}
              \item
                \(\ConstEnvZero(\possiblyWithSub\stageOmetaColor{p}) =  \openO{(} \possiblyWithSub\stageOmetaColor{x}_{{\mathrm{1}}}  \relO{:}  \possiblyWithSub\stageOmetaColor{T^{\superscriptO} }_{{\mathrm{1}}} \closeO{)} \relO{\to}   \cdots \relO{\to}   \openO{(} \possiblyWithSub\stageOmetaColor{x}_{\ottmv{k}}  \relO{:}  \possiblyWithSub\stageOmetaColor{T^{\superscriptO} }_{\ottmv{k}} \closeO{)} \relO{\to}  \possiblyWithSub\stageOmetaColor{T'^{\superscriptO} }   \), and
              \item
                \(\possiblyWithSub\stageOmetaColor{c}_{\ottmv{i}} \vDash [ \possiblyWithSub\stageOmetaColor{c}_{i - 1}/\possiblyWithSub\stageOmetaColor{x}_{i - 1} ] \cdots [ \possiblyWithSub\stageOmetaColor{c}_{{\mathrm{1}}}/\possiblyWithSub\stageOmetaColor{x}_{{\mathrm{1}}} ] \possiblyWithSub\stageOmetaColor{T^{\superscriptO} }_{\ottmv{i}}\)
                for each \(i \in \{1, \ldots, k\}\).
            \end{itemize}
        \end{itemize}
      \item
        \(\ConstEnvPers(c) =   \ttI{Bool}  \) implies
        \(c \in \{  \mathtt{true}  ,   \mathtt{false}  \}\).
      \item
        If \(\ConstEnvPers(c) =   \ttI{NatList}  \), then
        \(c\) is of the form~\(s\).
    \end{enumerate}
  \end{assumption}

\subsection{Soundness of Assertion Insertion}
  \begin{lemma}[Soundness of Assertive Cast Generation]\label{lem:cast-soundness}
    If \( \vdash  \mathit{\Gamma} \), \( \mathit{\Gamma}  \vdash^{0}  \possiblyWithSub\stageOmetaColor{T^{\superscriptO} }_{{\mathrm{1}}} \), \( \mathit{\Gamma}  \vdash^{0}  \possiblyWithSub\stageOmetaColor{T^{\superscriptO} }_{{\mathrm{2}}} \), and \( \mathit{\Gamma}  \vdash_{ L }  \possiblyWithSub\stageOmetaColor{T^{\superscriptO} }_{{\mathrm{1}}}  \CastArrow  \possiblyWithSub\stageOmetaColor{T^{\superscriptO} }_{{\mathrm{2}}}  \ElabArrow  \possiblyWithSub\stageOmetaColor{N^{\superscriptO} } \),
    then we have \( \mathit{\Gamma}  \vdash^{0}  \possiblyWithSub\stageOmetaColor{N^{\superscriptO} }  :   \openO{(} \possiblyWithSub\stageOmetaColor{x}  \relO{:}  \possiblyWithSub\stageOmetaColor{T^{\superscriptO} }_{{\mathrm{1}}} \closeO{)} \relO{\to}  \possiblyWithSub\stageOmetaColor{T^{\superscriptO} }_{{\mathrm{2}}}  \)
    for some \(\possiblyWithSub\stageOmetaColor{x}\) such that \(\possiblyWithSub\stageOmetaColor{x} \not\in \fv(\possiblyWithSub\stageOmetaColor{T^{\superscriptO} }_{{\mathrm{2}}})\).
  \end{lemma}
  \begin{proof}
    By induction on the derivation.
    \begin{itemize}
      \item Case \derive[I-Tensor]{}{%
         \mathit{\Gamma}  \vdash_{ L }   \ttO{Tensor}\  \possiblyWithSub\stageOmetaColor{s}   \CastArrow   \ttO{Tensor}\  \possiblyWithSub\stageOmetaColor{s}   \ElabArrow   \ordO{\lambda} \possiblyWithSub\stageOmetaColor{x}  \relO{:}   \ttO{Tensor}\  \possiblyWithSub\stageOmetaColor{s}  \punctO{.}\   \possiblyWithSub\stageOmetaColor{x}   
      }:
        We can immediately derive
        \begin{center}
          \infer[T0-Abs]{%
            \infer[T0-Var]{%
              \infer[WfEnv-Cons0]{%
                 \vdash  \mathit{\Gamma} 
              \andalso
                \infer[WfT0-Tensor]{}{
                   \mathit{\Gamma}  \vdash^{0}   \ttO{Tensor}\  \possiblyWithSub\stageOmetaColor{s}  
                }
              }{%
                 \vdash   \mathit{\Gamma} ,  \possiblyWithSub\stageOmetaColor{x}  : (  \ttO{Tensor}\  \possiblyWithSub\stageOmetaColor{s}  )^{0}  
              }
            \andalso
               (  \mathit{\Gamma} ,  \possiblyWithSub\stageOmetaColor{x}  : (  \ttO{Tensor}\  \possiblyWithSub\stageOmetaColor{s}  )^{0}  ) (\possiblyWithSub\stageOmetaColor{x}) = ( \ttO{Tensor}\  \possiblyWithSub\stageOmetaColor{s} )^{0}
            }{%
                \mathit{\Gamma} ,  \possiblyWithSub\stageOmetaColor{x}  : (  \ttO{Tensor}\  \possiblyWithSub\stageOmetaColor{s}  )^{0}   \vdash^{0}   \possiblyWithSub\stageOmetaColor{x}   :   \ttO{Tensor}\  \possiblyWithSub\stageOmetaColor{s}  
            }
          }{%
             \mathit{\Gamma}  \vdash^{0}   \ordO{\lambda} \possiblyWithSub\stageOmetaColor{x}  \relO{:}   \ttO{Tensor}\  \possiblyWithSub\stageOmetaColor{s}  \punctO{.}\   \possiblyWithSub\stageOmetaColor{x}    :   \openO{(} \possiblyWithSub\stageOmetaColor{x}  \relO{:}   \ttO{Tensor}\  \possiblyWithSub\stageOmetaColor{s}  \closeO{)} \relO{\to}   \ttO{Tensor}\  \possiblyWithSub\stageOmetaColor{s}   
          }.
        \end{center}
      \item Case \derive[I-Rfn]{}{%
         \mathit{\Gamma}  \vdash_{ L }    \openO{\{} \possiblyWithSub\stageOmetaColor{\nu}  \relO{:}  \possiblyWithSub\stageOmetaColor{B}  \relO{\mid}  \possiblyWithSub\stageOmetaColor{N^{\superscriptO} }_{{\mathrm{1}}} \closeO{\} }    \CastArrow    \openO{\{} \possiblyWithSub\stageOmetaColor{\nu}  \relO{:}  \possiblyWithSub\stageOmetaColor{B}  \relO{\mid}  \possiblyWithSub\stageOmetaColor{N^{\superscriptO} }_{{\mathrm{2}}} \closeO{\} }    \ElabArrow   \LeftAssertParen \relO{\CastArrow}   \openO{\{} \possiblyWithSub\stageOmetaColor{\nu}  \relO{:}  \possiblyWithSub\stageOmetaColor{B}  \relO{\mid}  \possiblyWithSub\stageOmetaColor{N^{\superscriptO} }_{{\mathrm{2}}} \closeO{\} }   \RightAssertParen^{ L }  
      }:
        We can immediately derive
        \begin{center}
          \derive[T0-Rfn]{%
             \mathit{\Gamma}  \vdash^{0}    \openO{\{} \possiblyWithSub\stageOmetaColor{\nu}  \relO{:}  \possiblyWithSub\stageOmetaColor{B}  \relO{\mid}  \possiblyWithSub\stageOmetaColor{N^{\superscriptO} }_{{\mathrm{2}}} \closeO{\} }   
          \andalso
             \mathit{\Gamma}  \vdash^{0}    \openO{\{} \possiblyWithSub\stageOmetaColor{\nu}  \relO{:}  \possiblyWithSub\stageOmetaColor{B}  \relO{\mid}  \possiblyWithSub\stageOmetaColor{N^{\superscriptO} }_{{\mathrm{1}}} \closeO{\} }   
          \andalso
            \possiblyWithSub\stageOmetaColor{x} \not\in \dom \mathit{\Gamma}
          }{%
             \mathit{\Gamma}  \vdash^{0}   \LeftAssertParen \relO{\CastArrow}   \openO{\{} \possiblyWithSub\stageOmetaColor{\nu}  \relO{:}  \possiblyWithSub\stageOmetaColor{B}  \relO{\mid}  \possiblyWithSub\stageOmetaColor{N^{\superscriptO} }_{{\mathrm{2}}} \closeO{\} }   \RightAssertParen^{ L }   :   \openO{(} \possiblyWithSub\stageOmetaColor{x}  \relO{:}    \openO{\{} \possiblyWithSub\stageOmetaColor{\nu}  \relO{:}  \possiblyWithSub\stageOmetaColor{B}  \relO{\mid}  \possiblyWithSub\stageOmetaColor{N^{\superscriptO} }_{{\mathrm{1}}} \closeO{\} }   \closeO{)} \relO{\to}    \openO{\{} \possiblyWithSub\stageOmetaColor{\nu}  \relO{:}  \possiblyWithSub\stageOmetaColor{B}  \relO{\mid}  \possiblyWithSub\stageOmetaColor{N^{\superscriptO} }_{{\mathrm{2}}} \closeO{\} }    
          }.
        \end{center}
        Since \(\possiblyWithSub\stageOmetaColor{x} \not\in \dom \mathit{\Gamma}\) and \( \mathit{\Gamma}  \vdash^{0}  \possiblyWithSub\stageOmetaColor{T^{\superscriptO} }_{{\mathrm{2}}} \),
        we clearly have \(\possiblyWithSub\stageOmetaColor{x} \not\in \fv(\possiblyWithSub\stageOmetaColor{T^{\superscriptO} }_{{\mathrm{2}}})\).
      \item Case \derive[I-Code]{%
      \andalso
         \possiblyWithSub\stageImetaColor{T^{\superscriptI} }_{{\mathrm{1}}}  \mathrel{||}^{1}  \possiblyWithSub\stageImetaColor{T^{\superscriptI} }_{{\mathrm{2}}} 
      }{%
         \mathit{\Gamma}  \vdash_{ L }   \openO{\langle} \possiblyWithSub\stageImetaColor{T^{\superscriptI} }_{{\mathrm{1}}} \closeO{\rangle}   \CastArrow   \openO{\langle} \possiblyWithSub\stageImetaColor{T^{\superscriptI} }_{{\mathrm{2}}} \closeO{\rangle}   \ElabArrow   \LeftAssertParen\openO{\langle} \possiblyWithSub\stageImetaColor{T^{\superscriptI} }_{{\mathrm{1}}} \closeO{\rangle} \relO{\CastArrow} \openO{\langle} \possiblyWithSub\stageImetaColor{T^{\superscriptI} }_{{\mathrm{2}}} \closeO{\rangle}\RightAssertParen^{ L }  
      }:
        By the assumptions \( \mathit{\Gamma}  \vdash^{0}   \openO{\langle} \possiblyWithSub\stageImetaColor{T^{\superscriptI} }_{{\mathrm{1}}} \closeO{\rangle}  \) and \( \mathit{\Gamma}  \vdash^{0}   \openO{\langle} \possiblyWithSub\stageImetaColor{T^{\superscriptI} }_{{\mathrm{2}}} \closeO{\rangle}  \),
        we have \( \mathit{\Gamma}  \vdash^{1}  \possiblyWithSub\stageImetaColor{T^{\superscriptI} }_{{\mathrm{1}}} \) and \( \mathit{\Gamma}  \vdash^{1}  \possiblyWithSub\stageImetaColor{T^{\superscriptI} }_{{\mathrm{2}}} \).
        We can thus derive the following immediately:
        \begin{center}
          \derive[T0-Ass]{%
             \mathit{\Gamma}  \vdash^{1}  \possiblyWithSub\stageImetaColor{T^{\superscriptI} }_{{\mathrm{1}}} 
          \andalso
             \mathit{\Gamma}  \vdash^{1}  \possiblyWithSub\stageImetaColor{T^{\superscriptI} }_{{\mathrm{2}}} 
          \andalso
             \possiblyWithSub\stageImetaColor{T^{\superscriptI} }_{{\mathrm{1}}}  \mathrel{||}^{1}  \possiblyWithSub\stageImetaColor{T^{\superscriptI} }_{{\mathrm{2}}} 
          \andalso
            \possiblyWithSub\stageOmetaColor{x}\ \not\in \dom(\mathit{\Gamma})
          }{%
             \mathit{\Gamma}  \vdash^{0}   \LeftAssertParen\openO{\langle} \possiblyWithSub\stageImetaColor{T^{\superscriptI} }_{{\mathrm{1}}} \closeO{\rangle} \relO{\CastArrow} \openO{\langle} \possiblyWithSub\stageImetaColor{T^{\superscriptI} }_{{\mathrm{2}}} \closeO{\rangle}\RightAssertParen^{ L }   :   \openO{(} \possiblyWithSub\stageOmetaColor{x}  \relO{:}   \openO{\langle} \possiblyWithSub\stageImetaColor{T^{\superscriptI} }_{{\mathrm{1}}} \closeO{\rangle}  \closeO{)} \relO{\to}   \openO{\langle} \possiblyWithSub\stageImetaColor{T^{\superscriptI} }_{{\mathrm{2}}} \closeO{\rangle}   
          }
        \end{center}
      \item Case \derive[I-Arr]{%
         \mathit{\Gamma}  \vdash_{  L .\mathbf{dom}  }  \possiblyWithSub\stageOmetaColor{T^{\superscriptO} }_{{\mathrm{21}}}  \CastArrow  \possiblyWithSub\stageOmetaColor{T^{\superscriptO} }_{{\mathrm{11}}}  \ElabArrow  \possiblyWithSub\stageOmetaColor{N^{\superscriptO} }_{{\mathrm{1}}} 
      \andalso
        \possiblyWithSub\stageOmetaColor{f} \not\in \dom(\mathit{\Gamma}) \uplus \{\possiblyWithSub\stageOmetaColor{x}\}
      \\
        \possiblyWithSub\stageOmetaColor{x'} \not\in \dom(\mathit{\Gamma}) \uplus \{\possiblyWithSub\stageOmetaColor{x}, \possiblyWithSub\stageOmetaColor{f}\}
      \andalso
          \mathit{\Gamma} ,  \possiblyWithSub\stageOmetaColor{x'}  : ( \possiblyWithSub\stageOmetaColor{T^{\superscriptO} }_{{\mathrm{11}}} )^{0}   \vdash_{  L .\mathbf{cod}  }    [   \possiblyWithSub\stageOmetaColor{x'}   /  \possiblyWithSub\stageOmetaColor{x}  ]    \possiblyWithSub\stageOmetaColor{T^{\superscriptO} }_{{\mathrm{12}}}   \CastArrow  \possiblyWithSub\stageOmetaColor{T^{\superscriptO} }_{{\mathrm{22}}}  \ElabArrow  \possiblyWithSub\stageOmetaColor{N^{\superscriptO} }_{{\mathrm{2}}} 
      }{%
        \begin{aligned}
          &
          \mathit{\Gamma} \vdash_{L}  \openO{(} \possiblyWithSub\stageOmetaColor{x}  \relO{:}  \possiblyWithSub\stageOmetaColor{T^{\superscriptO} }_{{\mathrm{11}}} \closeO{)} \relO{\to}  \possiblyWithSub\stageOmetaColor{T^{\superscriptO} }_{{\mathrm{12}}}  \Rightarrow  \openO{(} \possiblyWithSub\stageOmetaColor{x}  \relO{:}  \possiblyWithSub\stageOmetaColor{T^{\superscriptO} }_{{\mathrm{21}}} \closeO{)} \relO{\to}  \possiblyWithSub\stageOmetaColor{T^{\superscriptO} }_{{\mathrm{22}}} 
          \rightsquigarrow
        \\[-0.25em]
          &\quad
           \ordO{\lambda} \possiblyWithSub\stageOmetaColor{f}  \relO{:}   \openO{(} \possiblyWithSub\stageOmetaColor{x}  \relO{:}  \possiblyWithSub\stageOmetaColor{T^{\superscriptO} }_{{\mathrm{11}}} \closeO{)} \relO{\to}  \possiblyWithSub\stageOmetaColor{T^{\superscriptO} }_{{\mathrm{12}}}  \punctO{.}\    \ordO{\lambda} \possiblyWithSub\stageOmetaColor{x}  \relO{:}  \possiblyWithSub\stageOmetaColor{T^{\superscriptO} }_{{\mathrm{21}}} \punctO{.}\    \tokenO{let}\  \possiblyWithSub\stageOmetaColor{x'}  \relO{:}  \possiblyWithSub\stageOmetaColor{T^{\superscriptO} }_{{\mathrm{11}}}  \relO{=}   \possiblyWithSub\stageOmetaColor{N^{\superscriptO} }_{{\mathrm{1}}} \   \possiblyWithSub\stageOmetaColor{x}    \ \tokenO{in}\  \possiblyWithSub\stageOmetaColor{N^{\superscriptO} }_{{\mathrm{2}}}   \   \openO{(}   \possiblyWithSub\stageOmetaColor{f}  \   \possiblyWithSub\stageOmetaColor{x'}   \closeO{)}   
        \end{aligned}
      }:
        Since \( \mathit{\Gamma}  \vdash^{0}  \possiblyWithSub\stageOmetaColor{T^{\superscriptO} }_{{\mathrm{11}}} \) and \( \mathit{\Gamma}  \vdash^{0}  \possiblyWithSub\stageOmetaColor{T^{\superscriptO} }_{{\mathrm{21}}} \) hold from the assumption,
        by IH, we first have \( \mathit{\Gamma}  \vdash^{0}  \possiblyWithSub\stageOmetaColor{N^{\superscriptO} }_{{\mathrm{1}}}  :   \openO{(} \possiblyWithSub\stageOmetaColor{x}_{{\mathrm{1}}}  \relO{:}  \possiblyWithSub\stageOmetaColor{T^{\superscriptO} }_{{\mathrm{21}}} \closeO{)} \relO{\to}  \possiblyWithSub\stageOmetaColor{T^{\superscriptO} }_{{\mathrm{11}}}  \) for some fresh \(\possiblyWithSub\stageOmetaColor{x}_{{\mathrm{1}}}\).
        Let \(\mathit{\Gamma}'\), \(\mathit{\Gamma}_{{\mathrm{0}}}\), and \(\possiblyWithSub\stageOmetaColor{N^{\superscriptO} }_{{\mathrm{0}}}\) be defined as follows:
        \begin{gather*}
          \mathit{\Gamma}' :=  \mathit{\Gamma}_{{\mathrm{0}}} ,  \possiblyWithSub\stageOmetaColor{x'}  : ( \possiblyWithSub\stageOmetaColor{T^{\superscriptO} }_{{\mathrm{11}}} )^{0} 
        \qquad
          \mathit{\Gamma}_{{\mathrm{0}}} :=   \mathit{\Gamma} ,  \possiblyWithSub\stageOmetaColor{f}  : (  \openO{(} \possiblyWithSub\stageOmetaColor{x}  \relO{:}  \possiblyWithSub\stageOmetaColor{T^{\superscriptO} }_{{\mathrm{11}}} \closeO{)} \relO{\to}  \possiblyWithSub\stageOmetaColor{T^{\superscriptO} }_{{\mathrm{12}}}  )^{0}  ,  \possiblyWithSub\stageOmetaColor{x}  : ( \possiblyWithSub\stageOmetaColor{T^{\superscriptO} }_{{\mathrm{21}}} )^{0} 
        \\
          \possiblyWithSub\stageOmetaColor{N^{\superscriptO} }_{{\mathrm{0}}} :=  \openO{(}    \tokenO{let}\  \possiblyWithSub\stageOmetaColor{x'}  \relO{:}  \possiblyWithSub\stageOmetaColor{T^{\superscriptO} }_{{\mathrm{11}}}  \relO{=}   \possiblyWithSub\stageOmetaColor{N^{\superscriptO} }_{{\mathrm{1}}} \   \possiblyWithSub\stageOmetaColor{x}    \ \tokenO{in}\  \possiblyWithSub\stageOmetaColor{N^{\superscriptO} }_{{\mathrm{2}}}  \   \openO{(}   \possiblyWithSub\stageOmetaColor{f}  \   \possiblyWithSub\stageOmetaColor{x'}   \closeO{)}   \closeO{)} 
            =   \openO{(}   \ordO{\lambda} \possiblyWithSub\stageOmetaColor{x'}  \relO{:}  \possiblyWithSub\stageOmetaColor{T^{\superscriptO} }_{{\mathrm{11}}} \punctO{.}\  \possiblyWithSub\stageOmetaColor{N^{\superscriptO} }_{{\mathrm{2}}}  \   \openO{(}   \possiblyWithSub\stageOmetaColor{f}  \   \possiblyWithSub\stageOmetaColor{x'}   \closeO{)}   \closeO{)}  \   \openO{(}  \possiblyWithSub\stageOmetaColor{N^{\superscriptO} }_{{\mathrm{1}}} \   \possiblyWithSub\stageOmetaColor{x}   \closeO{)}  
        \end{gather*}
        From the assumptions, we also have \( \vdash  \mathit{\Gamma}' \), \( \mathit{\Gamma}'  \vdash^{0}  \possiblyWithSub\stageOmetaColor{T^{\superscriptO} }_{{\mathrm{12}}} \), and \( \mathit{\Gamma}'  \vdash^{0}  \possiblyWithSub\stageOmetaColor{T^{\superscriptO} }_{{\mathrm{22}}} \),
        and thus, by IH, \( \mathit{\Gamma}'  \vdash^{0}  \possiblyWithSub\stageOmetaColor{N^{\superscriptO} }_{{\mathrm{2}}}  :   \openO{(} \possiblyWithSub\stageOmetaColor{x}_{{\mathrm{2}}}  \relO{:}    [   \possiblyWithSub\stageOmetaColor{x'}   /  \possiblyWithSub\stageOmetaColor{x}  ]    \possiblyWithSub\stageOmetaColor{T^{\superscriptO} }_{{\mathrm{12}}}  \closeO{)} \relO{\to}  \possiblyWithSub\stageOmetaColor{T^{\superscriptO} }_{{\mathrm{22}}}  \) holds
        for some fresh \(\possiblyWithSub\stageOmetaColor{x}_{{\mathrm{2}}}\).
        Then, we can first derive
        \begin{center}
          \infer[T0-Abs]{%
            \infer[T0-App]{%
               \mathit{\Gamma}'  \vdash^{0}  \possiblyWithSub\stageOmetaColor{N^{\superscriptO} }_{{\mathrm{2}}}  :   \openO{(} \possiblyWithSub\stageOmetaColor{x}_{{\mathrm{2}}}  \relO{:}    [   \possiblyWithSub\stageOmetaColor{x'}   /  \possiblyWithSub\stageOmetaColor{x}  ]    \possiblyWithSub\stageOmetaColor{T^{\superscriptO} }_{{\mathrm{12}}}  \closeO{)} \relO{\to}  \possiblyWithSub\stageOmetaColor{T^{\superscriptO} }_{{\mathrm{22}}}  
            \andalso
              \infer[T0-App]{%
                \infer{}{
                   \mathit{\Gamma}'  \vdash^{0}   \possiblyWithSub\stageOmetaColor{f}   :   \openO{(} \possiblyWithSub\stageOmetaColor{x}  \relO{:}  \possiblyWithSub\stageOmetaColor{T^{\superscriptO} }_{{\mathrm{11}}} \closeO{)} \relO{\to}  \possiblyWithSub\stageOmetaColor{T^{\superscriptO} }_{{\mathrm{12}}}  
                }
              \andalso
                \infer{}{
                   \mathit{\Gamma}'  \vdash^{0}   \possiblyWithSub\stageOmetaColor{x'}   :  \possiblyWithSub\stageOmetaColor{T^{\superscriptO} }_{{\mathrm{11}}} 
                }
              }{%
                 \mathit{\Gamma}'  \vdash^{0}    \possiblyWithSub\stageOmetaColor{f}  \   \possiblyWithSub\stageOmetaColor{x'}    :    [   \possiblyWithSub\stageOmetaColor{x'}   /  \possiblyWithSub\stageOmetaColor{x}  ]    \possiblyWithSub\stageOmetaColor{T^{\superscriptO} }_{{\mathrm{12}}}  
              }
            }{
               \mathit{\Gamma}'  \vdash^{0}   \possiblyWithSub\stageOmetaColor{N^{\superscriptO} }_{{\mathrm{2}}} \   \openO{(}   \possiblyWithSub\stageOmetaColor{f}  \   \possiblyWithSub\stageOmetaColor{x'}   \closeO{)}    :  \possiblyWithSub\stageOmetaColor{T^{\superscriptO} }_{{\mathrm{22}}} 
            }
          }{%
             \mathit{\Gamma}_{{\mathrm{0}}}  \vdash^{0}    \ordO{\lambda} \possiblyWithSub\stageOmetaColor{x'}  \relO{:}  \possiblyWithSub\stageOmetaColor{T^{\superscriptO} }_{{\mathrm{11}}} \punctO{.}\  \possiblyWithSub\stageOmetaColor{N^{\superscriptO} }_{{\mathrm{2}}}  \   \openO{(}   \possiblyWithSub\stageOmetaColor{f}  \   \possiblyWithSub\stageOmetaColor{x'}   \closeO{)}    :   \openO{(} \possiblyWithSub\stageOmetaColor{x'}  \relO{:}  \possiblyWithSub\stageOmetaColor{T^{\superscriptO} }_{{\mathrm{11}}} \closeO{)} \relO{\to}  \possiblyWithSub\stageOmetaColor{T^{\superscriptO} }_{{\mathrm{22}}}  
          }
        \end{center}
        and
        \begin{center}
          \infer[T0-App]{%
             \mathit{\Gamma}_{{\mathrm{0}}}  \vdash^{0}  \possiblyWithSub\stageOmetaColor{N^{\superscriptO} }_{{\mathrm{1}}}  :   \openO{(} \possiblyWithSub\stageOmetaColor{x}_{{\mathrm{1}}}  \relO{:}  \possiblyWithSub\stageOmetaColor{T^{\superscriptO} }_{{\mathrm{21}}} \closeO{)} \relO{\to}  \possiblyWithSub\stageOmetaColor{T^{\superscriptO} }_{{\mathrm{11}}}  
          \andalso
            \infer[T0-Var]{}{%
               \mathit{\Gamma}_{{\mathrm{0}}}  \vdash^{0}   \possiblyWithSub\stageOmetaColor{x}   :  \possiblyWithSub\stageOmetaColor{T^{\superscriptO} }_{{\mathrm{21}}} 
            }
          }{
             \mathit{\Gamma}_{{\mathrm{0}}}  \vdash^{0}   \possiblyWithSub\stageOmetaColor{N^{\superscriptO} }_{{\mathrm{1}}} \   \possiblyWithSub\stageOmetaColor{x}    :  \possiblyWithSub\stageOmetaColor{T^{\superscriptO} }_{{\mathrm{11}}} 
          }
        \end{center}
        Therefore, we can derive
        \begin{center}
          \infer[T0-App]{%
             \mathit{\Gamma}_{{\mathrm{0}}}  \vdash^{0}    \ordO{\lambda} \possiblyWithSub\stageOmetaColor{x'}  \relO{:}  \possiblyWithSub\stageOmetaColor{T^{\superscriptO} }_{{\mathrm{11}}} \punctO{.}\  \possiblyWithSub\stageOmetaColor{N^{\superscriptO} }_{{\mathrm{2}}}  \   \openO{(}   \possiblyWithSub\stageOmetaColor{f}  \   \possiblyWithSub\stageOmetaColor{x'}   \closeO{)}    :   \openO{(} \possiblyWithSub\stageOmetaColor{x}  \relO{:}  \possiblyWithSub\stageOmetaColor{T^{\superscriptO} }_{{\mathrm{11}}} \closeO{)} \relO{\to}  \possiblyWithSub\stageOmetaColor{T^{\superscriptO} }_{{\mathrm{22}}}  
          \andalso
             \mathit{\Gamma}_{{\mathrm{0}}}  \vdash^{0}   \possiblyWithSub\stageOmetaColor{N^{\superscriptO} }_{{\mathrm{1}}} \   \possiblyWithSub\stageOmetaColor{x}    :  \possiblyWithSub\stageOmetaColor{T^{\superscriptO} }_{{\mathrm{11}}} 
          }{%
             \mathit{\Gamma}_{{\mathrm{0}}}  \vdash^{0}  \possiblyWithSub\stageOmetaColor{N^{\superscriptO} }_{{\mathrm{0}}}  :  \possiblyWithSub\stageOmetaColor{T^{\superscriptO} }_{{\mathrm{22}}} 
          }
        \end{center}
        and then we finally have
        \begin{center}
          \infer[T0-Abs]{%
             \mathit{\Gamma}  \vdash^{0}   \openO{(} \possiblyWithSub\stageOmetaColor{x}  \relO{:}  \possiblyWithSub\stageOmetaColor{T^{\superscriptO} }_{{\mathrm{11}}} \closeO{)} \relO{\to}  \possiblyWithSub\stageOmetaColor{T^{\superscriptO} }_{{\mathrm{12}}}  
          \andalso
            \infer[T0-Abs]{%
                \mathit{\Gamma} ,  \possiblyWithSub\stageOmetaColor{f}  : (  \openO{(} \possiblyWithSub\stageOmetaColor{x}  \relO{:}  \possiblyWithSub\stageOmetaColor{T^{\superscriptO} }_{{\mathrm{11}}} \closeO{)} \relO{\to}  \possiblyWithSub\stageOmetaColor{T^{\superscriptO} }_{{\mathrm{12}}}  )^{0}   \vdash^{0}  \possiblyWithSub\stageOmetaColor{T^{\superscriptO} }_{{\mathrm{11}}} 
            \andalso
               \mathit{\Gamma}_{{\mathrm{0}}}  \vdash^{0}  \possiblyWithSub\stageOmetaColor{N^{\superscriptO} }_{{\mathrm{0}}}  :  \possiblyWithSub\stageOmetaColor{T^{\superscriptO} }_{{\mathrm{22}}} 
            }{
                \mathit{\Gamma} ,  \possiblyWithSub\stageOmetaColor{f}  : (  \openO{(} \possiblyWithSub\stageOmetaColor{x}  \relO{:}  \possiblyWithSub\stageOmetaColor{T^{\superscriptO} }_{{\mathrm{11}}} \closeO{)} \relO{\to}  \possiblyWithSub\stageOmetaColor{T^{\superscriptO} }_{{\mathrm{12}}}  )^{0}   \vdash^{0}   \ordO{\lambda} \possiblyWithSub\stageOmetaColor{x}  \relO{:}  \possiblyWithSub\stageOmetaColor{T^{\superscriptO} }_{{\mathrm{21}}} \punctO{.}\  \possiblyWithSub\stageOmetaColor{N^{\superscriptO} }_{{\mathrm{0}}}   :   \openO{(} \possiblyWithSub\stageOmetaColor{x}  \relO{:}  \possiblyWithSub\stageOmetaColor{T^{\superscriptO} }_{{\mathrm{21}}} \closeO{)} \relO{\to}  \possiblyWithSub\stageOmetaColor{T^{\superscriptO} }_{{\mathrm{22}}}  
            }
          }{%
             \mathit{\Gamma}  \vdash^{0}   \ordO{\lambda} \possiblyWithSub\stageOmetaColor{f}  \relO{:}   \openO{(} \possiblyWithSub\stageOmetaColor{x}  \relO{:}  \possiblyWithSub\stageOmetaColor{T^{\superscriptO} }_{{\mathrm{11}}} \closeO{)} \relO{\to}  \possiblyWithSub\stageOmetaColor{T^{\superscriptO} }_{{\mathrm{12}}}  \punctO{.}\   \ordO{\lambda} \possiblyWithSub\stageOmetaColor{x}  \relO{:}  \possiblyWithSub\stageOmetaColor{T^{\superscriptO} }_{{\mathrm{21}}} \punctO{.}\  \possiblyWithSub\stageOmetaColor{N^{\superscriptO} }_{{\mathrm{0}}}    :   \openO{(} \possiblyWithSub\stageOmetaColor{f}  \relO{:}   \openO{(} \possiblyWithSub\stageOmetaColor{x}  \relO{:}  \possiblyWithSub\stageOmetaColor{T^{\superscriptO} }_{{\mathrm{11}}} \closeO{)} \relO{\to}  \possiblyWithSub\stageOmetaColor{T^{\superscriptO} }_{{\mathrm{12}}}  \closeO{)} \relO{\to}   \openO{(} \possiblyWithSub\stageOmetaColor{x}  \relO{:}  \possiblyWithSub\stageOmetaColor{T^{\superscriptO} }_{{\mathrm{21}}} \closeO{)} \relO{\to}  \possiblyWithSub\stageOmetaColor{T^{\superscriptO} }_{{\mathrm{22}}}   
          }
        \end{center}
    \end{itemize}
  \end{proof}
  \begin{lemma}[Substitution]\label{lem:subst}
    Suppose \( \mathit{\Gamma}  \vdash^{0}  \possiblyWithSub\stageOmetaColor{N'^{\superscriptO} }  :  \possiblyWithSub\stageOmetaColor{T'^{\superscriptO} } \).
    We have the following (under the Barendregt convention):
    \begin{enumerate}
      \item
        If \(   \mathit{\Gamma} ,  \possiblyWithSub\stageOmetaColor{x}  : ( \possiblyWithSub\stageOmetaColor{T'^{\superscriptO} } )^{0}  ,  \mathit{\Gamma}'   \vdash^{0}  \possiblyWithSub\stageOmetaColor{N^{\superscriptO} }  :  \possiblyWithSub\stageOmetaColor{T^{\superscriptO} } \),
        then \(  \mathit{\Gamma} ,    [  \possiblyWithSub\stageOmetaColor{N'^{\superscriptO} }  /  \possiblyWithSub\stageOmetaColor{x}  ]    \mathit{\Gamma}'    \vdash^{0}    [  \possiblyWithSub\stageOmetaColor{N'^{\superscriptO} }  /  \possiblyWithSub\stageOmetaColor{x}  ]    \possiblyWithSub\stageOmetaColor{N^{\superscriptO} }   :    [  \possiblyWithSub\stageOmetaColor{N'^{\superscriptO} }  /  \possiblyWithSub\stageOmetaColor{x}  ]    \possiblyWithSub\stageOmetaColor{T^{\superscriptO} }  \).
      \item
        If \(   \mathit{\Gamma} ,  \possiblyWithSub\stageOmetaColor{x}  : ( \possiblyWithSub\stageOmetaColor{T'^{\superscriptO} } )^{0}  ,  \mathit{\Gamma}'   \vdash^{1}  \possiblyWithSub\stageImetaColor{N^{\superscriptI} }  :  \possiblyWithSub\stageImetaColor{T^{\superscriptI} } \),
        then \(  \mathit{\Gamma} ,    [  \possiblyWithSub\stageOmetaColor{N'^{\superscriptO} }  /  \possiblyWithSub\stageOmetaColor{x}  ]    \mathit{\Gamma}'    \vdash^{1}    [  \possiblyWithSub\stageOmetaColor{N'^{\superscriptO} }  /  \possiblyWithSub\stageOmetaColor{x}  ]    \possiblyWithSub\stageImetaColor{N^{\superscriptI} }   :    [  \possiblyWithSub\stageOmetaColor{N'^{\superscriptO} }  /  \possiblyWithSub\stageOmetaColor{x}  ]    \possiblyWithSub\stageImetaColor{T^{\superscriptI} }  \).
      \item
        If \(   \mathit{\Gamma} ,  \possiblyWithSub\stageOmetaColor{x}  : ( \possiblyWithSub\stageOmetaColor{T'^{\superscriptO} } )^{0}  ,  \mathit{\Gamma}'   \vdash^{0}  \possiblyWithSub\stageOmetaColor{T^{\superscriptO} } \),
        then \(  \mathit{\Gamma} ,    [  \possiblyWithSub\stageOmetaColor{N'^{\superscriptO} }  /  \possiblyWithSub\stageOmetaColor{x}  ]    \mathit{\Gamma}'    \vdash^{0}    [  \possiblyWithSub\stageOmetaColor{N'^{\superscriptO} }  /  \possiblyWithSub\stageOmetaColor{x}  ]    \possiblyWithSub\stageOmetaColor{T^{\superscriptO} }  \).
      \item
        If \(   \mathit{\Gamma} ,  \possiblyWithSub\stageOmetaColor{x}  : ( \possiblyWithSub\stageOmetaColor{T'^{\superscriptO} } )^{0}  ,  \mathit{\Gamma}'   \vdash^{1}  \possiblyWithSub\stageImetaColor{T^{\superscriptI} } \),
        then \(  \mathit{\Gamma} ,    [  \possiblyWithSub\stageOmetaColor{N'^{\superscriptO} }  /  \possiblyWithSub\stageOmetaColor{x}  ]    \mathit{\Gamma}'    \vdash^{1}    [  \possiblyWithSub\stageOmetaColor{N'^{\superscriptO} }  /  \possiblyWithSub\stageOmetaColor{x}  ]    \possiblyWithSub\stageImetaColor{T^{\superscriptI} }  \).
      \item
        If \( \vdash    \mathit{\Gamma} ,  \possiblyWithSub\stageOmetaColor{x}  : ( \possiblyWithSub\stageOmetaColor{T'^{\superscriptO} } )^{0}  ,  \mathit{\Gamma}'  \),
        then \( \vdash   \mathit{\Gamma} ,    [  \possiblyWithSub\stageOmetaColor{N'^{\superscriptO} }  /  \possiblyWithSub\stageOmetaColor{x}  ]    \mathit{\Gamma}'   \).
    \end{enumerate}
  \end{lemma}
  \begin{proof}
    By induction on the derivation.
    \begin{enumerate}
      \item
        \begin{itemize}
          \item Case \derive[T0-Var]{%
             \vdash    \mathit{\Gamma} ,  \possiblyWithSub\stageOmetaColor{x}  : ( \possiblyWithSub\stageOmetaColor{T'^{\superscriptO} } )^{0}  ,  \mathit{\Gamma}'  
          \andalso
             (   \mathit{\Gamma} ,  \possiblyWithSub\stageOmetaColor{x}  : ( \possiblyWithSub\stageOmetaColor{T'^{\superscriptO} } )^{0}  ,  \mathit{\Gamma}'  ) (\possiblyWithSub\stageOmetaColor{x'}) = (\possiblyWithSub\stageOmetaColor{T^{\superscriptO} })^{0}
          }{%
               \mathit{\Gamma} ,  \possiblyWithSub\stageOmetaColor{x}  : ( \possiblyWithSub\stageOmetaColor{T'^{\superscriptO} } )^{0}  ,  \mathit{\Gamma}'   \vdash^{0}   \possiblyWithSub\stageOmetaColor{x'}   :  \possiblyWithSub\stageOmetaColor{T^{\superscriptO} } 
          }:
            By IH, from \( \vdash    \mathit{\Gamma} ,  \possiblyWithSub\stageOmetaColor{x}  : ( \possiblyWithSub\stageOmetaColor{T'^{\superscriptO} } )^{0}  ,  \mathit{\Gamma}'  \),
            we have \( \vdash   \mathit{\Gamma} ,    [  \possiblyWithSub\stageOmetaColor{N'^{\superscriptO} }  /  \possiblyWithSub\stageOmetaColor{x}  ]    \mathit{\Gamma}'   \).
            \begin{itemize}
              \item If \(\possiblyWithSub\stageOmetaColor{x'} \in \dom \mathit{\Gamma}'\):
                We can assume \(\possiblyWithSub\stageOmetaColor{x'} \neq \possiblyWithSub\stageOmetaColor{x}\) by the Barendregt convention,
                and we have \( (  \mathit{\Gamma} ,    [  \possiblyWithSub\stageOmetaColor{N'^{\superscriptO} }  /  \possiblyWithSub\stageOmetaColor{x}  ]    \mathit{\Gamma}'   ) (\possiblyWithSub\stageOmetaColor{x'}) = (  [  \possiblyWithSub\stageOmetaColor{N'^{\superscriptO} }  /  \possiblyWithSub\stageOmetaColor{x}  ]    \possiblyWithSub\stageOmetaColor{T^{\superscriptO} } )^{0}\).
                This enables us to derive
                \begin{center}
                  \derive[T0-Var]{%
                     \vdash   \mathit{\Gamma} ,    [  \possiblyWithSub\stageOmetaColor{N'^{\superscriptO} }  /  \possiblyWithSub\stageOmetaColor{x}  ]    \mathit{\Gamma}'   
                  \andalso
                     (  \mathit{\Gamma} ,    [  \possiblyWithSub\stageOmetaColor{N'^{\superscriptO} }  /  \possiblyWithSub\stageOmetaColor{x}  ]    \mathit{\Gamma}'   ) (\possiblyWithSub\stageOmetaColor{x'}) = (  [  \possiblyWithSub\stageOmetaColor{N'^{\superscriptO} }  /  \possiblyWithSub\stageOmetaColor{x}  ]    \possiblyWithSub\stageOmetaColor{T^{\superscriptO} } )^{0}
                  }{%
                      \mathit{\Gamma} ,    [  \possiblyWithSub\stageOmetaColor{N'^{\superscriptO} }  /  \possiblyWithSub\stageOmetaColor{x}  ]    \mathit{\Gamma}'    \vdash^{0}   \possiblyWithSub\stageOmetaColor{x'}   :    [  \possiblyWithSub\stageOmetaColor{N'^{\superscriptO} }  /  \possiblyWithSub\stageOmetaColor{x}  ]    \possiblyWithSub\stageOmetaColor{T^{\superscriptO} }  
                  }.
                \end{center}
              \item If \(\possiblyWithSub\stageOmetaColor{x'} = \possiblyWithSub\stageOmetaColor{x}\):
                We have \(\possiblyWithSub\stageOmetaColor{T^{\superscriptO} } = \possiblyWithSub\stageOmetaColor{T'^{\superscriptO} }\), and
                \(  [  \possiblyWithSub\stageOmetaColor{N'^{\superscriptO} }  /  \possiblyWithSub\stageOmetaColor{x}  ]    \possiblyWithSub\stageOmetaColor{T^{\superscriptO} }  =   [  \possiblyWithSub\stageOmetaColor{N'^{\superscriptO} }  /  \possiblyWithSub\stageOmetaColor{x}  ]    \possiblyWithSub\stageOmetaColor{T'^{\superscriptO} }  = \possiblyWithSub\stageOmetaColor{T'^{\superscriptO} }\) holds
                by the Barendregt convention.
                By Weakening, from \( \mathit{\Gamma}  \vdash^{0}  \possiblyWithSub\stageOmetaColor{N'^{\superscriptO} }  :  \possiblyWithSub\stageOmetaColor{T'^{\superscriptO} } \), we have
                \(  \mathit{\Gamma} ,    [  \possiblyWithSub\stageOmetaColor{N'^{\superscriptO} }  /  \possiblyWithSub\stageOmetaColor{x}  ]    \mathit{\Gamma}'    \vdash^{0}    [  \possiblyWithSub\stageOmetaColor{N'^{\superscriptO} }  /  \possiblyWithSub\stageOmetaColor{x}  ]     \possiblyWithSub\stageOmetaColor{x}    :  \possiblyWithSub\stageOmetaColor{T'^{\superscriptO} } \).
              \item If \(\possiblyWithSub\stageOmetaColor{x'} \in \dom \mathit{\Gamma}\):
                We have \(\mathit{\Gamma}(\possiblyWithSub\stageOmetaColor{x'}) = (\possiblyWithSub\stageOmetaColor{T^{\superscriptO} })^{0}\), and
                \(  [  \possiblyWithSub\stageOmetaColor{N'^{\superscriptO} }  /  \possiblyWithSub\stageOmetaColor{x}  ]    \possiblyWithSub\stageOmetaColor{T^{\superscriptO} }  = \possiblyWithSub\stageOmetaColor{T^{\superscriptO} }\) holds by the Barendregt convention.
                Since we clearly have \( \vdash  \mathit{\Gamma} \) from \( \vdash    \mathit{\Gamma} ,  \possiblyWithSub\stageOmetaColor{x}  : ( \possiblyWithSub\stageOmetaColor{T'^{\superscriptO} } )^{0}  ,  \mathit{\Gamma}'  \),
                we can derive
                \begin{center}
                  \infer[T0-Var]{%
                       \vdash  \mathit{\Gamma} 
                  \andalso
                    \mathit{\Gamma}(\possiblyWithSub\stageOmetaColor{x'}) = (\possiblyWithSub\stageOmetaColor{T^{\superscriptO} })^{0}
                  }{%
                     \mathit{\Gamma}  \vdash^{0}   \possiblyWithSub\stageOmetaColor{x'}   :  \possiblyWithSub\stageOmetaColor{T^{\superscriptO} } 
                  }.
                \end{center}
                Thus, by Weakening, we have \(  \mathit{\Gamma} ,    [  \possiblyWithSub\stageOmetaColor{N'^{\superscriptO} }  /  \possiblyWithSub\stageOmetaColor{x}  ]    \mathit{\Gamma}'    \vdash^{0}   \possiblyWithSub\stageOmetaColor{x'}   :  \possiblyWithSub\stageOmetaColor{T^{\superscriptO} } \).
            \end{itemize}
          \item Case \derive[T0-Abs]{%
               \mathit{\Gamma} ,  \possiblyWithSub\stageOmetaColor{x}  : ( \possiblyWithSub\stageOmetaColor{T'^{\superscriptO} } )^{0}  ,  \mathit{\Gamma}'   \vdash^{0}  \possiblyWithSub\stageOmetaColor{T^{\superscriptO} }_{{\mathrm{1}}} 
          \\
                \mathit{\Gamma} ,  \possiblyWithSub\stageOmetaColor{x}  : ( \possiblyWithSub\stageOmetaColor{T'^{\superscriptO} } )^{0}  ,  \mathit{\Gamma}'  ,  \possiblyWithSub\stageOmetaColor{x'}  : ( \possiblyWithSub\stageOmetaColor{T^{\superscriptO} }_{{\mathrm{1}}} )^{0}   \vdash^{0}  \possiblyWithSub\stageOmetaColor{N^{\superscriptO} }_{{\mathrm{2}}}  :  \possiblyWithSub\stageOmetaColor{T^{\superscriptO} }_{{\mathrm{2}}} 
          }{%
               \mathit{\Gamma} ,  \possiblyWithSub\stageOmetaColor{x}  : ( \possiblyWithSub\stageOmetaColor{T'^{\superscriptO} } )^{0}  ,  \mathit{\Gamma}'   \vdash^{0}   \openO{(}  \ordO{\lambda} \possiblyWithSub\stageOmetaColor{x'}  \relO{:}  \possiblyWithSub\stageOmetaColor{T^{\superscriptO} }_{{\mathrm{1}}} \punctO{.}\  \possiblyWithSub\stageOmetaColor{N^{\superscriptO} }_{{\mathrm{2}}}  \closeO{)}   :   \openO{(} \possiblyWithSub\stageOmetaColor{x'}  \relO{:}  \possiblyWithSub\stageOmetaColor{T^{\superscriptO} }_{{\mathrm{1}}} \closeO{)} \relO{\to}  \possiblyWithSub\stageOmetaColor{T^{\superscriptO} }_{{\mathrm{2}}}  
          }:
            By the Barendregt convention, w.l.o.g., we can assume \(\possiblyWithSub\stageOmetaColor{x'} \neq \possiblyWithSub\stageOmetaColor{x}\).
            By IH, from \(   \mathit{\Gamma} ,  \possiblyWithSub\stageOmetaColor{x}  : ( \possiblyWithSub\stageOmetaColor{T'^{\superscriptO} } )^{0}  ,  \mathit{\Gamma}'   \vdash^{0}  \possiblyWithSub\stageOmetaColor{T^{\superscriptO} }_{{\mathrm{1}}} \) and
            \(    \mathit{\Gamma} ,  \possiblyWithSub\stageOmetaColor{x}  : ( \possiblyWithSub\stageOmetaColor{T'^{\superscriptO} } )^{0}  ,  \mathit{\Gamma}'  ,  \possiblyWithSub\stageOmetaColor{x'}  : ( \possiblyWithSub\stageOmetaColor{T^{\superscriptO} }_{{\mathrm{1}}} )^{0}   \vdash^{0}  \possiblyWithSub\stageOmetaColor{N^{\superscriptO} }_{{\mathrm{2}}}  :  \possiblyWithSub\stageOmetaColor{T^{\superscriptO} }_{{\mathrm{2}}} \),
            we respectively have
            \(  \mathit{\Gamma} ,    [  \possiblyWithSub\stageOmetaColor{N'^{\superscriptO} }  /  \possiblyWithSub\stageOmetaColor{x}  ]    \mathit{\Gamma}'    \vdash^{0}    [  \possiblyWithSub\stageOmetaColor{N'^{\superscriptO} }  /  \possiblyWithSub\stageOmetaColor{x}  ]    \possiblyWithSub\stageOmetaColor{T^{\superscriptO} }_{{\mathrm{1}}}  \) and
            \(  \mathit{\Gamma} ,    [  \possiblyWithSub\stageOmetaColor{N'^{\superscriptO} }  /  \possiblyWithSub\stageOmetaColor{x}  ]     (  \mathit{\Gamma}' ,  \possiblyWithSub\stageOmetaColor{x'}  : ( \possiblyWithSub\stageOmetaColor{T^{\superscriptO} }_{{\mathrm{1}}} )^{0}  )     \vdash^{0}    [  \possiblyWithSub\stageOmetaColor{N'^{\superscriptO} }  /  \possiblyWithSub\stageOmetaColor{x}  ]    \possiblyWithSub\stageOmetaColor{N^{\superscriptO} }_{{\mathrm{2}}}   :    [  \possiblyWithSub\stageOmetaColor{N'^{\superscriptO} }  /  \possiblyWithSub\stageOmetaColor{x}  ]    \possiblyWithSub\stageOmetaColor{T^{\superscriptO} }_{{\mathrm{2}}}  \).
            Thus, we can derive
            \begin{center}
              \derive[T0-Abs]{%
                  \mathit{\Gamma} ,    [  \possiblyWithSub\stageOmetaColor{N'^{\superscriptO} }  /  \possiblyWithSub\stageOmetaColor{x}  ]    \mathit{\Gamma}'    \vdash^{0}    [  \possiblyWithSub\stageOmetaColor{N'^{\superscriptO} }  /  \possiblyWithSub\stageOmetaColor{x}  ]    \possiblyWithSub\stageOmetaColor{T^{\superscriptO} }_{{\mathrm{1}}}  
              \\
                   \mathit{\Gamma} ,    [  \possiblyWithSub\stageOmetaColor{N'^{\superscriptO} }  /  \possiblyWithSub\stageOmetaColor{x}  ]    \mathit{\Gamma}'   ,  \possiblyWithSub\stageOmetaColor{x'}  : (   [  \possiblyWithSub\stageOmetaColor{N'^{\superscriptO} }  /  \possiblyWithSub\stageOmetaColor{x}  ]    \possiblyWithSub\stageOmetaColor{T^{\superscriptO} }_{{\mathrm{1}}}  )^{0}   \vdash^{0}    [  \possiblyWithSub\stageOmetaColor{N'^{\superscriptO} }  /  \possiblyWithSub\stageOmetaColor{x}  ]    \possiblyWithSub\stageOmetaColor{N^{\superscriptO} }_{{\mathrm{2}}}   :    [  \possiblyWithSub\stageOmetaColor{N'^{\superscriptO} }  /  \possiblyWithSub\stageOmetaColor{x}  ]    \possiblyWithSub\stageOmetaColor{T^{\superscriptO} }_{{\mathrm{2}}}  
              }{%
                  \mathit{\Gamma} ,    [  \possiblyWithSub\stageOmetaColor{N'^{\superscriptO} }  /  \possiblyWithSub\stageOmetaColor{x}  ]    \mathit{\Gamma}'    \vdash^{0}    [  \possiblyWithSub\stageOmetaColor{N'^{\superscriptO} }  /  \possiblyWithSub\stageOmetaColor{x}  ]     \openO{(}  \ordO{\lambda} \possiblyWithSub\stageOmetaColor{x'}  \relO{:}  \possiblyWithSub\stageOmetaColor{T^{\superscriptO} }_{{\mathrm{1}}} \punctO{.}\  \possiblyWithSub\stageOmetaColor{N^{\superscriptO} }_{{\mathrm{2}}}  \closeO{)}    :    [  \possiblyWithSub\stageOmetaColor{N'^{\superscriptO} }  /  \possiblyWithSub\stageOmetaColor{x}  ]     (  \openO{(} \possiblyWithSub\stageOmetaColor{x'}  \relO{:}  \possiblyWithSub\stageOmetaColor{T^{\superscriptO} }_{{\mathrm{1}}} \closeO{)} \relO{\to}  \possiblyWithSub\stageOmetaColor{T^{\superscriptO} }_{{\mathrm{2}}}  )   
              }
            \end{center}
          \item The other cases are straightforward.
        \end{itemize}
      \item
        \begin{itemize}
          \item Case \rulename{T1-Var}: Immediate since substitution never happens.
          \item Case \rulename{T1-Abs}: Done in a way similar to \rulename{T0-Abs}.
          \item The other cases are straightforward.
        \end{itemize}
      \item
        \begin{itemize}
          \item Case \derive[Wf0-Arr]{%
               \mathit{\Gamma} ,  \possiblyWithSub\stageOmetaColor{x}  : ( \possiblyWithSub\stageOmetaColor{T'^{\superscriptO} } )^{0}  ,  \mathit{\Gamma}'   \vdash^{0}  \possiblyWithSub\stageOmetaColor{T^{\superscriptO} }_{{\mathrm{1}}} 
          \andalso
                \mathit{\Gamma} ,  \possiblyWithSub\stageOmetaColor{x}  : ( \possiblyWithSub\stageOmetaColor{T'^{\superscriptO} } )^{0}  ,  \mathit{\Gamma}'  ,  \possiblyWithSub\stageOmetaColor{x'}  : ( \possiblyWithSub\stageOmetaColor{T^{\superscriptO} }_{{\mathrm{1}}} )^{0}   \vdash^{0}  \possiblyWithSub\stageOmetaColor{T^{\superscriptO} }_{{\mathrm{2}}} 
          }{%
               \mathit{\Gamma} ,  \possiblyWithSub\stageOmetaColor{x}  : ( \possiblyWithSub\stageOmetaColor{T'^{\superscriptO} } )^{0}  ,  \mathit{\Gamma}'   \vdash^{0}   \openO{(} \possiblyWithSub\stageOmetaColor{x'}  \relO{:}  \possiblyWithSub\stageOmetaColor{T^{\superscriptO} }_{{\mathrm{1}}} \closeO{)} \relO{\to}  \possiblyWithSub\stageOmetaColor{T^{\superscriptO} }_{{\mathrm{2}}}  
          }:
            By IH, from \(   \mathit{\Gamma} ,  \possiblyWithSub\stageOmetaColor{x}  : ( \possiblyWithSub\stageOmetaColor{T'^{\superscriptO} } )^{0}  ,  \mathit{\Gamma}'   \vdash^{0}  \possiblyWithSub\stageOmetaColor{T^{\superscriptO} }_{{\mathrm{1}}} \)
            and \(    \mathit{\Gamma} ,  \possiblyWithSub\stageOmetaColor{x}  : ( \possiblyWithSub\stageOmetaColor{T'^{\superscriptO} } )^{0}  ,  \mathit{\Gamma}'  ,  \possiblyWithSub\stageOmetaColor{x'}  : ( \possiblyWithSub\stageOmetaColor{T^{\superscriptO} }_{{\mathrm{1}}} )^{0}   \vdash^{0}  \possiblyWithSub\stageOmetaColor{T^{\superscriptO} }_{{\mathrm{2}}} \),
            we respectively have \(  \mathit{\Gamma} ,  \mathit{\Gamma}'   \vdash^{0}    [  \possiblyWithSub\stageOmetaColor{N'^{\superscriptO} }  /  \possiblyWithSub\stageOmetaColor{x}  ]    \possiblyWithSub\stageOmetaColor{T^{\superscriptO} }_{{\mathrm{1}}}  \)
            and \(   \mathit{\Gamma} ,  \mathit{\Gamma}'  ,  \possiblyWithSub\stageOmetaColor{x'}  : ( \possiblyWithSub\stageOmetaColor{T^{\superscriptO} }_{{\mathrm{1}}} )^{0}   \vdash^{0}    [  \possiblyWithSub\stageOmetaColor{N'^{\superscriptO} }  /  \possiblyWithSub\stageOmetaColor{x}  ]    \possiblyWithSub\stageOmetaColor{T^{\superscriptO} }_{{\mathrm{2}}}  \).
            Thus, we can derive
            \begin{center}
              \derive[Wf0-Arr]{%
                  \mathit{\Gamma} ,  \mathit{\Gamma}'   \vdash^{0}    [  \possiblyWithSub\stageOmetaColor{N'^{\superscriptO} }  /  \possiblyWithSub\stageOmetaColor{x}  ]    \possiblyWithSub\stageOmetaColor{T^{\superscriptO} }_{{\mathrm{1}}}  
              \andalso
                   \mathit{\Gamma} ,  \mathit{\Gamma}'  ,  \possiblyWithSub\stageOmetaColor{x'}  : ( \possiblyWithSub\stageOmetaColor{T^{\superscriptO} }_{{\mathrm{1}}} )^{0}   \vdash^{0}    [  \possiblyWithSub\stageOmetaColor{N'^{\superscriptO} }  /  \possiblyWithSub\stageOmetaColor{x}  ]    \possiblyWithSub\stageOmetaColor{T^{\superscriptO} }_{{\mathrm{2}}}  
              }{%
                  \mathit{\Gamma} ,  \mathit{\Gamma}'   \vdash^{0}   \openO{(} \possiblyWithSub\stageOmetaColor{x'}  \relO{:}    [  \possiblyWithSub\stageOmetaColor{N'^{\superscriptO} }  /  \possiblyWithSub\stageOmetaColor{x}  ]    \possiblyWithSub\stageOmetaColor{T^{\superscriptO} }_{{\mathrm{1}}}  \closeO{)} \relO{\to}    [  \possiblyWithSub\stageOmetaColor{N'^{\superscriptO} }  /  \possiblyWithSub\stageOmetaColor{x}  ]    \possiblyWithSub\stageOmetaColor{T^{\superscriptO} }_{{\mathrm{2}}}   
              }
            \end{center}
          \item The other cases are straightforward.
        \end{itemize}
      \item
        \begin{itemize}
          \item Case \derive[WfT1-Tensor]{%
               \mathit{\Gamma} ,  \possiblyWithSub\stageOmetaColor{x}  : ( \possiblyWithSub\stageOmetaColor{T'^{\superscriptO} } )^{0}  ,  \mathit{\Gamma}'   \vdash^{0}  \possiblyWithSub\stageOmetaColor{N^{\superscriptO} }  :    \openO{\{} \possiblyWithSub\stageOmetaColor{\nu}  \relO{:}   \ttO{NatList}   \relO{\mid}     \ttO{true}    \closeO{\} }   
          }{%
               \mathit{\Gamma} ,  \possiblyWithSub\stageOmetaColor{x}  : ( \possiblyWithSub\stageOmetaColor{T'^{\superscriptO} } )^{0}  ,  \mathit{\Gamma}'   \vdash^{1}   \ttI{Tensor}\ \ordI{\%} \possiblyWithSub\stageOmetaColor{N^{\superscriptO} }  
          }:
            By IH, from \(   \mathit{\Gamma} ,  \possiblyWithSub\stageOmetaColor{x}  : ( \possiblyWithSub\stageOmetaColor{T'^{\superscriptO} } )^{0}  ,  \mathit{\Gamma}'   \vdash^{0}  \possiblyWithSub\stageOmetaColor{N^{\superscriptO} }  :    \openO{\{} \possiblyWithSub\stageOmetaColor{\nu}  \relO{:}   \ttO{NatList}   \relO{\mid}     \ttO{true}    \closeO{\} }   \),
            we have \(  \mathit{\Gamma} ,    [  \possiblyWithSub\stageOmetaColor{N'^{\superscriptO} }  /  \possiblyWithSub\stageOmetaColor{x}  ]    \mathit{\Gamma}'    \vdash^{0}    [  \possiblyWithSub\stageOmetaColor{N'^{\superscriptO} }  /  \possiblyWithSub\stageOmetaColor{x}  ]    \possiblyWithSub\stageOmetaColor{N^{\superscriptO} }   :    [  \possiblyWithSub\stageOmetaColor{N'^{\superscriptO} }  /  \possiblyWithSub\stageOmetaColor{x}  ]      \openO{\{} \possiblyWithSub\stageOmetaColor{\nu}  \relO{:}   \ttO{NatList}   \relO{\mid}     \ttO{true}    \closeO{\} }    \).
            This enables us to derive
            \begin{center}
              \derive[WfT1-Tensor]{%
                  \mathit{\Gamma} ,    [  \possiblyWithSub\stageOmetaColor{N'^{\superscriptO} }  /  \possiblyWithSub\stageOmetaColor{x}  ]    \mathit{\Gamma}'    \vdash^{0}    [  \possiblyWithSub\stageOmetaColor{N'^{\superscriptO} }  /  \possiblyWithSub\stageOmetaColor{x}  ]    \possiblyWithSub\stageOmetaColor{N^{\superscriptO} }   :    \openO{\{} \possiblyWithSub\stageOmetaColor{\nu}  \relO{:}   \ttO{NatList}   \relO{\mid}     \ttO{true}    \closeO{\} }   
              }{%
                  \mathit{\Gamma} ,    [  \possiblyWithSub\stageOmetaColor{N'^{\superscriptO} }  /  \possiblyWithSub\stageOmetaColor{x}  ]    \mathit{\Gamma}'    \vdash^{1}   \ttI{Tensor}\ \ordI{\%}  \openO{(}   [  \possiblyWithSub\stageOmetaColor{N'^{\superscriptO} }  /  \possiblyWithSub\stageOmetaColor{x}  ]    \possiblyWithSub\stageOmetaColor{N^{\superscriptO} }  \closeO{)}   
              }.
            \end{center}
          \item The other cases are straightforward.
        \end{itemize}
      \item
        \begin{itemize}
          \item Case \derive[WfEnv-Cons0]{%
             \vdash    \mathit{\Gamma} ,  \possiblyWithSub\stageOmetaColor{x}  : ( \possiblyWithSub\stageOmetaColor{T'^{\superscriptO} } )^{0}  ,  \mathit{\Gamma}''  
          \andalso
               \mathit{\Gamma} ,  \possiblyWithSub\stageOmetaColor{x}  : ( \possiblyWithSub\stageOmetaColor{T'^{\superscriptO} } )^{0}  ,  \mathit{\Gamma}''   \vdash^{0}  \possiblyWithSub\stageOmetaColor{T^{\superscriptO} } 
          }{%
             \vdash     \mathit{\Gamma} ,  \possiblyWithSub\stageOmetaColor{x}  : ( \possiblyWithSub\stageOmetaColor{T'^{\superscriptO} } )^{0}  ,  \mathit{\Gamma}''  ,  \possiblyWithSub\stageOmetaColor{x'}  : ( \possiblyWithSub\stageOmetaColor{T^{\superscriptO} } )^{0}  
          }:
            W.l.o.g., we can assume \(\possiblyWithSub\stageOmetaColor{x'} \neq \possiblyWithSub\stageOmetaColor{x}\) by the Barendregt convention.
            By IH, from \( \vdash    \mathit{\Gamma} ,  \possiblyWithSub\stageOmetaColor{x}  : ( \possiblyWithSub\stageOmetaColor{T'^{\superscriptO} } )^{0}  ,  \mathit{\Gamma}''  \)
            and \(   \mathit{\Gamma} ,  \possiblyWithSub\stageOmetaColor{x}  : ( \possiblyWithSub\stageOmetaColor{T'^{\superscriptO} } )^{0}  ,  \mathit{\Gamma}''   \vdash^{0}  \possiblyWithSub\stageOmetaColor{T^{\superscriptO} } \),
            we respectively have \( \vdash   \mathit{\Gamma} ,    [  \possiblyWithSub\stageOmetaColor{N'^{\superscriptO} }  /  \possiblyWithSub\stageOmetaColor{x}  ]    \mathit{\Gamma}''   \)
            and \(  \mathit{\Gamma} ,    [  \possiblyWithSub\stageOmetaColor{N'^{\superscriptO} }  /  \possiblyWithSub\stageOmetaColor{x}  ]    \mathit{\Gamma}''    \vdash^{0}    [  \possiblyWithSub\stageOmetaColor{N'^{\superscriptO} }  /  \possiblyWithSub\stageOmetaColor{x}  ]    \possiblyWithSub\stageOmetaColor{T^{\superscriptO} }  \).
            Thus, we can derive
            \begin{center}
              \derive[WfEnv-Cons0]{%
                 \vdash   \mathit{\Gamma} ,    [  \possiblyWithSub\stageOmetaColor{N'^{\superscriptO} }  /  \possiblyWithSub\stageOmetaColor{x}  ]    \mathit{\Gamma}''   
              \andalso
                  \mathit{\Gamma} ,    [  \possiblyWithSub\stageOmetaColor{N'^{\superscriptO} }  /  \possiblyWithSub\stageOmetaColor{x}  ]    \mathit{\Gamma}''    \vdash^{0}    [  \possiblyWithSub\stageOmetaColor{N'^{\superscriptO} }  /  \possiblyWithSub\stageOmetaColor{x}  ]    \possiblyWithSub\stageOmetaColor{T^{\superscriptO} }  
              }{%
                 \vdash    \mathit{\Gamma} ,    [  \possiblyWithSub\stageOmetaColor{N'^{\superscriptO} }  /  \possiblyWithSub\stageOmetaColor{x}  ]    \mathit{\Gamma}''   ,  \possiblyWithSub\stageOmetaColor{x'}  : (   [  \possiblyWithSub\stageOmetaColor{N'^{\superscriptO} }  /  \possiblyWithSub\stageOmetaColor{x}  ]    \possiblyWithSub\stageOmetaColor{T^{\superscriptO} }  )^{0}  
              }.
            \end{center}
          \item The other cases are similar.
        \end{itemize}
    \end{enumerate}
  \end{proof}
  \begin{lemma}[Target typing synthesizes only well-formed types]\label{lem:target-typing-implies-type-well-formedness}
    \noindent
    \begin{enumerate}
      \item \( \mathit{\Gamma}  \vdash^{0}  \possiblyWithSub\stageOmetaColor{N^{\superscriptO} }  :  \possiblyWithSub\stageOmetaColor{T^{\superscriptO} } \) implies \( \mathit{\Gamma}  \vdash^{0}  \possiblyWithSub\stageOmetaColor{T^{\superscriptO} } \).
      \item \( \mathit{\Gamma}  \vdash^{1}  \possiblyWithSub\stageImetaColor{N^{\superscriptI} }  :  \possiblyWithSub\stageImetaColor{T^{\superscriptI} } \) implies \( \mathit{\Gamma}  \vdash^{1}  \possiblyWithSub\stageImetaColor{T^{\superscriptI} } \).
    \end{enumerate}
  \end{lemma}
  \begin{proof}
    By induction on the derivation.
    \begin{enumerate}
      \item
        \begin{itemize}
          \item Case \rulename{T0-Cst0}:
            Straightforward from Assumption~\ref{assump:type-of-constants}.
          \item Case \derive[T0-App]{%
             \mathit{\Gamma}  \vdash^{0}  \possiblyWithSub\stageOmetaColor{N^{\superscriptO} }_{{\mathrm{1}}}  :   \openO{(} \possiblyWithSub\stageOmetaColor{x}  \relO{:}  \possiblyWithSub\stageOmetaColor{T^{\superscriptO} }_{{\mathrm{11}}} \closeO{)} \relO{\to}  \possiblyWithSub\stageOmetaColor{T^{\superscriptO} }_{{\mathrm{12}}}  
          \andalso
             \mathit{\Gamma}  \vdash^{0}  \possiblyWithSub\stageOmetaColor{N^{\superscriptO} }_{{\mathrm{2}}}  :  \possiblyWithSub\stageOmetaColor{T^{\superscriptO} }_{{\mathrm{11}}} 
          }{%
             \mathit{\Gamma}  \vdash^{0}   \possiblyWithSub\stageOmetaColor{N^{\superscriptO} }_{{\mathrm{1}}} \  \possiblyWithSub\stageOmetaColor{N^{\superscriptO} }_{{\mathrm{2}}}   :    [  \possiblyWithSub\stageOmetaColor{N^{\superscriptO} }_{{\mathrm{2}}}  /  \possiblyWithSub\stageOmetaColor{x}  ]    \possiblyWithSub\stageOmetaColor{T^{\superscriptO} }_{{\mathrm{12}}}  
          }:
            By IH, we have \( \mathit{\Gamma}  \vdash^{0}   \openO{(} \possiblyWithSub\stageOmetaColor{x}  \relO{:}  \possiblyWithSub\stageOmetaColor{T^{\superscriptO} }_{{\mathrm{11}}} \closeO{)} \relO{\to}  \possiblyWithSub\stageOmetaColor{T^{\superscriptO} }_{{\mathrm{12}}}  \) and thereby
            \(  \mathit{\Gamma} ,  \possiblyWithSub\stageOmetaColor{x}  : ( \possiblyWithSub\stageOmetaColor{T^{\superscriptO} }_{{\mathrm{11}}} )^{0}   \vdash^{0}  \possiblyWithSub\stageOmetaColor{T^{\superscriptO} }_{{\mathrm{12}}} \).
            Thus, by Lemma~\ref{lem:subst} and \( \mathit{\Gamma}  \vdash^{0}  \possiblyWithSub\stageOmetaColor{N^{\superscriptO} }_{{\mathrm{2}}}  :  \possiblyWithSub\stageOmetaColor{T^{\superscriptO} }_{{\mathrm{11}}} \),
            we have \( \mathit{\Gamma}  \vdash^{0}    [  \possiblyWithSub\stageOmetaColor{N^{\superscriptO} }_{{\mathrm{2}}}  /  \possiblyWithSub\stageOmetaColor{x}  ]    \possiblyWithSub\stageOmetaColor{T^{\superscriptO} }_{{\mathrm{12}}}  \).
          \item
            The other cases are straightforward.
        \end{itemize}
      \item
        \begin{itemize}
          \item Case \derive[T1-Abs]{%
             \mathit{\Gamma}  \vdash^{1}  \possiblyWithSub\stageImetaColor{T^{\superscriptI} }_{{\mathrm{1}}} 
          \andalso
              \mathit{\Gamma} ,  \possiblyWithSub\stageImetaColor{x}  : ( \possiblyWithSub\stageImetaColor{T^{\superscriptI} }_{{\mathrm{1}}} )^{1}   \vdash^{1}  \possiblyWithSub\stageImetaColor{N^{\superscriptI} }_{{\mathrm{2}}}  :  \possiblyWithSub\stageImetaColor{T^{\superscriptI} }_{{\mathrm{2}}} 
          \andalso
            \possiblyWithSub\stageImetaColor{x} \not\in \fv(\possiblyWithSub\stageImetaColor{T^{\superscriptI} }_{{\mathrm{2}}})
          }{%
             \mathit{\Gamma}  \vdash^{1}   \openI{(}  \ordI{\lambda} \possiblyWithSub\stageImetaColor{x}  \relI{:}  \possiblyWithSub\stageImetaColor{T^{\superscriptI} }_{{\mathrm{1}}} \punctI{.}\  \possiblyWithSub\stageImetaColor{N^{\superscriptI} }_{{\mathrm{2}}}  \closeI{)}   :   \possiblyWithSub\stageImetaColor{T^{\superscriptI} }_{{\mathrm{1}}}  \relI{\to}  \possiblyWithSub\stageImetaColor{T^{\superscriptI} }_{{\mathrm{2}}}  
          }:
            By IH, we have \(  \mathit{\Gamma} ,  \possiblyWithSub\stageImetaColor{x}  : ( \possiblyWithSub\stageImetaColor{T^{\superscriptI} }_{{\mathrm{1}}} )^{1}   \vdash^{1}  \possiblyWithSub\stageImetaColor{T^{\superscriptI} }_{{\mathrm{2}}} \).
            Since \(\possiblyWithSub\stageImetaColor{x} \not\in \fv(\possiblyWithSub\stageImetaColor{T^{\superscriptI} }_{{\mathrm{2}}})\) holds,
            we have \( \mathit{\Gamma}  \vdash^{1}  \possiblyWithSub\stageImetaColor{T^{\superscriptI} }_{{\mathrm{2}}} \) by evident contraction.
            This enables us to derive
              \derive[Wf1-Arr]{%
                 \mathit{\Gamma}  \vdash^{1}  \possiblyWithSub\stageImetaColor{T^{\superscriptI} }_{{\mathrm{1}}} 
              \andalso
                 \mathit{\Gamma}  \vdash^{1}  \possiblyWithSub\stageImetaColor{T^{\superscriptI} }_{{\mathrm{2}}} 
              }{%
                 \mathit{\Gamma}  \vdash^{1}   \possiblyWithSub\stageImetaColor{T^{\superscriptI} }_{{\mathrm{1}}}  \relI{\to}  \possiblyWithSub\stageImetaColor{T^{\superscriptI} }_{{\mathrm{2}}}  
              }.
          \item
            The other cases are all straightforward.
        \end{itemize}
    \end{enumerate}
  \end{proof}
  \recalltheorem[Soundness of Assertion Insertion]{thm:soundness-of-elaboration}{%
    Suppose \( \vdash  \mathit{\Gamma} \).
    \begin{enumerate}
      \item If \( \mathit{\Gamma}  \vdash^{0}  \possiblyWithSub\stageOmetaColor{M^{\scriptscriptstyle(0)} }  :  \possiblyWithSub\stageOmetaColor{T^{\superscriptO} }  \ElabArrow  \possiblyWithSub\stageOmetaColor{N^{\superscriptO} } \), then \( \mathit{\Gamma}  \vdash^{0}  \possiblyWithSub\stageOmetaColor{N^{\superscriptO} }  :  \possiblyWithSub\stageOmetaColor{T^{\superscriptO} } \).
      \item If \( \mathit{\Gamma}  \vdash^{1}  \possiblyWithSub\stageImetaColor{M^{\superscriptI} }  :  \possiblyWithSub\stageImetaColor{T^{\superscriptI} }  \ElabArrow  \possiblyWithSub\stageImetaColor{N^{\superscriptI} } \), then \( \mathit{\Gamma}  \vdash^{1}  \possiblyWithSub\stageImetaColor{N^{\superscriptI} }  :  \possiblyWithSub\stageImetaColor{T^{\superscriptI} } \).
      \item If \( \mathit{\Gamma}  \vdash^{0}  \possiblyWithSub\stageOmetaColor{S^{\superscriptO} }_{{\mathrm{1}}}  \ElabArrow  \possiblyWithSub\stageOmetaColor{T^{\superscriptO} }_{{\mathrm{1}}} \), then \( \mathit{\Gamma}  \vdash^{0}  \possiblyWithSub\stageOmetaColor{T^{\superscriptO} }_{{\mathrm{1}}} \).
      \item If \( \mathit{\Gamma}  \vdash^{1}  \possiblyWithSub\stageImetaColor{S^{\superscriptI} }_{{\mathrm{1}}}  \ElabArrow  \possiblyWithSub\stageImetaColor{T^{\superscriptI} }_{{\mathrm{1}}} \), then \( \mathit{\Gamma}  \vdash^{1}  \possiblyWithSub\stageImetaColor{T^{\superscriptI} }_{{\mathrm{1}}} \).
    \end{enumerate}
  }
  \begin{proof}
    By mutual induction on the derivation.
    \begin{enumerate}
      \item
        \begin{itemize}
          \item Case \derive[S0-App]{%
             \mathit{\Gamma}  \vdash^{0}  \possiblyWithSub\stageOmetaColor{M^{\scriptscriptstyle(0)} }_{{\mathrm{1}}}  :   \openO{(} \possiblyWithSub\stageOmetaColor{x}  \relO{:}  \possiblyWithSub\stageOmetaColor{T^{\superscriptO} }_{{\mathrm{11}}} \closeO{)} \relO{\to}  \possiblyWithSub\stageOmetaColor{T^{\superscriptO} }_{{\mathrm{12}}}   \ElabArrow  \possiblyWithSub\stageOmetaColor{N^{\superscriptO} }_{{\mathrm{1}}} 
          \\
             \mathit{\Gamma}  \vdash^{0}  \possiblyWithSub\stageOmetaColor{M^{\scriptscriptstyle(0)} }_{{\mathrm{2}}}  :  \possiblyWithSub\stageOmetaColor{T^{\superscriptO} }_{{\mathrm{2}}}  \ElabArrow  \possiblyWithSub\stageOmetaColor{N^{\superscriptO} }_{{\mathrm{2}}} 
          \andalso
             \mathit{\Gamma}  \vdash_{  \ell  }  \possiblyWithSub\stageOmetaColor{T^{\superscriptO} }_{{\mathrm{2}}}  \CastArrow  \possiblyWithSub\stageOmetaColor{T^{\superscriptO} }_{{\mathrm{11}}}  \ElabArrow  \possiblyWithSub\stageOmetaColor{N^{\superscriptO} }_{{\mathrm{0}}} 
          }{%
             \mathit{\Gamma}  \vdash^{0}   \openO{(} \possiblyWithSub\stageOmetaColor{M^{\scriptscriptstyle(0)} }_{{\mathrm{1}}} \  \possiblyWithSub\stageOmetaColor{M^{\scriptscriptstyle(0)} }_{{\mathrm{2}}} \closeO{)}_{ \ell }   :    [   \possiblyWithSub\stageOmetaColor{N^{\superscriptO} }_{{\mathrm{0}}} \  \possiblyWithSub\stageOmetaColor{N^{\superscriptO} }_{{\mathrm{2}}}   /  \possiblyWithSub\stageOmetaColor{x}  ]    \possiblyWithSub\stageOmetaColor{T^{\superscriptO} }_{{\mathrm{12}}}   \ElabArrow   \possiblyWithSub\stageOmetaColor{N^{\superscriptO} }_{{\mathrm{1}}} \   \openO{(}  \possiblyWithSub\stageOmetaColor{N^{\superscriptO} }_{{\mathrm{0}}} \  \possiblyWithSub\stageOmetaColor{N^{\superscriptO} }_{{\mathrm{2}}}  \closeO{)}   
          }:
            First, by IH, we have \( \mathit{\Gamma}  \vdash^{0}  \possiblyWithSub\stageOmetaColor{N^{\superscriptO} }_{{\mathrm{1}}}  :   \openO{(} \possiblyWithSub\stageOmetaColor{x}  \relO{:}  \possiblyWithSub\stageOmetaColor{T^{\superscriptO} }_{{\mathrm{11}}} \closeO{)} \relO{\to}  \possiblyWithSub\stageOmetaColor{T^{\superscriptO} }_{{\mathrm{12}}}  \)
            and \( \mathit{\Gamma}  \vdash^{0}  \possiblyWithSub\stageOmetaColor{N^{\superscriptO} }_{{\mathrm{2}}}  :  \possiblyWithSub\stageOmetaColor{T^{\superscriptO} }_{{\mathrm{2}}} \).
            By Lemma~\ref{lem:target-typing-implies-type-well-formedness},
            we also have \( \mathit{\Gamma}  \vdash^{0}   \openO{(} \possiblyWithSub\stageOmetaColor{x}  \relO{:}  \possiblyWithSub\stageOmetaColor{T^{\superscriptO} }_{{\mathrm{11}}} \closeO{)} \relO{\to}  \possiblyWithSub\stageOmetaColor{T^{\superscriptO} }_{{\mathrm{12}}}  \) (and thereby \( \mathit{\Gamma}  \vdash^{0}  \possiblyWithSub\stageOmetaColor{T^{\superscriptO} }_{{\mathrm{11}}} \))
            and \( \mathit{\Gamma}  \vdash^{0}  \possiblyWithSub\stageOmetaColor{T^{\superscriptO} }_{{\mathrm{2}}} \).
            Then, by Lemma~\ref{lem:cast-soundness}, from \( \mathit{\Gamma}  \vdash_{  \ell  }  \possiblyWithSub\stageOmetaColor{T^{\superscriptO} }_{{\mathrm{2}}}  \CastArrow  \possiblyWithSub\stageOmetaColor{T^{\superscriptO} }_{{\mathrm{11}}}  \ElabArrow  \possiblyWithSub\stageOmetaColor{N^{\superscriptO} }_{{\mathrm{0}}} \),
            we have \( \mathit{\Gamma}  \vdash^{0}  \possiblyWithSub\stageOmetaColor{N^{\superscriptO} }_{{\mathrm{0}}}  :   \openO{(} \possiblyWithSub\stageOmetaColor{x'}  \relO{:}  \possiblyWithSub\stageOmetaColor{T^{\superscriptO} }_{{\mathrm{2}}} \closeO{)} \relO{\to}  \possiblyWithSub\stageOmetaColor{T^{\superscriptO} }_{{\mathrm{11}}}  \)
            for some \(\possiblyWithSub\stageOmetaColor{x'} \not\in \fv(\possiblyWithSub\stageOmetaColor{T^{\superscriptO} }_{{\mathrm{11}}})\).
            Since \(  [  \possiblyWithSub\stageOmetaColor{N^{\superscriptO} }_{{\mathrm{2}}}  /  \possiblyWithSub\stageOmetaColor{x'}  ]    \possiblyWithSub\stageOmetaColor{T^{\superscriptO} }_{{\mathrm{11}}}  = \possiblyWithSub\stageOmetaColor{T^{\superscriptO} }_{{\mathrm{11}}}\), we can derive:
            \begin{center}
              \infer[T0-App]{%
                 \mathit{\Gamma}  \vdash^{0}  \possiblyWithSub\stageOmetaColor{N^{\superscriptO} }_{{\mathrm{1}}}  :   \openO{(} \possiblyWithSub\stageOmetaColor{x}  \relO{:}  \possiblyWithSub\stageOmetaColor{T^{\superscriptO} }_{{\mathrm{11}}} \closeO{)} \relO{\to}  \possiblyWithSub\stageOmetaColor{T^{\superscriptO} }_{{\mathrm{12}}}  
              \andalso
                \infer[T0-App]{%
                   \mathit{\Gamma}  \vdash^{0}  \possiblyWithSub\stageOmetaColor{N^{\superscriptO} }_{{\mathrm{0}}}  :   \openO{(} \possiblyWithSub\stageOmetaColor{x'}  \relO{:}  \possiblyWithSub\stageOmetaColor{T^{\superscriptO} }_{{\mathrm{2}}} \closeO{)} \relO{\to}  \possiblyWithSub\stageOmetaColor{T^{\superscriptO} }_{{\mathrm{11}}}  
                \andalso
                   \mathit{\Gamma}  \vdash^{0}  \possiblyWithSub\stageOmetaColor{N^{\superscriptO} }_{{\mathrm{2}}}  :  \possiblyWithSub\stageOmetaColor{T^{\superscriptO} }_{{\mathrm{2}}} 
                }{
                   \mathit{\Gamma}  \vdash^{0}   \possiblyWithSub\stageOmetaColor{N^{\superscriptO} }_{{\mathrm{0}}} \  \possiblyWithSub\stageOmetaColor{N^{\superscriptO} }_{{\mathrm{2}}}   :  \possiblyWithSub\stageOmetaColor{T^{\superscriptO} }_{{\mathrm{11}}} 
                }
              }{%
                 \mathit{\Gamma}  \vdash^{0}   \possiblyWithSub\stageOmetaColor{N^{\superscriptO} }_{{\mathrm{1}}} \   \openO{(}  \possiblyWithSub\stageOmetaColor{N^{\superscriptO} }_{{\mathrm{0}}} \  \possiblyWithSub\stageOmetaColor{N^{\superscriptO} }_{{\mathrm{2}}}  \closeO{)}    :    [   \possiblyWithSub\stageOmetaColor{N^{\superscriptO} }_{{\mathrm{0}}} \  \possiblyWithSub\stageOmetaColor{N^{\superscriptO} }_{{\mathrm{2}}}   /  \possiblyWithSub\stageOmetaColor{x}  ]    \possiblyWithSub\stageOmetaColor{T^{\superscriptO} }_{{\mathrm{12}}}  
              }
            \end{center}
          \item Case \derive[S0-Abs]{%
             \mathit{\Gamma}  \vdash^{0}  \possiblyWithSub\stageOmetaColor{S^{\superscriptO} }_{{\mathrm{1}}}  \ElabArrow  \possiblyWithSub\stageOmetaColor{T^{\superscriptO} }_{{\mathrm{1}}} 
          \andalso
              \mathit{\Gamma} ,  \possiblyWithSub\stageOmetaColor{x}  : ( \possiblyWithSub\stageOmetaColor{T^{\superscriptO} }_{{\mathrm{1}}} )^{0}   \vdash^{0}  \possiblyWithSub\stageOmetaColor{M^{\scriptscriptstyle(0)} }_{{\mathrm{2}}}  :  \possiblyWithSub\stageOmetaColor{T^{\superscriptO} }_{{\mathrm{2}}}  \ElabArrow  \possiblyWithSub\stageOmetaColor{N^{\superscriptO} }_{{\mathrm{2}}} 
          }{%
             \mathit{\Gamma}  \vdash^{0}   \openO{(}  \ordO{\lambda} \possiblyWithSub\stageOmetaColor{x}  \relO{:}  \possiblyWithSub\stageOmetaColor{S^{\superscriptO} }_{{\mathrm{1}}} \punctO{.}\  \possiblyWithSub\stageOmetaColor{M^{\scriptscriptstyle(0)} }_{{\mathrm{2}}}  \closeO{)}   :   \openO{(} \possiblyWithSub\stageOmetaColor{x}  \relO{:}  \possiblyWithSub\stageOmetaColor{T^{\superscriptO} }_{{\mathrm{1}}} \closeO{)} \relO{\to}  \possiblyWithSub\stageOmetaColor{T^{\superscriptO} }_{{\mathrm{2}}}   \ElabArrow   \openO{(}  \ordO{\lambda} \possiblyWithSub\stageOmetaColor{x}  \relO{:}  \possiblyWithSub\stageOmetaColor{T^{\superscriptO} }_{{\mathrm{1}}} \punctO{.}\  \possiblyWithSub\stageOmetaColor{N^{\superscriptO} }_{{\mathrm{2}}}  \closeO{)}  
          }:
            By Lemma~\ref{lem:target-typing-implies-type-well-formedness},
            from \( \mathit{\Gamma}  \vdash^{0}  \possiblyWithSub\stageOmetaColor{S^{\superscriptO} }_{{\mathrm{1}}}  \ElabArrow  \possiblyWithSub\stageOmetaColor{T^{\superscriptO} }_{{\mathrm{1}}} \), we have \( \mathit{\Gamma}  \vdash^{0}  \possiblyWithSub\stageOmetaColor{T^{\superscriptO} }_{{\mathrm{1}}} \),
            and thereby we can derive
            \begin{center}
              \derive[WfEnv-Cons0]{%
                 \vdash  \mathit{\Gamma} 
              \andalso
                 \mathit{\Gamma}  \vdash^{0}  \possiblyWithSub\stageOmetaColor{T^{\superscriptO} } 
              }{%
                 \vdash   \mathit{\Gamma} ,  \possiblyWithSub\stageOmetaColor{x}  : ( \possiblyWithSub\stageOmetaColor{T^{\superscriptO} }_{{\mathrm{1}}} )^{0}  
              }.
            \end{center}
            Then, by IH, we have \( \mathit{\Gamma}  \vdash^{0}  \possiblyWithSub\stageOmetaColor{T^{\superscriptO} }_{{\mathrm{1}}} \) and \(  \mathit{\Gamma} ,  \possiblyWithSub\stageOmetaColor{x}  : ( \possiblyWithSub\stageOmetaColor{T^{\superscriptO} }_{{\mathrm{1}}} )^{0}   \vdash^{0}  \possiblyWithSub\stageOmetaColor{N^{\superscriptO} }_{{\mathrm{2}}}  :  \possiblyWithSub\stageOmetaColor{T^{\superscriptO} }_{{\mathrm{2}}} \),
            and these enable us to derive
            \begin{center}
              \derive[T0-Abs]{%
                 \mathit{\Gamma}  \vdash^{0}  \possiblyWithSub\stageOmetaColor{T^{\superscriptO} }_{{\mathrm{1}}} 
              \andalso
                  \mathit{\Gamma} ,  \possiblyWithSub\stageOmetaColor{x}  : ( \possiblyWithSub\stageOmetaColor{T^{\superscriptO} }_{{\mathrm{1}}} )^{0}   \vdash^{0}  \possiblyWithSub\stageOmetaColor{N^{\superscriptO} }_{{\mathrm{2}}}  :  \possiblyWithSub\stageOmetaColor{T^{\superscriptO} }_{{\mathrm{2}}} 
              }{%
                 \mathit{\Gamma}  \vdash^{0}   \openO{(}  \ordO{\lambda} \possiblyWithSub\stageOmetaColor{x}  \relO{:}  \possiblyWithSub\stageOmetaColor{T^{\superscriptO} }_{{\mathrm{1}}} \punctO{.}\  \possiblyWithSub\stageOmetaColor{N^{\superscriptO} }_{{\mathrm{2}}}  \closeO{)}   :   \openO{(} \possiblyWithSub\stageOmetaColor{x}  \relO{:}  \possiblyWithSub\stageOmetaColor{T^{\superscriptO} }_{{\mathrm{1}}} \closeO{)} \relO{\to}  \possiblyWithSub\stageOmetaColor{T^{\superscriptO} }_{{\mathrm{2}}}  
              }.
            \end{center}
          \item
            The other cases are straightforward.
        \end{itemize}
      \item
        \begin{itemize}
          \item Case \derive[S1-App]{%
             \mathit{\Gamma}  \vdash^{1}  \possiblyWithSub\stageImetaColor{M^{\superscriptI} }_{{\mathrm{1}}}  :   \possiblyWithSub\stageImetaColor{T^{\superscriptI} }_{{\mathrm{11}}}  \relI{\to}  \possiblyWithSub\stageImetaColor{T^{\superscriptI} }_{{\mathrm{12}}}   \ElabArrow  \possiblyWithSub\stageImetaColor{N^{\superscriptI} }_{{\mathrm{1}}} 
          \\
             \mathit{\Gamma}  \vdash^{1}  \possiblyWithSub\stageImetaColor{M^{\superscriptI} }_{{\mathrm{2}}}  :  \possiblyWithSub\stageImetaColor{T^{\superscriptI} }_{{\mathrm{2}}}  \ElabArrow  \possiblyWithSub\stageImetaColor{N^{\superscriptI} }_{{\mathrm{2}}} 
          \andalso
             \possiblyWithSub\stageImetaColor{T^{\superscriptI} }_{{\mathrm{2}}}  \mathrel{||}^{1}  \possiblyWithSub\stageImetaColor{T^{\superscriptI} }_{{\mathrm{11}}} 
          }{%
             \mathit{\Gamma}  \vdash^{1}   \openI{(} \possiblyWithSub\stageImetaColor{M^{\superscriptI} }_{{\mathrm{1}}} \  \possiblyWithSub\stageImetaColor{M^{\superscriptI} }_{{\mathrm{2}}} \closeI{)}_{ \ell }   :  \possiblyWithSub\stageImetaColor{T^{\superscriptI} }_{{\mathrm{12}}}  \ElabArrow   \possiblyWithSub\stageImetaColor{N^{\superscriptI} }_{{\mathrm{1}}} \   \ordI{\sim}  \openO{(}   \LeftAssertParen\openO{\langle} \possiblyWithSub\stageImetaColor{T^{\superscriptI} }_{{\mathrm{2}}} \closeO{\rangle} \relO{\CastArrow} \openO{\langle} \possiblyWithSub\stageImetaColor{T^{\superscriptI} }_{{\mathrm{11}}} \closeO{\rangle}\RightAssertParen^{  \ell  }  \   \openO{\langle} \possiblyWithSub\stageImetaColor{N^{\superscriptI} }_{{\mathrm{2}}} \closeO{\rangle}   \closeO{)}    
          }:
            First, by IH, we have \( \mathit{\Gamma}  \vdash^{1}  \possiblyWithSub\stageImetaColor{N^{\superscriptI} }_{{\mathrm{1}}}  :   \possiblyWithSub\stageImetaColor{T^{\superscriptI} }_{{\mathrm{11}}}  \relI{\to}  \possiblyWithSub\stageImetaColor{T^{\superscriptI} }_{{\mathrm{12}}}  \)
            and \( \mathit{\Gamma}  \vdash^{1}  \possiblyWithSub\stageImetaColor{N^{\superscriptI} }_{{\mathrm{2}}}  :  \possiblyWithSub\stageImetaColor{T^{\superscriptI} }_{{\mathrm{2}}} \).
            Thus, we can derive:
            \begin{center}
              \infer{%
                \infer{%
                   \mathit{\Gamma}  \vdash^{1}  \possiblyWithSub\stageImetaColor{T^{\superscriptI} }_{{\mathrm{2}}} 
                \andalso
                   \mathit{\Gamma}  \vdash^{1}  \possiblyWithSub\stageImetaColor{T^{\superscriptI} }_{{\mathrm{11}}} 
                \andalso
                   \possiblyWithSub\stageImetaColor{T^{\superscriptI} }_{{\mathrm{2}}}  \mathrel{||}^{1}  \possiblyWithSub\stageImetaColor{T^{\superscriptI} }_{{\mathrm{11}}} 
                \andalso
                  \possiblyWithSub\stageOmetaColor{x} \not\in \dom(\mathit{\Gamma})
                }{
                   \mathit{\Gamma}  \vdash^{0}   \LeftAssertParen\openO{\langle} \possiblyWithSub\stageImetaColor{T^{\superscriptI} }_{{\mathrm{2}}} \closeO{\rangle} \relO{\CastArrow} \openO{\langle} \possiblyWithSub\stageImetaColor{T^{\superscriptI} }_{{\mathrm{11}}} \closeO{\rangle}\RightAssertParen^{  \ell  }   :   \openO{(} \possiblyWithSub\stageOmetaColor{x}  \relO{:}   \openO{\langle} \possiblyWithSub\stageImetaColor{T^{\superscriptI} }_{{\mathrm{2}}} \closeO{\rangle}  \closeO{)} \relO{\to}   \openO{\langle} \possiblyWithSub\stageImetaColor{T^{\superscriptI} }_{{\mathrm{11}}} \closeO{\rangle}   
                }
              \andalso
                \infer{%
                   \mathit{\Gamma}  \vdash^{1}  \possiblyWithSub\stageImetaColor{N^{\superscriptI} }_{{\mathrm{2}}}  :  \possiblyWithSub\stageImetaColor{T^{\superscriptI} }_{{\mathrm{2}}} 
                }{
                   \mathit{\Gamma}  \vdash^{0}   \openO{\langle} \possiblyWithSub\stageImetaColor{N^{\superscriptI} }_{{\mathrm{2}}} \closeO{\rangle}   :   \openO{\langle} \possiblyWithSub\stageImetaColor{T^{\superscriptI} }_{{\mathrm{2}}} \closeO{\rangle}  
                }
              }{%
                 \mathit{\Gamma}  \vdash^{0}    \LeftAssertParen\openO{\langle} \possiblyWithSub\stageImetaColor{T^{\superscriptI} }_{{\mathrm{2}}} \closeO{\rangle} \relO{\CastArrow} \openO{\langle} \possiblyWithSub\stageImetaColor{T^{\superscriptI} }_{{\mathrm{11}}} \closeO{\rangle}\RightAssertParen^{  \ell  }  \   \openO{\langle} \possiblyWithSub\stageImetaColor{N^{\superscriptI} }_{{\mathrm{2}}} \closeO{\rangle}    :   \openO{\langle} \possiblyWithSub\stageImetaColor{T^{\superscriptI} }_{{\mathrm{11}}} \closeO{\rangle}  
              }
            \end{center}
            and then have:
            \begin{center}
              \infer[T1-App]{%
                 \mathit{\Gamma}  \vdash^{1}  \possiblyWithSub\stageImetaColor{N^{\superscriptI} }_{{\mathrm{1}}}  :   \possiblyWithSub\stageImetaColor{T^{\superscriptI} }_{{\mathrm{11}}}  \relI{\to}  \possiblyWithSub\stageImetaColor{T^{\superscriptI} }_{{\mathrm{12}}}  
              \andalso
                \infer[T1-Esc]{%
                   \mathit{\Gamma}  \vdash^{0}    \LeftAssertParen\openO{\langle} \possiblyWithSub\stageImetaColor{T^{\superscriptI} }_{{\mathrm{2}}} \closeO{\rangle} \relO{\CastArrow} \openO{\langle} \possiblyWithSub\stageImetaColor{T^{\superscriptI} }_{{\mathrm{11}}} \closeO{\rangle}\RightAssertParen^{  \ell  }  \   \openO{\langle} \possiblyWithSub\stageImetaColor{N^{\superscriptI} }_{{\mathrm{2}}} \closeO{\rangle}    :   \openO{\langle} \possiblyWithSub\stageImetaColor{T^{\superscriptI} }_{{\mathrm{11}}} \closeO{\rangle}  
                }{
                   \mathit{\Gamma}  \vdash^{1}   \ordI{\sim}  \openO{(}   \LeftAssertParen\openO{\langle} \possiblyWithSub\stageImetaColor{T^{\superscriptI} }_{{\mathrm{2}}} \closeO{\rangle} \relO{\CastArrow} \openO{\langle} \possiblyWithSub\stageImetaColor{T^{\superscriptI} }_{{\mathrm{11}}} \closeO{\rangle}\RightAssertParen^{  \ell  }  \   \openO{\langle} \possiblyWithSub\stageImetaColor{N^{\superscriptI} }_{{\mathrm{2}}} \closeO{\rangle}   \closeO{)}    :  \possiblyWithSub\stageImetaColor{T^{\superscriptI} }_{{\mathrm{11}}} 
                }
              }{%
                 \mathit{\Gamma}  \vdash^{1}   \possiblyWithSub\stageImetaColor{N^{\superscriptI} }_{{\mathrm{1}}} \   \ordI{\sim}  \openO{(}   \LeftAssertParen\openO{\langle} \possiblyWithSub\stageImetaColor{T^{\superscriptI} }_{{\mathrm{2}}} \closeO{\rangle} \relO{\CastArrow} \openO{\langle} \possiblyWithSub\stageImetaColor{T^{\superscriptI} }_{{\mathrm{11}}} \closeO{\rangle}\RightAssertParen^{  \ell  }  \   \openO{\langle} \possiblyWithSub\stageImetaColor{N^{\superscriptI} }_{{\mathrm{2}}} \closeO{\rangle}   \closeO{)}     :  \possiblyWithSub\stageImetaColor{T^{\superscriptI} }_{{\mathrm{12}}} 
              }
            \end{center}
          \item
            The other cases are straightforward.
        \end{itemize}
      \item
        \begin{itemize}
          \item Case \derive[ST0-Base]{%
              \mathit{\Gamma} ,  \possiblyWithSub\stageOmetaColor{\nu}  : (   \openO{\{} \possiblyWithSub\stageOmetaColor{\nu}_{{\mathrm{1}}}  \relO{:}  \possiblyWithSub\stageOmetaColor{B}  \relO{\mid}     \ttO{true}    \closeO{\} }   )^{0}   \vdash^{0}  \possiblyWithSub\stageOmetaColor{M^{\scriptscriptstyle(0)} }  :    \openO{\{} \possiblyWithSub\stageOmetaColor{\nu}_{{\mathrm{2}}}  \relO{:}   \ttO{Bool}   \relO{\mid}  \possiblyWithSub\stageOmetaColor{N'^{\superscriptO} } \closeO{\} }    \ElabArrow  \possiblyWithSub\stageOmetaColor{N^{\superscriptO} } 
          }{%
             \mathit{\Gamma}  \vdash^{0}    \openO{\{} \possiblyWithSub\stageOmetaColor{\nu}  \relO{:}  \possiblyWithSub\stageOmetaColor{B}  \relO{\mid}  \possiblyWithSub\stageOmetaColor{M^{\scriptscriptstyle(0)} } \closeO{\} }    \ElabArrow    \openO{\{} \possiblyWithSub\stageOmetaColor{\nu}  \relO{:}  \possiblyWithSub\stageOmetaColor{B}  \relO{\mid}  \possiblyWithSub\stageOmetaColor{N^{\superscriptO} } \closeO{\} }   
          }:
            First, we can easily derive \( \vdash   \mathit{\Gamma} ,  \possiblyWithSub\stageOmetaColor{\nu}  : (   \openO{\{} \possiblyWithSub\stageOmetaColor{\nu}_{{\mathrm{1}}}  \relO{:}  \possiblyWithSub\stageOmetaColor{B}  \relO{\mid}     \ttO{true}    \closeO{\} }   )^{0}  \).
            Then, by IH, we have \(  \mathit{\Gamma} ,  \possiblyWithSub\stageOmetaColor{\nu}  : (   \openO{\{} \possiblyWithSub\stageOmetaColor{\nu}_{{\mathrm{1}}}  \relO{:}  \possiblyWithSub\stageOmetaColor{B}  \relO{\mid}     \ttO{true}    \closeO{\} }   )^{0}   \vdash^{0}  \possiblyWithSub\stageOmetaColor{N^{\superscriptO} }  :    \openO{\{} \possiblyWithSub\stageOmetaColor{\nu}_{{\mathrm{2}}}  \relO{:}   \ttO{Bool}   \relO{\mid}  \possiblyWithSub\stageOmetaColor{N'^{\superscriptO} } \closeO{\} }   \).
            This enables us to derive
            \begin{center}
              \derive[WfT0-Rfn]{%
                  \mathit{\Gamma} ,  \possiblyWithSub\stageOmetaColor{\nu}  : (   \openO{\{} \possiblyWithSub\stageOmetaColor{\nu}_{{\mathrm{1}}}  \relO{:}  \possiblyWithSub\stageOmetaColor{B}  \relO{\mid}     \ttO{true}    \closeO{\} }   )^{0}   \vdash^{0}  \possiblyWithSub\stageOmetaColor{N^{\superscriptO} }  :    \openO{\{} \possiblyWithSub\stageOmetaColor{\nu}_{{\mathrm{2}}}  \relO{:}   \ttO{Bool}   \relO{\mid}  \possiblyWithSub\stageOmetaColor{N'^{\superscriptO} } \closeO{\} }   
              }{%
                 \mathit{\Gamma}  \vdash^{0}    \openO{\{} \possiblyWithSub\stageOmetaColor{\nu}  \relO{:}  \possiblyWithSub\stageOmetaColor{B}  \relO{\mid}  \possiblyWithSub\stageOmetaColor{N^{\superscriptO} } \closeO{\} }   
              }.
            \end{center}
          \item The other cases are straightforward.
        \end{itemize}
      \item
        \begin{itemize}
          \item Case \derive[ST1-Tensor]{%
             \mathit{\Gamma}  \vdash^{0}  \possiblyWithSub\stageOmetaColor{M^{\scriptscriptstyle(0)} }  :    \openO{\{} \possiblyWithSub\stageOmetaColor{\nu}  \relO{:}   \ttO{NatList}   \relO{\mid}  \possiblyWithSub\stageOmetaColor{N'^{\superscriptO} } \closeO{\} }    \ElabArrow  \possiblyWithSub\stageOmetaColor{N^{\superscriptO} } 
          }{%
             \mathit{\Gamma}  \vdash^{1}   \ttI{Tensor}\ \ordI{\%} \possiblyWithSub\stageOmetaColor{M^{\scriptscriptstyle(0)} }   \ElabArrow   \ttI{Tensor}\ \ordI{\%} \possiblyWithSub\stageOmetaColor{N^{\superscriptO} }  
          }:
            By IH, we have \( \mathit{\Gamma}  \vdash^{0}  \possiblyWithSub\stageOmetaColor{N^{\superscriptO} }  :    \openO{\{} \possiblyWithSub\stageOmetaColor{\nu}  \relO{:}   \ttO{NatList}   \relO{\mid}  \possiblyWithSub\stageOmetaColor{N'^{\superscriptO} } \closeO{\} }   \).
            Thus, we can derive
            \begin{center}
              \derive[WfT1-Tensor]{%
                 \mathit{\Gamma}  \vdash^{0}  \possiblyWithSub\stageOmetaColor{N^{\superscriptO} }  :    \openO{\{} \possiblyWithSub\stageOmetaColor{\nu}  \relO{:}   \ttO{NatList}   \relO{\mid}  \possiblyWithSub\stageOmetaColor{N'^{\superscriptO} } \closeO{\} }   
              }{%
                 \mathit{\Gamma}  \vdash^{1}   \ttI{Tensor}\ \ordI{\%} \possiblyWithSub\stageOmetaColor{N^{\superscriptO} }  
              }.
            \end{center}
          \item The other cases are also straightforward.
        \end{itemize}
    \end{enumerate}
  \end{proof}

\subsection{Preservation and Progress}
  \begin{lemma}[Reduction conforms to CSR equivalence]\label{lem:reduction-implies-csr-equiv}
    \( \possiblyWithSub\stageImetaColor{T^{\superscriptI} }  \longrightarrow^{1}   \possiblyWithSub\stageImetaColor{T'^{\superscriptI} }  \) implies \( \possiblyWithSub\stageImetaColor{T^{\superscriptI} }  \equiv^{1}  \possiblyWithSub\stageImetaColor{T'^{\superscriptI} } \).
  \end{lemma}
  \begin{proof}
    By straightforward induction on the derivation.
  \end{proof}
  \begin{corollary}\label{cor:reduction-implies-csr-equiv}
    \( \possiblyWithSub\stageImetaColor{T^{\superscriptI} }  \longrightarrow^{1\,\ast}   \possiblyWithSub\stageImetaColor{T'^{\superscriptI} }  \) implies \( \possiblyWithSub\stageImetaColor{T^{\superscriptI} }  \equiv^{1}  \possiblyWithSub\stageImetaColor{T'^{\superscriptI} } \).
  \end{corollary}
  \begin{proof}
    By straightforward induction on the length of the reduction sequence.
  \end{proof}
  \begin{lemma}[Substitution preserves CSR equivalence]\label{lem:subst-preserves-csr-equiv}
    \noindent
    \begin{enumerate}
      \item \( \possiblyWithSub\stageImetaColor{T^{\superscriptI} }_{{\mathrm{1}}}  \equiv^{1}  \possiblyWithSub\stageImetaColor{T^{\superscriptI} }_{{\mathrm{1}}} \) implies \(   [  \possiblyWithSub\stageOmetaColor{N^{\superscriptO} }  /  \possiblyWithSub\stageOmetaColor{x}  ]    \possiblyWithSub\stageImetaColor{T^{\superscriptI} }_{{\mathrm{1}}}   \equiv^{1}    [  \possiblyWithSub\stageOmetaColor{N^{\superscriptO} }  /  \possiblyWithSub\stageOmetaColor{x}  ]    \possiblyWithSub\stageImetaColor{T^{\superscriptI} }_{{\mathrm{2}}}  \).
      \item \( \possiblyWithSub\stageOmetaColor{T^{\superscriptO} }_{{\mathrm{1}}}  \equiv^{0}  \possiblyWithSub\stageOmetaColor{T^{\superscriptO} }_{{\mathrm{2}}} \) implies \(   [  \possiblyWithSub\stageOmetaColor{N^{\superscriptO} }  /  \possiblyWithSub\stageOmetaColor{x}  ]    \possiblyWithSub\stageOmetaColor{T^{\superscriptO} }_{{\mathrm{1}}}   \equiv^{0}    [  \possiblyWithSub\stageOmetaColor{N^{\superscriptO} }  /  \possiblyWithSub\stageOmetaColor{x}  ]    \possiblyWithSub\stageOmetaColor{T^{\superscriptO} }_{{\mathrm{2}}}  \).
    \end{enumerate}
  \end{lemma}
  \begin{proof}
    \begin{enumerate}
      \item
        By induction on the derivation of \( \possiblyWithSub\stageImetaColor{T^{\superscriptI} }_{{\mathrm{1}}}  \equiv^{1}  \possiblyWithSub\stageImetaColor{T^{\superscriptI} }_{{\mathrm{2}}} \).
        \begin{itemize}
          \item Case \derive[CqT1-Tensor]{%
             \sigma_{{\mathrm{1}}}  \longrightarrow  \sigma_{{\mathrm{2}}} 
          }{%
              \ttI{Tensor}\ \ordI{\%}  \openO{(}  \sigma_{{\mathrm{1}}}   \possiblyWithSub\stageOmetaColor{N'^{\superscriptO} }  \closeO{)}    \equiv^{1}   \ttI{Tensor}\ \ordI{\%}  \openO{(}  \sigma_{{\mathrm{2}}}   \possiblyWithSub\stageOmetaColor{N'^{\superscriptO} }  \closeO{)}   
          }:
            W.l.o.g., we can assume \(\possiblyWithSub\stageOmetaColor{x} \not\in \dom \sigma_{{\mathrm{1}}} = \dom \sigma_{{\mathrm{2}}}\).
            Since only closed terms can reduce by \(\longrightarrow^0\),
            we have \(  [  \possiblyWithSub\stageOmetaColor{N^{\superscriptO} }  /  \possiblyWithSub\stageOmetaColor{x}  ]     \openO{(}  \sigma_{{\mathrm{1}}}   \possiblyWithSub\stageOmetaColor{N'^{\superscriptO} }  \closeO{)}   =  \sigma_{{\mathrm{1}}}    \openO{(}   [  \possiblyWithSub\stageOmetaColor{N^{\superscriptO} }  /  \possiblyWithSub\stageOmetaColor{x}  ]    \possiblyWithSub\stageOmetaColor{N'^{\superscriptO} }  \closeO{)}  \)
            and \(  [  \possiblyWithSub\stageOmetaColor{N^{\superscriptO} }  /  \possiblyWithSub\stageOmetaColor{x}  ]     \openO{(}  \sigma_{{\mathrm{2}}}   \possiblyWithSub\stageOmetaColor{N'^{\superscriptO} }  \closeO{)}   =  \sigma_{{\mathrm{2}}}    \openO{(}   [  \possiblyWithSub\stageOmetaColor{N^{\superscriptO} }  /  \possiblyWithSub\stageOmetaColor{x}  ]    \possiblyWithSub\stageOmetaColor{N'^{\superscriptO} }  \closeO{)}  \).
            Thus, we can derive
            \begin{center}
              \derive[CqT1-Tensor]{%
                 \sigma_{{\mathrm{1}}}  \longrightarrow  \sigma_{{\mathrm{2}}} 
              }{%
                  \ttI{Tensor}\ \ordI{\%}  \openO{(}  \sigma_{{\mathrm{1}}}    \openO{(}   [  \possiblyWithSub\stageOmetaColor{N^{\superscriptO} }  /  \possiblyWithSub\stageOmetaColor{x}  ]    \possiblyWithSub\stageOmetaColor{N'^{\superscriptO} }  \closeO{)}   \closeO{)}    \equiv^{1}   \ttI{Tensor}\ \ordI{\%}  \openO{(}  \sigma_{{\mathrm{2}}}    \openO{(}   [  \possiblyWithSub\stageOmetaColor{N^{\superscriptO} }  /  \possiblyWithSub\stageOmetaColor{x}  ]    \possiblyWithSub\stageOmetaColor{N'^{\superscriptO} }  \closeO{)}   \closeO{)}   
              }.
            \end{center}
          \item The other cases are all straightforward.
        \end{itemize}
      \item
        By induction on the derivation of \( \possiblyWithSub\stageOmetaColor{T^{\superscriptO} }_{{\mathrm{1}}}  \equiv^{0}  \possiblyWithSub\stageOmetaColor{T^{\superscriptO} }_{{\mathrm{2}}} \).
        \begin{itemize}
          \item Case \derive[CqT0-Rfn]{%
             \sigma_{{\mathrm{1}}}  \longrightarrow  \sigma_{{\mathrm{2}}} 
          }{%
               \openO{\{} \possiblyWithSub\stageOmetaColor{\nu}  \relO{:}  \possiblyWithSub\stageOmetaColor{B}  \relO{\mid}   \sigma_{{\mathrm{1}}}   \possiblyWithSub\stageOmetaColor{N'^{\superscriptO} }  \closeO{\} }    \equiv^{0}    \openO{\{} \possiblyWithSub\stageOmetaColor{\nu}  \relO{:}  \possiblyWithSub\stageOmetaColor{B}  \relO{\mid}   \sigma_{{\mathrm{2}}}   \possiblyWithSub\stageOmetaColor{N'^{\superscriptO} }  \closeO{\} }   
          }:
            Can be proved in the same manner as \rulename{CqT1-Tensor}.
          \item Case \derive[CqT0-Code]{%
             \possiblyWithSub\stageImetaColor{T^{\superscriptI} }_{{\mathrm{1}}}  \equiv^{1}  \possiblyWithSub\stageImetaColor{T^{\superscriptI} }_{{\mathrm{2}}} 
          }{%
              \openO{\langle} \possiblyWithSub\stageImetaColor{T^{\superscriptI} }_{{\mathrm{1}}} \closeO{\rangle}   \equiv^{0}   \openO{\langle} \possiblyWithSub\stageImetaColor{T^{\superscriptI} }_{{\mathrm{2}}} \closeO{\rangle}  
          }:
            Immediate by (1).
          \item Case \derive[CqT0-Arr]{%
             \possiblyWithSub\stageOmetaColor{T^{\superscriptO} }_{{\mathrm{11}}}  \equiv^{0}  \possiblyWithSub\stageOmetaColor{T^{\superscriptO} }_{{\mathrm{21}}} 
          \andalso
             \possiblyWithSub\stageOmetaColor{T^{\superscriptO} }_{{\mathrm{12}}}  \equiv^{0}  \possiblyWithSub\stageOmetaColor{T^{\superscriptO} }_{{\mathrm{22}}} 
          }{%
              \openO{(} \possiblyWithSub\stageOmetaColor{x'}  \relO{:}  \possiblyWithSub\stageOmetaColor{T^{\superscriptO} }_{{\mathrm{11}}} \closeO{)} \relO{\to}  \possiblyWithSub\stageOmetaColor{T^{\superscriptO} }_{{\mathrm{12}}}   \equiv^{0}   \openO{(} \possiblyWithSub\stageOmetaColor{x'}  \relO{:}  \possiblyWithSub\stageOmetaColor{T^{\superscriptO} }_{{\mathrm{21}}} \closeO{)} \relO{\to}  \possiblyWithSub\stageOmetaColor{T^{\superscriptO} }_{{\mathrm{22}}}  
          }:
            By IH and the Barendregt convention, which enables us to assume
            \(\possiblyWithSub\stageOmetaColor{x'} \neq \possiblyWithSub\stageOmetaColor{x}\) safely.
          \item The other cases are straightforward.
        \end{itemize}
    \end{enumerate}
  \end{proof}
  \begin{lemma}[Reduction of substitution satisfies CSR equivalence]\label{lem:reduction-of-subst-satisfies-csr-equiv}
    Suppose \( \possiblyWithSub\stageOmetaColor{N^{\superscriptO} }  \longrightarrow^{0}   \possiblyWithSub\stageOmetaColor{N'^{\superscriptO} }  \).
    \begin{enumerate}
      \item \(   [  \possiblyWithSub\stageOmetaColor{N^{\superscriptO} }  /  \possiblyWithSub\stageOmetaColor{x}  ]    \possiblyWithSub\stageImetaColor{T^{\superscriptI} }   \equiv^{1}    [  \possiblyWithSub\stageOmetaColor{N'^{\superscriptO} }  /  \possiblyWithSub\stageOmetaColor{x}  ]    \possiblyWithSub\stageImetaColor{T^{\superscriptI} }  \).
      \item \(   [  \possiblyWithSub\stageOmetaColor{N^{\superscriptO} }  /  \possiblyWithSub\stageOmetaColor{x}  ]    \possiblyWithSub\stageOmetaColor{T^{\superscriptO} }   \equiv^{0}    [  \possiblyWithSub\stageOmetaColor{N'^{\superscriptO} }  /  \possiblyWithSub\stageOmetaColor{x}  ]    \possiblyWithSub\stageOmetaColor{T^{\superscriptO} }  \).
    \end{enumerate}
  \end{lemma}
  \begin{proof}
    \begin{enumerate}
      \item
        By induction on the structure of \(\possiblyWithSub\stageImetaColor{T^{\superscriptI} }\).
        \begin{itemize}
          \item Case~\(\possiblyWithSub\stageImetaColor{T^{\superscriptI} } = \possiblyWithSub\stageImetaColor{B}\):
            Immediate by \rulename{CqT0-Refl}.
          \item Case~\(\possiblyWithSub\stageImetaColor{T^{\superscriptI} } =  \possiblyWithSub\stageImetaColor{T^{\superscriptI} }_{{\mathrm{1}}}  \relI{\to}  \possiblyWithSub\stageImetaColor{T^{\superscriptI} }_{{\mathrm{2}}} \):
            Straightforward by IH.
          \item Case~\(\possiblyWithSub\stageImetaColor{T^{\superscriptI} } =  \ttI{Tensor}\ \ordI{\%} \possiblyWithSub\stageOmetaColor{N^{\superscriptO} }_{{\mathrm{0}}} \):
            We can immediately derive
            \begin{center}
              \derive[CqT1-Tensor]{%
                  [  \possiblyWithSub\stageOmetaColor{N^{\superscriptO} }  /  \possiblyWithSub\stageOmetaColor{x}  ]   \longrightarrow   [  \possiblyWithSub\stageOmetaColor{N'^{\superscriptO} }  /  \possiblyWithSub\stageOmetaColor{x}  ]  
              }{%
                  \ttI{Tensor}\ \ordI{\%}  \openO{(}   [  \possiblyWithSub\stageOmetaColor{N^{\superscriptO} }  /  \possiblyWithSub\stageOmetaColor{x}  ]    \possiblyWithSub\stageOmetaColor{N^{\superscriptO} }_{{\mathrm{0}}}  \closeO{)}    \equiv^{1}   \ttI{Tensor}\ \ordI{\%}  \openO{(}   [  \possiblyWithSub\stageOmetaColor{N'^{\superscriptO} }  /  \possiblyWithSub\stageOmetaColor{x}  ]    \possiblyWithSub\stageOmetaColor{N^{\superscriptO} }_{{\mathrm{0}}}  \closeO{)}   
              }.
            \end{center}
        \end{itemize}
      \item
        By induction on the structure of \(\possiblyWithSub\stageOmetaColor{T^{\superscriptO} }\).
        \begin{itemize}
          \item Case~\(\possiblyWithSub\stageOmetaColor{T^{\superscriptO} } =  \ttO{Tensor}\  \possiblyWithSub\stageOmetaColor{s} \):
            Immediate by \rulename{CqT0-Refl}.
          \item Case~\(\possiblyWithSub\stageOmetaColor{T^{\superscriptO} } =  \openO{\langle} \possiblyWithSub\stageImetaColor{T^{\superscriptI} } \closeO{\rangle} \):
            Immediate by (1).
          \item Case~\(\possiblyWithSub\stageOmetaColor{T^{\superscriptO} } =  \openO{(} \possiblyWithSub\stageOmetaColor{x'}  \relO{:}  \possiblyWithSub\stageOmetaColor{T^{\superscriptO} }_{{\mathrm{1}}} \closeO{)} \relO{\to}  \possiblyWithSub\stageOmetaColor{T^{\superscriptO} }_{{\mathrm{2}}} \):
            By the Barendregt convention, we can safely assume \(\possiblyWithSub\stageOmetaColor{x'} \neq \possiblyWithSub\stageOmetaColor{x}\).
            By IH, we have \(   [  \possiblyWithSub\stageOmetaColor{N^{\superscriptO} }  /  \possiblyWithSub\stageOmetaColor{x}  ]    \possiblyWithSub\stageOmetaColor{T^{\superscriptO} }_{{\mathrm{1}}}   \equiv^{0}    [  \possiblyWithSub\stageOmetaColor{N'^{\superscriptO} }  /  \possiblyWithSub\stageOmetaColor{x}  ]    \possiblyWithSub\stageOmetaColor{T^{\superscriptO} }_{{\mathrm{1}}}  \)
            and \(   [  \possiblyWithSub\stageOmetaColor{N^{\superscriptO} }  /  \possiblyWithSub\stageOmetaColor{x}  ]    \possiblyWithSub\stageOmetaColor{T^{\superscriptO} }_{{\mathrm{2}}}   \equiv^{0}    [  \possiblyWithSub\stageOmetaColor{N'^{\superscriptO} }  /  \possiblyWithSub\stageOmetaColor{x}  ]    \possiblyWithSub\stageOmetaColor{T^{\superscriptO} }_{{\mathrm{2}}}  \).
            Thus, we can derive \(   [  \possiblyWithSub\stageOmetaColor{N^{\superscriptO} }  /  \possiblyWithSub\stageOmetaColor{x}  ]    \possiblyWithSub\stageOmetaColor{T^{\superscriptO} }   \equiv^{0}    [  \possiblyWithSub\stageOmetaColor{N'^{\superscriptO} }  /  \possiblyWithSub\stageOmetaColor{x}  ]    \possiblyWithSub\stageOmetaColor{T^{\superscriptO} }  \) by
            \begin{center}
              \derive[CqT0-Arr]{%
                   [  \possiblyWithSub\stageOmetaColor{N^{\superscriptO} }  /  \possiblyWithSub\stageOmetaColor{x}  ]    \possiblyWithSub\stageOmetaColor{T^{\superscriptO} }_{{\mathrm{1}}}   \equiv^{0}    [  \possiblyWithSub\stageOmetaColor{N'^{\superscriptO} }  /  \possiblyWithSub\stageOmetaColor{x}  ]    \possiblyWithSub\stageOmetaColor{T^{\superscriptO} }_{{\mathrm{1}}}  
              \andalso
                   [  \possiblyWithSub\stageOmetaColor{N^{\superscriptO} }  /  \possiblyWithSub\stageOmetaColor{x}  ]    \possiblyWithSub\stageOmetaColor{T^{\superscriptO} }_{{\mathrm{2}}}   \equiv^{0}    [  \possiblyWithSub\stageOmetaColor{N'^{\superscriptO} }  /  \possiblyWithSub\stageOmetaColor{x}  ]    \possiblyWithSub\stageOmetaColor{T^{\superscriptO} }_{{\mathrm{2}}}  
              }{%
                   \openO{(} \possiblyWithSub\stageOmetaColor{x'}  \relO{:}    [  \possiblyWithSub\stageOmetaColor{N^{\superscriptO} }  /  \possiblyWithSub\stageOmetaColor{x}  ]    \possiblyWithSub\stageOmetaColor{T^{\superscriptO} }_{{\mathrm{1}}}  \closeO{)} \relO{\to}    [  \possiblyWithSub\stageOmetaColor{N^{\superscriptO} }  /  \possiblyWithSub\stageOmetaColor{x}  ]    \possiblyWithSub\stageOmetaColor{T^{\superscriptO} }_{{\mathrm{2}}}    \equiv^{0}   \openO{(} \possiblyWithSub\stageOmetaColor{x'}  \relO{:}    [  \possiblyWithSub\stageOmetaColor{N'^{\superscriptO} }  /  \possiblyWithSub\stageOmetaColor{x}  ]    \possiblyWithSub\stageOmetaColor{T^{\superscriptO} }_{{\mathrm{1}}}  \closeO{)} \relO{\to}    [  \possiblyWithSub\stageOmetaColor{N'^{\superscriptO} }  /  \possiblyWithSub\stageOmetaColor{x}  ]    \possiblyWithSub\stageOmetaColor{T^{\superscriptO} }_{{\mathrm{2}}}   
              }.
            \end{center}
          \item Case~\(\possiblyWithSub\stageOmetaColor{T^{\superscriptO} } =  \openO{\{} \possiblyWithSub\stageOmetaColor{\nu}  \relO{:}  \possiblyWithSub\stageOmetaColor{B}  \relO{\mid}  \possiblyWithSub\stageOmetaColor{N^{\superscriptO} }_{{\mathrm{0}}} \closeO{\} } \):
            By the Barendregt convention, we can safely assume \(\possiblyWithSub\stageOmetaColor{\nu} \neq \possiblyWithSub\stageOmetaColor{x}\).
            Thus, we can derive
            \begin{center}
              \derive[CqT0-Rfn]{%
                  [  \possiblyWithSub\stageOmetaColor{N^{\superscriptO} }  /  \possiblyWithSub\stageOmetaColor{x}  ]   \longrightarrow   [  \possiblyWithSub\stageOmetaColor{N'^{\superscriptO} }  /  \possiblyWithSub\stageOmetaColor{x}  ]  
              }{%
                   \openO{\{} \possiblyWithSub\stageOmetaColor{\nu}  \relO{:}  \possiblyWithSub\stageOmetaColor{B}  \relO{\mid}    [  \possiblyWithSub\stageOmetaColor{N^{\superscriptO} }  /  \possiblyWithSub\stageOmetaColor{x}  ]    \possiblyWithSub\stageOmetaColor{N^{\superscriptO} }_{{\mathrm{0}}}  \closeO{\} }    \equiv^{0}    \openO{\{} \possiblyWithSub\stageOmetaColor{\nu}  \relO{:}  \possiblyWithSub\stageOmetaColor{B}  \relO{\mid}    [  \possiblyWithSub\stageOmetaColor{N'^{\superscriptO} }  /  \possiblyWithSub\stageOmetaColor{x}  ]    \possiblyWithSub\stageOmetaColor{N^{\superscriptO} }_{{\mathrm{0}}}  \closeO{\} }   
              }.
            \end{center}
        \end{itemize}
    \end{enumerate}
  \end{proof}
  \begin{lemma}[Lambda Inversion]\label{lem:lambda-inversion}
    If \( \mathit{\Gamma}  \vdash^{0}   \openO{(}  \ordO{\lambda} \possiblyWithSub\stageOmetaColor{x}  \relO{:}  \possiblyWithSub\stageOmetaColor{T^{\superscriptO} }_{{\mathrm{1}}} \punctO{.}\  \possiblyWithSub\stageOmetaColor{N^{\superscriptO} }_{{\mathrm{2}}}  \closeO{)}   :  \possiblyWithSub\stageOmetaColor{T^{\superscriptO} } \), then
    we have \( \mathit{\Gamma}  \vdash^{0}  \possiblyWithSub\stageOmetaColor{T^{\superscriptO} }_{{\mathrm{1}}} \) and there exists \(\possiblyWithSub\stageOmetaColor{T^{\superscriptO} }_{{\mathrm{2}}}\) such that
    \(  \mathit{\Gamma} ,  \possiblyWithSub\stageOmetaColor{x}  : ( \possiblyWithSub\stageOmetaColor{T^{\superscriptO} }_{{\mathrm{1}}} )^{0}   \vdash^{0}  \possiblyWithSub\stageOmetaColor{N^{\superscriptO} }_{{\mathrm{2}}}  :  \possiblyWithSub\stageOmetaColor{T^{\superscriptO} }_{{\mathrm{2}}} \) and \( \possiblyWithSub\stageOmetaColor{T^{\superscriptO} }  \equiv^{0}   \openO{(} \possiblyWithSub\stageOmetaColor{x}  \relO{:}  \possiblyWithSub\stageOmetaColor{T^{\superscriptO} }_{{\mathrm{1}}} \closeO{)} \relO{\to}  \possiblyWithSub\stageOmetaColor{T^{\superscriptO} }_{{\mathrm{2}}}  \).
  \end{lemma}
  \begin{proof}
    By induction on the derivation of \( \mathit{\Gamma}  \vdash^{0}   \openO{(}  \ordO{\lambda} \possiblyWithSub\stageOmetaColor{x}  \relO{:}  \possiblyWithSub\stageOmetaColor{T^{\superscriptO} }_{{\mathrm{1}}} \punctO{.}\  \possiblyWithSub\stageOmetaColor{N^{\superscriptO} }_{{\mathrm{2}}}  \closeO{)}   :  \possiblyWithSub\stageOmetaColor{T^{\superscriptO} } \).
    \begin{itemize}
      \item Case \derive[T0-TyEquiv]{%
         \mathit{\Gamma}  \vdash^{0}   \openO{(}  \ordO{\lambda} \possiblyWithSub\stageOmetaColor{x}  \relO{:}  \possiblyWithSub\stageOmetaColor{T^{\superscriptO} }_{{\mathrm{1}}} \punctO{.}\  \possiblyWithSub\stageOmetaColor{N^{\superscriptO} }_{{\mathrm{2}}}  \closeO{)}   :  \possiblyWithSub\stageOmetaColor{T'^{\superscriptO} } 
      \andalso
         \possiblyWithSub\stageOmetaColor{T'^{\superscriptO} }  \equiv^{0}  \possiblyWithSub\stageOmetaColor{T^{\superscriptO} } 
      \andalso
         \mathit{\Gamma}  \vdash^{0}  \possiblyWithSub\stageOmetaColor{T^{\superscriptO} } 
      }{%
         \mathit{\Gamma}  \vdash^{0}   \openO{(}  \ordO{\lambda} \possiblyWithSub\stageOmetaColor{x}  \relO{:}  \possiblyWithSub\stageOmetaColor{T^{\superscriptO} }_{{\mathrm{1}}} \punctO{.}\  \possiblyWithSub\stageOmetaColor{N^{\superscriptO} }_{{\mathrm{2}}}  \closeO{)}   :  \possiblyWithSub\stageOmetaColor{T^{\superscriptO} } 
      }:
        By IH on \( \mathit{\Gamma}  \vdash^{0}   \openO{(}  \ordO{\lambda} \possiblyWithSub\stageOmetaColor{x}  \relO{:}  \possiblyWithSub\stageOmetaColor{T^{\superscriptO} }_{{\mathrm{1}}} \punctO{.}\  \possiblyWithSub\stageOmetaColor{N^{\superscriptO} }_{{\mathrm{2}}}  \closeO{)}   :  \possiblyWithSub\stageOmetaColor{T'^{\superscriptO} } \),
        we have \( \mathit{\Gamma}  \vdash^{0}  \possiblyWithSub\stageOmetaColor{T^{\superscriptO} }_{{\mathrm{1}}} \) and there exists \(\possiblyWithSub\stageOmetaColor{T^{\superscriptO} }_{{\mathrm{2}}}\) such that
        \(  \mathit{\Gamma} ,  \possiblyWithSub\stageOmetaColor{x}  : ( \possiblyWithSub\stageOmetaColor{T^{\superscriptO} }_{{\mathrm{1}}} )^{0}   \vdash^{0}  \possiblyWithSub\stageOmetaColor{N^{\superscriptO} }_{{\mathrm{2}}}  :  \possiblyWithSub\stageOmetaColor{T^{\superscriptO} }_{{\mathrm{2}}} \) and \( \possiblyWithSub\stageOmetaColor{T'^{\superscriptO} }  \equiv^{0}   \openO{(} \possiblyWithSub\stageOmetaColor{x}  \relO{:}  \possiblyWithSub\stageOmetaColor{T^{\superscriptO} }_{{\mathrm{1}}} \closeO{)} \relO{\to}  \possiblyWithSub\stageOmetaColor{T^{\superscriptO} }_{{\mathrm{2}}}  \),
        which enables us to derive
        \begin{center}
          \infer[CqT0-Trans]{%
            \infer[CqT0-Sym]{%
               \possiblyWithSub\stageOmetaColor{T'^{\superscriptO} }  \equiv^{0}  \possiblyWithSub\stageOmetaColor{T^{\superscriptO} } 
            }{%
               \possiblyWithSub\stageOmetaColor{T^{\superscriptO} }  \equiv^{0}  \possiblyWithSub\stageOmetaColor{T'^{\superscriptO} } 
            }
          \andalso
             \possiblyWithSub\stageOmetaColor{T'^{\superscriptO} }  \equiv^{0}   \openO{(} \possiblyWithSub\stageOmetaColor{x}  \relO{:}  \possiblyWithSub\stageOmetaColor{T^{\superscriptO} }_{{\mathrm{1}}} \closeO{)} \relO{\to}  \possiblyWithSub\stageOmetaColor{T^{\superscriptO} }_{{\mathrm{2}}}  
          }{%
             \possiblyWithSub\stageOmetaColor{T^{\superscriptO} }  \equiv^{0}   \openO{(} \possiblyWithSub\stageOmetaColor{x}  \relO{:}  \possiblyWithSub\stageOmetaColor{T^{\superscriptO} }_{{\mathrm{1}}} \closeO{)} \relO{\to}  \possiblyWithSub\stageOmetaColor{T^{\superscriptO} }_{{\mathrm{2}}}  
          }.
        \end{center}
      \item Case \derive[T0-Abs]{%
         \mathit{\Gamma}  \vdash^{0}  \possiblyWithSub\stageOmetaColor{T^{\superscriptO} }_{{\mathrm{1}}} 
      \andalso
          \mathit{\Gamma} ,  \possiblyWithSub\stageOmetaColor{x}  : ( \possiblyWithSub\stageOmetaColor{T^{\superscriptO} }_{{\mathrm{1}}} )^{0}   \vdash^{0}  \possiblyWithSub\stageOmetaColor{N^{\superscriptO} }_{{\mathrm{2}}}  :  \possiblyWithSub\stageOmetaColor{T^{\superscriptO} }_{{\mathrm{2}}} 
      }{%
         \mathit{\Gamma}  \vdash^{0}   \openO{(}  \ordO{\lambda} \possiblyWithSub\stageOmetaColor{x}  \relO{:}  \possiblyWithSub\stageOmetaColor{T^{\superscriptO} }_{{\mathrm{1}}} \punctO{.}\  \possiblyWithSub\stageOmetaColor{N^{\superscriptO} }_{{\mathrm{2}}}  \closeO{)}   :   \openO{(} \possiblyWithSub\stageOmetaColor{x}  \relO{:}  \possiblyWithSub\stageOmetaColor{T^{\superscriptO} }_{{\mathrm{1}}} \closeO{)} \relO{\to}  \possiblyWithSub\stageOmetaColor{T^{\superscriptO} }_{{\mathrm{2}}}  
      }:
        Immediate from the premises and \rulename{CqT0-Refl}.
      \item
        The other cases contradict the assumption.
    \end{itemize}
  \end{proof}
  \begin{lemma}[Bracket Inversion]\label{lem:bracket-inversion}
    If \( \mathit{\Gamma}  \vdash^{0}   \openO{\langle} \possiblyWithSub\stageImetaColor{N^{\superscriptI} } \closeO{\rangle}   :  \possiblyWithSub\stageOmetaColor{T^{\superscriptO} } \), then
    there exists \(\possiblyWithSub\stageImetaColor{T^{\superscriptI} }\) such that
    \( \mathit{\Gamma}  \vdash^{1}  \possiblyWithSub\stageImetaColor{N^{\superscriptI} }  :  \possiblyWithSub\stageImetaColor{T^{\superscriptI} } \) and \( \possiblyWithSub\stageOmetaColor{T^{\superscriptO} }  \equiv^{0}   \openO{\langle} \possiblyWithSub\stageImetaColor{T^{\superscriptI} } \closeO{\rangle}  \).
  \end{lemma}
  \begin{proof}
    By induction on the derivation of \( \mathit{\Gamma}  \vdash^{0}   \openO{\langle} \possiblyWithSub\stageImetaColor{N^{\superscriptI} } \closeO{\rangle}   :  \possiblyWithSub\stageOmetaColor{T^{\superscriptO} } \).
    \begin{itemize}
      \item Case \derive[T0-TyEquiv]{%
         \mathit{\Gamma}  \vdash^{0}   \openO{\langle} \possiblyWithSub\stageImetaColor{N^{\superscriptI} } \closeO{\rangle}   :  \possiblyWithSub\stageOmetaColor{T'^{\superscriptO} } 
      \andalso
         \possiblyWithSub\stageOmetaColor{T'^{\superscriptO} }  \equiv^{0}  \possiblyWithSub\stageOmetaColor{T^{\superscriptO} } 
      \andalso
         \mathit{\Gamma}  \vdash^{0}  \possiblyWithSub\stageOmetaColor{T^{\superscriptO} } 
      }{%
         \mathit{\Gamma}  \vdash^{0}   \openO{\langle} \possiblyWithSub\stageImetaColor{N^{\superscriptI} } \closeO{\rangle}   :  \possiblyWithSub\stageOmetaColor{T^{\superscriptO} } 
      }:
        By IH on \( \mathit{\Gamma}  \vdash^{0}   \openO{\langle} \possiblyWithSub\stageImetaColor{N^{\superscriptI} } \closeO{\rangle}   :  \possiblyWithSub\stageOmetaColor{T'^{\superscriptO} } \),
        there exists \(\possiblyWithSub\stageImetaColor{T^{\superscriptI} }\) such that
        \( \mathit{\Gamma}  \vdash^{1}  \possiblyWithSub\stageImetaColor{N^{\superscriptI} }  :  \possiblyWithSub\stageImetaColor{T^{\superscriptI} } \) and \( \possiblyWithSub\stageOmetaColor{T'^{\superscriptO} }  \equiv^{0}   \openO{\langle} \possiblyWithSub\stageImetaColor{T^{\superscriptI} } \closeO{\rangle}  \),
        which enables us to derive
        \begin{center}
          \infer[CqT0-Trans]{%
            \infer[CqT0-Sym]{%
               \possiblyWithSub\stageOmetaColor{T'^{\superscriptO} }  \equiv^{0}  \possiblyWithSub\stageOmetaColor{T^{\superscriptO} } 
            }{
               \possiblyWithSub\stageOmetaColor{T^{\superscriptO} }  \equiv^{0}  \possiblyWithSub\stageOmetaColor{T'^{\superscriptO} } 
            }
          \andalso
             \possiblyWithSub\stageOmetaColor{T'^{\superscriptO} }  \equiv^{0}   \openO{\langle} \possiblyWithSub\stageImetaColor{T^{\superscriptI} } \closeO{\rangle}  
          }{%
             \possiblyWithSub\stageOmetaColor{T^{\superscriptO} }  \equiv^{0}   \openO{\langle} \possiblyWithSub\stageImetaColor{T^{\superscriptI} } \closeO{\rangle}  
          }.
        \end{center}
      \item Case \derive[T0-Brkt]{%
         \mathit{\Gamma}  \vdash^{1}  \possiblyWithSub\stageImetaColor{N^{\superscriptI} }  :  \possiblyWithSub\stageImetaColor{T^{\superscriptI} } 
      }{%
         \mathit{\Gamma}  \vdash^{0}   \openO{\langle} \possiblyWithSub\stageImetaColor{N^{\superscriptI} } \closeO{\rangle}   :   \openO{\langle} \possiblyWithSub\stageImetaColor{T^{\superscriptI} } \closeO{\rangle}  
      }:
        Immediate from the premise and \rulename{CqT0-Refl}.
      \item
        The other cases clearly contradict the assumption.
    \end{itemize}
  \end{proof}
  \begin{lemma}[CSR equivalence preserves the form of refinement types]\label{lem:rfn-type-csr-equiv-form}
    \noindent
    \begin{enumerate}
      \item If \(   \openO{\{} \possiblyWithSub\stageOmetaColor{\nu}  \relO{:}  \possiblyWithSub\stageOmetaColor{B}  \relO{\mid}  \possiblyWithSub\stageOmetaColor{N^{\superscriptO} }_{{\mathrm{1}}} \closeO{\} }    \equiv^{0}  \possiblyWithSub\stageOmetaColor{T^{\superscriptO} } \), then \(\possiblyWithSub\stageOmetaColor{T^{\superscriptO} }\) is of the form~\( \openO{\{} \possiblyWithSub\stageOmetaColor{\nu}  \relO{:}  \possiblyWithSub\stageOmetaColor{B}  \relO{\mid}  \possiblyWithSub\stageOmetaColor{N^{\superscriptO} }_{{\mathrm{2}}} \closeO{\} } \).
      \item If \( \possiblyWithSub\stageOmetaColor{T^{\superscriptO} }  \equiv^{0}    \openO{\{} \possiblyWithSub\stageOmetaColor{\nu}  \relO{:}  \possiblyWithSub\stageOmetaColor{B}  \relO{\mid}  \possiblyWithSub\stageOmetaColor{N^{\superscriptO} }_{{\mathrm{2}}} \closeO{\} }   \), then \(\possiblyWithSub\stageOmetaColor{T^{\superscriptO} }\) is of the form~\( \openO{\{} \possiblyWithSub\stageOmetaColor{\nu}  \relO{:}  \possiblyWithSub\stageOmetaColor{B}  \relO{\mid}  \possiblyWithSub\stageOmetaColor{N^{\superscriptO} }_{{\mathrm{1}}} \closeO{\} } \).
    \end{enumerate}
  \end{lemma}
  \begin{proof}
    By mutual induction on the derivation.
    \begin{enumerate}
      \item
        \begin{itemize}
          \item Case \rulename{CqT0-Rfn} is immediate.
          \item Case \rulename{CqT0-Refl} is also immediate.
          \item Case \rulename{CqT0-Sym} is straightforward by IH.
          \item Case \derive[CqT0-Trans]{%
               \openO{\{} \possiblyWithSub\stageOmetaColor{\nu}  \relO{:}  \possiblyWithSub\stageOmetaColor{B}  \relO{\mid}  \possiblyWithSub\stageOmetaColor{N^{\superscriptO} }_{{\mathrm{1}}} \closeO{\} }    \equiv^{0}  \possiblyWithSub\stageOmetaColor{T'^{\superscriptO} } 
          \andalso
             \possiblyWithSub\stageOmetaColor{T'^{\superscriptO} }  \equiv^{0}  \possiblyWithSub\stageOmetaColor{T^{\superscriptO} } 
          }{%
               \openO{\{} \possiblyWithSub\stageOmetaColor{\nu}  \relO{:}  \possiblyWithSub\stageOmetaColor{B}  \relO{\mid}  \possiblyWithSub\stageOmetaColor{N^{\superscriptO} }_{{\mathrm{1}}} \closeO{\} }    \equiv^{0}  \possiblyWithSub\stageOmetaColor{T^{\superscriptO} } 
          }:
            By IH, from \(   \openO{\{} \possiblyWithSub\stageOmetaColor{\nu}  \relO{:}  \possiblyWithSub\stageOmetaColor{B}  \relO{\mid}  \possiblyWithSub\stageOmetaColor{N^{\superscriptO} }_{{\mathrm{1}}} \closeO{\} }    \equiv^{0}  \possiblyWithSub\stageOmetaColor{T'^{\superscriptO} } \),
            \(\possiblyWithSub\stageOmetaColor{T'^{\superscriptO} }\) is of the form~\( \openO{\{} \possiblyWithSub\stageOmetaColor{\nu}  \relO{:}  \possiblyWithSub\stageOmetaColor{B}  \relO{\mid}  \possiblyWithSub\stageOmetaColor{N'^{\superscriptO} } \closeO{\} } \).
            Then, again by IH, from \( \possiblyWithSub\stageOmetaColor{T'^{\superscriptO} }  \equiv^{0}  \possiblyWithSub\stageOmetaColor{T^{\superscriptO} } \),
            \(\possiblyWithSub\stageOmetaColor{T^{\superscriptO} }\) is of the form~\( \openO{\{} \possiblyWithSub\stageOmetaColor{\nu}  \relO{:}  \possiblyWithSub\stageOmetaColor{B}  \relO{\mid}  \possiblyWithSub\stageOmetaColor{N^{\superscriptO} }_{{\mathrm{2}}} \closeO{\} } \).
          \item The other cases contradict the assumption.
        \end{itemize}
      \item
        Can be proved in the same manner as (1).
    \end{enumerate}
  \end{proof}
  \begin{lemma}[CSR equivalence preserves stage-\(0\) tensor types]\label{lem:tensor-type-csr-equiv-form}
    \noindent
    \begin{enumerate}
      \item If \(  \ttO{Tensor}\  \possiblyWithSub\stageOmetaColor{s}   \equiv^{0}  \possiblyWithSub\stageOmetaColor{T^{\superscriptO} } \), then \(\possiblyWithSub\stageOmetaColor{T^{\superscriptO} } =  \ttO{Tensor}\  \possiblyWithSub\stageOmetaColor{s} \).
      \item If \( \possiblyWithSub\stageOmetaColor{T^{\superscriptO} }  \equiv^{0}   \ttO{Tensor}\  \possiblyWithSub\stageOmetaColor{s}  \), then \(\possiblyWithSub\stageOmetaColor{T^{\superscriptO} } =  \ttO{Tensor}\  \possiblyWithSub\stageOmetaColor{s} \).
    \end{enumerate}
  \end{lemma}
  \begin{proof}
    By mutual induction on the derivation.
    \begin{enumerate}
      \item
        \begin{itemize}
          \item Case \rulename{CqT0-Refl} is immediate.
          \item Case \rulename{CqT0-Sym} is straightforward by IH.
          \item Case \derive[CqT0-Trans]{%
              \ttO{Tensor}\  \possiblyWithSub\stageOmetaColor{s}   \equiv^{0}  \possiblyWithSub\stageOmetaColor{T'^{\superscriptO} } 
          \andalso
             \possiblyWithSub\stageOmetaColor{T'^{\superscriptO} }  \equiv^{0}  \possiblyWithSub\stageOmetaColor{T^{\superscriptO} } 
          }{%
              \ttO{Tensor}\  \possiblyWithSub\stageOmetaColor{s}   \equiv^{0}  \possiblyWithSub\stageOmetaColor{T^{\superscriptO} } 
          }:
            By IH, from \(  \ttO{Tensor}\  \possiblyWithSub\stageOmetaColor{s}   \equiv^{0}  \possiblyWithSub\stageOmetaColor{T'^{\superscriptO} } \),
            we have \(\possiblyWithSub\stageOmetaColor{T'^{\superscriptO} } =  \ttO{Tensor}\  \possiblyWithSub\stageOmetaColor{s} \).
            Then, again by IH, from \( \possiblyWithSub\stageOmetaColor{T'^{\superscriptO} }  \equiv^{0}  \possiblyWithSub\stageOmetaColor{T^{\superscriptO} } \),
            we have \(\possiblyWithSub\stageOmetaColor{T^{\superscriptO} } =  \ttO{Tensor}\  \possiblyWithSub\stageOmetaColor{s} \).
          \item The other cases contradict the assumption.
        \end{itemize}
      \item
        Can be proved in the same manner as (1).
    \end{enumerate}
  \end{proof}
  \begin{lemma}[CSR equivalence preserves the form of stage-\(0\) function types]\label{lem:arrow-type-csr-equiv-form}
    \noindent
    \begin{enumerate}
      \item
        If \(  \openO{(} \possiblyWithSub\stageOmetaColor{x}  \relO{:}  \possiblyWithSub\stageOmetaColor{T^{\superscriptO} }_{{\mathrm{11}}} \closeO{)} \relO{\to}  \possiblyWithSub\stageOmetaColor{T^{\superscriptO} }_{{\mathrm{12}}}   \equiv^{0}  \possiblyWithSub\stageOmetaColor{T^{\superscriptO} } \), then
        \(\possiblyWithSub\stageOmetaColor{T^{\superscriptO} }\) is of the form~\( \openO{(} \possiblyWithSub\stageOmetaColor{x}  \relO{:}  \possiblyWithSub\stageOmetaColor{T^{\superscriptO} }_{{\mathrm{21}}} \closeO{)} \relO{\to}  \possiblyWithSub\stageOmetaColor{T^{\superscriptO} }_{{\mathrm{22}}} \).
      \item
        If \( \possiblyWithSub\stageOmetaColor{T^{\superscriptO} }  \equiv^{0}   \openO{(} \possiblyWithSub\stageOmetaColor{x}  \relO{:}  \possiblyWithSub\stageOmetaColor{T^{\superscriptO} }_{{\mathrm{21}}} \closeO{)} \relO{\to}  \possiblyWithSub\stageOmetaColor{T^{\superscriptO} }_{{\mathrm{22}}}  \), then
        \(\possiblyWithSub\stageOmetaColor{T^{\superscriptO} }\) is of the form~\( \openO{(} \possiblyWithSub\stageOmetaColor{x}  \relO{:}  \possiblyWithSub\stageOmetaColor{T^{\superscriptO} }_{{\mathrm{11}}} \closeO{)} \relO{\to}  \possiblyWithSub\stageOmetaColor{T^{\superscriptO} }_{{\mathrm{12}}} \).
    \end{enumerate}
  \end{lemma}
  \begin{proof}
    By mutual induction on the derivation.
    \begin{enumerate}
      \item
        \begin{itemize}
          \item Case \rulename{CqT0-Arr} is immediate.
          \item Case \rulename{CqT0-Refl} is also immediate.
          \item Case \rulename{CqT0-Sym} is straightforward by IH.
          \item Case \derive[CqT0-Trans]{%
              \openO{(} \possiblyWithSub\stageOmetaColor{x}  \relO{:}  \possiblyWithSub\stageOmetaColor{T^{\superscriptO} }_{{\mathrm{11}}} \closeO{)} \relO{\to}  \possiblyWithSub\stageOmetaColor{T^{\superscriptO} }_{{\mathrm{12}}}   \equiv^{0}  \possiblyWithSub\stageOmetaColor{T'^{\superscriptO} } 
          \andalso
             \possiblyWithSub\stageOmetaColor{T'^{\superscriptO} }  \equiv^{0}  \possiblyWithSub\stageOmetaColor{T^{\superscriptO} } 
          }{%
              \openO{(} \possiblyWithSub\stageOmetaColor{x}  \relO{:}  \possiblyWithSub\stageOmetaColor{T^{\superscriptO} }_{{\mathrm{11}}} \closeO{)} \relO{\to}  \possiblyWithSub\stageOmetaColor{T^{\superscriptO} }_{{\mathrm{12}}}   \equiv^{0}  \possiblyWithSub\stageOmetaColor{T^{\superscriptO} } 
          }:
            By IH, from \(  \openO{(} \possiblyWithSub\stageOmetaColor{x}  \relO{:}  \possiblyWithSub\stageOmetaColor{T^{\superscriptO} }_{{\mathrm{11}}} \closeO{)} \relO{\to}  \possiblyWithSub\stageOmetaColor{T^{\superscriptO} }_{{\mathrm{12}}}   \equiv^{0}  \possiblyWithSub\stageOmetaColor{T'^{\superscriptO} } \),
            \(\possiblyWithSub\stageOmetaColor{T'^{\superscriptO} }\) is of the form~\( \openO{(} \possiblyWithSub\stageOmetaColor{x}  \relO{:}  \possiblyWithSub\stageOmetaColor{T'^{\superscriptO} }_{{\mathrm{1}}} \closeO{)} \relO{\to}  \possiblyWithSub\stageOmetaColor{T'^{\superscriptO} }_{{\mathrm{2}}} \).
            Then, by IH again, from \(  \openO{(} \possiblyWithSub\stageOmetaColor{x}  \relO{:}  \possiblyWithSub\stageOmetaColor{T'^{\superscriptO} }_{{\mathrm{1}}} \closeO{)} \relO{\to}  \possiblyWithSub\stageOmetaColor{T'^{\superscriptO} }_{{\mathrm{2}}}   \equiv^{0}  \possiblyWithSub\stageOmetaColor{T^{\superscriptO} } \),
            \(\possiblyWithSub\stageOmetaColor{T^{\superscriptO} }\) is of the form~\( \openO{(} \possiblyWithSub\stageOmetaColor{x}  \relO{:}  \possiblyWithSub\stageOmetaColor{T'^{\superscriptO} }_{{\mathrm{21}}} \closeO{)} \relO{\to}  \possiblyWithSub\stageOmetaColor{T'^{\superscriptO} }_{{\mathrm{22}}} \).
          \item The other cases contradict the assumption.
        \end{itemize}
      \item
        Can be proved in the same manner as (1).
    \end{enumerate}
  \end{proof}
  \begin{lemma}[Arrow Type CSR Equivalence Inversion]\label{lem:arrow-type-csr-equiv-inversion}
    \(  \openO{(} \possiblyWithSub\stageOmetaColor{x}  \relO{:}  \possiblyWithSub\stageOmetaColor{T^{\superscriptO} }_{{\mathrm{11}}} \closeO{)} \relO{\to}  \possiblyWithSub\stageOmetaColor{T^{\superscriptO} }_{{\mathrm{12}}}   \equiv^{0}   \openO{(} \possiblyWithSub\stageOmetaColor{x}  \relO{:}  \possiblyWithSub\stageOmetaColor{T^{\superscriptO} }_{{\mathrm{21}}} \closeO{)} \relO{\to}  \possiblyWithSub\stageOmetaColor{T^{\superscriptO} }_{{\mathrm{22}}}  \) implies
    \( \possiblyWithSub\stageOmetaColor{T^{\superscriptO} }_{{\mathrm{11}}}  \equiv^{0}  \possiblyWithSub\stageOmetaColor{T^{\superscriptO} }_{{\mathrm{21}}} \) and \( \possiblyWithSub\stageOmetaColor{T^{\superscriptO} }_{{\mathrm{12}}}  \equiv^{0}  \possiblyWithSub\stageOmetaColor{T^{\superscriptO} }_{{\mathrm{22}}} \).
  \end{lemma}
  \begin{proof}
    By induction on the derivation of \(  \openO{(} \possiblyWithSub\stageOmetaColor{x}  \relO{:}  \possiblyWithSub\stageOmetaColor{T^{\superscriptO} }_{{\mathrm{11}}} \closeO{)} \relO{\to}  \possiblyWithSub\stageOmetaColor{T^{\superscriptO} }_{{\mathrm{12}}}   \equiv^{0}   \openO{(} \possiblyWithSub\stageOmetaColor{x}  \relO{:}  \possiblyWithSub\stageOmetaColor{T^{\superscriptO} }_{{\mathrm{21}}} \closeO{)} \relO{\to}  \possiblyWithSub\stageOmetaColor{T^{\superscriptO} }_{{\mathrm{22}}}  \).
    \begin{itemize}
      \item Case \derive[CqT0-Trans]{%
          \openO{(} \possiblyWithSub\stageOmetaColor{x}  \relO{:}  \possiblyWithSub\stageOmetaColor{T^{\superscriptO} }_{{\mathrm{11}}} \closeO{)} \relO{\to}  \possiblyWithSub\stageOmetaColor{T^{\superscriptO} }_{{\mathrm{12}}}   \equiv^{0}  \possiblyWithSub\stageOmetaColor{T'^{\superscriptO} } 
      \andalso
         \possiblyWithSub\stageOmetaColor{T'^{\superscriptO} }  \equiv^{0}   \openO{(} \possiblyWithSub\stageOmetaColor{x}  \relO{:}  \possiblyWithSub\stageOmetaColor{T^{\superscriptO} }_{{\mathrm{21}}} \closeO{)} \relO{\to}  \possiblyWithSub\stageOmetaColor{T^{\superscriptO} }_{{\mathrm{22}}}  
      }{%
          \openO{(} \possiblyWithSub\stageOmetaColor{x}  \relO{:}  \possiblyWithSub\stageOmetaColor{T^{\superscriptO} }_{{\mathrm{11}}} \closeO{)} \relO{\to}  \possiblyWithSub\stageOmetaColor{T^{\superscriptO} }_{{\mathrm{12}}}   \equiv^{0}   \openO{(} \possiblyWithSub\stageOmetaColor{x}  \relO{:}  \possiblyWithSub\stageOmetaColor{T^{\superscriptO} }_{{\mathrm{21}}} \closeO{)} \relO{\to}  \possiblyWithSub\stageOmetaColor{T^{\superscriptO} }_{{\mathrm{22}}}  
      }:
        By Lemma~\ref{lem:arrow-type-csr-equiv-form}, from \(  \openO{(} \possiblyWithSub\stageOmetaColor{x}  \relO{:}  \possiblyWithSub\stageOmetaColor{T^{\superscriptO} }_{{\mathrm{11}}} \closeO{)} \relO{\to}  \possiblyWithSub\stageOmetaColor{T^{\superscriptO} }_{{\mathrm{12}}}   \equiv^{0}  \possiblyWithSub\stageOmetaColor{T'^{\superscriptO} } \),
        \(\possiblyWithSub\stageOmetaColor{T'^{\superscriptO} }\) is of the form~\( \openO{(} \possiblyWithSub\stageOmetaColor{x}  \relO{:}  \possiblyWithSub\stageOmetaColor{T'^{\superscriptO} }_{{\mathrm{1}}} \closeO{)} \relO{\to}  \possiblyWithSub\stageOmetaColor{T'^{\superscriptO} }_{{\mathrm{2}}} \).
        Then, by IH, we have \( \possiblyWithSub\stageOmetaColor{T^{\superscriptO} }_{{\mathrm{11}}}  \equiv^{0}  \possiblyWithSub\stageOmetaColor{T'^{\superscriptO} }_{{\mathrm{1}}} \) and \( \possiblyWithSub\stageOmetaColor{T^{\superscriptO} }_{{\mathrm{12}}}  \equiv^{0}  \possiblyWithSub\stageOmetaColor{T'^{\superscriptO} }_{{\mathrm{2}}} \).
        Similarly, by IH, from \(  \openO{(} \possiblyWithSub\stageOmetaColor{x}  \relO{:}  \possiblyWithSub\stageOmetaColor{T'^{\superscriptO} }_{{\mathrm{1}}} \closeO{)} \relO{\to}  \possiblyWithSub\stageOmetaColor{T'^{\superscriptO} }_{{\mathrm{2}}}   \equiv^{0}   \openO{(} \possiblyWithSub\stageOmetaColor{x}  \relO{:}  \possiblyWithSub\stageOmetaColor{T^{\superscriptO} }_{{\mathrm{21}}} \closeO{)} \relO{\to}  \possiblyWithSub\stageOmetaColor{T^{\superscriptO} }_{{\mathrm{22}}}  \),
        we have \( \possiblyWithSub\stageOmetaColor{T'^{\superscriptO} }_{{\mathrm{1}}}  \equiv^{0}  \possiblyWithSub\stageOmetaColor{T^{\superscriptO} }_{{\mathrm{21}}} \) and \( \possiblyWithSub\stageOmetaColor{T'^{\superscriptO} }_{{\mathrm{2}}}  \equiv^{0}  \possiblyWithSub\stageOmetaColor{T^{\superscriptO} }_{{\mathrm{22}}} \).
        Therefore, we can derive
        \begin{center}
          \infer[CqT0-Trans]{%
             \possiblyWithSub\stageOmetaColor{T^{\superscriptO} }_{{\mathrm{11}}}  \equiv^{0}  \possiblyWithSub\stageOmetaColor{T'^{\superscriptO} }_{{\mathrm{1}}} 
          \andalso
             \possiblyWithSub\stageOmetaColor{T'^{\superscriptO} }_{{\mathrm{1}}}  \equiv^{0}  \possiblyWithSub\stageOmetaColor{T^{\superscriptO} }_{{\mathrm{21}}} 
          }{%
             \possiblyWithSub\stageOmetaColor{T^{\superscriptO} }_{{\mathrm{11}}}  \equiv^{0}  \possiblyWithSub\stageOmetaColor{T^{\superscriptO} }_{{\mathrm{21}}} 
          }
        \end{center}
        and
        \begin{center}
          \infer[CqT0-Trans]{%
             \possiblyWithSub\stageOmetaColor{T^{\superscriptO} }_{{\mathrm{12}}}  \equiv^{0}  \possiblyWithSub\stageOmetaColor{T'^{\superscriptO} }_{{\mathrm{2}}} 
          \andalso
             \possiblyWithSub\stageOmetaColor{T'^{\superscriptO} }_{{\mathrm{2}}}  \equiv^{0}  \possiblyWithSub\stageOmetaColor{T^{\superscriptO} }_{{\mathrm{22}}} 
          }{%
             \possiblyWithSub\stageOmetaColor{T^{\superscriptO} }_{{\mathrm{12}}}  \equiv^{0}  \possiblyWithSub\stageOmetaColor{T^{\superscriptO} }_{{\mathrm{22}}} 
          }.
        \end{center}
      \item The other cases are easy.
    \end{itemize}
  \end{proof}
  \begin{lemma}[CSR equivalence preserves unlifting relation]\label{lem:csr-equiv-preserves-unlifting-relation}
    If \( \possiblyWithSub\stageImetaColor{\tau^{\superscriptI} }  \gg  \possiblyWithSub\stageOmetaColor{T^{\superscriptO} } \) and \( \possiblyWithSub\stageOmetaColor{T^{\superscriptO} }  \equiv^{0}  \possiblyWithSub\stageOmetaColor{T'^{\superscriptO} } \), then \( \possiblyWithSub\stageImetaColor{\tau^{\superscriptI} }  \gg  \possiblyWithSub\stageOmetaColor{T'^{\superscriptO} } \).
  \end{lemma}
  \begin{proof}
    By induction on the structure of \(\possiblyWithSub\stageImetaColor{\tau^{\superscriptI} }\).
    \begin{itemize}
      \item Case \(\possiblyWithSub\stageImetaColor{\tau^{\superscriptI} } = \possiblyWithSub\stageImetaColor{B}\):
        By the definition of \(\gg\),
        \(\possiblyWithSub\stageOmetaColor{T^{\superscriptO} }\) is of the form~\( \openO{\{} \possiblyWithSub\stageOmetaColor{\nu}  \relO{:}  \possiblyWithSub\stageOmetaColor{B}  \relO{\mid}  \possiblyWithSub\stageOmetaColor{N^{\superscriptO} } \closeO{\} } \).
        Then, by Lemma~\ref{lem:rfn-type-csr-equiv-form},
        \(\possiblyWithSub\stageOmetaColor{T'^{\superscriptO} }\) is of the form~\( \openO{\{} \possiblyWithSub\stageOmetaColor{\nu}  \relO{:}  \possiblyWithSub\stageOmetaColor{B}  \relO{\mid}  \possiblyWithSub\stageOmetaColor{N'^{\superscriptO} } \closeO{\} } \),
        which enables us to have \( \possiblyWithSub\stageImetaColor{\tau^{\superscriptI} }  \gg  \possiblyWithSub\stageOmetaColor{T'^{\superscriptO} } \).
      \item Case \(\possiblyWithSub\stageImetaColor{\tau^{\superscriptI} } =  \ttI{Tensor}\ \ordI{\%} \possiblyWithSub\stageOmetaColor{s} \):
        By the definition of \(\gg\), \(\possiblyWithSub\stageOmetaColor{T^{\superscriptO} } =  \ttO{Tensor}\  \possiblyWithSub\stageOmetaColor{s} \).
        Then, by Lemma~\ref{lem:tensor-type-csr-equiv-form},
        we have \(\possiblyWithSub\stageOmetaColor{T'^{\superscriptO} } =  \ttO{Tensor}\  \possiblyWithSub\stageOmetaColor{s} \) and thereby \( \possiblyWithSub\stageImetaColor{\tau^{\superscriptI} }  \gg  \possiblyWithSub\stageOmetaColor{T'^{\superscriptO} } \).
      \item Case \(\possiblyWithSub\stageImetaColor{\tau^{\superscriptI} } =  \possiblyWithSub\stageImetaColor{\tau^{\superscriptI} }_{{\mathrm{1}}}  \relI{\to}  \possiblyWithSub\stageImetaColor{\tau^{\superscriptI} }_{{\mathrm{2}}} \):
        By tracing back the derivation of \( \possiblyWithSub\stageImetaColor{\tau^{\superscriptI} }  \gg  \possiblyWithSub\stageOmetaColor{T^{\superscriptO} } \), we have
        \begin{center}
          \derive{%
             \possiblyWithSub\stageImetaColor{\tau^{\superscriptI} }_{{\mathrm{1}}}  \gg  \possiblyWithSub\stageOmetaColor{T^{\superscriptO} }_{{\mathrm{1}}} 
          \andalso
             \possiblyWithSub\stageImetaColor{\tau^{\superscriptI} }_{{\mathrm{2}}}  \gg  \possiblyWithSub\stageOmetaColor{T^{\superscriptO} }_{{\mathrm{2}}} 
          }{%
              \possiblyWithSub\stageImetaColor{\tau^{\superscriptI} }_{{\mathrm{1}}}  \relI{\to}  \possiblyWithSub\stageImetaColor{\tau^{\superscriptI} }_{{\mathrm{2}}}   \gg   \openO{(} \possiblyWithSub\stageOmetaColor{x}  \relO{:}  \possiblyWithSub\stageOmetaColor{T^{\superscriptO} }_{{\mathrm{1}}} \closeO{)} \relO{\to}  \possiblyWithSub\stageOmetaColor{T^{\superscriptO} }_{{\mathrm{2}}}  
          }.
        \end{center}
        Since \(\possiblyWithSub\stageOmetaColor{T^{\superscriptO} } =  \openO{(} \possiblyWithSub\stageOmetaColor{x}  \relO{:}  \possiblyWithSub\stageOmetaColor{T^{\superscriptO} }_{{\mathrm{1}}} \closeO{)} \relO{\to}  \possiblyWithSub\stageOmetaColor{T^{\superscriptO} }_{{\mathrm{2}}} \),
        by Lemmata~\ref{lem:arrow-type-csr-equiv-form} and \ref{lem:arrow-type-csr-equiv-inversion},
        \(\possiblyWithSub\stageOmetaColor{T'^{\superscriptO} }\) is of the form~\( \openO{(} \possiblyWithSub\stageOmetaColor{x}  \relO{:}  \possiblyWithSub\stageOmetaColor{T'^{\superscriptO} }_{{\mathrm{1}}} \closeO{)} \relO{\to}  \possiblyWithSub\stageOmetaColor{T'^{\superscriptO} }_{{\mathrm{2}}} \),
        and we have \( \possiblyWithSub\stageOmetaColor{T^{\superscriptO} }_{{\mathrm{1}}}  \equiv^{0}  \possiblyWithSub\stageOmetaColor{T'^{\superscriptO} }_{{\mathrm{1}}} \) and \( \possiblyWithSub\stageOmetaColor{T^{\superscriptO} }_{{\mathrm{2}}}  \equiv^{0}  \possiblyWithSub\stageOmetaColor{T'^{\superscriptO} }_{{\mathrm{2}}} \).
        Then, by IH, we have \( \possiblyWithSub\stageImetaColor{\tau^{\superscriptI} }_{{\mathrm{1}}}  \gg  \possiblyWithSub\stageOmetaColor{T'^{\superscriptO} }_{{\mathrm{1}}} \) and \( \possiblyWithSub\stageImetaColor{\tau^{\superscriptI} }_{{\mathrm{2}}}  \gg  \possiblyWithSub\stageOmetaColor{T'^{\superscriptO} }_{{\mathrm{2}}} \).
        This enables us to derive
        \begin{center}
          \derive{%
             \possiblyWithSub\stageImetaColor{\tau^{\superscriptI} }_{{\mathrm{1}}}  \gg  \possiblyWithSub\stageOmetaColor{T'^{\superscriptO} }_{{\mathrm{1}}} 
          \andalso
             \possiblyWithSub\stageImetaColor{\tau^{\superscriptI} }_{{\mathrm{2}}}  \gg  \possiblyWithSub\stageOmetaColor{T'^{\superscriptO} }_{{\mathrm{2}}} 
          }{%
              \possiblyWithSub\stageImetaColor{\tau^{\superscriptI} }_{{\mathrm{1}}}  \relI{\to}  \possiblyWithSub\stageImetaColor{\tau^{\superscriptI} }_{{\mathrm{2}}}   \gg   \openO{(} \possiblyWithSub\stageOmetaColor{x}  \relO{:}  \possiblyWithSub\stageOmetaColor{T'^{\superscriptO} }_{{\mathrm{1}}} \closeO{)} \relO{\to}  \possiblyWithSub\stageOmetaColor{T'^{\superscriptO} }_{{\mathrm{2}}}  
          }.
        \end{center}
    \end{itemize}
  \end{proof}
  \begin{lemma}[Refinement Inversion]\label{lem:refinement-inversion}
    Suppose \( \mathit{\Gamma}  \vdash^{0}   \LeftAssertParen \relO{\CastArrow}   \openO{\{} \possiblyWithSub\stageOmetaColor{\nu}  \relO{:}  \possiblyWithSub\stageOmetaColor{B}  \relO{\mid}  \possiblyWithSub\stageOmetaColor{N^{\superscriptO} } \closeO{\} }   \RightAssertParen^{ L }   :   \openO{(} \possiblyWithSub\stageOmetaColor{x}  \relO{:}  \possiblyWithSub\stageOmetaColor{T^{\superscriptO} }_{{\mathrm{1}}} \closeO{)} \relO{\to}  \possiblyWithSub\stageOmetaColor{T^{\superscriptO} }_{{\mathrm{2}}}  \). Then,
    we have \(  \possiblyWithSub\stageImetaColor{B}   \gg  \possiblyWithSub\stageOmetaColor{T^{\superscriptO} }_{{\mathrm{1}}} \), \( \possiblyWithSub\stageOmetaColor{T^{\superscriptO} }_{{\mathrm{2}}}  \equiv^{0}    \openO{\{} \possiblyWithSub\stageOmetaColor{\nu}  \relO{:}  \possiblyWithSub\stageOmetaColor{B}  \relO{\mid}  \possiblyWithSub\stageOmetaColor{N^{\superscriptO} } \closeO{\} }   \), and \( \mathit{\Gamma}  \vdash^{0}    \openO{\{} \possiblyWithSub\stageOmetaColor{\nu}  \relO{:}  \possiblyWithSub\stageOmetaColor{B}  \relO{\mid}  \possiblyWithSub\stageOmetaColor{N^{\superscriptO} } \closeO{\} }   \).
  \end{lemma}
  \begin{proof}
    By induction on the derivation of \( \mathit{\Gamma}  \vdash^{0}   \LeftAssertParen \relO{\CastArrow}   \openO{\{} \possiblyWithSub\stageOmetaColor{\nu}  \relO{:}  \possiblyWithSub\stageOmetaColor{B}  \relO{\mid}  \possiblyWithSub\stageOmetaColor{N^{\superscriptO} } \closeO{\} }   \RightAssertParen^{ L }   :   \openO{(} \possiblyWithSub\stageOmetaColor{x}  \relO{:}  \possiblyWithSub\stageOmetaColor{T^{\superscriptO} }_{{\mathrm{1}}} \closeO{)} \relO{\to}  \possiblyWithSub\stageOmetaColor{T^{\superscriptO} }_{{\mathrm{2}}}  \).
    \begin{itemize}
      \item Case \derive[T0-Rfn]{%
         \mathit{\Gamma}  \vdash^{0}    \openO{\{} \possiblyWithSub\stageOmetaColor{\nu}  \relO{:}  \possiblyWithSub\stageOmetaColor{B}  \relO{\mid}  \possiblyWithSub\stageOmetaColor{N^{\superscriptO} } \closeO{\} }   
      \andalso
         \mathit{\Gamma}  \vdash^{0}    \openO{\{} \possiblyWithSub\stageOmetaColor{\nu}  \relO{:}  \possiblyWithSub\stageOmetaColor{B}  \relO{\mid}  \possiblyWithSub\stageOmetaColor{N'^{\superscriptO} } \closeO{\} }   
      \andalso
        \possiblyWithSub\stageOmetaColor{x} \not\in \dom \mathit{\Gamma}
      }{%
         \mathit{\Gamma}  \vdash^{0}   \LeftAssertParen \relO{\CastArrow}   \openO{\{} \possiblyWithSub\stageOmetaColor{\nu}  \relO{:}  \possiblyWithSub\stageOmetaColor{B}  \relO{\mid}  \possiblyWithSub\stageOmetaColor{N^{\superscriptO} } \closeO{\} }   \RightAssertParen^{ L }   :   \openO{(} \possiblyWithSub\stageOmetaColor{x}  \relO{:}    \openO{\{} \possiblyWithSub\stageOmetaColor{\nu}  \relO{:}  \possiblyWithSub\stageOmetaColor{B}  \relO{\mid}  \possiblyWithSub\stageOmetaColor{N'^{\superscriptO} } \closeO{\} }   \closeO{)} \relO{\to}    \openO{\{} \possiblyWithSub\stageOmetaColor{\nu}  \relO{:}  \possiblyWithSub\stageOmetaColor{B}  \relO{\mid}  \possiblyWithSub\stageOmetaColor{N^{\superscriptO} } \closeO{\} }    
      }:
        Immediate.
      \item Case \derive[T0-TyEquiv]{%
         \mathit{\Gamma}  \vdash^{0}   \LeftAssertParen \relO{\CastArrow}   \openO{\{} \possiblyWithSub\stageOmetaColor{\nu}  \relO{:}  \possiblyWithSub\stageOmetaColor{B}  \relO{\mid}  \possiblyWithSub\stageOmetaColor{N^{\superscriptO} } \closeO{\} }   \RightAssertParen^{ L }   :  \possiblyWithSub\stageOmetaColor{T'^{\superscriptO} } 
      \andalso
         \possiblyWithSub\stageOmetaColor{T'^{\superscriptO} }  \equiv^{0}   \openO{(} \possiblyWithSub\stageOmetaColor{x}  \relO{:}  \possiblyWithSub\stageOmetaColor{T^{\superscriptO} }_{{\mathrm{1}}} \closeO{)} \relO{\to}  \possiblyWithSub\stageOmetaColor{T^{\superscriptO} }_{{\mathrm{2}}}  
      \andalso
         \mathit{\Gamma}  \vdash^{0}  \possiblyWithSub\stageOmetaColor{T^{\superscriptO} } 
      }{%
         \mathit{\Gamma}  \vdash^{0}   \LeftAssertParen \relO{\CastArrow}   \openO{\{} \possiblyWithSub\stageOmetaColor{\nu}  \relO{:}  \possiblyWithSub\stageOmetaColor{B}  \relO{\mid}  \possiblyWithSub\stageOmetaColor{N^{\superscriptO} } \closeO{\} }   \RightAssertParen^{ L }   :   \openO{(} \possiblyWithSub\stageOmetaColor{x}  \relO{:}  \possiblyWithSub\stageOmetaColor{T^{\superscriptO} }_{{\mathrm{1}}} \closeO{)} \relO{\to}  \possiblyWithSub\stageOmetaColor{T^{\superscriptO} }_{{\mathrm{2}}}  
      }:
        By Lemma~\ref{lem:arrow-type-csr-equiv-form}, from \( \possiblyWithSub\stageOmetaColor{T'^{\superscriptO} }  \equiv^{0}   \openO{(} \possiblyWithSub\stageOmetaColor{x}  \relO{:}  \possiblyWithSub\stageOmetaColor{T^{\superscriptO} }_{{\mathrm{1}}} \closeO{)} \relO{\to}  \possiblyWithSub\stageOmetaColor{T^{\superscriptO} }_{{\mathrm{2}}}  \),
        there exist \(\possiblyWithSub\stageOmetaColor{T'^{\superscriptO} }_{{\mathrm{1}}}\) and \(\possiblyWithSub\stageOmetaColor{T'^{\superscriptO} }_{{\mathrm{2}}}\) such that
        \(\possiblyWithSub\stageOmetaColor{T'^{\superscriptO} } =  \openO{(} \possiblyWithSub\stageOmetaColor{x}  \relO{:}  \possiblyWithSub\stageOmetaColor{T'^{\superscriptO} }_{{\mathrm{1}}} \closeO{)} \relO{\to}  \possiblyWithSub\stageOmetaColor{T'^{\superscriptO} }_{{\mathrm{2}}} \), \( \possiblyWithSub\stageOmetaColor{T^{\superscriptO} }_{{\mathrm{1}}}  \equiv^{0}  \possiblyWithSub\stageOmetaColor{T'^{\superscriptO} }_{{\mathrm{1}}} \), and \( \possiblyWithSub\stageOmetaColor{T^{\superscriptO} }_{{\mathrm{2}}}  \equiv^{0}  \possiblyWithSub\stageOmetaColor{T'^{\superscriptO} }_{{\mathrm{2}}} \).
        Then, by IH, from \( \mathit{\Gamma}  \vdash^{0}   \LeftAssertParen \relO{\CastArrow}   \openO{\{} \possiblyWithSub\stageOmetaColor{\nu}  \relO{:}  \possiblyWithSub\stageOmetaColor{B}  \relO{\mid}  \possiblyWithSub\stageOmetaColor{N^{\superscriptO} } \closeO{\} }   \RightAssertParen^{ L }   :   \openO{(} \possiblyWithSub\stageOmetaColor{x}  \relO{:}  \possiblyWithSub\stageOmetaColor{T'^{\superscriptO} }_{{\mathrm{1}}} \closeO{)} \relO{\to}  \possiblyWithSub\stageOmetaColor{T'^{\superscriptO} }_{{\mathrm{2}}}  \),
        we have \(  \possiblyWithSub\stageImetaColor{B}   \gg  \possiblyWithSub\stageOmetaColor{T'^{\superscriptO} }_{{\mathrm{1}}} \), \( \possiblyWithSub\stageOmetaColor{T'^{\superscriptO} }_{{\mathrm{2}}}  \equiv^{0}    \openO{\{} \possiblyWithSub\stageOmetaColor{\nu}  \relO{:}  \possiblyWithSub\stageOmetaColor{B}  \relO{\mid}  \possiblyWithSub\stageOmetaColor{N^{\superscriptO} } \closeO{\} }   \), and \( \mathit{\Gamma}  \vdash^{0}    \openO{\{} \possiblyWithSub\stageOmetaColor{\nu}  \relO{:}  \possiblyWithSub\stageOmetaColor{B}  \relO{\mid}  \possiblyWithSub\stageOmetaColor{N^{\superscriptO} } \closeO{\} }   \).
        Thus, by \(  \possiblyWithSub\stageImetaColor{B}   \gg  \possiblyWithSub\stageOmetaColor{T'^{\superscriptO} }_{{\mathrm{1}}} \), \( \possiblyWithSub\stageOmetaColor{T^{\superscriptO} }_{{\mathrm{1}}}  \equiv^{0}  \possiblyWithSub\stageOmetaColor{T'^{\superscriptO} }_{{\mathrm{1}}} \),
        Lemma~\ref{lem:csr-equiv-preserves-unlifting-relation}, and \rulename{CqT0-Sym},
        we have \(  \possiblyWithSub\stageImetaColor{B}   \gg  \possiblyWithSub\stageOmetaColor{T^{\superscriptO} }_{{\mathrm{1}}} \).
        Also, we can derive
        \begin{center}
          \derive[CqT0-Trans]{%
             \possiblyWithSub\stageOmetaColor{T^{\superscriptO} }_{{\mathrm{2}}}  \equiv^{0}  \possiblyWithSub\stageOmetaColor{T'^{\superscriptO} }_{{\mathrm{2}}} 
          \andalso
             \possiblyWithSub\stageOmetaColor{T'^{\superscriptO} }_{{\mathrm{2}}}  \equiv^{0}    \openO{\{} \possiblyWithSub\stageOmetaColor{\nu}  \relO{:}  \possiblyWithSub\stageOmetaColor{B}  \relO{\mid}  \possiblyWithSub\stageOmetaColor{N^{\superscriptO} } \closeO{\} }   
          }{%
             \possiblyWithSub\stageOmetaColor{T^{\superscriptO} }_{{\mathrm{2}}}  \equiv^{0}    \openO{\{} \possiblyWithSub\stageOmetaColor{\nu}  \relO{:}  \possiblyWithSub\stageOmetaColor{B}  \relO{\mid}  \possiblyWithSub\stageOmetaColor{N^{\superscriptO} } \closeO{\} }   
          }.
        \end{center}
      \item
        The other cases contradict the assumption.
    \end{itemize}
  \end{proof}
  \begin{lemma}[CSR equivalence preserves the form of code types]\label{lem:code-type-csr-equiv-form}
    \noindent
    \begin{enumerate}
      \item If \(  \openO{\langle} \possiblyWithSub\stageImetaColor{T^{\superscriptI} }_{{\mathrm{1}}} \closeO{\rangle}   \equiv^{0}  \possiblyWithSub\stageOmetaColor{T^{\superscriptO} } \), then \(\possiblyWithSub\stageOmetaColor{T^{\superscriptO} }\) is of the form~\( \openO{\langle} \possiblyWithSub\stageImetaColor{T^{\superscriptI} }_{{\mathrm{2}}} \closeO{\rangle} \).
      \item If \( \possiblyWithSub\stageOmetaColor{T^{\superscriptO} }  \equiv^{0}   \openO{\langle} \possiblyWithSub\stageImetaColor{T^{\superscriptI} }_{{\mathrm{2}}} \closeO{\rangle}  \), then \(\possiblyWithSub\stageOmetaColor{T^{\superscriptO} }\) is of the form~\( \openO{\langle} \possiblyWithSub\stageImetaColor{T^{\superscriptI} }_{{\mathrm{1}}} \closeO{\rangle} \).
    \end{enumerate}
  \end{lemma}
  \begin{proof}
    By mutual induction on the derivation.
    \begin{enumerate}
      \item
        \begin{itemize}
          \item Case \rulename{CqT0-Code} is immediate.
          \item Case \rulename{CqT0-Refl} is also immediate.
          \item Case \rulename{CqT0-Sym} is straightforward by IH.
          \item Case \derive[CqT0-Trans]{%
              \openO{\langle} \possiblyWithSub\stageImetaColor{T^{\superscriptI} }_{{\mathrm{1}}} \closeO{\rangle}   \equiv^{0}  \possiblyWithSub\stageOmetaColor{T'^{\superscriptO} } 
          \andalso
             \possiblyWithSub\stageOmetaColor{T'^{\superscriptO} }  \equiv^{0}  \possiblyWithSub\stageOmetaColor{T^{\superscriptO} } 
          }{%
              \openO{\langle} \possiblyWithSub\stageImetaColor{T^{\superscriptI} }_{{\mathrm{1}}} \closeO{\rangle}   \equiv^{0}  \possiblyWithSub\stageOmetaColor{T^{\superscriptO} } 
          }:
            By IH, from \(  \openO{\langle} \possiblyWithSub\stageImetaColor{T^{\superscriptI} }_{{\mathrm{1}}} \closeO{\rangle}   \equiv^{0}  \possiblyWithSub\stageOmetaColor{T'^{\superscriptO} } \),
            \(\possiblyWithSub\stageOmetaColor{T'^{\superscriptO} }\) is of the form~\( \openO{\langle} \possiblyWithSub\stageImetaColor{T'^{\superscriptI} } \closeO{\rangle} \).
            Then, again by IH, from \(  \openO{\langle} \possiblyWithSub\stageImetaColor{T'^{\superscriptI} } \closeO{\rangle}   \equiv^{0}  \possiblyWithSub\stageOmetaColor{T^{\superscriptO} } \),
            \(\possiblyWithSub\stageOmetaColor{T^{\superscriptO} }\) is of the form~\( \openO{\langle} \possiblyWithSub\stageImetaColor{T^{\superscriptI} }_{{\mathrm{2}}} \closeO{\rangle} \).
          \item The other cases contradict the assumption.
        \end{itemize}
      \item
        Can be proved in the same manner as (1).
    \end{enumerate}
  \end{proof}
  \begin{lemma}[Code Type CSR Equivalence Inversion]\label{lem:code-type-csr-equiv-inversion}
    \(  \openO{\langle} \possiblyWithSub\stageImetaColor{T^{\superscriptI} }_{{\mathrm{1}}} \closeO{\rangle}   \equiv^{0}   \openO{\langle} \possiblyWithSub\stageImetaColor{T^{\superscriptI} }_{{\mathrm{2}}} \closeO{\rangle}  \) implies \( \possiblyWithSub\stageImetaColor{T^{\superscriptI} }_{{\mathrm{1}}}  \equiv^{1}  \possiblyWithSub\stageImetaColor{T^{\superscriptI} }_{{\mathrm{2}}} \).
  \end{lemma}
  \begin{proof}
    By induction on the derivation of \(  \openO{\langle} \possiblyWithSub\stageImetaColor{T^{\superscriptI} }_{{\mathrm{1}}} \closeO{\rangle}   \equiv^{0}   \openO{\langle} \possiblyWithSub\stageImetaColor{T^{\superscriptI} }_{{\mathrm{2}}} \closeO{\rangle}  \).
    \begin{itemize}
      \item Case \derive[CqT0-Trans]{%
          \openO{\langle} \possiblyWithSub\stageImetaColor{T^{\superscriptI} }_{{\mathrm{1}}} \closeO{\rangle}   \equiv^{0}  \possiblyWithSub\stageOmetaColor{T'^{\superscriptO} } 
      \andalso
         \possiblyWithSub\stageOmetaColor{T'^{\superscriptO} }  \equiv^{0}   \openO{\langle} \possiblyWithSub\stageImetaColor{T^{\superscriptI} }_{{\mathrm{2}}} \closeO{\rangle}  
      }{%
          \openO{\langle} \possiblyWithSub\stageImetaColor{T^{\superscriptI} }_{{\mathrm{1}}} \closeO{\rangle}   \equiv^{0}   \openO{\langle} \possiblyWithSub\stageImetaColor{T^{\superscriptI} }_{{\mathrm{2}}} \closeO{\rangle}  
      }:
        By Lemma~\ref{lem:code-type-csr-equiv-form}, from \(  \openO{\langle} \possiblyWithSub\stageImetaColor{T^{\superscriptI} }_{{\mathrm{1}}} \closeO{\rangle}   \equiv^{0}  \possiblyWithSub\stageOmetaColor{T'^{\superscriptO} } \),
        \(\possiblyWithSub\stageOmetaColor{T'^{\superscriptO} }\) is of the form~\( \openO{\langle} \possiblyWithSub\stageImetaColor{T'^{\superscriptI} } \closeO{\rangle} \).
        Then, by IH, \( \possiblyWithSub\stageImetaColor{T^{\superscriptI} }_{{\mathrm{1}}}  \equiv^{1}  \possiblyWithSub\stageImetaColor{T'^{\superscriptI} } \) holds.
        Again by IH, from \(  \openO{\langle} \possiblyWithSub\stageImetaColor{T'^{\superscriptI} } \closeO{\rangle}   \equiv^{0}   \openO{\langle} \possiblyWithSub\stageImetaColor{T^{\superscriptI} }_{{\mathrm{2}}} \closeO{\rangle}  \),
        we have \( \possiblyWithSub\stageImetaColor{T'^{\superscriptI} }  \equiv^{1}  \possiblyWithSub\stageImetaColor{T^{\superscriptI} }_{{\mathrm{2}}} \).
        Therefore, we can derive
        \begin{center}
          \infer[CqT1-Trans]{%
             \possiblyWithSub\stageImetaColor{T^{\superscriptI} }_{{\mathrm{1}}}  \equiv^{1}  \possiblyWithSub\stageImetaColor{T'^{\superscriptI} } 
          \andalso
             \possiblyWithSub\stageImetaColor{T'^{\superscriptI} }  \equiv^{1}  \possiblyWithSub\stageImetaColor{T^{\superscriptI} }_{{\mathrm{2}}} 
          }{%
             \possiblyWithSub\stageImetaColor{T^{\superscriptI} }_{{\mathrm{1}}}  \equiv^{1}  \possiblyWithSub\stageImetaColor{T^{\superscriptI} }_{{\mathrm{2}}} 
          }.
        \end{center}
      \item The other cases are straightforward.
    \end{itemize}
  \end{proof}
  \begin{lemma}[Reduction preserves compatibility]\label{lem:eval-preserves-compatibility}
    \noindent
    \begin{enumerate}
      \item If \( \possiblyWithSub\stageImetaColor{T^{\superscriptI} }_{{\mathrm{1}}}  \mathrel{||}^{1}  \possiblyWithSub\stageImetaColor{T^{\superscriptI} }_{{\mathrm{2}}} \) and \( \possiblyWithSub\stageImetaColor{T^{\superscriptI} }_{{\mathrm{1}}}  \longrightarrow^{1}   \possiblyWithSub\stageImetaColor{T'^{\superscriptI} }_{{\mathrm{1}}}  \), then \( \possiblyWithSub\stageImetaColor{T'^{\superscriptI} }_{{\mathrm{1}}}  \mathrel{||}^{1}  \possiblyWithSub\stageImetaColor{T^{\superscriptI} }_{{\mathrm{2}}} \).
      \item If \( \possiblyWithSub\stageImetaColor{T^{\superscriptI} }_{{\mathrm{1}}}  \mathrel{||}^{1}  \possiblyWithSub\stageImetaColor{T^{\superscriptI} }_{{\mathrm{2}}} \) and \( \possiblyWithSub\stageImetaColor{T^{\superscriptI} }_{{\mathrm{2}}}  \longrightarrow^{1}   \possiblyWithSub\stageImetaColor{T'^{\superscriptI} }_{{\mathrm{2}}}  \), then \( \possiblyWithSub\stageImetaColor{T^{\superscriptI} }_{{\mathrm{1}}}  \mathrel{||}^{1}  \possiblyWithSub\stageImetaColor{T'^{\superscriptI} }_{{\mathrm{2}}} \).
    \end{enumerate}
  \end{lemma}
  \begin{proof}
    By straightforward induction on the derivation.
  \end{proof}
  \begin{lemma}[CSR equivalence preserves the order of types]\label{lem:csr-equiv-preserves-order}
    \noindent
    \begin{enumerate}
      \item
        If \( \vdash^{0}_{\mathrm{dom} }  \possiblyWithSub\stageOmetaColor{T^{\superscriptO} } \) and \( \possiblyWithSub\stageOmetaColor{T^{\superscriptO} }  \equiv^{0}  \possiblyWithSub\stageOmetaColor{T'^{\superscriptO} } \), then \( \vdash^{0}_{\mathrm{dom} }  \possiblyWithSub\stageOmetaColor{T'^{\superscriptO} } \).
      \item
        If \( \vdash^{0}_{\mathrm{fo} }  \possiblyWithSub\stageOmetaColor{T^{\superscriptO} } \) and \( \possiblyWithSub\stageOmetaColor{T^{\superscriptO} }  \equiv^{0}  \possiblyWithSub\stageOmetaColor{T'^{\superscriptO} } \), then \( \vdash^{0}_{\mathrm{fo} }  \possiblyWithSub\stageOmetaColor{T'^{\superscriptO} } \).
    \end{enumerate}
  \end{lemma}
  \begin{proof}
    \begin{enumerate}
      \item
        By tracing back the derivation of \( \vdash^{0}_{\mathrm{dom} }  \possiblyWithSub\stageOmetaColor{T^{\superscriptO} } \),
        we have the following two cases:
        \begin{itemize}
          \item Case \(\possiblyWithSub\stageOmetaColor{T^{\superscriptO} } =  \openO{\{} \possiblyWithSub\stageOmetaColor{\nu}  \relO{:}  \possiblyWithSub\stageOmetaColor{B}  \relO{\mid}  \possiblyWithSub\stageOmetaColor{N^{\superscriptO} } \closeO{\} } \):
            By Lemma~\ref{lem:rfn-type-csr-equiv-form},
            \(\possiblyWithSub\stageOmetaColor{T'^{\superscriptO} }\) is of the form~\( \openO{\{} \possiblyWithSub\stageOmetaColor{\nu}  \relO{:}  \possiblyWithSub\stageOmetaColor{B}  \relO{\mid}  \possiblyWithSub\stageOmetaColor{N'^{\superscriptO} } \closeO{\} } \),
            which makes \( \vdash^{0}_{\mathrm{dom} }  \possiblyWithSub\stageOmetaColor{T'^{\superscriptO} } \) derivable.
          \item Case \(\possiblyWithSub\stageOmetaColor{T^{\superscriptO} } =  \ttO{Tensor}\  \possiblyWithSub\stageOmetaColor{s} \):
            By Lemma~\ref{lem:tensor-type-csr-equiv-form},
            we have \(\possiblyWithSub\stageOmetaColor{T'^{\superscriptO} } =  \ttO{Tensor}\  \possiblyWithSub\stageOmetaColor{s} \) and thereby \( \vdash^{0}_{\mathrm{dom} }  \possiblyWithSub\stageOmetaColor{T'^{\superscriptO} } \).
        \end{itemize}
      \item
        By induction on the derivation of \( \vdash^{0}_{\mathrm{fo} }  \possiblyWithSub\stageOmetaColor{T^{\superscriptO} } \).
        \begin{itemize}
          \item Case \derive{%
             \vdash^{0}_{\mathrm{dom} }  \possiblyWithSub\stageOmetaColor{T^{\superscriptO} }_{{\mathrm{1}}} 
          \andalso
             \vdash^{0}_{\mathrm{fo} }  \possiblyWithSub\stageOmetaColor{T^{\superscriptO} }_{{\mathrm{2}}} 
          }{%
             \vdash^{0}_{\mathrm{fo} }   \openO{(} \possiblyWithSub\stageOmetaColor{x}  \relO{:}  \possiblyWithSub\stageOmetaColor{T^{\superscriptO} }_{{\mathrm{1}}} \closeO{)} \relO{\to}  \possiblyWithSub\stageOmetaColor{T^{\superscriptO} }_{{\mathrm{2}}}  
          }:
            By Lemmata~\ref{lem:arrow-type-csr-equiv-form} and \ref{lem:arrow-type-csr-equiv-inversion},
            \(\possiblyWithSub\stageOmetaColor{T'^{\superscriptO} }\) is of the form~\( \openO{(} \possiblyWithSub\stageOmetaColor{x}  \relO{:}  \possiblyWithSub\stageOmetaColor{T'^{\superscriptO} }_{{\mathrm{1}}} \closeO{)} \relO{\to}  \possiblyWithSub\stageOmetaColor{T'^{\superscriptO} }_{{\mathrm{2}}} \),
            and we have \( \possiblyWithSub\stageOmetaColor{T^{\superscriptO} }_{{\mathrm{1}}}  \equiv^{0}  \possiblyWithSub\stageOmetaColor{T'^{\superscriptO} }_{{\mathrm{1}}} \) and \( \possiblyWithSub\stageOmetaColor{T^{\superscriptO} }_{{\mathrm{2}}}  \equiv^{0}  \possiblyWithSub\stageOmetaColor{T'^{\superscriptO} }_{{\mathrm{2}}} \).
            Then, by (1), we first have \( \vdash^{0}_{\mathrm{dom} }  \possiblyWithSub\stageOmetaColor{T'^{\superscriptO} }_{{\mathrm{1}}} \).
            Also, by IH, we have \( \vdash^{0}_{\mathrm{fo} }  \possiblyWithSub\stageOmetaColor{T'^{\superscriptO} }_{{\mathrm{2}}} \).
            Therefore, we can derive:
            \begin{center}
              \derive{%
                 \vdash^{0}_{\mathrm{dom} }  \possiblyWithSub\stageOmetaColor{T'^{\superscriptO} }_{{\mathrm{1}}} 
              \andalso
                 \vdash^{0}_{\mathrm{fo} }  \possiblyWithSub\stageOmetaColor{T'^{\superscriptO} }_{{\mathrm{2}}} 
              }{%
                 \vdash^{0}_{\mathrm{fo} }   \openO{(} \possiblyWithSub\stageOmetaColor{x}  \relO{:}  \possiblyWithSub\stageOmetaColor{T'^{\superscriptO} }_{{\mathrm{1}}} \closeO{)} \relO{\to}  \possiblyWithSub\stageOmetaColor{T'^{\superscriptO} }_{{\mathrm{2}}}  
              }.
            \end{center}
          \item Case \derive{%
             \vdash^{0}_{\mathrm{dom} }  \possiblyWithSub\stageOmetaColor{T^{\superscriptO} } 
          }{%
             \vdash^{0}_{\mathrm{fo} }  \possiblyWithSub\stageOmetaColor{T^{\superscriptO} } 
          }:
            Straightforward by IH.
          \item Case \derive{}{%
             \vdash^{0}_{\mathrm{fo} }   \openO{\langle} \possiblyWithSub\stageImetaColor{T^{\superscriptI} } \closeO{\rangle}  
          }:
            By Lemma~\ref{lem:code-type-csr-equiv-form},
            \(\possiblyWithSub\stageOmetaColor{T'^{\superscriptO} }\) is of the form~\( \openO{\langle} \possiblyWithSub\stageImetaColor{T'^{\superscriptI} } \closeO{\rangle} \),
            which makes \( \vdash^{0}_{\mathrm{fo} }  \possiblyWithSub\stageOmetaColor{T'^{\superscriptO} } \) hold.
        \end{itemize}
    \end{enumerate}
  \end{proof}
  \begin{lemma}[Substitution preserves the order of types]\label{lem:subst-preserves-order}
    \noindent
    \begin{enumerate}
      \item
        \( \vdash^{0}_{\mathrm{dom} }  \possiblyWithSub\stageOmetaColor{T^{\superscriptO} } \) implies \( \vdash^{0}_{\mathrm{dom} }    [  \possiblyWithSub\stageOmetaColor{N^{\superscriptO} }  /  \possiblyWithSub\stageOmetaColor{x}  ]    \possiblyWithSub\stageOmetaColor{T^{\superscriptO} }  \).
      \item
        \( \vdash^{0}_{\mathrm{fo} }  \possiblyWithSub\stageOmetaColor{T^{\superscriptO} } \) implies \( \vdash^{0}_{\mathrm{fo} }    [  \possiblyWithSub\stageOmetaColor{N^{\superscriptO} }  /  \possiblyWithSub\stageOmetaColor{x}  ]    \possiblyWithSub\stageOmetaColor{T^{\superscriptO} }  \).
    \end{enumerate}
  \end{lemma}
  \begin{proof}
    By straightforward induction on the derivation.
  \end{proof}
  \begin{lemma}\label{lem:non-code-order-zero-domain}
    If \( \mathit{\Gamma}  \vdash^{0}   \possiblyWithSub\stageOmetaColor{a}   :   \openO{(} \possiblyWithSub\stageOmetaColor{x}  \relO{:}  \possiblyWithSub\stageOmetaColor{T^{\superscriptO} }_{{\mathrm{1}}} \closeO{)} \relO{\to}  \possiblyWithSub\stageOmetaColor{T^{\superscriptO} }_{{\mathrm{2}}}  \),
    then we have \( \vdash^{0}_{\mathrm{dom} }  \possiblyWithSub\stageOmetaColor{T^{\superscriptO} }_{{\mathrm{1}}} \) and \( \vdash^{0}_{\mathrm{fo} }  \possiblyWithSub\stageOmetaColor{T^{\superscriptO} }_{{\mathrm{2}}} \).
  \end{lemma}
  \begin{proof}
    by induction on the derivation of \( \mathit{\Gamma}  \vdash^{0}   \possiblyWithSub\stageOmetaColor{a}   :   \openO{(} \possiblyWithSub\stageOmetaColor{x}  \relO{:}  \possiblyWithSub\stageOmetaColor{T^{\superscriptO} }_{{\mathrm{1}}} \closeO{)} \relO{\to}  \possiblyWithSub\stageOmetaColor{T^{\superscriptO} }_{{\mathrm{2}}}  \).
    \begin{itemize}
      \item Case \derive[T0-TyEquiv]{%
         \mathit{\Gamma}  \vdash^{0}   \possiblyWithSub\stageOmetaColor{a}   :  \possiblyWithSub\stageOmetaColor{T'^{\superscriptO} } 
      \andalso
         \possiblyWithSub\stageOmetaColor{T'^{\superscriptO} }  \equiv^{0}   \openO{(} \possiblyWithSub\stageOmetaColor{x}  \relO{:}  \possiblyWithSub\stageOmetaColor{T^{\superscriptO} }_{{\mathrm{1}}} \closeO{)} \relO{\to}  \possiblyWithSub\stageOmetaColor{T^{\superscriptO} }_{{\mathrm{2}}}  
      \andalso
         \mathit{\Gamma}  \vdash^{0}   \openO{(} \possiblyWithSub\stageOmetaColor{x}  \relO{:}  \possiblyWithSub\stageOmetaColor{T^{\superscriptO} }_{{\mathrm{1}}} \closeO{)} \relO{\to}  \possiblyWithSub\stageOmetaColor{T^{\superscriptO} }_{{\mathrm{2}}}  
      }{%
         \mathit{\Gamma}  \vdash^{0}   \possiblyWithSub\stageOmetaColor{a}   :   \openO{(} \possiblyWithSub\stageOmetaColor{x}  \relO{:}  \possiblyWithSub\stageOmetaColor{T^{\superscriptO} }_{{\mathrm{1}}} \closeO{)} \relO{\to}  \possiblyWithSub\stageOmetaColor{T^{\superscriptO} }_{{\mathrm{2}}}  
      }:
        By Lemma~\ref{lem:arrow-type-csr-equiv-form},
        \(\possiblyWithSub\stageOmetaColor{T'^{\superscriptO} }\) is of the form~\( \openO{(} \possiblyWithSub\stageOmetaColor{x}  \relO{:}  \possiblyWithSub\stageOmetaColor{T'^{\superscriptO} }_{{\mathrm{1}}} \closeO{)} \relO{\to}  \possiblyWithSub\stageOmetaColor{T'^{\superscriptO} }_{{\mathrm{2}}} \).
        Then, by IH on \( \mathit{\Gamma}  \vdash^{0}   \possiblyWithSub\stageOmetaColor{a}   :  \possiblyWithSub\stageOmetaColor{T'^{\superscriptO} } \),
        we have \( \vdash^{0}_{\mathrm{dom} }  \possiblyWithSub\stageOmetaColor{T'^{\superscriptO} }_{{\mathrm{1}}} \).
        By Lemma~\ref{lem:arrow-type-csr-equiv-inversion},
        we have \( \possiblyWithSub\stageOmetaColor{T'^{\superscriptO} }_{{\mathrm{1}}}  \equiv^{0}  \possiblyWithSub\stageOmetaColor{T^{\superscriptO} }_{{\mathrm{1}}} \) and \( \possiblyWithSub\stageOmetaColor{T'^{\superscriptO} }_{{\mathrm{2}}}  \equiv^{0}  \possiblyWithSub\stageOmetaColor{T^{\superscriptO} }_{{\mathrm{2}}} \),
        and thus, by Lemma~\ref{lem:csr-equiv-preserves-order},
        \( \vdash^{0}_{\mathrm{dom} }  \possiblyWithSub\stageOmetaColor{T^{\superscriptO} }_{{\mathrm{1}}} \) and \( \vdash^{0}_{\mathrm{fo} }  \possiblyWithSub\stageOmetaColor{T^{\superscriptO} }_{{\mathrm{2}}} \) also hold.
      \item Case \derive[T0-App]{%
         \mathit{\Gamma}  \vdash^{0}   \possiblyWithSub\stageOmetaColor{a}_{{\mathrm{1}}}   :   \openO{(} \possiblyWithSub\stageOmetaColor{x'}  \relO{:}  \possiblyWithSub\stageOmetaColor{T^{\superscriptO} }_{{\mathrm{11}}} \closeO{)} \relO{\to}  \possiblyWithSub\stageOmetaColor{T'^{\superscriptO} }  
      \andalso
         \mathit{\Gamma}  \vdash^{0}    \possiblyWithSub\stageOmetaColor{c}_{{\mathrm{2}}}    :  \possiblyWithSub\stageOmetaColor{T^{\superscriptO} }_{{\mathrm{11}}} 
      }{%
         \mathit{\Gamma}  \vdash^{0}    \possiblyWithSub\stageOmetaColor{a}_{{\mathrm{1}}}  \    \possiblyWithSub\stageOmetaColor{c}_{{\mathrm{2}}}     :    [    \possiblyWithSub\stageOmetaColor{c}_{{\mathrm{2}}}    /  \possiblyWithSub\stageOmetaColor{x'}  ]    \possiblyWithSub\stageOmetaColor{T'^{\superscriptO} }  
      }:
        Since \( \openO{(} \possiblyWithSub\stageOmetaColor{x}  \relO{:}  \possiblyWithSub\stageOmetaColor{T^{\superscriptO} }_{{\mathrm{1}}} \closeO{)} \relO{\to}  \possiblyWithSub\stageOmetaColor{T^{\superscriptO} }_{{\mathrm{2}}}  =   [    \possiblyWithSub\stageOmetaColor{c}_{{\mathrm{2}}}    /  \possiblyWithSub\stageOmetaColor{x'}  ]    \possiblyWithSub\stageOmetaColor{T'^{\superscriptO} } \) holds
        (and we can safely assume \(\possiblyWithSub\stageOmetaColor{x} \neq \possiblyWithSub\stageOmetaColor{x'}\) w.l.o.g. by the Barendregt convention),
        \(\possiblyWithSub\stageOmetaColor{T'^{\superscriptO} }\) is of the form~\( \openO{(} \possiblyWithSub\stageOmetaColor{x}  \relO{:}  \possiblyWithSub\stageOmetaColor{T'^{\superscriptO} }_{{\mathrm{1}}} \closeO{)} \relO{\to}  \possiblyWithSub\stageOmetaColor{T'^{\superscriptO} }_{{\mathrm{2}}} \),
        and we have \(\possiblyWithSub\stageOmetaColor{T^{\superscriptO} }_{{\mathrm{1}}} =   [    \possiblyWithSub\stageOmetaColor{c}_{{\mathrm{2}}}    /  \possiblyWithSub\stageOmetaColor{x'}  ]    \possiblyWithSub\stageOmetaColor{T'^{\superscriptO} }_{{\mathrm{1}}} \) and \(\possiblyWithSub\stageOmetaColor{T^{\superscriptO} }_{{\mathrm{2}}} =   [    \possiblyWithSub\stageOmetaColor{c}_{{\mathrm{2}}}    /  \possiblyWithSub\stageOmetaColor{x'}  ]    \possiblyWithSub\stageOmetaColor{T'^{\superscriptO} }_{{\mathrm{2}}} \).
        By IH, \( \vdash^{0}_{\mathrm{fo} }  \possiblyWithSub\stageOmetaColor{T'^{\superscriptO} } \) holds, and we thereby have
        \( \vdash^{0}_{\mathrm{dom} }  \possiblyWithSub\stageOmetaColor{T'^{\superscriptO} }_{{\mathrm{1}}} \) and \( \vdash^{0}_{\mathrm{fo} }  \possiblyWithSub\stageOmetaColor{T'^{\superscriptO} }_{{\mathrm{2}}} \).
        Therefore, by Lemma~\ref{lem:subst-preserves-order},
        we also have \( \vdash^{0}_{\mathrm{dom} }  \possiblyWithSub\stageOmetaColor{T^{\superscriptO} }_{{\mathrm{1}}} \) and \( \vdash^{0}_{\mathrm{fo} }  \possiblyWithSub\stageOmetaColor{T^{\superscriptO} }_{{\mathrm{2}}} \).
      \item Case \derive[T0-CstP]{%
         \vdash  \mathit{\Gamma} 
      \andalso
        \ConstEnvPers(c) = \possiblyWithSub\stageImetaColor{\tau^{\superscriptI} }
      }{%
         \mathit{\Gamma}  \vdash^{0}    \possiblyWithSub\stageOmetaColor{c}    :   \mathop{\downarrow}( \possiblyWithSub\stageImetaColor{\tau^{\superscriptI} } )  
      }:
        By Assumption~\ref{assump:type-of-constants} and \( \mathop{\downarrow}( \possiblyWithSub\stageImetaColor{\tau^{\superscriptI} } )  =  \openO{(} \possiblyWithSub\stageOmetaColor{x}  \relO{:}  \possiblyWithSub\stageOmetaColor{T^{\superscriptO} }_{{\mathrm{1}}} \closeO{)} \relO{\to}  \possiblyWithSub\stageOmetaColor{T^{\superscriptO} }_{{\mathrm{2}}} \),
        we clearly have \( \vdash^{0}_{\mathrm{dom} }  \possiblyWithSub\stageOmetaColor{T^{\superscriptO} }_{{\mathrm{1}}} \) and \( \vdash^{0}_{\mathrm{fo} }  \possiblyWithSub\stageOmetaColor{T^{\superscriptO} }_{{\mathrm{2}}} \).
      \item Case \derive[T0-Cst0]{%
         \vdash  \mathit{\Gamma} 
      \andalso
        \ConstEnvZero(\possiblyWithSub\stageOmetaColor{p}) =  \openO{(} \possiblyWithSub\stageOmetaColor{x}  \relO{:}  \possiblyWithSub\stageOmetaColor{T^{\superscriptO} }_{{\mathrm{1}}} \closeO{)} \relO{\to}  \possiblyWithSub\stageOmetaColor{T^{\superscriptO} }_{{\mathrm{2}}} 
      }{%
         \mathit{\Gamma}  \vdash^{0}    \possiblyWithSub\stageOmetaColor{p}    :   \openO{(} \possiblyWithSub\stageOmetaColor{x}  \relO{:}  \possiblyWithSub\stageOmetaColor{T^{\superscriptO} }_{{\mathrm{1}}} \closeO{)} \relO{\to}  \possiblyWithSub\stageOmetaColor{T^{\superscriptO} }_{{\mathrm{2}}}  
      }:
        Similarly to the previous case,
        by Assumption~\ref{assump:type-of-constants},
        we have \( \vdash^{0}_{\mathrm{dom} }  \possiblyWithSub\stageOmetaColor{T^{\superscriptO} }_{{\mathrm{1}}} \) and \( \vdash^{0}_{\mathrm{fo} }  \possiblyWithSub\stageOmetaColor{T^{\superscriptO} }_{{\mathrm{2}}} \).
      \item The other cases contradict the assumption.
    \end{itemize}
  \end{proof}
  \begin{lemma}\label{lem:inversion-of-cpo}
    If \( \mathit{\Gamma}  \vdash^{0}    \possiblyWithSub\stageOmetaColor{c}    :  \possiblyWithSub\stageOmetaColor{T^{\superscriptO} } \), then there exists \(\possiblyWithSub\stageImetaColor{\tau^{\superscriptI} }\) such that
    \(\ConstEnvPers(c) = \possiblyWithSub\stageImetaColor{\tau^{\superscriptI} }\) and \( \possiblyWithSub\stageImetaColor{\tau^{\superscriptI} }  \gg  \possiblyWithSub\stageOmetaColor{T^{\superscriptO} } \).
  \end{lemma}
  \begin{proof}
    By induction on the derivation of \( \mathit{\Gamma}  \vdash^{0}    \possiblyWithSub\stageOmetaColor{c}    :  \possiblyWithSub\stageOmetaColor{T^{\superscriptO} } \).
    \begin{itemize}
      \item Case \derive[T0-TyEquiv]{%
         \mathit{\Gamma}  \vdash^{0}    \possiblyWithSub\stageOmetaColor{c}    :  \possiblyWithSub\stageOmetaColor{T'^{\superscriptO} } 
      \andalso
         \possiblyWithSub\stageOmetaColor{T'^{\superscriptO} }  \equiv^{0}  \possiblyWithSub\stageOmetaColor{T^{\superscriptO} } 
      \andalso
         \mathit{\Gamma}  \vdash^{0}  \possiblyWithSub\stageOmetaColor{T^{\superscriptO} } 
      }{%
         \mathit{\Gamma}  \vdash^{0}    \possiblyWithSub\stageOmetaColor{c}    :  \possiblyWithSub\stageOmetaColor{T^{\superscriptO} } 
      }:
        By IH on \( \mathit{\Gamma}  \vdash^{0}    \possiblyWithSub\stageOmetaColor{c}    :  \possiblyWithSub\stageOmetaColor{T'^{\superscriptO} } \), there exists \(\possiblyWithSub\stageImetaColor{\tau^{\superscriptI} }\) such that
        \(\ConstEnvPers(\possiblyWithSub\stageOmetaColor{c}) = \possiblyWithSub\stageImetaColor{\tau^{\superscriptI} }\) and \( \possiblyWithSub\stageImetaColor{\tau^{\superscriptI} }  \gg  \possiblyWithSub\stageOmetaColor{T'^{\superscriptO} } \).
        Thus, by Lemma~\ref{lem:csr-equiv-preserves-unlifting-relation},
        we have \( \possiblyWithSub\stageImetaColor{\tau^{\superscriptI} }  \gg  \possiblyWithSub\stageOmetaColor{T^{\superscriptO} } \).
      \item Case \derive[T0-RfnPred]{%
         \mathit{\Gamma}  \vdash^{0}    \openO{\{} \possiblyWithSub\stageOmetaColor{\nu}  \relO{:}  \possiblyWithSub\stageOmetaColor{B}  \relO{\mid}  \possiblyWithSub\stageOmetaColor{N^{\superscriptO} } \closeO{\} }   
      \andalso
        \ConstEnvPers(c) = \possiblyWithSub\stageImetaColor{B}
      \andalso
           [    \possiblyWithSub\stageOmetaColor{c}    /  \possiblyWithSub\stageOmetaColor{\nu}  ]    \possiblyWithSub\stageOmetaColor{N^{\superscriptO} }   \longrightarrow^{0\,\ast}      \ttO{true}     
      }{%
         \mathit{\Gamma}  \vdash^{0}    \possiblyWithSub\stageOmetaColor{c}    :    \openO{\{} \possiblyWithSub\stageOmetaColor{\nu}  \relO{:}  \possiblyWithSub\stageOmetaColor{B}  \relO{\mid}  \possiblyWithSub\stageOmetaColor{N^{\superscriptO} } \closeO{\} }   
      }:
        Immediate.
      \item Case \derive[T0-CstP]{%
         \vdash  \mathit{\Gamma} 
      \andalso
        \ConstEnvPers(c) = \possiblyWithSub\stageImetaColor{\tau^{\superscriptI} }
      }{%
         \mathit{\Gamma}  \vdash^{0}    \possiblyWithSub\stageOmetaColor{c}    :   \mathop{\downarrow}( \possiblyWithSub\stageImetaColor{\tau^{\superscriptI} } )  
      }:
        This is also immediate.
      \item The other cases contradict the assumption.
    \end{itemize}
  \end{proof}
  \begin{lemma}\label{lem:type-value-equal-by-stage-0-type}
    If \( \possiblyWithSub\stageImetaColor{\tau^{\superscriptI} }_{{\mathrm{1}}}  \gg  \possiblyWithSub\stageOmetaColor{T^{\superscriptO} } \) and \( \possiblyWithSub\stageImetaColor{\tau^{\superscriptI} }_{{\mathrm{2}}}  \gg  \possiblyWithSub\stageOmetaColor{T^{\superscriptO} } \), then \(\possiblyWithSub\stageImetaColor{\tau^{\superscriptI} }_{{\mathrm{1}}} = \possiblyWithSub\stageImetaColor{\tau^{\superscriptI} }_{{\mathrm{2}}}\).
  \end{lemma}
  \begin{proof}
    By induction on the structure of \(\possiblyWithSub\stageOmetaColor{T^{\superscriptO} }\).
    \begin{itemize}
      \item Case \(\possiblyWithSub\stageOmetaColor{T^{\superscriptO} } =  \openO{\{} \possiblyWithSub\stageOmetaColor{\nu}  \relO{:}  \possiblyWithSub\stageOmetaColor{B}  \relO{\mid}  \possiblyWithSub\stageOmetaColor{N^{\superscriptO} } \closeO{\} } \):
        By the definition of \(\gg\),
        we have \(\possiblyWithSub\stageImetaColor{\tau^{\superscriptI} }_{{\mathrm{1}}} = \possiblyWithSub\stageImetaColor{B}\) and \(\possiblyWithSub\stageImetaColor{\tau^{\superscriptI} }_{{\mathrm{2}}} = \possiblyWithSub\stageImetaColor{B}\).
      \item Case \(\possiblyWithSub\stageOmetaColor{T^{\superscriptO} } =  \ttI{Tensor}\ \ordI{\%} \possiblyWithSub\stageOmetaColor{N^{\superscriptO} } \):
        By the definition of \(\gg\),
        \(\possiblyWithSub\stageOmetaColor{N^{\superscriptO} }\) is of the form~\(\possiblyWithSub\stageOmetaColor{s}\),
        and we have \(\possiblyWithSub\stageImetaColor{\tau^{\superscriptI} }_{{\mathrm{1}}} =  \ttO{Tensor}\  \possiblyWithSub\stageOmetaColor{s} \) and \(\possiblyWithSub\stageImetaColor{\tau^{\superscriptI} }_{{\mathrm{2}}} =  \ttO{Tensor}\  \possiblyWithSub\stageOmetaColor{s} \).
      \item Case \(\possiblyWithSub\stageOmetaColor{T^{\superscriptO} } =  \openO{(} \possiblyWithSub\stageOmetaColor{x}  \relO{:}  \possiblyWithSub\stageOmetaColor{T^{\superscriptO} }_{{\mathrm{1}}} \closeO{)} \relO{\to}  \possiblyWithSub\stageOmetaColor{T^{\superscriptO} }_{{\mathrm{2}}} \):
        By tracing back the derivations of \( \possiblyWithSub\stageImetaColor{\tau^{\superscriptI} }_{{\mathrm{1}}}  \gg  \possiblyWithSub\stageOmetaColor{T^{\superscriptO} } \) and \( \possiblyWithSub\stageImetaColor{\tau^{\superscriptI} }_{{\mathrm{2}}}  \gg  \possiblyWithSub\stageOmetaColor{T^{\superscriptO} } \) as follows,
        it turns out that \(\possiblyWithSub\stageImetaColor{\tau^{\superscriptI} }_{{\mathrm{1}}}\) and \(\possiblyWithSub\stageImetaColor{\tau^{\superscriptI} }_{{\mathrm{2}}}\) are
        of the forms~\( \possiblyWithSub\stageImetaColor{\tau^{\superscriptI} }_{{\mathrm{11}}}  \relI{\to}  \possiblyWithSub\stageImetaColor{\tau^{\superscriptI} }_{{\mathrm{12}}} \) and \( \possiblyWithSub\stageImetaColor{\tau^{\superscriptI} }_{{\mathrm{21}}}  \relI{\to}  \possiblyWithSub\stageImetaColor{\tau^{\superscriptI} }_{{\mathrm{22}}} \), respectively:
        \begin{center}
          \derive{%
             \possiblyWithSub\stageImetaColor{\tau^{\superscriptI} }_{{\mathrm{11}}}  \gg  \possiblyWithSub\stageOmetaColor{T^{\superscriptO} }_{{\mathrm{1}}} 
          \andalso
             \possiblyWithSub\stageImetaColor{\tau^{\superscriptI} }_{{\mathrm{12}}}  \gg  \possiblyWithSub\stageOmetaColor{T^{\superscriptO} }_{{\mathrm{2}}} 
          }{%
             \possiblyWithSub\stageImetaColor{\tau^{\superscriptI} }_{{\mathrm{1}}}  \gg   \openO{(} \possiblyWithSub\stageOmetaColor{x}  \relO{:}  \possiblyWithSub\stageOmetaColor{T^{\superscriptO} }_{{\mathrm{1}}} \closeO{)} \relO{\to}  \possiblyWithSub\stageOmetaColor{T^{\superscriptO} }_{{\mathrm{2}}}  
          },
        \qquad
          \derive{%
             \possiblyWithSub\stageImetaColor{\tau^{\superscriptI} }_{{\mathrm{21}}}  \gg  \possiblyWithSub\stageOmetaColor{T^{\superscriptO} }_{{\mathrm{1}}} 
          \andalso
             \possiblyWithSub\stageImetaColor{\tau^{\superscriptI} }_{{\mathrm{22}}}  \gg  \possiblyWithSub\stageOmetaColor{T^{\superscriptO} }_{{\mathrm{2}}} 
          }{%
             \possiblyWithSub\stageImetaColor{\tau^{\superscriptI} }_{{\mathrm{2}}}  \gg   \openO{(} \possiblyWithSub\stageOmetaColor{x}  \relO{:}  \possiblyWithSub\stageOmetaColor{T^{\superscriptO} }_{{\mathrm{1}}} \closeO{)} \relO{\to}  \possiblyWithSub\stageOmetaColor{T^{\superscriptO} }_{{\mathrm{2}}}  
          }.
        \end{center}
        Then, by IH, we have \(\possiblyWithSub\stageImetaColor{\tau^{\superscriptI} }_{{\mathrm{11}}} = \possiblyWithSub\stageImetaColor{\tau^{\superscriptI} }_{{\mathrm{21}}}\) and \(\possiblyWithSub\stageImetaColor{\tau^{\superscriptI} }_{{\mathrm{12}}} = \possiblyWithSub\stageImetaColor{\tau^{\superscriptI} }_{{\mathrm{22}}}\),
        which make \(\possiblyWithSub\stageImetaColor{\tau^{\superscriptI} }_{{\mathrm{1}}} = \possiblyWithSub\stageImetaColor{\tau^{\superscriptI} }_{{\mathrm{2}}}\) hold.
      \item Case \(\possiblyWithSub\stageOmetaColor{T^{\superscriptO} } =  \openO{\langle} \possiblyWithSub\stageImetaColor{T^{\superscriptI} } \closeO{\rangle} \) contradicts the assumption.
    \end{itemize}
  \end{proof}
  \begin{lemma}\label{lem:persistent-const-inversion}
    If \( \mathit{\Gamma}  \vdash^{0}    \possiblyWithSub\stageOmetaColor{c}    :  \possiblyWithSub\stageOmetaColor{T^{\superscriptO} } \) and \( \possiblyWithSub\stageImetaColor{\tau^{\superscriptI} }  \gg  \possiblyWithSub\stageOmetaColor{T^{\superscriptO} } \),
    then we have \(\ConstEnvPers(c) = \possiblyWithSub\stageImetaColor{\tau^{\superscriptI} }\).
  \end{lemma}
  \begin{proof}
    By induction on the derivation of \( \mathit{\Gamma}  \vdash^{0}    \possiblyWithSub\stageOmetaColor{c}    :  \possiblyWithSub\stageOmetaColor{T^{\superscriptO} } \).
    \begin{itemize}
      \item Case \derive[T0-CstP]{%
         \vdash  \mathit{\Gamma} 
      \andalso
        \ConstEnvPers(c) = \possiblyWithSub\stageImetaColor{\tau'^{\superscriptI} }
      }{%
         \mathit{\Gamma}  \vdash^{0}    \possiblyWithSub\stageOmetaColor{c}    :   \mathop{\downarrow}( \possiblyWithSub\stageImetaColor{\tau'^{\superscriptI} } )  
      }:
        Since \(\possiblyWithSub\stageOmetaColor{T^{\superscriptO} } =  \mathop{\downarrow}( \possiblyWithSub\stageImetaColor{\tau'^{\superscriptI} } ) \), we have \( \possiblyWithSub\stageImetaColor{\tau^{\superscriptI} }  \gg   \mathop{\downarrow}( \possiblyWithSub\stageImetaColor{\tau'^{\superscriptI} } )  \).
        Then, by \( \possiblyWithSub\stageImetaColor{\tau'^{\superscriptI} }  \gg   \mathop{\downarrow}( \possiblyWithSub\stageImetaColor{\tau'^{\superscriptI} } )  \) and Lemma~\ref{lem:type-value-equal-by-stage-0-type},
        we have \(\possiblyWithSub\stageImetaColor{\tau^{\superscriptI} } = \possiblyWithSub\stageImetaColor{\tau'^{\superscriptI} }\).
      \item Case \derive[T0-RfnPred]{%
         \mathit{\Gamma}  \vdash^{0}    \openO{\{} \possiblyWithSub\stageOmetaColor{\nu}  \relO{:}  \possiblyWithSub\stageOmetaColor{B}  \relO{\mid}  \possiblyWithSub\stageOmetaColor{N^{\superscriptO} } \closeO{\} }   
      \andalso
        \ConstEnvPers(c) = \possiblyWithSub\stageImetaColor{B}
      \andalso
           [    \possiblyWithSub\stageOmetaColor{c}    /  \possiblyWithSub\stageOmetaColor{\nu}  ]    \possiblyWithSub\stageOmetaColor{N^{\superscriptO} }   \longrightarrow^{0\,\ast}      \ttO{true}     
      }{%
         \mathit{\Gamma}  \vdash^{0}    \possiblyWithSub\stageOmetaColor{c}    :    \openO{\{} \possiblyWithSub\stageOmetaColor{\nu}  \relO{:}  \possiblyWithSub\stageOmetaColor{B}  \relO{\mid}  \possiblyWithSub\stageOmetaColor{N^{\superscriptO} } \closeO{\} }   
      }:
        Since \(\possiblyWithSub\stageOmetaColor{T^{\superscriptO} } =  \openO{\{} \possiblyWithSub\stageOmetaColor{\nu}  \relO{:}  \possiblyWithSub\stageOmetaColor{B}  \relO{\mid}  \possiblyWithSub\stageOmetaColor{N^{\superscriptO} } \closeO{\} } \), we clearly have \(\possiblyWithSub\stageImetaColor{\tau^{\superscriptI} } = \possiblyWithSub\stageImetaColor{B}\).
        Thus, we can finish the proof by \(\ConstEnvPers(c) = \possiblyWithSub\stageImetaColor{B}\).
      \item Case \derive[T0-TyEquiv]{%
         \mathit{\Gamma}  \vdash^{0}    \possiblyWithSub\stageOmetaColor{c}    :  \possiblyWithSub\stageOmetaColor{T'^{\superscriptO} } 
      \andalso
         \possiblyWithSub\stageOmetaColor{T'^{\superscriptO} }  \equiv^{0}  \possiblyWithSub\stageOmetaColor{T^{\superscriptO} } 
      \andalso
         \mathit{\Gamma}  \vdash^{0}  \possiblyWithSub\stageOmetaColor{T^{\superscriptO} } 
      }{%
         \mathit{\Gamma}  \vdash^{0}    \possiblyWithSub\stageOmetaColor{c}    :  \possiblyWithSub\stageOmetaColor{T^{\superscriptO} } 
      }:
        By Lemma~\ref{lem:csr-equiv-preserves-unlifting-relation} and \rulename{CqT0-Sym},
        from \( \possiblyWithSub\stageImetaColor{\tau^{\superscriptI} }  \gg  \possiblyWithSub\stageOmetaColor{T^{\superscriptO} } \) and \( \possiblyWithSub\stageOmetaColor{T'^{\superscriptO} }  \equiv^{0}  \possiblyWithSub\stageOmetaColor{T^{\superscriptO} } \),
        we have \( \possiblyWithSub\stageImetaColor{\tau^{\superscriptI} }  \gg  \possiblyWithSub\stageOmetaColor{T'^{\superscriptO} } \).
        Thus, by IH, we have \(\ConstEnvPers(c) = \possiblyWithSub\stageImetaColor{\tau^{\superscriptI} }\).
      \item The other cases contradict the assumption.
    \end{itemize}
  \end{proof}
  \begin{lemma}[Determinism of \(\longrightarrow\)]\label{lem:determinism}
    \noindent
    \begin{enumerate}
      \item If \( \possiblyWithSub\stageOmetaColor{N^{\superscriptO} }  \longrightarrow^{0}   \possiblyWithSub\stageOmetaColor{N'^{\superscriptO} }_{{\mathrm{1}}}  \) and \( \possiblyWithSub\stageOmetaColor{N^{\superscriptO} }  \longrightarrow^{0}   \possiblyWithSub\stageOmetaColor{N'^{\superscriptO} }_{{\mathrm{2}}}  \), then \(\possiblyWithSub\stageOmetaColor{N'^{\superscriptO} }_{{\mathrm{1}}} = \possiblyWithSub\stageOmetaColor{N'^{\superscriptO} }_{{\mathrm{2}}}\).
      \item If \( \possiblyWithSub\stageImetaColor{N^{\superscriptI} }  \longrightarrow^{1}   \possiblyWithSub\stageImetaColor{N'^{\superscriptI} }_{{\mathrm{1}}}  \) and \( \possiblyWithSub\stageImetaColor{N^{\superscriptI} }  \longrightarrow^{1}   \possiblyWithSub\stageImetaColor{N'^{\superscriptI} }_{{\mathrm{2}}}  \), then \(\possiblyWithSub\stageImetaColor{N'^{\superscriptI} }_{{\mathrm{1}}} = \possiblyWithSub\stageImetaColor{N'^{\superscriptI} }_{{\mathrm{2}}}\).
      \item If \( \possiblyWithSub\stageImetaColor{T^{\superscriptI} }  \longrightarrow^{1}   \possiblyWithSub\stageImetaColor{T'^{\superscriptI} }_{{\mathrm{1}}}  \) and \( \possiblyWithSub\stageImetaColor{T^{\superscriptI} }  \longrightarrow^{1}   \possiblyWithSub\stageImetaColor{T'^{\superscriptI} }_{{\mathrm{2}}}  \), then \(\possiblyWithSub\stageImetaColor{T'^{\superscriptI} }_{{\mathrm{1}}} = \possiblyWithSub\stageImetaColor{T'^{\superscriptI} }_{{\mathrm{2}}}\).
    \end{enumerate}
  \end{lemma}
  \begin{proof}
    By straightforward mutual induction on the structure of \(\possiblyWithSub\stageOmetaColor{N^{\superscriptO} }\), \(\possiblyWithSub\stageImetaColor{N^{\superscriptI} }\), and \(\possiblyWithSub\stageImetaColor{T^{\superscriptI} }\).
  \end{proof}
  \begin{lemma}\label{lem:nonvalue-substitution-inversion}
    Suppose that \(\possiblyWithSub\stageOmetaColor{N^{\superscriptO} }_{{\mathrm{0}}}\) is a closed term and is not a value.
    \begin{enumerate}
      \item \(  [  \possiblyWithSub\stageOmetaColor{N^{\superscriptO} }_{{\mathrm{0}}}  /  \possiblyWithSub\stageOmetaColor{x}  ]    \possiblyWithSub\stageOmetaColor{N^{\superscriptO} }  = \possiblyWithSub\stageOmetaColor{c}\) implies \(\possiblyWithSub\stageOmetaColor{N^{\superscriptO} } = \possiblyWithSub\stageOmetaColor{c}\).
      \item \(  [  \possiblyWithSub\stageOmetaColor{N^{\superscriptO} }_{{\mathrm{0}}}  /  \possiblyWithSub\stageOmetaColor{x}  ]    \possiblyWithSub\stageOmetaColor{N^{\superscriptO} }  = \possiblyWithSub\stageOmetaColor{p}\) implies \(\possiblyWithSub\stageOmetaColor{N^{\superscriptO} } = \possiblyWithSub\stageOmetaColor{p}\).
      \item \(  [  \possiblyWithSub\stageOmetaColor{N^{\superscriptO} }_{{\mathrm{0}}}  /  \possiblyWithSub\stageOmetaColor{x}  ]    \possiblyWithSub\stageOmetaColor{N^{\superscriptO} }  = \possiblyWithSub\stageOmetaColor{a}\) implies \(\possiblyWithSub\stageOmetaColor{N^{\superscriptO} } = \possiblyWithSub\stageOmetaColor{a}\).
    \end{enumerate}
  \end{lemma}
  \begin{proof}
    \begin{enumerate}
      \item
        By case analysis on the structure of \(\possiblyWithSub\stageOmetaColor{N^{\superscriptO} }\).
        \begin{itemize}
          \item Case \(\possiblyWithSub\stageOmetaColor{N^{\superscriptO} } = \possiblyWithSub\stageOmetaColor{c'}\):
            Since \(  [  \possiblyWithSub\stageOmetaColor{N^{\superscriptO} }_{{\mathrm{0}}}  /  \possiblyWithSub\stageOmetaColor{x}  ]    \possiblyWithSub\stageOmetaColor{N^{\superscriptO} }  = \possiblyWithSub\stageOmetaColor{c'}\),
            we have \(\possiblyWithSub\stageOmetaColor{c'} = \possiblyWithSub\stageOmetaColor{c}\), which finishes the proof.
          \item Case \(\possiblyWithSub\stageOmetaColor{N^{\superscriptO} } = \possiblyWithSub\stageOmetaColor{x}\):
            Since \(  [  \possiblyWithSub\stageOmetaColor{N^{\superscriptO} }_{{\mathrm{0}}}  /  \possiblyWithSub\stageOmetaColor{x}  ]    \possiblyWithSub\stageOmetaColor{N^{\superscriptO} }  = \possiblyWithSub\stageOmetaColor{N^{\superscriptO} }_{{\mathrm{0}}}\),
            we have \(\possiblyWithSub\stageOmetaColor{N^{\superscriptO} }_{{\mathrm{0}}} = \possiblyWithSub\stageOmetaColor{c}\).
            However, this contradicts the assumption
            because \(\possiblyWithSub\stageOmetaColor{N^{\superscriptO} }_{{\mathrm{0}}}\) is assumed not to be a value
            while \(\possiblyWithSub\stageOmetaColor{c}\) is.
          \item The other cases clearly contradict the assumption.
        \end{itemize}
      \item
        Can be proved in the same manner as (1).
      \item
        By induction on the structure of \(\possiblyWithSub\stageOmetaColor{a}\).
        \begin{itemize}
          \item Case \(\possiblyWithSub\stageOmetaColor{a} = \possiblyWithSub\stageOmetaColor{c}\):
            Immediate from (1).
          \item Case \(\possiblyWithSub\stageOmetaColor{a} = \possiblyWithSub\stageOmetaColor{p}\):
            This is also immediate from (2).
          \item Case \(\possiblyWithSub\stageOmetaColor{a} =   \possiblyWithSub\stageOmetaColor{a}_{{\mathrm{1}}}  \    \possiblyWithSub\stageOmetaColor{c}_{{\mathrm{2}}}   \):
            We do the following case analysis on \(\possiblyWithSub\stageOmetaColor{N^{\superscriptO} }\):
            \begin{itemize}
              \item Case~\(\possiblyWithSub\stageOmetaColor{N^{\superscriptO} } = \possiblyWithSub\stageOmetaColor{x}\):
                This contradicts the assumption that \(\possiblyWithSub\stageOmetaColor{N^{\superscriptO} }_{{\mathrm{0}}}\) is not a value
                since we have \(\possiblyWithSub\stageOmetaColor{a} =   [  \possiblyWithSub\stageOmetaColor{N^{\superscriptO} }_{{\mathrm{0}}}  /  \possiblyWithSub\stageOmetaColor{x}  ]    \possiblyWithSub\stageOmetaColor{N^{\superscriptO} }  = \possiblyWithSub\stageOmetaColor{N^{\superscriptO} }_{{\mathrm{0}}}\).
              \item Case~\(\possiblyWithSub\stageOmetaColor{N^{\superscriptO} } = \possiblyWithSub\stageOmetaColor{x'}\) such that \(\possiblyWithSub\stageOmetaColor{x'} \neq \possiblyWithSub\stageOmetaColor{x}\):
                This contradicts the assumption by
                \(\possiblyWithSub\stageOmetaColor{a} =   [  \possiblyWithSub\stageOmetaColor{N^{\superscriptO} }_{{\mathrm{0}}}  /  \possiblyWithSub\stageOmetaColor{x}  ]    \possiblyWithSub\stageOmetaColor{N^{\superscriptO} }  = \possiblyWithSub\stageOmetaColor{x'}\).
              \item Case~\(\possiblyWithSub\stageOmetaColor{N^{\superscriptO} } =  \possiblyWithSub\stageOmetaColor{N^{\superscriptO} }_{{\mathrm{1}}} \  \possiblyWithSub\stageOmetaColor{N^{\superscriptO} }_{{\mathrm{2}}} \):
                We have \(  [  \possiblyWithSub\stageOmetaColor{N^{\superscriptO} }_{{\mathrm{0}}}  /  \possiblyWithSub\stageOmetaColor{x}  ]    \possiblyWithSub\stageOmetaColor{N^{\superscriptO} }_{{\mathrm{1}}}  = \possiblyWithSub\stageOmetaColor{a}_{{\mathrm{1}}}\)
                and \(  [  \possiblyWithSub\stageOmetaColor{N^{\superscriptO} }_{{\mathrm{0}}}  /  \possiblyWithSub\stageOmetaColor{x}  ]    \possiblyWithSub\stageOmetaColor{N^{\superscriptO} }_{{\mathrm{2}}}  = \possiblyWithSub\stageOmetaColor{c}_{{\mathrm{2}}}\).
                By IH, we have \(\possiblyWithSub\stageOmetaColor{N^{\superscriptO} }_{{\mathrm{1}}} = \possiblyWithSub\stageOmetaColor{a}_{{\mathrm{1}}}\),
                and by (1), we have \(\possiblyWithSub\stageOmetaColor{N^{\superscriptO} }_{{\mathrm{2}}} = \possiblyWithSub\stageOmetaColor{c}_{{\mathrm{2}}}\).
                Therefore, we have
                \(\possiblyWithSub\stageOmetaColor{N^{\superscriptO} } =  \possiblyWithSub\stageOmetaColor{N^{\superscriptO} }_{{\mathrm{1}}} \  \possiblyWithSub\stageOmetaColor{N^{\superscriptO} }_{{\mathrm{2}}}  =   \possiblyWithSub\stageOmetaColor{a}_{{\mathrm{1}}}  \    \possiblyWithSub\stageOmetaColor{c}_{{\mathrm{2}}}    = \possiblyWithSub\stageOmetaColor{a}\).
              \item The other cases clearly contradict
              the assumption~\(  [  \possiblyWithSub\stageOmetaColor{N^{\superscriptO} }_{{\mathrm{0}}}  /  \possiblyWithSub\stageOmetaColor{x}  ]    \possiblyWithSub\stageOmetaColor{N^{\superscriptO} }  =   \possiblyWithSub\stageOmetaColor{a}_{{\mathrm{1}}}  \    \possiblyWithSub\stageOmetaColor{c}_{{\mathrm{2}}}   \).
            \end{itemize}
        \end{itemize}
    \end{enumerate}
  \end{proof}
  \begin{lemma}\label{lem:subst-valueness}
    Suppose \(\possiblyWithSub\stageOmetaColor{N^{\superscriptO} }_{{\mathrm{0}}}\) is a closed term and is not a value.
    \begin{enumerate}
      \item
        If \(  [  \possiblyWithSub\stageOmetaColor{N^{\superscriptO} }_{{\mathrm{0}}}  /  \possiblyWithSub\stageOmetaColor{x}  ]    \possiblyWithSub\stageOmetaColor{N^{\superscriptO} } \) is a stage-\(0\) value,
        then \(\possiblyWithSub\stageOmetaColor{N^{\superscriptO} }\) is also a stage-\(0\) value.
      \item
        If \(  [  \possiblyWithSub\stageOmetaColor{N^{\superscriptO} }_{{\mathrm{0}}}  /  \possiblyWithSub\stageOmetaColor{x}  ]    \possiblyWithSub\stageImetaColor{N^{\superscriptI} } \) is a stage-\(1\) value,
        then \(\possiblyWithSub\stageImetaColor{N^{\superscriptI} }\) is also a stage-\(1\) value.
      \item
        If \(  [  \possiblyWithSub\stageOmetaColor{N^{\superscriptO} }_{{\mathrm{0}}}  /  \possiblyWithSub\stageOmetaColor{x}  ]    \possiblyWithSub\stageImetaColor{T^{\superscriptI} }  \revdefeq \possiblyWithSub\stageImetaColor{\tau^{\superscriptI} }\) is a type value,
        then \(\possiblyWithSub\stageImetaColor{T^{\superscriptI} } = \possiblyWithSub\stageImetaColor{\tau^{\superscriptI} }\).
    \end{enumerate}
  \end{lemma}
  \begin{proof}
    By mutual induction on the structure of \(\possiblyWithSub\stageOmetaColor{N^{\superscriptO} }\), \(\possiblyWithSub\stageImetaColor{N^{\superscriptI} }\), and \(\possiblyWithSub\stageImetaColor{T^{\superscriptI} }\).
    \begin{enumerate}
      \item
        \begin{itemize}
          \item Case where we have \(\possiblyWithSub\stageOmetaColor{N^{\superscriptO} } = \possiblyWithSub\stageOmetaColor{p}\), \(\possiblyWithSub\stageOmetaColor{N^{\superscriptO} } = \possiblyWithSub\stageOmetaColor{c}\),
          \(\possiblyWithSub\stageOmetaColor{N^{\superscriptO} } =  \openO{(}  \ordO{\lambda} \possiblyWithSub\stageOmetaColor{x'}  \relO{:}  \possiblyWithSub\stageOmetaColor{T^{\superscriptO} }_{{\mathrm{1}}} \punctO{.}\  \possiblyWithSub\stageOmetaColor{N^{\superscriptO} }_{{\mathrm{2}}}  \closeO{)} \), or
          \(\possiblyWithSub\stageOmetaColor{N^{\superscriptO} } =  \LeftAssertParen \relO{\CastArrow}   \openO{\{} \possiblyWithSub\stageOmetaColor{\nu}  \relO{:}  \possiblyWithSub\stageOmetaColor{B}  \relO{\mid}  \possiblyWithSub\stageOmetaColor{N^{\superscriptO} }_{{\mathrm{1}}} \closeO{\} }   \RightAssertParen^{ L } \)
            is immediate.
          \item Case~\(\possiblyWithSub\stageOmetaColor{N^{\superscriptO} } = \possiblyWithSub\stageOmetaColor{x}\) contradicts the assumption
            because we have \(  [  \possiblyWithSub\stageOmetaColor{N^{\superscriptO} }_{{\mathrm{0}}}  /  \possiblyWithSub\stageOmetaColor{x}  ]    \possiblyWithSub\stageOmetaColor{N^{\superscriptO} }  = \possiblyWithSub\stageOmetaColor{N^{\superscriptO} }_{{\mathrm{0}}}\),
            the left-hand side (resp.~the right-hand side) of which
            is a value (resp.~is not a value).
          \item Case~\(\possiblyWithSub\stageOmetaColor{N^{\superscriptO} } = \possiblyWithSub\stageOmetaColor{x'}\) such that \(\possiblyWithSub\stageOmetaColor{x'} \neq \possiblyWithSub\stageOmetaColor{x}\)
            also contradicts the assumption
            since \(  [  \possiblyWithSub\stageOmetaColor{N^{\superscriptO} }_{{\mathrm{0}}}  /  \possiblyWithSub\stageOmetaColor{x}  ]    \possiblyWithSub\stageOmetaColor{N^{\superscriptO} }  = \possiblyWithSub\stageOmetaColor{x'}\) is not a value.
          \item Case where \(\possiblyWithSub\stageOmetaColor{N^{\superscriptO} } =  \LeftAssertParen   \openO{\{} \possiblyWithSub\stageOmetaColor{\nu}  \relO{:}  \possiblyWithSub\stageOmetaColor{B}  \relO{\mid}  \possiblyWithSub\stageOmetaColor{N^{\superscriptO} }_{{\mathrm{1}}} \closeO{\} }  \punctO{,}  \possiblyWithSub\stageOmetaColor{N^{\superscriptO} }_{{\mathrm{2}}} \punctO{,}  \possiblyWithSub\stageOmetaColor{c}  \RightAssertParen^{ L } \)
            or \(\possiblyWithSub\stageOmetaColor{N^{\superscriptO} } =  \LeftAssertParen\openO{\langle} \possiblyWithSub\stageImetaColor{T^{\superscriptI} }_{{\mathrm{1}}} \closeO{\rangle} \relO{\CastArrow} \openO{\langle} \possiblyWithSub\stageImetaColor{T^{\superscriptI} }_{{\mathrm{2}}} \closeO{\rangle}\RightAssertParen^{ L } \):
            Clearly contradicts the assumption that
            \(  [  \possiblyWithSub\stageOmetaColor{N^{\superscriptO} }_{{\mathrm{0}}}  /  \possiblyWithSub\stageOmetaColor{x}  ]    \possiblyWithSub\stageOmetaColor{N^{\superscriptO} } \) is a value.
          \item Case~\(\possiblyWithSub\stageOmetaColor{N^{\superscriptO} } =  \openO{\langle} \possiblyWithSub\stageImetaColor{N^{\superscriptI} } \closeO{\rangle} \):
            By the definition of stage-\(0\) values,
            \(  [  \possiblyWithSub\stageOmetaColor{N^{\superscriptO} }_{{\mathrm{0}}}  /  \possiblyWithSub\stageOmetaColor{x}  ]    \possiblyWithSub\stageImetaColor{N^{\superscriptI} } \) is a stage-\(1\) value.
            Then, by IH, \(\possiblyWithSub\stageImetaColor{N^{\superscriptI} }\) is also a stage-\(1\) value,
            and thereby \(\possiblyWithSub\stageOmetaColor{N^{\superscriptO} } =  \openO{\langle} \possiblyWithSub\stageImetaColor{N^{\superscriptI} } \closeO{\rangle} \) is a stage-\(0\) value.
          \item Case~\(\possiblyWithSub\stageOmetaColor{N^{\superscriptO} } =  \possiblyWithSub\stageOmetaColor{N^{\superscriptO} }_{{\mathrm{1}}} \  \possiblyWithSub\stageOmetaColor{N^{\superscriptO} }_{{\mathrm{2}}} \):
            Since \(  [  \possiblyWithSub\stageOmetaColor{N^{\superscriptO} }_{{\mathrm{0}}}  /  \possiblyWithSub\stageOmetaColor{x}  ]    \possiblyWithSub\stageOmetaColor{N^{\superscriptO} } \) is a value,
            We only have the case where
            \(  [  \possiblyWithSub\stageOmetaColor{N^{\superscriptO} }_{{\mathrm{0}}}  /  \possiblyWithSub\stageOmetaColor{x}  ]    \possiblyWithSub\stageOmetaColor{N^{\superscriptO} }_{{\mathrm{1}}}  = \possiblyWithSub\stageOmetaColor{a}_{{\mathrm{1}}}\)
            and \(  [  \possiblyWithSub\stageOmetaColor{N^{\superscriptO} }_{{\mathrm{0}}}  /  \possiblyWithSub\stageOmetaColor{x}  ]    \possiblyWithSub\stageOmetaColor{N^{\superscriptO} }_{{\mathrm{2}}}  = \possiblyWithSub\stageOmetaColor{c}_{{\mathrm{2}}}\).
            Then, by Lemma~\ref{lem:nonvalue-substitution-inversion},
            we have \(\possiblyWithSub\stageOmetaColor{N^{\superscriptO} }_{{\mathrm{1}}} = \possiblyWithSub\stageOmetaColor{a}_{{\mathrm{1}}}\) and \(\possiblyWithSub\stageOmetaColor{N^{\superscriptO} }_{{\mathrm{2}}} = \possiblyWithSub\stageOmetaColor{c}_{{\mathrm{2}}}\).
            Thus, \(\possiblyWithSub\stageOmetaColor{N^{\superscriptO} } =   \possiblyWithSub\stageOmetaColor{a}_{{\mathrm{1}}}  \    \possiblyWithSub\stageOmetaColor{c}_{{\mathrm{2}}}   \) is a value.
        \end{itemize}
      \item
        \begin{itemize}
          \item Case~\(\possiblyWithSub\stageImetaColor{N^{\superscriptI} } =  \openI{(}  \ordI{\lambda} \possiblyWithSub\stageImetaColor{x}  \relI{:}  \possiblyWithSub\stageImetaColor{T^{\superscriptI} }_{{\mathrm{1}}} \punctI{.}\  \possiblyWithSub\stageImetaColor{N^{\superscriptI} }_{{\mathrm{2}}}  \closeI{)} \):
            We have
            \(  [  \possiblyWithSub\stageOmetaColor{N^{\superscriptO} }_{{\mathrm{0}}}  /  \possiblyWithSub\stageOmetaColor{x}  ]    \possiblyWithSub\stageImetaColor{N^{\superscriptI} } 
              =  \openI{(}  \ordI{\lambda} \possiblyWithSub\stageImetaColor{x}  \relI{:}    [  \possiblyWithSub\stageOmetaColor{N^{\superscriptO} }_{{\mathrm{0}}}  /  \possiblyWithSub\stageOmetaColor{x}  ]    \possiblyWithSub\stageImetaColor{T^{\superscriptI} }_{{\mathrm{1}}}  \punctI{.}\    [  \possiblyWithSub\stageOmetaColor{N^{\superscriptO} }_{{\mathrm{0}}}  /  \possiblyWithSub\stageOmetaColor{x}  ]    \possiblyWithSub\stageImetaColor{N^{\superscriptI} }_{{\mathrm{2}}}   \closeI{)} \).
            By the definition of stage-\(1\) values,
            \(  [  \possiblyWithSub\stageOmetaColor{N^{\superscriptO} }_{{\mathrm{0}}}  /  \possiblyWithSub\stageOmetaColor{x}  ]    \possiblyWithSub\stageImetaColor{T^{\superscriptI} }_{{\mathrm{1}}}  \revdefeq \possiblyWithSub\stageImetaColor{\tau^{\superscriptI} }_{{\mathrm{1}}}\)
            and \(  [  \possiblyWithSub\stageOmetaColor{N^{\superscriptO} }_{{\mathrm{0}}}  /  \possiblyWithSub\stageOmetaColor{x}  ]    \possiblyWithSub\stageImetaColor{N^{\superscriptI} }_{{\mathrm{2}}} \)
            are a stage-\(1\) type value and a stage-\(1\) value, respectively.
            Then, by IH, we have \(\possiblyWithSub\stageImetaColor{T^{\superscriptI} }_{{\mathrm{1}}} = \possiblyWithSub\stageImetaColor{\tau^{\superscriptI} }_{{\mathrm{1}}}\),
            and \(\possiblyWithSub\stageImetaColor{N^{\superscriptI} }_{{\mathrm{2}}}\) is also a stage-\(1\) value.
            Therefore,
            \(\possiblyWithSub\stageImetaColor{N^{\superscriptI} } =  \openI{(}  \ordI{\lambda} \possiblyWithSub\stageImetaColor{x}  \relI{:}   \possiblyWithSub\stageImetaColor{\tau^{\superscriptI} }_{{\mathrm{1}}}  \punctI{.}\  \possiblyWithSub\stageImetaColor{N^{\superscriptI} }_{{\mathrm{2}}}  \closeI{)} \)
            is a stage-\(1\) value.
          \item Case~\(\possiblyWithSub\stageImetaColor{N^{\superscriptI} } =  \ordI{\sim} \possiblyWithSub\stageOmetaColor{N^{\superscriptO} }_{{\mathrm{1}}} \):
            This contradicts the assumption that
            \(  [  \possiblyWithSub\stageOmetaColor{N^{\superscriptO} }_{{\mathrm{0}}}  /  \possiblyWithSub\stageOmetaColor{x}  ]    \possiblyWithSub\stageImetaColor{N^{\superscriptI} } \) is a value
            since \(  [  \possiblyWithSub\stageOmetaColor{N^{\superscriptO} }_{{\mathrm{0}}}  /  \possiblyWithSub\stageOmetaColor{x}  ]    \possiblyWithSub\stageImetaColor{N^{\superscriptI} }  =  \ordI{\sim}  \openO{(}   [  \possiblyWithSub\stageOmetaColor{N^{\superscriptO} }_{{\mathrm{0}}}  /  \possiblyWithSub\stageOmetaColor{x}  ]    \possiblyWithSub\stageOmetaColor{N^{\superscriptO} }_{{\mathrm{1}}}  \closeO{)}  \) holds.
          \item The other cases are all straightforward.
        \end{itemize}
      \item
        \begin{itemize}
          \item Case~\(\possiblyWithSub\stageImetaColor{T^{\superscriptI} } =  \ttI{Tensor}\ \ordI{\%} \possiblyWithSub\stageOmetaColor{N^{\superscriptO} }_{{\mathrm{1}}} \):
            Since
            \(\possiblyWithSub\stageImetaColor{\tau^{\superscriptI} }
              =   [  \possiblyWithSub\stageOmetaColor{N^{\superscriptO} }_{{\mathrm{0}}}  /  \possiblyWithSub\stageOmetaColor{x}  ]     \openI{(}  \ttI{Tensor}\ \ordI{\%} \possiblyWithSub\stageOmetaColor{N^{\superscriptO} }_{{\mathrm{1}}}  \closeI{)}  
              =  \ttI{Tensor}\ \ordI{\%}  \openO{(}   [  \possiblyWithSub\stageOmetaColor{N^{\superscriptO} }_{{\mathrm{0}}}  /  \possiblyWithSub\stageOmetaColor{x}  ]    \possiblyWithSub\stageOmetaColor{N^{\superscriptO} }_{{\mathrm{1}}}  \closeO{)}  \)
            holds,
            by the definition of stage-\(1\) type values,
            \(  [  \possiblyWithSub\stageOmetaColor{N^{\superscriptO} }_{{\mathrm{0}}}  /  \possiblyWithSub\stageOmetaColor{x}  ]    \possiblyWithSub\stageOmetaColor{N^{\superscriptO} }_{{\mathrm{1}}} \) is of the form~\(  \possiblyWithSub\stageOmetaColor{s}  \).
            Then, by Lemma~\ref{lem:nonvalue-substitution-inversion},
            we have \(\possiblyWithSub\stageOmetaColor{N^{\superscriptO} }_{{\mathrm{1}}} = \possiblyWithSub\stageOmetaColor{s}\).
            Thus, we have
            \(\possiblyWithSub\stageImetaColor{T^{\superscriptI} } =  \ttI{Tensor}\ \ordI{\%} \possiblyWithSub\stageOmetaColor{s}  = \possiblyWithSub\stageImetaColor{\tau^{\superscriptI} }\).
          \item Case~\(\possiblyWithSub\stageImetaColor{T^{\superscriptI} } = \possiblyWithSub\stageImetaColor{B}\):
            This is immediate.
          \item Case~\(\possiblyWithSub\stageImetaColor{T^{\superscriptI} } =  \possiblyWithSub\stageImetaColor{T^{\superscriptI} }_{{\mathrm{1}}}  \relI{\to}  \possiblyWithSub\stageImetaColor{T^{\superscriptI} }_{{\mathrm{2}}} \):
            Straightforward by IH.
        \end{itemize}
    \end{enumerate}
  \end{proof}
  \begin{lemma}\label{lem:weak-bisimulation-app-pre}
    Suppose that \(\stageOmetaColor{N_{\mathrm{A} }^{\superscriptO} }\) and \(\stageOmetaColor{N_{\mathrm{B} }^{\superscriptO} }\) are closed terms
    and that \(\possiblyWithSub\stageOmetaColor{v^{\superscriptO} }_{{\mathrm{1}}}\), \(  [  \stageOmetaColor{N_{\mathrm{A} }^{\superscriptO} }  /  \possiblyWithSub\stageOmetaColor{x}  ]    \possiblyWithSub\stageOmetaColor{N^{\superscriptO} }_{{\mathrm{2}}} \), and \(  [  \stageOmetaColor{N_{\mathrm{B} }^{\superscriptO} }  /  \possiblyWithSub\stageOmetaColor{x}  ]    \possiblyWithSub\stageOmetaColor{N^{\superscriptO} }_{{\mathrm{2}}} \) are all values.
    If \(   [  \stageOmetaColor{N_{\mathrm{A} }^{\superscriptO} }  /  \possiblyWithSub\stageOmetaColor{x}  ]     \openO{(}   \possiblyWithSub\stageOmetaColor{v^{\superscriptO} }_{{\mathrm{1}}}  \  \possiblyWithSub\stageOmetaColor{N^{\superscriptO} }_{{\mathrm{2}}}  \closeO{)}    \longrightarrow^{0}   \possiblyWithSub\stageOmetaColor{N'^{\superscriptO} }  \), then
    there exists \(\possiblyWithSub\stageOmetaColor{\Hat{N}^{\superscriptO} }\) such that
    \(   [  \stageOmetaColor{N_{\mathrm{B} }^{\superscriptO} }  /  \possiblyWithSub\stageOmetaColor{x}  ]     \openO{(}   \possiblyWithSub\stageOmetaColor{v^{\superscriptO} }_{{\mathrm{1}}}  \  \possiblyWithSub\stageOmetaColor{N^{\superscriptO} }_{{\mathrm{2}}}  \closeO{)}    \longrightarrow^{0}     [  \stageOmetaColor{N_{\mathrm{B} }^{\superscriptO} }  /  \possiblyWithSub\stageOmetaColor{x}  ]    \possiblyWithSub\stageOmetaColor{\Hat{N}^{\superscriptO} }   \)
    and \(\possiblyWithSub\stageOmetaColor{N'^{\superscriptO} } =   [  \stageOmetaColor{N_{\mathrm{A} }^{\superscriptO} }  /  \possiblyWithSub\stageOmetaColor{x}  ]    \possiblyWithSub\stageOmetaColor{\Hat{N}^{\superscriptO} } \).
  \end{lemma}
  \begin{proof}
    By case analysis on \(\possiblyWithSub\stageOmetaColor{v^{\superscriptO} }_{{\mathrm{1}}}\):
    \begin{itemize}
      \item Case~\(\possiblyWithSub\stageOmetaColor{v^{\superscriptO} }_{{\mathrm{1}}} =  \openO{(}  \ordO{\lambda} \possiblyWithSub\stageOmetaColor{x'}  \relO{:}  \possiblyWithSub\stageOmetaColor{T^{\superscriptO} }_{{\mathrm{11}}} \punctO{.}\  \possiblyWithSub\stageOmetaColor{N^{\superscriptO} }_{{\mathrm{12}}}  \closeO{)} \):
        We have only the following derivation for
        \(   [  \stageOmetaColor{N_{\mathrm{A} }^{\superscriptO} }  /  \possiblyWithSub\stageOmetaColor{x}  ]     \openO{(}   \possiblyWithSub\stageOmetaColor{v^{\superscriptO} }_{{\mathrm{1}}}  \  \possiblyWithSub\stageOmetaColor{N^{\superscriptO} }_{{\mathrm{2}}}  \closeO{)}    \longrightarrow^{0}   \possiblyWithSub\stageOmetaColor{N'^{\superscriptO} }  \),
        assuming \(\possiblyWithSub\stageOmetaColor{x'} \neq \possiblyWithSub\stageOmetaColor{x}\) w.l.o.g. by the Barendregt convention:
        \begin{center}
          \derive[E0-Beta]{}{%
               \openO{(}  \ordO{\lambda} \possiblyWithSub\stageOmetaColor{x'}  \relO{:}    [  \stageOmetaColor{N_{\mathrm{A} }^{\superscriptO} }  /  \possiblyWithSub\stageOmetaColor{x}  ]    \possiblyWithSub\stageOmetaColor{T^{\superscriptO} }_{{\mathrm{11}}}  \punctO{.}\    [  \stageOmetaColor{N_{\mathrm{A} }^{\superscriptO} }  /  \possiblyWithSub\stageOmetaColor{x}  ]    \possiblyWithSub\stageOmetaColor{N^{\superscriptO} }_{{\mathrm{12}}}   \closeO{)}  \   \openO{(}   [  \stageOmetaColor{N_{\mathrm{A} }^{\superscriptO} }  /  \possiblyWithSub\stageOmetaColor{x}  ]    \possiblyWithSub\stageOmetaColor{N^{\superscriptO} }_{{\mathrm{2}}}  \closeO{)}    \longrightarrow^{0}     [    [  \stageOmetaColor{N_{\mathrm{A} }^{\superscriptO} }  /  \possiblyWithSub\stageOmetaColor{x}  ]    \possiblyWithSub\stageOmetaColor{N^{\superscriptO} }_{{\mathrm{2}}}   /  \possiblyWithSub\stageOmetaColor{x'}  ]      [  \stageOmetaColor{N_{\mathrm{A} }^{\superscriptO} }  /  \possiblyWithSub\stageOmetaColor{x}  ]    \possiblyWithSub\stageOmetaColor{N^{\superscriptO} }_{{\mathrm{12}}}    
          }.
        \end{center}
        We will show that
        \(\possiblyWithSub\stageOmetaColor{\Hat{N}^{\superscriptO} } \defeq   [  \possiblyWithSub\stageOmetaColor{N^{\superscriptO} }_{{\mathrm{2}}}  /  \possiblyWithSub\stageOmetaColor{x'}  ]    \possiblyWithSub\stageOmetaColor{N^{\superscriptO} }_{{\mathrm{12}}} \)
        satisfies the desired properties.
        First, we have
        \(\possiblyWithSub\stageOmetaColor{N'^{\superscriptO} } =   [    [  \stageOmetaColor{N_{\mathrm{A} }^{\superscriptO} }  /  \possiblyWithSub\stageOmetaColor{x}  ]    \possiblyWithSub\stageOmetaColor{N^{\superscriptO} }_{{\mathrm{2}}}   /  \possiblyWithSub\stageOmetaColor{x'}  ]      [  \stageOmetaColor{N_{\mathrm{A} }^{\superscriptO} }  /  \possiblyWithSub\stageOmetaColor{x}  ]    \possiblyWithSub\stageOmetaColor{N^{\superscriptO} }_{{\mathrm{12}}}  
          =   [  \stageOmetaColor{N_{\mathrm{A} }^{\superscriptO} }  /  \possiblyWithSub\stageOmetaColor{x}  ]      [  \possiblyWithSub\stageOmetaColor{N^{\superscriptO} }_{{\mathrm{2}}}  /  \possiblyWithSub\stageOmetaColor{x'}  ]    \possiblyWithSub\stageOmetaColor{N^{\superscriptO} }_{{\mathrm{12}}}  
          =   [  \stageOmetaColor{N_{\mathrm{A} }^{\superscriptO} }  /  \possiblyWithSub\stageOmetaColor{x}  ]    \possiblyWithSub\stageOmetaColor{\Hat{N}^{\superscriptO} } \).
        Then,
        since \(  [  \stageOmetaColor{N_{\mathrm{A} }^{\superscriptO} }  /  \possiblyWithSub\stageOmetaColor{x}  ]    \possiblyWithSub\stageOmetaColor{N^{\superscriptO} }_{{\mathrm{2}}} \) is a value while \(\stageOmetaColor{N_{\mathrm{A} }^{\superscriptO} }\) is not,
        \(\possiblyWithSub\stageOmetaColor{N^{\superscriptO} }_{{\mathrm{2}}}\) is a value by Lemma~\ref{lem:subst-valueness}.
        Therefore, \(  [  \stageOmetaColor{N_{\mathrm{B} }^{\superscriptO} }  /  \possiblyWithSub\stageOmetaColor{x}  ]    \possiblyWithSub\stageOmetaColor{N^{\superscriptO} }_{{\mathrm{2}}} \) is clearly a value,
        and we can derive \(   [  \stageOmetaColor{N_{\mathrm{B} }^{\superscriptO} }  /  \possiblyWithSub\stageOmetaColor{x}  ]    \possiblyWithSub\stageOmetaColor{N^{\superscriptO} }   \longrightarrow^{0}     [  \stageOmetaColor{N_{\mathrm{B} }^{\superscriptO} }  /  \possiblyWithSub\stageOmetaColor{x}  ]    \possiblyWithSub\stageOmetaColor{\Hat{N}^{\superscriptO} }   \)
        as follows:
        \begin{center}
          \derive[E0-Beta]{}{%
               \openO{(}  \ordO{\lambda} \possiblyWithSub\stageOmetaColor{x'}  \relO{:}    [  \stageOmetaColor{N_{\mathrm{B} }^{\superscriptO} }  /  \possiblyWithSub\stageOmetaColor{x}  ]    \possiblyWithSub\stageOmetaColor{T^{\superscriptO} }_{{\mathrm{11}}}  \punctO{.}\    [  \stageOmetaColor{N_{\mathrm{B} }^{\superscriptO} }  /  \possiblyWithSub\stageOmetaColor{x}  ]    \possiblyWithSub\stageOmetaColor{N^{\superscriptO} }_{{\mathrm{12}}}   \closeO{)}  \   \openO{(}   [  \stageOmetaColor{N_{\mathrm{B} }^{\superscriptO} }  /  \possiblyWithSub\stageOmetaColor{x}  ]    \possiblyWithSub\stageOmetaColor{N^{\superscriptO} }_{{\mathrm{2}}}  \closeO{)}    \longrightarrow^{0}     [    [  \stageOmetaColor{N_{\mathrm{B} }^{\superscriptO} }  /  \possiblyWithSub\stageOmetaColor{x}  ]    \possiblyWithSub\stageOmetaColor{N^{\superscriptO} }_{{\mathrm{2}}}   /  \possiblyWithSub\stageOmetaColor{x'}  ]      [  \stageOmetaColor{N_{\mathrm{B} }^{\superscriptO} }  /  \possiblyWithSub\stageOmetaColor{x}  ]    \possiblyWithSub\stageOmetaColor{N^{\superscriptO} }_{{\mathrm{12}}}    
          }.
        \end{center}
      \item Case~\(\possiblyWithSub\stageOmetaColor{v^{\superscriptO} }_{{\mathrm{1}}} = \possiblyWithSub\stageOmetaColor{a}_{{\mathrm{1}}}\):
        Since \(  [  \stageOmetaColor{N_{\mathrm{A} }^{\superscriptO} }  /  \possiblyWithSub\stageOmetaColor{x}  ]     \possiblyWithSub\stageOmetaColor{v^{\superscriptO} }_{{\mathrm{1}}}   = \possiblyWithSub\stageOmetaColor{a}_{{\mathrm{1}}}\),
        we have the following as the sole derivation of
        \(   [  \stageOmetaColor{N_{\mathrm{A} }^{\superscriptO} }  /  \possiblyWithSub\stageOmetaColor{x}  ]     \openO{(}   \possiblyWithSub\stageOmetaColor{v^{\superscriptO} }_{{\mathrm{1}}}  \  \possiblyWithSub\stageOmetaColor{N^{\superscriptO} }_{{\mathrm{2}}}  \closeO{)}    \longrightarrow^{0}   \possiblyWithSub\stageOmetaColor{N'^{\superscriptO} }  \):
        \begin{center}
          \derive[E0-Delta]{%
            \delta(  \possiblyWithSub\stageOmetaColor{a}_{{\mathrm{1}}}  \    \possiblyWithSub\stageOmetaColor{c}_{{\mathrm{2}}}   ) = \possiblyWithSub\stageOmetaColor{q}
          }{%
               \possiblyWithSub\stageOmetaColor{a}_{{\mathrm{1}}}  \    \possiblyWithSub\stageOmetaColor{c}_{{\mathrm{2}}}     \longrightarrow^{0}    \possiblyWithSub\stageOmetaColor{q}   
          }.
        \end{center}
        We will show that \(\possiblyWithSub\stageOmetaColor{\Hat{N}^{\superscriptO} } \defeq \possiblyWithSub\stageOmetaColor{q}\) satisfies the desired properties.
        First, by \(  [  \stageOmetaColor{N_{\mathrm{A} }^{\superscriptO} }  /  \possiblyWithSub\stageOmetaColor{x}  ]    \possiblyWithSub\stageOmetaColor{N^{\superscriptO} }_{{\mathrm{2}}}  = \possiblyWithSub\stageOmetaColor{c}_{{\mathrm{2}}}\)
        and Lemma~\ref{lem:nonvalue-substitution-inversion},
        we have \(\possiblyWithSub\stageOmetaColor{N^{\superscriptO} }_{{\mathrm{2}}} = \possiblyWithSub\stageOmetaColor{c}_{{\mathrm{2}}}\).
        Since we have \(  [  \stageOmetaColor{N_{\mathrm{B} }^{\superscriptO} }  /  \possiblyWithSub\stageOmetaColor{x}  ]     \possiblyWithSub\stageOmetaColor{v^{\superscriptO} }_{{\mathrm{1}}}   = \possiblyWithSub\stageOmetaColor{a}_{{\mathrm{1}}}\)
        and \(  [  \stageOmetaColor{N_{\mathrm{B} }^{\superscriptO} }  /  \possiblyWithSub\stageOmetaColor{x}  ]    \possiblyWithSub\stageOmetaColor{N^{\superscriptO} }_{{\mathrm{2}}}  = \possiblyWithSub\stageOmetaColor{c}_{{\mathrm{2}}}\),
        we can immediately derive
        \(   [  \stageOmetaColor{N_{\mathrm{B} }^{\superscriptO} }  /  \possiblyWithSub\stageOmetaColor{x}  ]    \possiblyWithSub\stageOmetaColor{N^{\superscriptO} }   \longrightarrow^{0}     [  \stageOmetaColor{N_{\mathrm{B} }^{\superscriptO} }  /  \possiblyWithSub\stageOmetaColor{x}  ]    \possiblyWithSub\stageOmetaColor{\Hat{N}^{\superscriptO} }   \)
        as follows:
        \begin{center}
          \derive[E0-Delta]{%
            \delta(  \possiblyWithSub\stageOmetaColor{a}_{{\mathrm{1}}}  \    \possiblyWithSub\stageOmetaColor{c}_{{\mathrm{2}}}   ) = \possiblyWithSub\stageOmetaColor{q}
          }{%
               \possiblyWithSub\stageOmetaColor{a}_{{\mathrm{1}}}  \    \possiblyWithSub\stageOmetaColor{c}_{{\mathrm{2}}}     \longrightarrow^{0}    \possiblyWithSub\stageOmetaColor{q}   
          }.
        \end{center}
        Also, we clearly have \(\possiblyWithSub\stageOmetaColor{N'^{\superscriptO} } =   [  \stageOmetaColor{N_{\mathrm{A} }^{\superscriptO} }  /  \possiblyWithSub\stageOmetaColor{x}  ]    \possiblyWithSub\stageOmetaColor{\Hat{N}^{\superscriptO} } \)
        since \(  [  \stageOmetaColor{N_{\mathrm{A} }^{\superscriptO} }  /  \possiblyWithSub\stageOmetaColor{x}  ]     \possiblyWithSub\stageOmetaColor{q}   = \possiblyWithSub\stageOmetaColor{q}\).
      \item Case~\(\possiblyWithSub\stageOmetaColor{v^{\superscriptO} }_{{\mathrm{1}}} =  \LeftAssertParen \relO{\CastArrow}   \openO{\{} \possiblyWithSub\stageOmetaColor{\nu}  \relO{:}  \possiblyWithSub\stageOmetaColor{B}  \relO{\mid}  \possiblyWithSub\stageOmetaColor{N^{\superscriptO} }_{{\mathrm{11}}} \closeO{\} }   \RightAssertParen^{ L } \):
        The sole possible derivation of
        \(   [  \stageOmetaColor{N_{\mathrm{A} }^{\superscriptO} }  /  \possiblyWithSub\stageOmetaColor{x}  ]     \openO{(}   \possiblyWithSub\stageOmetaColor{v^{\superscriptO} }_{{\mathrm{1}}}  \  \possiblyWithSub\stageOmetaColor{N^{\superscriptO} }_{{\mathrm{2}}}  \closeO{)}    \longrightarrow^{0}   \possiblyWithSub\stageOmetaColor{N'^{\superscriptO} }  \)
        is the following,
        assuming \(\possiblyWithSub\stageOmetaColor{\nu} \neq \possiblyWithSub\stageOmetaColor{x}\) w.l.o.g. by the Barendregt convention:
        \begin{center}
          \derive[E0-RfnStart]{}{%
               \LeftAssertParen \relO{\CastArrow}   \openO{\{} \possiblyWithSub\stageOmetaColor{\nu}  \relO{:}  \possiblyWithSub\stageOmetaColor{B}  \relO{\mid}    [  \stageOmetaColor{N_{\mathrm{A} }^{\superscriptO} }  /  \possiblyWithSub\stageOmetaColor{x}  ]    \possiblyWithSub\stageOmetaColor{N^{\superscriptO} }_{{\mathrm{11}}}  \closeO{\} }   \RightAssertParen^{ L }  \    \possiblyWithSub\stageOmetaColor{c}_{{\mathrm{2}}}     \longrightarrow^{0}    \LeftAssertParen   \openO{\{} \possiblyWithSub\stageOmetaColor{\nu}  \relO{:}  \possiblyWithSub\stageOmetaColor{B}  \relO{\mid}    [  \stageOmetaColor{N_{\mathrm{A} }^{\superscriptO} }  /  \possiblyWithSub\stageOmetaColor{x}  ]    \possiblyWithSub\stageOmetaColor{N^{\superscriptO} }_{{\mathrm{11}}}  \closeO{\} }  \punctO{,}    [    \possiblyWithSub\stageOmetaColor{c}_{{\mathrm{2}}}    /  \possiblyWithSub\stageOmetaColor{\nu}  ]      [  \stageOmetaColor{N_{\mathrm{A} }^{\superscriptO} }  /  \possiblyWithSub\stageOmetaColor{x}  ]    \possiblyWithSub\stageOmetaColor{N^{\superscriptO} }_{{\mathrm{11}}}   \punctO{,}  \possiblyWithSub\stageOmetaColor{c}_{{\mathrm{2}}}  \RightAssertParen^{ L }   
          }.
        \end{center}
        Similarly to the previous case, we first have \(\possiblyWithSub\stageOmetaColor{N^{\superscriptO} }_{{\mathrm{2}}} = \possiblyWithSub\stageOmetaColor{c}_{{\mathrm{2}}}\)
        by \(  [  \stageOmetaColor{N_{\mathrm{A} }^{\superscriptO} }  /  \possiblyWithSub\stageOmetaColor{x}  ]    \possiblyWithSub\stageOmetaColor{N^{\superscriptO} }_{{\mathrm{2}}}  = \possiblyWithSub\stageOmetaColor{c}_{{\mathrm{2}}}\) and Lemma~\ref{lem:nonvalue-substitution-inversion}.
        We will show that
        \(\possiblyWithSub\stageOmetaColor{\Hat{N}^{\superscriptO} } \defeq  \LeftAssertParen   \openO{\{} \possiblyWithSub\stageOmetaColor{\nu}  \relO{:}  \possiblyWithSub\stageOmetaColor{B}  \relO{\mid}  \possiblyWithSub\stageOmetaColor{N^{\superscriptO} }_{{\mathrm{11}}} \closeO{\} }  \punctO{,}    [    \possiblyWithSub\stageOmetaColor{c}_{{\mathrm{2}}}    /  \possiblyWithSub\stageOmetaColor{\nu}  ]    \possiblyWithSub\stageOmetaColor{N^{\superscriptO} }_{{\mathrm{11}}}  \punctO{,}  \possiblyWithSub\stageOmetaColor{c}_{{\mathrm{2}}}  \RightAssertParen^{ L } \)
        satisfies the desired properties.
        First, since \(\stageOmetaColor{N_{\mathrm{A} }^{\superscriptO} }\) is a closed term, we have
        \begin{align*}
          \possiblyWithSub\stageOmetaColor{N'^{\superscriptO} }
          &=  \LeftAssertParen   \openO{\{} \possiblyWithSub\stageOmetaColor{\nu}  \relO{:}  \possiblyWithSub\stageOmetaColor{B}  \relO{\mid}    [  \stageOmetaColor{N_{\mathrm{A} }^{\superscriptO} }  /  \possiblyWithSub\stageOmetaColor{x}  ]    \possiblyWithSub\stageOmetaColor{N^{\superscriptO} }_{{\mathrm{11}}}  \closeO{\} }  \punctO{,}    [    \possiblyWithSub\stageOmetaColor{c}_{{\mathrm{2}}}    /  \possiblyWithSub\stageOmetaColor{\nu}  ]      [  \stageOmetaColor{N_{\mathrm{A} }^{\superscriptO} }  /  \possiblyWithSub\stageOmetaColor{x}  ]    \possiblyWithSub\stageOmetaColor{N^{\superscriptO} }_{{\mathrm{11}}}   \punctO{,}  \possiblyWithSub\stageOmetaColor{c}_{{\mathrm{2}}}  \RightAssertParen^{ L } 
        \\&=  \LeftAssertParen   \openO{\{} \possiblyWithSub\stageOmetaColor{\nu}  \relO{:}  \possiblyWithSub\stageOmetaColor{B}  \relO{\mid}    [  \stageOmetaColor{N_{\mathrm{A} }^{\superscriptO} }  /  \possiblyWithSub\stageOmetaColor{x}  ]    \possiblyWithSub\stageOmetaColor{N^{\superscriptO} }_{{\mathrm{11}}}  \closeO{\} }  \punctO{,}    [  \stageOmetaColor{N_{\mathrm{A} }^{\superscriptO} }  /  \possiblyWithSub\stageOmetaColor{x}  ]      [    \possiblyWithSub\stageOmetaColor{c}_{{\mathrm{2}}}    /  \possiblyWithSub\stageOmetaColor{\nu}  ]    \possiblyWithSub\stageOmetaColor{N^{\superscriptO} }_{{\mathrm{11}}}   \punctO{,}  \possiblyWithSub\stageOmetaColor{c}_{{\mathrm{2}}}  \RightAssertParen^{ L } 
        \\&=   [  \stageOmetaColor{N_{\mathrm{A} }^{\superscriptO} }  /  \possiblyWithSub\stageOmetaColor{x}  ]     \LeftAssertParen   \openO{\{} \possiblyWithSub\stageOmetaColor{\nu}  \relO{:}  \possiblyWithSub\stageOmetaColor{B}  \relO{\mid}  \possiblyWithSub\stageOmetaColor{N^{\superscriptO} }_{{\mathrm{11}}} \closeO{\} }  \punctO{,}    [    \possiblyWithSub\stageOmetaColor{c}_{{\mathrm{2}}}    /  \possiblyWithSub\stageOmetaColor{\nu}  ]    \possiblyWithSub\stageOmetaColor{N^{\superscriptO} }_{{\mathrm{11}}}  \punctO{,}  \possiblyWithSub\stageOmetaColor{c}_{{\mathrm{2}}}  \RightAssertParen^{ L }  
           =   [  \stageOmetaColor{N_{\mathrm{A} }^{\superscriptO} }  /  \possiblyWithSub\stageOmetaColor{x}  ]    \possiblyWithSub\stageOmetaColor{\Hat{N}^{\superscriptO} } .
        \end{align*}
        Also, since \(  [  \stageOmetaColor{N_{\mathrm{B} }^{\superscriptO} }  /  \possiblyWithSub\stageOmetaColor{x}  ]    \possiblyWithSub\stageOmetaColor{N^{\superscriptO} }_{{\mathrm{2}}}  = \possiblyWithSub\stageOmetaColor{c}_{{\mathrm{2}}}\) holds
        and \(\stageOmetaColor{N_{\mathrm{B} }^{\superscriptO} }\) is a closed term,
        we can derive \(   [  \stageOmetaColor{N_{\mathrm{B} }^{\superscriptO} }  /  \possiblyWithSub\stageOmetaColor{x}  ]    \possiblyWithSub\stageOmetaColor{N^{\superscriptO} }   \longrightarrow^{0}     [  \stageOmetaColor{N_{\mathrm{B} }^{\superscriptO} }  /  \possiblyWithSub\stageOmetaColor{x}  ]    \possiblyWithSub\stageOmetaColor{\Hat{N}^{\superscriptO} }   \)
        as follows:
        \begin{center}
          \derive[E0-RfnStart]{}{%
               \LeftAssertParen \relO{\CastArrow}   \openO{\{} \possiblyWithSub\stageOmetaColor{\nu}  \relO{:}  \possiblyWithSub\stageOmetaColor{B}  \relO{\mid}    [  \stageOmetaColor{N_{\mathrm{B} }^{\superscriptO} }  /  \possiblyWithSub\stageOmetaColor{x}  ]    \possiblyWithSub\stageOmetaColor{N^{\superscriptO} }_{{\mathrm{11}}}  \closeO{\} }   \RightAssertParen^{ L }  \    \possiblyWithSub\stageOmetaColor{c}_{{\mathrm{2}}}     \longrightarrow^{0}    \LeftAssertParen   \openO{\{} \possiblyWithSub\stageOmetaColor{\nu}  \relO{:}  \possiblyWithSub\stageOmetaColor{B}  \relO{\mid}    [  \stageOmetaColor{N_{\mathrm{B} }^{\superscriptO} }  /  \possiblyWithSub\stageOmetaColor{x}  ]    \possiblyWithSub\stageOmetaColor{N^{\superscriptO} }_{{\mathrm{11}}}  \closeO{\} }  \punctO{,}    [    \possiblyWithSub\stageOmetaColor{c}_{{\mathrm{2}}}    /  \possiblyWithSub\stageOmetaColor{\nu}  ]      [  \stageOmetaColor{N_{\mathrm{B} }^{\superscriptO} }  /  \possiblyWithSub\stageOmetaColor{x}  ]    \possiblyWithSub\stageOmetaColor{N^{\superscriptO} }_{{\mathrm{11}}}   \punctO{,}  \possiblyWithSub\stageOmetaColor{c}_{{\mathrm{2}}}  \RightAssertParen^{ L }   
          }.
        \end{center}
      \item The other cases contradict the assumption~\(   [  \stageOmetaColor{N_{\mathrm{A} }^{\superscriptO} }  /  \possiblyWithSub\stageOmetaColor{x}  ]     \openO{(}   \possiblyWithSub\stageOmetaColor{v^{\superscriptO} }_{{\mathrm{1}}}  \  \possiblyWithSub\stageOmetaColor{N^{\superscriptO} }_{{\mathrm{2}}}  \closeO{)}    \longrightarrow^{0}   \possiblyWithSub\stageOmetaColor{N'^{\superscriptO} }  \).
    \end{itemize}
  \end{proof}
  \begin{lemma}[Bisimulation on essential reductions of applications]\label{lem:weak-bisimulation-app-left-and-right}
    Suppose \( \possiblyWithSub\stageOmetaColor{N^{\superscriptO} }_{{\mathrm{0}}}  \longrightarrow^{0}   \possiblyWithSub\stageOmetaColor{N'^{\superscriptO} }_{{\mathrm{0}}}  \) and that
    \(  [  \possiblyWithSub\stageOmetaColor{N^{\superscriptO} }_{{\mathrm{0}}}  /  \possiblyWithSub\stageOmetaColor{x}  ]    \possiblyWithSub\stageOmetaColor{N^{\superscriptO} }_{{\mathrm{1}}} \), \(  [  \possiblyWithSub\stageOmetaColor{N^{\superscriptO} }_{{\mathrm{0}}}  /  \possiblyWithSub\stageOmetaColor{x}  ]    \possiblyWithSub\stageOmetaColor{N^{\superscriptO} }_{{\mathrm{2}}} \), and \(  [  \possiblyWithSub\stageOmetaColor{N'^{\superscriptO} }_{{\mathrm{0}}}  /  \possiblyWithSub\stageOmetaColor{x}  ]    \possiblyWithSub\stageOmetaColor{N^{\superscriptO} }_{{\mathrm{2}}} \) are all values.
    \begin{enumerate}
      \item Left-hand side:
        If \(   [  \possiblyWithSub\stageOmetaColor{N^{\superscriptO} }_{{\mathrm{0}}}  /  \possiblyWithSub\stageOmetaColor{x}  ]     \openO{(}  \possiblyWithSub\stageOmetaColor{N^{\superscriptO} }_{{\mathrm{1}}} \  \possiblyWithSub\stageOmetaColor{N^{\superscriptO} }_{{\mathrm{2}}}  \closeO{)}    \longrightarrow^{0}   \possiblyWithSub\stageOmetaColor{N'^{\superscriptO} }  \), then
        there exists \(\possiblyWithSub\stageOmetaColor{\Hat{N}^{\superscriptO} }\) such that
        \(   [  \possiblyWithSub\stageOmetaColor{N'^{\superscriptO} }_{{\mathrm{0}}}  /  \possiblyWithSub\stageOmetaColor{x}  ]     \openO{(}  \possiblyWithSub\stageOmetaColor{N^{\superscriptO} }_{{\mathrm{1}}} \  \possiblyWithSub\stageOmetaColor{N^{\superscriptO} }_{{\mathrm{2}}}  \closeO{)}    \longrightarrow^{0}     [  \possiblyWithSub\stageOmetaColor{N'^{\superscriptO} }_{{\mathrm{0}}}  /  \possiblyWithSub\stageOmetaColor{x}  ]    \possiblyWithSub\stageOmetaColor{\Hat{N}^{\superscriptO} }   \)
        and \(\possiblyWithSub\stageOmetaColor{N'^{\superscriptO} } =   [  \possiblyWithSub\stageOmetaColor{N^{\superscriptO} }_{{\mathrm{0}}}  /  \possiblyWithSub\stageOmetaColor{x}  ]    \possiblyWithSub\stageOmetaColor{\Hat{N}^{\superscriptO} } \).
      \item Right-hand side:
        If \(   [  \possiblyWithSub\stageOmetaColor{N'^{\superscriptO} }_{{\mathrm{0}}}  /  \possiblyWithSub\stageOmetaColor{x}  ]     \openO{(}  \possiblyWithSub\stageOmetaColor{N^{\superscriptO} }_{{\mathrm{1}}} \  \possiblyWithSub\stageOmetaColor{N^{\superscriptO} }_{{\mathrm{2}}}  \closeO{)}    \longrightarrow^{0}   \possiblyWithSub\stageOmetaColor{N'^{\superscriptO} }  \), then
        there exists \(\possiblyWithSub\stageOmetaColor{\Hat{N}^{\superscriptO} }\) such that
        \(   [  \possiblyWithSub\stageOmetaColor{N^{\superscriptO} }_{{\mathrm{0}}}  /  \possiblyWithSub\stageOmetaColor{x}  ]     \openO{(}  \possiblyWithSub\stageOmetaColor{N^{\superscriptO} }_{{\mathrm{1}}} \  \possiblyWithSub\stageOmetaColor{N^{\superscriptO} }_{{\mathrm{2}}}  \closeO{)}    \longrightarrow^{0}     [  \possiblyWithSub\stageOmetaColor{N^{\superscriptO} }_{{\mathrm{0}}}  /  \possiblyWithSub\stageOmetaColor{x}  ]    \possiblyWithSub\stageOmetaColor{\Hat{N}^{\superscriptO} }   \)
        and \(\possiblyWithSub\stageOmetaColor{N'^{\superscriptO} } =   [  \possiblyWithSub\stageOmetaColor{N'^{\superscriptO} }_{{\mathrm{0}}}  /  \possiblyWithSub\stageOmetaColor{x}  ]    \possiblyWithSub\stageOmetaColor{\Hat{N}^{\superscriptO} } \).
    \end{enumerate}
  \end{lemma}
  \begin{proof}
    Since \(  [  \possiblyWithSub\stageOmetaColor{N^{\superscriptO} }_{{\mathrm{0}}}  /  \possiblyWithSub\stageOmetaColor{x}  ]    \possiblyWithSub\stageOmetaColor{N^{\superscriptO} }_{{\mathrm{1}}} \) is a value while \(\possiblyWithSub\stageOmetaColor{N^{\superscriptO} }_{{\mathrm{0}}}\) is not (by \( \possiblyWithSub\stageOmetaColor{N^{\superscriptO} }_{{\mathrm{0}}}  \longrightarrow^{0}   \possiblyWithSub\stageOmetaColor{N'^{\superscriptO} }_{{\mathrm{0}}}  \)),
    \(\possiblyWithSub\stageOmetaColor{N^{\superscriptO} }_{{\mathrm{1}}}\) is also a value by Lemma~\ref{lem:subst-valueness}.
    Therefore, we can prove (1) and (2) as follows:
    \begin{enumerate}
      \item
        We can finish by Lemma~\ref{lem:weak-bisimulation-app-pre}
        (where \(\stageOmetaColor{N_{\mathrm{A} }^{\superscriptO} } \defeq \possiblyWithSub\stageOmetaColor{N^{\superscriptO} }_{{\mathrm{0}}}\) and \(\stageOmetaColor{N_{\mathrm{B} }^{\superscriptO} } \defeq \possiblyWithSub\stageOmetaColor{N'^{\superscriptO} }_{{\mathrm{0}}}\)).
      \item
        We can finish by Lemma~\ref{lem:weak-bisimulation-app-pre}
        (where \(\stageOmetaColor{N_{\mathrm{A} }^{\superscriptO} } \defeq \possiblyWithSub\stageOmetaColor{N'^{\superscriptO} }_{{\mathrm{0}}}\) and \(\stageOmetaColor{N_{\mathrm{B} }^{\superscriptO} } \defeq \possiblyWithSub\stageOmetaColor{N^{\superscriptO} }_{{\mathrm{0}}}\)).
    \end{enumerate}
  \end{proof}
  \begin{lemma}[Weak bisimulation, left-hand side]\label{lem:weak-bisimulation-left}
    Suppose \( \possiblyWithSub\stageOmetaColor{N^{\superscriptO} }_{{\mathrm{0}}}  \longrightarrow^{0}   \possiblyWithSub\stageOmetaColor{N'^{\superscriptO} }_{{\mathrm{0}}}  \).
    \begin{enumerate}
      \item
        If \(   [  \possiblyWithSub\stageOmetaColor{N^{\superscriptO} }_{{\mathrm{0}}}  /  \possiblyWithSub\stageOmetaColor{x}  ]    \possiblyWithSub\stageOmetaColor{N^{\superscriptO} }   \longrightarrow^{0}   \possiblyWithSub\stageOmetaColor{N'^{\superscriptO} }  \),
        then there exists \(\possiblyWithSub\stageOmetaColor{\Hat{N}^{\superscriptO} }\) such that
        \(   [  \possiblyWithSub\stageOmetaColor{N'^{\superscriptO} }_{{\mathrm{0}}}  /  \possiblyWithSub\stageOmetaColor{x}  ]    \possiblyWithSub\stageOmetaColor{N^{\superscriptO} }   \longrightarrow^{0\,\ast}     [  \possiblyWithSub\stageOmetaColor{N'^{\superscriptO} }_{{\mathrm{0}}}  /  \possiblyWithSub\stageOmetaColor{x}  ]    \possiblyWithSub\stageOmetaColor{\Hat{N}^{\superscriptO} }   \)
        and \(\possiblyWithSub\stageOmetaColor{N'^{\superscriptO} } =   [  \possiblyWithSub\stageOmetaColor{N^{\superscriptO} }_{{\mathrm{0}}}  /  \possiblyWithSub\stageOmetaColor{x}  ]    \possiblyWithSub\stageOmetaColor{\Hat{N}^{\superscriptO} } \).
      \item
        If \(   [  \possiblyWithSub\stageOmetaColor{N^{\superscriptO} }_{{\mathrm{0}}}  /  \possiblyWithSub\stageOmetaColor{x}  ]    \possiblyWithSub\stageImetaColor{N^{\superscriptI} }   \longrightarrow^{1}   \possiblyWithSub\stageImetaColor{N'^{\superscriptI} }  \),
        then there exists \(\possiblyWithSub\stageImetaColor{\Hat{N}^{\superscriptI} }\) such that
        \(   [  \possiblyWithSub\stageOmetaColor{N'^{\superscriptO} }_{{\mathrm{0}}}  /  \possiblyWithSub\stageOmetaColor{x}  ]    \possiblyWithSub\stageImetaColor{N^{\superscriptI} }   \longrightarrow^{1\,\ast}     [  \possiblyWithSub\stageOmetaColor{N'^{\superscriptO} }_{{\mathrm{0}}}  /  \possiblyWithSub\stageOmetaColor{x}  ]    \possiblyWithSub\stageImetaColor{\Hat{N}^{\superscriptI} }   \)
        and \(\possiblyWithSub\stageImetaColor{N'^{\superscriptI} } =   [  \possiblyWithSub\stageOmetaColor{N^{\superscriptO} }_{{\mathrm{0}}}  /  \possiblyWithSub\stageOmetaColor{x}  ]    \possiblyWithSub\stageImetaColor{\Hat{N}^{\superscriptI} } \).
      \item
        If \(   [  \possiblyWithSub\stageOmetaColor{N^{\superscriptO} }_{{\mathrm{0}}}  /  \possiblyWithSub\stageOmetaColor{x}  ]    \possiblyWithSub\stageImetaColor{T^{\superscriptI} }   \longrightarrow^{1}   \possiblyWithSub\stageImetaColor{T'^{\superscriptI} }  \),
        then there exists \(\possiblyWithSub\stageImetaColor{\Hat{T}^{\superscriptI} }\) such that
        \(   [  \possiblyWithSub\stageOmetaColor{N'^{\superscriptO} }_{{\mathrm{0}}}  /  \possiblyWithSub\stageOmetaColor{x}  ]    \possiblyWithSub\stageImetaColor{T^{\superscriptI} }   \longrightarrow^{1\,\ast}     [  \possiblyWithSub\stageOmetaColor{N'^{\superscriptO} }_{{\mathrm{0}}}  /  \possiblyWithSub\stageOmetaColor{x}  ]    \possiblyWithSub\stageImetaColor{\Hat{T}^{\superscriptI} }   \)
        and \(\possiblyWithSub\stageImetaColor{T'^{\superscriptI} } =   [  \possiblyWithSub\stageOmetaColor{N^{\superscriptO} }_{{\mathrm{0}}}  /  \possiblyWithSub\stageOmetaColor{x}  ]    \possiblyWithSub\stageImetaColor{\Hat{T}^{\superscriptI} } \).
    \end{enumerate}
  \end{lemma}
  \begin{proof}
    By mutual induction on the structure of \(\possiblyWithSub\stageOmetaColor{N^{\superscriptO} }\), \(\possiblyWithSub\stageImetaColor{N^{\superscriptI} }\), and \(\possiblyWithSub\stageImetaColor{T^{\superscriptI} }\).
    \begin{enumerate}
      \item
        \begin{itemize}
          \item Case \(\possiblyWithSub\stageOmetaColor{N^{\superscriptO} } = \possiblyWithSub\stageOmetaColor{x}\):
            Since \(  [  \possiblyWithSub\stageOmetaColor{N^{\superscriptO} }_{{\mathrm{0}}}  /  \possiblyWithSub\stageOmetaColor{x}  ]    \possiblyWithSub\stageOmetaColor{N^{\superscriptO} }  = \possiblyWithSub\stageOmetaColor{N^{\superscriptO} }_{{\mathrm{0}}}\),
            we have \(\possiblyWithSub\stageOmetaColor{N'^{\superscriptO} } = \possiblyWithSub\stageOmetaColor{N'^{\superscriptO} }_{{\mathrm{0}}}\) by Lemma~\ref{lem:determinism}.
            Thus, \(\possiblyWithSub\stageOmetaColor{\Hat{N}^{\superscriptO} } \defeq \possiblyWithSub\stageOmetaColor{N'^{\superscriptO} }_{{\mathrm{0}}}\) satisfies the desired properties
            since \(\possiblyWithSub\stageOmetaColor{N'^{\superscriptO} }_{{\mathrm{0}}}\) is a closed term (by \( \possiblyWithSub\stageOmetaColor{N^{\superscriptO} }_{{\mathrm{0}}}  \longrightarrow^{0}   \possiblyWithSub\stageOmetaColor{N'^{\superscriptO} }_{{\mathrm{0}}}  \)) and thereby
            we have \(  [  \possiblyWithSub\stageOmetaColor{N'^{\superscriptO} }_{{\mathrm{0}}}  /  \possiblyWithSub\stageOmetaColor{x}  ]    \possiblyWithSub\stageOmetaColor{N'^{\superscriptO} }_{{\mathrm{0}}}  =   [  \possiblyWithSub\stageOmetaColor{N^{\superscriptO} }_{{\mathrm{0}}}  /  \possiblyWithSub\stageOmetaColor{x}  ]    \possiblyWithSub\stageOmetaColor{N'^{\superscriptO} }_{{\mathrm{0}}}  = \possiblyWithSub\stageOmetaColor{N'^{\superscriptO} }_{{\mathrm{0}}}\).
          \item Case \(\possiblyWithSub\stageOmetaColor{N^{\superscriptO} } = \possiblyWithSub\stageOmetaColor{x'}\) such that \(\possiblyWithSub\stageOmetaColor{x'} \neq \possiblyWithSub\stageOmetaColor{x}\):
            Since \(  [  \possiblyWithSub\stageOmetaColor{N^{\superscriptO} }_{{\mathrm{0}}}  /  \possiblyWithSub\stageOmetaColor{x}  ]    \possiblyWithSub\stageOmetaColor{N^{\superscriptO} }  = \possiblyWithSub\stageOmetaColor{x'}\),
            this case contradicts the assumption~\(   [  \possiblyWithSub\stageOmetaColor{N^{\superscriptO} }_{{\mathrm{0}}}  /  \possiblyWithSub\stageOmetaColor{x}  ]    \possiblyWithSub\stageOmetaColor{N^{\superscriptO} }   \longrightarrow^{0}   \possiblyWithSub\stageOmetaColor{N'^{\superscriptO} }  \).
          \item Case where we have \(\possiblyWithSub\stageOmetaColor{N^{\superscriptO} } =  \openO{(}  \ordO{\lambda} \possiblyWithSub\stageOmetaColor{x'}  \relO{:}  \possiblyWithSub\stageOmetaColor{T^{\superscriptO} }_{{\mathrm{1}}} \punctO{.}\  \possiblyWithSub\stageOmetaColor{N^{\superscriptO} }_{{\mathrm{2}}}  \closeO{)} \)
          or \(\possiblyWithSub\stageOmetaColor{N^{\superscriptO} } =  \LeftAssertParen \relO{\CastArrow}   \openO{\{} \possiblyWithSub\stageOmetaColor{x'}  \relO{:}  \possiblyWithSub\stageOmetaColor{B}  \relO{\mid}  \possiblyWithSub\stageOmetaColor{N^{\superscriptO} }_{{\mathrm{1}}} \closeO{\} }   \RightAssertParen^{ L } \) also contradicts
          the assumption~\(   [  \possiblyWithSub\stageOmetaColor{N^{\superscriptO} }_{{\mathrm{0}}}  /  \possiblyWithSub\stageOmetaColor{x}  ]    \possiblyWithSub\stageOmetaColor{N^{\superscriptO} }   \longrightarrow^{0}   \possiblyWithSub\stageOmetaColor{N'^{\superscriptO} }  \).
          \item Case \(\possiblyWithSub\stageOmetaColor{N^{\superscriptO} } =  \possiblyWithSub\stageOmetaColor{N^{\superscriptO} }_{{\mathrm{1}}} \  \possiblyWithSub\stageOmetaColor{N^{\superscriptO} }_{{\mathrm{2}}} \):
            We have one of the following cases:
            \begin{itemize}
              \item Case where \(  [  \possiblyWithSub\stageOmetaColor{N^{\superscriptO} }_{{\mathrm{0}}}  /  \possiblyWithSub\stageOmetaColor{x}  ]    \possiblyWithSub\stageOmetaColor{N^{\superscriptO} }_{{\mathrm{1}}} \) is not a value:
                We have only the following form
                as the derivation of \(   \openO{(}   [  \possiblyWithSub\stageOmetaColor{N^{\superscriptO} }_{{\mathrm{0}}}  /  \possiblyWithSub\stageOmetaColor{x}  ]    \possiblyWithSub\stageOmetaColor{N^{\superscriptO} }_{{\mathrm{1}}}  \closeO{)}  \   \openO{(}   [  \possiblyWithSub\stageOmetaColor{N^{\superscriptO} }_{{\mathrm{0}}}  /  \possiblyWithSub\stageOmetaColor{x}  ]    \possiblyWithSub\stageOmetaColor{N^{\superscriptO} }_{{\mathrm{2}}}  \closeO{)}    \longrightarrow^{0}   \possiblyWithSub\stageOmetaColor{N'^{\superscriptO} }  \):
                \begin{center}
                  \derive[E0-App1]{%
                       [  \possiblyWithSub\stageOmetaColor{N^{\superscriptO} }_{{\mathrm{0}}}  /  \possiblyWithSub\stageOmetaColor{x}  ]    \possiblyWithSub\stageOmetaColor{N^{\superscriptO} }_{{\mathrm{1}}}   \longrightarrow^{0}   \possiblyWithSub\stageOmetaColor{N'^{\superscriptO} }_{{\mathrm{1}}}  
                  }{%
                       \openO{(}   [  \possiblyWithSub\stageOmetaColor{N^{\superscriptO} }_{{\mathrm{0}}}  /  \possiblyWithSub\stageOmetaColor{x}  ]    \possiblyWithSub\stageOmetaColor{N^{\superscriptO} }_{{\mathrm{1}}}  \closeO{)}  \   \openO{(}   [  \possiblyWithSub\stageOmetaColor{N^{\superscriptO} }_{{\mathrm{0}}}  /  \possiblyWithSub\stageOmetaColor{x}  ]    \possiblyWithSub\stageOmetaColor{N^{\superscriptO} }_{{\mathrm{2}}}  \closeO{)}    \longrightarrow^{0}    \possiblyWithSub\stageOmetaColor{N'^{\superscriptO} }_{{\mathrm{1}}} \   \openO{(}   [  \possiblyWithSub\stageOmetaColor{N^{\superscriptO} }_{{\mathrm{0}}}  /  \possiblyWithSub\stageOmetaColor{x}  ]    \possiblyWithSub\stageOmetaColor{N^{\superscriptO} }_{{\mathrm{2}}}  \closeO{)}    
                  }.
                \end{center}
                By IH, there exists \(\possiblyWithSub\stageOmetaColor{\Hat{N}^{\superscriptO} }_{{\mathrm{1}}}\) such that
                \(   [  \possiblyWithSub\stageOmetaColor{N'^{\superscriptO} }_{{\mathrm{0}}}  /  \possiblyWithSub\stageOmetaColor{x}  ]    \possiblyWithSub\stageOmetaColor{N^{\superscriptO} }_{{\mathrm{1}}}   \longrightarrow^{0\,\ast}     [  \possiblyWithSub\stageOmetaColor{N'^{\superscriptO} }_{{\mathrm{0}}}  /  \possiblyWithSub\stageOmetaColor{x}  ]    \possiblyWithSub\stageOmetaColor{\Hat{N}^{\superscriptO} }_{{\mathrm{1}}}   \)
                and \(\possiblyWithSub\stageOmetaColor{N'^{\superscriptO} }_{{\mathrm{1}}} =   [  \possiblyWithSub\stageOmetaColor{N^{\superscriptO} }_{{\mathrm{0}}}  /  \possiblyWithSub\stageOmetaColor{x}  ]    \possiblyWithSub\stageOmetaColor{\Hat{N}^{\superscriptO} }_{{\mathrm{1}}} \).
                Thus, \(\possiblyWithSub\stageOmetaColor{\Hat{N}^{\superscriptO} } \defeq  \possiblyWithSub\stageOmetaColor{\Hat{N}^{\superscriptO} }_{{\mathrm{1}}} \  \possiblyWithSub\stageOmetaColor{N^{\superscriptO} }_{{\mathrm{2}}} \) satisfies the desired properties.
              \item Case where \(  [  \possiblyWithSub\stageOmetaColor{N^{\superscriptO} }_{{\mathrm{0}}}  /  \possiblyWithSub\stageOmetaColor{x}  ]    \possiblyWithSub\stageOmetaColor{N^{\superscriptO} }_{{\mathrm{1}}}  \revdefeq \possiblyWithSub\stageOmetaColor{v^{\superscriptO} }_{{\mathrm{1}}}\) is a value
              while \(  [  \possiblyWithSub\stageOmetaColor{N^{\superscriptO} }_{{\mathrm{0}}}  /  \possiblyWithSub\stageOmetaColor{x}  ]    \possiblyWithSub\stageOmetaColor{N^{\superscriptO} }_{{\mathrm{2}}} \) is not:
                We have only the following form
                as the derivation of \(   \openO{(}   [  \possiblyWithSub\stageOmetaColor{N^{\superscriptO} }_{{\mathrm{0}}}  /  \possiblyWithSub\stageOmetaColor{x}  ]    \possiblyWithSub\stageOmetaColor{N^{\superscriptO} }_{{\mathrm{1}}}  \closeO{)}  \   \openO{(}   [  \possiblyWithSub\stageOmetaColor{N^{\superscriptO} }_{{\mathrm{0}}}  /  \possiblyWithSub\stageOmetaColor{x}  ]    \possiblyWithSub\stageOmetaColor{N^{\superscriptO} }_{{\mathrm{2}}}  \closeO{)}    \longrightarrow^{0}   \possiblyWithSub\stageOmetaColor{N'^{\superscriptO} }  \):
                \begin{center}
                  \derive[E0-App2]{%
                       [  \possiblyWithSub\stageOmetaColor{N^{\superscriptO} }_{{\mathrm{0}}}  /  \possiblyWithSub\stageOmetaColor{x}  ]    \possiblyWithSub\stageOmetaColor{N^{\superscriptO} }_{{\mathrm{2}}}   \longrightarrow^{0}   \possiblyWithSub\stageOmetaColor{N'^{\superscriptO} }_{{\mathrm{2}}}  
                  }{%
                       \possiblyWithSub\stageOmetaColor{v^{\superscriptO} }_{{\mathrm{1}}}  \   \openO{(}   [  \possiblyWithSub\stageOmetaColor{N^{\superscriptO} }_{{\mathrm{0}}}  /  \possiblyWithSub\stageOmetaColor{x}  ]    \possiblyWithSub\stageOmetaColor{N^{\superscriptO} }_{{\mathrm{2}}}  \closeO{)}    \longrightarrow^{0}     \possiblyWithSub\stageOmetaColor{v^{\superscriptO} }_{{\mathrm{1}}}  \  \possiblyWithSub\stageOmetaColor{N'^{\superscriptO} }_{{\mathrm{2}}}   
                  }.
                \end{center}
                This case can be proved in the same manner as the previous one.
              \item Case where both \(  [  \possiblyWithSub\stageOmetaColor{N^{\superscriptO} }_{{\mathrm{0}}}  /  \possiblyWithSub\stageOmetaColor{x}  ]    \possiblyWithSub\stageOmetaColor{N^{\superscriptO} }_{{\mathrm{1}}} \) and \(  [  \possiblyWithSub\stageOmetaColor{N^{\superscriptO} }_{{\mathrm{0}}}  /  \possiblyWithSub\stageOmetaColor{x}  ]    \possiblyWithSub\stageOmetaColor{N^{\superscriptO} }_{{\mathrm{2}}} \) are values:
                Since \(  [  \possiblyWithSub\stageOmetaColor{N^{\superscriptO} }_{{\mathrm{0}}}  /  \possiblyWithSub\stageOmetaColor{x}  ]    \possiblyWithSub\stageOmetaColor{N^{\superscriptO} }_{{\mathrm{2}}} \) is a value while \(\possiblyWithSub\stageOmetaColor{N^{\superscriptO} }_{{\mathrm{0}}}\) is not,
                \(\possiblyWithSub\stageOmetaColor{N^{\superscriptO} }_{{\mathrm{2}}}\) is also a value by Lemma~\ref{lem:subst-valueness}.
                Thus, \(  [  \possiblyWithSub\stageOmetaColor{N'^{\superscriptO} }_{{\mathrm{0}}}  /  \possiblyWithSub\stageOmetaColor{x}  ]    \possiblyWithSub\stageOmetaColor{N^{\superscriptO} }_{{\mathrm{2}}} \) is clearly a value,
                and we can finish the proof by
                Lemma~\ref{lem:weak-bisimulation-app-left-and-right}~(1).
            \end{itemize}
          \item Case \(\possiblyWithSub\stageOmetaColor{N^{\superscriptO} } =  \openO{\langle} \possiblyWithSub\stageImetaColor{N^{\superscriptI} } \closeO{\rangle} \):
            By tracing back the derivation, we have
            \begin{center}
              \derive[E0-Brkt]{%
                   [  \possiblyWithSub\stageOmetaColor{N^{\superscriptO} }_{{\mathrm{0}}}  /  \possiblyWithSub\stageOmetaColor{x}  ]    \possiblyWithSub\stageImetaColor{N^{\superscriptI} }   \longrightarrow^{1}   \possiblyWithSub\stageImetaColor{N'^{\superscriptI} }  
              }{%
                  \openO{\langle}   [  \possiblyWithSub\stageOmetaColor{N^{\superscriptO} }_{{\mathrm{0}}}  /  \possiblyWithSub\stageOmetaColor{x}  ]    \possiblyWithSub\stageImetaColor{N^{\superscriptI} }  \closeO{\rangle}   \longrightarrow^{0}    \openO{\langle} \possiblyWithSub\stageImetaColor{N'^{\superscriptI} } \closeO{\rangle}   
              }.
            \end{center}
            Then, by IH, there exists \(\possiblyWithSub\stageImetaColor{\Hat{N}^{\superscriptI} }\) such that
            \(   [  \possiblyWithSub\stageOmetaColor{N'^{\superscriptO} }_{{\mathrm{0}}}  /  \possiblyWithSub\stageOmetaColor{x}  ]    \possiblyWithSub\stageImetaColor{N^{\superscriptI} }   \longrightarrow^{1\,\ast}     [  \possiblyWithSub\stageOmetaColor{N'^{\superscriptO} }_{{\mathrm{0}}}  /  \possiblyWithSub\stageOmetaColor{x}  ]    \possiblyWithSub\stageImetaColor{\Hat{N}^{\superscriptI} }   \)
            and \(\possiblyWithSub\stageImetaColor{N'^{\superscriptI} } =   [  \possiblyWithSub\stageOmetaColor{N^{\superscriptO} }_{{\mathrm{0}}}  /  \possiblyWithSub\stageOmetaColor{x}  ]    \possiblyWithSub\stageImetaColor{\Hat{N}^{\superscriptI} } \).
            Thus, letting \(\possiblyWithSub\stageOmetaColor{\Hat{N}^{\superscriptO} } \defeq  \openO{\langle} \possiblyWithSub\stageImetaColor{\Hat{N}^{\superscriptI} } \closeO{\rangle} \),
            we have \(\possiblyWithSub\stageOmetaColor{N'^{\superscriptO} } =   [  \possiblyWithSub\stageOmetaColor{N^{\superscriptO} }_{{\mathrm{0}}}  /  \possiblyWithSub\stageOmetaColor{x}  ]    \possiblyWithSub\stageOmetaColor{\Hat{N}^{\superscriptO} } \) and can clearly derive
            \(   [  \possiblyWithSub\stageOmetaColor{N'^{\superscriptO} }_{{\mathrm{0}}}  /  \possiblyWithSub\stageOmetaColor{x}  ]    \possiblyWithSub\stageOmetaColor{N^{\superscriptO} }   \longrightarrow^{0\,\ast}     [  \possiblyWithSub\stageOmetaColor{N'^{\superscriptO} }_{{\mathrm{0}}}  /  \possiblyWithSub\stageOmetaColor{x}  ]    \possiblyWithSub\stageOmetaColor{\Hat{N}^{\superscriptO} }   \)
            by reflexivity and the repeated use of \rulename{E0-Brkt}.
          \item Case \(\possiblyWithSub\stageOmetaColor{N^{\superscriptO} } =  \LeftAssertParen\openO{\langle} \possiblyWithSub\stageImetaColor{T^{\superscriptI} }_{{\mathrm{1}}} \closeO{\rangle} \relO{\CastArrow} \openO{\langle} \possiblyWithSub\stageImetaColor{T^{\superscriptI} }_{{\mathrm{2}}} \closeO{\rangle}\RightAssertParen^{ L } \):
            We have one of the following cases with the derivation of
            \(  \LeftAssertParen\openO{\langle}   [  \possiblyWithSub\stageOmetaColor{N^{\superscriptO} }_{{\mathrm{0}}}  /  \possiblyWithSub\stageOmetaColor{x}  ]    \possiblyWithSub\stageImetaColor{T^{\superscriptI} }_{{\mathrm{1}}}  \closeO{\rangle} \relO{\CastArrow} \openO{\langle}   [  \possiblyWithSub\stageOmetaColor{N^{\superscriptO} }_{{\mathrm{0}}}  /  \possiblyWithSub\stageOmetaColor{x}  ]    \possiblyWithSub\stageImetaColor{T^{\superscriptI} }_{{\mathrm{2}}}  \closeO{\rangle}\RightAssertParen^{ L }   \longrightarrow^{0}   \possiblyWithSub\stageOmetaColor{N'^{\superscriptO} }  \):
            \begin{itemize}
              \item Case \derive[E0-Ass1]{%
                   [  \possiblyWithSub\stageOmetaColor{N^{\superscriptO} }_{{\mathrm{0}}}  /  \possiblyWithSub\stageOmetaColor{x}  ]    \possiblyWithSub\stageImetaColor{T^{\superscriptI} }_{{\mathrm{1}}}   \longrightarrow^{1}   \possiblyWithSub\stageImetaColor{T'^{\superscriptI} }_{{\mathrm{1}}}  
              }{%
                  \LeftAssertParen\openO{\langle}   [  \possiblyWithSub\stageOmetaColor{N^{\superscriptO} }_{{\mathrm{0}}}  /  \possiblyWithSub\stageOmetaColor{x}  ]    \possiblyWithSub\stageImetaColor{T^{\superscriptI} }_{{\mathrm{1}}}  \closeO{\rangle} \relO{\CastArrow} \openO{\langle}   [  \possiblyWithSub\stageOmetaColor{N^{\superscriptO} }_{{\mathrm{0}}}  /  \possiblyWithSub\stageOmetaColor{x}  ]    \possiblyWithSub\stageImetaColor{T^{\superscriptI} }_{{\mathrm{2}}}  \closeO{\rangle}\RightAssertParen^{ L }   \longrightarrow^{0}    \LeftAssertParen\openO{\langle} \possiblyWithSub\stageImetaColor{T'^{\superscriptI} }_{{\mathrm{1}}} \closeO{\rangle} \relO{\CastArrow} \openO{\langle}   [  \possiblyWithSub\stageOmetaColor{N^{\superscriptO} }_{{\mathrm{0}}}  /  \possiblyWithSub\stageOmetaColor{x}  ]    \possiblyWithSub\stageImetaColor{T^{\superscriptI} }_{{\mathrm{2}}}  \closeO{\rangle}\RightAssertParen^{ L }   
              }:
                By IH, there exists \(\possiblyWithSub\stageImetaColor{\Hat{T}^{\superscriptI} }_{{\mathrm{1}}}\) such that
                \(   [  \possiblyWithSub\stageOmetaColor{N'^{\superscriptO} }_{{\mathrm{0}}}  /  \possiblyWithSub\stageOmetaColor{x}  ]    \possiblyWithSub\stageImetaColor{T^{\superscriptI} }_{{\mathrm{1}}}   \longrightarrow^{1\,\ast}     [  \possiblyWithSub\stageOmetaColor{N'^{\superscriptO} }_{{\mathrm{0}}}  /  \possiblyWithSub\stageOmetaColor{x}  ]    \possiblyWithSub\stageImetaColor{\Hat{T}^{\superscriptI} }_{{\mathrm{1}}}   \)
                and \(\possiblyWithSub\stageImetaColor{T'^{\superscriptI} }_{{\mathrm{1}}} =   [  \possiblyWithSub\stageOmetaColor{N^{\superscriptO} }_{{\mathrm{0}}}  /  \possiblyWithSub\stageOmetaColor{x}  ]    \possiblyWithSub\stageImetaColor{\Hat{T}^{\superscriptI} }_{{\mathrm{1}}} \).
                We will show that
                \(\possiblyWithSub\stageOmetaColor{\Hat{T}^{\superscriptO} } \defeq  \LeftAssertParen\openO{\langle} \possiblyWithSub\stageImetaColor{\Hat{T}^{\superscriptI} }_{{\mathrm{1}}} \closeO{\rangle} \relO{\CastArrow} \openO{\langle} \possiblyWithSub\stageImetaColor{T^{\superscriptI} }_{{\mathrm{2}}} \closeO{\rangle}\RightAssertParen^{ L } \)
                satisfies the desired properties.
                First, we have
                \begin{align*}
                  \possiblyWithSub\stageOmetaColor{N'^{\superscriptO} }
                  &=  \LeftAssertParen\openO{\langle} \possiblyWithSub\stageImetaColor{T'^{\superscriptI} }_{{\mathrm{1}}} \closeO{\rangle} \relO{\CastArrow} \openO{\langle}   [  \possiblyWithSub\stageOmetaColor{N^{\superscriptO} }_{{\mathrm{0}}}  /  \possiblyWithSub\stageOmetaColor{x}  ]    \possiblyWithSub\stageImetaColor{T^{\superscriptI} }_{{\mathrm{2}}}  \closeO{\rangle}\RightAssertParen^{ L } 
                \\&=  \LeftAssertParen\openO{\langle}   [  \possiblyWithSub\stageOmetaColor{N^{\superscriptO} }_{{\mathrm{0}}}  /  \possiblyWithSub\stageOmetaColor{x}  ]    \possiblyWithSub\stageImetaColor{\Hat{T}^{\superscriptI} }_{{\mathrm{1}}}  \closeO{\rangle} \relO{\CastArrow} \openO{\langle}   [  \possiblyWithSub\stageOmetaColor{N^{\superscriptO} }_{{\mathrm{0}}}  /  \possiblyWithSub\stageOmetaColor{x}  ]    \possiblyWithSub\stageImetaColor{T^{\superscriptI} }_{{\mathrm{2}}}  \closeO{\rangle}\RightAssertParen^{ L } 
                \\&=   [  \possiblyWithSub\stageOmetaColor{N^{\superscriptO} }_{{\mathrm{0}}}  /  \possiblyWithSub\stageOmetaColor{x}  ]     \LeftAssertParen\openO{\langle} \possiblyWithSub\stageImetaColor{\Hat{T}^{\superscriptI} }_{{\mathrm{1}}} \closeO{\rangle} \relO{\CastArrow} \openO{\langle} \possiblyWithSub\stageImetaColor{T^{\superscriptI} }_{{\mathrm{2}}} \closeO{\rangle}\RightAssertParen^{ L }  
                  =   [  \possiblyWithSub\stageOmetaColor{N^{\superscriptO} }_{{\mathrm{0}}}  /  \possiblyWithSub\stageOmetaColor{x}  ]    \possiblyWithSub\stageOmetaColor{\Hat{T}^{\superscriptO} } .
                \end{align*}
                Also, by reflexivity and the repeated use of \rulename{E0-Ass1},
                we can derive \(   [  \possiblyWithSub\stageOmetaColor{N'^{\superscriptO} }_{{\mathrm{0}}}  /  \possiblyWithSub\stageOmetaColor{x}  ]    \possiblyWithSub\stageOmetaColor{N^{\superscriptO} }   \longrightarrow^{0\,\ast}     [  \possiblyWithSub\stageOmetaColor{N'^{\superscriptO} }_{{\mathrm{0}}}  /  \possiblyWithSub\stageOmetaColor{x}  ]    \possiblyWithSub\stageOmetaColor{\Hat{N}^{\superscriptO} }   \)
                from \(   [  \possiblyWithSub\stageOmetaColor{N'^{\superscriptO} }_{{\mathrm{0}}}  /  \possiblyWithSub\stageOmetaColor{x}  ]    \possiblyWithSub\stageImetaColor{T^{\superscriptI} }_{{\mathrm{1}}}   \longrightarrow^{1\,\ast}     [  \possiblyWithSub\stageOmetaColor{N'^{\superscriptO} }_{{\mathrm{0}}}  /  \possiblyWithSub\stageOmetaColor{x}  ]    \possiblyWithSub\stageImetaColor{\Hat{T}^{\superscriptI} }_{{\mathrm{1}}}   \).
              \item Case \derive[E0-Ass2]{%
                   [  \possiblyWithSub\stageOmetaColor{N^{\superscriptO} }_{{\mathrm{0}}}  /  \possiblyWithSub\stageOmetaColor{x}  ]    \possiblyWithSub\stageImetaColor{T^{\superscriptI} }_{{\mathrm{2}}}   \longrightarrow^{1}   \possiblyWithSub\stageImetaColor{T'^{\superscriptI} }_{{\mathrm{2}}}  
              }{%
                  \LeftAssertParen\openO{\langle}  \possiblyWithSub\stageImetaColor{\tau^{\superscriptI} }_{{\mathrm{1}}}  \closeO{\rangle} \relO{\CastArrow} \openO{\langle}   [  \possiblyWithSub\stageOmetaColor{N^{\superscriptO} }_{{\mathrm{0}}}  /  \possiblyWithSub\stageOmetaColor{x}  ]    \possiblyWithSub\stageImetaColor{T^{\superscriptI} }_{{\mathrm{2}}}  \closeO{\rangle}\RightAssertParen^{ L }   \longrightarrow^{0}    \LeftAssertParen\openO{\langle}  \possiblyWithSub\stageImetaColor{\tau^{\superscriptI} }_{{\mathrm{1}}}  \closeO{\rangle} \relO{\CastArrow} \openO{\langle}   [  \possiblyWithSub\stageOmetaColor{N^{\superscriptO} }_{{\mathrm{0}}}  /  \possiblyWithSub\stageOmetaColor{x}  ]    \possiblyWithSub\stageImetaColor{T'^{\superscriptI} }_{{\mathrm{2}}}  \closeO{\rangle}\RightAssertParen^{ L }   
              }, where \(\possiblyWithSub\stageImetaColor{\tau^{\superscriptI} }_{{\mathrm{1}}} =   [  \possiblyWithSub\stageOmetaColor{N^{\superscriptO} }_{{\mathrm{0}}}  /  \possiblyWithSub\stageOmetaColor{x}  ]    \possiblyWithSub\stageImetaColor{T^{\superscriptI} }_{{\mathrm{1}}} \):
                Can be proved in the same manner as the previous case.
              \item Case \derive[E0-AssPass]{}{%
                  \LeftAssertParen\openO{\langle}  \possiblyWithSub\stageImetaColor{\tau^{\superscriptI} }  \closeO{\rangle} \relO{\CastArrow} \openO{\langle}  \possiblyWithSub\stageImetaColor{\tau^{\superscriptI} }  \closeO{\rangle}\RightAssertParen^{ L }   \longrightarrow^{0}    \ordO{\lambda} \possiblyWithSub\stageOmetaColor{x'}  \relO{:}   \openO{\langle}  \possiblyWithSub\stageImetaColor{\tau^{\superscriptI} }  \closeO{\rangle}  \punctO{.}\   \possiblyWithSub\stageOmetaColor{x'}    
              }, where \(\possiblyWithSub\stageImetaColor{\tau^{\superscriptI} } =   [  \possiblyWithSub\stageOmetaColor{N^{\superscriptO} }_{{\mathrm{0}}}  /  \possiblyWithSub\stageOmetaColor{x}  ]    \possiblyWithSub\stageImetaColor{T^{\superscriptI} }_{{\mathrm{1}}}  =   [  \possiblyWithSub\stageOmetaColor{N^{\superscriptO} }_{{\mathrm{0}}}  /  \possiblyWithSub\stageOmetaColor{x}  ]    \possiblyWithSub\stageImetaColor{T^{\superscriptI} }_{{\mathrm{2}}} \):
                Since \(\possiblyWithSub\stageOmetaColor{N^{\superscriptO} }_{{\mathrm{0}}}\) is not a value,
                we have \(\possiblyWithSub\stageImetaColor{\tau^{\superscriptI} } = \possiblyWithSub\stageImetaColor{T^{\superscriptI} }_{{\mathrm{1}}} = \possiblyWithSub\stageImetaColor{T^{\superscriptI} }_{{\mathrm{2}}}\) by Lemma~\ref{lem:subst-valueness}.
                Therefore, \(\possiblyWithSub\stageOmetaColor{\Hat{N}^{\superscriptO} } \defeq  \openO{(}  \ordO{\lambda} \possiblyWithSub\stageOmetaColor{x'}  \relO{:}   \openO{\langle}  \possiblyWithSub\stageImetaColor{\tau^{\superscriptI} }  \closeO{\rangle}  \punctO{.}\   \possiblyWithSub\stageOmetaColor{x'}   \closeO{)} \)
                clearly satisfies the desired properties.
            \end{itemize}
          \item Case \(\possiblyWithSub\stageOmetaColor{N^{\superscriptO} } =  \LeftAssertParen   \openO{\{} \possiblyWithSub\stageOmetaColor{\nu}  \relO{:}  \possiblyWithSub\stageOmetaColor{B}  \relO{\mid}  \possiblyWithSub\stageOmetaColor{N^{\superscriptO} }_{{\mathrm{1}}} \closeO{\} }  \punctO{,}  \possiblyWithSub\stageOmetaColor{N^{\superscriptO} }_{{\mathrm{2}}} \punctO{,}  \possiblyWithSub\stageOmetaColor{c}  \RightAssertParen^{ L } \):
            We can assume \(\possiblyWithSub\stageOmetaColor{\nu} \neq \possiblyWithSub\stageOmetaColor{x}\) w.l.o.g. by the Barendregt convention
            and thereby have
            \(  [  \possiblyWithSub\stageOmetaColor{N^{\superscriptO} }_{{\mathrm{0}}}  /  \possiblyWithSub\stageOmetaColor{x}  ]    \possiblyWithSub\stageOmetaColor{N^{\superscriptO} } 
              =  \LeftAssertParen   \openO{\{} \possiblyWithSub\stageOmetaColor{\nu}  \relO{:}  \possiblyWithSub\stageOmetaColor{B}  \relO{\mid}    [  \possiblyWithSub\stageOmetaColor{N^{\superscriptO} }_{{\mathrm{0}}}  /  \possiblyWithSub\stageOmetaColor{x}  ]    \possiblyWithSub\stageOmetaColor{N^{\superscriptO} }_{{\mathrm{1}}}  \closeO{\} }  \punctO{,}    [  \possiblyWithSub\stageOmetaColor{N^{\superscriptO} }_{{\mathrm{0}}}  /  \possiblyWithSub\stageOmetaColor{x}  ]    \possiblyWithSub\stageOmetaColor{N^{\superscriptO} }_{{\mathrm{2}}}  \punctO{,}  \possiblyWithSub\stageOmetaColor{c}  \RightAssertParen^{ L } \).
            Then, we have one of the following cases with the derivation of
            \(  \LeftAssertParen   \openO{\{} \possiblyWithSub\stageOmetaColor{\nu}  \relO{:}  \possiblyWithSub\stageOmetaColor{B}  \relO{\mid}    [  \possiblyWithSub\stageOmetaColor{N^{\superscriptO} }_{{\mathrm{0}}}  /  \possiblyWithSub\stageOmetaColor{x}  ]    \possiblyWithSub\stageOmetaColor{N^{\superscriptO} }_{{\mathrm{1}}}  \closeO{\} }  \punctO{,}    [  \possiblyWithSub\stageOmetaColor{N^{\superscriptO} }_{{\mathrm{0}}}  /  \possiblyWithSub\stageOmetaColor{x}  ]    \possiblyWithSub\stageOmetaColor{N^{\superscriptO} }_{{\mathrm{2}}}  \punctO{,}  \possiblyWithSub\stageOmetaColor{c}  \RightAssertParen^{ L }   \longrightarrow^{0}   \possiblyWithSub\stageOmetaColor{N'^{\superscriptO} }  \):
            \begin{itemize}
              \item Case \derive[E0-RfnAct]{%
                   [  \possiblyWithSub\stageOmetaColor{N^{\superscriptO} }_{{\mathrm{0}}}  /  \possiblyWithSub\stageOmetaColor{x}  ]    \possiblyWithSub\stageOmetaColor{N^{\superscriptO} }_{{\mathrm{2}}}   \longrightarrow^{0}   \possiblyWithSub\stageOmetaColor{N'^{\superscriptO} }_{{\mathrm{2}}}  
              }{%
                  \LeftAssertParen   \openO{\{} \possiblyWithSub\stageOmetaColor{\nu}  \relO{:}  \possiblyWithSub\stageOmetaColor{B}  \relO{\mid}    [  \possiblyWithSub\stageOmetaColor{N^{\superscriptO} }_{{\mathrm{0}}}  /  \possiblyWithSub\stageOmetaColor{x}  ]    \possiblyWithSub\stageOmetaColor{N^{\superscriptO} }_{{\mathrm{1}}}  \closeO{\} }  \punctO{,}    [  \possiblyWithSub\stageOmetaColor{N^{\superscriptO} }_{{\mathrm{0}}}  /  \possiblyWithSub\stageOmetaColor{x}  ]    \possiblyWithSub\stageOmetaColor{N^{\superscriptO} }_{{\mathrm{2}}}  \punctO{,}  \possiblyWithSub\stageOmetaColor{c}  \RightAssertParen^{ L }   \longrightarrow^{0}    \LeftAssertParen   \openO{\{} \possiblyWithSub\stageOmetaColor{\nu}  \relO{:}  \possiblyWithSub\stageOmetaColor{B}  \relO{\mid}    [  \possiblyWithSub\stageOmetaColor{N^{\superscriptO} }_{{\mathrm{0}}}  /  \possiblyWithSub\stageOmetaColor{x}  ]    \possiblyWithSub\stageOmetaColor{N^{\superscriptO} }_{{\mathrm{1}}}  \closeO{\} }  \punctO{,}  \possiblyWithSub\stageOmetaColor{N'^{\superscriptO} }_{{\mathrm{2}}} \punctO{,}  \possiblyWithSub\stageOmetaColor{c}  \RightAssertParen^{ L }   
              }:
                By IH, there exists \(\possiblyWithSub\stageOmetaColor{\Hat{N}^{\superscriptO} }_{{\mathrm{2}}}\) such that
                \(   [  \possiblyWithSub\stageOmetaColor{N'^{\superscriptO} }_{{\mathrm{0}}}  /  \possiblyWithSub\stageOmetaColor{x}  ]    \possiblyWithSub\stageOmetaColor{N^{\superscriptO} }_{{\mathrm{2}}}   \longrightarrow^{0\,\ast}     [  \possiblyWithSub\stageOmetaColor{N'^{\superscriptO} }_{{\mathrm{0}}}  /  \possiblyWithSub\stageOmetaColor{x}  ]    \possiblyWithSub\stageOmetaColor{\Hat{N}^{\superscriptO} }_{{\mathrm{2}}}   \)
                and \(\possiblyWithSub\stageOmetaColor{N'^{\superscriptO} }_{{\mathrm{2}}} =   [  \possiblyWithSub\stageOmetaColor{N^{\superscriptO} }_{{\mathrm{0}}}  /  \possiblyWithSub\stageOmetaColor{x}  ]    \possiblyWithSub\stageOmetaColor{\Hat{N}^{\superscriptO} }_{{\mathrm{2}}} \).
                We can easily show that
                \(\possiblyWithSub\stageOmetaColor{\Hat{N}^{\superscriptO} } \defeq  \LeftAssertParen   \openO{\{} \possiblyWithSub\stageOmetaColor{\nu}  \relO{:}  \possiblyWithSub\stageOmetaColor{B}  \relO{\mid}  \possiblyWithSub\stageOmetaColor{N^{\superscriptO} }_{{\mathrm{1}}} \closeO{\} }  \punctO{,}  \possiblyWithSub\stageOmetaColor{\Hat{N}^{\superscriptO} }_{{\mathrm{2}}} \punctO{,}  \possiblyWithSub\stageOmetaColor{c}  \RightAssertParen^{ L } \)
                satisfies the desired properties.
              \item Case \derive[E0-RfnPass]{}{%
                  \LeftAssertParen   \openO{\{} \possiblyWithSub\stageOmetaColor{\nu}  \relO{:}  \possiblyWithSub\stageOmetaColor{B}  \relO{\mid}    [  \possiblyWithSub\stageOmetaColor{N^{\superscriptO} }_{{\mathrm{0}}}  /  \possiblyWithSub\stageOmetaColor{x}  ]    \possiblyWithSub\stageOmetaColor{N^{\superscriptO} }_{{\mathrm{1}}}  \closeO{\} }  \punctO{,}     \ttO{true}    \punctO{,}  \possiblyWithSub\stageOmetaColor{c}  \RightAssertParen^{ L }   \longrightarrow^{0}     \possiblyWithSub\stageOmetaColor{c}    
              }:
                Since \(  [  \possiblyWithSub\stageOmetaColor{N^{\superscriptO} }_{{\mathrm{0}}}  /  \possiblyWithSub\stageOmetaColor{x}  ]    \possiblyWithSub\stageOmetaColor{N^{\superscriptO} }_{{\mathrm{2}}}  =   \ttO{true}  \)
                and \(\possiblyWithSub\stageOmetaColor{N^{\superscriptO} }_{{\mathrm{0}}}\) is not a value,
                we have \(\possiblyWithSub\stageOmetaColor{N^{\superscriptO} }_{{\mathrm{2}}} =   \ttO{true}  \).
                Thus, \(\possiblyWithSub\stageOmetaColor{\Hat{N}^{\superscriptO} } \defeq \possiblyWithSub\stageOmetaColor{c}\)
                clearly satisfies the desired properties.
            \end{itemize}
        \end{itemize}
      \item
        \begin{itemize}
          \item Case~\(\possiblyWithSub\stageImetaColor{N^{\superscriptI} } =  \ordI{\sim} \possiblyWithSub\stageOmetaColor{N^{\superscriptO} } \):
            We have one of the following two cases with the derivation of
            \(   [  \possiblyWithSub\stageOmetaColor{N^{\superscriptO} }_{{\mathrm{0}}}  /  \possiblyWithSub\stageOmetaColor{x}  ]    \possiblyWithSub\stageImetaColor{N^{\superscriptI} }   \longrightarrow^{1}   \possiblyWithSub\stageImetaColor{N'^{\superscriptI} }  \):
            \begin{itemize}
              \item Case \derive[E1-Esc]{%
                   [  \possiblyWithSub\stageOmetaColor{N^{\superscriptO} }_{{\mathrm{0}}}  /  \possiblyWithSub\stageOmetaColor{x}  ]    \possiblyWithSub\stageOmetaColor{N^{\superscriptO} }   \longrightarrow^{0}   \possiblyWithSub\stageOmetaColor{N'^{\superscriptO} }  
              }{%
                  \ordI{\sim}  \openO{(}   [  \possiblyWithSub\stageOmetaColor{N^{\superscriptO} }_{{\mathrm{0}}}  /  \possiblyWithSub\stageOmetaColor{x}  ]    \possiblyWithSub\stageOmetaColor{N^{\superscriptO} }  \closeO{)}    \longrightarrow^{1}    \ordI{\sim} \possiblyWithSub\stageOmetaColor{N'^{\superscriptO} }   
              }:
                By IH, there exists \(\possiblyWithSub\stageOmetaColor{\Hat{N}^{\superscriptO} }\) such that
                \(   [  \possiblyWithSub\stageOmetaColor{N'^{\superscriptO} }_{{\mathrm{0}}}  /  \possiblyWithSub\stageOmetaColor{x}  ]    \possiblyWithSub\stageOmetaColor{N^{\superscriptO} }   \longrightarrow^{0\,\ast}     [  \possiblyWithSub\stageOmetaColor{N'^{\superscriptO} }_{{\mathrm{0}}}  /  \possiblyWithSub\stageOmetaColor{x}  ]    \possiblyWithSub\stageOmetaColor{\Hat{N}^{\superscriptO} }   \)
                and \(\possiblyWithSub\stageOmetaColor{N'^{\superscriptO} } =   [  \possiblyWithSub\stageOmetaColor{N^{\superscriptO} }_{{\mathrm{0}}}  /  \possiblyWithSub\stageOmetaColor{x}  ]    \possiblyWithSub\stageOmetaColor{\Hat{N}^{\superscriptO} } \).
                We can easily show that \(\possiblyWithSub\stageImetaColor{\Hat{N}^{\superscriptI} } \defeq  \ordI{\sim} \possiblyWithSub\stageOmetaColor{\Hat{N}^{\superscriptO} } \)
                satisfies the desired properties.
              \item Case \derive[E1-Cancel]{}{%
                  \ordI{\sim}  \openO{\langle}  \possiblyWithSub\stageImetaColor{v^{\superscriptI} }_{{\mathrm{1}}}  \closeO{\rangle}    \longrightarrow^{1}    \possiblyWithSub\stageImetaColor{v^{\superscriptI} }_{{\mathrm{1}}}   
              }, where \(\possiblyWithSub\stageImetaColor{v^{\superscriptI} }_{{\mathrm{1}}} =   [  \possiblyWithSub\stageOmetaColor{N^{\superscriptO} }_{{\mathrm{0}}}  /  \possiblyWithSub\stageOmetaColor{x}  ]    \possiblyWithSub\stageImetaColor{N^{\superscriptI} }_{{\mathrm{1}}} \):
                \(\possiblyWithSub\stageImetaColor{\Hat{N}^{\superscriptI} } \defeq \possiblyWithSub\stageImetaColor{N^{\superscriptI} }_{{\mathrm{1}}}\) clearly satisfies the desired properties.
            \end{itemize}
          \item The other cases are more straightforward.
        \end{itemize}
      \item
        \begin{itemize}
          \item Case~\(\possiblyWithSub\stageImetaColor{T^{\superscriptI} } =  \ttI{Tensor}\ \ordI{\%} \possiblyWithSub\stageOmetaColor{N^{\superscriptO} }_{{\mathrm{1}}} \):
            Tracing back the derivation of \(   [  \possiblyWithSub\stageOmetaColor{N^{\superscriptO} }_{{\mathrm{0}}}  /  \possiblyWithSub\stageOmetaColor{x}  ]    \possiblyWithSub\stageImetaColor{T^{\superscriptI} }   \longrightarrow^{1}   \possiblyWithSub\stageImetaColor{T'^{\superscriptI} }  \),
            we have
            \begin{center}
              \derive[ET1-Tensor]{%
                   [  \possiblyWithSub\stageOmetaColor{N^{\superscriptO} }_{{\mathrm{0}}}  /  \possiblyWithSub\stageOmetaColor{x}  ]    \possiblyWithSub\stageOmetaColor{N^{\superscriptO} }_{{\mathrm{1}}}   \longrightarrow^{0}   \possiblyWithSub\stageOmetaColor{N'^{\superscriptO} }_{{\mathrm{1}}}  
              }{%
                  \ttI{Tensor}\ \ordI{\%}  \openO{(}   [  \possiblyWithSub\stageOmetaColor{N^{\superscriptO} }_{{\mathrm{0}}}  /  \possiblyWithSub\stageOmetaColor{x}  ]    \possiblyWithSub\stageOmetaColor{N^{\superscriptO} }_{{\mathrm{1}}}  \closeO{)}    \longrightarrow^{1}    \ttI{Tensor}\ \ordI{\%} \possiblyWithSub\stageOmetaColor{N'^{\superscriptO} }_{{\mathrm{1}}}   
              }.
            \end{center}
            By IH, there exists \(\possiblyWithSub\stageOmetaColor{\Hat{N}^{\superscriptO} }_{{\mathrm{1}}}\) such that
            \(   [  \possiblyWithSub\stageOmetaColor{N'^{\superscriptO} }_{{\mathrm{0}}}  /  \possiblyWithSub\stageOmetaColor{x}  ]    \possiblyWithSub\stageOmetaColor{N^{\superscriptO} }_{{\mathrm{1}}}   \longrightarrow^{0\,\ast}     [  \possiblyWithSub\stageOmetaColor{N'^{\superscriptO} }_{{\mathrm{0}}}  /  \possiblyWithSub\stageOmetaColor{x}  ]    \possiblyWithSub\stageOmetaColor{\Hat{N}^{\superscriptO} }_{{\mathrm{1}}}   \)
            and \(\possiblyWithSub\stageOmetaColor{N'^{\superscriptO} }_{{\mathrm{1}}} =   [  \possiblyWithSub\stageOmetaColor{N^{\superscriptO} }_{{\mathrm{0}}}  /  \possiblyWithSub\stageOmetaColor{x}  ]    \possiblyWithSub\stageOmetaColor{\Hat{N}^{\superscriptO} }_{{\mathrm{1}}} \).
            We can easily show that \(\possiblyWithSub\stageImetaColor{\Hat{T}^{\superscriptI} } \defeq  \ttI{Tensor}\ \ordI{\%} \possiblyWithSub\stageOmetaColor{\Hat{N}^{\superscriptO} }_{{\mathrm{1}}} \)
            satisfies the desired properties.
          \item The other cases are also straightforward.
        \end{itemize}
    \end{enumerate}
  \end{proof}
  \begin{lemma}\label{lem:weak-bisimulation-right-pre}
    Suppose \( \possiblyWithSub\stageOmetaColor{N^{\superscriptO} }_{{\mathrm{0}}}  \longrightarrow^{0}   \possiblyWithSub\stageOmetaColor{N'^{\superscriptO} }_{{\mathrm{0}}}  \).
    \begin{enumerate}
      \item
        If \(  [  \possiblyWithSub\stageOmetaColor{N'^{\superscriptO} }_{{\mathrm{0}}}  /  \possiblyWithSub\stageOmetaColor{x}  ]    \possiblyWithSub\stageOmetaColor{N^{\superscriptO} } \) is a stage-\(0\) value,
        then there exists \(\possiblyWithSub\stageOmetaColor{N'^{\superscriptO} }\) such that
        \begin{itemize}
          \item[(A)] \(   [  \possiblyWithSub\stageOmetaColor{N^{\superscriptO} }_{{\mathrm{0}}}  /  \possiblyWithSub\stageOmetaColor{x}  ]    \possiblyWithSub\stageOmetaColor{N^{\superscriptO} }   \longrightarrow^{0\,\ast}     [  \possiblyWithSub\stageOmetaColor{N^{\superscriptO} }_{{\mathrm{0}}}  /  \possiblyWithSub\stageOmetaColor{x}  ]    \possiblyWithSub\stageOmetaColor{N'^{\superscriptO} }   \),
          \item[(B)] \(  [  \possiblyWithSub\stageOmetaColor{N^{\superscriptO} }_{{\mathrm{0}}}  /  \possiblyWithSub\stageOmetaColor{x}  ]    \possiblyWithSub\stageOmetaColor{N'^{\superscriptO} } \) is a stage-\(0\) value, and
          \item[(C)] \(  [  \possiblyWithSub\stageOmetaColor{N'^{\superscriptO} }_{{\mathrm{0}}}  /  \possiblyWithSub\stageOmetaColor{x}  ]    \possiblyWithSub\stageOmetaColor{N^{\superscriptO} }  =   [  \possiblyWithSub\stageOmetaColor{N'^{\superscriptO} }_{{\mathrm{0}}}  /  \possiblyWithSub\stageOmetaColor{x}  ]    \possiblyWithSub\stageOmetaColor{N'^{\superscriptO} } \).
        \end{itemize}
      \item
        If \(  [  \possiblyWithSub\stageOmetaColor{N'^{\superscriptO} }_{{\mathrm{0}}}  /  \possiblyWithSub\stageOmetaColor{x}  ]    \possiblyWithSub\stageImetaColor{N^{\superscriptI} } \) is a stage-\(1\) value,
        then there exists \(\possiblyWithSub\stageImetaColor{N'^{\superscriptI} }\) such that
        \begin{itemize}
          \item[(A)] \(   [  \possiblyWithSub\stageOmetaColor{N^{\superscriptO} }_{{\mathrm{0}}}  /  \possiblyWithSub\stageOmetaColor{x}  ]    \possiblyWithSub\stageImetaColor{N^{\superscriptI} }   \longrightarrow^{1\,\ast}     [  \possiblyWithSub\stageOmetaColor{N^{\superscriptO} }_{{\mathrm{0}}}  /  \possiblyWithSub\stageOmetaColor{x}  ]    \possiblyWithSub\stageImetaColor{N'^{\superscriptI} }   \),
          \item[(B)] \(  [  \possiblyWithSub\stageOmetaColor{N^{\superscriptO} }_{{\mathrm{0}}}  /  \possiblyWithSub\stageOmetaColor{x}  ]    \possiblyWithSub\stageImetaColor{N'^{\superscriptI} } \) is a stage-\(1\) value, and
          \item[(C)] \(  [  \possiblyWithSub\stageOmetaColor{N'^{\superscriptO} }_{{\mathrm{0}}}  /  \possiblyWithSub\stageOmetaColor{x}  ]    \possiblyWithSub\stageImetaColor{N^{\superscriptI} }  =   [  \possiblyWithSub\stageOmetaColor{N'^{\superscriptO} }_{{\mathrm{0}}}  /  \possiblyWithSub\stageOmetaColor{x}  ]    \possiblyWithSub\stageImetaColor{N'^{\superscriptI} } \).
        \end{itemize}
      \item
        If \(  [  \possiblyWithSub\stageOmetaColor{N'^{\superscriptO} }_{{\mathrm{0}}}  /  \possiblyWithSub\stageOmetaColor{x}  ]    \possiblyWithSub\stageImetaColor{T^{\superscriptI} }  \revdefeq \possiblyWithSub\stageImetaColor{\tau^{\superscriptI} }\) is a type value, then
        \begin{itemize}
          \item[(X)] \(   [  \possiblyWithSub\stageOmetaColor{N^{\superscriptO} }_{{\mathrm{0}}}  /  \possiblyWithSub\stageOmetaColor{x}  ]    \possiblyWithSub\stageImetaColor{T^{\superscriptI} }   \longrightarrow^{1\,\ast}    \possiblyWithSub\stageImetaColor{\tau^{\superscriptI} }   \).
        \end{itemize}
    \end{enumerate}
  \end{lemma}
  \begin{proof}
    By induction on the structure of \(\possiblyWithSub\stageOmetaColor{N^{\superscriptO} }\), \(\possiblyWithSub\stageImetaColor{N^{\superscriptI} }\), and \(\possiblyWithSub\stageImetaColor{T^{\superscriptI} }\):
    \begin{enumerate}
      \item
        \begin{itemize}
          \item Case~\(\possiblyWithSub\stageOmetaColor{N^{\superscriptO} } = \possiblyWithSub\stageOmetaColor{x}\):
            We will show that \(\possiblyWithSub\stageOmetaColor{N'^{\superscriptO} } \defeq \possiblyWithSub\stageOmetaColor{N'^{\superscriptO} }_{{\mathrm{0}}}\) satisfies the desired properties.
            First, since \(\possiblyWithSub\stageOmetaColor{N'^{\superscriptO} }_{{\mathrm{0}}}\) is a closed term by \( \possiblyWithSub\stageOmetaColor{N^{\superscriptO} }_{{\mathrm{0}}}  \longrightarrow^{0}   \possiblyWithSub\stageOmetaColor{N'^{\superscriptO} }_{{\mathrm{0}}}  \),
            we have \(  [  \possiblyWithSub\stageOmetaColor{N^{\superscriptO} }_{{\mathrm{0}}}  /  \possiblyWithSub\stageOmetaColor{x}  ]    \possiblyWithSub\stageOmetaColor{N'^{\superscriptO} }  = \possiblyWithSub\stageOmetaColor{N'^{\superscriptO} }_{{\mathrm{0}}}\).
            Therefore, we have (A) by \(  [  \possiblyWithSub\stageOmetaColor{N^{\superscriptO} }_{{\mathrm{0}}}  /  \possiblyWithSub\stageOmetaColor{x}  ]    \possiblyWithSub\stageOmetaColor{N^{\superscriptO} }  = \possiblyWithSub\stageOmetaColor{N^{\superscriptO} }_{{\mathrm{0}}}\) and \( \possiblyWithSub\stageOmetaColor{N^{\superscriptO} }_{{\mathrm{0}}}  \longrightarrow^{0}   \possiblyWithSub\stageOmetaColor{N'^{\superscriptO} }_{{\mathrm{0}}}  \).
            Also, (B) immediately holds by the assumption that
            \(  [  \possiblyWithSub\stageOmetaColor{N'^{\superscriptO} }_{{\mathrm{0}}}  /  \possiblyWithSub\stageOmetaColor{x}  ]    \possiblyWithSub\stageOmetaColor{N^{\superscriptO} }  = \possiblyWithSub\stageOmetaColor{N'^{\superscriptO} }_{{\mathrm{0}}}\) is a stage-\(0\) value.
            Finally, we have (C) by \(  [  \possiblyWithSub\stageOmetaColor{N'^{\superscriptO} }_{{\mathrm{0}}}  /  \possiblyWithSub\stageOmetaColor{x}  ]    \possiblyWithSub\stageOmetaColor{N^{\superscriptO} }  = \possiblyWithSub\stageOmetaColor{N'^{\superscriptO} }_{{\mathrm{0}}} =   [  \possiblyWithSub\stageOmetaColor{N'^{\superscriptO} }_{{\mathrm{0}}}  /  \possiblyWithSub\stageOmetaColor{x}  ]    \possiblyWithSub\stageOmetaColor{N'^{\superscriptO} } \).
          \item Case~\(\possiblyWithSub\stageOmetaColor{N^{\superscriptO} } = \possiblyWithSub\stageOmetaColor{x'}\) such that \(\possiblyWithSub\stageOmetaColor{x'} \neq \possiblyWithSub\stageOmetaColor{x}\):
            This contradicts the assumption that \(  [  \possiblyWithSub\stageOmetaColor{N'^{\superscriptO} }_{{\mathrm{0}}}  /  \possiblyWithSub\stageOmetaColor{x}  ]    \possiblyWithSub\stageOmetaColor{N^{\superscriptO} } \) is a stage-\(0\) value.
          \item Case where \(\possiblyWithSub\stageOmetaColor{N^{\superscriptO} } =  \LeftAssertParen\openO{\langle} \possiblyWithSub\stageImetaColor{T^{\superscriptI} } \closeO{\rangle} \relO{\CastArrow} \openO{\langle} \possiblyWithSub\stageImetaColor{T^{\superscriptI} } \closeO{\rangle}\RightAssertParen^{ L } \)
          or \(\possiblyWithSub\stageOmetaColor{N^{\superscriptO} } =  \LeftAssertParen   \openO{\{} \possiblyWithSub\stageOmetaColor{x}  \relO{:}  \possiblyWithSub\stageOmetaColor{B}  \relO{\mid}  \possiblyWithSub\stageOmetaColor{N^{\superscriptO} } \closeO{\} }  \punctO{,}  \possiblyWithSub\stageOmetaColor{N^{\superscriptO} } \punctO{,}  \possiblyWithSub\stageOmetaColor{c}  \RightAssertParen^{ L } \):
            This also contradicts the assumption.
          \item Case where \(\possiblyWithSub\stageOmetaColor{N^{\superscriptO} } = \possiblyWithSub\stageOmetaColor{p}\), \(\possiblyWithSub\stageOmetaColor{N^{\superscriptO} } = \possiblyWithSub\stageOmetaColor{c}\), \(\possiblyWithSub\stageOmetaColor{N^{\superscriptO} } =  \openO{(}  \ordO{\lambda} \possiblyWithSub\stageOmetaColor{x}  \relO{:}  \possiblyWithSub\stageOmetaColor{T^{\superscriptO} }_{{\mathrm{1}}} \punctO{.}\  \possiblyWithSub\stageOmetaColor{N^{\superscriptO} }_{{\mathrm{2}}}  \closeO{)} \),
          or \( \LeftAssertParen \relO{\CastArrow}   \openO{\{} \possiblyWithSub\stageOmetaColor{x}  \relO{:}  \possiblyWithSub\stageOmetaColor{B}  \relO{\mid}  \possiblyWithSub\stageOmetaColor{N^{\superscriptO} }_{{\mathrm{1}}} \closeO{\} }   \RightAssertParen^{ L } \):
            Since \(\possiblyWithSub\stageOmetaColor{N^{\superscriptO} }\) is a stage-\(0\) value,
            \(\possiblyWithSub\stageOmetaColor{N'^{\superscriptO} } \defeq \possiblyWithSub\stageOmetaColor{N^{\superscriptO} }\) clearly satisfies (A), (B), and (C).
          \item Case~\(\possiblyWithSub\stageOmetaColor{N^{\superscriptO} } =  \possiblyWithSub\stageOmetaColor{N^{\superscriptO} }_{{\mathrm{1}}} \  \possiblyWithSub\stageOmetaColor{N^{\superscriptO} }_{{\mathrm{2}}} \):
            By the assumption that \(  [  \possiblyWithSub\stageOmetaColor{N'^{\superscriptO} }_{{\mathrm{0}}}  /  \possiblyWithSub\stageOmetaColor{x}  ]    \possiblyWithSub\stageOmetaColor{N^{\superscriptO} } \) is a stage-\(0\) value,
            there exist \(\possiblyWithSub\stageOmetaColor{a}_{{\mathrm{1}}}\) and \(\possiblyWithSub\stageOmetaColor{c}_{{\mathrm{2}}}\) such that
            \(\possiblyWithSub\stageOmetaColor{a}_{{\mathrm{1}}} =   [  \possiblyWithSub\stageOmetaColor{N'^{\superscriptO} }_{{\mathrm{0}}}  /  \possiblyWithSub\stageOmetaColor{x}  ]    \possiblyWithSub\stageOmetaColor{N^{\superscriptO} }_{{\mathrm{1}}} \) and \(\possiblyWithSub\stageOmetaColor{c}_{{\mathrm{2}}} =   [  \possiblyWithSub\stageOmetaColor{N'^{\superscriptO} }_{{\mathrm{0}}}  /  \possiblyWithSub\stageOmetaColor{x}  ]    \possiblyWithSub\stageOmetaColor{N^{\superscriptO} }_{{\mathrm{2}}} \).
            Then, by Lemma~\ref{lem:nonvalue-substitution-inversion},
            we have \(\possiblyWithSub\stageOmetaColor{N^{\superscriptO} }_{{\mathrm{1}}} = \possiblyWithSub\stageOmetaColor{a}_{{\mathrm{1}}}\) and \(\possiblyWithSub\stageOmetaColor{N^{\superscriptO} }_{{\mathrm{2}}} = \possiblyWithSub\stageOmetaColor{c}_{{\mathrm{2}}}\).
            We can easily show that \(\possiblyWithSub\stageOmetaColor{N'^{\superscriptO} } \defeq  \openO{(}   \possiblyWithSub\stageOmetaColor{a}_{{\mathrm{1}}}  \    \possiblyWithSub\stageOmetaColor{c}_{{\mathrm{2}}}    \closeO{)}  = \possiblyWithSub\stageOmetaColor{N^{\superscriptO} }\) satisfies
            (A), (B), and (C).
          \item Case~\(\possiblyWithSub\stageOmetaColor{N^{\superscriptO} } =  \openO{\langle} \possiblyWithSub\stageImetaColor{N^{\superscriptI} } \closeO{\rangle} \):
            By the assumption that \(  [  \possiblyWithSub\stageOmetaColor{N'^{\superscriptO} }_{{\mathrm{0}}}  /  \possiblyWithSub\stageOmetaColor{x}  ]    \possiblyWithSub\stageOmetaColor{N^{\superscriptO} } \) is a stage-\(0\) value,
            \(  [  \possiblyWithSub\stageOmetaColor{N'^{\superscriptO} }_{{\mathrm{0}}}  /  \possiblyWithSub\stageOmetaColor{x}  ]    \possiblyWithSub\stageImetaColor{N^{\superscriptI} } \) is a stage-\(1\) value.
            Then, by IH, there exists \(\possiblyWithSub\stageImetaColor{N'^{\superscriptI} }\) such that
            \begin{itemize}
              \item[(\Aprime)] \(   [  \possiblyWithSub\stageOmetaColor{N^{\superscriptO} }_{{\mathrm{0}}}  /  \possiblyWithSub\stageOmetaColor{x}  ]    \possiblyWithSub\stageImetaColor{N^{\superscriptI} }   \longrightarrow^{1\,\ast}     [  \possiblyWithSub\stageOmetaColor{N^{\superscriptO} }_{{\mathrm{0}}}  /  \possiblyWithSub\stageOmetaColor{x}  ]    \possiblyWithSub\stageImetaColor{N'^{\superscriptI} }   \),
              \item[(\Bprime)] \(  [  \possiblyWithSub\stageOmetaColor{N^{\superscriptO} }_{{\mathrm{0}}}  /  \possiblyWithSub\stageOmetaColor{x}  ]    \possiblyWithSub\stageImetaColor{N'^{\superscriptI} } \) is a stage-\(1\) value, and
              \item[(\Cprime)] \(  [  \possiblyWithSub\stageOmetaColor{N'^{\superscriptO} }_{{\mathrm{0}}}  /  \possiblyWithSub\stageOmetaColor{x}  ]    \possiblyWithSub\stageImetaColor{N^{\superscriptI} }  =   [  \possiblyWithSub\stageOmetaColor{N'^{\superscriptO} }_{{\mathrm{0}}}  /  \possiblyWithSub\stageOmetaColor{x}  ]    \possiblyWithSub\stageImetaColor{N'^{\superscriptI} } \).
            \end{itemize}
            We will show that \(\possiblyWithSub\stageOmetaColor{N'^{\superscriptO} } \defeq  \openO{\langle} \possiblyWithSub\stageImetaColor{N'^{\superscriptI} } \closeO{\rangle} \) satisfies the desired properties.
            First, by (\Aprime) and the repeated use of \rulename{E0-Brkt},
            we can derive
            \begin{itemize}
              \item[(A)] \(   [  \possiblyWithSub\stageOmetaColor{N^{\superscriptO} }_{{\mathrm{0}}}  /  \possiblyWithSub\stageOmetaColor{x}  ]     \openO{\langle} \possiblyWithSub\stageImetaColor{N^{\superscriptI} } \closeO{\rangle}    \longrightarrow^{0\,\ast}     [  \possiblyWithSub\stageOmetaColor{N^{\superscriptO} }_{{\mathrm{0}}}  /  \possiblyWithSub\stageOmetaColor{x}  ]     \openO{\langle} \possiblyWithSub\stageImetaColor{N'^{\superscriptI} } \closeO{\rangle}    \).
            \end{itemize}
            Also, by (\Bprime) and (\Cprime), we immediately have the following:
            \begin{itemize}
              \item[(B)] \(  [  \possiblyWithSub\stageOmetaColor{N^{\superscriptO} }_{{\mathrm{0}}}  /  \possiblyWithSub\stageOmetaColor{x}  ]     \openO{\langle} \possiblyWithSub\stageImetaColor{N'^{\superscriptI} } \closeO{\rangle}  \) is a stage-\(0\) value, and
              \item[(C)] \(  [  \possiblyWithSub\stageOmetaColor{N'^{\superscriptO} }_{{\mathrm{0}}}  /  \possiblyWithSub\stageOmetaColor{x}  ]     \openO{\langle} \possiblyWithSub\stageImetaColor{N^{\superscriptI} } \closeO{\rangle}   =   [  \possiblyWithSub\stageOmetaColor{N'^{\superscriptO} }_{{\mathrm{0}}}  /  \possiblyWithSub\stageOmetaColor{x}  ]     \openO{\langle} \possiblyWithSub\stageImetaColor{N'^{\superscriptI} } \closeO{\rangle}  \).
            \end{itemize}
        \end{itemize}
      \item
        \begin{itemize}
          \item Case where \(\possiblyWithSub\stageImetaColor{N^{\superscriptI} } = \possiblyWithSub\stageImetaColor{c}\) or \(\possiblyWithSub\stageImetaColor{N^{\superscriptI} } = \possiblyWithSub\stageImetaColor{x'}\):
            \(\possiblyWithSub\stageImetaColor{N'^{\superscriptI} } \defeq \possiblyWithSub\stageImetaColor{N'^{\superscriptI} }\) clearly satisfies (A), (B), and (C).
          \item Case where \(\possiblyWithSub\stageImetaColor{N^{\superscriptI} } =  \openI{(}  \ordI{\lambda} \possiblyWithSub\stageImetaColor{x'}  \relI{:}  \possiblyWithSub\stageImetaColor{T^{\superscriptI} }_{{\mathrm{1}}} \punctI{.}\  \possiblyWithSub\stageImetaColor{N^{\superscriptI} }_{{\mathrm{2}}}  \closeI{)} \):
            Since \(  [  \possiblyWithSub\stageOmetaColor{N'^{\superscriptO} }_{{\mathrm{0}}}  /  \possiblyWithSub\stageOmetaColor{x}  ]    \possiblyWithSub\stageImetaColor{N^{\superscriptI} } \) is a stage-\(1\) value,
            \(  [  \possiblyWithSub\stageOmetaColor{N'^{\superscriptO} }_{{\mathrm{0}}}  /  \possiblyWithSub\stageOmetaColor{x}  ]    \possiblyWithSub\stageImetaColor{T^{\superscriptI} }_{{\mathrm{1}}}  \revdefeq \possiblyWithSub\stageImetaColor{\tau^{\superscriptI} }_{{\mathrm{1}}}\) and \(  [  \possiblyWithSub\stageOmetaColor{N'^{\superscriptO} }_{{\mathrm{0}}}  /  \possiblyWithSub\stageOmetaColor{x}  ]    \possiblyWithSub\stageImetaColor{N^{\superscriptI} }_{{\mathrm{2}}} \) are also
            a type value and a stage-\(1\) value, respectively.
            Then, by IH, there exists \(\possiblyWithSub\stageImetaColor{N'^{\superscriptI} }_{{\mathrm{2}}}\) such that
            \begin{itemize}
              \item[(1X)] \(   [  \possiblyWithSub\stageOmetaColor{N^{\superscriptO} }_{{\mathrm{0}}}  /  \possiblyWithSub\stageOmetaColor{x}  ]    \possiblyWithSub\stageImetaColor{T^{\superscriptI} }_{{\mathrm{1}}}   \longrightarrow^{1\,\ast}    \possiblyWithSub\stageImetaColor{\tau^{\superscriptI} }_{{\mathrm{1}}}   \),
              \item[(2A)] \(   [  \possiblyWithSub\stageOmetaColor{N^{\superscriptO} }_{{\mathrm{0}}}  /  \possiblyWithSub\stageOmetaColor{x}  ]    \possiblyWithSub\stageImetaColor{N^{\superscriptI} }_{{\mathrm{2}}}   \longrightarrow^{1\,\ast}     [  \possiblyWithSub\stageOmetaColor{N^{\superscriptO} }_{{\mathrm{0}}}  /  \possiblyWithSub\stageOmetaColor{x}  ]    \possiblyWithSub\stageImetaColor{N'^{\superscriptI} }_{{\mathrm{2}}}   \),
              \item[(2B)] \(  [  \possiblyWithSub\stageOmetaColor{N^{\superscriptO} }_{{\mathrm{0}}}  /  \possiblyWithSub\stageOmetaColor{x}  ]    \possiblyWithSub\stageImetaColor{N'^{\superscriptI} }_{{\mathrm{2}}} \) is a stage-\(1\) value, and
              \item[(2C)] \(  [  \possiblyWithSub\stageOmetaColor{N'^{\superscriptO} }_{{\mathrm{0}}}  /  \possiblyWithSub\stageOmetaColor{x}  ]    \possiblyWithSub\stageImetaColor{N^{\superscriptI} }_{{\mathrm{2}}}  =   [  \possiblyWithSub\stageOmetaColor{N'^{\superscriptO} }_{{\mathrm{0}}}  /  \possiblyWithSub\stageOmetaColor{x}  ]    \possiblyWithSub\stageImetaColor{N'^{\superscriptI} }_{{\mathrm{2}}} \).
            \end{itemize}
            We will show that \(\possiblyWithSub\stageImetaColor{N'^{\superscriptI} } \defeq  \openI{(}  \ordI{\lambda} \possiblyWithSub\stageImetaColor{x}  \relI{:}   \possiblyWithSub\stageImetaColor{\tau^{\superscriptI} }_{{\mathrm{1}}}  \punctI{.}\  \possiblyWithSub\stageImetaColor{N'^{\superscriptI} }_{{\mathrm{2}}}  \closeI{)} \) satisfies the desired properties.
            Indeed, by (1X) and (2A), we can clearly derive the following,
            using \rulename{E1-Abs1} and \rulename{E1-Abs2} repeatedly:
            \begin{itemize}
              \item[(A)] \(   [  \possiblyWithSub\stageOmetaColor{N^{\superscriptO} }_{{\mathrm{0}}}  /  \possiblyWithSub\stageOmetaColor{x}  ]     \openI{(}  \ordI{\lambda} \possiblyWithSub\stageImetaColor{x'}  \relI{:}  \possiblyWithSub\stageImetaColor{T^{\superscriptI} }_{{\mathrm{1}}} \punctI{.}\  \possiblyWithSub\stageImetaColor{N^{\superscriptI} }_{{\mathrm{2}}}  \closeI{)}    \longrightarrow^{1\,\ast}     [  \possiblyWithSub\stageOmetaColor{N^{\superscriptO} }_{{\mathrm{0}}}  /  \possiblyWithSub\stageOmetaColor{x}  ]     \openI{(}  \ordI{\lambda} \possiblyWithSub\stageImetaColor{x'}  \relI{:}   \possiblyWithSub\stageImetaColor{\tau^{\superscriptI} }_{{\mathrm{1}}}  \punctI{.}\  \possiblyWithSub\stageImetaColor{N'^{\superscriptI} }_{{\mathrm{2}}}  \closeI{)}    \).
            \end{itemize}
            We can also have the following immediately by (1X), (2B), and (2C):
            \begin{itemize}
              \item[(B)] \(  [  \possiblyWithSub\stageOmetaColor{N^{\superscriptO} }_{{\mathrm{0}}}  /  \possiblyWithSub\stageOmetaColor{x}  ]     \openI{(}  \ordI{\lambda} \possiblyWithSub\stageImetaColor{x}  \relI{:}   \possiblyWithSub\stageImetaColor{\tau^{\superscriptI} }_{{\mathrm{1}}}  \punctI{.}\  \possiblyWithSub\stageImetaColor{N'^{\superscriptI} }_{{\mathrm{2}}}  \closeI{)}  \) is a stage-\(1\) value, and
              \item[(C)] \(  [  \possiblyWithSub\stageOmetaColor{N'^{\superscriptO} }_{{\mathrm{0}}}  /  \possiblyWithSub\stageOmetaColor{x}  ]     \openI{(}  \ordI{\lambda} \possiblyWithSub\stageImetaColor{x}  \relI{:}  \possiblyWithSub\stageImetaColor{T^{\superscriptI} }_{{\mathrm{1}}} \punctI{.}\  \possiblyWithSub\stageImetaColor{N^{\superscriptI} }_{{\mathrm{2}}}  \closeI{)}   =   [  \possiblyWithSub\stageOmetaColor{N'^{\superscriptO} }_{{\mathrm{0}}}  /  \possiblyWithSub\stageOmetaColor{x}  ]     \openI{(}  \ordI{\lambda} \possiblyWithSub\stageImetaColor{x}  \relI{:}   \possiblyWithSub\stageImetaColor{\tau^{\superscriptI} }_{{\mathrm{1}}}  \punctI{.}\  \possiblyWithSub\stageImetaColor{N'^{\superscriptI} }_{{\mathrm{2}}}  \closeI{)}  \).
            \end{itemize}
          \item Case~\(\possiblyWithSub\stageImetaColor{N^{\superscriptI} } =  \possiblyWithSub\stageImetaColor{N^{\superscriptI} }_{{\mathrm{1}}} \  \possiblyWithSub\stageImetaColor{N^{\superscriptI} }_{{\mathrm{2}}} \):
            Straightforward by IH and the repeated use of \rulename{E1-App1} and \rulename{E1-App2},
            similarly to the previous case.
          \item Case~\(\possiblyWithSub\stageImetaColor{N^{\superscriptI} } =  \ordI{\sim} \possiblyWithSub\stageOmetaColor{N^{\superscriptO} } \):
            This contradicts the assumption that \(  [  \possiblyWithSub\stageOmetaColor{N'^{\superscriptO} }_{{\mathrm{0}}}  /  \possiblyWithSub\stageOmetaColor{x}  ]    \possiblyWithSub\stageImetaColor{N^{\superscriptI} } \) is a stage-\(1\) value.
        \end{itemize}
      \item
        \begin{itemize}
          \item Case~\(\possiblyWithSub\stageImetaColor{T^{\superscriptI} } = \possiblyWithSub\stageImetaColor{B}\):
            Since \(\possiblyWithSub\stageImetaColor{\tau^{\superscriptI} } = \possiblyWithSub\stageImetaColor{B}\), we clearly have (X).
          \item Case~\(\possiblyWithSub\stageImetaColor{T^{\superscriptI} } =  \ttI{Tensor}\ \ordI{\%} \possiblyWithSub\stageOmetaColor{N^{\superscriptO} } \):
            Since \(  [  \possiblyWithSub\stageOmetaColor{N'^{\superscriptO} }_{{\mathrm{0}}}  /  \possiblyWithSub\stageOmetaColor{x}  ]    \possiblyWithSub\stageImetaColor{T^{\superscriptI} }  \revdefeq \possiblyWithSub\stageImetaColor{\tau^{\superscriptI} }\) is a type value,
            \(  [  \possiblyWithSub\stageOmetaColor{N'^{\superscriptO} }_{{\mathrm{0}}}  /  \possiblyWithSub\stageOmetaColor{x}  ]    \possiblyWithSub\stageOmetaColor{N^{\superscriptO} } \) must be of the form~\(\possiblyWithSub\stageOmetaColor{s}\).
            Then, by the definition of substitution, we have the following two cases:
            \begin{itemize}
              \item Case where \(\possiblyWithSub\stageOmetaColor{N^{\superscriptO} } = \possiblyWithSub\stageOmetaColor{x}\) and \(\possiblyWithSub\stageOmetaColor{N'^{\superscriptO} }_{{\mathrm{0}}} = \possiblyWithSub\stageOmetaColor{s}\):
                By the assumption~\( \possiblyWithSub\stageOmetaColor{N^{\superscriptO} }_{{\mathrm{0}}}  \longrightarrow^{0}   \possiblyWithSub\stageOmetaColor{N'^{\superscriptO} }_{{\mathrm{0}}}   = \possiblyWithSub\stageOmetaColor{s}\),
                we can derive (X)~\(   [  \possiblyWithSub\stageOmetaColor{N^{\superscriptO} }_{{\mathrm{0}}}  /  \possiblyWithSub\stageOmetaColor{x}  ]     \openI{(}  \ttI{Tensor}\ \ordI{\%}  \possiblyWithSub\stageOmetaColor{x}   \closeI{)}    \longrightarrow^{1\,\ast}    \possiblyWithSub\stageImetaColor{\tau^{\superscriptI} }   \) as follows:
                \begin{center}
                  \derive[ET1-Tensor]{%
                     \possiblyWithSub\stageOmetaColor{N^{\superscriptO} }_{{\mathrm{0}}}  \longrightarrow^{0}      \possiblyWithSub\stageOmetaColor{s}     
                  }{%
                      \ttI{Tensor}\ \ordI{\%} \possiblyWithSub\stageOmetaColor{N^{\superscriptO} }_{{\mathrm{0}}}   \longrightarrow^{1}    \ttI{Tensor}\ \ordI{\%}    \possiblyWithSub\stageOmetaColor{s}      
                  }.
                \end{center}
              \item Case where \(\possiblyWithSub\stageOmetaColor{N^{\superscriptO} } = \possiblyWithSub\stageOmetaColor{s}\):
                we have \(  [  \possiblyWithSub\stageOmetaColor{N^{\superscriptO} }_{{\mathrm{0}}}  /  \possiblyWithSub\stageOmetaColor{x}  ]     \openI{(}  \ttI{Tensor}\ \ordI{\%}    \possiblyWithSub\stageOmetaColor{s}     \closeI{)}   =  \ttI{Tensor}\ \ordI{\%} \possiblyWithSub\stageOmetaColor{s}  = \possiblyWithSub\stageImetaColor{\tau^{\superscriptI} }\),
                which implies (X).
            \end{itemize}
          \item Case~\(\possiblyWithSub\stageImetaColor{T^{\superscriptI} } =  \possiblyWithSub\stageImetaColor{T^{\superscriptI} }_{{\mathrm{1}}}  \relI{\to}  \possiblyWithSub\stageImetaColor{T^{\superscriptI} }_{{\mathrm{2}}} \):
            Since \(  [  \possiblyWithSub\stageOmetaColor{N'^{\superscriptO} }_{{\mathrm{0}}}  /  \possiblyWithSub\stageOmetaColor{x}  ]    \possiblyWithSub\stageImetaColor{T^{\superscriptI} }  \revdefeq \possiblyWithSub\stageImetaColor{\tau^{\superscriptI} }\) is a type value,
            there exist \(\possiblyWithSub\stageImetaColor{\tau^{\superscriptI} }_{{\mathrm{1}}}\) and \(\possiblyWithSub\stageImetaColor{\tau^{\superscriptI} }_{{\mathrm{2}}}\) such that \(\possiblyWithSub\stageImetaColor{\tau^{\superscriptI} } =  \possiblyWithSub\stageImetaColor{\tau^{\superscriptI} }_{{\mathrm{1}}}  \relI{\to}  \possiblyWithSub\stageImetaColor{\tau^{\superscriptI} }_{{\mathrm{2}}} \),
            \(  [  \possiblyWithSub\stageOmetaColor{N'^{\superscriptO} }_{{\mathrm{0}}}  /  \possiblyWithSub\stageOmetaColor{x}  ]    \possiblyWithSub\stageImetaColor{T^{\superscriptI} }_{{\mathrm{1}}}  = \possiblyWithSub\stageImetaColor{\tau^{\superscriptI} }_{{\mathrm{1}}}\), and \(  [  \possiblyWithSub\stageOmetaColor{N'^{\superscriptO} }_{{\mathrm{0}}}  /  \possiblyWithSub\stageOmetaColor{x}  ]    \possiblyWithSub\stageImetaColor{T^{\superscriptI} }_{{\mathrm{2}}}  = \possiblyWithSub\stageImetaColor{\tau^{\superscriptI} }_{{\mathrm{2}}}\).
            Then, by IH, we have
            \begin{itemize}
              \item[(1X)] \(   [  \possiblyWithSub\stageOmetaColor{N^{\superscriptO} }_{{\mathrm{0}}}  /  \possiblyWithSub\stageOmetaColor{x}  ]    \possiblyWithSub\stageImetaColor{T^{\superscriptI} }_{{\mathrm{1}}}   \longrightarrow^{1\,\ast}    \possiblyWithSub\stageImetaColor{\tau^{\superscriptI} }_{{\mathrm{1}}}   \) and
              \item[(2X)] \(   [  \possiblyWithSub\stageOmetaColor{N^{\superscriptO} }_{{\mathrm{0}}}  /  \possiblyWithSub\stageOmetaColor{x}  ]    \possiblyWithSub\stageImetaColor{T^{\superscriptI} }_{{\mathrm{2}}}   \longrightarrow^{1\,\ast}    \possiblyWithSub\stageImetaColor{\tau^{\superscriptI} }_{{\mathrm{2}}}   \).
            \end{itemize}
            Thus, using \rulename{ET1-Arr1} and \rulename{ET1-Arr2} repeatedly,
            we can derive (X)~\(   [  \possiblyWithSub\stageOmetaColor{N^{\superscriptO} }_{{\mathrm{0}}}  /  \possiblyWithSub\stageOmetaColor{x}  ]     \openI{(}  \possiblyWithSub\stageImetaColor{T^{\superscriptI} }_{{\mathrm{1}}}  \relI{\to}  \possiblyWithSub\stageImetaColor{T^{\superscriptI} }_{{\mathrm{2}}}  \closeI{)}    \longrightarrow^{1\,\ast}     \possiblyWithSub\stageImetaColor{\tau^{\superscriptI} }_{{\mathrm{1}}}   \relI{\to}   \possiblyWithSub\stageImetaColor{\tau^{\superscriptI} }_{{\mathrm{2}}}    \).
        \end{itemize}
    \end{enumerate}
  \end{proof}
  \begin{lemma}[Weak bisimulation, right-hand side]\label{lem:weak-bisimulation-right}
    Suppose \( \possiblyWithSub\stageOmetaColor{N^{\superscriptO} }_{{\mathrm{0}}}  \longrightarrow^{0}   \possiblyWithSub\stageOmetaColor{N'^{\superscriptO} }_{{\mathrm{0}}}  \).
    \begin{enumerate}
      \item
        If \(   [  \possiblyWithSub\stageOmetaColor{N'^{\superscriptO} }_{{\mathrm{0}}}  /  \possiblyWithSub\stageOmetaColor{x}  ]    \possiblyWithSub\stageOmetaColor{N^{\superscriptO} }   \longrightarrow^{0}   \possiblyWithSub\stageOmetaColor{N'^{\superscriptO} }  \),
        then there exists \(\possiblyWithSub\stageOmetaColor{\Hat{N}^{\superscriptO} }\) such that
        \(   [  \possiblyWithSub\stageOmetaColor{N^{\superscriptO} }_{{\mathrm{0}}}  /  \possiblyWithSub\stageOmetaColor{x}  ]    \possiblyWithSub\stageOmetaColor{N^{\superscriptO} }   \longrightarrow^{0\,\ast}     [  \possiblyWithSub\stageOmetaColor{N^{\superscriptO} }_{{\mathrm{0}}}  /  \possiblyWithSub\stageOmetaColor{x}  ]    \possiblyWithSub\stageOmetaColor{\Hat{N}^{\superscriptO} }   \)
        and \(\possiblyWithSub\stageOmetaColor{N'^{\superscriptO} } =   [  \possiblyWithSub\stageOmetaColor{N'^{\superscriptO} }_{{\mathrm{0}}}  /  \possiblyWithSub\stageOmetaColor{x}  ]    \possiblyWithSub\stageOmetaColor{\Hat{N}^{\superscriptO} } \).
      \item
        If \(   [  \possiblyWithSub\stageOmetaColor{N'^{\superscriptO} }_{{\mathrm{0}}}  /  \possiblyWithSub\stageOmetaColor{x}  ]    \possiblyWithSub\stageImetaColor{N^{\superscriptI} }   \longrightarrow^{1}   \possiblyWithSub\stageImetaColor{N'^{\superscriptI} }  \),
        then there exists \(\possiblyWithSub\stageImetaColor{\Hat{N}^{\superscriptI} }\) such that
        \(   [  \possiblyWithSub\stageOmetaColor{N^{\superscriptO} }_{{\mathrm{0}}}  /  \possiblyWithSub\stageOmetaColor{x}  ]    \possiblyWithSub\stageImetaColor{N^{\superscriptI} }   \longrightarrow^{1\,\ast}     [  \possiblyWithSub\stageOmetaColor{N^{\superscriptO} }_{{\mathrm{0}}}  /  \possiblyWithSub\stageOmetaColor{x}  ]    \possiblyWithSub\stageImetaColor{\Hat{N}^{\superscriptI} }   \)
        and \(\possiblyWithSub\stageImetaColor{N'^{\superscriptI} } =   [  \possiblyWithSub\stageOmetaColor{N'^{\superscriptO} }_{{\mathrm{0}}}  /  \possiblyWithSub\stageOmetaColor{x}  ]    \possiblyWithSub\stageImetaColor{\Hat{N}^{\superscriptI} } \).
      \item
        If \(   [  \possiblyWithSub\stageOmetaColor{N'^{\superscriptO} }_{{\mathrm{0}}}  /  \possiblyWithSub\stageOmetaColor{x}  ]    \possiblyWithSub\stageImetaColor{T^{\superscriptI} }   \longrightarrow^{1}   \possiblyWithSub\stageImetaColor{T'^{\superscriptI} }  \),
        then there exists \(\possiblyWithSub\stageImetaColor{\Hat{T}^{\superscriptI} }\) such that
        \(   [  \possiblyWithSub\stageOmetaColor{N^{\superscriptO} }_{{\mathrm{0}}}  /  \possiblyWithSub\stageOmetaColor{x}  ]    \possiblyWithSub\stageImetaColor{T^{\superscriptI} }   \longrightarrow^{1\,\ast}     [  \possiblyWithSub\stageOmetaColor{N^{\superscriptO} }_{{\mathrm{0}}}  /  \possiblyWithSub\stageOmetaColor{x}  ]    \possiblyWithSub\stageImetaColor{\Hat{T}^{\superscriptI} }   \)
        and \(\possiblyWithSub\stageImetaColor{T'^{\superscriptI} } =   [  \possiblyWithSub\stageOmetaColor{N'^{\superscriptO} }_{{\mathrm{0}}}  /  \possiblyWithSub\stageOmetaColor{x}  ]    \possiblyWithSub\stageImetaColor{\Hat{T}^{\superscriptI} } \).
    \end{enumerate}
  \end{lemma}
  \begin{proof}
    By mutual induction on the structure of \(\possiblyWithSub\stageOmetaColor{N^{\superscriptO} }\), \(\possiblyWithSub\stageImetaColor{N^{\superscriptI} }\), and \(\possiblyWithSub\stageImetaColor{T^{\superscriptI} }\)
    similar to the one for Lemma~\ref{lem:weak-bisimulation-left} (i.e.~the left-hand side version).
    \begin{enumerate}
      \item
        \begin{itemize}
          \item Case~\(\possiblyWithSub\stageOmetaColor{N^{\superscriptO} } = \possiblyWithSub\stageOmetaColor{x}\):
            We will show that \(\possiblyWithSub\stageOmetaColor{\Hat{N}^{\superscriptO} } \defeq \possiblyWithSub\stageOmetaColor{N'^{\superscriptO} }\)
            satisfies the desired properties.
            First, since \(\possiblyWithSub\stageOmetaColor{N'^{\superscriptO} }\) is a closed term (by \( \possiblyWithSub\stageOmetaColor{N^{\superscriptO} }_{{\mathrm{0}}}  \longrightarrow^{0}   \possiblyWithSub\stageOmetaColor{N'^{\superscriptO} }_{{\mathrm{0}}}  \)),
            we have \(  [  \possiblyWithSub\stageOmetaColor{N'^{\superscriptO} }_{{\mathrm{0}}}  /  \possiblyWithSub\stageOmetaColor{x}  ]    \possiblyWithSub\stageOmetaColor{\Hat{N}^{\superscriptO} }  =   [  \possiblyWithSub\stageOmetaColor{N'^{\superscriptO} }_{{\mathrm{0}}}  /  \possiblyWithSub\stageOmetaColor{x}  ]    \possiblyWithSub\stageOmetaColor{N'^{\superscriptO} }  = \possiblyWithSub\stageOmetaColor{N'^{\superscriptO} }\).
            Also, since \(  [  \possiblyWithSub\stageOmetaColor{N'^{\superscriptO} }_{{\mathrm{0}}}  /  \possiblyWithSub\stageOmetaColor{x}  ]    \possiblyWithSub\stageOmetaColor{N^{\superscriptO} }  = \possiblyWithSub\stageOmetaColor{N'^{\superscriptO} }_{{\mathrm{0}}}\) holds,
            we have \( \possiblyWithSub\stageOmetaColor{N'^{\superscriptO} }_{{\mathrm{0}}}  \longrightarrow^{0}   \possiblyWithSub\stageOmetaColor{N'^{\superscriptO} }  \).
            Therefore, by
            \(  [  \possiblyWithSub\stageOmetaColor{N^{\superscriptO} }_{{\mathrm{0}}}  /  \possiblyWithSub\stageOmetaColor{x}  ]    \possiblyWithSub\stageOmetaColor{N^{\superscriptO} }  = \possiblyWithSub\stageOmetaColor{N^{\superscriptO} }_{{\mathrm{0}}}\),
            \(  [  \possiblyWithSub\stageOmetaColor{N^{\superscriptO} }_{{\mathrm{0}}}  /  \possiblyWithSub\stageOmetaColor{x}  ]    \possiblyWithSub\stageOmetaColor{\Hat{N}^{\superscriptO} }  = \possiblyWithSub\stageOmetaColor{N'^{\superscriptO} }\),
            and the assumption~\( \possiblyWithSub\stageOmetaColor{N^{\superscriptO} }_{{\mathrm{0}}}  \longrightarrow^{0}   \possiblyWithSub\stageOmetaColor{N'^{\superscriptO} }_{{\mathrm{0}}}  \),
            we have
            \(   [  \possiblyWithSub\stageOmetaColor{N^{\superscriptO} }_{{\mathrm{0}}}  /  \possiblyWithSub\stageOmetaColor{x}  ]    \possiblyWithSub\stageOmetaColor{N^{\superscriptO} }   \longrightarrow^{0\,\ast}     [  \possiblyWithSub\stageOmetaColor{N^{\superscriptO} }_{{\mathrm{0}}}  /  \possiblyWithSub\stageOmetaColor{x}  ]    \possiblyWithSub\stageOmetaColor{\Hat{N}^{\superscriptO} }   \).
          \item Case~\(\possiblyWithSub\stageOmetaColor{N^{\superscriptO} } = \possiblyWithSub\stageOmetaColor{x'}\) such that \(\possiblyWithSub\stageOmetaColor{x'} \neq \possiblyWithSub\stageOmetaColor{x}\):
            Since \(  [  \possiblyWithSub\stageOmetaColor{N'^{\superscriptO} }_{{\mathrm{0}}}  /  \possiblyWithSub\stageOmetaColor{x}  ]    \possiblyWithSub\stageOmetaColor{N^{\superscriptO} }  = \possiblyWithSub\stageOmetaColor{x'}\),
            this case contradicts the assumption~\(   [  \possiblyWithSub\stageOmetaColor{N'^{\superscriptO} }_{{\mathrm{0}}}  /  \possiblyWithSub\stageOmetaColor{x}  ]    \possiblyWithSub\stageOmetaColor{N^{\superscriptO} }   \longrightarrow^{0}   \possiblyWithSub\stageOmetaColor{N'^{\superscriptO} }  \).
          \item Case where we have \(\possiblyWithSub\stageOmetaColor{N^{\superscriptO} } = \possiblyWithSub\stageOmetaColor{p}\), \(\possiblyWithSub\stageOmetaColor{N^{\superscriptO} } = \possiblyWithSub\stageOmetaColor{c}\),
          \(\possiblyWithSub\stageOmetaColor{N^{\superscriptO} } =  \openO{(}  \ordO{\lambda} \possiblyWithSub\stageOmetaColor{x'}  \relO{:}  \possiblyWithSub\stageOmetaColor{T^{\superscriptO} }_{{\mathrm{1}}} \punctO{.}\  \possiblyWithSub\stageOmetaColor{N^{\superscriptO} }_{{\mathrm{2}}}  \closeO{)} \), or
          \(\possiblyWithSub\stageOmetaColor{N^{\superscriptO} } =  \LeftAssertParen \relO{\CastArrow}   \openO{\{} \possiblyWithSub\stageOmetaColor{x'}  \relO{:}  \possiblyWithSub\stageOmetaColor{B}  \relO{\mid}  \possiblyWithSub\stageOmetaColor{N^{\superscriptO} }_{{\mathrm{1}}} \closeO{\} }   \RightAssertParen^{ L } \)
          also contradicts the assumption~\(   [  \possiblyWithSub\stageOmetaColor{N'^{\superscriptO} }_{{\mathrm{0}}}  /  \possiblyWithSub\stageOmetaColor{x}  ]    \possiblyWithSub\stageOmetaColor{N^{\superscriptO} }   \longrightarrow^{0}   \possiblyWithSub\stageOmetaColor{N'^{\superscriptO} }  \).
          \item Case \(\possiblyWithSub\stageOmetaColor{N^{\superscriptO} } =  \possiblyWithSub\stageOmetaColor{N^{\superscriptO} }_{{\mathrm{1}}} \  \possiblyWithSub\stageOmetaColor{N^{\superscriptO} }_{{\mathrm{2}}} \):
            We have one of the following cases:
            \begin{itemize}
              \item Case where \(  [  \possiblyWithSub\stageOmetaColor{N'^{\superscriptO} }_{{\mathrm{0}}}  /  \possiblyWithSub\stageOmetaColor{x}  ]    \possiblyWithSub\stageOmetaColor{N^{\superscriptO} }_{{\mathrm{1}}} \) is not a value:
                We can uniquely trace back the derivation of
                \(   [  \possiblyWithSub\stageOmetaColor{N'^{\superscriptO} }_{{\mathrm{0}}}  /  \possiblyWithSub\stageOmetaColor{x}  ]    \possiblyWithSub\stageOmetaColor{N^{\superscriptO} }   \longrightarrow^{0}   \possiblyWithSub\stageOmetaColor{N'^{\superscriptO} }  \) as follows:
                \begin{center}
                  \derive[E0-App1]{%
                       [  \possiblyWithSub\stageOmetaColor{N'^{\superscriptO} }_{{\mathrm{0}}}  /  \possiblyWithSub\stageOmetaColor{x}  ]    \possiblyWithSub\stageOmetaColor{N^{\superscriptO} }_{{\mathrm{1}}}   \longrightarrow^{0}   \possiblyWithSub\stageOmetaColor{N'^{\superscriptO} }_{{\mathrm{1}}}  
                  }{%
                       \openO{(}   [  \possiblyWithSub\stageOmetaColor{N'^{\superscriptO} }_{{\mathrm{0}}}  /  \possiblyWithSub\stageOmetaColor{x}  ]    \possiblyWithSub\stageOmetaColor{N^{\superscriptO} }_{{\mathrm{1}}}  \closeO{)}  \   \openO{(}   [  \possiblyWithSub\stageOmetaColor{N'^{\superscriptO} }_{{\mathrm{0}}}  /  \possiblyWithSub\stageOmetaColor{x}  ]    \possiblyWithSub\stageOmetaColor{N^{\superscriptO} }_{{\mathrm{2}}}  \closeO{)}    \longrightarrow^{0}    \possiblyWithSub\stageOmetaColor{N'^{\superscriptO} }_{{\mathrm{1}}} \   \openO{(}   [  \possiblyWithSub\stageOmetaColor{N'^{\superscriptO} }_{{\mathrm{0}}}  /  \possiblyWithSub\stageOmetaColor{x}  ]    \possiblyWithSub\stageOmetaColor{N^{\superscriptO} }_{{\mathrm{2}}}  \closeO{)}    
                  }.
                \end{center}
                By IH, there exists \(\possiblyWithSub\stageOmetaColor{\Hat{N}^{\superscriptO} }_{{\mathrm{1}}}\) such that
                \(   [  \possiblyWithSub\stageOmetaColor{N^{\superscriptO} }_{{\mathrm{0}}}  /  \possiblyWithSub\stageOmetaColor{x}  ]    \possiblyWithSub\stageOmetaColor{N^{\superscriptO} }_{{\mathrm{1}}}   \longrightarrow^{0\,\ast}     [  \possiblyWithSub\stageOmetaColor{N^{\superscriptO} }_{{\mathrm{0}}}  /  \possiblyWithSub\stageOmetaColor{x}  ]    \possiblyWithSub\stageOmetaColor{\Hat{N}^{\superscriptO} }_{{\mathrm{1}}}   \)
                and \(\possiblyWithSub\stageOmetaColor{N'^{\superscriptO} }_{{\mathrm{1}}} =   [  \possiblyWithSub\stageOmetaColor{N'^{\superscriptO} }_{{\mathrm{0}}}  /  \possiblyWithSub\stageOmetaColor{x}  ]    \possiblyWithSub\stageOmetaColor{\Hat{N}^{\superscriptO} }_{{\mathrm{1}}} \).
                Thus, \(\possiblyWithSub\stageOmetaColor{\Hat{N}^{\superscriptO} } \defeq  \possiblyWithSub\stageOmetaColor{\Hat{N}^{\superscriptO} }_{{\mathrm{1}}} \  \possiblyWithSub\stageOmetaColor{N^{\superscriptO} }_{{\mathrm{2}}} \) satisfies the desired properties
                by the repeated use of \rulename{E0-App1}.
              \item Case where \(  [  \possiblyWithSub\stageOmetaColor{N'^{\superscriptO} }_{{\mathrm{0}}}  /  \possiblyWithSub\stageOmetaColor{x}  ]    \possiblyWithSub\stageOmetaColor{N^{\superscriptO} }_{{\mathrm{1}}}  \revdefeq \possiblyWithSub\stageOmetaColor{v^{\superscriptO} }_{{\mathrm{1}}}\) is a value
              while \(  [  \possiblyWithSub\stageOmetaColor{N'^{\superscriptO} }_{{\mathrm{0}}}  /  \possiblyWithSub\stageOmetaColor{x}  ]    \possiblyWithSub\stageOmetaColor{N^{\superscriptO} }_{{\mathrm{2}}} \) is not:
                The sole possible derivation of \(   [  \possiblyWithSub\stageOmetaColor{N'^{\superscriptO} }_{{\mathrm{0}}}  /  \possiblyWithSub\stageOmetaColor{x}  ]    \possiblyWithSub\stageOmetaColor{N^{\superscriptO} }   \longrightarrow^{0}   \possiblyWithSub\stageOmetaColor{N'^{\superscriptO} }  \) is as follows:
                \begin{center}
                  \derive[E0-App2]{%
                       [  \possiblyWithSub\stageOmetaColor{N'^{\superscriptO} }_{{\mathrm{0}}}  /  \possiblyWithSub\stageOmetaColor{x}  ]    \possiblyWithSub\stageOmetaColor{N^{\superscriptO} }_{{\mathrm{2}}}   \longrightarrow^{0}   \possiblyWithSub\stageOmetaColor{N'^{\superscriptO} }_{{\mathrm{2}}}  
                  }{%
                       \possiblyWithSub\stageOmetaColor{v^{\superscriptO} }_{{\mathrm{1}}}  \   \openO{(}   [  \possiblyWithSub\stageOmetaColor{N'^{\superscriptO} }_{{\mathrm{0}}}  /  \possiblyWithSub\stageOmetaColor{x}  ]    \possiblyWithSub\stageOmetaColor{N^{\superscriptO} }_{{\mathrm{2}}}  \closeO{)}    \longrightarrow^{0}     \possiblyWithSub\stageOmetaColor{v^{\superscriptO} }_{{\mathrm{1}}}  \  \possiblyWithSub\stageOmetaColor{N'^{\superscriptO} }_{{\mathrm{2}}}   
                  }.
                \end{center}
                By IH, there exists \(\possiblyWithSub\stageOmetaColor{\Hat{N}^{\superscriptO} }_{{\mathrm{2}}}\) such that
                \(   [  \possiblyWithSub\stageOmetaColor{N^{\superscriptO} }_{{\mathrm{0}}}  /  \possiblyWithSub\stageOmetaColor{x}  ]    \possiblyWithSub\stageOmetaColor{N^{\superscriptO} }_{{\mathrm{2}}}   \longrightarrow^{0\,\ast}     [  \possiblyWithSub\stageOmetaColor{N^{\superscriptO} }_{{\mathrm{0}}}  /  \possiblyWithSub\stageOmetaColor{x}  ]    \possiblyWithSub\stageOmetaColor{\Hat{N}^{\superscriptO} }_{{\mathrm{2}}}   \)
                and \(\possiblyWithSub\stageOmetaColor{N'^{\superscriptO} }_{{\mathrm{2}}} =   [  \possiblyWithSub\stageOmetaColor{N'^{\superscriptO} }_{{\mathrm{0}}}  /  \possiblyWithSub\stageOmetaColor{x}  ]    \possiblyWithSub\stageOmetaColor{\Hat{N}^{\superscriptO} }_{{\mathrm{2}}} \).
                Thus, \(\possiblyWithSub\stageOmetaColor{\Hat{N}^{\superscriptO} } \defeq  \possiblyWithSub\stageOmetaColor{N^{\superscriptO} }_{{\mathrm{1}}} \  \possiblyWithSub\stageOmetaColor{\Hat{N}^{\superscriptO} }_{{\mathrm{2}}} \) clearly satisfies the desired properties
                by the repeated use of \rulename{E0-App2}.
              \item Case where both \(  [  \possiblyWithSub\stageOmetaColor{N'^{\superscriptO} }_{{\mathrm{0}}}  /  \possiblyWithSub\stageOmetaColor{x}  ]    \possiblyWithSub\stageOmetaColor{N^{\superscriptO} }_{{\mathrm{1}}} \) and \(  [  \possiblyWithSub\stageOmetaColor{N'^{\superscriptO} }_{{\mathrm{0}}}  /  \possiblyWithSub\stageOmetaColor{x}  ]    \possiblyWithSub\stageOmetaColor{N^{\superscriptO} }_{{\mathrm{2}}} \) are values:
                First, by using Lemma~\ref{lem:weak-bisimulation-right-pre} twice,
                there exist \(\possiblyWithSub\stageOmetaColor{N'^{\superscriptO} }_{{\mathrm{1}}}\) and \(\possiblyWithSub\stageOmetaColor{N'^{\superscriptO} }_{{\mathrm{2}}}\) such that
                \begin{enumerate}
                  \item[(1A)] \(   [  \possiblyWithSub\stageOmetaColor{N^{\superscriptO} }_{{\mathrm{0}}}  /  \possiblyWithSub\stageOmetaColor{x}  ]    \possiblyWithSub\stageOmetaColor{N^{\superscriptO} }_{{\mathrm{1}}}   \longrightarrow^{0\,\ast}     [  \possiblyWithSub\stageOmetaColor{N^{\superscriptO} }_{{\mathrm{0}}}  /  \possiblyWithSub\stageOmetaColor{x}  ]    \possiblyWithSub\stageOmetaColor{N'^{\superscriptO} }_{{\mathrm{1}}}   \),
                  \item[(1B)] \(  [  \possiblyWithSub\stageOmetaColor{N^{\superscriptO} }_{{\mathrm{0}}}  /  \possiblyWithSub\stageOmetaColor{x}  ]    \possiblyWithSub\stageOmetaColor{N'^{\superscriptO} }_{{\mathrm{1}}} \) is a stage-\(0\) value,
                  \item[(1C)] \(  [  \possiblyWithSub\stageOmetaColor{N'^{\superscriptO} }_{{\mathrm{0}}}  /  \possiblyWithSub\stageOmetaColor{x}  ]    \possiblyWithSub\stageOmetaColor{N^{\superscriptO} }_{{\mathrm{1}}}  =   [  \possiblyWithSub\stageOmetaColor{N'^{\superscriptO} }_{{\mathrm{0}}}  /  \possiblyWithSub\stageOmetaColor{x}  ]    \possiblyWithSub\stageOmetaColor{N'^{\superscriptO} }_{{\mathrm{1}}} \),
                  \item[(2A)] \(   [  \possiblyWithSub\stageOmetaColor{N^{\superscriptO} }_{{\mathrm{0}}}  /  \possiblyWithSub\stageOmetaColor{x}  ]    \possiblyWithSub\stageOmetaColor{N^{\superscriptO} }_{{\mathrm{2}}}   \longrightarrow^{0\,\ast}     [  \possiblyWithSub\stageOmetaColor{N^{\superscriptO} }_{{\mathrm{0}}}  /  \possiblyWithSub\stageOmetaColor{x}  ]    \possiblyWithSub\stageOmetaColor{N'^{\superscriptO} }_{{\mathrm{2}}}   \),
                  \item[(2B)] \(  [  \possiblyWithSub\stageOmetaColor{N^{\superscriptO} }_{{\mathrm{0}}}  /  \possiblyWithSub\stageOmetaColor{x}  ]    \possiblyWithSub\stageOmetaColor{N'^{\superscriptO} }_{{\mathrm{2}}} \) is a stage-\(0\) value, and
                  \item[(2C)] \(  [  \possiblyWithSub\stageOmetaColor{N'^{\superscriptO} }_{{\mathrm{0}}}  /  \possiblyWithSub\stageOmetaColor{x}  ]    \possiblyWithSub\stageOmetaColor{N^{\superscriptO} }_{{\mathrm{2}}}  =   [  \possiblyWithSub\stageOmetaColor{N'^{\superscriptO} }_{{\mathrm{0}}}  /  \possiblyWithSub\stageOmetaColor{x}  ]    \possiblyWithSub\stageOmetaColor{N'^{\superscriptO} }_{{\mathrm{2}}} \).
                \end{enumerate}
                Then, by (1A), (1B), and (2A), we can clearly derive the following,
                using \rulename{E0-App1} and \rulename{E0-App2} repeatedly:
                \begin{enumerate}
                  \item[(P)] \(   [  \possiblyWithSub\stageOmetaColor{N^{\superscriptO} }_{{\mathrm{0}}}  /  \possiblyWithSub\stageOmetaColor{x}  ]     \openO{(}  \possiblyWithSub\stageOmetaColor{N^{\superscriptO} }_{{\mathrm{1}}} \  \possiblyWithSub\stageOmetaColor{N^{\superscriptO} }_{{\mathrm{2}}}  \closeO{)}    \longrightarrow^{0\,\ast}     [  \possiblyWithSub\stageOmetaColor{N^{\superscriptO} }_{{\mathrm{0}}}  /  \possiblyWithSub\stageOmetaColor{x}  ]     \openO{(}  \possiblyWithSub\stageOmetaColor{N'^{\superscriptO} }_{{\mathrm{1}}} \  \possiblyWithSub\stageOmetaColor{N'^{\superscriptO} }_{{\mathrm{2}}}  \closeO{)}    \).
                \end{enumerate}
                Also, by (1B), (2B), (2C), and the assumption that
                \(  [  \possiblyWithSub\stageOmetaColor{N'^{\superscriptO} }_{{\mathrm{0}}}  /  \possiblyWithSub\stageOmetaColor{x}  ]    \possiblyWithSub\stageOmetaColor{N^{\superscriptO} }_{{\mathrm{2}}} \) is a stage-\(0\) value,
                \(  [  \possiblyWithSub\stageOmetaColor{N^{\superscriptO} }_{{\mathrm{0}}}  /  \possiblyWithSub\stageOmetaColor{x}  ]    \possiblyWithSub\stageOmetaColor{N'^{\superscriptO} }_{{\mathrm{1}}} \), \(  [  \possiblyWithSub\stageOmetaColor{N^{\superscriptO} }_{{\mathrm{0}}}  /  \possiblyWithSub\stageOmetaColor{x}  ]    \possiblyWithSub\stageOmetaColor{N'^{\superscriptO} }_{{\mathrm{2}}} \), and \(  [  \possiblyWithSub\stageOmetaColor{N'^{\superscriptO} }_{{\mathrm{0}}}  /  \possiblyWithSub\stageOmetaColor{x}  ]    \possiblyWithSub\stageOmetaColor{N'^{\superscriptO} }_{{\mathrm{2}}} \)
                are all stage-\(0\) values.
                Thus, by Lemma~\ref{lem:weak-bisimulation-app-left-and-right}~(2),
                there exists \(\possiblyWithSub\stageOmetaColor{\Hat{N}^{\superscriptO} }\) such that
                \begin{enumerate}
                  \item[(Q)] \(   [  \possiblyWithSub\stageOmetaColor{N^{\superscriptO} }_{{\mathrm{0}}}  /  \possiblyWithSub\stageOmetaColor{x}  ]     \openO{(}  \possiblyWithSub\stageOmetaColor{N'^{\superscriptO} }_{{\mathrm{1}}} \  \possiblyWithSub\stageOmetaColor{N'^{\superscriptO} }_{{\mathrm{2}}}  \closeO{)}    \longrightarrow^{0}     [  \possiblyWithSub\stageOmetaColor{N^{\superscriptO} }_{{\mathrm{0}}}  /  \possiblyWithSub\stageOmetaColor{x}  ]    \possiblyWithSub\stageOmetaColor{\Hat{N}^{\superscriptO} }   \) and
                  \item[(R)] \(\possiblyWithSub\stageOmetaColor{N'^{\superscriptO} } =   [  \possiblyWithSub\stageOmetaColor{N^{\superscriptO} }_{{\mathrm{0}}}  /  \possiblyWithSub\stageOmetaColor{x}  ]    \possiblyWithSub\stageOmetaColor{\Hat{N}^{\superscriptO} } \).
                \end{enumerate}
                Thus, by (P), (Q), and (R), we have proved the desired properties.
            \end{itemize}
          \item Case~\(\possiblyWithSub\stageOmetaColor{N^{\superscriptO} } =  \openO{\langle} \possiblyWithSub\stageImetaColor{N^{\superscriptI} } \closeO{\rangle} \):
            By tracing back the derivation of \(   [  \possiblyWithSub\stageOmetaColor{N'^{\superscriptO} }_{{\mathrm{0}}}  /  \possiblyWithSub\stageOmetaColor{x}  ]    \possiblyWithSub\stageOmetaColor{N^{\superscriptO} }   \longrightarrow^{0}   \possiblyWithSub\stageOmetaColor{N'^{\superscriptO} }  \),
            we have
            \begin{center}
              \derive[E0-Brkt]{%
                   [  \possiblyWithSub\stageOmetaColor{N'^{\superscriptO} }_{{\mathrm{0}}}  /  \possiblyWithSub\stageOmetaColor{x}  ]    \possiblyWithSub\stageImetaColor{N^{\superscriptI} }   \longrightarrow^{1}   \possiblyWithSub\stageImetaColor{N'^{\superscriptI} }  
              }{%
                  \openO{\langle}   [  \possiblyWithSub\stageOmetaColor{N'^{\superscriptO} }_{{\mathrm{0}}}  /  \possiblyWithSub\stageOmetaColor{x}  ]    \possiblyWithSub\stageImetaColor{N^{\superscriptI} }  \closeO{\rangle}   \longrightarrow^{0}    \openO{\langle} \possiblyWithSub\stageImetaColor{N'^{\superscriptI} } \closeO{\rangle}   
              }.
            \end{center}
            Then, by IH, there exists \(\possiblyWithSub\stageImetaColor{\Hat{N}^{\superscriptI} }\) such that
            \(   [  \possiblyWithSub\stageOmetaColor{N^{\superscriptO} }_{{\mathrm{0}}}  /  \possiblyWithSub\stageOmetaColor{x}  ]    \possiblyWithSub\stageImetaColor{N^{\superscriptI} }   \longrightarrow^{1\,\ast}     [  \possiblyWithSub\stageOmetaColor{N^{\superscriptO} }_{{\mathrm{0}}}  /  \possiblyWithSub\stageOmetaColor{x}  ]    \possiblyWithSub\stageImetaColor{\Hat{N}^{\superscriptI} }   \)
            and \(\possiblyWithSub\stageImetaColor{N'^{\superscriptI} } =   [  \possiblyWithSub\stageOmetaColor{N'^{\superscriptO} }_{{\mathrm{0}}}  /  \possiblyWithSub\stageOmetaColor{x}  ]    \possiblyWithSub\stageImetaColor{\Hat{N}^{\superscriptI} } \).
            We will show that \(\possiblyWithSub\stageOmetaColor{\Hat{N}^{\superscriptO} } \defeq  \openO{\langle} \possiblyWithSub\stageImetaColor{\Hat{N}^{\superscriptI} } \closeO{\rangle} \) satisfies the desired properties.
            Indeed, using \rulename{E0-Brkt} repeatedly, we can derive
            \(   [  \possiblyWithSub\stageOmetaColor{N^{\superscriptO} }_{{\mathrm{0}}}  /  \possiblyWithSub\stageOmetaColor{x}  ]     \openO{\langle} \possiblyWithSub\stageImetaColor{N^{\superscriptI} } \closeO{\rangle}    \longrightarrow^{0\,\ast}     [  \possiblyWithSub\stageOmetaColor{N^{\superscriptO} }_{{\mathrm{0}}}  /  \possiblyWithSub\stageOmetaColor{x}  ]     \openO{\langle} \possiblyWithSub\stageImetaColor{\Hat{N}^{\superscriptI} } \closeO{\rangle}    \)
            from \(   [  \possiblyWithSub\stageOmetaColor{N^{\superscriptO} }_{{\mathrm{0}}}  /  \possiblyWithSub\stageOmetaColor{x}  ]    \possiblyWithSub\stageImetaColor{N^{\superscriptI} }   \longrightarrow^{1\,\ast}     [  \possiblyWithSub\stageOmetaColor{N^{\superscriptO} }_{{\mathrm{0}}}  /  \possiblyWithSub\stageOmetaColor{x}  ]    \possiblyWithSub\stageImetaColor{\Hat{N}^{\superscriptI} }   \).
            We also have
            \(\possiblyWithSub\stageOmetaColor{N'^{\superscriptO} }
              =  \openO{\langle} \possiblyWithSub\stageImetaColor{N'^{\superscriptI} } \closeO{\rangle} 
              =  \openO{\langle}   [  \possiblyWithSub\stageOmetaColor{N'^{\superscriptO} }_{{\mathrm{0}}}  /  \possiblyWithSub\stageOmetaColor{x}  ]    \possiblyWithSub\stageImetaColor{\Hat{N}^{\superscriptI} }  \closeO{\rangle} 
              =   [  \possiblyWithSub\stageOmetaColor{N'^{\superscriptO} }_{{\mathrm{0}}}  /  \possiblyWithSub\stageOmetaColor{x}  ]     \openO{\langle} \possiblyWithSub\stageImetaColor{\Hat{N}^{\superscriptI} } \closeO{\rangle}  
              =   [  \possiblyWithSub\stageOmetaColor{N'^{\superscriptO} }_{{\mathrm{0}}}  /  \possiblyWithSub\stageOmetaColor{x}  ]    \possiblyWithSub\stageOmetaColor{\Hat{N}^{\superscriptO} } \).
          \item Case~\(\possiblyWithSub\stageOmetaColor{N^{\superscriptO} } =  \LeftAssertParen   \openO{\{} \possiblyWithSub\stageOmetaColor{x}  \relO{:}  \possiblyWithSub\stageOmetaColor{B}  \relO{\mid}  \possiblyWithSub\stageOmetaColor{N^{\superscriptO} }_{{\mathrm{1}}} \closeO{\} }  \punctO{,}  \possiblyWithSub\stageOmetaColor{N^{\superscriptO} }_{{\mathrm{2}}} \punctO{,}  \possiblyWithSub\stageOmetaColor{c}  \RightAssertParen^{ L } \):
            By tracing back the derivation of \(   [  \possiblyWithSub\stageOmetaColor{N'^{\superscriptO} }_{{\mathrm{0}}}  /  \possiblyWithSub\stageOmetaColor{x}  ]    \possiblyWithSub\stageOmetaColor{N^{\superscriptO} }   \longrightarrow^{0}   \possiblyWithSub\stageOmetaColor{N'^{\superscriptO} }  \),
            we have one of the following:
            \begin{itemize}
              \item Case \derive[E0-RfnAct]{%
                   [  \possiblyWithSub\stageOmetaColor{N'^{\superscriptO} }_{{\mathrm{0}}}  /  \possiblyWithSub\stageOmetaColor{x}  ]    \possiblyWithSub\stageOmetaColor{N^{\superscriptO} }_{{\mathrm{2}}}   \longrightarrow^{0}   \possiblyWithSub\stageOmetaColor{N'^{\superscriptO} }_{{\mathrm{2}}}  
              }{%
                  \LeftAssertParen   \openO{\{} \possiblyWithSub\stageOmetaColor{\nu}  \relO{:}  \possiblyWithSub\stageOmetaColor{B}  \relO{\mid}    [  \possiblyWithSub\stageOmetaColor{N'^{\superscriptO} }_{{\mathrm{0}}}  /  \possiblyWithSub\stageOmetaColor{x}  ]    \possiblyWithSub\stageOmetaColor{N^{\superscriptO} }_{{\mathrm{1}}}  \closeO{\} }  \punctO{,}    [  \possiblyWithSub\stageOmetaColor{N'^{\superscriptO} }_{{\mathrm{0}}}  /  \possiblyWithSub\stageOmetaColor{x}  ]    \possiblyWithSub\stageOmetaColor{N^{\superscriptO} }_{{\mathrm{2}}}  \punctO{,}  \possiblyWithSub\stageOmetaColor{c}  \RightAssertParen^{ L }   \longrightarrow^{0}    \LeftAssertParen   \openO{\{} \possiblyWithSub\stageOmetaColor{\nu}  \relO{:}  \possiblyWithSub\stageOmetaColor{B}  \relO{\mid}    [  \possiblyWithSub\stageOmetaColor{N'^{\superscriptO} }_{{\mathrm{0}}}  /  \possiblyWithSub\stageOmetaColor{x}  ]    \possiblyWithSub\stageOmetaColor{N^{\superscriptO} }_{{\mathrm{1}}}  \closeO{\} }  \punctO{,}  \possiblyWithSub\stageOmetaColor{N'^{\superscriptO} }_{{\mathrm{2}}} \punctO{,}  \possiblyWithSub\stageOmetaColor{c}  \RightAssertParen^{ L }   
              }, where we can assume \(\possiblyWithSub\stageOmetaColor{\nu} \neq \possiblyWithSub\stageOmetaColor{x}\) w.l.o.g.:
                By IH, there exists \(\possiblyWithSub\stageOmetaColor{\Hat{N}^{\superscriptO} }_{{\mathrm{2}}}\) such that
                \(   [  \possiblyWithSub\stageOmetaColor{N^{\superscriptO} }_{{\mathrm{0}}}  /  \possiblyWithSub\stageOmetaColor{x}  ]    \possiblyWithSub\stageOmetaColor{N^{\superscriptO} }_{{\mathrm{2}}}   \longrightarrow^{0\,\ast}     [  \possiblyWithSub\stageOmetaColor{N^{\superscriptO} }_{{\mathrm{0}}}  /  \possiblyWithSub\stageOmetaColor{x}  ]    \possiblyWithSub\stageOmetaColor{\Hat{N}^{\superscriptO} }_{{\mathrm{2}}}   \)
                and \(\possiblyWithSub\stageOmetaColor{N'^{\superscriptO} }_{{\mathrm{2}}} =   [  \possiblyWithSub\stageOmetaColor{N'^{\superscriptO} }_{{\mathrm{0}}}  /  \possiblyWithSub\stageOmetaColor{x}  ]    \possiblyWithSub\stageOmetaColor{\Hat{N}^{\superscriptO} }_{{\mathrm{2}}} \).
                We will show that \(\possiblyWithSub\stageOmetaColor{\Hat{N}^{\superscriptO} } \defeq  \LeftAssertParen   \openO{\{} \possiblyWithSub\stageOmetaColor{\nu}  \relO{:}  \possiblyWithSub\stageOmetaColor{B}  \relO{\mid}  \possiblyWithSub\stageOmetaColor{N^{\superscriptO} }_{{\mathrm{1}}} \closeO{\} }  \punctO{,}  \possiblyWithSub\stageOmetaColor{\Hat{N}^{\superscriptO} }_{{\mathrm{2}}} \punctO{,}  \possiblyWithSub\stageOmetaColor{c}  \RightAssertParen^{ L } \)
                satisfies the desired properties.
                Indeed, we first have
                \begin{align*}
                  \possiblyWithSub\stageOmetaColor{N'^{\superscriptO} }
                  &=  \LeftAssertParen   \openO{\{} \possiblyWithSub\stageOmetaColor{\nu}  \relO{:}  \possiblyWithSub\stageOmetaColor{B}  \relO{\mid}    [  \possiblyWithSub\stageOmetaColor{N'^{\superscriptO} }_{{\mathrm{0}}}  /  \possiblyWithSub\stageOmetaColor{x}  ]    \possiblyWithSub\stageOmetaColor{N^{\superscriptO} }_{{\mathrm{1}}}  \closeO{\} }  \punctO{,}  \possiblyWithSub\stageOmetaColor{N'^{\superscriptO} }_{{\mathrm{2}}} \punctO{,}  \possiblyWithSub\stageOmetaColor{c}  \RightAssertParen^{ L } 
                \\&=  \LeftAssertParen   \openO{\{} \possiblyWithSub\stageOmetaColor{\nu}  \relO{:}  \possiblyWithSub\stageOmetaColor{B}  \relO{\mid}    [  \possiblyWithSub\stageOmetaColor{N'^{\superscriptO} }_{{\mathrm{0}}}  /  \possiblyWithSub\stageOmetaColor{x}  ]    \possiblyWithSub\stageOmetaColor{N^{\superscriptO} }_{{\mathrm{1}}}  \closeO{\} }  \punctO{,}    [  \possiblyWithSub\stageOmetaColor{N'^{\superscriptO} }_{{\mathrm{0}}}  /  \possiblyWithSub\stageOmetaColor{x}  ]    \possiblyWithSub\stageOmetaColor{\Hat{N}^{\superscriptO} }_{{\mathrm{2}}}  \punctO{,}  \possiblyWithSub\stageOmetaColor{c}  \RightAssertParen^{ L } 
                \\&=   [  \possiblyWithSub\stageOmetaColor{N'^{\superscriptO} }_{{\mathrm{0}}}  /  \possiblyWithSub\stageOmetaColor{x}  ]     \LeftAssertParen   \openO{\{} \possiblyWithSub\stageOmetaColor{\nu}  \relO{:}  \possiblyWithSub\stageOmetaColor{B}  \relO{\mid}  \possiblyWithSub\stageOmetaColor{N^{\superscriptO} }_{{\mathrm{1}}} \closeO{\} }  \punctO{,}  \possiblyWithSub\stageOmetaColor{\Hat{N}^{\superscriptO} }_{{\mathrm{2}}} \punctO{,}  \possiblyWithSub\stageOmetaColor{c}  \RightAssertParen^{ L }  
                  =   [  \possiblyWithSub\stageOmetaColor{N'^{\superscriptO} }_{{\mathrm{0}}}  /  \possiblyWithSub\stageOmetaColor{x}  ]    \possiblyWithSub\stageOmetaColor{\Hat{N}^{\superscriptO} } .
                \end{align*}
                We can also derive
                \(   [  \possiblyWithSub\stageOmetaColor{N^{\superscriptO} }_{{\mathrm{0}}}  /  \possiblyWithSub\stageOmetaColor{x}  ]    \possiblyWithSub\stageOmetaColor{N^{\superscriptO} }   \longrightarrow^{0}     [  \possiblyWithSub\stageOmetaColor{N^{\superscriptO} }_{{\mathrm{0}}}  /  \possiblyWithSub\stageOmetaColor{x}  ]    \possiblyWithSub\stageOmetaColor{\Hat{N}^{\superscriptO} }   \)
                as follows:
                \begin{center}
                  \derive[E0-RfnAct]{%
                       [  \possiblyWithSub\stageOmetaColor{N^{\superscriptO} }_{{\mathrm{0}}}  /  \possiblyWithSub\stageOmetaColor{x}  ]    \possiblyWithSub\stageOmetaColor{N^{\superscriptO} }_{{\mathrm{2}}}   \longrightarrow^{0}     [  \possiblyWithSub\stageOmetaColor{N^{\superscriptO} }_{{\mathrm{0}}}  /  \possiblyWithSub\stageOmetaColor{x}  ]    \possiblyWithSub\stageOmetaColor{\Hat{N}^{\superscriptO} }_{{\mathrm{2}}}   
                  }{%
                      \LeftAssertParen   \openO{\{} \possiblyWithSub\stageOmetaColor{\nu}  \relO{:}  \possiblyWithSub\stageOmetaColor{B}  \relO{\mid}    [  \possiblyWithSub\stageOmetaColor{N^{\superscriptO} }_{{\mathrm{0}}}  /  \possiblyWithSub\stageOmetaColor{x}  ]    \possiblyWithSub\stageOmetaColor{N^{\superscriptO} }_{{\mathrm{1}}}  \closeO{\} }  \punctO{,}    [  \possiblyWithSub\stageOmetaColor{N^{\superscriptO} }_{{\mathrm{0}}}  /  \possiblyWithSub\stageOmetaColor{x}  ]    \possiblyWithSub\stageOmetaColor{N^{\superscriptO} }_{{\mathrm{2}}}  \punctO{,}  \possiblyWithSub\stageOmetaColor{c}  \RightAssertParen^{ L }   \longrightarrow^{0}    \LeftAssertParen   \openO{\{} \possiblyWithSub\stageOmetaColor{\nu}  \relO{:}  \possiblyWithSub\stageOmetaColor{B}  \relO{\mid}    [  \possiblyWithSub\stageOmetaColor{N^{\superscriptO} }_{{\mathrm{0}}}  /  \possiblyWithSub\stageOmetaColor{x}  ]    \possiblyWithSub\stageOmetaColor{N^{\superscriptO} }_{{\mathrm{1}}}  \closeO{\} }  \punctO{,}    [  \possiblyWithSub\stageOmetaColor{N^{\superscriptO} }_{{\mathrm{0}}}  /  \possiblyWithSub\stageOmetaColor{x}  ]    \possiblyWithSub\stageOmetaColor{\Hat{N}^{\superscriptO} }_{{\mathrm{2}}}  \punctO{,}  \possiblyWithSub\stageOmetaColor{c}  \RightAssertParen^{ L }   
                  }.
                \end{center}
              \item Case \derive[E0-RfnPass]{}{%
                  \LeftAssertParen   \openO{\{} \possiblyWithSub\stageOmetaColor{\nu}  \relO{:}  \possiblyWithSub\stageOmetaColor{B}  \relO{\mid}    [  \possiblyWithSub\stageOmetaColor{N'^{\superscriptO} }_{{\mathrm{0}}}  /  \possiblyWithSub\stageOmetaColor{x}  ]    \possiblyWithSub\stageOmetaColor{N^{\superscriptO} }_{{\mathrm{1}}}  \closeO{\} }  \punctO{,}     \ttO{true}    \punctO{,}  \possiblyWithSub\stageOmetaColor{c}_{{\mathrm{2}}}  \RightAssertParen^{ L }   \longrightarrow^{0}     \possiblyWithSub\stageOmetaColor{c}_{{\mathrm{2}}}    
              }:
                Since \(  [  \possiblyWithSub\stageOmetaColor{N'^{\superscriptO} }_{{\mathrm{0}}}  /  \possiblyWithSub\stageOmetaColor{x}  ]    \possiblyWithSub\stageOmetaColor{N^{\superscriptO} }_{{\mathrm{2}}}  =   \ttO{true}  \) holds,
                we have one of the following:
                \begin{itemize}
                  \item Case~\(\possiblyWithSub\stageOmetaColor{N^{\superscriptO} }_{{\mathrm{2}}} = \possiblyWithSub\stageOmetaColor{x}\) and \(\possiblyWithSub\stageOmetaColor{N'^{\superscriptO} }_{{\mathrm{0}}} =   \ttO{true}  \):
                    We will show that \(\possiblyWithSub\stageOmetaColor{\Hat{N}^{\superscriptO} } \defeq  \possiblyWithSub\stageOmetaColor{c}_{{\mathrm{2}}} \) satisfies the desired properties.
                    Indeed, we first have \(  [  \possiblyWithSub\stageOmetaColor{N'^{\superscriptO} }_{{\mathrm{0}}}  /  \possiblyWithSub\stageOmetaColor{x}  ]    \possiblyWithSub\stageOmetaColor{\Hat{N}^{\superscriptO} }  =  \possiblyWithSub\stageOmetaColor{c}_{{\mathrm{2}}}  = \possiblyWithSub\stageOmetaColor{N'^{\superscriptO} }\).
                    Also, by the assumption~\( \possiblyWithSub\stageOmetaColor{N^{\superscriptO} }_{{\mathrm{0}}}  \longrightarrow^{0}   \possiblyWithSub\stageOmetaColor{N'^{\superscriptO} }_{{\mathrm{0}}}   =   \ttO{true}  \),
                    we can derive
                    \(   [  \possiblyWithSub\stageOmetaColor{N^{\superscriptO} }_{{\mathrm{0}}}  /  \possiblyWithSub\stageOmetaColor{x}  ]    \possiblyWithSub\stageOmetaColor{N^{\superscriptO} }   \longrightarrow^{0}    \LeftAssertParen   \openO{\{} \possiblyWithSub\stageOmetaColor{\nu}  \relO{:}  \possiblyWithSub\stageOmetaColor{B}  \relO{\mid}    [  \possiblyWithSub\stageOmetaColor{N^{\superscriptO} }_{{\mathrm{0}}}  /  \possiblyWithSub\stageOmetaColor{x}  ]    \possiblyWithSub\stageOmetaColor{N^{\superscriptO} }_{{\mathrm{1}}}  \closeO{\} }  \punctO{,}     \ttO{true}    \punctO{,}  \possiblyWithSub\stageOmetaColor{c}_{{\mathrm{2}}}  \RightAssertParen^{ L }   \)
                    as follows:
                    \begin{center}
                      \derive[E0-RfnAct]{%
                         \possiblyWithSub\stageOmetaColor{N^{\superscriptO} }_{{\mathrm{0}}}  \longrightarrow^{0}      \ttO{true}     
                      }{%
                          \LeftAssertParen   \openO{\{} \possiblyWithSub\stageOmetaColor{\nu}  \relO{:}  \possiblyWithSub\stageOmetaColor{B}  \relO{\mid}    [  \possiblyWithSub\stageOmetaColor{N^{\superscriptO} }_{{\mathrm{0}}}  /  \possiblyWithSub\stageOmetaColor{x}  ]    \possiblyWithSub\stageOmetaColor{N^{\superscriptO} }_{{\mathrm{1}}}  \closeO{\} }  \punctO{,}  \possiblyWithSub\stageOmetaColor{N^{\superscriptO} }_{{\mathrm{0}}} \punctO{,}  \possiblyWithSub\stageOmetaColor{c}_{{\mathrm{2}}}  \RightAssertParen^{ L }   \longrightarrow^{0}    \LeftAssertParen   \openO{\{} \possiblyWithSub\stageOmetaColor{\nu}  \relO{:}  \possiblyWithSub\stageOmetaColor{B}  \relO{\mid}    [  \possiblyWithSub\stageOmetaColor{N^{\superscriptO} }_{{\mathrm{0}}}  /  \possiblyWithSub\stageOmetaColor{x}  ]    \possiblyWithSub\stageOmetaColor{N^{\superscriptO} }_{{\mathrm{1}}}  \closeO{\} }  \punctO{,}     \ttO{true}    \punctO{,}  \possiblyWithSub\stageOmetaColor{c}_{{\mathrm{2}}}  \RightAssertParen^{ L }   
                      }.
                    \end{center}
                    In addition, we can immediately derive
                    \begin{center}
                      \derive[E0-RfnPass]{}{%
                          \LeftAssertParen   \openO{\{} \possiblyWithSub\stageOmetaColor{\nu}  \relO{:}  \possiblyWithSub\stageOmetaColor{B}  \relO{\mid}    [  \possiblyWithSub\stageOmetaColor{N^{\superscriptO} }_{{\mathrm{0}}}  /  \possiblyWithSub\stageOmetaColor{x}  ]    \possiblyWithSub\stageOmetaColor{N^{\superscriptO} }_{{\mathrm{1}}}  \closeO{\} }  \punctO{,}     \ttO{true}    \punctO{,}  \possiblyWithSub\stageOmetaColor{c}_{{\mathrm{2}}}  \RightAssertParen^{ L }   \longrightarrow^{0}     \possiblyWithSub\stageOmetaColor{c}_{{\mathrm{2}}}    
                      }.
                    \end{center}
                    Therefore, we finally have \(   [  \possiblyWithSub\stageOmetaColor{N^{\superscriptO} }_{{\mathrm{0}}}  /  \possiblyWithSub\stageOmetaColor{x}  ]    \possiblyWithSub\stageOmetaColor{N^{\superscriptO} }   \longrightarrow^{0}     [  \possiblyWithSub\stageOmetaColor{N^{\superscriptO} }_{{\mathrm{0}}}  /  \possiblyWithSub\stageOmetaColor{x}  ]    \possiblyWithSub\stageOmetaColor{\Hat{N}^{\superscriptO} }   \).
                  \item Case~\(\possiblyWithSub\stageOmetaColor{N^{\superscriptO} }_{{\mathrm{2}}} =   \ttO{true}  \):
                    We can easily show that \(\possiblyWithSub\stageOmetaColor{\Hat{N}^{\superscriptO} } \defeq  \possiblyWithSub\stageOmetaColor{c}_{{\mathrm{2}}} \)
                    satisfies the desired properties as well.
                \end{itemize}
            \end{itemize}
          \item Case~\(\possiblyWithSub\stageOmetaColor{N^{\superscriptO} } =  \LeftAssertParen\openO{\langle} \possiblyWithSub\stageImetaColor{T^{\superscriptI} }_{{\mathrm{1}}} \closeO{\rangle} \relO{\CastArrow} \openO{\langle} \possiblyWithSub\stageImetaColor{T^{\superscriptI} }_{{\mathrm{2}}} \closeO{\rangle}\RightAssertParen^{ L } \):
            We have one of the following cases:
            \begin{itemize}
              \item Case where \(  [  \possiblyWithSub\stageOmetaColor{N'^{\superscriptO} }_{{\mathrm{0}}}  /  \possiblyWithSub\stageOmetaColor{x}  ]    \possiblyWithSub\stageImetaColor{T^{\superscriptI} }_{{\mathrm{1}}} \) is not a type value:
                We can uniquely trace back the derivation of
                \(   [  \possiblyWithSub\stageOmetaColor{N'^{\superscriptO} }_{{\mathrm{0}}}  /  \possiblyWithSub\stageOmetaColor{x}  ]    \possiblyWithSub\stageOmetaColor{N^{\superscriptO} }   \longrightarrow^{0}   \possiblyWithSub\stageOmetaColor{N'^{\superscriptO} }  \) as follows:
                \begin{center}
                  \derive[E0-Ass1]{%
                       [  \possiblyWithSub\stageOmetaColor{N'^{\superscriptO} }_{{\mathrm{0}}}  /  \possiblyWithSub\stageOmetaColor{x}  ]    \possiblyWithSub\stageImetaColor{T^{\superscriptI} }_{{\mathrm{1}}}   \longrightarrow^{1}   \possiblyWithSub\stageImetaColor{T'^{\superscriptI} }_{{\mathrm{1}}}  
                  }{%
                      \LeftAssertParen\openO{\langle}   [  \possiblyWithSub\stageOmetaColor{N'^{\superscriptO} }_{{\mathrm{0}}}  /  \possiblyWithSub\stageOmetaColor{x}  ]    \possiblyWithSub\stageImetaColor{T^{\superscriptI} }_{{\mathrm{1}}}  \closeO{\rangle} \relO{\CastArrow} \openO{\langle}   [  \possiblyWithSub\stageOmetaColor{N'^{\superscriptO} }_{{\mathrm{0}}}  /  \possiblyWithSub\stageOmetaColor{x}  ]    \possiblyWithSub\stageImetaColor{T^{\superscriptI} }_{{\mathrm{2}}}  \closeO{\rangle}\RightAssertParen^{ L }   \longrightarrow^{0}    \LeftAssertParen\openO{\langle} \possiblyWithSub\stageImetaColor{T'^{\superscriptI} }_{{\mathrm{1}}} \closeO{\rangle} \relO{\CastArrow} \openO{\langle}   [  \possiblyWithSub\stageOmetaColor{N'^{\superscriptO} }_{{\mathrm{0}}}  /  \possiblyWithSub\stageOmetaColor{x}  ]    \possiblyWithSub\stageImetaColor{T^{\superscriptI} }_{{\mathrm{2}}}  \closeO{\rangle}\RightAssertParen^{ L }   
                  }.
                \end{center}
                By IH, there exists \(\possiblyWithSub\stageImetaColor{\Hat{T}^{\superscriptI} }_{{\mathrm{1}}}\) such that
                \(   [  \possiblyWithSub\stageOmetaColor{N^{\superscriptO} }_{{\mathrm{0}}}  /  \possiblyWithSub\stageOmetaColor{x}  ]    \possiblyWithSub\stageImetaColor{T^{\superscriptI} }_{{\mathrm{1}}}   \longrightarrow^{1\,\ast}     [  \possiblyWithSub\stageOmetaColor{N^{\superscriptO} }_{{\mathrm{0}}}  /  \possiblyWithSub\stageOmetaColor{x}  ]    \possiblyWithSub\stageImetaColor{\Hat{T}^{\superscriptI} }_{{\mathrm{1}}}   \)
                and \(\possiblyWithSub\stageImetaColor{T'^{\superscriptI} }_{{\mathrm{1}}} =   [  \possiblyWithSub\stageOmetaColor{N'^{\superscriptO} }_{{\mathrm{0}}}  /  \possiblyWithSub\stageOmetaColor{x}  ]    \possiblyWithSub\stageImetaColor{\Hat{T}^{\superscriptI} }_{{\mathrm{1}}} \).
                Thus, \(\possiblyWithSub\stageOmetaColor{\Hat{N}^{\superscriptO} } \defeq  \LeftAssertParen\openO{\langle} \possiblyWithSub\stageImetaColor{\Hat{T}^{\superscriptI} }_{{\mathrm{1}}} \closeO{\rangle} \relO{\CastArrow} \openO{\langle} \possiblyWithSub\stageImetaColor{T^{\superscriptI} }_{{\mathrm{2}}} \closeO{\rangle}\RightAssertParen^{ L } \) satisfies the desired properties
                by the repeated use of \rulename{E0-Ass1}.
              \item Case where \(  [  \possiblyWithSub\stageOmetaColor{N'^{\superscriptO} }_{{\mathrm{0}}}  /  \possiblyWithSub\stageOmetaColor{x}  ]    \possiblyWithSub\stageImetaColor{T^{\superscriptI} }_{{\mathrm{1}}}  \revdefeq \possiblyWithSub\stageImetaColor{\tau^{\superscriptI} }_{{\mathrm{1}}}\) is a type value
              while \(  [  \possiblyWithSub\stageOmetaColor{N'^{\superscriptO} }_{{\mathrm{0}}}  /  \possiblyWithSub\stageOmetaColor{x}  ]    \possiblyWithSub\stageImetaColor{T^{\superscriptI} }_{{\mathrm{2}}} \) is not:
                The sole possible derivation of \(   [  \possiblyWithSub\stageOmetaColor{N'^{\superscriptO} }_{{\mathrm{0}}}  /  \possiblyWithSub\stageOmetaColor{x}  ]    \possiblyWithSub\stageOmetaColor{N^{\superscriptO} }   \longrightarrow^{0}   \possiblyWithSub\stageOmetaColor{N'^{\superscriptO} }  \) is as follows:
                \begin{center}
                  \derive[E0-Ass2]{%
                       [  \possiblyWithSub\stageOmetaColor{N'^{\superscriptO} }_{{\mathrm{0}}}  /  \possiblyWithSub\stageOmetaColor{x}  ]    \possiblyWithSub\stageImetaColor{T^{\superscriptI} }_{{\mathrm{2}}}   \longrightarrow^{1}   \possiblyWithSub\stageImetaColor{T'^{\superscriptI} }_{{\mathrm{2}}}  
                  }{%
                      \LeftAssertParen\openO{\langle}  \possiblyWithSub\stageImetaColor{\tau^{\superscriptI} }_{{\mathrm{1}}}  \closeO{\rangle} \relO{\CastArrow} \openO{\langle}   [  \possiblyWithSub\stageOmetaColor{N'^{\superscriptO} }_{{\mathrm{0}}}  /  \possiblyWithSub\stageOmetaColor{x}  ]    \possiblyWithSub\stageImetaColor{T^{\superscriptI} }_{{\mathrm{2}}}  \closeO{\rangle}\RightAssertParen^{ L }   \longrightarrow^{0}    \LeftAssertParen\openO{\langle}  \possiblyWithSub\stageImetaColor{\tau^{\superscriptI} }_{{\mathrm{1}}}  \closeO{\rangle} \relO{\CastArrow} \openO{\langle} \possiblyWithSub\stageImetaColor{T'^{\superscriptI} }_{{\mathrm{2}}} \closeO{\rangle}\RightAssertParen^{ L }   
                  }.
                \end{center}
                By IH, there exists \(\possiblyWithSub\stageImetaColor{\Hat{T}^{\superscriptI} }_{{\mathrm{2}}}\) such that
                \(   [  \possiblyWithSub\stageOmetaColor{N^{\superscriptO} }_{{\mathrm{0}}}  /  \possiblyWithSub\stageOmetaColor{x}  ]    \possiblyWithSub\stageImetaColor{T^{\superscriptI} }_{{\mathrm{2}}}   \longrightarrow^{1\,\ast}     [  \possiblyWithSub\stageOmetaColor{N^{\superscriptO} }_{{\mathrm{0}}}  /  \possiblyWithSub\stageOmetaColor{x}  ]    \possiblyWithSub\stageImetaColor{\Hat{T}^{\superscriptI} }_{{\mathrm{2}}}   \)
                and \(\possiblyWithSub\stageImetaColor{T'^{\superscriptI} }_{{\mathrm{2}}} =   [  \possiblyWithSub\stageOmetaColor{N'^{\superscriptO} }_{{\mathrm{0}}}  /  \possiblyWithSub\stageOmetaColor{x}  ]    \possiblyWithSub\stageImetaColor{\Hat{T}^{\superscriptI} }_{{\mathrm{2}}} \).
                Thus, \(\possiblyWithSub\stageOmetaColor{\Hat{N}^{\superscriptO} } \defeq  \LeftAssertParen\openO{\langle} \possiblyWithSub\stageImetaColor{T^{\superscriptI} }_{{\mathrm{1}}} \closeO{\rangle} \relO{\CastArrow} \openO{\langle} \possiblyWithSub\stageImetaColor{\Hat{T}^{\superscriptI} }_{{\mathrm{2}}} \closeO{\rangle}\RightAssertParen^{ L } \)
                clearly satisfies the desired properties
                by the repeated use of \rulename{E0-Ass2}.
              \item Case where both \(  [  \possiblyWithSub\stageOmetaColor{N'^{\superscriptO} }_{{\mathrm{0}}}  /  \possiblyWithSub\stageOmetaColor{x}  ]    \possiblyWithSub\stageImetaColor{T^{\superscriptI} }_{{\mathrm{1}}} \)
              and \(  [  \possiblyWithSub\stageOmetaColor{N'^{\superscriptO} }_{{\mathrm{0}}}  /  \possiblyWithSub\stageOmetaColor{x}  ]    \possiblyWithSub\stageImetaColor{T^{\superscriptI} }_{{\mathrm{2}}} \) are type values:
                We can uniquely track back the derivation of
                \(   [  \possiblyWithSub\stageOmetaColor{N'^{\superscriptO} }_{{\mathrm{0}}}  /  \possiblyWithSub\stageOmetaColor{x}  ]    \possiblyWithSub\stageOmetaColor{N^{\superscriptO} }   \longrightarrow^{0}   \possiblyWithSub\stageOmetaColor{N'^{\superscriptO} }  \) as follows:
                \begin{center}
                  \derive[E0-AssPass]{}{%
                      \LeftAssertParen\openO{\langle}  \possiblyWithSub\stageImetaColor{\tau^{\superscriptI} }  \closeO{\rangle} \relO{\CastArrow} \openO{\langle}  \possiblyWithSub\stageImetaColor{\tau^{\superscriptI} }  \closeO{\rangle}\RightAssertParen^{ L }   \longrightarrow^{0}    \ordO{\lambda} \possiblyWithSub\stageOmetaColor{x'}  \relO{:}   \openO{\langle}  \possiblyWithSub\stageImetaColor{\tau^{\superscriptI} }  \closeO{\rangle}  \punctO{.}\   \possiblyWithSub\stageOmetaColor{x'}    
                  }.
                \end{center}
                Thus, \(  [  \possiblyWithSub\stageOmetaColor{N'^{\superscriptO} }_{{\mathrm{0}}}  /  \possiblyWithSub\stageOmetaColor{x}  ]    \possiblyWithSub\stageImetaColor{T^{\superscriptI} }_{{\mathrm{1}}}  =   [  \possiblyWithSub\stageOmetaColor{N'^{\superscriptO} }_{{\mathrm{0}}}  /  \possiblyWithSub\stageOmetaColor{x}  ]    \possiblyWithSub\stageImetaColor{T^{\superscriptI} }_{{\mathrm{2}}}  = \possiblyWithSub\stageImetaColor{\tau^{\superscriptI} }\) holds.
                We will show that \(\possiblyWithSub\stageOmetaColor{\Hat{N}^{\superscriptO} } \defeq  \openO{(}  \ordO{\lambda} \possiblyWithSub\stageOmetaColor{x'}  \relO{:}   \openO{\langle}  \possiblyWithSub\stageImetaColor{\tau^{\superscriptI} }  \closeO{\rangle}  \punctO{.}\   \possiblyWithSub\stageOmetaColor{x'}   \closeO{)} \)
                satisfies the desired properties.
                Indeed, we first have
                \(  [  \possiblyWithSub\stageOmetaColor{N'^{\superscriptO} }_{{\mathrm{0}}}  /  \possiblyWithSub\stageOmetaColor{x}  ]    \possiblyWithSub\stageOmetaColor{\Hat{N}^{\superscriptO} }  =  \openO{(}  \ordO{\lambda} \possiblyWithSub\stageOmetaColor{x'}  \relO{:}   \openO{\langle}  \possiblyWithSub\stageImetaColor{\tau^{\superscriptI} }  \closeO{\rangle}  \punctO{.}\   \possiblyWithSub\stageOmetaColor{x'}   \closeO{)}  = \possiblyWithSub\stageOmetaColor{N'^{\superscriptO} }\).
                Also, by using Lemma~\ref{lem:weak-bisimulation-right-pre} twice, we have
                (1X)~\(   [  \possiblyWithSub\stageOmetaColor{N^{\superscriptO} }_{{\mathrm{0}}}  /  \possiblyWithSub\stageOmetaColor{x}  ]    \possiblyWithSub\stageImetaColor{T^{\superscriptI} }_{{\mathrm{1}}}   \longrightarrow^{1\,\ast}    \possiblyWithSub\stageImetaColor{\tau^{\superscriptI} }   \) and
                (2X)~\(   [  \possiblyWithSub\stageOmetaColor{N^{\superscriptO} }_{{\mathrm{0}}}  /  \possiblyWithSub\stageOmetaColor{x}  ]    \possiblyWithSub\stageImetaColor{T^{\superscriptI} }_{{\mathrm{2}}}   \longrightarrow^{1\,\ast}    \possiblyWithSub\stageImetaColor{\tau^{\superscriptI} }   \).
                Then, by (1X) and (2X), we can derive
                \(   [  \possiblyWithSub\stageOmetaColor{N^{\superscriptO} }_{{\mathrm{0}}}  /  \possiblyWithSub\stageOmetaColor{x}  ]     \LeftAssertParen\openO{\langle} \possiblyWithSub\stageImetaColor{T^{\superscriptI} }_{{\mathrm{1}}} \closeO{\rangle} \relO{\CastArrow} \openO{\langle} \possiblyWithSub\stageImetaColor{T^{\superscriptI} }_{{\mathrm{2}}} \closeO{\rangle}\RightAssertParen^{ L }    \longrightarrow^{0\,\ast}    \LeftAssertParen\openO{\langle}  \possiblyWithSub\stageImetaColor{\tau^{\superscriptI} }  \closeO{\rangle} \relO{\CastArrow} \openO{\langle}  \possiblyWithSub\stageImetaColor{\tau^{\superscriptI} }  \closeO{\rangle}\RightAssertParen^{ L }   \),
                using \rulename{E0-Ass1} and \rulename{E0-Ass2} repeatedly.
                In addition, we can derive
                \begin{center}
                  \derive[E0-AssPass]{}{%
                      \LeftAssertParen\openO{\langle}  \possiblyWithSub\stageImetaColor{\tau^{\superscriptI} }  \closeO{\rangle} \relO{\CastArrow} \openO{\langle}  \possiblyWithSub\stageImetaColor{\tau^{\superscriptI} }  \closeO{\rangle}\RightAssertParen^{ L }   \longrightarrow^{0}    \ordO{\lambda} \possiblyWithSub\stageOmetaColor{x'}  \relO{:}   \openO{\langle}  \possiblyWithSub\stageImetaColor{\tau^{\superscriptI} }  \closeO{\rangle}  \punctO{.}\   \possiblyWithSub\stageOmetaColor{x'}    
                  }.
                \end{center}
                Therefore, we finally have
                \(   [  \possiblyWithSub\stageOmetaColor{N^{\superscriptO} }_{{\mathrm{0}}}  /  \possiblyWithSub\stageOmetaColor{x}  ]     \LeftAssertParen\openO{\langle} \possiblyWithSub\stageImetaColor{T^{\superscriptI} }_{{\mathrm{1}}} \closeO{\rangle} \relO{\CastArrow} \openO{\langle} \possiblyWithSub\stageImetaColor{T^{\superscriptI} }_{{\mathrm{2}}} \closeO{\rangle}\RightAssertParen^{ L }    \longrightarrow^{0\,\ast}     [  \possiblyWithSub\stageOmetaColor{N^{\superscriptO} }_{{\mathrm{0}}}  /  \possiblyWithSub\stageOmetaColor{x}  ]     \openO{(}  \ordO{\lambda} \possiblyWithSub\stageOmetaColor{x'}  \relO{:}   \openO{\langle}  \possiblyWithSub\stageImetaColor{\tau^{\superscriptI} }  \closeO{\rangle}  \punctO{.}\   \possiblyWithSub\stageOmetaColor{x'}   \closeO{)}    \).
            \end{itemize}
        \end{itemize}
      \item
        \begin{itemize}
          \item Case where \(\possiblyWithSub\stageImetaColor{N^{\superscriptI} } = \possiblyWithSub\stageImetaColor{c}\) or \(\possiblyWithSub\stageImetaColor{N^{\superscriptI} } = \possiblyWithSub\stageImetaColor{x'}\):
            This contradicts the assumption~\(   [  \possiblyWithSub\stageOmetaColor{N'^{\superscriptO} }_{{\mathrm{0}}}  /  \possiblyWithSub\stageOmetaColor{x}  ]    \possiblyWithSub\stageImetaColor{N^{\superscriptI} }   \longrightarrow^{1}   \possiblyWithSub\stageImetaColor{N'^{\superscriptI} }  \).
          \item Case~\(\possiblyWithSub\stageImetaColor{N^{\superscriptI} } =  \openI{(}  \ordI{\lambda} \possiblyWithSub\stageImetaColor{x'}  \relI{:}  \possiblyWithSub\stageImetaColor{T^{\superscriptI} }_{{\mathrm{1}}} \punctI{.}\  \possiblyWithSub\stageImetaColor{N^{\superscriptI} }_{{\mathrm{2}}}  \closeI{)} \):
            We have one of the following cases:
            \begin{itemize}
              \item Case where \(  [  \possiblyWithSub\stageOmetaColor{N'^{\superscriptO} }_{{\mathrm{0}}}  /  \possiblyWithSub\stageOmetaColor{x}  ]    \possiblyWithSub\stageImetaColor{T^{\superscriptI} }_{{\mathrm{1}}} \) is not a type value:
                We can uniquely trace back the derivation of
                \(   [  \possiblyWithSub\stageOmetaColor{N'^{\superscriptO} }_{{\mathrm{0}}}  /  \possiblyWithSub\stageOmetaColor{x}  ]    \possiblyWithSub\stageImetaColor{N^{\superscriptI} }   \longrightarrow^{1}   \possiblyWithSub\stageImetaColor{N'^{\superscriptI} }  \) as follows:
                \begin{center}
                  \derive[E1-Abs1]{%
                       [  \possiblyWithSub\stageOmetaColor{N'^{\superscriptO} }_{{\mathrm{0}}}  /  \possiblyWithSub\stageOmetaColor{x}  ]    \possiblyWithSub\stageImetaColor{T^{\superscriptI} }_{{\mathrm{1}}}   \longrightarrow^{1}   \possiblyWithSub\stageImetaColor{T'^{\superscriptI} }_{{\mathrm{1}}}  
                  }{%
                      \ordI{\lambda} \possiblyWithSub\stageImetaColor{x'}  \relI{:}    [  \possiblyWithSub\stageOmetaColor{N'^{\superscriptO} }_{{\mathrm{0}}}  /  \possiblyWithSub\stageOmetaColor{x}  ]    \possiblyWithSub\stageImetaColor{T^{\superscriptI} }_{{\mathrm{1}}}  \punctI{.}\    [  \possiblyWithSub\stageOmetaColor{N'^{\superscriptO} }_{{\mathrm{0}}}  /  \possiblyWithSub\stageOmetaColor{x}  ]    \possiblyWithSub\stageImetaColor{N^{\superscriptI} }_{{\mathrm{2}}}    \longrightarrow^{1}    \ordI{\lambda} \possiblyWithSub\stageImetaColor{x'}  \relI{:}  \possiblyWithSub\stageImetaColor{T'^{\superscriptI} }_{{\mathrm{1}}} \punctI{.}\    [  \possiblyWithSub\stageOmetaColor{N'^{\superscriptO} }_{{\mathrm{0}}}  /  \possiblyWithSub\stageOmetaColor{x}  ]    \possiblyWithSub\stageImetaColor{N^{\superscriptI} }_{{\mathrm{2}}}    
                  }.
                \end{center}
                By IH, there exists \(\possiblyWithSub\stageImetaColor{\Hat{T}^{\superscriptI} }_{{\mathrm{1}}}\) such that
                \(   [  \possiblyWithSub\stageOmetaColor{N^{\superscriptO} }_{{\mathrm{0}}}  /  \possiblyWithSub\stageOmetaColor{x}  ]    \possiblyWithSub\stageImetaColor{T^{\superscriptI} }_{{\mathrm{1}}}   \longrightarrow^{1\,\ast}     [  \possiblyWithSub\stageOmetaColor{N^{\superscriptO} }_{{\mathrm{0}}}  /  \possiblyWithSub\stageOmetaColor{x}  ]    \possiblyWithSub\stageImetaColor{\Hat{T}^{\superscriptI} }_{{\mathrm{1}}}   \)
                and \(\possiblyWithSub\stageImetaColor{T'^{\superscriptI} }_{{\mathrm{1}}} =   [  \possiblyWithSub\stageOmetaColor{N'^{\superscriptO} }_{{\mathrm{0}}}  /  \possiblyWithSub\stageOmetaColor{x}  ]    \possiblyWithSub\stageImetaColor{\Hat{T}^{\superscriptI} }_{{\mathrm{1}}} \).
                Thus, \(\possiblyWithSub\stageImetaColor{\Hat{N}^{\superscriptI} } \defeq  \openI{(}  \ordI{\lambda} \possiblyWithSub\stageImetaColor{x'}  \relI{:}  \possiblyWithSub\stageImetaColor{\Hat{T}^{\superscriptI} }_{{\mathrm{1}}} \punctI{.}\  \possiblyWithSub\stageImetaColor{N^{\superscriptI} }_{{\mathrm{2}}}  \closeI{)} \) clearly satisfies the desired properties
                by the repeated use of \rulename{E1-Abs1}.
              \item Case where \(  [  \possiblyWithSub\stageOmetaColor{N'^{\superscriptO} }_{{\mathrm{0}}}  /  \possiblyWithSub\stageOmetaColor{x}  ]    \possiblyWithSub\stageImetaColor{T^{\superscriptI} }_{{\mathrm{1}}}  \revdefeq \possiblyWithSub\stageImetaColor{\tau^{\superscriptI} }_{{\mathrm{1}}}\) is a type value
              while \(  [  \possiblyWithSub\stageOmetaColor{N'^{\superscriptO} }_{{\mathrm{0}}}  /  \possiblyWithSub\stageOmetaColor{x}  ]    \possiblyWithSub\stageImetaColor{N^{\superscriptI} }_{{\mathrm{2}}} \) is not a stage-\(1\) value:
                The sole possible derivation of \(   [  \possiblyWithSub\stageOmetaColor{N'^{\superscriptO} }_{{\mathrm{0}}}  /  \possiblyWithSub\stageOmetaColor{x}  ]    \possiblyWithSub\stageImetaColor{N^{\superscriptI} }   \longrightarrow^{1}   \possiblyWithSub\stageImetaColor{N'^{\superscriptI} }  \) is as follows:
                \begin{center}
                  \derive[E1-Abs2]{%
                       [  \possiblyWithSub\stageOmetaColor{N'^{\superscriptO} }_{{\mathrm{0}}}  /  \possiblyWithSub\stageOmetaColor{x}  ]    \possiblyWithSub\stageImetaColor{N^{\superscriptI} }_{{\mathrm{2}}}   \longrightarrow^{1}   \possiblyWithSub\stageImetaColor{N'^{\superscriptI} }_{{\mathrm{2}}}  
                  }{%
                      \ordI{\lambda} \possiblyWithSub\stageImetaColor{x'}  \relI{:}   \possiblyWithSub\stageImetaColor{\tau^{\superscriptI} }_{{\mathrm{1}}}  \punctI{.}\    [  \possiblyWithSub\stageOmetaColor{N'^{\superscriptO} }_{{\mathrm{0}}}  /  \possiblyWithSub\stageOmetaColor{x}  ]    \possiblyWithSub\stageImetaColor{N^{\superscriptI} }_{{\mathrm{2}}}    \longrightarrow^{1}    \ordI{\lambda} \possiblyWithSub\stageImetaColor{x'}  \relI{:}   \possiblyWithSub\stageImetaColor{\tau^{\superscriptI} }_{{\mathrm{1}}}  \punctI{.}\  \possiblyWithSub\stageImetaColor{N'^{\superscriptI} }_{{\mathrm{2}}}   
                  }.
                \end{center}
                By IH, there exists \(\possiblyWithSub\stageImetaColor{\Hat{N}^{\superscriptI} }_{{\mathrm{2}}}\) such that
                \(   [  \possiblyWithSub\stageOmetaColor{N^{\superscriptO} }_{{\mathrm{0}}}  /  \possiblyWithSub\stageOmetaColor{x}  ]    \possiblyWithSub\stageImetaColor{N^{\superscriptI} }_{{\mathrm{2}}}   \longrightarrow^{1\,\ast}     [  \possiblyWithSub\stageOmetaColor{N^{\superscriptO} }_{{\mathrm{0}}}  /  \possiblyWithSub\stageOmetaColor{x}  ]    \possiblyWithSub\stageImetaColor{\Hat{N}^{\superscriptI} }_{{\mathrm{2}}}   \)
                and \(\possiblyWithSub\stageImetaColor{N'^{\superscriptI} }_{{\mathrm{2}}} =   [  \possiblyWithSub\stageOmetaColor{N'^{\superscriptO} }_{{\mathrm{0}}}  /  \possiblyWithSub\stageOmetaColor{x}  ]    \possiblyWithSub\stageImetaColor{\Hat{N}^{\superscriptI} }_{{\mathrm{2}}} \).
                Thus, \(\possiblyWithSub\stageImetaColor{\Hat{N}^{\superscriptI} } \defeq  \openI{(}  \ordI{\lambda} \possiblyWithSub\stageImetaColor{x'}  \relI{:}  \possiblyWithSub\stageImetaColor{T^{\superscriptI} }_{{\mathrm{1}}} \punctI{.}\  \possiblyWithSub\stageImetaColor{N'^{\superscriptI} }_{{\mathrm{2}}}  \closeI{)} \) clearly satisfies the desired properties
                by the repeated use of \rulename{E1-Abs2}.
              \item Case where \(  [  \possiblyWithSub\stageOmetaColor{N'^{\superscriptO} }_{{\mathrm{0}}}  /  \possiblyWithSub\stageOmetaColor{x}  ]    \possiblyWithSub\stageImetaColor{T^{\superscriptI} }_{{\mathrm{1}}} \) is a type value
              and \(  [  \possiblyWithSub\stageOmetaColor{N'^{\superscriptO} }_{{\mathrm{0}}}  /  \possiblyWithSub\stageOmetaColor{x}  ]    \possiblyWithSub\stageImetaColor{N^{\superscriptI} }_{{\mathrm{2}}} \) is a stage-\(1\) value:
                This case contradicts the assumption~\(   [  \possiblyWithSub\stageOmetaColor{N'^{\superscriptO} }_{{\mathrm{0}}}  /  \possiblyWithSub\stageOmetaColor{x}  ]    \possiblyWithSub\stageImetaColor{N^{\superscriptI} }   \longrightarrow^{1}   \possiblyWithSub\stageImetaColor{N'^{\superscriptI} }  \).
            \end{itemize}
          \item Case~\(\possiblyWithSub\stageImetaColor{N^{\superscriptI} } =  \possiblyWithSub\stageImetaColor{N^{\superscriptI} }_{{\mathrm{1}}} \  \possiblyWithSub\stageImetaColor{N^{\superscriptI} }_{{\mathrm{2}}} \):
            We have one of the following cases:
            \begin{itemize}
              \item Case where \(  [  \possiblyWithSub\stageOmetaColor{N'^{\superscriptO} }_{{\mathrm{0}}}  /  \possiblyWithSub\stageOmetaColor{x}  ]    \possiblyWithSub\stageImetaColor{N^{\superscriptI} }_{{\mathrm{1}}} \) is not a stage-\(1\) value:
                We can uniquely trace back the derivation of
                \(   [  \possiblyWithSub\stageOmetaColor{N'^{\superscriptO} }_{{\mathrm{0}}}  /  \possiblyWithSub\stageOmetaColor{x}  ]    \possiblyWithSub\stageImetaColor{N^{\superscriptI} }   \longrightarrow^{1}   \possiblyWithSub\stageImetaColor{N'^{\superscriptI} }  \) as follows:
                \begin{center}
                  \derive[E1-App1]{%
                       [  \possiblyWithSub\stageOmetaColor{N'^{\superscriptO} }_{{\mathrm{0}}}  /  \possiblyWithSub\stageOmetaColor{x}  ]    \possiblyWithSub\stageImetaColor{N^{\superscriptI} }_{{\mathrm{1}}}   \longrightarrow^{1}   \possiblyWithSub\stageImetaColor{N'^{\superscriptI} }_{{\mathrm{1}}}  
                  }{%
                       \openI{(}   [  \possiblyWithSub\stageOmetaColor{N'^{\superscriptO} }_{{\mathrm{0}}}  /  \possiblyWithSub\stageOmetaColor{x}  ]    \possiblyWithSub\stageImetaColor{N^{\superscriptI} }_{{\mathrm{1}}}  \closeI{)}  \   \openI{(}   [  \possiblyWithSub\stageOmetaColor{N'^{\superscriptO} }_{{\mathrm{0}}}  /  \possiblyWithSub\stageOmetaColor{x}  ]    \possiblyWithSub\stageImetaColor{N^{\superscriptI} }_{{\mathrm{2}}}  \closeI{)}    \longrightarrow^{1}    \possiblyWithSub\stageImetaColor{N'^{\superscriptI} }_{{\mathrm{1}}} \   \openI{(}   [  \possiblyWithSub\stageOmetaColor{N'^{\superscriptO} }_{{\mathrm{0}}}  /  \possiblyWithSub\stageOmetaColor{x}  ]    \possiblyWithSub\stageImetaColor{N^{\superscriptI} }_{{\mathrm{2}}}  \closeI{)}    
                  }.
                \end{center}
                By IH, there exists \(\possiblyWithSub\stageImetaColor{\Hat{N}^{\superscriptI} }_{{\mathrm{1}}}\) such that
                \(   [  \possiblyWithSub\stageOmetaColor{N^{\superscriptO} }_{{\mathrm{0}}}  /  \possiblyWithSub\stageOmetaColor{x}  ]    \possiblyWithSub\stageImetaColor{N^{\superscriptI} }_{{\mathrm{1}}}   \longrightarrow^{1\,\ast}     [  \possiblyWithSub\stageOmetaColor{N^{\superscriptO} }_{{\mathrm{0}}}  /  \possiblyWithSub\stageOmetaColor{x}  ]    \possiblyWithSub\stageImetaColor{\Hat{N}^{\superscriptI} }_{{\mathrm{1}}}   \)
                and \(\possiblyWithSub\stageImetaColor{N'^{\superscriptI} }_{{\mathrm{1}}} =   [  \possiblyWithSub\stageOmetaColor{N'^{\superscriptO} }_{{\mathrm{0}}}  /  \possiblyWithSub\stageOmetaColor{x}  ]    \possiblyWithSub\stageImetaColor{\Hat{N}^{\superscriptI} }_{{\mathrm{1}}} \).
                Thus, \(\possiblyWithSub\stageImetaColor{\Hat{N}^{\superscriptI} } \defeq  \possiblyWithSub\stageImetaColor{\Hat{N}^{\superscriptI} }_{{\mathrm{1}}} \  \possiblyWithSub\stageImetaColor{N^{\superscriptI} }_{{\mathrm{2}}} \) satisfies the desired properties
                by the repeated use of \rulename{E1-App1}.
              \item Case where \(  [  \possiblyWithSub\stageOmetaColor{N'^{\superscriptO} }_{{\mathrm{0}}}  /  \possiblyWithSub\stageOmetaColor{x}  ]    \possiblyWithSub\stageImetaColor{N^{\superscriptI} }_{{\mathrm{1}}}  \revdefeq \possiblyWithSub\stageImetaColor{v^{\superscriptI} }_{{\mathrm{1}}}\) is a stage-\(1\) value
              while \(  [  \possiblyWithSub\stageOmetaColor{N'^{\superscriptO} }_{{\mathrm{0}}}  /  \possiblyWithSub\stageOmetaColor{x}  ]    \possiblyWithSub\stageImetaColor{N^{\superscriptI} }_{{\mathrm{2}}} \) is not:
                The sole possible derivation of \(   [  \possiblyWithSub\stageOmetaColor{N'^{\superscriptO} }_{{\mathrm{0}}}  /  \possiblyWithSub\stageOmetaColor{x}  ]    \possiblyWithSub\stageImetaColor{N^{\superscriptI} }   \longrightarrow^{1}   \possiblyWithSub\stageImetaColor{N'^{\superscriptI} }  \) is as follows:
                \begin{center}
                  \derive[E1-App2]{%
                       [  \possiblyWithSub\stageOmetaColor{N'^{\superscriptO} }_{{\mathrm{0}}}  /  \possiblyWithSub\stageOmetaColor{x}  ]    \possiblyWithSub\stageImetaColor{N^{\superscriptI} }_{{\mathrm{2}}}   \longrightarrow^{1}   \possiblyWithSub\stageImetaColor{N'^{\superscriptI} }_{{\mathrm{2}}}  
                  }{%
                       \possiblyWithSub\stageImetaColor{v^{\superscriptI} }_{{\mathrm{1}}}  \   \openI{(}   [  \possiblyWithSub\stageOmetaColor{N'^{\superscriptO} }_{{\mathrm{0}}}  /  \possiblyWithSub\stageOmetaColor{x}  ]    \possiblyWithSub\stageImetaColor{N^{\superscriptI} }_{{\mathrm{2}}}  \closeI{)}    \longrightarrow^{1}     \possiblyWithSub\stageImetaColor{v^{\superscriptI} }_{{\mathrm{1}}}  \  \possiblyWithSub\stageImetaColor{N'^{\superscriptI} }_{{\mathrm{2}}}   
                  }.
                \end{center}
                By IH, there exists \(\possiblyWithSub\stageImetaColor{\Hat{N}^{\superscriptI} }_{{\mathrm{2}}}\) such that
                \(   [  \possiblyWithSub\stageOmetaColor{N^{\superscriptO} }_{{\mathrm{0}}}  /  \possiblyWithSub\stageOmetaColor{x}  ]    \possiblyWithSub\stageImetaColor{N^{\superscriptI} }_{{\mathrm{2}}}   \longrightarrow^{1\,\ast}     [  \possiblyWithSub\stageOmetaColor{N^{\superscriptO} }_{{\mathrm{0}}}  /  \possiblyWithSub\stageOmetaColor{x}  ]    \possiblyWithSub\stageImetaColor{\Hat{N}^{\superscriptI} }_{{\mathrm{2}}}   \)
                and \(\possiblyWithSub\stageImetaColor{N'^{\superscriptI} }_{{\mathrm{2}}} =   [  \possiblyWithSub\stageOmetaColor{N'^{\superscriptO} }_{{\mathrm{0}}}  /  \possiblyWithSub\stageOmetaColor{x}  ]    \possiblyWithSub\stageImetaColor{\Hat{N}^{\superscriptI} }_{{\mathrm{2}}} \).
                Thus, \(\possiblyWithSub\stageImetaColor{\Hat{N}^{\superscriptI} } \defeq  \possiblyWithSub\stageImetaColor{N^{\superscriptI} }_{{\mathrm{1}}} \  \possiblyWithSub\stageImetaColor{\Hat{N}^{\superscriptI} }_{{\mathrm{2}}} \) clearly satisfies the desired properties
                by the repeated use of \rulename{E1-App2}.
              \item Case where both \(  [  \possiblyWithSub\stageOmetaColor{N'^{\superscriptO} }_{{\mathrm{0}}}  /  \possiblyWithSub\stageOmetaColor{x}  ]    \possiblyWithSub\stageImetaColor{N^{\superscriptI} }_{{\mathrm{1}}} \)
              and \(  [  \possiblyWithSub\stageOmetaColor{N'^{\superscriptO} }_{{\mathrm{0}}}  /  \possiblyWithSub\stageOmetaColor{x}  ]    \possiblyWithSub\stageImetaColor{N^{\superscriptI} }_{{\mathrm{2}}} \) are stage-\(1\) values:
                This case contradicts the assumption~\(   [  \possiblyWithSub\stageOmetaColor{N'^{\superscriptO} }_{{\mathrm{0}}}  /  \possiblyWithSub\stageOmetaColor{x}  ]    \possiblyWithSub\stageImetaColor{N^{\superscriptI} }   \longrightarrow^{1}   \possiblyWithSub\stageImetaColor{N'^{\superscriptI} }  \).
            \end{itemize}
          \item Case~\(\possiblyWithSub\stageImetaColor{N^{\superscriptI} } =  \ordI{\sim} \possiblyWithSub\stageOmetaColor{N^{\superscriptO} }_{{\mathrm{1}}} \):
            By tracing back the derivation of \(   [  \possiblyWithSub\stageOmetaColor{N'^{\superscriptO} }_{{\mathrm{0}}}  /  \possiblyWithSub\stageOmetaColor{x}  ]    \possiblyWithSub\stageImetaColor{N^{\superscriptI} }   \longrightarrow^{1}   \possiblyWithSub\stageImetaColor{N'^{\superscriptI} }  \),
            we have one of the following:
            \begin{itemize}
              \item Case \derive[E1-Esc]{%
                   [  \possiblyWithSub\stageOmetaColor{N'^{\superscriptO} }_{{\mathrm{0}}}  /  \possiblyWithSub\stageOmetaColor{x}  ]    \possiblyWithSub\stageOmetaColor{N^{\superscriptO} }_{{\mathrm{1}}}   \longrightarrow^{0}   \possiblyWithSub\stageOmetaColor{N'^{\superscriptO} }_{{\mathrm{1}}}  
              }{%
                  \ordI{\sim}  \openO{(}   [  \possiblyWithSub\stageOmetaColor{N'^{\superscriptO} }_{{\mathrm{0}}}  /  \possiblyWithSub\stageOmetaColor{x}  ]    \possiblyWithSub\stageOmetaColor{N^{\superscriptO} }_{{\mathrm{1}}}  \closeO{)}    \longrightarrow^{1}    \ordI{\sim} \possiblyWithSub\stageOmetaColor{N'^{\superscriptO} }_{{\mathrm{1}}}   
              }:
                By IH, there exists \(\possiblyWithSub\stageOmetaColor{\Hat{N}^{\superscriptO} }_{{\mathrm{1}}}\) such that
                \(   [  \possiblyWithSub\stageOmetaColor{N^{\superscriptO} }_{{\mathrm{0}}}  /  \possiblyWithSub\stageOmetaColor{x}  ]    \possiblyWithSub\stageOmetaColor{N^{\superscriptO} }_{{\mathrm{1}}}   \longrightarrow^{0\,\ast}     [  \possiblyWithSub\stageOmetaColor{N^{\superscriptO} }_{{\mathrm{0}}}  /  \possiblyWithSub\stageOmetaColor{x}  ]    \possiblyWithSub\stageOmetaColor{\Hat{N}^{\superscriptO} }_{{\mathrm{1}}}   \)
                and \(\possiblyWithSub\stageOmetaColor{N'^{\superscriptO} }_{{\mathrm{1}}} =   [  \possiblyWithSub\stageOmetaColor{N'^{\superscriptO} }_{{\mathrm{0}}}  /  \possiblyWithSub\stageOmetaColor{x}  ]    \possiblyWithSub\stageOmetaColor{\Hat{N}^{\superscriptO} }_{{\mathrm{1}}} \).
                Thus, \(\possiblyWithSub\stageImetaColor{\Hat{N}^{\superscriptI} } \defeq  \ordI{\sim} \possiblyWithSub\stageOmetaColor{\Hat{N}^{\superscriptO} }_{{\mathrm{1}}} \) clearly satisfies the desired properties
                by the repeated use of \rulename{E1-Esc}.
              \item Case \derive[E1-Cancel]{}{%
                  \ordI{\sim}  \openO{\langle}  \possiblyWithSub\stageImetaColor{v'^{\superscriptI} }  \closeO{\rangle}    \longrightarrow^{1}    \possiblyWithSub\stageImetaColor{v'^{\superscriptI} }   
              }:
                Since \(\possiblyWithSub\stageImetaColor{v'^{\superscriptI} } = \possiblyWithSub\stageImetaColor{N'^{\superscriptI} }\) is a closed stage-\(1\) value,
                \(\possiblyWithSub\stageImetaColor{\Hat{N}^{\superscriptI} } \defeq \possiblyWithSub\stageImetaColor{v'^{\superscriptI} }\) clearly satisfies the desired properties.
            \end{itemize}
        \end{itemize}
      \item
        \begin{itemize}
          \item Case~\(\possiblyWithSub\stageImetaColor{T^{\superscriptI} } = \possiblyWithSub\stageImetaColor{B}\):
            This contradicts the assumption~\(   [  \possiblyWithSub\stageOmetaColor{N'^{\superscriptO} }_{{\mathrm{0}}}  /  \possiblyWithSub\stageOmetaColor{x}  ]    \possiblyWithSub\stageImetaColor{T^{\superscriptI} }   \longrightarrow^{1}   \possiblyWithSub\stageImetaColor{T'^{\superscriptI} }  \).
          \item Case~\(\possiblyWithSub\stageImetaColor{T^{\superscriptI} } =  \ttI{Tensor}\ \ordI{\%} \possiblyWithSub\stageOmetaColor{N^{\superscriptO} } \):
            We can trace back the derivation of \(   [  \possiblyWithSub\stageOmetaColor{N'^{\superscriptO} }_{{\mathrm{0}}}  /  \possiblyWithSub\stageOmetaColor{x}  ]    \possiblyWithSub\stageImetaColor{T^{\superscriptI} }   \longrightarrow^{1}   \possiblyWithSub\stageImetaColor{T'^{\superscriptI} }  \) as follows:
            \begin{center}
              \derive[ET1-Tensor]{%
                   [  \possiblyWithSub\stageOmetaColor{N'^{\superscriptO} }_{{\mathrm{0}}}  /  \possiblyWithSub\stageOmetaColor{x}  ]    \possiblyWithSub\stageOmetaColor{N^{\superscriptO} }_{{\mathrm{1}}}   \longrightarrow^{0}   \possiblyWithSub\stageOmetaColor{N'^{\superscriptO} }_{{\mathrm{1}}}  
              }{%
                  \ttI{Tensor}\ \ordI{\%}  \openO{(}   [  \possiblyWithSub\stageOmetaColor{N'^{\superscriptO} }_{{\mathrm{0}}}  /  \possiblyWithSub\stageOmetaColor{x}  ]    \possiblyWithSub\stageOmetaColor{N^{\superscriptO} }_{{\mathrm{1}}}  \closeO{)}    \longrightarrow^{1}    \ttI{Tensor}\ \ordI{\%} \possiblyWithSub\stageOmetaColor{N'^{\superscriptO} }_{{\mathrm{1}}}   
              }.
            \end{center}
            By IH, there exists \(\possiblyWithSub\stageOmetaColor{\Hat{N}^{\superscriptO} }_{{\mathrm{1}}}\) such that
            \(   [  \possiblyWithSub\stageOmetaColor{N^{\superscriptO} }_{{\mathrm{0}}}  /  \possiblyWithSub\stageOmetaColor{x}  ]    \possiblyWithSub\stageOmetaColor{N^{\superscriptO} }_{{\mathrm{1}}}   \longrightarrow^{0\,\ast}     [  \possiblyWithSub\stageOmetaColor{N^{\superscriptO} }_{{\mathrm{0}}}  /  \possiblyWithSub\stageOmetaColor{x}  ]    \possiblyWithSub\stageOmetaColor{\Hat{N}^{\superscriptO} }_{{\mathrm{1}}}   \)
            and \(\possiblyWithSub\stageOmetaColor{N'^{\superscriptO} }_{{\mathrm{1}}} =   [  \possiblyWithSub\stageOmetaColor{N'^{\superscriptO} }_{{\mathrm{0}}}  /  \possiblyWithSub\stageOmetaColor{x}  ]    \possiblyWithSub\stageOmetaColor{\Hat{N}^{\superscriptO} }_{{\mathrm{1}}} \).
            Thus, \(\possiblyWithSub\stageImetaColor{\Hat{T}^{\superscriptI} } \defeq  \ttI{Tensor}\ \ordI{\%} \possiblyWithSub\stageOmetaColor{\Hat{N}^{\superscriptO} }_{{\mathrm{1}}} \) satisfies the desired properties
            by the repeated use of \rulename{ET1-Tensor}.
          \item Case~\(\possiblyWithSub\stageImetaColor{T^{\superscriptI} } =  \possiblyWithSub\stageImetaColor{T^{\superscriptI} }_{{\mathrm{1}}}  \relI{\to}  \possiblyWithSub\stageImetaColor{T^{\superscriptI} }_{{\mathrm{2}}} \):
            Can be proved by using IH in a manner similar to
            the case on (2) where \(\possiblyWithSub\stageImetaColor{N^{\superscriptI} } =  \possiblyWithSub\stageImetaColor{N^{\superscriptI} }_{{\mathrm{1}}} \  \possiblyWithSub\stageImetaColor{N^{\superscriptI} }_{{\mathrm{2}}} \).
        \end{itemize}
    \end{enumerate}
  \end{proof}
  \begin{lemma}[One-step, one-variable cotermination]\label{lem:cotermination-one-step-one-var}
    \noindent
    \begin{enumerate}
      \item
        If \( \possiblyWithSub\stageOmetaColor{N^{\superscriptO} }_{{\mathrm{1}}}  \longrightarrow^{0}   \possiblyWithSub\stageOmetaColor{N^{\superscriptO} }_{{\mathrm{2}}}  \) and \(   [  \possiblyWithSub\stageOmetaColor{N^{\superscriptO} }_{{\mathrm{1}}}  /  \possiblyWithSub\stageOmetaColor{x}  ]    \possiblyWithSub\stageOmetaColor{N^{\superscriptO} }   \longrightarrow^{0\,\ast}     \possiblyWithSub\stageOmetaColor{c}    \),
        then \(   [  \possiblyWithSub\stageOmetaColor{N^{\superscriptO} }_{{\mathrm{2}}}  /  \possiblyWithSub\stageOmetaColor{x}  ]    \possiblyWithSub\stageOmetaColor{N^{\superscriptO} }   \longrightarrow^{0\,\ast}     \possiblyWithSub\stageOmetaColor{c}    \).
      \item
        If \( \possiblyWithSub\stageOmetaColor{N^{\superscriptO} }_{{\mathrm{1}}}  \longrightarrow^{0}   \possiblyWithSub\stageOmetaColor{N^{\superscriptO} }_{{\mathrm{2}}}  \) and \(   [  \possiblyWithSub\stageOmetaColor{N^{\superscriptO} }_{{\mathrm{2}}}  /  \possiblyWithSub\stageOmetaColor{x}  ]    \possiblyWithSub\stageOmetaColor{N^{\superscriptO} }   \longrightarrow^{0\,\ast}     \possiblyWithSub\stageOmetaColor{c}    \),
        then \(   [  \possiblyWithSub\stageOmetaColor{N^{\superscriptO} }_{{\mathrm{1}}}  /  \possiblyWithSub\stageOmetaColor{x}  ]    \possiblyWithSub\stageOmetaColor{N^{\superscriptO} }   \longrightarrow^{0\,\ast}     \possiblyWithSub\stageOmetaColor{c}    \).
    \end{enumerate}
  \end{lemma}
  \begin{proof}
    \begin{enumerate}
      \item
        By induction on the number of steps of \(   [  \possiblyWithSub\stageOmetaColor{N^{\superscriptO} }_{{\mathrm{1}}}  /  \possiblyWithSub\stageOmetaColor{x}  ]    \possiblyWithSub\stageOmetaColor{N^{\superscriptO} }   \longrightarrow^{0\,\ast}     \possiblyWithSub\stageOmetaColor{c}    \).
        \begin{itemize}
          \item Case \(  [  \possiblyWithSub\stageOmetaColor{N^{\superscriptO} }_{{\mathrm{1}}}  /  \possiblyWithSub\stageOmetaColor{x}  ]    \possiblyWithSub\stageOmetaColor{N^{\superscriptO} }  = \possiblyWithSub\stageOmetaColor{c}\):
            We have the following two cases:
            \begin{itemize}
              \item Case where \(\possiblyWithSub\stageOmetaColor{N^{\superscriptO} } = \possiblyWithSub\stageOmetaColor{x}\) and \(\possiblyWithSub\stageOmetaColor{N^{\superscriptO} }_{{\mathrm{1}}} = \possiblyWithSub\stageOmetaColor{c}\):
                This contradicts the assumption~\( \possiblyWithSub\stageOmetaColor{N^{\superscriptO} }_{{\mathrm{1}}}  \longrightarrow^{0}   \possiblyWithSub\stageOmetaColor{N^{\superscriptO} }_{{\mathrm{2}}}  \).
              \item Case \(\possiblyWithSub\stageOmetaColor{N^{\superscriptO} } = \possiblyWithSub\stageOmetaColor{c}\):
                We clearly have \(   [  \possiblyWithSub\stageOmetaColor{N^{\superscriptO} }_{{\mathrm{2}}}  /  \possiblyWithSub\stageOmetaColor{x}  ]    \possiblyWithSub\stageOmetaColor{N^{\superscriptO} }   \longrightarrow^{0\,\ast}     \possiblyWithSub\stageOmetaColor{c}    \).
            \end{itemize}
          \item Case where there exists \(\possiblyWithSub\stageOmetaColor{N'^{\superscriptO} }\) such that
            \(   [  \possiblyWithSub\stageOmetaColor{N^{\superscriptO} }_{{\mathrm{1}}}  /  \possiblyWithSub\stageOmetaColor{x}  ]    \possiblyWithSub\stageOmetaColor{N^{\superscriptO} }   \longrightarrow^{0}   \possiblyWithSub\stageOmetaColor{N'^{\superscriptO} }  \) and \( \possiblyWithSub\stageOmetaColor{N'^{\superscriptO} }  \longrightarrow^{0\,\ast}     \possiblyWithSub\stageOmetaColor{c}    \):
            By Lemma~\ref{lem:weak-bisimulation-left},
            there exists \(\possiblyWithSub\stageOmetaColor{N^{\superscriptO} }_{{\mathrm{0}}}\) such that
            \(   [  \possiblyWithSub\stageOmetaColor{N^{\superscriptO} }_{{\mathrm{2}}}  /  \possiblyWithSub\stageOmetaColor{x}  ]    \possiblyWithSub\stageOmetaColor{N^{\superscriptO} }   \longrightarrow^{0\,\ast}     [  \possiblyWithSub\stageOmetaColor{N^{\superscriptO} }_{{\mathrm{2}}}  /  \possiblyWithSub\stageOmetaColor{x}  ]    \possiblyWithSub\stageOmetaColor{N^{\superscriptO} }_{{\mathrm{0}}}   \) and \(\possiblyWithSub\stageOmetaColor{N'^{\superscriptO} } =   [  \possiblyWithSub\stageOmetaColor{N^{\superscriptO} }_{{\mathrm{1}}}  /  \possiblyWithSub\stageOmetaColor{x}  ]    \possiblyWithSub\stageOmetaColor{N^{\superscriptO} }_{{\mathrm{0}}} \).
            Then, by IH on \(   [  \possiblyWithSub\stageOmetaColor{N^{\superscriptO} }_{{\mathrm{1}}}  /  \possiblyWithSub\stageOmetaColor{x}  ]    \possiblyWithSub\stageOmetaColor{N^{\superscriptO} }_{{\mathrm{0}}}   \longrightarrow^{0\,\ast}     \possiblyWithSub\stageOmetaColor{c}    \),
            which has a strictly shorter evaluation sequence,
            we have \(   [  \possiblyWithSub\stageOmetaColor{N^{\superscriptO} }_{{\mathrm{2}}}  /  \possiblyWithSub\stageOmetaColor{x}  ]    \possiblyWithSub\stageOmetaColor{N^{\superscriptO} }_{{\mathrm{0}}}   \longrightarrow^{0\,\ast}     \possiblyWithSub\stageOmetaColor{c}    \).
            Therefore, we have
            \(  [  \possiblyWithSub\stageOmetaColor{N^{\superscriptO} }_{{\mathrm{2}}}  /  \possiblyWithSub\stageOmetaColor{x}  ]    \possiblyWithSub\stageOmetaColor{N^{\superscriptO} }  \longrightarrow^{\ast\,0}    [  \possiblyWithSub\stageOmetaColor{N^{\superscriptO} }_{{\mathrm{2}}}  /  \possiblyWithSub\stageOmetaColor{x}  ]    \possiblyWithSub\stageOmetaColor{N^{\superscriptO} }_{{\mathrm{0}}}   \longrightarrow^{0\,\ast}     \possiblyWithSub\stageOmetaColor{c}    \).
        \end{itemize}
      \item
        By induction on the number of steps of \(   [  \possiblyWithSub\stageOmetaColor{N^{\superscriptO} }_{{\mathrm{1}}}  /  \possiblyWithSub\stageOmetaColor{x}  ]    \possiblyWithSub\stageOmetaColor{N^{\superscriptO} }   \longrightarrow^{0\,\ast}     \possiblyWithSub\stageOmetaColor{c}    \).
        \begin{itemize}
          \item Case \(  [  \possiblyWithSub\stageOmetaColor{N^{\superscriptO} }_{{\mathrm{2}}}  /  \possiblyWithSub\stageOmetaColor{x}  ]    \possiblyWithSub\stageOmetaColor{N^{\superscriptO} }  = \possiblyWithSub\stageOmetaColor{c}\):
            We have the following two cases:
            \begin{itemize}
              \item Case where \(\possiblyWithSub\stageOmetaColor{N^{\superscriptO} } = \possiblyWithSub\stageOmetaColor{x}\) and \(\possiblyWithSub\stageOmetaColor{N^{\superscriptO} }_{{\mathrm{2}}} = \possiblyWithSub\stageOmetaColor{c}\):
                Since \( \possiblyWithSub\stageOmetaColor{N^{\superscriptO} }_{{\mathrm{1}}}  \longrightarrow^{0}     \possiblyWithSub\stageOmetaColor{c}    \),
                we clearly have \(   [  \possiblyWithSub\stageOmetaColor{N^{\superscriptO} }_{{\mathrm{1}}}  /  \possiblyWithSub\stageOmetaColor{x}  ]     \possiblyWithSub\stageOmetaColor{x}    \longrightarrow^{0\,\ast}     \possiblyWithSub\stageOmetaColor{c}    \).
              \item Case \(\possiblyWithSub\stageOmetaColor{N^{\superscriptO} } = \possiblyWithSub\stageOmetaColor{c}\):
                We clearly have \(   [  \possiblyWithSub\stageOmetaColor{N^{\superscriptO} }_{{\mathrm{1}}}  /  \possiblyWithSub\stageOmetaColor{x}  ]    \possiblyWithSub\stageOmetaColor{N^{\superscriptO} }   \longrightarrow^{0\,\ast}     \possiblyWithSub\stageOmetaColor{c}    \).
            \end{itemize}
          \item Case where there exists \(\possiblyWithSub\stageOmetaColor{N'^{\superscriptO} }\) such that
          \(   [  \possiblyWithSub\stageOmetaColor{N^{\superscriptO} }_{{\mathrm{2}}}  /  \possiblyWithSub\stageOmetaColor{x}  ]    \possiblyWithSub\stageOmetaColor{N^{\superscriptO} }   \longrightarrow^{0}   \possiblyWithSub\stageOmetaColor{N'^{\superscriptO} }  \) and \( \possiblyWithSub\stageOmetaColor{N'^{\superscriptO} }  \longrightarrow^{0\,\ast}     \possiblyWithSub\stageOmetaColor{c}    \):
            By Lemma~\ref{lem:weak-bisimulation-right},
            there exists \(\possiblyWithSub\stageOmetaColor{N^{\superscriptO} }_{{\mathrm{0}}}\) such that
            \(   [  \possiblyWithSub\stageOmetaColor{N^{\superscriptO} }_{{\mathrm{1}}}  /  \possiblyWithSub\stageOmetaColor{x}  ]    \possiblyWithSub\stageOmetaColor{N^{\superscriptO} }   \longrightarrow^{0\,\ast}     [  \possiblyWithSub\stageOmetaColor{N^{\superscriptO} }_{{\mathrm{1}}}  /  \possiblyWithSub\stageOmetaColor{x}  ]    \possiblyWithSub\stageOmetaColor{N^{\superscriptO} }_{{\mathrm{0}}}   \) and \(\possiblyWithSub\stageOmetaColor{N'^{\superscriptO} } =   [  \possiblyWithSub\stageOmetaColor{N^{\superscriptO} }_{{\mathrm{2}}}  /  \possiblyWithSub\stageOmetaColor{x}  ]    \possiblyWithSub\stageOmetaColor{N^{\superscriptO} }_{{\mathrm{0}}} \).
            Then, by IH on \(   [  \possiblyWithSub\stageOmetaColor{N^{\superscriptO} }_{{\mathrm{2}}}  /  \possiblyWithSub\stageOmetaColor{x}  ]    \possiblyWithSub\stageOmetaColor{N^{\superscriptO} }_{{\mathrm{0}}}   \longrightarrow^{0\,\ast}     \possiblyWithSub\stageOmetaColor{c}    \),
            which has a strictly shorter evaluation sequence,
            we have \(   [  \possiblyWithSub\stageOmetaColor{N^{\superscriptO} }_{{\mathrm{1}}}  /  \possiblyWithSub\stageOmetaColor{x}  ]    \possiblyWithSub\stageOmetaColor{N^{\superscriptO} }_{{\mathrm{0}}}   \longrightarrow^{0\,\ast}     \possiblyWithSub\stageOmetaColor{c}    \).
            Therefore, we have
            \(  [  \possiblyWithSub\stageOmetaColor{N^{\superscriptO} }_{{\mathrm{1}}}  /  \possiblyWithSub\stageOmetaColor{x}  ]    \possiblyWithSub\stageOmetaColor{N^{\superscriptO} }  \longrightarrow^{\ast\,0}    [  \possiblyWithSub\stageOmetaColor{N^{\superscriptO} }_{{\mathrm{1}}}  /  \possiblyWithSub\stageOmetaColor{x}  ]    \possiblyWithSub\stageOmetaColor{N^{\superscriptO} }_{{\mathrm{0}}}   \longrightarrow^{0\,\ast}     \possiblyWithSub\stageOmetaColor{c}    \).
        \end{itemize}
    \end{enumerate}
  \end{proof}
  \begin{lemma}[One-variable cotermination]\label{lem:cotermination-one-var}
    \noindent
    \begin{enumerate}
      \item
        If \( \possiblyWithSub\stageOmetaColor{N^{\superscriptO} }_{{\mathrm{1}}}  \longrightarrow^{0\,\ast}   \possiblyWithSub\stageOmetaColor{N^{\superscriptO} }_{{\mathrm{2}}}  \) and \(   [  \possiblyWithSub\stageOmetaColor{N^{\superscriptO} }_{{\mathrm{1}}}  /  \possiblyWithSub\stageOmetaColor{x}  ]    \possiblyWithSub\stageOmetaColor{N^{\superscriptO} }   \longrightarrow^{0\,\ast}     \possiblyWithSub\stageOmetaColor{c}    \),
        then \(   [  \possiblyWithSub\stageOmetaColor{N^{\superscriptO} }_{{\mathrm{2}}}  /  \possiblyWithSub\stageOmetaColor{x}  ]    \possiblyWithSub\stageOmetaColor{N^{\superscriptO} }   \longrightarrow^{0\,\ast}     \possiblyWithSub\stageOmetaColor{c}    \).
      \item
        If \( \possiblyWithSub\stageOmetaColor{N^{\superscriptO} }_{{\mathrm{1}}}  \longrightarrow^{0\,\ast}   \possiblyWithSub\stageOmetaColor{N^{\superscriptO} }_{{\mathrm{2}}}  \) and \(   [  \possiblyWithSub\stageOmetaColor{N^{\superscriptO} }_{{\mathrm{2}}}  /  \possiblyWithSub\stageOmetaColor{x}  ]    \possiblyWithSub\stageOmetaColor{N^{\superscriptO} }   \longrightarrow^{0\,\ast}     \possiblyWithSub\stageOmetaColor{c}    \),
        then \(   [  \possiblyWithSub\stageOmetaColor{N^{\superscriptO} }_{{\mathrm{1}}}  /  \possiblyWithSub\stageOmetaColor{x}  ]    \possiblyWithSub\stageOmetaColor{N^{\superscriptO} }   \longrightarrow^{0\,\ast}     \possiblyWithSub\stageOmetaColor{c}    \).
    \end{enumerate}
  \end{lemma}
  \begin{proof}
    By induction on the number of steps of \( \possiblyWithSub\stageOmetaColor{N^{\superscriptO} }_{{\mathrm{1}}}  \longrightarrow^{0\,\ast}   \possiblyWithSub\stageOmetaColor{N^{\superscriptO} }_{{\mathrm{2}}}  \)
    using Lemma~\ref{lem:cotermination-one-step-one-var}.
  \end{proof}
  \begin{lemma}[Cotermination]\label{lem:cotermination}
    \noindent
    \begin{enumerate}
      \item
        If \( \sigma_{{\mathrm{1}}}  \longrightarrow  \sigma_{{\mathrm{2}}} \) and \(  \sigma_{{\mathrm{1}}}   \possiblyWithSub\stageOmetaColor{N^{\superscriptO} }   \longrightarrow^{0\,\ast}     \possiblyWithSub\stageOmetaColor{c}    \),
        then \(  \sigma_{{\mathrm{2}}}   \possiblyWithSub\stageOmetaColor{N^{\superscriptO} }   \longrightarrow^{0\,\ast}     \possiblyWithSub\stageOmetaColor{c}    \).
      \item
        If \( \sigma_{{\mathrm{1}}}  \longrightarrow  \sigma_{{\mathrm{2}}} \) and \(  \sigma_{{\mathrm{2}}}   \possiblyWithSub\stageOmetaColor{N^{\superscriptO} }   \longrightarrow^{0\,\ast}     \possiblyWithSub\stageOmetaColor{c}    \),
        then \(  \sigma_{{\mathrm{1}}}   \possiblyWithSub\stageOmetaColor{N^{\superscriptO} }   \longrightarrow^{0\,\ast}     \possiblyWithSub\stageOmetaColor{c}    \).
    \end{enumerate}
  \end{lemma}
  \begin{proof}
    By induction on the size of \(\dom \sigma_{{\mathrm{1}}}\) using Lemma~\ref{lem:cotermination-one-var}.
  \end{proof}
  \begin{lemma}[Equivalence preserves reduction of predicates]\label{lem:equiv-preserves-predicate-reduction}
    \noindent
    \begin{enumerate}
      \item
        If \(   \openO{\{} \possiblyWithSub\stageOmetaColor{\nu}  \relO{:}  \possiblyWithSub\stageOmetaColor{B}  \relO{\mid}  \possiblyWithSub\stageOmetaColor{N^{\superscriptO} }_{{\mathrm{1}}} \closeO{\} }    \equiv^{0}    \openO{\{} \possiblyWithSub\stageOmetaColor{\nu}  \relO{:}  \possiblyWithSub\stageOmetaColor{B}  \relO{\mid}  \possiblyWithSub\stageOmetaColor{N^{\superscriptO} }_{{\mathrm{2}}} \closeO{\} }   \) and \(   [    \possiblyWithSub\stageOmetaColor{c}    /  \possiblyWithSub\stageOmetaColor{\nu}  ]    \possiblyWithSub\stageOmetaColor{N^{\superscriptO} }_{{\mathrm{1}}}   \longrightarrow^{0\,\ast}      \ttO{true}     \),
        then \(   [    \possiblyWithSub\stageOmetaColor{c}    /  \possiblyWithSub\stageOmetaColor{\nu}  ]    \possiblyWithSub\stageOmetaColor{N^{\superscriptO} }_{{\mathrm{2}}}   \longrightarrow^{0\,\ast}      \ttO{true}     \).
      \item
        If \(   \openO{\{} \possiblyWithSub\stageOmetaColor{\nu}  \relO{:}  \possiblyWithSub\stageOmetaColor{B}  \relO{\mid}  \possiblyWithSub\stageOmetaColor{N^{\superscriptO} }_{{\mathrm{1}}} \closeO{\} }    \equiv^{0}    \openO{\{} \possiblyWithSub\stageOmetaColor{\nu}  \relO{:}  \possiblyWithSub\stageOmetaColor{B}  \relO{\mid}  \possiblyWithSub\stageOmetaColor{N^{\superscriptO} }_{{\mathrm{2}}} \closeO{\} }   \) and \(   [    \possiblyWithSub\stageOmetaColor{c}    /  \possiblyWithSub\stageOmetaColor{\nu}  ]    \possiblyWithSub\stageOmetaColor{N^{\superscriptO} }_{{\mathrm{2}}}   \longrightarrow^{0\,\ast}      \ttO{true}     \),
        then \(   [    \possiblyWithSub\stageOmetaColor{c}    /  \possiblyWithSub\stageOmetaColor{\nu}  ]    \possiblyWithSub\stageOmetaColor{N^{\superscriptO} }_{{\mathrm{1}}}   \longrightarrow^{0\,\ast}      \ttO{true}     \).
    \end{enumerate}
  \end{lemma}
  \begin{proof}
    By mutual induction on the derivation of \(   \openO{\{} \possiblyWithSub\stageOmetaColor{\nu}  \relO{:}  \possiblyWithSub\stageOmetaColor{B}  \relO{\mid}  \possiblyWithSub\stageOmetaColor{N^{\superscriptO} }_{{\mathrm{1}}} \closeO{\} }    \equiv^{0}    \openO{\{} \possiblyWithSub\stageOmetaColor{\nu}  \relO{:}  \possiblyWithSub\stageOmetaColor{B}  \relO{\mid}  \possiblyWithSub\stageOmetaColor{N^{\superscriptO} }_{{\mathrm{2}}} \closeO{\} }   \).
    \begin{enumerate}
      \item
        \begin{itemize}
          \item Case~\rulename{CqT0-Refl} is immediate.
          \item Case~\rulename{CqT0-Sym} is straightforward by IH.
          \item Case \derive[CqT0-Trans]{%
               \openO{\{} \possiblyWithSub\stageOmetaColor{\nu}  \relO{:}  \possiblyWithSub\stageOmetaColor{B}  \relO{\mid}  \possiblyWithSub\stageOmetaColor{N^{\superscriptO} }_{{\mathrm{1}}} \closeO{\} }    \equiv^{0}  \possiblyWithSub\stageOmetaColor{T'^{\superscriptO} } 
          \andalso
             \possiblyWithSub\stageOmetaColor{T'^{\superscriptO} }  \equiv^{0}    \openO{\{} \possiblyWithSub\stageOmetaColor{\nu}  \relO{:}  \possiblyWithSub\stageOmetaColor{B}  \relO{\mid}  \possiblyWithSub\stageOmetaColor{N^{\superscriptO} }_{{\mathrm{2}}} \closeO{\} }   
          }{%
               \openO{\{} \possiblyWithSub\stageOmetaColor{\nu}  \relO{:}  \possiblyWithSub\stageOmetaColor{B}  \relO{\mid}  \possiblyWithSub\stageOmetaColor{N^{\superscriptO} }_{{\mathrm{1}}} \closeO{\} }    \equiv^{0}    \openO{\{} \possiblyWithSub\stageOmetaColor{\nu}  \relO{:}  \possiblyWithSub\stageOmetaColor{B}  \relO{\mid}  \possiblyWithSub\stageOmetaColor{N^{\superscriptO} }_{{\mathrm{2}}} \closeO{\} }   
          }:
            By Lemma~\ref{lem:rfn-type-csr-equiv-form}, \(\possiblyWithSub\stageOmetaColor{T'^{\superscriptO} }\) is of the form \( \openO{\{} \possiblyWithSub\stageOmetaColor{\nu}  \relO{:}  \possiblyWithSub\stageOmetaColor{B}  \relO{\mid}  \possiblyWithSub\stageOmetaColor{N'^{\superscriptO} } \closeO{\} } \).
            Then, by IH on \(   \openO{\{} \possiblyWithSub\stageOmetaColor{\nu}  \relO{:}  \possiblyWithSub\stageOmetaColor{B}  \relO{\mid}  \possiblyWithSub\stageOmetaColor{N^{\superscriptO} }_{{\mathrm{1}}} \closeO{\} }    \equiv^{0}  \possiblyWithSub\stageOmetaColor{T'^{\superscriptO} } \),
            we have \(   [    \possiblyWithSub\stageOmetaColor{c}    /  \possiblyWithSub\stageOmetaColor{\nu}  ]    \possiblyWithSub\stageOmetaColor{N'^{\superscriptO} }   \longrightarrow^{0\,\ast}      \ttO{true}     \).
            Thus, again by IH on \( \possiblyWithSub\stageOmetaColor{T'^{\superscriptO} }  \equiv^{0}    \openO{\{} \possiblyWithSub\stageOmetaColor{\nu}  \relO{:}  \possiblyWithSub\stageOmetaColor{B}  \relO{\mid}  \possiblyWithSub\stageOmetaColor{N^{\superscriptO} }_{{\mathrm{2}}} \closeO{\} }   \),
            we have \(   [    \possiblyWithSub\stageOmetaColor{c}    /  \possiblyWithSub\stageOmetaColor{\nu}  ]    \possiblyWithSub\stageOmetaColor{N^{\superscriptO} }_{{\mathrm{2}}}   \longrightarrow^{0\,\ast}      \ttO{true}     \).
          \item Case \derive[CqT0-Rfn]{%
             \sigma_{{\mathrm{1}}}  \longrightarrow  \sigma_{{\mathrm{2}}} 
          }{%
               \openO{\{} \possiblyWithSub\stageOmetaColor{\nu}  \relO{:}  \possiblyWithSub\stageOmetaColor{B}  \relO{\mid}   \sigma_{{\mathrm{1}}}   \possiblyWithSub\stageOmetaColor{N^{\superscriptO} }  \closeO{\} }    \equiv^{0}    \openO{\{} \possiblyWithSub\stageOmetaColor{\nu}  \relO{:}  \possiblyWithSub\stageOmetaColor{B}  \relO{\mid}   \sigma_{{\mathrm{2}}}   \possiblyWithSub\stageOmetaColor{N^{\superscriptO} }  \closeO{\} }   
          }:
            Since \(\possiblyWithSub\stageOmetaColor{N^{\superscriptO} }_{{\mathrm{1}}} =  \sigma_{{\mathrm{1}}}   \possiblyWithSub\stageOmetaColor{N^{\superscriptO} } \),
            we have \(   (   [    \possiblyWithSub\stageOmetaColor{c}    /  \possiblyWithSub\stageOmetaColor{\nu}  ]   \circ  \sigma_{{\mathrm{1}}}  )    \possiblyWithSub\stageOmetaColor{N^{\superscriptO} }   \longrightarrow^{0\,\ast}      \ttO{true}     \).
            Then, by Lemma~\ref{lem:cotermination},
            we have \(   (   [    \possiblyWithSub\stageOmetaColor{c}    /  \possiblyWithSub\stageOmetaColor{\nu}  ]   \circ  \sigma_{{\mathrm{2}}}  )    \possiblyWithSub\stageOmetaColor{N^{\superscriptO} }   \longrightarrow^{0\,\ast}      \ttO{true}     \),
            which is equivalent to \(   [    \possiblyWithSub\stageOmetaColor{c}    /  \possiblyWithSub\stageOmetaColor{\nu}  ]    \possiblyWithSub\stageOmetaColor{N^{\superscriptO} }_{{\mathrm{2}}}   \longrightarrow^{0\,\ast}      \ttO{true}     \).
          \item The other cases contradict the assumption.
        \end{itemize}
      \item
        Can be proved in mostly the same way as (1) by using Cotermination.
    \end{enumerate}
  \end{proof}
  \begin{lemma}\label{lem:equiv-preserves-rfn}
    If \(  \possiblyWithSub\stageOmetaColor{c}   \vDash  \possiblyWithSub\stageOmetaColor{T^{\superscriptO} }_{{\mathrm{1}}} \) and \( \possiblyWithSub\stageOmetaColor{T^{\superscriptO} }_{{\mathrm{1}}}  \equiv^{0}  \possiblyWithSub\stageOmetaColor{T^{\superscriptO} }_{{\mathrm{2}}} \), then \(  \possiblyWithSub\stageOmetaColor{c}   \vDash  \possiblyWithSub\stageOmetaColor{T^{\superscriptO} }_{{\mathrm{2}}} \).
  \end{lemma}
  \begin{proof}
    Immediate from Lemmata~\ref{lem:rfn-type-csr-equiv-form} and \ref{lem:equiv-preserves-predicate-reduction}.
  \end{proof}
  \begin{lemma}\label{lem:const-typing-implies-predicate-satisfaction}
    \( \mathit{\Gamma}  \vdash^{0}    \possiblyWithSub\stageOmetaColor{c}    :  \possiblyWithSub\stageOmetaColor{T^{\superscriptO} } \) and \( \vdash^{0}_{\mathrm{dom} }  \possiblyWithSub\stageOmetaColor{T^{\superscriptO} } \) implies \(  \possiblyWithSub\stageOmetaColor{c}   \vDash  \possiblyWithSub\stageOmetaColor{T^{\superscriptO} } \).
  \end{lemma}
  \begin{proof}
    By induction on the derivation of \( \mathit{\Gamma}  \vdash^{0}    \possiblyWithSub\stageOmetaColor{c}    :  \possiblyWithSub\stageOmetaColor{T^{\superscriptO} } \).
    \begin{itemize}
      \item Case \derive[T0-RfnPred]{%
         \mathit{\Gamma}  \vdash^{0}    \openO{\{} \possiblyWithSub\stageOmetaColor{\nu}  \relO{:}  \possiblyWithSub\stageOmetaColor{B}  \relO{\mid}  \possiblyWithSub\stageOmetaColor{N^{\superscriptO} } \closeO{\} }   
      \andalso
        \ConstEnvPers(c) = \possiblyWithSub\stageImetaColor{B}
      \andalso
           [    \possiblyWithSub\stageOmetaColor{c}    /  \possiblyWithSub\stageOmetaColor{\nu}  ]    \possiblyWithSub\stageOmetaColor{N^{\superscriptO} }   \longrightarrow^{0\,\ast}      \ttO{true}     
      }{%
         \mathit{\Gamma}  \vdash^{0}    \possiblyWithSub\stageOmetaColor{c}    :    \openO{\{} \possiblyWithSub\stageOmetaColor{\nu}  \relO{:}  \possiblyWithSub\stageOmetaColor{B}  \relO{\mid}  \possiblyWithSub\stageOmetaColor{N^{\superscriptO} } \closeO{\} }   
      }:
        We can immediately derive
        \begin{center}
          \derive{%
            \ConstEnvPers(c) = \possiblyWithSub\stageImetaColor{B}
          \andalso
               [    \possiblyWithSub\stageOmetaColor{c}    /  \possiblyWithSub\stageOmetaColor{\nu}  ]    \possiblyWithSub\stageOmetaColor{N^{\superscriptO} }   \longrightarrow^{0\,\ast}      \ttO{true}     
          }{%
              \possiblyWithSub\stageOmetaColor{c}   \vDash    \openO{\{} \possiblyWithSub\stageOmetaColor{\nu}  \relO{:}  \possiblyWithSub\stageOmetaColor{B}  \relO{\mid}  \possiblyWithSub\stageOmetaColor{N^{\superscriptO} } \closeO{\} }   
          }
        \end{center}
      \item Case \derive[T0-TyEquiv]{%
         \mathit{\Gamma}  \vdash^{0}    \possiblyWithSub\stageOmetaColor{c}    :  \possiblyWithSub\stageOmetaColor{T'^{\superscriptO} } 
      \andalso
         \possiblyWithSub\stageOmetaColor{T'^{\superscriptO} }  \equiv^{0}  \possiblyWithSub\stageOmetaColor{T^{\superscriptO} } 
      \andalso
         \mathit{\Gamma}  \vdash^{0}  \possiblyWithSub\stageOmetaColor{T^{\superscriptO} } 
      }{%
         \mathit{\Gamma}  \vdash^{0}    \possiblyWithSub\stageOmetaColor{c}    :  \possiblyWithSub\stageOmetaColor{T^{\superscriptO} } 
      }:
        By Lemma~\ref{lem:csr-equiv-preserves-order}, we have \( \vdash^{0}_{\mathrm{dom} }  \possiblyWithSub\stageOmetaColor{T'^{\superscriptO} } \).
        Then, by IH, we have \(  \possiblyWithSub\stageOmetaColor{c}   \vDash  \possiblyWithSub\stageOmetaColor{T'^{\superscriptO} } \).
        Thus, by Lemma~\ref{lem:equiv-preserves-rfn}, we have \(  \possiblyWithSub\stageOmetaColor{c}   \vDash  \possiblyWithSub\stageOmetaColor{T^{\superscriptO} } \).
      \item Case \derive[T0-CstP]{%
         \vdash  \mathit{\Gamma} 
      \andalso
        \ConstEnvPers(c) = \possiblyWithSub\stageImetaColor{\tau^{\superscriptI} }
      }{%
         \mathit{\Gamma}  \vdash^{0}    \possiblyWithSub\stageOmetaColor{c}    :   \mathop{\downarrow}( \possiblyWithSub\stageImetaColor{\tau^{\superscriptI} } )  
      }:
        By \( \vdash^{0}_{\mathrm{dom} }  \possiblyWithSub\stageOmetaColor{T^{\superscriptO} } \), we have the following two cases:
        \begin{itemize}
          \item Case \(\possiblyWithSub\stageOmetaColor{T^{\superscriptO} } =  \ttO{Tensor}\  \possiblyWithSub\stageOmetaColor{s} \):
            Since \( \mathop{\downarrow}( \possiblyWithSub\stageImetaColor{\tau^{\superscriptI} } )  = \possiblyWithSub\stageOmetaColor{T^{\superscriptO} }\), we have \(\possiblyWithSub\stageImetaColor{\tau^{\superscriptI} } =  \ttI{Tensor}\ \ordI{\%} \possiblyWithSub\stageOmetaColor{s} \).
            Thus, we can derive
            \begin{center}
              \derive{%
                \ConstEnvPers(c) =  \ttI{Tensor}\ \ordI{\%} \possiblyWithSub\stageOmetaColor{s} 
              }{%
                  \possiblyWithSub\stageOmetaColor{c}   \vDash   \ttO{Tensor}\  \possiblyWithSub\stageOmetaColor{s}  
              }.
            \end{center}
          \item Case \(\possiblyWithSub\stageOmetaColor{T^{\superscriptO} } =  \openO{\{} \possiblyWithSub\stageOmetaColor{\nu}  \relO{:}  \possiblyWithSub\stageOmetaColor{B}  \relO{\mid}  \possiblyWithSub\stageOmetaColor{N^{\superscriptO} } \closeO{\} } \):
            Since \( \mathop{\downarrow}( \possiblyWithSub\stageImetaColor{\tau^{\superscriptI} } )  = \possiblyWithSub\stageOmetaColor{T^{\superscriptO} }\),
            we have \(\possiblyWithSub\stageImetaColor{\tau^{\superscriptI} } =  \possiblyWithSub\stageImetaColor{B} \) and \(\possiblyWithSub\stageOmetaColor{N^{\superscriptO} } =   \ttO{true}  \).
            Therefore, we can clearly derive
            \begin{center}
              \derive{%
                \ConstEnvPers(c) =  \possiblyWithSub\stageImetaColor{B} 
              \andalso
                   [    \possiblyWithSub\stageOmetaColor{c}    /  \possiblyWithSub\stageOmetaColor{\nu}  ]       \ttO{true}      \longrightarrow^{0\,\ast}      \ttO{true}     
              }{%
                  \possiblyWithSub\stageOmetaColor{c}   \vDash    \openO{\{} \possiblyWithSub\stageOmetaColor{\nu}  \relO{:}  \possiblyWithSub\stageOmetaColor{B}  \relO{\mid}  \possiblyWithSub\stageOmetaColor{N^{\superscriptO} } \closeO{\} }   
              }.
            \end{center}
        \end{itemize}
    \end{itemize}
  \end{proof}
  \begin{lemma}\label{lem:csr-equiv-unlift-inversion}
    \( \possiblyWithSub\stageOmetaColor{T^{\superscriptO} }  \equiv^{0}   \mathop{\downarrow}( \possiblyWithSub\stageImetaColor{\tau^{\superscriptI} } )  \) implies \( \possiblyWithSub\stageImetaColor{\tau^{\superscriptI} }  \gg  \possiblyWithSub\stageOmetaColor{T^{\superscriptO} } \).
  \end{lemma}
  \begin{proof}
    By induction on the structure of \(\possiblyWithSub\stageImetaColor{\tau^{\superscriptI} }\).
    \begin{itemize}
      \item Case~\(\possiblyWithSub\stageImetaColor{\tau^{\superscriptI} } = \possiblyWithSub\stageImetaColor{B}\):
        We have \( \possiblyWithSub\stageOmetaColor{T^{\superscriptO} }  \equiv^{0}    \openO{\{} \possiblyWithSub\stageOmetaColor{\nu}  \relO{:}  \possiblyWithSub\stageOmetaColor{B}  \relO{\mid}     \ttO{true}    \closeO{\} }   \),
        and by Lemma~\ref{lem:rfn-type-csr-equiv-form},
        \(\possiblyWithSub\stageOmetaColor{T^{\superscriptO} }\) is of the form~\( \openO{\{} \possiblyWithSub\stageOmetaColor{\nu}  \relO{:}  \possiblyWithSub\stageOmetaColor{B}  \relO{\mid}  \possiblyWithSub\stageOmetaColor{N^{\superscriptO} } \closeO{\} } \).
        Thus, we have \( \possiblyWithSub\stageImetaColor{\tau^{\superscriptI} }  \gg  \possiblyWithSub\stageOmetaColor{T^{\superscriptO} } \).
      \item Case~\(\possiblyWithSub\stageImetaColor{\tau^{\superscriptI} } =  \ttI{Tensor}\ \ordI{\%} \possiblyWithSub\stageOmetaColor{s} \):
        We have \( \possiblyWithSub\stageOmetaColor{T^{\superscriptO} }  \equiv^{0}   \ttO{Tensor}\  \possiblyWithSub\stageOmetaColor{s}  \),
        and by Lemma~\ref{lem:tensor-type-csr-equiv-form},
        \(\possiblyWithSub\stageOmetaColor{T^{\superscriptO} } =  \ttO{Tensor}\  \possiblyWithSub\stageOmetaColor{s} \).
        Therefore, we have \( \possiblyWithSub\stageImetaColor{\tau^{\superscriptI} }  \gg  \possiblyWithSub\stageOmetaColor{T^{\superscriptO} } \).
      \item Case~\(\possiblyWithSub\stageImetaColor{\tau^{\superscriptI} } =  \possiblyWithSub\stageImetaColor{\tau^{\superscriptI} }_{{\mathrm{1}}}  \relI{\to}  \possiblyWithSub\stageImetaColor{\tau^{\superscriptI} }_{{\mathrm{2}}} \):
        We have \( \possiblyWithSub\stageOmetaColor{T^{\superscriptO} }  \equiv^{0}   \openO{(} \possiblyWithSub\stageOmetaColor{x}  \relO{:}   \mathop{\downarrow}( \possiblyWithSub\stageImetaColor{\tau^{\superscriptI} }_{{\mathrm{1}}} )  \closeO{)} \relO{\to}   \mathop{\downarrow}( \possiblyWithSub\stageImetaColor{\tau^{\superscriptI} }_{{\mathrm{2}}} )   \),
        and by Lemma~\ref{lem:arrow-type-csr-equiv-form},
        \(\possiblyWithSub\stageOmetaColor{T^{\superscriptO} }\) is of the form~\( \openO{(} \possiblyWithSub\stageOmetaColor{x}  \relO{:}  \possiblyWithSub\stageOmetaColor{T^{\superscriptO} }_{{\mathrm{1}}} \closeO{)} \relO{\to}  \possiblyWithSub\stageOmetaColor{T^{\superscriptO} }_{{\mathrm{2}}} \).
        Then, by Lemma~\ref{lem:arrow-type-csr-equiv-inversion},
        we have \( \possiblyWithSub\stageOmetaColor{T^{\superscriptO} }_{{\mathrm{1}}}  \equiv^{0}   \mathop{\downarrow}( \possiblyWithSub\stageImetaColor{\tau^{\superscriptI} }_{{\mathrm{1}}} )  \) and \( \possiblyWithSub\stageOmetaColor{T^{\superscriptO} }_{{\mathrm{2}}}  \equiv^{0}   \mathop{\downarrow}( \possiblyWithSub\stageImetaColor{\tau^{\superscriptI} }_{{\mathrm{2}}} )  \).
        By IH, we have \( \possiblyWithSub\stageImetaColor{\tau^{\superscriptI} }_{{\mathrm{1}}}  \gg  \possiblyWithSub\stageOmetaColor{T^{\superscriptO} }_{{\mathrm{1}}} \) and \( \possiblyWithSub\stageImetaColor{\tau^{\superscriptI} }_{{\mathrm{2}}}  \gg  \possiblyWithSub\stageOmetaColor{T^{\superscriptO} }_{{\mathrm{2}}} \).
        Thus, we can derive \( \possiblyWithSub\stageImetaColor{\tau^{\superscriptI} }  \gg  \possiblyWithSub\stageOmetaColor{T^{\superscriptO} } \) by
        \begin{center}
          \derive{%
             \possiblyWithSub\stageImetaColor{\tau^{\superscriptI} }_{{\mathrm{1}}}  \gg  \possiblyWithSub\stageOmetaColor{T^{\superscriptO} }_{{\mathrm{1}}} 
          \andalso
             \possiblyWithSub\stageImetaColor{\tau^{\superscriptI} }_{{\mathrm{2}}}  \gg  \possiblyWithSub\stageOmetaColor{T^{\superscriptO} }_{{\mathrm{2}}} 
          }{%
              \possiblyWithSub\stageImetaColor{\tau^{\superscriptI} }_{{\mathrm{1}}}  \relI{\to}  \possiblyWithSub\stageImetaColor{\tau^{\superscriptI} }_{{\mathrm{2}}}   \gg   \openO{(} \possiblyWithSub\stageOmetaColor{x}  \relO{:}  \possiblyWithSub\stageOmetaColor{T^{\superscriptO} }_{{\mathrm{1}}} \closeO{)} \relO{\to}  \possiblyWithSub\stageOmetaColor{T^{\superscriptO} }_{{\mathrm{2}}}  
          }.
        \end{center}
    \end{itemize}
  \end{proof}
  \begin{lemma}\label{lem:partial-app-persistent}
    If \( \mathit{\Gamma}  \vdash^{0}   \openO{(}    \possiblyWithSub\stageOmetaColor{\Hat{c} }   \    \possiblyWithSub\stageOmetaColor{c}_{{\mathrm{1}}}   \ \cdots\    \possiblyWithSub\stageOmetaColor{c}_{\ottmv{k}}    \closeO{)}   :  \possiblyWithSub\stageOmetaColor{T^{\superscriptO} } \),
    \(m \defeq \arity{\Hat{c}} \geq 1\), and \(k \leq m\),
    then we have
    \begin{itemize}
      \item
        \(\ConstEnvPers(c_i) = \possiblyWithSub\stageImetaColor{\Hat{\tau}^{\superscriptI} }_i\) for each \(i \in \{1, \ldots, k\}\) and
      \item
        \(\possiblyWithSub\stageOmetaColor{T^{\superscriptO} } \equiv^0
          \mathord{\downarrow}(\possiblyWithSub\stageImetaColor{\Hat{\tau}^{\superscriptI} }_{k + 1} \relI{\to} \cdots \relI{\to}  \possiblyWithSub\stageImetaColor{\Hat{\tau}^{\superscriptI} }_{\ottmv{m}}  \relI{\to}  \possiblyWithSub\stageImetaColor{\Hat{\tau}^{\superscriptI} } )\),
    \end{itemize}
    where \(\ConstEnvPers(\Hat{c}) =   \possiblyWithSub\stageImetaColor{\Hat{\tau}^{\superscriptI} }_{{\mathrm{1}}}  \relI{\to}   \cdots \relI{\to}  \possiblyWithSub\stageImetaColor{\Hat{\tau}^{\superscriptI} }_{\ottmv{m}}    \relI{\to}  \possiblyWithSub\stageImetaColor{\Hat{\tau}^{\superscriptI} } \).
  \end{lemma}
  \begin{proof}
    By induction on the derivation of \( \mathit{\Gamma}  \vdash^{0}   \openO{(}    \possiblyWithSub\stageOmetaColor{\Hat{c} }   \    \possiblyWithSub\stageOmetaColor{c}_{{\mathrm{1}}}   \ \cdots\    \possiblyWithSub\stageOmetaColor{c}_{\ottmv{k}}    \closeO{)}   :  \possiblyWithSub\stageOmetaColor{T^{\superscriptO} } \).
    \begin{itemize}
      \item Case \derive[T0-CstP]{%
         \vdash  \mathit{\Gamma} 
      \andalso
        \ConstEnvPers(\Hat{c}) = \possiblyWithSub\stageImetaColor{\tau^{\superscriptI} }
      }{%
         \mathit{\Gamma}  \vdash^{0}    \possiblyWithSub\stageOmetaColor{\Hat{c} }    :   \mathop{\downarrow}( \possiblyWithSub\stageImetaColor{\tau^{\superscriptI} } )  
      }:
        We have \(k = 0\), and since \(\possiblyWithSub\stageOmetaColor{T^{\superscriptO} } =  \mathop{\downarrow}( \possiblyWithSub\stageImetaColor{\tau^{\superscriptI} } ) \),
        we can immediately finish the proof by deriving
        \derive[CqT0-Refl]{}{%
            \mathop{\downarrow}( \possiblyWithSub\stageImetaColor{\tau^{\superscriptI} } )   \equiv^{0}   \mathop{\downarrow}( \possiblyWithSub\stageImetaColor{\tau^{\superscriptI} } )  
        }.
      \item Case \rulename{T0-RfnPred} contradicts the assumption~\(\arity{\Hat{c}} \geq 1\).
      \item Case \derive[T0-TyEquiv]{%
         \mathit{\Gamma}  \vdash^{0}    \possiblyWithSub\stageOmetaColor{\Hat{c} }    :  \possiblyWithSub\stageOmetaColor{T'^{\superscriptO} } 
      \andalso
         \possiblyWithSub\stageOmetaColor{T'^{\superscriptO} }  \equiv^{0}  \possiblyWithSub\stageOmetaColor{T^{\superscriptO} } 
      \andalso
         \mathit{\Gamma}  \vdash^{0}  \possiblyWithSub\stageOmetaColor{T^{\superscriptO} } 
      }{%
         \mathit{\Gamma}  \vdash^{0}    \possiblyWithSub\stageOmetaColor{\Hat{c} }    :  \possiblyWithSub\stageOmetaColor{T^{\superscriptO} } 
      }:
        Immediate by IH on \( \mathit{\Gamma}  \vdash^{0}    \possiblyWithSub\stageOmetaColor{\Hat{c} }    :  \possiblyWithSub\stageOmetaColor{T'^{\superscriptO} } \) and \rulename{CqT0-Trans}.
      \item Case \derive[T0-App]{%
        \mathit{\Gamma} \vdash^0 \openO{(}\possiblyWithSub\stageOmetaColor{\Hat{c} }\ \possiblyWithSub\stageOmetaColor{c}_{{\mathrm{1}}}\ \cdots\ \possiblyWithSub\stageOmetaColor{c}_{k - 1}\closeO{)} :  \openO{(} \possiblyWithSub\stageOmetaColor{x}  \relO{:}  \possiblyWithSub\stageOmetaColor{T'^{\superscriptO} }_{{\mathrm{1}}} \closeO{)} \relO{\to}  \possiblyWithSub\stageOmetaColor{T'^{\superscriptO} }_{{\mathrm{2}}} 
      \andalso
         \mathit{\Gamma}  \vdash^{0}    \possiblyWithSub\stageOmetaColor{c}_{\ottmv{k}}    :  \possiblyWithSub\stageOmetaColor{T'^{\superscriptO} }_{{\mathrm{1}}} 
      }{%
         \mathit{\Gamma}  \vdash^{0}   \openO{(}    \possiblyWithSub\stageOmetaColor{\Hat{c} }   \    \possiblyWithSub\stageOmetaColor{c}_{{\mathrm{1}}}   \ \cdots\    \possiblyWithSub\stageOmetaColor{c}_{\ottmv{k}}    \closeO{)}   :    [    \possiblyWithSub\stageOmetaColor{c}_{\ottmv{k}}    /  \possiblyWithSub\stageOmetaColor{x}  ]    \possiblyWithSub\stageOmetaColor{T'^{\superscriptO} }_{{\mathrm{2}}}  
      }:
        By IH, we have
        \begin{itemize}
          \item
            \(\ConstEnvPers(c_i) = \possiblyWithSub\stageImetaColor{\Hat{\tau}^{\superscriptI} }_i\) for each \(i \in \{1, \ldots, k - 1\}\) and
          \item
            \( \openO{(} \possiblyWithSub\stageOmetaColor{x}  \relO{:}  \possiblyWithSub\stageOmetaColor{T'^{\superscriptO} }_{{\mathrm{1}}} \closeO{)} \relO{\to}  \possiblyWithSub\stageOmetaColor{T'^{\superscriptO} }_{{\mathrm{2}}}  \equiv^0
              \mathord{\downarrow}(\possiblyWithSub\stageImetaColor{\Hat{\tau}^{\superscriptI} }_{\ottmv{k}} \relI{\to} \possiblyWithSub\stageImetaColor{\Hat{\tau}^{\superscriptI} }_{k + 1} \relI{\to}
                \cdots \relI{\to}  \possiblyWithSub\stageImetaColor{\Hat{\tau}^{\superscriptI} }_{\ottmv{m}}  \relI{\to}  \possiblyWithSub\stageImetaColor{\Hat{\tau}^{\superscriptI} } )\).
        \end{itemize}
        Then, by Lemma~\ref{lem:arrow-type-csr-equiv-inversion},
        we have \( \possiblyWithSub\stageOmetaColor{T'^{\superscriptO} }_{{\mathrm{1}}}  \equiv^{0}   \mathop{\downarrow}( \possiblyWithSub\stageImetaColor{\Hat{\tau}^{\superscriptI} }_{\ottmv{k}} )  \) and
        \(\possiblyWithSub\stageOmetaColor{T'^{\superscriptO} }_{{\mathrm{2}}} \equiv^0
          \mathord{\downarrow}(\possiblyWithSub\stageImetaColor{\Hat{\tau}^{\superscriptI} }_{k + 1} \relI{\to} \cdots \relI{\to}  \possiblyWithSub\stageImetaColor{\Hat{\tau}^{\superscriptI} }_{\ottmv{m}}  \relI{\to}  \possiblyWithSub\stageImetaColor{\Hat{\tau}^{\superscriptI} } )\).
        Here, by \( \possiblyWithSub\stageOmetaColor{T'^{\superscriptO} }_{{\mathrm{1}}}  \equiv^{0}   \mathop{\downarrow}( \possiblyWithSub\stageImetaColor{\Hat{\tau}^{\superscriptI} }_{\ottmv{k}} )  \) and Lemma~\ref{lem:csr-equiv-unlift-inversion},
        we have \( \possiblyWithSub\stageImetaColor{\Hat{\tau}^{\superscriptI} }_{\ottmv{k}}  \gg  \possiblyWithSub\stageOmetaColor{T'^{\superscriptO} }_{{\mathrm{1}}} \),
        and then by \( \mathit{\Gamma}  \vdash^{0}    \possiblyWithSub\stageOmetaColor{c}_{\ottmv{k}}    :  \possiblyWithSub\stageOmetaColor{T'^{\superscriptO} }_{{\mathrm{1}}} \) and Lemma~\ref{lem:persistent-const-inversion},
        we have \(\ConstEnvPers(c_{\ottmv{k}}) = \possiblyWithSub\stageImetaColor{\Hat{\tau}^{\superscriptI} }_{\ottmv{k}}\).
        Therefore, we first have
        \(\ConstEnvPers(c_i) = \possiblyWithSub\stageImetaColor{\Hat{\tau}^{\superscriptI} }_i\) for each \(i \in \{1, \ldots, k\}\).
        Also, by Lemma~\ref{lem:subst-preserves-csr-equiv},
        we have
        \(\possiblyWithSub\stageOmetaColor{T^{\superscriptO} } =   [    \possiblyWithSub\stageOmetaColor{c}_{\ottmv{k}}    /  \possiblyWithSub\stageOmetaColor{x}  ]    \possiblyWithSub\stageOmetaColor{T'^{\superscriptO} }_{{\mathrm{2}}}  \equiv^0
          [ \possiblyWithSub\stageOmetaColor{c}_{\ottmv{k}} / \possiblyWithSub\stageOmetaColor{x} ]\mathord{\downarrow}(\possiblyWithSub\stageImetaColor{\Hat{\tau}^{\superscriptI} }_{k + 1} \relI{\to}
            \cdots \relI{\to}  \possiblyWithSub\stageImetaColor{\Hat{\tau}^{\superscriptI} }_{\ottmv{m}}  \relI{\to}  \possiblyWithSub\stageImetaColor{\Hat{\tau}^{\superscriptI} } )
          = \mathord{\downarrow}(\possiblyWithSub\stageImetaColor{\Hat{\tau}^{\superscriptI} }_{k + 1} \relI{\to} \cdots \relI{\to}  \possiblyWithSub\stageImetaColor{\Hat{\tau}^{\superscriptI} }_{\ottmv{m}}  \relI{\to}  \possiblyWithSub\stageImetaColor{\Hat{\tau}^{\superscriptI} } )\).
      \item The other cases contradict the form~\( \openO{(}    \possiblyWithSub\stageOmetaColor{\Hat{c} }   \    \possiblyWithSub\stageOmetaColor{c}_{{\mathrm{1}}}   \ \cdots\    \possiblyWithSub\stageOmetaColor{c}_{\ottmv{k}}    \closeO{)} \).
    \end{itemize}
  \end{proof}
  \begin{lemma}\label{lem:partial-app-zero}
    If \( \mathit{\Gamma}  \vdash^{0}   \openO{(}    \possiblyWithSub\stageOmetaColor{p}   \    \possiblyWithSub\stageOmetaColor{c}_{{\mathrm{1}}}   \ \cdots\    \possiblyWithSub\stageOmetaColor{c}_{\ottmv{k}}    \closeO{)}   :  \possiblyWithSub\stageOmetaColor{T^{\superscriptO} } \) and \(k \leq m \defeq \arity{\possiblyWithSub\stageOmetaColor{p}}\),
    then we have
    \begin{itemize}
      \item
        \(\possiblyWithSub\stageOmetaColor{c}_{\ottmv{i}} \vDash [ \possiblyWithSub\stageOmetaColor{c}_{i - 1} / \possiblyWithSub\stageOmetaColor{x}_{i - 1} ] \cdots [ \possiblyWithSub\stageOmetaColor{c}_{{\mathrm{1}}} / \possiblyWithSub\stageOmetaColor{x}_{{\mathrm{1}}} ] \possiblyWithSub\stageOmetaColor{\Hat{T}^{\superscriptO} }_{\ottmv{i}}\)
        for each \(i \in \{1, \ldots, k\}\) and
      \item
        \(\possiblyWithSub\stageOmetaColor{T^{\superscriptO} } \equiv^0
          [ \possiblyWithSub\stageOmetaColor{c}_{\ottmv{k}} / \possiblyWithSub\stageOmetaColor{x}_{\ottmv{k}} ] \cdots [ \possiblyWithSub\stageOmetaColor{c}_{{\mathrm{1}}} / \possiblyWithSub\stageOmetaColor{x}_{{\mathrm{1}}} ]\openO{(}
            \openO{(}\possiblyWithSub\stageOmetaColor{x}_{k + 1} \relO{:} \possiblyWithSub\stageOmetaColor{\Hat{T}^{\superscriptO} }_{k + 1}\closeO{)} \relO{\to}
            \cdots \relO{\to}
             \openO{(} \possiblyWithSub\stageOmetaColor{x}_{\ottmv{m}}  \relO{:}  \possiblyWithSub\stageOmetaColor{\Hat{T}^{\superscriptO} }_{\ottmv{m}} \closeO{)} \relO{\to}  \possiblyWithSub\stageOmetaColor{\Hat{T}^{\superscriptO} } 
          \closeO{)}\),
    \end{itemize}
    where \(\ConstEnvZero(\possiblyWithSub\stageOmetaColor{p}) =  \openO{(} \possiblyWithSub\stageOmetaColor{x}_{{\mathrm{1}}}  \relO{:}  \possiblyWithSub\stageOmetaColor{\Hat{T}^{\superscriptO} }_{{\mathrm{1}}} \closeO{)} \relO{\to}   \cdots \relO{\to}   \openO{(} \possiblyWithSub\stageOmetaColor{x}_{\ottmv{m}}  \relO{:}  \possiblyWithSub\stageOmetaColor{\Hat{T}^{\superscriptO} }_{\ottmv{m}} \closeO{)} \relO{\to}  \possiblyWithSub\stageOmetaColor{\Hat{T}^{\superscriptO} }   \).
  \end{lemma}
  \begin{proof}
    By induction on the derivation of \( \mathit{\Gamma}  \vdash^{0}   \openO{(}    \possiblyWithSub\stageOmetaColor{p}   \    \possiblyWithSub\stageOmetaColor{c}_{{\mathrm{1}}}   \ \cdots\    \possiblyWithSub\stageOmetaColor{c}_{\ottmv{k}}    \closeO{)}   :  \possiblyWithSub\stageOmetaColor{T^{\superscriptO} } \).
    \begin{itemize}
      \item Case \derive[T0-Cst0]{%
         \vdash  \mathit{\Gamma} 
      \andalso
        \ConstEnvZero(\possiblyWithSub\stageOmetaColor{p}) = \possiblyWithSub\stageOmetaColor{T^{\superscriptO} }
      }{%
         \mathit{\Gamma}  \vdash^{0}    \possiblyWithSub\stageOmetaColor{p}    :  \possiblyWithSub\stageOmetaColor{T^{\superscriptO} } 
      }:
        We have \(k = 0\), and since \(\possiblyWithSub\stageOmetaColor{\Hat{T}^{\superscriptO} } = \possiblyWithSub\stageOmetaColor{T^{\superscriptO} }\),
        we can immediately finish the proof by deriving
        \derive[CqT0-Refl]{}{%
           \possiblyWithSub\stageOmetaColor{\Hat{T}^{\superscriptO} }  \equiv^{0}  \possiblyWithSub\stageOmetaColor{\Hat{T}^{\superscriptO} } .
        }
      \item Case \derive[T0-TyEquiv]{%
         \mathit{\Gamma}  \vdash^{0}   \openO{(}    \possiblyWithSub\stageOmetaColor{p}   \    \possiblyWithSub\stageOmetaColor{c}_{{\mathrm{1}}}   \ \cdots\    \possiblyWithSub\stageOmetaColor{c}_{\ottmv{k}}    \closeO{)}   :  \possiblyWithSub\stageOmetaColor{T'^{\superscriptO} } 
      \andalso
         \possiblyWithSub\stageOmetaColor{T'^{\superscriptO} }  \equiv^{0}  \possiblyWithSub\stageOmetaColor{T^{\superscriptO} } 
      \andalso
         \mathit{\Gamma}  \vdash^{0}  \possiblyWithSub\stageOmetaColor{T^{\superscriptO} } 
      }{%
         \mathit{\Gamma}  \vdash^{0}   \openO{(}    \possiblyWithSub\stageOmetaColor{p}   \    \possiblyWithSub\stageOmetaColor{c}_{{\mathrm{1}}}   \ \cdots\    \possiblyWithSub\stageOmetaColor{c}_{\ottmv{k}}    \closeO{)}   :  \possiblyWithSub\stageOmetaColor{T^{\superscriptO} } 
      }:
        Immediate by IH and \rulename{CqT0-Trans}.
      \item Case \derive[T0-App]{%
        \mathit{\Gamma} \vdash^0 \openO{(}\possiblyWithSub\stageOmetaColor{p}\ \possiblyWithSub\stageOmetaColor{c}_{{\mathrm{1}}}\ \cdots\ \possiblyWithSub\stageOmetaColor{c}_{k - 1}\closeO{)} :  \openO{(} \possiblyWithSub\stageOmetaColor{x}  \relO{:}  \possiblyWithSub\stageOmetaColor{T'^{\superscriptO} }_{{\mathrm{1}}} \closeO{)} \relO{\to}  \possiblyWithSub\stageOmetaColor{T'^{\superscriptO} }_{{\mathrm{2}}} 
      \andalso
         \mathit{\Gamma}  \vdash^{0}    \possiblyWithSub\stageOmetaColor{c}_{\ottmv{k}}    :  \possiblyWithSub\stageOmetaColor{T'^{\superscriptO} }_{{\mathrm{1}}} 
      }{%
         \mathit{\Gamma}  \vdash^{0}   \openO{(}    \possiblyWithSub\stageOmetaColor{\Hat{c} }   \    \possiblyWithSub\stageOmetaColor{c}_{{\mathrm{1}}}   \ \cdots\    \possiblyWithSub\stageOmetaColor{c}_{\ottmv{k}}    \closeO{)}   :    [    \possiblyWithSub\stageOmetaColor{c}_{\ottmv{k}}    /  \possiblyWithSub\stageOmetaColor{x}  ]    \possiblyWithSub\stageOmetaColor{T'^{\superscriptO} }_{{\mathrm{2}}}  
      }:
        By IH, we have
        \begin{itemize}
          \item
            \(\possiblyWithSub\stageOmetaColor{c}_{\ottmv{i}} \vDash [ \possiblyWithSub\stageOmetaColor{c}_{i - 1} / \possiblyWithSub\stageOmetaColor{x}_{i - 1} ] \cdots [ \possiblyWithSub\stageOmetaColor{c}_{{\mathrm{1}}} / \possiblyWithSub\stageOmetaColor{x}_{{\mathrm{1}}} ] \possiblyWithSub\stageOmetaColor{\Hat{T}^{\superscriptO} }_{\ottmv{i}}\)
            for each \(i \in \{1, \ldots, k - 1\}\) and
          \item
            \( \openO{(} \possiblyWithSub\stageOmetaColor{x}  \relO{:}  \possiblyWithSub\stageOmetaColor{T'^{\superscriptO} }_{{\mathrm{1}}} \closeO{)} \relO{\to}  \possiblyWithSub\stageOmetaColor{T'^{\superscriptO} }_{{\mathrm{2}}}  \equiv^0
              [ \possiblyWithSub\stageOmetaColor{c}_{k - 1} / \possiblyWithSub\stageOmetaColor{x}_{k - 1} ] \cdots [ \possiblyWithSub\stageOmetaColor{c}_{{\mathrm{1}}} / \possiblyWithSub\stageOmetaColor{x}_{{\mathrm{1}}} ]\openO{(}
                 \openO{(} \possiblyWithSub\stageOmetaColor{x}_{\ottmv{k}}  \relO{:}  \possiblyWithSub\stageOmetaColor{\Hat{T}^{\superscriptO} }_{\ottmv{k}} \closeO{)} \relO{\to}   \cdots \relO{\to}   \openO{(} \possiblyWithSub\stageOmetaColor{x}_{\ottmv{m}}  \relO{:}  \possiblyWithSub\stageOmetaColor{\Hat{T}^{\superscriptO} }_{\ottmv{m}} \closeO{)} \relO{\to}  \possiblyWithSub\stageOmetaColor{\Hat{T}^{\superscriptO} }   
              \closeO{)}\).
        \end{itemize}
        Then, by Lemma~\ref{lem:arrow-type-csr-equiv-inversion}
        and the Barendregt convention, which enables us to assume \(\possiblyWithSub\stageOmetaColor{x} = \possiblyWithSub\stageOmetaColor{x}_{\ottmv{k}}\) safely,
        we have
        \begin{itemize}
          \item
            \(\possiblyWithSub\stageOmetaColor{T'^{\superscriptO} }_{{\mathrm{1}}} \equiv^0
              [ \possiblyWithSub\stageOmetaColor{c}_{k - 1} / \possiblyWithSub\stageOmetaColor{x}_{k - 1} ] \cdots [ \possiblyWithSub\stageOmetaColor{c}_{{\mathrm{1}}} / \possiblyWithSub\stageOmetaColor{x}_{{\mathrm{1}}} ] \possiblyWithSub\stageOmetaColor{\Hat{T}^{\superscriptO} }_{\ottmv{k}}\)
            and
          \item
            \(\possiblyWithSub\stageOmetaColor{T'^{\superscriptO} }_{{\mathrm{2}}} \equiv^0
              [ \possiblyWithSub\stageOmetaColor{c}_{k - 1} / \possiblyWithSub\stageOmetaColor{x}_{k - 1} ] \cdots [ \possiblyWithSub\stageOmetaColor{c}_{{\mathrm{1}}} / \possiblyWithSub\stageOmetaColor{x}_{{\mathrm{1}}} ]\openO{(}
                \openO{(}\possiblyWithSub\stageOmetaColor{x}_{k + 1} \relO{:} \possiblyWithSub\stageOmetaColor{\Hat{T}^{\superscriptO} }_{k + 1}\closeO{)} \relO{\to}
                \cdots \relO{\to}
                 \openO{(} \possiblyWithSub\stageOmetaColor{x}_{\ottmv{m}}  \relO{:}  \possiblyWithSub\stageOmetaColor{\Hat{T}^{\superscriptO} }_{\ottmv{m}} \closeO{)} \relO{\to}  \possiblyWithSub\stageOmetaColor{\Hat{T}^{\superscriptO} } 
              \closeO{)}\).
        \end{itemize}
        Here, by \( \mathit{\Gamma}  \vdash^{0}    \possiblyWithSub\stageOmetaColor{c}_{\ottmv{k}}    :  \possiblyWithSub\stageOmetaColor{T'^{\superscriptO} }_{{\mathrm{1}}} \) and Lemma~\ref{lem:const-typing-implies-predicate-satisfaction},
        we have \(  \possiblyWithSub\stageOmetaColor{c}_{\ottmv{k}}   \vDash  \possiblyWithSub\stageOmetaColor{T'^{\superscriptO} }_{{\mathrm{1}}} \), and then by Lemma~\ref{lem:equiv-preserves-rfn},
        we have
        \(\possiblyWithSub\stageOmetaColor{c}_{\ottmv{k}} \vDash [ \possiblyWithSub\stageOmetaColor{c}_{k - 1} / \possiblyWithSub\stageOmetaColor{x}_{k - 1} ] \cdots [ \possiblyWithSub\stageOmetaColor{c}_{{\mathrm{1}}} / \possiblyWithSub\stageOmetaColor{x}_{{\mathrm{1}}} ] \possiblyWithSub\stageOmetaColor{\Hat{T}^{\superscriptO} }_{\ottmv{k}}\).
        Therefore, we first have
        \(\possiblyWithSub\stageOmetaColor{c}_{\ottmv{i}} \vDash [ \possiblyWithSub\stageOmetaColor{c}_{i - 1} / \possiblyWithSub\stageOmetaColor{x}_{i - 1} ] \cdots [ \possiblyWithSub\stageOmetaColor{c}_{{\mathrm{1}}} / \possiblyWithSub\stageOmetaColor{x}_{{\mathrm{1}}} ] \possiblyWithSub\stageOmetaColor{\Hat{T}^{\superscriptO} }_i\)
        for each \(i \in \{1, \ldots, k\}\).
        Also, by Lemma~\ref{lem:subst-preserves-csr-equiv},
        we have
        \(\possiblyWithSub\stageOmetaColor{T^{\superscriptO} } =   [    \possiblyWithSub\stageOmetaColor{c}_{\ottmv{k}}    /  \possiblyWithSub\stageOmetaColor{x}  ]    \possiblyWithSub\stageOmetaColor{T'^{\superscriptO} }_{{\mathrm{2}}} 
          \equiv^0
            [ \possiblyWithSub\stageOmetaColor{c}_{\ottmv{k}} / \possiblyWithSub\stageOmetaColor{x}_{\ottmv{k}} ]
              [ \possiblyWithSub\stageOmetaColor{c}_{k - 1} / \possiblyWithSub\stageOmetaColor{x}_{k - 1} ] \cdots [ \possiblyWithSub\stageOmetaColor{c}_{{\mathrm{1}}} / \possiblyWithSub\stageOmetaColor{x}_{{\mathrm{1}}} ]\openO{(}
                \openO{(}\possiblyWithSub\stageOmetaColor{x}_{k + 1} \relO{:} \possiblyWithSub\stageOmetaColor{\Hat{T}^{\superscriptO} }_{k + 1}\closeO{)} \relO{\to}
                \cdots \relO{\to}
                 \openO{(} \possiblyWithSub\stageOmetaColor{x}_{\ottmv{m}}  \relO{:}  \possiblyWithSub\stageOmetaColor{\Hat{T}^{\superscriptO} }_{\ottmv{m}} \closeO{)} \relO{\to}  \possiblyWithSub\stageOmetaColor{\Hat{T}^{\superscriptO} } 
              \closeO{)}
        \).
      \item The other cases contradict the form~\( \openO{(}    \possiblyWithSub\stageOmetaColor{p}   \    \possiblyWithSub\stageOmetaColor{c}_{{\mathrm{1}}}   \ \cdots\    \possiblyWithSub\stageOmetaColor{c}_{\ottmv{k}}    \closeO{)} \).
    \end{itemize}
  \end{proof}
  \begin{lemma}[Preservation on \(\delta\)-reduction of persistent built-in functions]\label{lem:preservation-of-delta-persistent}
    If \( \mathit{\Gamma}  \vdash^{0}   \openO{(}    \possiblyWithSub\stageOmetaColor{\Hat{c} }   \    \possiblyWithSub\stageOmetaColor{c}_{{\mathrm{1}}}   \ \cdots\    \possiblyWithSub\stageOmetaColor{c}_{\ottmv{m}}    \closeO{)}   :  \possiblyWithSub\stageOmetaColor{T^{\superscriptO} } \) and
    \(\delta(\possiblyWithSub\stageOmetaColor{\Hat{c} }, (\possiblyWithSub\stageOmetaColor{c}_{{\mathrm{1}}}, \ldots, \possiblyWithSub\stageOmetaColor{c}_{\ottmv{m}})) = \possiblyWithSub\stageOmetaColor{q}\),
    then \( \mathit{\Gamma}  \vdash^{0}   \possiblyWithSub\stageOmetaColor{q}   :  \possiblyWithSub\stageOmetaColor{T^{\superscriptO} } \).
  \end{lemma}
  \begin{proof}
    By Assumption~\ref{assump:type-of-constants}, we have \(m = \arity{\Hat{c}} \geq 1\).
    Let \(\ConstEnvPers(\Hat{c}) \revdefeq   \possiblyWithSub\stageImetaColor{\Hat{\tau}^{\superscriptI} }_{{\mathrm{1}}}  \relI{\to}   \cdots \relI{\to}  \possiblyWithSub\stageImetaColor{\Hat{\tau}^{\superscriptI} }_{\ottmv{m}}    \relI{\to}  \possiblyWithSub\stageImetaColor{\Hat{\tau}^{\superscriptI} } \).
    Then, by Lemma~\ref{lem:partial-app-persistent} with \(k \defeq m\), we have
    \(\ConstEnvPers(c_i) = \possiblyWithSub\stageImetaColor{\Hat{\tau}^{\superscriptI} }_i\) for each \(i \in \{1, \ldots, m\}\) and
    \(  \mathop{\downarrow}( \possiblyWithSub\stageImetaColor{\Hat{\tau}^{\superscriptI} } )   \equiv^{0}  \possiblyWithSub\stageOmetaColor{T^{\superscriptO} } \).
    Thus, again by Assumption~\ref{assump:type-of-constants},
    \(\possiblyWithSub\stageOmetaColor{q}\) is of the form~\(\possiblyWithSub\stageOmetaColor{c}\),
    and this \(c\) satisfies \(\ConstEnvPers(c) = \possiblyWithSub\stageImetaColor{\Hat{\tau}^{\superscriptI} }\).
    Since \( \mathit{\Gamma}  \vdash^{0}  \possiblyWithSub\stageOmetaColor{T^{\superscriptO} } \) holds by \( \mathit{\Gamma}  \vdash^{0}   \openO{(}    \possiblyWithSub\stageOmetaColor{\Hat{c} }   \    \possiblyWithSub\stageOmetaColor{c}_{{\mathrm{1}}}   \ \cdots\    \possiblyWithSub\stageOmetaColor{c}_{\ottmv{m}}    \closeO{)}   :  \possiblyWithSub\stageOmetaColor{T^{\superscriptO} } \) and
    Lemma~\ref{lem:target-typing-implies-type-well-formedness},
    we can derive
    \begin{center}
      \infer[T0-TyEquiv]{
        \infer[T0-CstP]{%
          \ConstEnvPers(c) = \possiblyWithSub\stageImetaColor{\Hat{\tau}^{\superscriptI} }
        }{%
           \mathit{\Gamma}  \vdash^{0}    \possiblyWithSub\stageOmetaColor{c}    :   \mathop{\downarrow}( \possiblyWithSub\stageImetaColor{\Hat{\tau}^{\superscriptI} } )  
        }
      \andalso
          \mathop{\downarrow}( \possiblyWithSub\stageImetaColor{\Hat{\tau}^{\superscriptI} } )   \equiv^{0}  \possiblyWithSub\stageOmetaColor{T^{\superscriptO} } 
      \andalso
         \mathit{\Gamma}  \vdash^{0}  \possiblyWithSub\stageOmetaColor{T^{\superscriptO} } 
      }{%
         \mathit{\Gamma}  \vdash^{0}    \possiblyWithSub\stageOmetaColor{c}    :  \possiblyWithSub\stageOmetaColor{T^{\superscriptO} } 
      }.
    \end{center}
  \end{proof}
  \begin{lemma}[Preservation on \(\delta\)-reduction of stage-0 built-in functions]\label{lem:preservation-of-delta-zero}
    If \( \mathit{\Gamma}  \vdash^{0}   \openO{(}    \possiblyWithSub\stageOmetaColor{p}   \    \possiblyWithSub\stageOmetaColor{c}_{{\mathrm{1}}}   \ \cdots\    \possiblyWithSub\stageOmetaColor{c}_{\ottmv{m}}    \closeO{)}   :  \possiblyWithSub\stageOmetaColor{T^{\superscriptO} } \) and
    \(\delta(\possiblyWithSub\stageOmetaColor{p}, (\possiblyWithSub\stageOmetaColor{c}_{{\mathrm{1}}}, \ldots, \possiblyWithSub\stageOmetaColor{c}_{\ottmv{m}})) = \possiblyWithSub\stageOmetaColor{q}\),
    then \( \mathit{\Gamma}  \vdash^{0}   \possiblyWithSub\stageOmetaColor{q}   :  \possiblyWithSub\stageOmetaColor{T^{\superscriptO} } \).
  \end{lemma}
  \begin{proof}
    We first have \( \mathit{\Gamma}  \vdash^{0}  \possiblyWithSub\stageOmetaColor{T^{\superscriptO} } \) by Lemma~\ref{lem:target-typing-implies-type-well-formedness}.
    By Assumption~\ref{assump:type-of-constants}, we have \(m = \arity{\possiblyWithSub\stageOmetaColor{p}}\).
    Let \(\ConstEnvZero(\possiblyWithSub\stageOmetaColor{p}) \revdefeq  \openO{(} \possiblyWithSub\stageOmetaColor{x}_{{\mathrm{1}}}  \relO{:}  \possiblyWithSub\stageOmetaColor{\Hat{T}^{\superscriptO} }_{{\mathrm{1}}} \closeO{)} \relO{\to}   \cdots \relO{\to}   \openO{(} \possiblyWithSub\stageOmetaColor{x}_{\ottmv{m}}  \relO{:}  \possiblyWithSub\stageOmetaColor{\Hat{T}^{\superscriptO} }_{\ottmv{m}} \closeO{)} \relO{\to}  \possiblyWithSub\stageOmetaColor{\Hat{T}^{\superscriptO} }   \).
    By Lemma~\ref{lem:partial-app-zero} with \(k \defeq m\),
    we have
    \(\possiblyWithSub\stageOmetaColor{c}_{\ottmv{i}} \vDash [ \possiblyWithSub\stageOmetaColor{c}_{i - 1} / \possiblyWithSub\stageOmetaColor{x}_{i - 1} ] \cdots [ \possiblyWithSub\stageOmetaColor{c}_{{\mathrm{1}}} / \possiblyWithSub\stageOmetaColor{x}_{{\mathrm{1}}} ] \possiblyWithSub\stageOmetaColor{\Hat{T}^{\superscriptO} }_{\ottmv{i}}\)
    for each \(i \in \{1, \ldots, m\}\) and
    \(\possiblyWithSub\stageOmetaColor{T^{\superscriptO} } \equiv^0 [ \possiblyWithSub\stageOmetaColor{c}_{\ottmv{m}} / \possiblyWithSub\stageOmetaColor{x}_{\ottmv{m}} ] \cdots [ \possiblyWithSub\stageOmetaColor{c}_{{\mathrm{1}}} / \possiblyWithSub\stageOmetaColor{x}_{{\mathrm{1}}} ] \possiblyWithSub\stageOmetaColor{\Hat{T}^{\superscriptO} }\).
    Then, by Assumption~\ref{assump:type-of-constants} and repeated use of Lemma~\ref{lem:subst},
    we have \(\mathit{\Gamma} \vdash^0 [ \possiblyWithSub\stageOmetaColor{c}_{\ottmv{m}} / \possiblyWithSub\stageOmetaColor{x}_{\ottmv{m}} ] \cdots [ \possiblyWithSub\stageOmetaColor{c}_{{\mathrm{1}}} / \possiblyWithSub\stageOmetaColor{x}_{{\mathrm{1}}} ] \possiblyWithSub\stageOmetaColor{\Hat{T}^{\superscriptO} }\).
    Also by Assumption~\ref{assump:type-of-constants},
    we have \(\possiblyWithSub\stageOmetaColor{q} \vDash [ \possiblyWithSub\stageOmetaColor{c}_{\ottmv{m}} / \possiblyWithSub\stageOmetaColor{x}_{\ottmv{m}} ] \cdots [ \possiblyWithSub\stageOmetaColor{c}_{{\mathrm{1}}} / \possiblyWithSub\stageOmetaColor{x}_{{\mathrm{1}}} ] \possiblyWithSub\stageOmetaColor{\Hat{T}^{\superscriptO} }\).
    We do a case analysis on this derivation:
    \begin{itemize}
      \item Case \derive{%
        \ConstEnvPers(c) =  \ttI{Tensor}\ \ordI{\%} \possiblyWithSub\stageOmetaColor{s} 
      }{%
          \possiblyWithSub\stageOmetaColor{c}   \vDash   \ttO{Tensor}\  \possiblyWithSub\stageOmetaColor{s}  
      }:
        Since \([ \possiblyWithSub\stageOmetaColor{c}_{\ottmv{m}} / \possiblyWithSub\stageOmetaColor{x}_{\ottmv{m}} ] \cdots [ \possiblyWithSub\stageOmetaColor{c}_{{\mathrm{1}}} / \possiblyWithSub\stageOmetaColor{x}_{{\mathrm{1}}} ] \possiblyWithSub\stageOmetaColor{\Hat{T}^{\superscriptO} } =  \ttO{Tensor}\  \possiblyWithSub\stageOmetaColor{s} \),
        we clearly have \(\possiblyWithSub\stageOmetaColor{\Hat{T}^{\superscriptO} } =  \ttO{Tensor}\  \possiblyWithSub\stageOmetaColor{s} \).
        Then, by \( \possiblyWithSub\stageOmetaColor{T^{\superscriptO} }  \equiv^{0}   \ttO{Tensor}\  \possiblyWithSub\stageOmetaColor{s}  \) and Lemma~\ref{lem:tensor-type-csr-equiv-form},
        we have \(\possiblyWithSub\stageOmetaColor{T^{\superscriptO} } =  \ttO{Tensor}\  \possiblyWithSub\stageOmetaColor{s} \).
        Therefore, we can immediately derive \( \mathit{\Gamma}  \vdash^{0}   \possiblyWithSub\stageOmetaColor{q}   :  \possiblyWithSub\stageOmetaColor{T^{\superscriptO} } \) by
        \derive[T0-CstP]{%
          \ConstEnvPers(c) =  \ttI{Tensor}\ \ordI{\%} \possiblyWithSub\stageOmetaColor{s} 
        }{%
           \mathit{\Gamma}  \vdash^{0}    \possiblyWithSub\stageOmetaColor{c}    :   \ttO{Tensor}\  \possiblyWithSub\stageOmetaColor{s}  
        }.
      \item Case \derive{%
        \ConstEnvPers(c) = \possiblyWithSub\stageImetaColor{B}
      \andalso
           [    \possiblyWithSub\stageOmetaColor{c}    /  \possiblyWithSub\stageOmetaColor{\nu}  ]    \possiblyWithSub\stageOmetaColor{N^{\superscriptO} }   \longrightarrow^{0\,\ast}      \ttO{true}     
      }{%
          \possiblyWithSub\stageOmetaColor{c}   \vDash    \openO{\{} \possiblyWithSub\stageOmetaColor{\nu}  \relO{:}  \possiblyWithSub\stageOmetaColor{B}  \relO{\mid}  \possiblyWithSub\stageOmetaColor{N^{\superscriptO} } \closeO{\} }   
      }:
        Since \( \possiblyWithSub\stageOmetaColor{T^{\superscriptO} }  \equiv^{0}    \openO{\{} \possiblyWithSub\stageOmetaColor{\nu}  \relO{:}  \possiblyWithSub\stageOmetaColor{B}  \relO{\mid}  \possiblyWithSub\stageOmetaColor{N^{\superscriptO} } \closeO{\} }   \) holds,
        we can derive \( \mathit{\Gamma}  \vdash^{0}   \possiblyWithSub\stageOmetaColor{q}   :  \possiblyWithSub\stageOmetaColor{T^{\superscriptO} } \) by the following two steps:
        \begin{center}
          \derive[T0-RfnPred]{%
             \mathit{\Gamma}  \vdash^{0}    \openO{\{} \possiblyWithSub\stageOmetaColor{\nu}  \relO{:}  \possiblyWithSub\stageOmetaColor{B}  \relO{\mid}  \possiblyWithSub\stageOmetaColor{N^{\superscriptO} } \closeO{\} }   
          \andalso
            \ConstEnvPers(c) = \possiblyWithSub\stageImetaColor{B}
          \andalso
               [    \possiblyWithSub\stageOmetaColor{c}    /  \possiblyWithSub\stageOmetaColor{\nu}  ]    \possiblyWithSub\stageOmetaColor{N^{\superscriptO} }   \longrightarrow^{0\,\ast}      \ttO{true}     
          }{%
             \mathit{\Gamma}  \vdash^{0}    \possiblyWithSub\stageOmetaColor{c}    :    \openO{\{} \possiblyWithSub\stageOmetaColor{\nu}  \relO{:}  \possiblyWithSub\stageOmetaColor{B}  \relO{\mid}  \possiblyWithSub\stageOmetaColor{N^{\superscriptO} } \closeO{\} }   
          }
        \end{center}
        and then
        \begin{center}
          \infer[T0-TyEquiv]{%
             \mathit{\Gamma}  \vdash^{0}    \possiblyWithSub\stageOmetaColor{c}    :    \openO{\{} \possiblyWithSub\stageOmetaColor{\nu}  \relO{:}  \possiblyWithSub\stageOmetaColor{B}  \relO{\mid}  \possiblyWithSub\stageOmetaColor{N^{\superscriptO} } \closeO{\} }   
          \andalso
            \infer[CqT0-Sym]{%
               \possiblyWithSub\stageOmetaColor{T^{\superscriptO} }  \equiv^{0}    \openO{\{} \possiblyWithSub\stageOmetaColor{\nu}  \relO{:}  \possiblyWithSub\stageOmetaColor{B}  \relO{\mid}  \possiblyWithSub\stageOmetaColor{N^{\superscriptO} } \closeO{\} }   
            }{%
                 \openO{\{} \possiblyWithSub\stageOmetaColor{\nu}  \relO{:}  \possiblyWithSub\stageOmetaColor{B}  \relO{\mid}  \possiblyWithSub\stageOmetaColor{N^{\superscriptO} } \closeO{\} }    \equiv^{0}  \possiblyWithSub\stageOmetaColor{T^{\superscriptO} } 
            }
          \andalso
             \mathit{\Gamma}  \vdash^{0}  \possiblyWithSub\stageOmetaColor{T^{\superscriptO} } 
          }{%
             \mathit{\Gamma}  \vdash^{0}    \possiblyWithSub\stageOmetaColor{c}    :  \possiblyWithSub\stageOmetaColor{T^{\superscriptO} } 
          }.
        \end{center}
      \item Case \derive{%
        \ConstEnvPers(c) = \possiblyWithSub\stageImetaColor{\tau^{\superscriptI} }
      \andalso
         \possiblyWithSub\stageImetaColor{T^{\superscriptI} }  \longrightarrow^{1\,\ast}    \possiblyWithSub\stageImetaColor{\tau^{\superscriptI} }   
      }{%
          \openO{\langle} \possiblyWithSub\stageImetaColor{c} \closeO{\rangle}   \vDash   \openO{\langle} \possiblyWithSub\stageImetaColor{T^{\superscriptI} } \closeO{\rangle}  
      }:
        We have \( \possiblyWithSub\stageOmetaColor{T^{\superscriptO} }  \equiv^{0}   \openO{\langle} \possiblyWithSub\stageImetaColor{T^{\superscriptI} } \closeO{\rangle}  \).
        By \( \possiblyWithSub\stageImetaColor{T^{\superscriptI} }  \longrightarrow^{1\,\ast}    \possiblyWithSub\stageImetaColor{\tau^{\superscriptI} }   \) and Corollary~\ref{cor:reduction-implies-csr-equiv},
        \( \possiblyWithSub\stageImetaColor{T^{\superscriptI} }  \equiv^{1}   \possiblyWithSub\stageImetaColor{\tau^{\superscriptI} }  \) holds.
        Also, by the inversion of \( \mathit{\Gamma}  \vdash^{0}   \openO{\langle} \possiblyWithSub\stageImetaColor{T^{\superscriptI} } \closeO{\rangle}  \), we have \( \mathit{\Gamma}  \vdash^{1}  \possiblyWithSub\stageImetaColor{T^{\superscriptI} } \).
        Therefore, we can derive
        \begin{center}
          \infer[T0-Brkt]{%
            \infer[T1-TyEquiv]{%
              \infer[T1-CstP]{%
                \ConstEnvPers(c) = \possiblyWithSub\stageImetaColor{\tau^{\superscriptI} }
              }{%
                 \mathit{\Gamma}  \vdash^{1}   \possiblyWithSub\stageImetaColor{c}   :   \possiblyWithSub\stageImetaColor{\tau^{\superscriptI} }  
              }
            \andalso
              \infer[CqT0-Sym]{%
                 \possiblyWithSub\stageImetaColor{T^{\superscriptI} }  \equiv^{1}   \possiblyWithSub\stageImetaColor{\tau^{\superscriptI} }  
              }{%
                  \possiblyWithSub\stageImetaColor{\tau^{\superscriptI} }   \equiv^{1}  \possiblyWithSub\stageImetaColor{T^{\superscriptI} } 
              }
            \andalso
               \mathit{\Gamma}  \vdash^{1}  \possiblyWithSub\stageImetaColor{T^{\superscriptI} } 
            }{%
               \mathit{\Gamma}  \vdash^{1}   \possiblyWithSub\stageImetaColor{c}   :  \possiblyWithSub\stageImetaColor{T^{\superscriptI} } 
            }
          }{%
             \mathit{\Gamma}  \vdash^{0}   \openO{\langle}  \possiblyWithSub\stageImetaColor{c}  \closeO{\rangle}   :   \openO{\langle} \possiblyWithSub\stageImetaColor{T^{\superscriptI} } \closeO{\rangle}  
          }
        \end{center}
        and then
        \begin{center}
          \infer[T0-TyEquiv]{%
             \mathit{\Gamma}  \vdash^{0}   \openO{\langle}  \possiblyWithSub\stageImetaColor{c}  \closeO{\rangle}   :   \openO{\langle} \possiblyWithSub\stageImetaColor{T^{\superscriptI} } \closeO{\rangle}  
          \andalso
            \infer[CqT0-Sym]{%
               \possiblyWithSub\stageOmetaColor{T^{\superscriptO} }  \equiv^{0}   \openO{\langle} \possiblyWithSub\stageImetaColor{T^{\superscriptI} } \closeO{\rangle}  
            }{%
                \openO{\langle} \possiblyWithSub\stageImetaColor{T^{\superscriptI} } \closeO{\rangle}   \equiv^{0}  \possiblyWithSub\stageOmetaColor{T^{\superscriptO} } 
            }
          \andalso
             \mathit{\Gamma}  \vdash^{0}  \possiblyWithSub\stageOmetaColor{T^{\superscriptO} } 
          }{%
             \mathit{\Gamma}  \vdash^{0}   \openO{\langle}  \possiblyWithSub\stageImetaColor{c}  \closeO{\rangle}   :  \possiblyWithSub\stageOmetaColor{T^{\superscriptO} } 
          }.
        \end{center}
    \end{itemize}
  \end{proof}
  \recalltheorem[Preservation]{thm:preservation}{%
    \noindent
    \begin{enumerate}
      \item If \( \mathit{\Gamma}  \vdash^{0}  \possiblyWithSub\stageOmetaColor{N^{\superscriptO} }  :  \possiblyWithSub\stageOmetaColor{T^{\superscriptO} } \) and \( \possiblyWithSub\stageOmetaColor{N^{\superscriptO} }  \longrightarrow^{0}   \possiblyWithSub\stageOmetaColor{N'^{\superscriptO} }  \), then \( \mathit{\Gamma}  \vdash^{0}  \possiblyWithSub\stageOmetaColor{N'^{\superscriptO} }  :  \possiblyWithSub\stageOmetaColor{T^{\superscriptO} } \).
      \item If \( \mathit{\Gamma}  \vdash^{1}  \possiblyWithSub\stageImetaColor{N^{\superscriptI} }  :  \possiblyWithSub\stageImetaColor{T^{\superscriptI} } \) and \( \possiblyWithSub\stageImetaColor{N^{\superscriptI} }  \longrightarrow^{1}   \possiblyWithSub\stageImetaColor{N'^{\superscriptI} }  \), then \( \mathit{\Gamma}  \vdash^{1}  \possiblyWithSub\stageImetaColor{N'^{\superscriptI} }  :  \possiblyWithSub\stageImetaColor{T^{\superscriptI} } \).
      \item If \( \mathit{\Gamma}  \vdash^{1}  \possiblyWithSub\stageImetaColor{T^{\superscriptI} } \) and \( \possiblyWithSub\stageImetaColor{T^{\superscriptI} }  \longrightarrow^{1}   \possiblyWithSub\stageImetaColor{T'^{\superscriptI} }  \), then \( \mathit{\Gamma}  \vdash^{1}  \possiblyWithSub\stageImetaColor{T'^{\superscriptI} } \).
    \end{enumerate}
  }
  \begin{proof}
    By mutual induction on the derivation of type judgments.
    \begin{enumerate}
      \item
        \begin{itemize}
          \item Case \derive[T0-TyEquiv]{%
             \mathit{\Gamma}  \vdash^{0}  \possiblyWithSub\stageOmetaColor{N^{\superscriptO} }  :  \possiblyWithSub\stageOmetaColor{T'^{\superscriptO} } 
          \andalso
             \possiblyWithSub\stageOmetaColor{T'^{\superscriptO} }  \equiv^{0}  \possiblyWithSub\stageOmetaColor{T^{\superscriptO} } 
          \andalso
             \mathit{\Gamma}  \vdash^{0}  \possiblyWithSub\stageOmetaColor{T^{\superscriptO} } 
          }{%
             \mathit{\Gamma}  \vdash^{0}  \possiblyWithSub\stageOmetaColor{N^{\superscriptO} }  :  \possiblyWithSub\stageOmetaColor{T^{\superscriptO} } 
          }:
            Straightforward by IH.
          \item Case \derive[T0-App]{%
             \mathit{\Gamma}  \vdash^{0}  \possiblyWithSub\stageOmetaColor{N^{\superscriptO} }_{{\mathrm{1}}}  :   \openO{(} \possiblyWithSub\stageOmetaColor{x}  \relO{:}  \possiblyWithSub\stageOmetaColor{T^{\superscriptO} }_{{\mathrm{11}}} \closeO{)} \relO{\to}  \possiblyWithSub\stageOmetaColor{T^{\superscriptO} }_{{\mathrm{12}}}  
          \andalso
             \mathit{\Gamma}  \vdash^{0}  \possiblyWithSub\stageOmetaColor{N^{\superscriptO} }_{{\mathrm{2}}}  :  \possiblyWithSub\stageOmetaColor{T^{\superscriptO} }_{{\mathrm{11}}} 
          }{%
             \mathit{\Gamma}  \vdash^{0}   \possiblyWithSub\stageOmetaColor{N^{\superscriptO} }_{{\mathrm{1}}} \  \possiblyWithSub\stageOmetaColor{N^{\superscriptO} }_{{\mathrm{2}}}   :    [  \possiblyWithSub\stageOmetaColor{N^{\superscriptO} }_{{\mathrm{2}}}  /  \possiblyWithSub\stageOmetaColor{x}  ]    \possiblyWithSub\stageOmetaColor{T^{\superscriptO} }_{{\mathrm{12}}}  
          }:
            By case analysis on the derivation of \( \possiblyWithSub\stageOmetaColor{N^{\superscriptO} }  \longrightarrow^{0}   \possiblyWithSub\stageOmetaColor{N'^{\superscriptO} }  \).
            \begin{itemize}
              \item Case \rulename{E0-App1}:
                Straightforward by IH.
              \item Case \derive[E0-App2]{%
                 \possiblyWithSub\stageOmetaColor{N^{\superscriptO} }_{{\mathrm{2}}}  \longrightarrow^{0}   \possiblyWithSub\stageOmetaColor{N'^{\superscriptO} }_{{\mathrm{2}}}  
              }{%
                  \possiblyWithSub\stageOmetaColor{N^{\superscriptO} }_{{\mathrm{1}}} \  \possiblyWithSub\stageOmetaColor{N^{\superscriptO} }_{{\mathrm{2}}}   \longrightarrow^{0}    \possiblyWithSub\stageOmetaColor{N^{\superscriptO} }_{{\mathrm{1}}} \  \possiblyWithSub\stageOmetaColor{N'^{\superscriptO} }_{{\mathrm{2}}}   
              }:
                By IH, from \( \mathit{\Gamma}  \vdash^{0}  \possiblyWithSub\stageOmetaColor{N^{\superscriptO} }_{{\mathrm{2}}}  :  \possiblyWithSub\stageOmetaColor{T^{\superscriptO} }_{{\mathrm{11}}} \) and \( \possiblyWithSub\stageOmetaColor{N^{\superscriptO} }_{{\mathrm{2}}}  \longrightarrow^{0}   \possiblyWithSub\stageOmetaColor{N'^{\superscriptO} }_{{\mathrm{2}}}  \),
                we have \( \mathit{\Gamma}  \vdash^{0}  \possiblyWithSub\stageOmetaColor{N'^{\superscriptO} }_{{\mathrm{2}}}  :  \possiblyWithSub\stageOmetaColor{T^{\superscriptO} }_{{\mathrm{11}}} \).
                This enables us to first derive
                \begin{center}
                  \derive[T0-App]{%
                     \mathit{\Gamma}  \vdash^{0}  \possiblyWithSub\stageOmetaColor{N^{\superscriptO} }_{{\mathrm{1}}}  :   \openO{(} \possiblyWithSub\stageOmetaColor{x}  \relO{:}  \possiblyWithSub\stageOmetaColor{T^{\superscriptO} }_{{\mathrm{11}}} \closeO{)} \relO{\to}  \possiblyWithSub\stageOmetaColor{T^{\superscriptO} }_{{\mathrm{12}}}  
                  \andalso
                     \mathit{\Gamma}  \vdash^{0}  \possiblyWithSub\stageOmetaColor{N'^{\superscriptO} }_{{\mathrm{2}}}  :  \possiblyWithSub\stageOmetaColor{T^{\superscriptO} }_{{\mathrm{11}}} 
                  }{%
                     \mathit{\Gamma}  \vdash^{0}   \possiblyWithSub\stageOmetaColor{N^{\superscriptO} }_{{\mathrm{1}}} \  \possiblyWithSub\stageOmetaColor{N'^{\superscriptO} }_{{\mathrm{2}}}   :    [  \possiblyWithSub\stageOmetaColor{N'^{\superscriptO} }_{{\mathrm{2}}}  /  \possiblyWithSub\stageOmetaColor{x}  ]    \possiblyWithSub\stageOmetaColor{T^{\superscriptO} }_{{\mathrm{12}}}  
                  }.
                \end{center}
                By Lemma~\ref{lem:reduction-of-subst-satisfies-csr-equiv},
                from \( \possiblyWithSub\stageOmetaColor{N^{\superscriptO} }_{{\mathrm{2}}}  \longrightarrow^{0}   \possiblyWithSub\stageOmetaColor{N'^{\superscriptO} }_{{\mathrm{2}}}  \),
                we have \(   [  \possiblyWithSub\stageOmetaColor{N'^{\superscriptO} }_{{\mathrm{2}}}  /  \possiblyWithSub\stageOmetaColor{x}  ]    \possiblyWithSub\stageOmetaColor{T^{\superscriptO} }_{{\mathrm{12}}}   \equiv^{0}    [  \possiblyWithSub\stageOmetaColor{N^{\superscriptO} }_{{\mathrm{2}}}  /  \possiblyWithSub\stageOmetaColor{x}  ]    \possiblyWithSub\stageOmetaColor{T^{\superscriptO} }_{{\mathrm{12}}}  \).
                Also, by Lemma~\ref{lem:target-typing-implies-type-well-formedness},
                we have \( \mathit{\Gamma}  \vdash^{0}    [  \possiblyWithSub\stageOmetaColor{N^{\superscriptO} }_{{\mathrm{2}}}  /  \possiblyWithSub\stageOmetaColor{x}  ]    \possiblyWithSub\stageOmetaColor{T^{\superscriptO} }_{{\mathrm{12}}}  \).
                Therefore, we have
                \begin{center}
                  \derive[T0-TyEquiv]{%
                     \mathit{\Gamma}  \vdash^{0}   \possiblyWithSub\stageOmetaColor{N^{\superscriptO} }_{{\mathrm{1}}} \  \possiblyWithSub\stageOmetaColor{N'^{\superscriptO} }_{{\mathrm{2}}}   :    [  \possiblyWithSub\stageOmetaColor{N'^{\superscriptO} }_{{\mathrm{2}}}  /  \possiblyWithSub\stageOmetaColor{x}  ]    \possiblyWithSub\stageOmetaColor{T^{\superscriptO} }_{{\mathrm{12}}}  
                  \\\empty 
                       [  \possiblyWithSub\stageOmetaColor{N'^{\superscriptO} }_{{\mathrm{2}}}  /  \possiblyWithSub\stageOmetaColor{x}  ]    \possiblyWithSub\stageOmetaColor{T^{\superscriptO} }_{{\mathrm{12}}}   \equiv^{0}    [  \possiblyWithSub\stageOmetaColor{N^{\superscriptO} }_{{\mathrm{2}}}  /  \possiblyWithSub\stageOmetaColor{x}  ]    \possiblyWithSub\stageOmetaColor{T^{\superscriptO} }_{{\mathrm{12}}}  
                  \andalso
                     \mathit{\Gamma}  \vdash^{0}    [  \possiblyWithSub\stageOmetaColor{N^{\superscriptO} }_{{\mathrm{2}}}  /  \possiblyWithSub\stageOmetaColor{x}  ]    \possiblyWithSub\stageOmetaColor{T^{\superscriptO} }_{{\mathrm{12}}}  
                  }{%
                     \mathit{\Gamma}  \vdash^{0}   \possiblyWithSub\stageOmetaColor{N^{\superscriptO} }_{{\mathrm{1}}} \  \possiblyWithSub\stageOmetaColor{N'^{\superscriptO} }_{{\mathrm{2}}}   :    [  \possiblyWithSub\stageOmetaColor{N^{\superscriptO} }_{{\mathrm{2}}}  /  \possiblyWithSub\stageOmetaColor{x}  ]    \possiblyWithSub\stageOmetaColor{T^{\superscriptO} }_{{\mathrm{12}}}  
                  }.
                \end{center}
              \item Case \derive[E0-Beta]{}{%
                   \openO{(}  \ordO{\lambda} \possiblyWithSub\stageOmetaColor{x'}  \relO{:}  \possiblyWithSub\stageOmetaColor{T'^{\superscriptO} }_{{\mathrm{11}}} \punctO{.}\  \possiblyWithSub\stageOmetaColor{N^{\superscriptO} }_{{\mathrm{12}}}  \closeO{)}  \   \possiblyWithSub\stageOmetaColor{v^{\superscriptO} }_{{\mathrm{2}}}    \longrightarrow^{0}     [   \possiblyWithSub\stageOmetaColor{v^{\superscriptO} }_{{\mathrm{2}}}   /  \possiblyWithSub\stageOmetaColor{x}  ]    \possiblyWithSub\stageOmetaColor{N^{\superscriptO} }_{{\mathrm{12}}}   
              }:
                We have \(\possiblyWithSub\stageOmetaColor{N^{\superscriptO} }_{{\mathrm{1}}} =  \ordO{\lambda} \possiblyWithSub\stageOmetaColor{x'}  \relO{:}  \possiblyWithSub\stageOmetaColor{T'^{\superscriptO} }_{{\mathrm{11}}} \punctO{.}\  \possiblyWithSub\stageOmetaColor{N^{\superscriptO} }_{{\mathrm{12}}} \) and \(\possiblyWithSub\stageOmetaColor{N^{\superscriptO} }_{{\mathrm{2}}} =  \possiblyWithSub\stageOmetaColor{v^{\superscriptO} }_{{\mathrm{2}}} \), i.e.,
                \( \mathit{\Gamma}  \vdash^{0}   \openO{(}  \ordO{\lambda} \possiblyWithSub\stageOmetaColor{x'}  \relO{:}  \possiblyWithSub\stageOmetaColor{T'^{\superscriptO} }_{{\mathrm{11}}} \punctO{.}\  \possiblyWithSub\stageOmetaColor{N^{\superscriptO} }_{{\mathrm{12}}}  \closeO{)}   :   \openO{(} \possiblyWithSub\stageOmetaColor{x}  \relO{:}  \possiblyWithSub\stageOmetaColor{T^{\superscriptO} }_{{\mathrm{11}}} \closeO{)} \relO{\to}  \possiblyWithSub\stageOmetaColor{T^{\superscriptO} }_{{\mathrm{12}}}  \) and
                \( \mathit{\Gamma}  \vdash^{0}   \possiblyWithSub\stageOmetaColor{v^{\superscriptO} }_{{\mathrm{2}}}   :  \possiblyWithSub\stageOmetaColor{T^{\superscriptO} }_{{\mathrm{11}}} \) hold.
                Then, by Lemma~\ref{lem:lambda-inversion},
                we have \( \mathit{\Gamma}  \vdash^{0}  \possiblyWithSub\stageOmetaColor{T'^{\superscriptO} }_{{\mathrm{11}}} \) and
                there exists \(\possiblyWithSub\stageOmetaColor{T'^{\superscriptO} }_{{\mathrm{12}}}\) such that
                \(  \mathit{\Gamma} ,  \possiblyWithSub\stageOmetaColor{x'}  : ( \possiblyWithSub\stageOmetaColor{T'^{\superscriptO} }_{{\mathrm{11}}} )^{0}   \vdash^{0}  \possiblyWithSub\stageOmetaColor{N^{\superscriptO} }_{{\mathrm{12}}}  :  \possiblyWithSub\stageOmetaColor{T'^{\superscriptO} }_{{\mathrm{12}}} \) and
                \(  \openO{(} \possiblyWithSub\stageOmetaColor{x}  \relO{:}  \possiblyWithSub\stageOmetaColor{T^{\superscriptO} }_{{\mathrm{11}}} \closeO{)} \relO{\to}  \possiblyWithSub\stageOmetaColor{T^{\superscriptO} }_{{\mathrm{12}}}   \equiv^{0}   \openO{(} \possiblyWithSub\stageOmetaColor{x'}  \relO{:}  \possiblyWithSub\stageOmetaColor{T'^{\superscriptO} }_{{\mathrm{11}}} \closeO{)} \relO{\to}  \possiblyWithSub\stageOmetaColor{T'^{\superscriptO} }_{{\mathrm{12}}}  \).
                Here, w.l.o.g., we can regard \(\possiblyWithSub\stageOmetaColor{x'} = \possiblyWithSub\stageOmetaColor{x}\),
                and by Lemma~\ref{lem:arrow-type-csr-equiv-inversion},
                we have \( \possiblyWithSub\stageOmetaColor{T^{\superscriptO} }_{{\mathrm{11}}}  \equiv^{0}  \possiblyWithSub\stageOmetaColor{T'^{\superscriptO} }_{{\mathrm{11}}} \) and \( \possiblyWithSub\stageOmetaColor{T^{\superscriptO} }_{{\mathrm{12}}}  \equiv^{0}  \possiblyWithSub\stageOmetaColor{T'^{\superscriptO} }_{{\mathrm{12}}} \).
                Also,
                by Lemma~\ref{lem:target-typing-implies-type-well-formedness},
                and \( \mathit{\Gamma}  \vdash^{0}   \openO{(}  \ordO{\lambda} \possiblyWithSub\stageOmetaColor{x'}  \relO{:}  \possiblyWithSub\stageOmetaColor{T'^{\superscriptO} }_{{\mathrm{11}}} \punctO{.}\  \possiblyWithSub\stageOmetaColor{N^{\superscriptO} }_{{\mathrm{12}}}  \closeO{)}   :   \openO{(} \possiblyWithSub\stageOmetaColor{x}  \relO{:}  \possiblyWithSub\stageOmetaColor{T^{\superscriptO} }_{{\mathrm{11}}} \closeO{)} \relO{\to}  \possiblyWithSub\stageOmetaColor{T^{\superscriptO} }_{{\mathrm{12}}}  \),
                we have \( \mathit{\Gamma}  \vdash^{0}   \openO{(} \possiblyWithSub\stageOmetaColor{x}  \relO{:}  \possiblyWithSub\stageOmetaColor{T^{\superscriptO} }_{{\mathrm{11}}} \closeO{)} \relO{\to}  \possiblyWithSub\stageOmetaColor{T^{\superscriptO} }_{{\mathrm{12}}}  \)
                and thereby \( \mathit{\Gamma}  \vdash^{0}  \possiblyWithSub\stageOmetaColor{T^{\superscriptO} }_{{\mathrm{12}}} \) by inversion.
                Thus, we can derive
                \begin{center}
                  \derive[T0-TyEquiv]{%
                      \mathit{\Gamma} ,  \possiblyWithSub\stageOmetaColor{x'}  : ( \possiblyWithSub\stageOmetaColor{T'^{\superscriptO} }_{{\mathrm{11}}} )^{0}   \vdash^{0}  \possiblyWithSub\stageOmetaColor{N^{\superscriptO} }_{{\mathrm{12}}}  :  \possiblyWithSub\stageOmetaColor{T'^{\superscriptO} }_{{\mathrm{12}}} 
                  \andalso
                     \possiblyWithSub\stageOmetaColor{T'^{\superscriptO} }_{{\mathrm{12}}}  \equiv^{0}  \possiblyWithSub\stageOmetaColor{T^{\superscriptO} }_{{\mathrm{12}}} 
                  \andalso
                     \mathit{\Gamma}  \vdash^{0}  \possiblyWithSub\stageOmetaColor{T^{\superscriptO} }_{{\mathrm{12}}} 
                  }{%
                      \mathit{\Gamma} ,  \possiblyWithSub\stageOmetaColor{x}  : ( \possiblyWithSub\stageOmetaColor{T'^{\superscriptO} }_{{\mathrm{11}}} )^{0}   \vdash^{0}  \possiblyWithSub\stageOmetaColor{N^{\superscriptO} }_{{\mathrm{12}}}  :  \possiblyWithSub\stageOmetaColor{T^{\superscriptO} }_{{\mathrm{12}}} 
                  }
                \end{center}
                and
                \begin{center}
                  \derive[T0-TyEquiv]{%
                     \mathit{\Gamma}  \vdash^{0}   \possiblyWithSub\stageOmetaColor{v^{\superscriptO} }_{{\mathrm{2}}}   :  \possiblyWithSub\stageOmetaColor{T^{\superscriptO} }_{{\mathrm{11}}} 
                  \andalso
                     \possiblyWithSub\stageOmetaColor{T^{\superscriptO} }_{{\mathrm{11}}}  \equiv^{0}  \possiblyWithSub\stageOmetaColor{T'^{\superscriptO} }_{{\mathrm{11}}} 
                  \andalso
                     \mathit{\Gamma}  \vdash^{0}  \possiblyWithSub\stageOmetaColor{T'^{\superscriptO} }_{{\mathrm{11}}} 
                  }{%
                     \mathit{\Gamma}  \vdash^{0}   \possiblyWithSub\stageOmetaColor{v^{\superscriptO} }_{{\mathrm{2}}}   :  \possiblyWithSub\stageOmetaColor{T'^{\superscriptO} }_{{\mathrm{11}}} 
                  }.
                \end{center}
                Finally, by Lemma~\ref{lem:subst},
                from the two judgments above,
                we have \( \mathit{\Gamma}  \vdash^{0}    [   \possiblyWithSub\stageOmetaColor{v^{\superscriptO} }_{{\mathrm{2}}}   /  \possiblyWithSub\stageOmetaColor{x}  ]    \possiblyWithSub\stageOmetaColor{N^{\superscriptO} }_{{\mathrm{12}}}   :    [   \possiblyWithSub\stageOmetaColor{v^{\superscriptO} }_{{\mathrm{2}}}   /  \possiblyWithSub\stageOmetaColor{x}  ]    \possiblyWithSub\stageOmetaColor{T^{\superscriptO} }_{{\mathrm{12}}}  \).
              \item Case \derive[E0-Delta]{%
                \delta(  \possiblyWithSub\stageOmetaColor{a}_{{\mathrm{1}}}  \    \possiblyWithSub\stageOmetaColor{c}_{{\mathrm{2}}}   ) = \possiblyWithSub\stageOmetaColor{q}
              }{%
                   \possiblyWithSub\stageOmetaColor{a}_{{\mathrm{1}}}  \    \possiblyWithSub\stageOmetaColor{c}_{{\mathrm{2}}}     \longrightarrow^{0}    \possiblyWithSub\stageOmetaColor{q}   
              }:
                Immediate from Lemmata~\ref{lem:preservation-of-delta-persistent}
                and \ref{lem:preservation-of-delta-zero}.
              \item Case \derive[E0-RfnStart]{}{%
                   \LeftAssertParen \relO{\CastArrow}   \openO{\{} \possiblyWithSub\stageOmetaColor{\nu}  \relO{:}  \possiblyWithSub\stageOmetaColor{B}  \relO{\mid}  \possiblyWithSub\stageOmetaColor{N^{\superscriptO} }_{{\mathrm{1}}} \closeO{\} }   \RightAssertParen^{ L }  \    \possiblyWithSub\stageOmetaColor{c}_{{\mathrm{2}}}     \longrightarrow^{0}    \LeftAssertParen   \openO{\{} \possiblyWithSub\stageOmetaColor{\nu}  \relO{:}  \possiblyWithSub\stageOmetaColor{B}  \relO{\mid}  \possiblyWithSub\stageOmetaColor{N^{\superscriptO} }_{{\mathrm{1}}} \closeO{\} }  \punctO{,}    [    \possiblyWithSub\stageOmetaColor{c}_{{\mathrm{2}}}    /  \possiblyWithSub\stageOmetaColor{\nu}  ]    \possiblyWithSub\stageOmetaColor{N^{\superscriptO} }_{{\mathrm{1}}}  \punctO{,}  \possiblyWithSub\stageOmetaColor{c}_{{\mathrm{2}}}  \RightAssertParen^{ L }   
              }:
                We have \(\possiblyWithSub\stageOmetaColor{N^{\superscriptO} }_{{\mathrm{1}}} =  \LeftAssertParen \relO{\CastArrow}   \openO{\{} \possiblyWithSub\stageOmetaColor{\nu}  \relO{:}  \possiblyWithSub\stageOmetaColor{B}  \relO{\mid}  \possiblyWithSub\stageOmetaColor{N^{\superscriptO} }_{{\mathrm{1}}} \closeO{\} }   \RightAssertParen^{ L } \) and \(\possiblyWithSub\stageOmetaColor{N^{\superscriptO} }_{{\mathrm{2}}} = \possiblyWithSub\stageOmetaColor{c}_{{\mathrm{2}}}\).
                By \( \mathit{\Gamma}  \vdash^{0}   \LeftAssertParen \relO{\CastArrow}   \openO{\{} \possiblyWithSub\stageOmetaColor{\nu}  \relO{:}  \possiblyWithSub\stageOmetaColor{B}  \relO{\mid}  \possiblyWithSub\stageOmetaColor{N^{\superscriptO} }_{{\mathrm{1}}} \closeO{\} }   \RightAssertParen^{ L }   :   \openO{(} \possiblyWithSub\stageOmetaColor{x}  \relO{:}  \possiblyWithSub\stageOmetaColor{T^{\superscriptO} }_{{\mathrm{11}}} \closeO{)} \relO{\to}  \possiblyWithSub\stageOmetaColor{T^{\superscriptO} }_{{\mathrm{12}}}  \)
                and Lemma~\ref{lem:refinement-inversion},
                we have \(  \possiblyWithSub\stageImetaColor{B}   \gg  \possiblyWithSub\stageOmetaColor{T^{\superscriptO} }_{{\mathrm{11}}} \), \( \possiblyWithSub\stageOmetaColor{T^{\superscriptO} }_{{\mathrm{12}}}  \equiv^{0}    \openO{\{} \possiblyWithSub\stageOmetaColor{\nu}  \relO{:}  \possiblyWithSub\stageOmetaColor{B}  \relO{\mid}  \possiblyWithSub\stageOmetaColor{N^{\superscriptO} }_{{\mathrm{1}}} \closeO{\} }   \),
                and \( \mathit{\Gamma}  \vdash^{0}    \openO{\{} \possiblyWithSub\stageOmetaColor{\nu}  \relO{:}  \possiblyWithSub\stageOmetaColor{B}  \relO{\mid}  \possiblyWithSub\stageOmetaColor{N^{\superscriptO} }_{{\mathrm{1}}} \closeO{\} }   \).
                Thus, by Lemma~\ref{lem:persistent-const-inversion},
                from \( \mathit{\Gamma}  \vdash^{0}    \possiblyWithSub\stageOmetaColor{c}_{{\mathrm{2}}}    :  \possiblyWithSub\stageOmetaColor{T^{\superscriptO} }_{{\mathrm{11}}} \) and \(  \possiblyWithSub\stageImetaColor{B}   \gg  \possiblyWithSub\stageOmetaColor{T^{\superscriptO} }_{{\mathrm{11}}} \),
                we have \(\ConstEnvPers(c_{{\mathrm{2}}}) = \possiblyWithSub\stageImetaColor{B}\).
                Since \( \vdash  \mathit{\Gamma} \) holds by Lemma~\ref{lem:target-typing-implies-type-well-formedness},
                we can derive
                \begin{center}
                  \derive[T0-CstP]{%
                     \vdash  \mathit{\Gamma} 
                  \andalso
                    \ConstEnvPers(c_{{\mathrm{2}}}) = \possiblyWithSub\stageImetaColor{B}
                  }{
                     \mathit{\Gamma}  \vdash^{0}    \possiblyWithSub\stageOmetaColor{c}_{{\mathrm{2}}}    :    \openO{\{} \possiblyWithSub\stageOmetaColor{\nu}  \relO{:}  \possiblyWithSub\stageOmetaColor{B}  \relO{\mid}     \ttO{true}    \closeO{\} }   
                  }.
                \end{center}
                Also, by tracing back the derivation of \( \mathit{\Gamma}  \vdash^{0}    \openO{\{} \possiblyWithSub\stageOmetaColor{\nu}  \relO{:}  \possiblyWithSub\stageOmetaColor{B}  \relO{\mid}  \possiblyWithSub\stageOmetaColor{N^{\superscriptO} }_{{\mathrm{1}}} \closeO{\} }   \),
                we have the following:
                \begin{center}
                  \derive[WfT0-Rfn]{%
                      \mathit{\Gamma} ,  \possiblyWithSub\stageOmetaColor{\nu}  : (   \openO{\{} \possiblyWithSub\stageOmetaColor{x}  \relO{:}  \possiblyWithSub\stageOmetaColor{B}  \relO{\mid}     \ttO{true}    \closeO{\} }   )^{0}   \vdash^{0}  \possiblyWithSub\stageOmetaColor{N^{\superscriptO} }_{{\mathrm{1}}}  :    \openO{\{} \possiblyWithSub\stageOmetaColor{\nu}_{{\mathrm{0}}}  \relO{:}   \ttO{Bool}   \relO{\mid}     \ttO{true}    \closeO{\} }   
                  }{%
                     \mathit{\Gamma}  \vdash^{0}    \openO{\{} \possiblyWithSub\stageOmetaColor{\nu}  \relO{:}  \possiblyWithSub\stageOmetaColor{B}  \relO{\mid}  \possiblyWithSub\stageOmetaColor{N^{\superscriptO} }_{{\mathrm{1}}} \closeO{\} }   
                  }.
                \end{center}
                Then, by Lemma~\ref{lem:subst},
                from \(  \mathit{\Gamma} ,  \possiblyWithSub\stageOmetaColor{\nu}  : (   \openO{\{} \possiblyWithSub\stageOmetaColor{x}  \relO{:}  \possiblyWithSub\stageOmetaColor{B}  \relO{\mid}     \ttO{true}    \closeO{\} }   )^{0}   \vdash^{0}  \possiblyWithSub\stageOmetaColor{N^{\superscriptO} }_{{\mathrm{1}}}  :    \openO{\{} \possiblyWithSub\stageOmetaColor{\nu}_{{\mathrm{1}}}  \relO{:}   \ttO{Bool}   \relO{\mid}     \ttO{true}    \closeO{\} }   \)
                and \( \mathit{\Gamma}  \vdash^{0}    \possiblyWithSub\stageOmetaColor{c}_{{\mathrm{2}}}    :    \openO{\{} \possiblyWithSub\stageOmetaColor{\nu}  \relO{:}  \possiblyWithSub\stageOmetaColor{B}  \relO{\mid}     \ttO{true}    \closeO{\} }   \),
                we have \( \mathit{\Gamma}  \vdash^{0}    [    \possiblyWithSub\stageOmetaColor{c}_{{\mathrm{2}}}    /  \possiblyWithSub\stageOmetaColor{\nu}  ]    \possiblyWithSub\stageOmetaColor{N^{\superscriptO} }_{{\mathrm{1}}}   :    \openO{\{} \possiblyWithSub\stageOmetaColor{\nu}_{{\mathrm{1}}}  \relO{:}   \ttO{Bool}   \relO{\mid}     \ttO{true}    \closeO{\} }   \).
                Since \(   [    \possiblyWithSub\stageOmetaColor{c}_{{\mathrm{2}}}    /  \possiblyWithSub\stageOmetaColor{\nu}  ]    \possiblyWithSub\stageOmetaColor{N^{\superscriptO} }_{{\mathrm{1}}}   \longrightarrow^{0\,\ast}     [    \possiblyWithSub\stageOmetaColor{c}_{{\mathrm{2}}}    /  \possiblyWithSub\stageOmetaColor{\nu}  ]    \possiblyWithSub\stageOmetaColor{N^{\superscriptO} }_{{\mathrm{1}}}   \) clearly holds,
                we can finally derive
                \begin{center}
                  \derive[T0-RfnAct]{%
                     \mathit{\Gamma}  \vdash^{0}    \openO{\{} \possiblyWithSub\stageOmetaColor{\nu}  \relO{:}  \possiblyWithSub\stageOmetaColor{B}  \relO{\mid}  \possiblyWithSub\stageOmetaColor{N^{\superscriptO} }_{{\mathrm{1}}} \closeO{\} }   
                  \andalso
                     \mathit{\Gamma}  \vdash^{0}    [    \possiblyWithSub\stageOmetaColor{c}_{{\mathrm{2}}}    /  \possiblyWithSub\stageOmetaColor{\nu}  ]    \possiblyWithSub\stageOmetaColor{N^{\superscriptO} }_{{\mathrm{1}}}   :    \openO{\{} \possiblyWithSub\stageOmetaColor{\nu}_{{\mathrm{1}}}  \relO{:}   \ttO{Bool}   \relO{\mid}     \ttO{true}    \closeO{\} }   
                  \\
                    \ConstEnvPers(c_{{\mathrm{2}}}) = \possiblyWithSub\stageImetaColor{B}
                  \andalso
                       [    \possiblyWithSub\stageOmetaColor{c}_{{\mathrm{2}}}    /  \possiblyWithSub\stageOmetaColor{\nu}  ]    \possiblyWithSub\stageOmetaColor{N^{\superscriptO} }_{{\mathrm{1}}}   \longrightarrow^{0\,\ast}     [    \possiblyWithSub\stageOmetaColor{c}_{{\mathrm{2}}}    /  \possiblyWithSub\stageOmetaColor{\nu}  ]    \possiblyWithSub\stageOmetaColor{N^{\superscriptO} }_{{\mathrm{1}}}   
                  }{%
                     \mathit{\Gamma}  \vdash^{0}   \LeftAssertParen   \openO{\{} \possiblyWithSub\stageOmetaColor{\nu}  \relO{:}  \possiblyWithSub\stageOmetaColor{B}  \relO{\mid}  \possiblyWithSub\stageOmetaColor{N^{\superscriptO} }_{{\mathrm{1}}} \closeO{\} }  \punctO{,}    [    \possiblyWithSub\stageOmetaColor{c}_{{\mathrm{2}}}    /  \possiblyWithSub\stageOmetaColor{\nu}  ]    \possiblyWithSub\stageOmetaColor{N^{\superscriptO} }_{{\mathrm{1}}}  \punctO{,}  \possiblyWithSub\stageOmetaColor{c}_{{\mathrm{2}}}  \RightAssertParen^{ L }   :    \openO{\{} \possiblyWithSub\stageOmetaColor{\nu}  \relO{:}  \possiblyWithSub\stageOmetaColor{B}  \relO{\mid}  \possiblyWithSub\stageOmetaColor{N^{\superscriptO} }_{{\mathrm{1}}} \closeO{\} }   
                  }.
                \end{center}
              \item The other cases contradict the form~\( \possiblyWithSub\stageOmetaColor{N^{\superscriptO} }_{{\mathrm{1}}} \  \possiblyWithSub\stageOmetaColor{N^{\superscriptO} }_{{\mathrm{2}}} \).
            \end{itemize}
          \item Case \derive[T0-Brkt]{%
             \mathit{\Gamma}  \vdash^{1}  \possiblyWithSub\stageImetaColor{N^{\superscriptI} }  :  \possiblyWithSub\stageImetaColor{T^{\superscriptI} } 
          }{%
             \mathit{\Gamma}  \vdash^{0}   \openO{\langle} \possiblyWithSub\stageImetaColor{N^{\superscriptI} } \closeO{\rangle}   :   \openO{\langle} \possiblyWithSub\stageImetaColor{T^{\superscriptI} } \closeO{\rangle}  
          }:
            Straightforward by IH;
            The last rule used for deriving the reduction is \rulename{E0-Brkt}.
          \item \derive[T0-Ass]{%
             \mathit{\Gamma}  \vdash^{1}  \possiblyWithSub\stageImetaColor{T^{\superscriptI} }_{{\mathrm{1}}} 
          \andalso
             \mathit{\Gamma}  \vdash^{1}  \possiblyWithSub\stageImetaColor{T^{\superscriptI} }_{{\mathrm{2}}} 
          \andalso
             \possiblyWithSub\stageImetaColor{T^{\superscriptI} }_{{\mathrm{1}}}  \mathrel{||}^{1}  \possiblyWithSub\stageImetaColor{T^{\superscriptI} }_{{\mathrm{2}}} 
          \andalso
            \possiblyWithSub\stageOmetaColor{x} \not\in \dom(\mathit{\Gamma})
          }{%
             \mathit{\Gamma}  \vdash^{0}   \LeftAssertParen\openO{\langle} \possiblyWithSub\stageImetaColor{T^{\superscriptI} }_{{\mathrm{1}}} \closeO{\rangle} \relO{\CastArrow} \openO{\langle} \possiblyWithSub\stageImetaColor{T^{\superscriptI} }_{{\mathrm{2}}} \closeO{\rangle}\RightAssertParen^{ L }   :   \openO{(} \possiblyWithSub\stageOmetaColor{x}  \relO{:}   \openO{\langle} \possiblyWithSub\stageImetaColor{T^{\superscriptI} }_{{\mathrm{1}}} \closeO{\rangle}  \closeO{)} \relO{\to}   \openO{\langle} \possiblyWithSub\stageImetaColor{T^{\superscriptI} }_{{\mathrm{2}}} \closeO{\rangle}   
          }:
            \begin{itemize}
              \item \infer[E0-Ass1]{%
                 \possiblyWithSub\stageImetaColor{T^{\superscriptI} }_{{\mathrm{1}}}  \longrightarrow^{1}   \possiblyWithSub\stageImetaColor{T'^{\superscriptI} }_{{\mathrm{1}}}  
              }{%
                  \LeftAssertParen\openO{\langle} \possiblyWithSub\stageImetaColor{T^{\superscriptI} }_{{\mathrm{1}}} \closeO{\rangle} \relO{\CastArrow} \openO{\langle} \possiblyWithSub\stageImetaColor{T^{\superscriptI} }_{{\mathrm{2}}} \closeO{\rangle}\RightAssertParen^{ L }   \longrightarrow^{0}    \LeftAssertParen\openO{\langle} \possiblyWithSub\stageImetaColor{T'^{\superscriptI} }_{{\mathrm{1}}} \closeO{\rangle} \relO{\CastArrow} \openO{\langle} \possiblyWithSub\stageImetaColor{T^{\superscriptI} }_{{\mathrm{2}}} \closeO{\rangle}\RightAssertParen^{ L }   
              }:
                By IH, from \( \mathit{\Gamma}  \vdash^{1}  \possiblyWithSub\stageImetaColor{T^{\superscriptI} }_{{\mathrm{1}}} \) and \( \possiblyWithSub\stageImetaColor{T^{\superscriptI} }_{{\mathrm{1}}}  \longrightarrow^{1}   \possiblyWithSub\stageImetaColor{T'^{\superscriptI} }_{{\mathrm{1}}}  \),
                we have \( \mathit{\Gamma}  \vdash^{1}  \possiblyWithSub\stageImetaColor{T'^{\superscriptI} }_{{\mathrm{1}}} \).
                Then, by Lemma~\ref{lem:eval-preserves-compatibility},
                from \( \possiblyWithSub\stageImetaColor{T^{\superscriptI} }_{{\mathrm{1}}}  \mathrel{||}^{1}  \possiblyWithSub\stageImetaColor{T^{\superscriptI} }_{{\mathrm{2}}} \) and \( \possiblyWithSub\stageImetaColor{T^{\superscriptI} }_{{\mathrm{1}}}  \longrightarrow^{1}   \possiblyWithSub\stageImetaColor{T'^{\superscriptI} }_{{\mathrm{1}}}  \),
                we have \( \possiblyWithSub\stageImetaColor{T'^{\superscriptI} }_{{\mathrm{1}}}  \mathrel{||}^{1}  \possiblyWithSub\stageImetaColor{T^{\superscriptI} }_{{\mathrm{2}}} \).
                Thus, we can derive
                \begin{center}
                  \derive[T0-Ass]{%
                     \mathit{\Gamma}  \vdash^{1}  \possiblyWithSub\stageImetaColor{T'^{\superscriptI} }_{{\mathrm{1}}} 
                  \andalso
                     \mathit{\Gamma}  \vdash^{1}  \possiblyWithSub\stageImetaColor{T^{\superscriptI} }_{{\mathrm{2}}} 
                  \andalso
                     \possiblyWithSub\stageImetaColor{T'^{\superscriptI} }_{{\mathrm{1}}}  \mathrel{||}^{1}  \possiblyWithSub\stageImetaColor{T^{\superscriptI} }_{{\mathrm{2}}} 
                  \andalso
                    \possiblyWithSub\stageOmetaColor{x}\ \text{fresh}
                  }{%
                     \mathit{\Gamma}  \vdash^{0}   \LeftAssertParen\openO{\langle} \possiblyWithSub\stageImetaColor{T'^{\superscriptI} }_{{\mathrm{1}}} \closeO{\rangle} \relO{\CastArrow} \openO{\langle} \possiblyWithSub\stageImetaColor{T^{\superscriptI} }_{{\mathrm{2}}} \closeO{\rangle}\RightAssertParen^{ L }   :   \openO{(} \possiblyWithSub\stageOmetaColor{x}  \relO{:}   \openO{\langle} \possiblyWithSub\stageImetaColor{T'^{\superscriptI} }_{{\mathrm{1}}} \closeO{\rangle}  \closeO{)} \relO{\to}   \openO{\langle} \possiblyWithSub\stageImetaColor{T^{\superscriptI} }_{{\mathrm{2}}} \closeO{\rangle}   
                  }.
                \end{center}
                Then, by Lemma~\ref{lem:reduction-implies-csr-equiv}, from \( \possiblyWithSub\stageImetaColor{T^{\superscriptI} }_{{\mathrm{1}}}  \longrightarrow^{1}   \possiblyWithSub\stageImetaColor{T'^{\superscriptI} }_{{\mathrm{1}}}  \),
                we have \( \possiblyWithSub\stageImetaColor{T'^{\superscriptI} }_{{\mathrm{1}}}  \equiv^{1}  \possiblyWithSub\stageImetaColor{T^{\superscriptI} }_{{\mathrm{1}}} \), and can thereby derive
                \begin{center}
                  \infer[CqT0-Arr]{%
                     \possiblyWithSub\stageImetaColor{T'^{\superscriptI} }_{{\mathrm{1}}}  \equiv^{1}  \possiblyWithSub\stageImetaColor{T^{\superscriptI} }_{{\mathrm{1}}} 
                  \andalso
                    \infer[CqT1-Refl]{}{%
                       \possiblyWithSub\stageImetaColor{T^{\superscriptI} }_{{\mathrm{2}}}  \equiv^{1}  \possiblyWithSub\stageImetaColor{T^{\superscriptI} }_{{\mathrm{2}}} 
                    }
                  }{%
                      \openO{(} \possiblyWithSub\stageOmetaColor{x}  \relO{:}   \openO{\langle} \possiblyWithSub\stageImetaColor{T'^{\superscriptI} }_{{\mathrm{1}}} \closeO{\rangle}  \closeO{)} \relO{\to}   \openO{\langle} \possiblyWithSub\stageImetaColor{T^{\superscriptI} }_{{\mathrm{2}}} \closeO{\rangle}    \equiv^{0}   \openO{(} \possiblyWithSub\stageOmetaColor{x}  \relO{:}   \openO{\langle} \possiblyWithSub\stageImetaColor{T^{\superscriptI} }_{{\mathrm{1}}} \closeO{\rangle}  \closeO{)} \relO{\to}   \openO{\langle} \possiblyWithSub\stageImetaColor{T^{\superscriptI} }_{{\mathrm{2}}} \closeO{\rangle}   
                  }
                \end{center}
                Also, \(  \mathit{\Gamma} ,  \possiblyWithSub\stageOmetaColor{x}  : (  \openO{\langle} \possiblyWithSub\stageImetaColor{T^{\superscriptI} }_{{\mathrm{1}}} \closeO{\rangle}  )^{0}   \vdash^{1}  \possiblyWithSub\stageImetaColor{T^{\superscriptI} }_{{\mathrm{2}}} \) holds
                by evident weakening from \( \mathit{\Gamma}  \vdash^{1}  \possiblyWithSub\stageImetaColor{T^{\superscriptI} }_{{\mathrm{2}}} \), and enables us to derive
                \begin{center}
                  \infer[Wf0-Arr]{%
                    \infer[Wf0-Code]{%
                       \mathit{\Gamma}  \vdash^{1}  \possiblyWithSub\stageImetaColor{T^{\superscriptI} }_{{\mathrm{1}}} 
                    }{%
                       \mathit{\Gamma}  \vdash^{0}   \openO{\langle} \possiblyWithSub\stageImetaColor{T^{\superscriptI} }_{{\mathrm{1}}} \closeO{\rangle}  
                    }
                  \andalso
                    \infer[Wf0-Code]{%
                        \mathit{\Gamma} ,  \possiblyWithSub\stageOmetaColor{x}  : (  \openO{\langle} \possiblyWithSub\stageImetaColor{T^{\superscriptI} }_{{\mathrm{1}}} \closeO{\rangle}  )^{0}   \vdash^{1}  \possiblyWithSub\stageImetaColor{T^{\superscriptI} }_{{\mathrm{2}}} 
                    }{%
                        \mathit{\Gamma} ,  \possiblyWithSub\stageOmetaColor{x}  : (  \openO{\langle} \possiblyWithSub\stageImetaColor{T^{\superscriptI} }_{{\mathrm{1}}} \closeO{\rangle}  )^{0}   \vdash^{0}   \openO{\langle} \possiblyWithSub\stageImetaColor{T^{\superscriptI} }_{{\mathrm{2}}} \closeO{\rangle}  
                    }
                  }{%
                     \mathit{\Gamma}  \vdash^{0}   \openO{(} \possiblyWithSub\stageOmetaColor{x}  \relO{:}   \openO{\langle} \possiblyWithSub\stageImetaColor{T^{\superscriptI} }_{{\mathrm{1}}} \closeO{\rangle}  \closeO{)} \relO{\to}   \openO{\langle} \possiblyWithSub\stageImetaColor{T^{\superscriptI} }_{{\mathrm{2}}} \closeO{\rangle}   
                  }
                \end{center}
                Therefore, we can derive
                \begin{center}
                  \derive[T0-TyEquiv]{%
                     \mathit{\Gamma}  \vdash^{0}   \LeftAssertParen\openO{\langle} \possiblyWithSub\stageImetaColor{T'^{\superscriptI} }_{{\mathrm{1}}} \closeO{\rangle} \relO{\CastArrow} \openO{\langle} \possiblyWithSub\stageImetaColor{T^{\superscriptI} }_{{\mathrm{2}}} \closeO{\rangle}\RightAssertParen^{ L }   :   \openO{(} \possiblyWithSub\stageOmetaColor{x}  \relO{:}   \openO{\langle} \possiblyWithSub\stageImetaColor{T'^{\superscriptI} }_{{\mathrm{1}}} \closeO{\rangle}  \closeO{)} \relO{\to}   \openO{\langle} \possiblyWithSub\stageImetaColor{T^{\superscriptI} }_{{\mathrm{2}}} \closeO{\rangle}   
                  \\
                      \openO{(} \possiblyWithSub\stageOmetaColor{x}  \relO{:}   \openO{\langle} \possiblyWithSub\stageImetaColor{T'^{\superscriptI} }_{{\mathrm{1}}} \closeO{\rangle}  \closeO{)} \relO{\to}   \openO{\langle} \possiblyWithSub\stageImetaColor{T^{\superscriptI} }_{{\mathrm{2}}} \closeO{\rangle}    \equiv^{0}   \openO{(} \possiblyWithSub\stageOmetaColor{x}  \relO{:}   \openO{\langle} \possiblyWithSub\stageImetaColor{T^{\superscriptI} }_{{\mathrm{1}}} \closeO{\rangle}  \closeO{)} \relO{\to}   \openO{\langle} \possiblyWithSub\stageImetaColor{T^{\superscriptI} }_{{\mathrm{2}}} \closeO{\rangle}   
                  \andalso
                     \mathit{\Gamma}  \vdash^{0}   \openO{(} \possiblyWithSub\stageOmetaColor{x}  \relO{:}   \openO{\langle} \possiblyWithSub\stageImetaColor{T^{\superscriptI} }_{{\mathrm{1}}} \closeO{\rangle}  \closeO{)} \relO{\to}   \openO{\langle} \possiblyWithSub\stageImetaColor{T^{\superscriptI} }_{{\mathrm{2}}} \closeO{\rangle}   
                  }{%
                     \mathit{\Gamma}  \vdash^{0}   \LeftAssertParen\openO{\langle} \possiblyWithSub\stageImetaColor{T'^{\superscriptI} }_{{\mathrm{1}}} \closeO{\rangle} \relO{\CastArrow} \openO{\langle} \possiblyWithSub\stageImetaColor{T^{\superscriptI} }_{{\mathrm{2}}} \closeO{\rangle}\RightAssertParen^{ L }   :   \openO{(} \possiblyWithSub\stageOmetaColor{x}  \relO{:}   \openO{\langle} \possiblyWithSub\stageImetaColor{T^{\superscriptI} }_{{\mathrm{1}}} \closeO{\rangle}  \closeO{)} \relO{\to}   \openO{\langle} \possiblyWithSub\stageImetaColor{T^{\superscriptI} }_{{\mathrm{2}}} \closeO{\rangle}   
                  }
                \end{center}
              \item Case \derive[E0-Ass2]{%
                 \possiblyWithSub\stageImetaColor{T^{\superscriptI} }_{{\mathrm{2}}}  \longrightarrow^{1}   \possiblyWithSub\stageImetaColor{T'^{\superscriptI} }_{{\mathrm{2}}}  
              }{%
                  \LeftAssertParen\openO{\langle}  \possiblyWithSub\stageImetaColor{\tau^{\superscriptI} }_{{\mathrm{1}}}  \closeO{\rangle} \relO{\CastArrow} \openO{\langle} \possiblyWithSub\stageImetaColor{T^{\superscriptI} }_{{\mathrm{2}}} \closeO{\rangle}\RightAssertParen^{ L }   \longrightarrow^{0}    \LeftAssertParen\openO{\langle}  \possiblyWithSub\stageImetaColor{\tau^{\superscriptI} }_{{\mathrm{1}}}  \closeO{\rangle} \relO{\CastArrow} \openO{\langle} \possiblyWithSub\stageImetaColor{T'^{\superscriptI} }_{{\mathrm{2}}} \closeO{\rangle}\RightAssertParen^{ L }   
              }:
                We have \(\possiblyWithSub\stageImetaColor{T^{\superscriptI} }_{{\mathrm{1}}} =  \possiblyWithSub\stageImetaColor{\tau^{\superscriptI} }_{{\mathrm{1}}} \) and \( \mathit{\Gamma}  \vdash^{1}   \possiblyWithSub\stageImetaColor{\tau^{\superscriptI} }_{{\mathrm{1}}}  \).
                By IH, from \( \mathit{\Gamma}  \vdash^{1}  \possiblyWithSub\stageImetaColor{T^{\superscriptI} }_{{\mathrm{2}}} \) and \( \possiblyWithSub\stageImetaColor{T^{\superscriptI} }_{{\mathrm{2}}}  \longrightarrow^{1}   \possiblyWithSub\stageImetaColor{T'^{\superscriptI} }_{{\mathrm{2}}}  \),
                we have \( \mathit{\Gamma}  \vdash^{1}  \possiblyWithSub\stageImetaColor{T'^{\superscriptI} }_{{\mathrm{2}}} \).
                Then, by Lemma~\ref{lem:eval-preserves-compatibility},
                from \(  \possiblyWithSub\stageImetaColor{\tau^{\superscriptI} }_{{\mathrm{1}}}   \mathrel{||}^{1}  \possiblyWithSub\stageImetaColor{T^{\superscriptI} }_{{\mathrm{2}}} \) and \( \possiblyWithSub\stageImetaColor{T^{\superscriptI} }_{{\mathrm{2}}}  \longrightarrow^{1}   \possiblyWithSub\stageImetaColor{T'^{\superscriptI} }_{{\mathrm{2}}}  \),
                we have \(  \possiblyWithSub\stageImetaColor{\tau^{\superscriptI} }_{{\mathrm{1}}}   \mathrel{||}^{1}  \possiblyWithSub\stageImetaColor{T'^{\superscriptI} }_{{\mathrm{2}}} \).
                Thus, we can derive
                \begin{center}
                  \infer[T0-Ass]{%
                     \mathit{\Gamma}  \vdash^{1}   \possiblyWithSub\stageImetaColor{\tau^{\superscriptI} }_{{\mathrm{1}}}  
                  \andalso
                     \mathit{\Gamma}  \vdash^{1}  \possiblyWithSub\stageImetaColor{T'^{\superscriptI} }_{{\mathrm{2}}} 
                  \andalso
                      \possiblyWithSub\stageImetaColor{\tau^{\superscriptI} }_{{\mathrm{1}}}   \mathrel{||}^{1}  \possiblyWithSub\stageImetaColor{T'^{\superscriptI} }_{{\mathrm{2}}} 
                  \andalso
                    \possiblyWithSub\stageOmetaColor{x} \not\in \dom(\mathit{\Gamma})
                  }{%
                     \mathit{\Gamma}  \vdash^{0}   \LeftAssertParen\openO{\langle}  \possiblyWithSub\stageImetaColor{\tau^{\superscriptI} }_{{\mathrm{1}}}  \closeO{\rangle} \relO{\CastArrow} \openO{\langle} \possiblyWithSub\stageImetaColor{T'^{\superscriptI} }_{{\mathrm{2}}} \closeO{\rangle}\RightAssertParen^{ L }   :   \openO{(} \possiblyWithSub\stageOmetaColor{x}  \relO{:}   \openO{\langle}  \possiblyWithSub\stageImetaColor{\tau^{\superscriptI} }_{{\mathrm{1}}}  \closeO{\rangle}  \closeO{)} \relO{\to}   \openO{\langle} \possiblyWithSub\stageImetaColor{T'^{\superscriptI} }_{{\mathrm{2}}} \closeO{\rangle}   
                  }
                \end{center}
                By Lemma~\ref{lem:reduction-implies-csr-equiv},
                from \( \possiblyWithSub\stageImetaColor{T^{\superscriptI} }_{{\mathrm{2}}}  \longrightarrow^{1}   \possiblyWithSub\stageImetaColor{T'^{\superscriptI} }_{{\mathrm{2}}}  \), we have \( \possiblyWithSub\stageImetaColor{T'^{\superscriptI} }_{{\mathrm{2}}}  \equiv^{1}  \possiblyWithSub\stageImetaColor{T^{\superscriptI} }_{{\mathrm{2}}} \),
                and can thereby derive
                \begin{center}
                  \infer[CqT0-Arr]{%
                    \infer[CqT0-Code]{
                      \infer[CqT0-Refl]{}{
                          \possiblyWithSub\stageImetaColor{\tau^{\superscriptI} }_{{\mathrm{1}}}   \equiv^{1}   \possiblyWithSub\stageImetaColor{\tau^{\superscriptI} }_{{\mathrm{1}}}  
                      }
                    }{%
                        \openO{\langle}  \possiblyWithSub\stageImetaColor{\tau^{\superscriptI} }_{{\mathrm{1}}}  \closeO{\rangle}   \equiv^{0}   \openO{\langle}  \possiblyWithSub\stageImetaColor{\tau^{\superscriptI} }_{{\mathrm{1}}}  \closeO{\rangle}  
                    }
                  \andalso
                    \infer[CqT0-Code]{%
                       \possiblyWithSub\stageImetaColor{T'^{\superscriptI} }_{{\mathrm{2}}}  \equiv^{1}  \possiblyWithSub\stageImetaColor{T^{\superscriptI} }_{{\mathrm{2}}} 
                    }{
                        \openO{\langle} \possiblyWithSub\stageImetaColor{T'^{\superscriptI} }_{{\mathrm{2}}} \closeO{\rangle}   \equiv^{0}   \openO{\langle} \possiblyWithSub\stageImetaColor{T^{\superscriptI} }_{{\mathrm{2}}} \closeO{\rangle}  
                    }
                  }{%
                      \openO{(} \possiblyWithSub\stageOmetaColor{x}  \relO{:}   \openO{\langle}  \possiblyWithSub\stageImetaColor{\tau^{\superscriptI} }_{{\mathrm{1}}}  \closeO{\rangle}  \closeO{)} \relO{\to}   \openO{\langle} \possiblyWithSub\stageImetaColor{T'^{\superscriptI} }_{{\mathrm{2}}} \closeO{\rangle}    \equiv^{0}   \openO{(} \possiblyWithSub\stageOmetaColor{x}  \relO{:}   \openO{\langle}  \possiblyWithSub\stageImetaColor{\tau^{\superscriptI} }_{{\mathrm{1}}}  \closeO{\rangle}  \closeO{)} \relO{\to}   \openO{\langle} \possiblyWithSub\stageImetaColor{T^{\superscriptI} }_{{\mathrm{2}}} \closeO{\rangle}   
                  }
                \end{center}
                At the same time, we have \(  \mathit{\Gamma} ,  \possiblyWithSub\stageOmetaColor{x}  : (  \openO{\langle} \possiblyWithSub\stageImetaColor{T^{\superscriptI} }_{{\mathrm{1}}} \closeO{\rangle}  )^{0}   \vdash^{1}  \possiblyWithSub\stageImetaColor{T^{\superscriptI} }_{{\mathrm{2}}} \)
                by evident weakening from \( \mathit{\Gamma}  \vdash^{1}  \possiblyWithSub\stageImetaColor{T^{\superscriptI} }_{{\mathrm{2}}} \), and thereby have
                \begin{center}
                  \infer[Wf0-Arr]{%
                     \mathit{\Gamma}  \vdash^{0}   \openO{\langle}  \possiblyWithSub\stageImetaColor{\tau^{\superscriptI} }_{{\mathrm{1}}}  \closeO{\rangle}  
                  \andalso
                    \infer[Wf0-Code]{%
                        \mathit{\Gamma} ,  \possiblyWithSub\stageOmetaColor{x}  : (  \openO{\langle}  \possiblyWithSub\stageImetaColor{\tau^{\superscriptI} }_{{\mathrm{1}}}  \closeO{\rangle}  )^{0}   \vdash^{1}  \possiblyWithSub\stageImetaColor{T^{\superscriptI} }_{{\mathrm{2}}} 
                    }{
                        \mathit{\Gamma} ,  \possiblyWithSub\stageOmetaColor{x}  : (  \openO{\langle}  \possiblyWithSub\stageImetaColor{\tau^{\superscriptI} }_{{\mathrm{1}}}  \closeO{\rangle}  )^{0}   \vdash^{0}   \openO{\langle} \possiblyWithSub\stageImetaColor{T^{\superscriptI} }_{{\mathrm{2}}} \closeO{\rangle}  
                    }
                  }{%
                     \mathit{\Gamma}  \vdash^{0}   \openO{(} \possiblyWithSub\stageOmetaColor{x}  \relO{:}   \openO{\langle}  \possiblyWithSub\stageImetaColor{\tau^{\superscriptI} }_{{\mathrm{1}}}  \closeO{\rangle}  \closeO{)} \relO{\to}   \openO{\langle} \possiblyWithSub\stageImetaColor{T^{\superscriptI} }_{{\mathrm{2}}} \closeO{\rangle}   
                  }
                \end{center}
                Therefore, we can derive
                \begin{center}
                  \derive[T0-TyEquiv]{%
                     \mathit{\Gamma}  \vdash^{0}   \LeftAssertParen\openO{\langle}  \possiblyWithSub\stageImetaColor{\tau^{\superscriptI} }_{{\mathrm{1}}}  \closeO{\rangle} \relO{\CastArrow} \openO{\langle} \possiblyWithSub\stageImetaColor{T'^{\superscriptI} }_{{\mathrm{2}}} \closeO{\rangle}\RightAssertParen^{ L }   :   \openO{(} \possiblyWithSub\stageOmetaColor{x}  \relO{:}   \openO{\langle}  \possiblyWithSub\stageImetaColor{\tau^{\superscriptI} }_{{\mathrm{1}}}  \closeO{\rangle}  \closeO{)} \relO{\to}   \openO{\langle} \possiblyWithSub\stageImetaColor{T'^{\superscriptI} }_{{\mathrm{2}}} \closeO{\rangle}   
                  \\
                      \openO{(} \possiblyWithSub\stageOmetaColor{x}  \relO{:}   \openO{\langle}  \possiblyWithSub\stageImetaColor{\tau^{\superscriptI} }_{{\mathrm{1}}}  \closeO{\rangle}  \closeO{)} \relO{\to}   \openO{\langle} \possiblyWithSub\stageImetaColor{T'^{\superscriptI} }_{{\mathrm{2}}} \closeO{\rangle}    \equiv^{0}   \openO{(} \possiblyWithSub\stageOmetaColor{x}  \relO{:}   \openO{\langle}  \possiblyWithSub\stageImetaColor{\tau^{\superscriptI} }_{{\mathrm{1}}}  \closeO{\rangle}  \closeO{)} \relO{\to}   \openO{\langle} \possiblyWithSub\stageImetaColor{T^{\superscriptI} }_{{\mathrm{2}}} \closeO{\rangle}   
                  \andalso
                     \mathit{\Gamma}  \vdash^{0}   \openO{(} \possiblyWithSub\stageOmetaColor{x}  \relO{:}   \openO{\langle}  \possiblyWithSub\stageImetaColor{\tau^{\superscriptI} }_{{\mathrm{1}}}  \closeO{\rangle}  \closeO{)} \relO{\to}   \openO{\langle} \possiblyWithSub\stageImetaColor{T^{\superscriptI} }_{{\mathrm{2}}} \closeO{\rangle}   
                  }{%
                     \mathit{\Gamma}  \vdash^{0}   \LeftAssertParen\openO{\langle}  \possiblyWithSub\stageImetaColor{\tau^{\superscriptI} }_{{\mathrm{1}}}  \closeO{\rangle} \relO{\CastArrow} \openO{\langle} \possiblyWithSub\stageImetaColor{T'^{\superscriptI} }_{{\mathrm{2}}} \closeO{\rangle}\RightAssertParen^{ L }   :   \openO{(} \possiblyWithSub\stageOmetaColor{x}  \relO{:}   \openO{\langle}  \possiblyWithSub\stageImetaColor{\tau^{\superscriptI} }_{{\mathrm{1}}}  \closeO{\rangle}  \closeO{)} \relO{\to}   \openO{\langle} \possiblyWithSub\stageImetaColor{T^{\superscriptI} }_{{\mathrm{2}}} \closeO{\rangle}   
                  }
                \end{center}
              \item Case \derive[E0-AssPass]{}{%
                  \LeftAssertParen\openO{\langle}  \possiblyWithSub\stageImetaColor{\tau^{\superscriptI} }  \closeO{\rangle} \relO{\CastArrow} \openO{\langle}  \possiblyWithSub\stageImetaColor{\tau^{\superscriptI} }  \closeO{\rangle}\RightAssertParen^{ L }   \longrightarrow^{0}    \ordO{\lambda} \possiblyWithSub\stageOmetaColor{x}  \relO{:}   \openO{\langle}  \possiblyWithSub\stageImetaColor{\tau^{\superscriptI} }  \closeO{\rangle}  \punctO{.}\   \possiblyWithSub\stageOmetaColor{x}    
              }:
                Straightforward.
            \end{itemize}
          \item Case \derive[T0-RfnAct]{%
             \mathit{\Gamma}  \vdash^{0}    \openO{\{} \possiblyWithSub\stageOmetaColor{\nu}  \relO{:}  \possiblyWithSub\stageOmetaColor{B}  \relO{\mid}  \possiblyWithSub\stageOmetaColor{N^{\superscriptO} }_{{\mathrm{1}}} \closeO{\} }   
          \andalso
             \mathit{\Gamma}  \vdash^{0}  \possiblyWithSub\stageOmetaColor{N^{\superscriptO} }_{{\mathrm{0}}}  :    \openO{\{} \possiblyWithSub\stageOmetaColor{\nu}_{{\mathrm{0}}}  \relO{:}   \ttO{Bool}   \relO{\mid}     \ttO{true}    \closeO{\} }   
          \\
            \ConstEnvPers(c_{{\mathrm{2}}}) = \possiblyWithSub\stageImetaColor{B}
          \andalso
               [    \possiblyWithSub\stageOmetaColor{c}_{{\mathrm{2}}}    /  \possiblyWithSub\stageOmetaColor{\nu}  ]    \possiblyWithSub\stageOmetaColor{N^{\superscriptO} }_{{\mathrm{1}}}   \longrightarrow^{0\,\ast}   \possiblyWithSub\stageOmetaColor{N^{\superscriptO} }_{{\mathrm{0}}}  
          }{%
             \mathit{\Gamma}  \vdash^{0}   \LeftAssertParen   \openO{\{} \possiblyWithSub\stageOmetaColor{\nu}  \relO{:}  \possiblyWithSub\stageOmetaColor{B}  \relO{\mid}  \possiblyWithSub\stageOmetaColor{N^{\superscriptO} }_{{\mathrm{1}}} \closeO{\} }  \punctO{,}  \possiblyWithSub\stageOmetaColor{N^{\superscriptO} }_{{\mathrm{0}}} \punctO{,}  \possiblyWithSub\stageOmetaColor{c}_{{\mathrm{2}}}  \RightAssertParen^{ L }   :    \openO{\{} \possiblyWithSub\stageOmetaColor{\nu}  \relO{:}  \possiblyWithSub\stageOmetaColor{B}  \relO{\mid}  \possiblyWithSub\stageOmetaColor{N^{\superscriptO} }_{{\mathrm{1}}} \closeO{\} }   
          }:
            By case analysis for the derivation of \( \possiblyWithSub\stageOmetaColor{N^{\superscriptO} }  \longrightarrow^{0}   \possiblyWithSub\stageOmetaColor{N'^{\superscriptO} }  \):
            \begin{itemize}
              \item Case \derive[E0-RfnAct]{%
                 \possiblyWithSub\stageOmetaColor{N^{\superscriptO} }_{{\mathrm{0}}}  \longrightarrow^{0}   \possiblyWithSub\stageOmetaColor{N'^{\superscriptO} }_{{\mathrm{0}}}  
              }{%
                  \LeftAssertParen   \openO{\{} \possiblyWithSub\stageOmetaColor{\nu}  \relO{:}  \possiblyWithSub\stageOmetaColor{B}  \relO{\mid}  \possiblyWithSub\stageOmetaColor{N^{\superscriptO} }_{{\mathrm{1}}} \closeO{\} }  \punctO{,}  \possiblyWithSub\stageOmetaColor{N^{\superscriptO} }_{{\mathrm{0}}} \punctO{,}  \possiblyWithSub\stageOmetaColor{c}_{{\mathrm{2}}}  \RightAssertParen^{ L }   \longrightarrow^{0}    \LeftAssertParen   \openO{\{} \possiblyWithSub\stageOmetaColor{\nu}  \relO{:}  \possiblyWithSub\stageOmetaColor{B}  \relO{\mid}  \possiblyWithSub\stageOmetaColor{N^{\superscriptO} }_{{\mathrm{1}}} \closeO{\} }  \punctO{,}  \possiblyWithSub\stageOmetaColor{N'^{\superscriptO} }_{{\mathrm{0}}} \punctO{,}  \possiblyWithSub\stageOmetaColor{c}_{{\mathrm{2}}}  \RightAssertParen^{ L }   
              }:
                By IH, from \( \mathit{\Gamma}  \vdash^{0}  \possiblyWithSub\stageOmetaColor{N^{\superscriptO} }_{{\mathrm{0}}}  :    \openO{\{} \possiblyWithSub\stageOmetaColor{\nu}_{{\mathrm{0}}}  \relO{:}   \ttO{Bool}   \relO{\mid}     \ttO{true}    \closeO{\} }   \) and \( \possiblyWithSub\stageOmetaColor{N^{\superscriptO} }_{{\mathrm{0}}}  \longrightarrow^{0}   \possiblyWithSub\stageOmetaColor{N'^{\superscriptO} }_{{\mathrm{0}}}  \),
                we have \( \mathit{\Gamma}  \vdash^{0}  \possiblyWithSub\stageOmetaColor{N'^{\superscriptO} }_{{\mathrm{0}}}  :    \openO{\{} \possiblyWithSub\stageOmetaColor{\nu}_{{\mathrm{0}}}  \relO{:}   \ttO{Bool}   \relO{\mid}     \ttO{true}    \closeO{\} }   \).
                Since \(   [    \possiblyWithSub\stageOmetaColor{c}_{{\mathrm{2}}}    /  \possiblyWithSub\stageOmetaColor{\nu}  ]    \possiblyWithSub\stageOmetaColor{N^{\superscriptO} }_{{\mathrm{1}}}   \longrightarrow^{0\,\ast}   \possiblyWithSub\stageOmetaColor{N^{\superscriptO} }_{{\mathrm{0}}}   \longrightarrow^{0} \possiblyWithSub\stageOmetaColor{N'^{\superscriptO} }_{{\mathrm{0}}}\) holds,
                we can derive
                \begin{center}
                  \derive[T0-RfnAct]{%
                     \mathit{\Gamma}  \vdash^{0}    \openO{\{} \possiblyWithSub\stageOmetaColor{\nu}  \relO{:}  \possiblyWithSub\stageOmetaColor{B}  \relO{\mid}  \possiblyWithSub\stageOmetaColor{N^{\superscriptO} }_{{\mathrm{1}}} \closeO{\} }   
                  \andalso
                     \mathit{\Gamma}  \vdash^{0}  \possiblyWithSub\stageOmetaColor{N'^{\superscriptO} }_{{\mathrm{0}}}  :    \openO{\{} \possiblyWithSub\stageOmetaColor{\nu}_{{\mathrm{0}}}  \relO{:}   \ttO{Bool}   \relO{\mid}     \ttO{true}    \closeO{\} }   
                  \\
                    \ConstEnvPers(c_{{\mathrm{2}}}) = \possiblyWithSub\stageImetaColor{B}
                  \andalso
                       [    \possiblyWithSub\stageOmetaColor{c}_{{\mathrm{2}}}    /  \possiblyWithSub\stageOmetaColor{\nu}  ]    \possiblyWithSub\stageOmetaColor{N^{\superscriptO} }_{{\mathrm{1}}}   \longrightarrow^{0\,\ast}   \possiblyWithSub\stageOmetaColor{N'^{\superscriptO} }_{{\mathrm{0}}}  
                  }{%
                     \mathit{\Gamma}  \vdash^{0}   \LeftAssertParen   \openO{\{} \possiblyWithSub\stageOmetaColor{\nu}  \relO{:}  \possiblyWithSub\stageOmetaColor{B}  \relO{\mid}  \possiblyWithSub\stageOmetaColor{N^{\superscriptO} }_{{\mathrm{1}}} \closeO{\} }  \punctO{,}  \possiblyWithSub\stageOmetaColor{N'^{\superscriptO} }_{{\mathrm{0}}} \punctO{,}  \possiblyWithSub\stageOmetaColor{c}_{{\mathrm{2}}}  \RightAssertParen^{ L }   :    \openO{\{} \possiblyWithSub\stageOmetaColor{\nu}  \relO{:}  \possiblyWithSub\stageOmetaColor{B}  \relO{\mid}  \possiblyWithSub\stageOmetaColor{N^{\superscriptO} }_{{\mathrm{1}}} \closeO{\} }   
                  }.
                \end{center}
              \item Case \derive[E0-RfnPass]{}{%
                  \LeftAssertParen   \openO{\{} \possiblyWithSub\stageOmetaColor{\nu}  \relO{:}  \possiblyWithSub\stageOmetaColor{B}  \relO{\mid}  \possiblyWithSub\stageOmetaColor{N^{\superscriptO} }_{{\mathrm{1}}} \closeO{\} }  \punctO{,}     \ttO{true}    \punctO{,}  \possiblyWithSub\stageOmetaColor{c}_{{\mathrm{2}}}  \RightAssertParen^{ L }   \longrightarrow^{0}     \possiblyWithSub\stageOmetaColor{c}_{{\mathrm{2}}}    
              }:
                Since \(\possiblyWithSub\stageOmetaColor{N^{\superscriptO} }_{{\mathrm{0}}} =   \ttO{true}  \), we have \(   [    \possiblyWithSub\stageOmetaColor{c}_{{\mathrm{2}}}    /  \possiblyWithSub\stageOmetaColor{\nu}  ]    \possiblyWithSub\stageOmetaColor{N^{\superscriptO} }_{{\mathrm{1}}}   \longrightarrow^{0\,\ast}      \ttO{true}     \).
                This enables us to derive
                \begin{center}
                  \derive[T0-RfnPred]{%
                     \mathit{\Gamma}  \vdash^{0}    \openO{\{} \possiblyWithSub\stageOmetaColor{\nu}  \relO{:}  \possiblyWithSub\stageOmetaColor{B}  \relO{\mid}  \possiblyWithSub\stageOmetaColor{N^{\superscriptO} }_{{\mathrm{1}}} \closeO{\} }   
                  \andalso
                    \ConstEnvPers(c_{{\mathrm{2}}}) = \possiblyWithSub\stageImetaColor{B}
                  \andalso
                       [    \possiblyWithSub\stageOmetaColor{c}_{{\mathrm{2}}}    /  \possiblyWithSub\stageOmetaColor{\nu}  ]    \possiblyWithSub\stageOmetaColor{N^{\superscriptO} }_{{\mathrm{1}}}   \longrightarrow^{0\,\ast}      \ttO{true}     
                  }{%
                     \mathit{\Gamma}  \vdash^{0}    \possiblyWithSub\stageOmetaColor{c}_{{\mathrm{2}}}    :    \openO{\{} \possiblyWithSub\stageOmetaColor{\nu}  \relO{:}  \possiblyWithSub\stageOmetaColor{B}  \relO{\mid}  \possiblyWithSub\stageOmetaColor{N^{\superscriptO} }_{{\mathrm{1}}} \closeO{\} }   
                  }.
                \end{center}
              \item The other cases contradict the form~\( \LeftAssertParen   \openO{\{} \possiblyWithSub\stageOmetaColor{\nu}  \relO{:}  \possiblyWithSub\stageOmetaColor{B}  \relO{\mid}  \possiblyWithSub\stageOmetaColor{N^{\superscriptO} }_{{\mathrm{1}}} \closeO{\} }  \punctO{,}  \possiblyWithSub\stageOmetaColor{N^{\superscriptO} }_{{\mathrm{0}}} \punctO{,}  \possiblyWithSub\stageOmetaColor{c}_{{\mathrm{2}}}  \RightAssertParen^{ L } \).
            \end{itemize}
          \item The other cases contradict the assumption.
        \end{itemize}
      \item
        \begin{itemize}
          \item Case \derive[T1-TyEquiv]{%
             \mathit{\Gamma}  \vdash^{1}  \possiblyWithSub\stageImetaColor{N^{\superscriptI} }  :  \possiblyWithSub\stageImetaColor{T'^{\superscriptI} } 
          \andalso
             \possiblyWithSub\stageImetaColor{T'^{\superscriptI} }  \equiv^{1}  \possiblyWithSub\stageImetaColor{T^{\superscriptI} } 
          \andalso
             \mathit{\Gamma}  \vdash^{1}  \possiblyWithSub\stageImetaColor{T^{\superscriptI} } 
          }{%
             \mathit{\Gamma}  \vdash^{1}  \possiblyWithSub\stageImetaColor{N^{\superscriptI} }  :  \possiblyWithSub\stageImetaColor{T^{\superscriptI} } 
          }:
            Straightforward by IH.
          \item Case \derive[T1-Esc]{%
             \mathit{\Gamma}  \vdash^{0}  \possiblyWithSub\stageOmetaColor{N^{\superscriptO} }  :   \openO{\langle} \possiblyWithSub\stageImetaColor{T^{\superscriptI} } \closeO{\rangle}  
          }{%
             \mathit{\Gamma}  \vdash^{1}   \ordI{\sim} \possiblyWithSub\stageOmetaColor{N^{\superscriptO} }   :  \possiblyWithSub\stageImetaColor{T^{\superscriptI} } 
          }:
            By case analysis of the last rule used for deriving the reduction.
            \begin{itemize}
              \item Case \derive[E1-Esc]{%
                 \possiblyWithSub\stageOmetaColor{N^{\superscriptO} }  \longrightarrow^{0}   \possiblyWithSub\stageOmetaColor{N'^{\superscriptO} }  
              }{%
                  \ordI{\sim} \possiblyWithSub\stageOmetaColor{N^{\superscriptO} }   \longrightarrow^{1}    \ordI{\sim} \possiblyWithSub\stageOmetaColor{N'^{\superscriptO} }   
              }:
                Straightforward by IH.
              \item Case \derive[E1-Cancel]{}{%
                  \ordI{\sim}  \openO{\langle} \possiblyWithSub\stageImetaColor{N'^{\superscriptI} } \closeO{\rangle}    \longrightarrow^{1}   \possiblyWithSub\stageImetaColor{N'^{\superscriptI} }  
              }:
                By Lemma~\ref{lem:bracket-inversion},
                from \( \mathit{\Gamma}  \vdash^{0}   \openO{\langle} \possiblyWithSub\stageImetaColor{N'^{\superscriptI} } \closeO{\rangle}   :   \openO{\langle} \possiblyWithSub\stageImetaColor{T^{\superscriptI} } \closeO{\rangle}  \),
                there exists \(\possiblyWithSub\stageImetaColor{T'^{\superscriptI} }\) such that
                \( \mathit{\Gamma}  \vdash^{1}  \possiblyWithSub\stageImetaColor{N'^{\superscriptI} }  :  \possiblyWithSub\stageImetaColor{T'^{\superscriptI} } \) and \(  \openO{\langle} \possiblyWithSub\stageImetaColor{T'^{\superscriptI} } \closeO{\rangle}   \equiv^{0}   \openO{\langle} \possiblyWithSub\stageImetaColor{T^{\superscriptI} } \closeO{\rangle}  \).
                Therefore, by Lemma~\ref{lem:code-type-csr-equiv-inversion},
                we have \( \possiblyWithSub\stageImetaColor{T'^{\superscriptI} }  \equiv^{1}  \possiblyWithSub\stageImetaColor{T^{\superscriptI} } \), and we can derive
                \begin{center}
                  \derive[T1-TyEquiv]{%
                     \mathit{\Gamma}  \vdash^{1}  \possiblyWithSub\stageImetaColor{N'^{\superscriptI} }  :  \possiblyWithSub\stageImetaColor{T'^{\superscriptI} } 
                  \andalso
                     \possiblyWithSub\stageImetaColor{T'^{\superscriptI} }  \equiv^{1}  \possiblyWithSub\stageImetaColor{T^{\superscriptI} } 
                  \andalso
                     \mathit{\Gamma}  \vdash^{1}  \possiblyWithSub\stageImetaColor{T^{\superscriptI} } 
                  }{%
                     \mathit{\Gamma}  \vdash^{1}  \possiblyWithSub\stageImetaColor{N'^{\superscriptI} }  :  \possiblyWithSub\stageImetaColor{T^{\superscriptI} } 
                  }.
                \end{center}
            \end{itemize}
          \item The other cases are straightforward by IH or contradiction.
        \end{itemize}
      \item
        All the cases are straightforward by IH or contradiction.
    \end{enumerate}
  \end{proof}
  \begin{lemma}[Value Inversion for Arrow Types]\label{lem:value-inversion-of-arrow}
    If \( \mathit{\Gamma}  \vdash^{0}   \possiblyWithSub\stageOmetaColor{v^{\superscriptO} }   :   \openO{(} \possiblyWithSub\stageOmetaColor{x}  \relO{:}  \possiblyWithSub\stageOmetaColor{T^{\superscriptO} }_{{\mathrm{1}}} \closeO{)} \relO{\to}  \possiblyWithSub\stageOmetaColor{T^{\superscriptO} }_{{\mathrm{2}}}  \), then
    one of the following holds:
    \begin{enumerate}
      \item
        there exist \(\possiblyWithSub\stageOmetaColor{T'^{\superscriptO} }_{{\mathrm{1}}}\) and \(\possiblyWithSub\stageOmetaColor{N^{\superscriptO} }_{{\mathrm{2}}}\) such that \(\possiblyWithSub\stageOmetaColor{v^{\superscriptO} } =  \openO{(}  \ordO{\lambda} \possiblyWithSub\stageOmetaColor{x}  \relO{:}  \possiblyWithSub\stageOmetaColor{T'^{\superscriptO} }_{{\mathrm{1}}} \punctO{.}\  \possiblyWithSub\stageOmetaColor{N^{\superscriptO} }_{{\mathrm{2}}}  \closeO{)} \);
      \item
        there exist \(\possiblyWithSub\stageOmetaColor{\nu}\), \(\possiblyWithSub\stageOmetaColor{B}\), \(\possiblyWithSub\stageOmetaColor{N'^{\superscriptO} }\), and \(L\) such that
        \(\possiblyWithSub\stageOmetaColor{v^{\superscriptO} } =  \LeftAssertParen \relO{\CastArrow}   \openO{\{} \possiblyWithSub\stageOmetaColor{\nu}  \relO{:}  \possiblyWithSub\stageOmetaColor{B}  \relO{\mid}  \possiblyWithSub\stageOmetaColor{N'^{\superscriptO} } \closeO{\} }   \RightAssertParen^{ L } \) and \(  \possiblyWithSub\stageImetaColor{B}   \gg  \possiblyWithSub\stageOmetaColor{T^{\superscriptO} }_{{\mathrm{1}}} \); or
      \item
        \(\possiblyWithSub\stageOmetaColor{v^{\superscriptO} }\) is of the form~\(\possiblyWithSub\stageOmetaColor{a}\).
    \end{enumerate}
  \end{lemma}
  \begin{proof}
    By induction on the derivation of \( \mathit{\Gamma}  \vdash^{0}   \possiblyWithSub\stageOmetaColor{v^{\superscriptO} }   :   \openO{(} \possiblyWithSub\stageOmetaColor{x}  \relO{:}  \possiblyWithSub\stageOmetaColor{T^{\superscriptO} }_{{\mathrm{1}}} \closeO{)} \relO{\to}  \possiblyWithSub\stageOmetaColor{T^{\superscriptO} }_{{\mathrm{2}}}  \).
    \begin{itemize}
      \item Cases \rulename{T0-Abs}, \rulename{T0-Rfn}, \rulename{T0-Cst0}, and \rulename{T0-CstP} are immediate.
      \item Case \derive[T0-TyEquiv]{%
         \mathit{\Gamma}  \vdash^{0}   \possiblyWithSub\stageOmetaColor{v^{\superscriptO} }   :  \possiblyWithSub\stageOmetaColor{T'^{\superscriptO} } 
      \andalso
         \possiblyWithSub\stageOmetaColor{T'^{\superscriptO} }  \equiv^{0}   \openO{(} \possiblyWithSub\stageOmetaColor{x}  \relO{:}  \possiblyWithSub\stageOmetaColor{T^{\superscriptO} }_{{\mathrm{1}}} \closeO{)} \relO{\to}  \possiblyWithSub\stageOmetaColor{T^{\superscriptO} }_{{\mathrm{2}}}  
      \andalso
         \mathit{\Gamma}  \vdash^{0}   \openO{(} \possiblyWithSub\stageOmetaColor{x}  \relO{:}  \possiblyWithSub\stageOmetaColor{T^{\superscriptO} }_{{\mathrm{1}}} \closeO{)} \relO{\to}  \possiblyWithSub\stageOmetaColor{T^{\superscriptO} }_{{\mathrm{2}}}  
      }{%
         \mathit{\Gamma}  \vdash^{0}   \possiblyWithSub\stageOmetaColor{v^{\superscriptO} }   :   \openO{(} \possiblyWithSub\stageOmetaColor{x}  \relO{:}  \possiblyWithSub\stageOmetaColor{T^{\superscriptO} }_{{\mathrm{1}}} \closeO{)} \relO{\to}  \possiblyWithSub\stageOmetaColor{T^{\superscriptO} }_{{\mathrm{2}}}  
      }:
        By \( \possiblyWithSub\stageOmetaColor{T'^{\superscriptO} }  \equiv^{0}   \openO{(} \possiblyWithSub\stageOmetaColor{x}  \relO{:}  \possiblyWithSub\stageOmetaColor{T^{\superscriptO} }_{{\mathrm{1}}} \closeO{)} \relO{\to}  \possiblyWithSub\stageOmetaColor{T^{\superscriptO} }_{{\mathrm{2}}}  \) and Lemma~\ref{lem:arrow-type-csr-equiv-form},
        \(\possiblyWithSub\stageOmetaColor{T'^{\superscriptO} }\) is of the form~\( \openO{(} \possiblyWithSub\stageOmetaColor{x}  \relO{:}  \possiblyWithSub\stageOmetaColor{T'^{\superscriptO} }_{{\mathrm{1}}} \closeO{)} \relO{\to}  \possiblyWithSub\stageOmetaColor{T'^{\superscriptO} }_{{\mathrm{2}}} \).
        This enables us to use IH on \( \mathit{\Gamma}  \vdash^{0}   \possiblyWithSub\stageOmetaColor{v^{\superscriptO} }   :   \openO{(} \possiblyWithSub\stageOmetaColor{x}  \relO{:}  \possiblyWithSub\stageOmetaColor{T'^{\superscriptO} }_{{\mathrm{1}}} \closeO{)} \relO{\to}  \possiblyWithSub\stageOmetaColor{T'^{\superscriptO} }_{{\mathrm{2}}}  \).
        \begin{itemize}
          \item Case where there exist \(\possiblyWithSub\stageOmetaColor{\nu}\), \(\possiblyWithSub\stageOmetaColor{B}\), \(\possiblyWithSub\stageOmetaColor{N'^{\superscriptO} }\), and \(L\) such that
          \(\possiblyWithSub\stageOmetaColor{v^{\superscriptO} } =  \LeftAssertParen \relO{\CastArrow}   \openO{\{} \possiblyWithSub\stageOmetaColor{\nu}  \relO{:}  \possiblyWithSub\stageOmetaColor{B}  \relO{\mid}  \possiblyWithSub\stageOmetaColor{N'^{\superscriptO} } \closeO{\} }   \RightAssertParen^{ L } \) and \(  \possiblyWithSub\stageImetaColor{B}   \gg  \possiblyWithSub\stageOmetaColor{T'^{\superscriptO} }_{{\mathrm{1}}} \):
            By Lemma~\ref{lem:csr-equiv-preserves-unlifting-relation}, we have \(  \possiblyWithSub\stageImetaColor{B}   \gg  \possiblyWithSub\stageOmetaColor{T^{\superscriptO} }_{{\mathrm{1}}} \).
          \item The other two cases are immediate.
        \end{itemize}
      \item Case \rulename{To-App}:
        By the definition of values, \(\possiblyWithSub\stageOmetaColor{v^{\superscriptO} }\) is necessarily of the form~\(\possiblyWithSub\stageOmetaColor{a}\).
      \item The other cases contradict the assumption.
    \end{itemize}
  \end{proof}
  \begin{lemma}[Value Inversion for Code Types]\label{lem:value-inversion-of-code}
    If \( \mathit{\Gamma}  \vdash^{0}   \possiblyWithSub\stageOmetaColor{v^{\superscriptO} }   :   \openO{\langle} \possiblyWithSub\stageImetaColor{T^{\superscriptI} } \closeO{\rangle}  \), then
    there exists a stage-1 value \(\possiblyWithSub\stageImetaColor{v^{\superscriptI} }\) such that
    \(\possiblyWithSub\stageOmetaColor{v^{\superscriptO} } =  \openO{\langle} \possiblyWithSub\stageImetaColor{v^{\superscriptI} } \closeO{\rangle} \).
  \end{lemma}
  \begin{proof}
    By induction on the derivation of \( \mathit{\Gamma}  \vdash^{0}   \possiblyWithSub\stageOmetaColor{v^{\superscriptO} }   :   \openO{\langle} \possiblyWithSub\stageImetaColor{T^{\superscriptI} } \closeO{\rangle}  \).
    \begin{itemize}
      \item Case \rulename{T0-Brkt} is immediate.
      \item Case \derive[T0-TyEquiv]{%
         \mathit{\Gamma}  \vdash^{0}   \possiblyWithSub\stageOmetaColor{v^{\superscriptO} }   :  \possiblyWithSub\stageOmetaColor{T'^{\superscriptO} } 
      \andalso
         \possiblyWithSub\stageOmetaColor{T'^{\superscriptO} }  \equiv^{0}   \openO{\langle} \possiblyWithSub\stageImetaColor{T^{\superscriptI} } \closeO{\rangle}  
      \andalso
         \mathit{\Gamma}  \vdash^{0}   \openO{\langle} \possiblyWithSub\stageImetaColor{T^{\superscriptI} } \closeO{\rangle}  
      }{%
         \mathit{\Gamma}  \vdash^{0}   \possiblyWithSub\stageOmetaColor{v^{\superscriptO} }   :   \openO{\langle} \possiblyWithSub\stageImetaColor{T^{\superscriptI} } \closeO{\rangle}  
      }:
        By \( \possiblyWithSub\stageOmetaColor{T'^{\superscriptO} }  \equiv^{0}   \openO{\langle} \possiblyWithSub\stageImetaColor{T^{\superscriptI} } \closeO{\rangle}  \) and Lemma~\ref{lem:code-type-csr-equiv-form},
        \(\possiblyWithSub\stageOmetaColor{T'^{\superscriptO} }\) is of the form~\( \openO{\langle} \possiblyWithSub\stageImetaColor{T'^{\superscriptI} } \closeO{\rangle} \).
        Then, by IH, from \( \mathit{\Gamma}  \vdash^{0}   \possiblyWithSub\stageOmetaColor{v^{\superscriptO} }   :   \openO{\langle} \possiblyWithSub\stageImetaColor{T'^{\superscriptI} } \closeO{\rangle}  \),
        we have \(\possiblyWithSub\stageOmetaColor{v^{\superscriptO} } =  \openO{\langle} \possiblyWithSub\stageImetaColor{v^{\superscriptI} } \closeO{\rangle} \) for some stage-1 value \(\possiblyWithSub\stageImetaColor{v^{\superscriptI} }\).
      \item Case \rulename{T0-Cst0}:
        Since the arity of partial built-in functions~\(\possiblyWithSub\stageOmetaColor{p}\) cannot be zero
        by Assumption~\ref{assump:type-of-constants},
        this contradicts the assumption of the present lemma.
      \item The other cases also contradict the assumption.
    \end{itemize}
  \end{proof}
  \begin{lemma}[Value Inversion for Non-Code Order-\(0\) Types]\label{lem:value-inversion-of-non-code-order-zero}
    If \( \mathit{\Gamma}  \vdash^{0}   \possiblyWithSub\stageOmetaColor{v^{\superscriptO} }   :  \possiblyWithSub\stageOmetaColor{T^{\superscriptO} } \) and \( \vdash^{0}_{\mathrm{dom} }  \possiblyWithSub\stageOmetaColor{T^{\superscriptO} } \),
    then \(\possiblyWithSub\stageOmetaColor{v^{\superscriptO} }\) is of the form~\(\possiblyWithSub\stageOmetaColor{c}\),
    and there exists \(\possiblyWithSub\stageImetaColor{\tau^{\superscriptI} }\) such that \( \vdash^{1}_{\mathrm{oz} }  \possiblyWithSub\stageImetaColor{\tau^{\superscriptI} } \),
    \(\ConstEnvPers(c) = \possiblyWithSub\stageImetaColor{\tau^{\superscriptI} }\), and \( \possiblyWithSub\stageImetaColor{\tau^{\superscriptI} }  \gg  \possiblyWithSub\stageOmetaColor{T^{\superscriptO} } \).
  \end{lemma}
  \begin{proof}
    By induction on the derivation of \( \mathit{\Gamma}  \vdash^{0}   \possiblyWithSub\stageOmetaColor{v^{\superscriptO} }   :  \possiblyWithSub\stageOmetaColor{T^{\superscriptO} } \).
    \begin{itemize}
      \item Case \derive[T0-CstP]{%
         \vdash  \mathit{\Gamma} 
      \andalso
        \ConstEnvPers(c) = \possiblyWithSub\stageImetaColor{\tau^{\superscriptI} }
      }{%
         \mathit{\Gamma}  \vdash^{0}    \possiblyWithSub\stageOmetaColor{c}    :   \mathop{\downarrow}( \possiblyWithSub\stageImetaColor{\tau^{\superscriptI} } )  
      }:
        The desired properties immediately hold.
      \item Case \derive[T0-TyEquiv]{%
         \mathit{\Gamma}  \vdash^{0}   \possiblyWithSub\stageOmetaColor{v^{\superscriptO} }   :  \possiblyWithSub\stageOmetaColor{T'^{\superscriptO} } 
      \andalso
         \possiblyWithSub\stageOmetaColor{T'^{\superscriptO} }  \equiv^{0}  \possiblyWithSub\stageOmetaColor{T^{\superscriptO} } 
      \andalso
         \mathit{\Gamma}  \vdash^{0}  \possiblyWithSub\stageOmetaColor{T^{\superscriptO} } 
      }{%
         \mathit{\Gamma}  \vdash^{0}   \possiblyWithSub\stageOmetaColor{v^{\superscriptO} }   :  \possiblyWithSub\stageOmetaColor{T^{\superscriptO} } 
      }:
        By Lemma~\ref{lem:csr-equiv-preserves-order},
        we have \( \vdash^{0}_{\mathrm{dom} }  \possiblyWithSub\stageOmetaColor{T'^{\superscriptO} } \).
        Then, by IH, \(\possiblyWithSub\stageOmetaColor{v^{\superscriptO} }\) is of the form~\(\possiblyWithSub\stageOmetaColor{c}\),
        and there exists \(\possiblyWithSub\stageImetaColor{\tau^{\superscriptI} }\) such that \( \vdash^{1}_{\mathrm{oz} }  \possiblyWithSub\stageImetaColor{\tau^{\superscriptI} } \),
        \(\ConstEnvPers(\possiblyWithSub\stageOmetaColor{c}) = \possiblyWithSub\stageImetaColor{\tau^{\superscriptI} }\), and \( \possiblyWithSub\stageImetaColor{\tau^{\superscriptI} }  \gg  \possiblyWithSub\stageOmetaColor{T'^{\superscriptO} } \).
        By Lemma~\ref{lem:csr-equiv-preserves-unlifting-relation},
        we also have \( \possiblyWithSub\stageImetaColor{\tau^{\superscriptI} }  \gg  \possiblyWithSub\stageOmetaColor{T^{\superscriptO} } \).
      \item Case \derive[T0-RfnPred]{%
         \mathit{\Gamma}  \vdash^{0}    \openO{\{} \possiblyWithSub\stageOmetaColor{\nu}  \relO{:}  \possiblyWithSub\stageOmetaColor{B}  \relO{\mid}  \possiblyWithSub\stageOmetaColor{N^{\superscriptO} } \closeO{\} }   
      \andalso
        \ConstEnvPers(c) = \possiblyWithSub\stageImetaColor{B}
      \andalso
           [    \possiblyWithSub\stageOmetaColor{c}    /  \possiblyWithSub\stageOmetaColor{\nu}  ]    \possiblyWithSub\stageOmetaColor{N^{\superscriptO} }   \longrightarrow^{0\,\ast}      \ttO{true}     
      }{%
         \mathit{\Gamma}  \vdash^{0}    \possiblyWithSub\stageOmetaColor{c}    :    \openO{\{} \possiblyWithSub\stageOmetaColor{\nu}  \relO{:}  \possiblyWithSub\stageOmetaColor{B}  \relO{\mid}  \possiblyWithSub\stageOmetaColor{N^{\superscriptO} } \closeO{\} }   
      }:
        Again, the desired properties immediately hold.
      \item Cases~\rulename{T0-Var}, \rulename{T0-Ass}, and \rulename{T0-RfnAct}
      clearly contradict the form~\(\possiblyWithSub\stageOmetaColor{v^{\superscriptO} }\).
      \item Case~\rulename{T0-App} also contradicts the form~\(\possiblyWithSub\stageOmetaColor{v^{\superscriptO} }\);
      due to the definition of values,
      \(\possiblyWithSub\stageOmetaColor{v^{\superscriptO} }\) can only be a partial application of a built-in function,
      but this cannot be the case by \( \vdash^{0}_{\mathrm{dom} }  \possiblyWithSub\stageOmetaColor{T^{\superscriptO} } \) and Assumption~\ref{assump:type-of-constants}.
      \item Cases~\rulename{T0-Brkt}, \rulename{T0-Rfn}
      clearly contradict the assumption~\( \vdash^{0}_{\mathrm{dom} }  \possiblyWithSub\stageOmetaColor{T^{\superscriptO} } \).
      \item Case \rulename{T0-Cst0} also contradicts \( \vdash^{0}_{\mathrm{dom} }  \possiblyWithSub\stageOmetaColor{T^{\superscriptO} } \);
      stage-\(0\)-specific built-in functions are of positive arity
      by Assumption~\ref{assump:type-of-constants}.
    \end{itemize}
  \end{proof}
  \begin{lemma}[Progress by \(\delta\)-reduction of persistent built-in functions]\label{lem:progress-of-delta-persistent}
    If \( \mathit{\Gamma}  \vdash^{0}   \openO{(}    \possiblyWithSub\stageOmetaColor{\Hat{c} }   \    \possiblyWithSub\stageOmetaColor{c}_{{\mathrm{1}}}   \ \cdots\    \possiblyWithSub\stageOmetaColor{c}_{\ottmv{m}}    \closeO{)}   :  \possiblyWithSub\stageOmetaColor{T^{\superscriptO} } \) and \(m = \arity{\Hat{c}} \geq 1\),
    then there exists \(c\) such that
    \(\delta(\possiblyWithSub\stageOmetaColor{\Hat{c} }, (\possiblyWithSub\stageOmetaColor{c}_{{\mathrm{1}}}, \ldots, \possiblyWithSub\stageOmetaColor{c}_{\ottmv{m}})) = \possiblyWithSub\stageOmetaColor{c'}\).
  \end{lemma}
  \begin{proof}
    Let \(\ConstEnvPers(\Hat{c}) \revdefeq   \possiblyWithSub\stageImetaColor{\Hat{\tau}^{\superscriptI} }_{{\mathrm{1}}}  \relI{\to}   \cdots \relI{\to}  \possiblyWithSub\stageImetaColor{\Hat{\tau}^{\superscriptI} }_{\ottmv{m}}    \relI{\to}  \possiblyWithSub\stageImetaColor{\Hat{\tau}^{\superscriptI} } \).
    By Lemma~\ref{lem:partial-app-persistent} with \(k \defeq m\), we have
    \(\ConstEnvPers(c_i) = \possiblyWithSub\stageImetaColor{\Hat{\tau}^{\superscriptI} }_i\) for each \(i \in \{1, \ldots, m\}\).
    Then, by Assumption~\ref{assump:type-of-constants},
    there exists \(c\) such that
    \(\delta(\possiblyWithSub\stageOmetaColor{\Hat{c} }, (\possiblyWithSub\stageOmetaColor{c}_{{\mathrm{1}}}, \ldots, \possiblyWithSub\stageOmetaColor{c}_{\ottmv{m}})) = \possiblyWithSub\stageOmetaColor{c}\).
  \end{proof}
  \begin{lemma}[Progress by \(\delta\)-reduction of stage-0 built-in functions]\label{lem:progress-of-delta-zero}
    If \( \mathit{\Gamma}  \vdash^{0}   \openO{(}    \possiblyWithSub\stageOmetaColor{p}   \    \possiblyWithSub\stageOmetaColor{c}_{{\mathrm{1}}}   \ \cdots\    \possiblyWithSub\stageOmetaColor{c}_{\ottmv{m}}    \closeO{)}   :  \possiblyWithSub\stageOmetaColor{T^{\superscriptO} } \) and \(m = \arity{\possiblyWithSub\stageOmetaColor{p}}\),
    then there exists \(\possiblyWithSub\stageOmetaColor{v^{\superscriptO} }\) such that
    \(\delta(\possiblyWithSub\stageOmetaColor{p}, (\possiblyWithSub\stageOmetaColor{c}_{{\mathrm{1}}}, \ldots, \possiblyWithSub\stageOmetaColor{c}_{\ottmv{m}})) = \possiblyWithSub\stageOmetaColor{v^{\superscriptO} }\).
  \end{lemma}
  \begin{proof}
    Let \(\ConstEnvZero(\possiblyWithSub\stageOmetaColor{p}) \revdefeq  \openO{(} \possiblyWithSub\stageOmetaColor{x}_{{\mathrm{1}}}  \relO{:}  \possiblyWithSub\stageOmetaColor{\Hat{T}^{\superscriptO} }_{{\mathrm{1}}} \closeO{)} \relO{\to}   \cdots \relO{\to}   \openO{(} \possiblyWithSub\stageOmetaColor{x}_{\ottmv{m}}  \relO{:}  \possiblyWithSub\stageOmetaColor{\Hat{T}^{\superscriptO} }_{\ottmv{m}} \closeO{)} \relO{\to}  \possiblyWithSub\stageOmetaColor{\Hat{T}^{\superscriptO} }   \).
    By Lemma~\ref{lem:partial-app-zero} with \(k \defeq m\), we have
    \(\possiblyWithSub\stageOmetaColor{c}_{\ottmv{i}} \vDash [ \possiblyWithSub\stageOmetaColor{c}_{i - 1} / \possiblyWithSub\stageOmetaColor{x}_{i - 1} ] \cdots [ \possiblyWithSub\stageOmetaColor{c}_{{\mathrm{1}}} / \possiblyWithSub\stageOmetaColor{x}_{{\mathrm{1}}} ] \possiblyWithSub\stageOmetaColor{\Hat{T}^{\superscriptO} }_{\ottmv{i}}\)
    for each \(i \in \{1, \ldots, m\}\).
    Then, by Assumption~\ref{assump:type-of-constants},
    there exists \(\possiblyWithSub\stageOmetaColor{q}\) such that
    \(\delta(\possiblyWithSub\stageOmetaColor{p}, (\possiblyWithSub\stageOmetaColor{c}_{{\mathrm{1}}}, \ldots, \possiblyWithSub\stageOmetaColor{c}_{\ottmv{m}})) = \possiblyWithSub\stageOmetaColor{q}\).
  \end{proof}
  \recalltheorem[Progress]{thm:progress}{%
    Suppose \( \vdash^{1}  \mathit{\Gamma} \).
    \begin{enumerate}
      \item If \( \mathit{\Gamma}  \vdash^{0}  \possiblyWithSub\stageOmetaColor{N^{\superscriptO} }  :  \possiblyWithSub\stageOmetaColor{T^{\superscriptO} } \), then one of the following holds:
        \begin{itemize}
          \item \( \possiblyWithSub\stageOmetaColor{N^{\superscriptO} }  \longrightarrow^{0}   \BlameSign^{ L }  \) for some \(L\).
          \item there exists \(\possiblyWithSub\stageOmetaColor{N'^{\superscriptO} }\) such that \( \possiblyWithSub\stageOmetaColor{N^{\superscriptO} }  \longrightarrow^{0}   \possiblyWithSub\stageOmetaColor{N'^{\superscriptO} }  \); or
          \item \(\possiblyWithSub\stageOmetaColor{N^{\superscriptO} }\) is a stage-0 value;
        \end{itemize}
      \item If \( \mathit{\Gamma}  \vdash^{1}  \possiblyWithSub\stageImetaColor{N^{\superscriptI} }  :  \possiblyWithSub\stageImetaColor{T^{\superscriptI} } \), then one of the following holds:
        \begin{itemize}
          \item \( \possiblyWithSub\stageImetaColor{N^{\superscriptI} }  \longrightarrow^{1}   \BlameSign^{ L }  \) for some \(L\).
          \item there exists \(\possiblyWithSub\stageImetaColor{N'^{\superscriptI} }\) such that \( \possiblyWithSub\stageImetaColor{N^{\superscriptI} }  \longrightarrow^{1}   \possiblyWithSub\stageImetaColor{N'^{\superscriptI} }  \); or
          \item \(\possiblyWithSub\stageImetaColor{N^{\superscriptI} }\) is a stage-1 value;
        \end{itemize}
      \item If \( \mathit{\Gamma}  \vdash^{1}  \possiblyWithSub\stageImetaColor{T^{\superscriptI} } \), then one of the following holds:
        \begin{itemize}
          \item \( \possiblyWithSub\stageImetaColor{T^{\superscriptI} }  \longrightarrow^{1}   \BlameSign^{ L }  \) for some \(L\).
          \item there exists \(\possiblyWithSub\stageImetaColor{T'^{\superscriptI} }\) such that \( \possiblyWithSub\stageImetaColor{T^{\superscriptI} }  \longrightarrow^{1}   \possiblyWithSub\stageImetaColor{T'^{\superscriptI} }  \); or
          \item \(\possiblyWithSub\stageImetaColor{T^{\superscriptI} }\) is a stage-1 value type;
        \end{itemize}
    \end{enumerate}
  }
  \begin{proof}
    By induction on the derivation.
    \begin{enumerate}
      \item
        \begin{itemize}
          \item Case \rulename{T0-Var} contradicts the assumption \( \vdash^{1}  \mathit{\Gamma} \).
          \item Case \rulename{T0-Abs} is immediate.
          \item Case \derive[T0-App]{%
             \mathit{\Gamma}  \vdash^{0}  \possiblyWithSub\stageOmetaColor{N^{\superscriptO} }_{{\mathrm{1}}}  :   \openO{(} \possiblyWithSub\stageOmetaColor{x}  \relO{:}  \possiblyWithSub\stageOmetaColor{T^{\superscriptO} }_{{\mathrm{11}}} \closeO{)} \relO{\to}  \possiblyWithSub\stageOmetaColor{T^{\superscriptO} }_{{\mathrm{12}}}  
          \andalso
             \mathit{\Gamma}  \vdash^{0}  \possiblyWithSub\stageOmetaColor{N^{\superscriptO} }_{{\mathrm{2}}}  :  \possiblyWithSub\stageOmetaColor{T^{\superscriptO} }_{{\mathrm{11}}} 
          }{%
             \mathit{\Gamma}  \vdash^{0}   \possiblyWithSub\stageOmetaColor{N^{\superscriptO} }_{{\mathrm{1}}} \  \possiblyWithSub\stageOmetaColor{N^{\superscriptO} }_{{\mathrm{2}}}   :    [  \possiblyWithSub\stageOmetaColor{N^{\superscriptO} }_{{\mathrm{2}}}  /  \possiblyWithSub\stageOmetaColor{x}  ]    \possiblyWithSub\stageOmetaColor{T^{\superscriptO} }_{{\mathrm{12}}}  
          }:
            By IH on \( \mathit{\Gamma}  \vdash^{0}  \possiblyWithSub\stageOmetaColor{N^{\superscriptO} }_{{\mathrm{1}}}  :   \openO{(} \possiblyWithSub\stageOmetaColor{x}  \relO{:}  \possiblyWithSub\stageOmetaColor{T^{\superscriptO} }_{{\mathrm{11}}} \closeO{)} \relO{\to}  \possiblyWithSub\stageOmetaColor{T^{\superscriptO} }_{{\mathrm{12}}}  \),
            we have one of the following cases:
            \begin{itemize}
              \item Case \( \possiblyWithSub\stageOmetaColor{N^{\superscriptO} }_{{\mathrm{1}}}  \longrightarrow^{0}   \BlameSign^{ L }  \):
                We have
                  \derive[E0-App1F]{%
                     \possiblyWithSub\stageOmetaColor{N^{\superscriptO} }_{{\mathrm{1}}}  \longrightarrow^{0}   \BlameSign^{ L }  
                  }{%
                      \possiblyWithSub\stageOmetaColor{N^{\superscriptO} }_{{\mathrm{1}}} \  \possiblyWithSub\stageOmetaColor{N^{\superscriptO} }_{{\mathrm{2}}}   \longrightarrow^{0}   \BlameSign^{ L }  
                  }.
              \item Case where \( \possiblyWithSub\stageOmetaColor{N^{\superscriptO} }_{{\mathrm{1}}}  \longrightarrow^{0}   \possiblyWithSub\stageOmetaColor{N'^{\superscriptO} }_{{\mathrm{1}}}  \) for some \(\possiblyWithSub\stageOmetaColor{N'^{\superscriptO} }_{{\mathrm{1}}}\):
                Similarly to the previous case, we have
                  \derive[E0-App1]{%
                     \possiblyWithSub\stageOmetaColor{N^{\superscriptO} }_{{\mathrm{1}}}  \longrightarrow^{0}   \possiblyWithSub\stageOmetaColor{N'^{\superscriptO} }_{{\mathrm{1}}}  
                  }{%
                      \possiblyWithSub\stageOmetaColor{N^{\superscriptO} }_{{\mathrm{1}}} \  \possiblyWithSub\stageOmetaColor{N^{\superscriptO} }_{{\mathrm{2}}}   \longrightarrow^{0}    \possiblyWithSub\stageOmetaColor{N'^{\superscriptO} }_{{\mathrm{1}}} \  \possiblyWithSub\stageOmetaColor{N^{\superscriptO} }_{{\mathrm{2}}}   
                  }.
              \item Case where \(\possiblyWithSub\stageOmetaColor{N^{\superscriptO} }_{{\mathrm{1}}} \revdefeq \possiblyWithSub\stageOmetaColor{v^{\superscriptO} }_{{\mathrm{1}}}\) is a stage-0 value:
                By IH on \( \mathit{\Gamma}  \vdash^{0}  \possiblyWithSub\stageOmetaColor{N^{\superscriptO} }_{{\mathrm{2}}}  :  \possiblyWithSub\stageOmetaColor{T^{\superscriptO} }_{{\mathrm{11}}} \), we have one of the following:
                \begin{itemize}
                  \item Case \( \possiblyWithSub\stageOmetaColor{N^{\superscriptO} }_{{\mathrm{2}}}  \longrightarrow^{0}   \BlameSign^{ L }  \):
                    We have
                      \derive[E0-App2Fail]{%
                         \possiblyWithSub\stageOmetaColor{N^{\superscriptO} }_{{\mathrm{2}}}  \longrightarrow^{0}   \BlameSign^{ L }  
                      }{%
                           \possiblyWithSub\stageOmetaColor{v^{\superscriptO} }_{{\mathrm{1}}}  \  \possiblyWithSub\stageOmetaColor{N^{\superscriptO} }_{{\mathrm{2}}}   \longrightarrow^{0}   \BlameSign^{ L }  
                      }.
                  \item Case where \( \possiblyWithSub\stageOmetaColor{N^{\superscriptO} }_{{\mathrm{2}}}  \longrightarrow^{0}   \possiblyWithSub\stageOmetaColor{N'^{\superscriptO} }_{{\mathrm{2}}}  \) for some \(\possiblyWithSub\stageOmetaColor{N'^{\superscriptO} }_{{\mathrm{2}}}\):
                    Similarly to the previous case, we have
                      \derive[E0-App2]{%
                         \possiblyWithSub\stageOmetaColor{N^{\superscriptO} }_{{\mathrm{2}}}  \longrightarrow^{0}   \possiblyWithSub\stageOmetaColor{N'^{\superscriptO} }_{{\mathrm{2}}}  
                      }{%
                           \possiblyWithSub\stageOmetaColor{v^{\superscriptO} }_{{\mathrm{1}}}  \  \possiblyWithSub\stageOmetaColor{N^{\superscriptO} }_{{\mathrm{2}}}   \longrightarrow^{0}     \possiblyWithSub\stageOmetaColor{v^{\superscriptO} }_{{\mathrm{1}}}  \  \possiblyWithSub\stageOmetaColor{N'^{\superscriptO} }_{{\mathrm{2}}}   
                      }.
                  \item Case where \(\possiblyWithSub\stageOmetaColor{N^{\superscriptO} }_{{\mathrm{2}}} \revdefeq \possiblyWithSub\stageOmetaColor{v^{\superscriptO} }_{{\mathrm{2}}}\) is a stage-0 value:
                    By \( \mathit{\Gamma}  \vdash^{0}   \possiblyWithSub\stageOmetaColor{v^{\superscriptO} }_{{\mathrm{1}}}   :   \openO{(} \possiblyWithSub\stageOmetaColor{x}  \relO{:}  \possiblyWithSub\stageOmetaColor{T^{\superscriptO} }_{{\mathrm{11}}} \closeO{)} \relO{\to}  \possiblyWithSub\stageOmetaColor{T^{\superscriptO} }_{{\mathrm{12}}}  \)
                    and Lemma~\ref{lem:value-inversion-of-arrow},
                    we have one of the following:
                    \begin{itemize}
                      \item Case where \(\possiblyWithSub\stageOmetaColor{v^{\superscriptO} }_{{\mathrm{1}}}\) is of the form~\( \openO{(}  \ordO{\lambda} \possiblyWithSub\stageOmetaColor{x}  \relO{:}  \possiblyWithSub\stageOmetaColor{T'^{\superscriptO} }_{{\mathrm{11}}} \punctO{.}\  \possiblyWithSub\stageOmetaColor{N^{\superscriptO} }_{{\mathrm{12}}}  \closeO{)} \):
                        We can derive
                        \begin{center}
                          \derive[E0-Beta]{}{%
                               \openO{(}  \ordO{\lambda} \possiblyWithSub\stageOmetaColor{x}  \relO{:}  \possiblyWithSub\stageOmetaColor{T'^{\superscriptO} }_{{\mathrm{11}}} \punctO{.}\  \possiblyWithSub\stageOmetaColor{N^{\superscriptO} }_{{\mathrm{12}}}  \closeO{)}  \   \possiblyWithSub\stageOmetaColor{v^{\superscriptO} }_{{\mathrm{2}}}    \longrightarrow^{0}     [   \possiblyWithSub\stageOmetaColor{v^{\superscriptO} }_{{\mathrm{2}}}   /  \possiblyWithSub\stageOmetaColor{x}  ]    \possiblyWithSub\stageOmetaColor{N^{\superscriptO} }_{{\mathrm{12}}}   
                          }.
                        \end{center}
                      \item Case where \(\possiblyWithSub\stageOmetaColor{v^{\superscriptO} }_{{\mathrm{1}}}\) is of the form~\( \LeftAssertParen \relO{\CastArrow}   \openO{\{} \possiblyWithSub\stageOmetaColor{\nu}  \relO{:}  \possiblyWithSub\stageOmetaColor{B}  \relO{\mid}  \possiblyWithSub\stageOmetaColor{N'^{\superscriptO} }_{{\mathrm{1}}} \closeO{\} }   \RightAssertParen^{ L } \),
                      and \(  \possiblyWithSub\stageImetaColor{B}   \gg  \possiblyWithSub\stageOmetaColor{T^{\superscriptO} }_{{\mathrm{11}}} \) holds:
                        By \(  \possiblyWithSub\stageImetaColor{B}   \gg  \possiblyWithSub\stageOmetaColor{T^{\superscriptO} }_{{\mathrm{11}}} \), we clearly have \( \vdash^{0}_{\mathrm{dom} }  \possiblyWithSub\stageOmetaColor{T^{\superscriptO} }_{{\mathrm{11}}} \),
                        and thereby by \( \mathit{\Gamma}  \vdash^{0}   \possiblyWithSub\stageOmetaColor{v^{\superscriptO} }_{{\mathrm{2}}}   :  \possiblyWithSub\stageOmetaColor{T^{\superscriptO} }_{{\mathrm{11}}} \)
                        and Lemma~\ref{lem:value-inversion-of-non-code-order-zero},
                        \(\possiblyWithSub\stageOmetaColor{v^{\superscriptO} }_{{\mathrm{2}}}\) is of the form~\(\possiblyWithSub\stageOmetaColor{c'}\).
                        This enables us to derive
                        \begin{center}
                          \derive[E0-RfnStart]{}{%
                               \LeftAssertParen \relO{\CastArrow}   \openO{\{} \possiblyWithSub\stageOmetaColor{\nu}  \relO{:}  \possiblyWithSub\stageOmetaColor{B}  \relO{\mid}  \possiblyWithSub\stageOmetaColor{N'^{\superscriptO} }_{{\mathrm{1}}} \closeO{\} }   \RightAssertParen^{ L }  \    \possiblyWithSub\stageOmetaColor{c'}     \longrightarrow^{0}    \LeftAssertParen   \openO{\{} \possiblyWithSub\stageOmetaColor{\nu}  \relO{:}  \possiblyWithSub\stageOmetaColor{B}  \relO{\mid}  \possiblyWithSub\stageOmetaColor{N'^{\superscriptO} }_{{\mathrm{1}}} \closeO{\} }  \punctO{,}    [    \possiblyWithSub\stageOmetaColor{c'}    /  \possiblyWithSub\stageOmetaColor{\nu}  ]    \possiblyWithSub\stageOmetaColor{N'^{\superscriptO} }_{{\mathrm{1}}}  \punctO{,}  \possiblyWithSub\stageOmetaColor{c'}  \RightAssertParen^{ L }   
                          }.
                        \end{center}
                      \item Case where \(\possiblyWithSub\stageOmetaColor{v^{\superscriptO} }_{{\mathrm{1}}}\) is of the form~\(\possiblyWithSub\stageOmetaColor{a}_{{\mathrm{1}}}\):
                        Since \( \mathit{\Gamma}  \vdash^{0}   \possiblyWithSub\stageOmetaColor{a}_{{\mathrm{1}}}   :   \openO{(} \possiblyWithSub\stageOmetaColor{x}  \relO{:}  \possiblyWithSub\stageOmetaColor{T^{\superscriptO} }_{{\mathrm{11}}} \closeO{)} \relO{\to}  \possiblyWithSub\stageOmetaColor{T^{\superscriptO} }_{{\mathrm{12}}}  \) holds,
                        by Lemma~\ref{lem:non-code-order-zero-domain},
                        we have \( \vdash^{0}_{\mathrm{dom} }  \possiblyWithSub\stageOmetaColor{T^{\superscriptO} }_{{\mathrm{11}}} \).
                        Then, by Lemma~\ref{lem:value-inversion-of-non-code-order-zero},
                        \(\possiblyWithSub\stageOmetaColor{v^{\superscriptO} }_{{\mathrm{2}}}\) is of the form~\(\possiblyWithSub\stageOmetaColor{c'}\).
                        Here, we can do the following case analysis about the form of \(\possiblyWithSub\stageOmetaColor{a}_{{\mathrm{1}}}\):
                        \begin{enumerate}
                          \item Case where \(\possiblyWithSub\stageOmetaColor{a}_{{\mathrm{1}}}\) is of the form~\( \openO{(}    \possiblyWithSub\stageOmetaColor{\Hat{c} }   \    \possiblyWithSub\stageOmetaColor{c}_{{\mathrm{1}}}   \ \cdots\    \possiblyWithSub\stageOmetaColor{c}_{\ottmv{k}}    \closeO{)} \)
                          with some \(k < \arity{\Hat{c}}\):
                            We further have the following two cases:
                            (i)~Case~\(k + 1 < \arity{\Hat{c}}\):
                              \(\possiblyWithSub\stageOmetaColor{N^{\superscriptO} } =  \openO{(}   \possiblyWithSub\stageOmetaColor{a}_{{\mathrm{1}}}  \    \possiblyWithSub\stageOmetaColor{c'}    \closeO{)} \) is a partial application
                              and thereby is already a value.
                            (ii)~Case~\(k + 1 = \arity{\Hat{c}}\):
                              By Lemma~\ref{lem:progress-of-delta-persistent},
                              there exists \(c''\) such that
                              \(\delta(\possiblyWithSub\stageOmetaColor{\Hat{c} }, (\possiblyWithSub\stageOmetaColor{c}_{{\mathrm{1}}}, \ldots, \possiblyWithSub\stageOmetaColor{c}_{\ottmv{k}}, \possiblyWithSub\stageOmetaColor{c'})) = \possiblyWithSub\stageOmetaColor{c''}\)
                              This enables us to derive:
                              \begin{center}
                                \derive[E0-Delta]{%
                                  \delta(  \possiblyWithSub\stageOmetaColor{a}_{{\mathrm{1}}}  \    \possiblyWithSub\stageOmetaColor{c'}   ) = \possiblyWithSub\stageOmetaColor{c''}
                                }{%
                                     \possiblyWithSub\stageOmetaColor{a}_{{\mathrm{1}}}  \    \possiblyWithSub\stageOmetaColor{c'}     \longrightarrow^{0}     \possiblyWithSub\stageOmetaColor{c''}    
                                }.
                              \end{center}
                          \item Case where \(\possiblyWithSub\stageOmetaColor{a}_{{\mathrm{1}}}\) is of the form~\( \openO{(}    \possiblyWithSub\stageOmetaColor{p}   \    \possiblyWithSub\stageOmetaColor{c'}_{{\mathrm{1}}}   \ \cdots\    \possiblyWithSub\stageOmetaColor{c'}_{\ottmv{k}}    \closeO{)} \)
                            with some \(k < \arity{\possiblyWithSub\stageOmetaColor{p}}\):
                            We further have the following two cases:
                            (i)~Case~\(k + 1 < \arity{\possiblyWithSub\stageOmetaColor{p}}\):
                              The application \( \openO{(}   \possiblyWithSub\stageOmetaColor{a}_{{\mathrm{1}}}  \    \possiblyWithSub\stageOmetaColor{c'}    \closeO{)} \) is already a value.
                            (ii)~Case~\(k + 1 = \arity{\possiblyWithSub\stageOmetaColor{p}}\):
                              By Lemma~\ref{lem:progress-of-delta-zero},
                              there exists \(\possiblyWithSub\stageOmetaColor{q}\) such that
                              \(\delta(\possiblyWithSub\stageOmetaColor{p}, (\possiblyWithSub\stageOmetaColor{c}_{{\mathrm{1}}}, \ldots, \possiblyWithSub\stageOmetaColor{c}_{\ottmv{k}}, \possiblyWithSub\stageOmetaColor{c'})) = \possiblyWithSub\stageOmetaColor{q}\).
                              This enables us to derive:
                              \begin{center}
                                \derive[E0-Delta]{%
                                  \delta(  \possiblyWithSub\stageOmetaColor{a}_{{\mathrm{1}}}  \    \possiblyWithSub\stageOmetaColor{c'}   ) = \possiblyWithSub\stageOmetaColor{v^{\superscriptO} }
                                }{%
                                     \possiblyWithSub\stageOmetaColor{a}_{{\mathrm{1}}}  \    \possiblyWithSub\stageOmetaColor{c'}     \longrightarrow^{0}    \possiblyWithSub\stageOmetaColor{v^{\superscriptO} }   
                                }.
                              \end{center}
                        \end{enumerate}
                    \end{itemize}
                \end{itemize}
            \end{itemize}
          \item \derive[T0-Ass]{%
             \mathit{\Gamma}  \vdash^{1}  \possiblyWithSub\stageImetaColor{T^{\superscriptI} }_{{\mathrm{1}}} 
          \andalso
             \mathit{\Gamma}  \vdash^{1}  \possiblyWithSub\stageImetaColor{T^{\superscriptI} }_{{\mathrm{2}}} 
          \andalso
             \possiblyWithSub\stageImetaColor{T^{\superscriptI} }_{{\mathrm{1}}}  \mathrel{||}^{1}  \possiblyWithSub\stageImetaColor{T^{\superscriptI} }_{{\mathrm{2}}} 
          \andalso
            \possiblyWithSub\stageOmetaColor{x} \not\in \dom(\mathit{\Gamma})
          }{%
             \mathit{\Gamma}  \vdash^{0}   \LeftAssertParen\openO{\langle} \possiblyWithSub\stageImetaColor{T^{\superscriptI} }_{{\mathrm{1}}} \closeO{\rangle} \relO{\CastArrow} \openO{\langle} \possiblyWithSub\stageImetaColor{T^{\superscriptI} }_{{\mathrm{2}}} \closeO{\rangle}\RightAssertParen^{ L' }   :   \openO{(} \possiblyWithSub\stageOmetaColor{x}  \relO{:}   \openO{\langle} \possiblyWithSub\stageImetaColor{T^{\superscriptI} }_{{\mathrm{1}}} \closeO{\rangle}  \closeO{)} \relO{\to}   \openO{\langle} \possiblyWithSub\stageImetaColor{T^{\superscriptI} }_{{\mathrm{2}}} \closeO{\rangle}   
          }:
            By IH, from \( \mathit{\Gamma}  \vdash^{1}  \possiblyWithSub\stageImetaColor{T^{\superscriptI} }_{{\mathrm{1}}} \), we have one of the following cases:
            \begin{itemize}
              \item Case \( \possiblyWithSub\stageImetaColor{T^{\superscriptI} }_{{\mathrm{1}}}  \longrightarrow^{1}   \BlameSign^{ L }  \):
                We have
                \begin{center}
                  \derive[E0-Ass1F]{%
                     \possiblyWithSub\stageImetaColor{T^{\superscriptI} }_{{\mathrm{1}}}  \longrightarrow^{1}   \BlameSign^{ L }  
                  }{%
                      \LeftAssertParen\openO{\langle} \possiblyWithSub\stageImetaColor{T^{\superscriptI} }_{{\mathrm{1}}} \closeO{\rangle} \relO{\CastArrow} \openO{\langle} \possiblyWithSub\stageImetaColor{T^{\superscriptI} }_{{\mathrm{2}}} \closeO{\rangle}\RightAssertParen^{ L' }   \longrightarrow^{0}   \BlameSign^{ L }  
                  }.
                \end{center}
              \item Case where \( \possiblyWithSub\stageImetaColor{T^{\superscriptI} }_{{\mathrm{1}}}  \longrightarrow^{1}   \possiblyWithSub\stageImetaColor{T'^{\superscriptI} }_{{\mathrm{1}}}  \) for some \(\possiblyWithSub\stageImetaColor{T'^{\superscriptI} }_{{\mathrm{1}}}\):
                We have
                \begin{center}
                  \derive[E0-Ass1]{%
                     \possiblyWithSub\stageImetaColor{T^{\superscriptI} }_{{\mathrm{1}}}  \longrightarrow^{1}   \possiblyWithSub\stageImetaColor{T'^{\superscriptI} }_{{\mathrm{1}}}  
                  }{%
                      \LeftAssertParen\openO{\langle} \possiblyWithSub\stageImetaColor{T^{\superscriptI} }_{{\mathrm{1}}} \closeO{\rangle} \relO{\CastArrow} \openO{\langle} \possiblyWithSub\stageImetaColor{T^{\superscriptI} }_{{\mathrm{2}}} \closeO{\rangle}\RightAssertParen^{ L }   \longrightarrow^{0}    \LeftAssertParen\openO{\langle} \possiblyWithSub\stageImetaColor{T'^{\superscriptI} }_{{\mathrm{1}}} \closeO{\rangle} \relO{\CastArrow} \openO{\langle} \possiblyWithSub\stageImetaColor{T^{\superscriptI} }_{{\mathrm{2}}} \closeO{\rangle}\RightAssertParen^{ L }   
                  }.
                \end{center}
              \item Case where \(\possiblyWithSub\stageImetaColor{T^{\superscriptI} }_{{\mathrm{1}}} = \possiblyWithSub\stageImetaColor{\tau^{\superscriptI} }_{{\mathrm{1}}}\) is a stage-1 value type:
                By IH, from \( \mathit{\Gamma}  \vdash^{1}  \possiblyWithSub\stageImetaColor{T^{\superscriptI} }_{{\mathrm{2}}} \), we have one of the following cases:
                \begin{itemize}
                  \item Case \( \possiblyWithSub\stageImetaColor{T^{\superscriptI} }_{{\mathrm{1}}}  \longrightarrow^{1}   \BlameSign^{ L }  \):
                    We have
                    \begin{center}
                      \derive[E0-Ass2F]{%
                         \possiblyWithSub\stageImetaColor{T^{\superscriptI} }_{{\mathrm{2}}}  \longrightarrow^{1}   \BlameSign^{ L }  
                      }{%
                          \LeftAssertParen\openO{\langle}  \possiblyWithSub\stageImetaColor{\tau^{\superscriptI} }_{{\mathrm{1}}}  \closeO{\rangle} \relO{\CastArrow} \openO{\langle} \possiblyWithSub\stageImetaColor{T^{\superscriptI} }_{{\mathrm{2}}} \closeO{\rangle}\RightAssertParen^{ L' }   \longrightarrow^{0}   \BlameSign^{ L }  
                      }
                    \end{center}
                  \item Case where \( \possiblyWithSub\stageImetaColor{T^{\superscriptI} }_{{\mathrm{2}}}  \longrightarrow^{1}   \possiblyWithSub\stageImetaColor{T'^{\superscriptI} }_{{\mathrm{2}}}  \) for some \(\possiblyWithSub\stageImetaColor{T'^{\superscriptI} }_{{\mathrm{2}}}\):
                    We have
                    \begin{center}
                      \derive[E0-Ass2]{%
                         \possiblyWithSub\stageImetaColor{T^{\superscriptI} }_{{\mathrm{2}}}  \longrightarrow^{1}   \possiblyWithSub\stageImetaColor{T'^{\superscriptI} }_{{\mathrm{2}}}  
                      }{%
                          \LeftAssertParen\openO{\langle}  \possiblyWithSub\stageImetaColor{\tau^{\superscriptI} }_{{\mathrm{1}}}  \closeO{\rangle} \relO{\CastArrow} \openO{\langle} \possiblyWithSub\stageImetaColor{T^{\superscriptI} }_{{\mathrm{2}}} \closeO{\rangle}\RightAssertParen^{ L }   \longrightarrow^{0}    \LeftAssertParen\openO{\langle}  \possiblyWithSub\stageImetaColor{\tau^{\superscriptI} }_{{\mathrm{1}}}  \closeO{\rangle} \relO{\CastArrow} \openO{\langle} \possiblyWithSub\stageImetaColor{T'^{\superscriptI} }_{{\mathrm{2}}} \closeO{\rangle}\RightAssertParen^{ L }   
                      }.
                    \end{center}
                  \item Case where \(\possiblyWithSub\stageImetaColor{T^{\superscriptI} }_{{\mathrm{2}}} = \possiblyWithSub\stageImetaColor{\tau^{\superscriptI} }_{{\mathrm{2}}}\) is a stage-1 value type:
                    \begin{itemize}
                      \item If \(\possiblyWithSub\stageImetaColor{\tau^{\superscriptI} }_{{\mathrm{2}}} = \possiblyWithSub\stageImetaColor{\tau^{\superscriptI} }_{{\mathrm{1}}}\):
                        we have
                        \begin{center}
                          \derive[E0-AssPass]{}{%
                              \LeftAssertParen\openO{\langle}  \possiblyWithSub\stageImetaColor{\tau^{\superscriptI} }_{{\mathrm{1}}}  \closeO{\rangle} \relO{\CastArrow} \openO{\langle}  \possiblyWithSub\stageImetaColor{\tau^{\superscriptI} }_{{\mathrm{1}}}  \closeO{\rangle}\RightAssertParen^{ L }   \longrightarrow^{0}    \ordO{\lambda} \possiblyWithSub\stageOmetaColor{x}  \relO{:}   \openO{\langle}  \possiblyWithSub\stageImetaColor{\tau^{\superscriptI} }_{{\mathrm{1}}}  \closeO{\rangle}  \punctO{.}\   \possiblyWithSub\stageOmetaColor{x}    
                          }.
                        \end{center}
                      \item If \(\possiblyWithSub\stageImetaColor{\tau^{\superscriptI} }_{{\mathrm{2}}} \neq \possiblyWithSub\stageImetaColor{\tau^{\superscriptI} }_{{\mathrm{1}}}\):
                        we have
                        \begin{center}
                          \derive[E0-AssFail]{%
                            \possiblyWithSub\stageImetaColor{\tau^{\superscriptI} }_{{\mathrm{1}}} \neq \possiblyWithSub\stageImetaColor{\tau^{\superscriptI} }_{{\mathrm{2}}}
                          }{%
                              \LeftAssertParen\openO{\langle}  \possiblyWithSub\stageImetaColor{\tau^{\superscriptI} }_{{\mathrm{1}}}  \closeO{\rangle} \relO{\CastArrow} \openO{\langle}  \possiblyWithSub\stageImetaColor{\tau^{\superscriptI} }_{{\mathrm{2}}}  \closeO{\rangle}\RightAssertParen^{ L }   \longrightarrow^{0}   \BlameSign^{ L }  
                          }.
                        \end{center}
                    \end{itemize}
                \end{itemize}
            \end{itemize}
          \item Case \derive[T0-RfnAct]{%
             \mathit{\Gamma}  \vdash^{0}    \openO{\{} \possiblyWithSub\stageOmetaColor{\nu}  \relO{:}  \possiblyWithSub\stageOmetaColor{B}  \relO{\mid}  \possiblyWithSub\stageOmetaColor{N^{\superscriptO} }_{{\mathrm{1}}} \closeO{\} }   
          \andalso
             \mathit{\Gamma}  \vdash^{0}  \possiblyWithSub\stageOmetaColor{N^{\superscriptO} }_{{\mathrm{2}}}  :    \openO{\{} \possiblyWithSub\stageOmetaColor{\nu}_{{\mathrm{0}}}  \relO{:}   \ttO{Bool}   \relO{\mid}     \ttO{true}    \closeO{\} }   
          \\
            \ConstEnvPers(c) = \possiblyWithSub\stageImetaColor{B}
          \andalso
               [    \possiblyWithSub\stageOmetaColor{c}    /  \possiblyWithSub\stageOmetaColor{\nu}  ]    \possiblyWithSub\stageOmetaColor{N^{\superscriptO} }_{{\mathrm{1}}}   \longrightarrow^{0\,\ast}   \possiblyWithSub\stageOmetaColor{N^{\superscriptO} }_{{\mathrm{2}}}  
          }{%
             \mathit{\Gamma}  \vdash^{0}   \LeftAssertParen   \openO{\{} \possiblyWithSub\stageOmetaColor{\nu}  \relO{:}  \possiblyWithSub\stageOmetaColor{B}  \relO{\mid}  \possiblyWithSub\stageOmetaColor{N^{\superscriptO} }_{{\mathrm{1}}} \closeO{\} }  \punctO{,}  \possiblyWithSub\stageOmetaColor{N^{\superscriptO} }_{{\mathrm{2}}} \punctO{,}  \possiblyWithSub\stageOmetaColor{c}  \RightAssertParen^{ L }   :    \openO{\{} \possiblyWithSub\stageOmetaColor{\nu}  \relO{:}  \possiblyWithSub\stageOmetaColor{B}  \relO{\mid}  \possiblyWithSub\stageOmetaColor{N^{\superscriptO} }_{{\mathrm{1}}} \closeO{\} }   
          }:
            By IH on \( \mathit{\Gamma}  \vdash^{0}  \possiblyWithSub\stageOmetaColor{N^{\superscriptO} }_{{\mathrm{2}}}  :    \openO{\{} \possiblyWithSub\stageOmetaColor{\nu}_{{\mathrm{0}}}  \relO{:}   \ttO{Bool}   \relO{\mid}     \ttO{true}    \closeO{\} }   \),
            we have one of the following:
            \begin{itemize}
              \item Case \( \possiblyWithSub\stageOmetaColor{N^{\superscriptO} }_{{\mathrm{2}}}  \longrightarrow^{0}   \BlameSign^{ L }  \):
                We can immediately derive
                \begin{center}
                  \derive[E0-RfnActF]{%
                     \possiblyWithSub\stageOmetaColor{N^{\superscriptO} }_{{\mathrm{2}}}  \longrightarrow^{0}   \BlameSign^{ L' }  
                  }{%
                      \LeftAssertParen   \openO{\{} \possiblyWithSub\stageOmetaColor{\nu}  \relO{:}  \possiblyWithSub\stageOmetaColor{B}  \relO{\mid}  \possiblyWithSub\stageOmetaColor{N^{\superscriptO} }_{{\mathrm{1}}} \closeO{\} }  \punctO{,}  \possiblyWithSub\stageOmetaColor{N^{\superscriptO} }_{{\mathrm{2}}} \punctO{,}  \possiblyWithSub\stageOmetaColor{c}  \RightAssertParen^{ L }   \longrightarrow^{0}   \BlameSign^{ L' }  
                  }.
                \end{center}
              \item Case \( \possiblyWithSub\stageOmetaColor{N^{\superscriptO} }_{{\mathrm{2}}}  \longrightarrow^{0}   \possiblyWithSub\stageOmetaColor{N'^{\superscriptO} }_{{\mathrm{2}}}  \) for some \(\possiblyWithSub\stageOmetaColor{N'^{\superscriptO} }_{{\mathrm{2}}}\):
                This is also immediate:
                \begin{center}
                  \derive[E0-RfnAct]{%
                     \possiblyWithSub\stageOmetaColor{N^{\superscriptO} }_{{\mathrm{1}}}  \longrightarrow^{0}   \possiblyWithSub\stageOmetaColor{N'^{\superscriptO} }_{{\mathrm{1}}}  
                  }{%
                      \LeftAssertParen   \openO{\{} \possiblyWithSub\stageOmetaColor{\nu}  \relO{:}  \possiblyWithSub\stageOmetaColor{B}  \relO{\mid}  \possiblyWithSub\stageOmetaColor{N^{\superscriptO} }_{{\mathrm{1}}} \closeO{\} }  \punctO{,}  \possiblyWithSub\stageOmetaColor{N^{\superscriptO} }_{{\mathrm{2}}} \punctO{,}  \possiblyWithSub\stageOmetaColor{c}  \RightAssertParen^{ L }   \longrightarrow^{0}    \LeftAssertParen   \openO{\{} \possiblyWithSub\stageOmetaColor{\nu}  \relO{:}  \possiblyWithSub\stageOmetaColor{B}  \relO{\mid}  \possiblyWithSub\stageOmetaColor{N^{\superscriptO} }_{{\mathrm{1}}} \closeO{\} }  \punctO{,}  \possiblyWithSub\stageOmetaColor{N'^{\superscriptO} }_{{\mathrm{2}}} \punctO{,}  \possiblyWithSub\stageOmetaColor{c}  \RightAssertParen^{ L }   
                  }.
                \end{center}
              \item Case \(\possiblyWithSub\stageOmetaColor{N^{\superscriptO} }_{{\mathrm{2}}} = \possiblyWithSub\stageOmetaColor{v^{\superscriptO} }_{{\mathrm{2}}}\) is a stage-\(0\) value:
                By Lemma~\ref{lem:value-inversion-of-non-code-order-zero},
                \(\possiblyWithSub\stageOmetaColor{v^{\superscriptO} }_{{\mathrm{2}}}\) is of the form~\(\possiblyWithSub\stageOmetaColor{c}_{{\mathrm{2}}}\), and
                we clearly have \(\ConstEnvPers(c_{{\mathrm{2}}}) =   \ttI{Bool}  \).
                By Assumption~\ref{assump:type-of-constants},
                \(\possiblyWithSub\stageOmetaColor{c}_{{\mathrm{2}}}\) is \(  \ttO{true}  \) or \(  \ttO{false}  \).
                \begin{itemize}
                  \item Case \(\possiblyWithSub\stageOmetaColor{c}_{{\mathrm{2}}} =   \ttO{true}  \):
                    We can immediately derive
                    \begin{center}
                      \derive[E0-RfnPass]{}{%
                          \LeftAssertParen   \openO{\{} \possiblyWithSub\stageOmetaColor{\nu}  \relO{:}  \possiblyWithSub\stageOmetaColor{B}  \relO{\mid}  \possiblyWithSub\stageOmetaColor{N^{\superscriptO} }_{{\mathrm{1}}} \closeO{\} }  \punctO{,}     \ttO{true}    \punctO{,}  \possiblyWithSub\stageOmetaColor{c}  \RightAssertParen^{ L }   \longrightarrow^{0}     \possiblyWithSub\stageOmetaColor{c}    
                      }.
                    \end{center}
                  \item Case \(\possiblyWithSub\stageOmetaColor{c}_{{\mathrm{2}}} =   \ttO{false}  \):
                    We can immediately derive the following as well:
                    \begin{center}
                      \derive[E0-RfnFail]{}{%
                          \LeftAssertParen   \openO{\{} \possiblyWithSub\stageOmetaColor{\nu}  \relO{:}  \possiblyWithSub\stageOmetaColor{B}  \relO{\mid}  \possiblyWithSub\stageOmetaColor{N^{\superscriptO} }_{{\mathrm{1}}} \closeO{\} }  \punctO{,}     \ttO{false}    \punctO{,}  \possiblyWithSub\stageOmetaColor{c}  \RightAssertParen^{ L }   \longrightarrow^{0}   \BlameSign^{ L }  
                      }.
                    \end{center}
                \end{itemize}
            \end{itemize}
          \item Cases~\rulename{T0-Brkt} and \rulename{T0-TyEquiv} are straightforward by IH.
          \item Cases~\rulename{T0-CstP}, \rulename{T0-Cst0},
          \rulename{T0-Rfn}, \rulename{T0-RfnPred} are immediate
          since \(\possiblyWithSub\stageOmetaColor{N^{\superscriptO} }\) is already a value.
        \end{itemize}
      \item
        \begin{itemize}
          \item Case \rulename{T1-Var} is immediate since \(\possiblyWithSub\stageImetaColor{x}\) is a stage-1 value.
          \item Case \derive[T1-Esc]{%
             \mathit{\Gamma}  \vdash^{0}  \possiblyWithSub\stageOmetaColor{N^{\superscriptO} }  :   \openO{\langle} \possiblyWithSub\stageImetaColor{T^{\superscriptI} } \closeO{\rangle}  
          }{%
             \mathit{\Gamma}  \vdash^{1}   \ordI{\sim} \possiblyWithSub\stageOmetaColor{N^{\superscriptO} }   :  \possiblyWithSub\stageImetaColor{T^{\superscriptI} } 
          }:
            By IH, from \( \mathit{\Gamma}  \vdash^{0}  \possiblyWithSub\stageOmetaColor{N^{\superscriptO} }  :   \openO{\langle} \possiblyWithSub\stageImetaColor{T^{\superscriptI} } \closeO{\rangle}  \), we have one of the following cases:
            \begin{itemize}
              \item Case \( \possiblyWithSub\stageOmetaColor{N^{\superscriptO} }  \longrightarrow^{0}   \BlameSign^{ L }  \):
                We have
                  \derive[E1-EscF]{%
                     \possiblyWithSub\stageOmetaColor{N^{\superscriptO} }  \longrightarrow^{0}   \BlameSign^{ L }  
                  }{%
                      \ordI{\sim} \possiblyWithSub\stageOmetaColor{N^{\superscriptO} }   \longrightarrow^{1}   \BlameSign^{ L }  
                  }.
              \item Case where \( \possiblyWithSub\stageOmetaColor{N^{\superscriptO} }  \longrightarrow^{0}   \possiblyWithSub\stageOmetaColor{N'^{\superscriptO} }  \) for some \(\possiblyWithSub\stageOmetaColor{N'^{\superscriptO} }\):
                We have
                  \derive[E1-Esc]{%
                     \possiblyWithSub\stageOmetaColor{N^{\superscriptO} }  \longrightarrow^{0}   \possiblyWithSub\stageOmetaColor{N'^{\superscriptO} }  
                  }{%
                      \ordI{\sim} \possiblyWithSub\stageOmetaColor{N^{\superscriptO} }   \longrightarrow^{1}    \ordI{\sim} \possiblyWithSub\stageOmetaColor{N'^{\superscriptO} }   
                  }
              \item Case where \(\possiblyWithSub\stageOmetaColor{N^{\superscriptO} } = \possiblyWithSub\stageOmetaColor{v^{\superscriptO} }\) is a stage-0 value:
                By Lemma~\ref{lem:value-inversion-of-code},
                from \( \mathit{\Gamma}  \vdash^{0}   \possiblyWithSub\stageOmetaColor{v^{\superscriptO} }   :   \openO{\langle} \possiblyWithSub\stageImetaColor{T^{\superscriptI} } \closeO{\rangle}  \),
                the value \(\possiblyWithSub\stageOmetaColor{v^{\superscriptO} }\) is of the form~\( \openO{\langle} \possiblyWithSub\stageImetaColor{v^{\superscriptI} } \closeO{\rangle} \).
                This enables us to derive
                  \derive[E1-Cancel]{}{%
                      \ordI{\sim}  \openO{\langle} \possiblyWithSub\stageImetaColor{N^{\superscriptI} } \closeO{\rangle}    \longrightarrow^{1}   \possiblyWithSub\stageImetaColor{N^{\superscriptI} }  
                  }.
            \end{itemize}
          \item The other cases are straightforward by IH.
        \end{itemize}
      \item
        \begin{itemize}
          \item Case \derive[WfT1-Tensor]{%
             \mathit{\Gamma}  \vdash^{0}  \possiblyWithSub\stageOmetaColor{N^{\superscriptO} }  :    \openO{\{} \possiblyWithSub\stageOmetaColor{\nu}  \relO{:}   \ttO{NatList}   \relO{\mid}     \ttO{true}    \closeO{\} }   
          }{%
             \mathit{\Gamma}  \vdash^{1}   \ttI{Tensor}\ \ordI{\%} \possiblyWithSub\stageOmetaColor{N^{\superscriptO} }  
          }:
            By IH, from \( \mathit{\Gamma}  \vdash^{0}  \possiblyWithSub\stageOmetaColor{N^{\superscriptO} }  :    \openO{\{} \possiblyWithSub\stageOmetaColor{\nu}  \relO{:}   \ttO{NatList}   \relO{\mid}     \ttO{true}    \closeO{\} }   \),
            we have one of the following cases:
            \begin{itemize}
              \item Case \( \possiblyWithSub\stageOmetaColor{N^{\superscriptO} }  \longrightarrow^{0}   \BlameSign^{ L }  \):
                We have
                  \derive[ET1-TensorF]{%
                     \possiblyWithSub\stageOmetaColor{N^{\superscriptO} }  \longrightarrow^{0}   \BlameSign^{ L }  
                  }{%
                      \ttI{Tensor}\ \ordI{\%} \possiblyWithSub\stageOmetaColor{N^{\superscriptO} }   \longrightarrow^{1}   \BlameSign^{ L }  
                  }.
              \item Case where \( \possiblyWithSub\stageOmetaColor{N^{\superscriptO} }  \longrightarrow^{0}   \possiblyWithSub\stageOmetaColor{N'^{\superscriptO} }  \) for some \(\possiblyWithSub\stageOmetaColor{N'^{\superscriptO} }\):
                Similarly to the previous case, we have
                  \derive[ET1-Tensor]{%
                     \possiblyWithSub\stageOmetaColor{N^{\superscriptO} }  \longrightarrow^{0}   \possiblyWithSub\stageOmetaColor{N'^{\superscriptO} }  
                  }{%
                      \ttI{Tensor}\ \ordI{\%} \possiblyWithSub\stageOmetaColor{N^{\superscriptO} }   \longrightarrow^{1}    \ttI{Tensor}\ \ordI{\%} \possiblyWithSub\stageOmetaColor{N'^{\superscriptO} }   
                  }.
              \item Case where \(\possiblyWithSub\stageOmetaColor{N^{\superscriptO} } = \possiblyWithSub\stageOmetaColor{v^{\superscriptO} }\) is a stage-0 value:
                By Lemma~\ref{lem:value-inversion-of-non-code-order-zero},
                \(\possiblyWithSub\stageOmetaColor{v^{\superscriptO} }\) is of the form~\(\possiblyWithSub\stageOmetaColor{c}\), and
                we clearly have \(\ConstEnvPers(c) =   \ttI{NatList}  \).
                By Assumption~\ref{assump:type-of-constants},
                \(\possiblyWithSub\stageOmetaColor{c}\) is of the form~\(\possiblyWithSub\stageOmetaColor{s}\).
                Therefore, \(\possiblyWithSub\stageImetaColor{T^{\superscriptI} }\) is a stage-1 value type~\( \ttI{Tensor}\ \ordI{\%}    \possiblyWithSub\stageOmetaColor{s}    \).
            \end{itemize}
          \item The other cases are all straightforward.
        \end{itemize}
    \end{enumerate}
  \end{proof}

\subsection{Safety of Generated Code}
\begin{figure}[tbp]
  \begin{flushleft}
    \fbox{\( \possiblyWithSub\stageOmetaColor{T^{\superscriptO} }_{{\mathrm{1}}}  \cong^{0}  \possiblyWithSub\stageOmetaColor{T^{\superscriptO} }_{{\mathrm{2}}} \)}
  \end{flushleft}
  \vspace{-2.5em}
  \begin{center}
  \hspace{5em}%
    \derive[BqT0-Refl]{}{%
       \possiblyWithSub\stageOmetaColor{T^{\superscriptO} }  \cong^{0}  \possiblyWithSub\stageOmetaColor{T^{\superscriptO} } 
    }
  \qquad
    \derive[BqT0-Sym]{%
       \possiblyWithSub\stageOmetaColor{T^{\superscriptO} }_{{\mathrm{1}}}  \cong^{0}  \possiblyWithSub\stageOmetaColor{T^{\superscriptO} }_{{\mathrm{2}}} 
    }{%
       \possiblyWithSub\stageOmetaColor{T^{\superscriptO} }_{{\mathrm{2}}}  \cong^{0}  \possiblyWithSub\stageOmetaColor{T^{\superscriptO} }_{{\mathrm{1}}} 
    }
  \qquad
    \derive[BqT0-Code]{%
       \possiblyWithSub\stageImetaColor{T^{\superscriptI} }_{{\mathrm{1}}}  \cong^{1}  \possiblyWithSub\stageImetaColor{T^{\superscriptI} }_{{\mathrm{2}}} 
    }{%
        \openO{\langle} \possiblyWithSub\stageImetaColor{T^{\superscriptI} }_{{\mathrm{1}}} \closeO{\rangle}   \cong^{0}   \openO{\langle} \possiblyWithSub\stageImetaColor{T^{\superscriptI} }_{{\mathrm{2}}} \closeO{\rangle}  
    }
  \\[0.7em]
    \derive[BqT0-Trans]{%
       \possiblyWithSub\stageOmetaColor{T^{\superscriptO} }_{{\mathrm{1}}}  \cong^{0}  \possiblyWithSub\stageOmetaColor{T^{\superscriptO} }_{{\mathrm{2}}} 
    \andalso
       \possiblyWithSub\stageOmetaColor{T^{\superscriptO} }_{{\mathrm{2}}}  \cong^{0}  \possiblyWithSub\stageOmetaColor{T^{\superscriptO} }_{{\mathrm{3}}} 
    }{%
       \possiblyWithSub\stageOmetaColor{T^{\superscriptO} }_{{\mathrm{1}}}  \cong^{0}  \possiblyWithSub\stageOmetaColor{T^{\superscriptO} }_{{\mathrm{3}}} 
    }
  \qquad
    \derive[BqT0-Arr]{%
       \possiblyWithSub\stageOmetaColor{T^{\superscriptO} }_{{\mathrm{11}}}  \cong^{0}  \possiblyWithSub\stageOmetaColor{T^{\superscriptO} }_{{\mathrm{21}}} 
    \andalso
       \possiblyWithSub\stageOmetaColor{T^{\superscriptO} }_{{\mathrm{12}}}  \cong^{0}  \possiblyWithSub\stageOmetaColor{T^{\superscriptO} }_{{\mathrm{22}}} 
    }{%
        \openO{(} \possiblyWithSub\stageOmetaColor{x}  \relO{:}  \possiblyWithSub\stageOmetaColor{T^{\superscriptO} }_{{\mathrm{11}}} \closeO{)} \relO{\to}  \possiblyWithSub\stageOmetaColor{T^{\superscriptO} }_{{\mathrm{12}}}   \cong^{0}   \openO{(} \possiblyWithSub\stageOmetaColor{x}  \relO{:}  \possiblyWithSub\stageOmetaColor{T^{\superscriptO} }_{{\mathrm{21}}} \closeO{)} \relO{\to}  \possiblyWithSub\stageOmetaColor{T^{\superscriptO} }_{{\mathrm{22}}}  
    }
  \\[0.7em]
    \derive[BqT0-Rfn]{%
       \possiblyWithSub\stageOmetaColor{N^{\superscriptO} }_{{\mathrm{1}}}  \cong^{0}  \possiblyWithSub\stageOmetaColor{N^{\superscriptO} }_{{\mathrm{2}}} 
    }{%
         \openO{\{} \possiblyWithSub\stageOmetaColor{\nu}  \relO{:}  \possiblyWithSub\stageOmetaColor{B}  \relO{\mid}  \possiblyWithSub\stageOmetaColor{N^{\superscriptO} }_{{\mathrm{1}}} \closeO{\} }    \cong^{0}    \openO{\{} \possiblyWithSub\stageOmetaColor{\nu}  \relO{:}  \possiblyWithSub\stageOmetaColor{B}  \relO{\mid}  \possiblyWithSub\stageOmetaColor{N^{\superscriptO} }_{{\mathrm{2}}} \closeO{\} }   
    }
  \end{center}
  \vspace{1em}
  \begin{flushleft}
    \fbox{\( \possiblyWithSub\stageImetaColor{T^{\superscriptI} }_{{\mathrm{1}}}  \cong^{1}  \possiblyWithSub\stageImetaColor{T^{\superscriptI} }_{{\mathrm{2}}} \)}
  \end{flushleft}
  \vspace{-2.5em}
  \begin{center}
    \derive[BqT1-Refl]{}{%
       \possiblyWithSub\stageImetaColor{T^{\superscriptI} }  \cong^{1}  \possiblyWithSub\stageImetaColor{T^{\superscriptI} } 
    }
  \qquad
    \derive[BqT1-Sym]{%
       \possiblyWithSub\stageImetaColor{T^{\superscriptI} }_{{\mathrm{1}}}  \cong^{1}  \possiblyWithSub\stageImetaColor{T^{\superscriptI} }_{{\mathrm{2}}} 
    }{%
       \possiblyWithSub\stageImetaColor{T^{\superscriptI} }_{{\mathrm{2}}}  \cong^{1}  \possiblyWithSub\stageImetaColor{T^{\superscriptI} }_{{\mathrm{1}}} 
    }
  \\[0.7em]
    \derive[BqT1-Trans]{%
       \possiblyWithSub\stageImetaColor{T^{\superscriptI} }_{{\mathrm{1}}}  \cong^{1}  \possiblyWithSub\stageImetaColor{T^{\superscriptI} }_{{\mathrm{2}}} 
    \andalso
       \possiblyWithSub\stageImetaColor{T^{\superscriptI} }_{{\mathrm{2}}}  \cong^{1}  \possiblyWithSub\stageImetaColor{T^{\superscriptI} }_{{\mathrm{3}}} 
    }{%
       \possiblyWithSub\stageImetaColor{T^{\superscriptI} }_{{\mathrm{1}}}  \cong^{1}  \possiblyWithSub\stageImetaColor{T^{\superscriptI} }_{{\mathrm{3}}} 
    }
  \qquad
    \derive[BqT1-Arr]{%
       \possiblyWithSub\stageImetaColor{T^{\superscriptI} }_{{\mathrm{11}}}  \cong^{1}  \possiblyWithSub\stageImetaColor{T^{\superscriptI} }_{{\mathrm{21}}} 
    \andalso
       \possiblyWithSub\stageImetaColor{T^{\superscriptI} }_{{\mathrm{12}}}  \cong^{1}  \possiblyWithSub\stageImetaColor{T^{\superscriptI} }_{{\mathrm{22}}} 
    }{%
        \possiblyWithSub\stageImetaColor{T^{\superscriptI} }_{{\mathrm{11}}}  \relI{\to}  \possiblyWithSub\stageImetaColor{T^{\superscriptI} }_{{\mathrm{12}}}   \cong^{1}   \possiblyWithSub\stageImetaColor{T^{\superscriptI} }_{{\mathrm{21}}}  \relI{\to}  \possiblyWithSub\stageImetaColor{T^{\superscriptI} }_{{\mathrm{22}}}  
    }
  \\[0.7em]
    \derive[BqT1-Tensor]{%
       \possiblyWithSub\stageOmetaColor{N^{\superscriptO} }_{{\mathrm{1}}}  \cong^{0}  \possiblyWithSub\stageOmetaColor{N^{\superscriptO} }_{{\mathrm{2}}} 
    }{%
        \ttI{Tensor}\ \ordI{\%} \possiblyWithSub\stageOmetaColor{N^{\superscriptO} }_{{\mathrm{1}}}   \cong^{1}   \ttI{Tensor}\ \ordI{\%} \possiblyWithSub\stageOmetaColor{N^{\superscriptO} }_{{\mathrm{2}}}  
    }
  \end{center}
  \caption{\(\beta\)-equivalence relations on types}
  \label{fig:beta-type-equivalence}
\end{figure}
\begin{figure}[tbp]
  \begin{flushleft}
    \fbox{\( \possiblyWithSub\stageOmetaColor{N^{\superscriptO} }_{{\mathrm{1}}}  \cong^{0}  \possiblyWithSub\stageOmetaColor{N^{\superscriptO} }_{{\mathrm{2}}} \)}
  \end{flushleft}
  \vspace{-3em}
  \begin{center}
  \hspace{6em}%
    \derive[Bq0-Brkt]{%
       \possiblyWithSub\stageImetaColor{N^{\superscriptI} }_{{\mathrm{1}}}  \cong^{1}  \possiblyWithSub\stageImetaColor{N^{\superscriptI} }_{{\mathrm{2}}} 
    }{%
        \openO{\langle} \possiblyWithSub\stageImetaColor{N^{\superscriptI} }_{{\mathrm{1}}} \closeO{\rangle}   \cong^{0}   \openO{\langle} \possiblyWithSub\stageImetaColor{N^{\superscriptI} }_{{\mathrm{2}}} \closeO{\rangle}  
    }
  \qquad
    \derive[Bq0-Ass]{%
       \possiblyWithSub\stageImetaColor{T^{\superscriptI} }_{{\mathrm{11}}}  \cong^{1}  \possiblyWithSub\stageImetaColor{T^{\superscriptI} }_{{\mathrm{21}}} 
    \andalso
       \possiblyWithSub\stageImetaColor{T^{\superscriptI} }_{{\mathrm{12}}}  \cong^{1}  \possiblyWithSub\stageImetaColor{T^{\superscriptI} }_{{\mathrm{22}}} 
    }{%
        \LeftAssertParen\openO{\langle} \possiblyWithSub\stageImetaColor{T^{\superscriptI} }_{{\mathrm{11}}} \closeO{\rangle} \relO{\CastArrow} \openO{\langle} \possiblyWithSub\stageImetaColor{T^{\superscriptI} }_{{\mathrm{12}}} \closeO{\rangle}\RightAssertParen^{ L }   \cong^{0}   \LeftAssertParen\openO{\langle} \possiblyWithSub\stageImetaColor{T^{\superscriptI} }_{{\mathrm{21}}} \closeO{\rangle} \relO{\CastArrow} \openO{\langle} \possiblyWithSub\stageImetaColor{T^{\superscriptI} }_{{\mathrm{22}}} \closeO{\rangle}\RightAssertParen^{ L }  
    }
  \\[0.7em]
    \derive[Bq0-AssPass]{}{%
        \LeftAssertParen\openO{\langle}  \possiblyWithSub\stageImetaColor{\tau^{\superscriptI} }  \closeO{\rangle} \relO{\CastArrow} \openO{\langle}  \possiblyWithSub\stageImetaColor{\tau^{\superscriptI} }  \closeO{\rangle}\RightAssertParen^{ L }   \cong^{0}   \ordO{\lambda} \possiblyWithSub\stageOmetaColor{x}  \relO{:}   \openO{\langle}  \possiblyWithSub\stageImetaColor{\tau^{\superscriptI} }  \closeO{\rangle}  \punctO{.}\   \possiblyWithSub\stageOmetaColor{x}   
    }
  \qquad
    \derive[Bq0-Abs]{%
       \possiblyWithSub\stageOmetaColor{T^{\superscriptO} }_{{\mathrm{1}}}  \cong^{0}  \possiblyWithSub\stageOmetaColor{T^{\superscriptO} }_{{\mathrm{2}}} 
    \andalso
       \possiblyWithSub\stageOmetaColor{N^{\superscriptO} }_{{\mathrm{1}}}  \cong^{0}  \possiblyWithSub\stageOmetaColor{N^{\superscriptO} }_{{\mathrm{2}}} 
    }{%
        \ordO{\lambda} \possiblyWithSub\stageOmetaColor{x}  \relO{:}  \possiblyWithSub\stageOmetaColor{T^{\superscriptO} }_{{\mathrm{1}}} \punctO{.}\  \possiblyWithSub\stageOmetaColor{N^{\superscriptO} }_{{\mathrm{1}}}   \cong^{0}   \ordO{\lambda} \possiblyWithSub\stageOmetaColor{x}  \relO{:}  \possiblyWithSub\stageOmetaColor{T^{\superscriptO} }_{{\mathrm{2}}} \punctO{.}\  \possiblyWithSub\stageOmetaColor{N^{\superscriptO} }_{{\mathrm{2}}}  
    }
  \\[0.7em]
    \derive[Bq0-Refl]{}{%
       \possiblyWithSub\stageOmetaColor{N^{\superscriptO} }  \cong^{0}  \possiblyWithSub\stageOmetaColor{N^{\superscriptO} } 
    }
  \qquad
    \derive[Bq0-Trans]{%
       \possiblyWithSub\stageOmetaColor{N^{\superscriptO} }_{{\mathrm{1}}}  \cong^{0}  \possiblyWithSub\stageOmetaColor{N^{\superscriptO} }_{{\mathrm{2}}} 
    \andalso
       \possiblyWithSub\stageOmetaColor{N^{\superscriptO} }_{{\mathrm{2}}}  \cong^{0}  \possiblyWithSub\stageOmetaColor{N^{\superscriptO} }_{{\mathrm{3}}} 
    }{%
       \possiblyWithSub\stageOmetaColor{N^{\superscriptO} }_{{\mathrm{1}}}  \cong^{0}  \possiblyWithSub\stageOmetaColor{N^{\superscriptO} }_{{\mathrm{3}}} 
    }
  \qquad
    \derive[Bq0-Sym]{%
       \possiblyWithSub\stageOmetaColor{N^{\superscriptO} }_{{\mathrm{1}}}  \cong^{0}  \possiblyWithSub\stageOmetaColor{N^{\superscriptO} }_{{\mathrm{2}}} 
    }{%
       \possiblyWithSub\stageOmetaColor{N^{\superscriptO} }_{{\mathrm{2}}}  \cong^{0}  \possiblyWithSub\stageOmetaColor{N^{\superscriptO} }_{{\mathrm{1}}} 
    }
  \\[0.7em]
    \derive[Bq0-App]{%
       \possiblyWithSub\stageOmetaColor{N^{\superscriptO} }_{{\mathrm{11}}}  \cong^{0}  \possiblyWithSub\stageOmetaColor{N^{\superscriptO} }_{{\mathrm{21}}} 
    \andalso
       \possiblyWithSub\stageOmetaColor{N^{\superscriptO} }_{{\mathrm{12}}}  \cong^{0}  \possiblyWithSub\stageOmetaColor{N^{\superscriptO} }_{{\mathrm{22}}} 
    }{%
        \possiblyWithSub\stageOmetaColor{N^{\superscriptO} }_{{\mathrm{11}}} \  \possiblyWithSub\stageOmetaColor{N^{\superscriptO} }_{{\mathrm{12}}}   \cong^{0}   \possiblyWithSub\stageOmetaColor{N^{\superscriptO} }_{{\mathrm{21}}} \  \possiblyWithSub\stageOmetaColor{N^{\superscriptO} }_{{\mathrm{22}}}  
    }
  \qquad
    \derive[Bq0-Beta]{}{%
         \openO{(}  \ordO{\lambda} \possiblyWithSub\stageOmetaColor{x}  \relO{:}  \possiblyWithSub\stageOmetaColor{T^{\superscriptO} } \punctO{.}\  \possiblyWithSub\stageOmetaColor{N^{\superscriptO} }_{{\mathrm{1}}}  \closeO{)}  \  \possiblyWithSub\stageOmetaColor{N^{\superscriptO} }_{{\mathrm{2}}}   \cong^{0}    [  \possiblyWithSub\stageOmetaColor{N^{\superscriptO} }_{{\mathrm{2}}}  /  \possiblyWithSub\stageOmetaColor{x}  ]    \possiblyWithSub\stageOmetaColor{N^{\superscriptO} }_{{\mathrm{1}}}  
    }
  \\[0.7em]
    \derive[Bq0-Delta]{%
      \delta(  \possiblyWithSub\stageOmetaColor{a}_{{\mathrm{1}}}  \    \possiblyWithSub\stageOmetaColor{c}_{{\mathrm{2}}}   ) = \possiblyWithSub\stageOmetaColor{q}
    }{%
         \possiblyWithSub\stageOmetaColor{a}_{{\mathrm{1}}}  \    \possiblyWithSub\stageOmetaColor{c}_{{\mathrm{2}}}     \cong^{0}   \possiblyWithSub\stageOmetaColor{q}  
    }
  \qquad
    \derive[Bq0-Rfn]{%
       \possiblyWithSub\stageOmetaColor{N^{\superscriptO} }_{{\mathrm{1}}}  \cong^{0}  \possiblyWithSub\stageOmetaColor{N^{\superscriptO} }_{{\mathrm{2}}} 
    }{%
        \LeftAssertParen \relO{\CastArrow}   \openO{\{} \possiblyWithSub\stageOmetaColor{\nu}  \relO{:}  \possiblyWithSub\stageOmetaColor{B}  \relO{\mid}  \possiblyWithSub\stageOmetaColor{N^{\superscriptO} }_{{\mathrm{1}}} \closeO{\} }   \RightAssertParen^{ L }   \cong^{0}   \LeftAssertParen \relO{\CastArrow}   \openO{\{} \possiblyWithSub\stageOmetaColor{\nu}  \relO{:}  \possiblyWithSub\stageOmetaColor{B}  \relO{\mid}  \possiblyWithSub\stageOmetaColor{N^{\superscriptO} }_{{\mathrm{2}}} \closeO{\} }   \RightAssertParen^{ L }  
    }
  \\[0.7em]
    \derive[Bq0-RfnStart]{}{%
         \LeftAssertParen \relO{\CastArrow}   \openO{\{} \possiblyWithSub\stageOmetaColor{\nu}  \relO{:}  \possiblyWithSub\stageOmetaColor{B}  \relO{\mid}  \possiblyWithSub\stageOmetaColor{N^{\superscriptO} } \closeO{\} }   \RightAssertParen^{ L }  \    \possiblyWithSub\stageOmetaColor{c}     \cong^{0}   \LeftAssertParen   \openO{\{} \possiblyWithSub\stageOmetaColor{\nu}  \relO{:}  \possiblyWithSub\stageOmetaColor{B}  \relO{\mid}  \possiblyWithSub\stageOmetaColor{N^{\superscriptO} } \closeO{\} }  \punctO{,}    [    \possiblyWithSub\stageOmetaColor{c}    /  \possiblyWithSub\stageOmetaColor{\nu}  ]    \possiblyWithSub\stageOmetaColor{N^{\superscriptO} }  \punctO{,}  \possiblyWithSub\stageOmetaColor{c}  \RightAssertParen^{ L }  
    }
  \\[0.7em]
    \derive[Bq0-RfnAct]{%
       \possiblyWithSub\stageOmetaColor{N^{\superscriptO} }_{{\mathrm{11}}}  \cong^{0}  \possiblyWithSub\stageOmetaColor{N^{\superscriptO} }_{{\mathrm{21}}} 
    \andalso
       \possiblyWithSub\stageOmetaColor{N^{\superscriptO} }_{{\mathrm{12}}}  \cong^{0}  \possiblyWithSub\stageOmetaColor{N^{\superscriptO} }_{{\mathrm{22}}} 
    }{%
        \LeftAssertParen   \openO{\{} \possiblyWithSub\stageOmetaColor{\nu}  \relO{:}  \possiblyWithSub\stageOmetaColor{B}  \relO{\mid}  \possiblyWithSub\stageOmetaColor{N^{\superscriptO} }_{{\mathrm{11}}} \closeO{\} }  \punctO{,}  \possiblyWithSub\stageOmetaColor{N^{\superscriptO} }_{{\mathrm{12}}} \punctO{,}  \possiblyWithSub\stageOmetaColor{c}  \RightAssertParen^{ L }   \cong^{0}   \LeftAssertParen   \openO{\{} \possiblyWithSub\stageOmetaColor{\nu}  \relO{:}  \possiblyWithSub\stageOmetaColor{B}  \relO{\mid}  \possiblyWithSub\stageOmetaColor{N^{\superscriptO} }_{{\mathrm{21}}} \closeO{\} }  \punctO{,}  \possiblyWithSub\stageOmetaColor{N^{\superscriptO} }_{{\mathrm{22}}} \punctO{,}  \possiblyWithSub\stageOmetaColor{c}  \RightAssertParen^{ L }  
    }
  \\[0.7em]
    \derive[Bq0-RfnPass]{}{%
        \LeftAssertParen   \openO{\{} \possiblyWithSub\stageOmetaColor{\nu}  \relO{:}  \possiblyWithSub\stageOmetaColor{B}  \relO{\mid}  \possiblyWithSub\stageOmetaColor{N^{\superscriptO} } \closeO{\} }  \punctO{,}     \ttO{true}    \punctO{,}  \possiblyWithSub\stageOmetaColor{c}  \RightAssertParen^{ L }   \cong^{0}    \possiblyWithSub\stageOmetaColor{c}   
    }
  \end{center}
  \vspace{0.5em}
  \begin{flushleft}
    \fbox{\( \possiblyWithSub\stageImetaColor{N^{\superscriptI} }_{{\mathrm{1}}}  \cong^{1}  \possiblyWithSub\stageImetaColor{N^{\superscriptI} }_{{\mathrm{2}}} \)}
  \end{flushleft}
  \vspace{-2.5em}
  \begin{center}
  \hspace{2em}%
    \derive[Bq1-Esc]{%
       \possiblyWithSub\stageOmetaColor{N^{\superscriptO} }_{{\mathrm{1}}}  \cong^{0}  \possiblyWithSub\stageOmetaColor{N^{\superscriptO} }_{{\mathrm{2}}} 
    }{%
        \ordI{\sim} \possiblyWithSub\stageOmetaColor{N^{\superscriptO} }_{{\mathrm{1}}}   \cong^{1}   \ordI{\sim} \possiblyWithSub\stageOmetaColor{N^{\superscriptO} }_{{\mathrm{2}}}  
    }
  \qquad
    \derive[Bq1-Cancel]{}{%
        \ordI{\sim}  \openO{\langle} \possiblyWithSub\stageImetaColor{N^{\superscriptI} } \closeO{\rangle}    \cong^{1}  \possiblyWithSub\stageImetaColor{N^{\superscriptI} } 
    }
  \\[0.7em]
    \derive[Bq1-Refl]{}{%
       \possiblyWithSub\stageImetaColor{N^{\superscriptI} }  \cong^{1}  \possiblyWithSub\stageImetaColor{N^{\superscriptI} } 
    }
  \quad
    \derive[Bq1-Sym]{%
       \possiblyWithSub\stageImetaColor{N^{\superscriptI} }_{{\mathrm{1}}}  \cong^{1}  \possiblyWithSub\stageImetaColor{N^{\superscriptI} }_{{\mathrm{2}}} 
    }{%
       \possiblyWithSub\stageImetaColor{N^{\superscriptI} }_{{\mathrm{2}}}  \cong^{1}  \possiblyWithSub\stageImetaColor{N^{\superscriptI} }_{{\mathrm{1}}} 
    }
  \quad
    \derive[Bq1-Trans]{%
       \possiblyWithSub\stageImetaColor{N^{\superscriptI} }_{{\mathrm{1}}}  \cong^{1}  \possiblyWithSub\stageImetaColor{N^{\superscriptI} }_{{\mathrm{2}}} 
    \andalso
       \possiblyWithSub\stageImetaColor{N^{\superscriptI} }_{{\mathrm{2}}}  \cong^{1}  \possiblyWithSub\stageImetaColor{N^{\superscriptI} }_{{\mathrm{3}}} 
    }{%
       \possiblyWithSub\stageImetaColor{N^{\superscriptI} }_{{\mathrm{1}}}  \cong^{1}  \possiblyWithSub\stageImetaColor{N^{\superscriptI} }_{{\mathrm{3}}} 
    }
  \\[0.7em]
    \derive[Bq1-App]{%
       \possiblyWithSub\stageImetaColor{N^{\superscriptI} }_{{\mathrm{11}}}  \cong^{1}  \possiblyWithSub\stageImetaColor{N^{\superscriptI} }_{{\mathrm{21}}} 
    \andalso
       \possiblyWithSub\stageImetaColor{N^{\superscriptI} }_{{\mathrm{12}}}  \cong^{1}  \possiblyWithSub\stageImetaColor{N^{\superscriptI} }_{{\mathrm{22}}} 
    }{%
        \possiblyWithSub\stageImetaColor{N^{\superscriptI} }_{{\mathrm{11}}} \  \possiblyWithSub\stageImetaColor{N^{\superscriptI} }_{{\mathrm{12}}}   \cong^{1}   \possiblyWithSub\stageImetaColor{N^{\superscriptI} }_{{\mathrm{21}}} \  \possiblyWithSub\stageImetaColor{N^{\superscriptI} }_{{\mathrm{22}}}  
    }
  \qquad
    \derive[Bq1-Abs]{%
       \possiblyWithSub\stageImetaColor{T^{\superscriptI} }_{{\mathrm{11}}}  \cong^{1}  \possiblyWithSub\stageImetaColor{T^{\superscriptI} }_{{\mathrm{21}}} 
    \andalso
       \possiblyWithSub\stageImetaColor{N^{\superscriptI} }_{{\mathrm{12}}}  \cong^{1}  \possiblyWithSub\stageImetaColor{N^{\superscriptI} }_{{\mathrm{22}}} 
    }{%
        \ordI{\lambda} \possiblyWithSub\stageImetaColor{x}  \relI{:}  \possiblyWithSub\stageImetaColor{T^{\superscriptI} }_{{\mathrm{11}}} \punctI{.}\  \possiblyWithSub\stageImetaColor{N^{\superscriptI} }_{{\mathrm{12}}}   \cong^{1}   \ordI{\lambda} \possiblyWithSub\stageImetaColor{x}  \relI{:}  \possiblyWithSub\stageImetaColor{T^{\superscriptI} }_{{\mathrm{21}}} \punctI{.}\  \possiblyWithSub\stageImetaColor{N^{\superscriptI} }_{{\mathrm{22}}}  
    }
  \end{center}
  \caption{\(\beta\)-equivalence relations on terms}
  \label{fig:beta-term-equivalence}
\end{figure}
\begin{figure}[tbp]
\small
  \begin{flushleft}
    \fbox{\( \possiblyWithSub\stageOmetaColor{N^{\superscriptO} }  \Longrightarrow^{0}  \possiblyWithSub\stageOmetaColor{N'^{\superscriptO} } \)}
  \end{flushleft}
  \vspace{-2.75em}
  \begin{center}
    \hspace{5em}%
    \derive[P0-Cst0]{}{%
         \possiblyWithSub\stageOmetaColor{p}    \Longrightarrow^{0}    \possiblyWithSub\stageOmetaColor{p}   
    }
  \qquad
    \derive[P0-CstP]{}{%
         \possiblyWithSub\stageOmetaColor{c}    \Longrightarrow^{0}    \possiblyWithSub\stageOmetaColor{c}   
    }
  \qquad
    \derive[P0-Var]{}{%
        \possiblyWithSub\stageOmetaColor{x}   \Longrightarrow^{0}   \possiblyWithSub\stageOmetaColor{x}  
    }
  \\[0.7em]
    \derive[P0-Abs]{%
       \possiblyWithSub\stageOmetaColor{T^{\superscriptO} }  \Longrightarrow^{0}  \possiblyWithSub\stageOmetaColor{T'^{\superscriptO} } 
    \andalso
       \possiblyWithSub\stageOmetaColor{N^{\superscriptO} }  \Longrightarrow^{0}  \possiblyWithSub\stageOmetaColor{N'^{\superscriptO} } 
    }{%
        \ordO{\lambda} \possiblyWithSub\stageOmetaColor{x}  \relO{:}  \possiblyWithSub\stageOmetaColor{T^{\superscriptO} } \punctO{.}\  \possiblyWithSub\stageOmetaColor{N^{\superscriptO} }   \Longrightarrow^{0}   \ordO{\lambda} \possiblyWithSub\stageOmetaColor{x}  \relO{:}  \possiblyWithSub\stageOmetaColor{T'^{\superscriptO} } \punctO{.}\  \possiblyWithSub\stageOmetaColor{N'^{\superscriptO} }  
    }
  \qquad
    \derive[P0-App]{%
       \possiblyWithSub\stageOmetaColor{N^{\superscriptO} }_{{\mathrm{1}}}  \Longrightarrow^{0}  \possiblyWithSub\stageOmetaColor{N'^{\superscriptO} }_{{\mathrm{1}}} 
    \andalso
       \possiblyWithSub\stageOmetaColor{N^{\superscriptO} }_{{\mathrm{2}}}  \Longrightarrow^{0}  \possiblyWithSub\stageOmetaColor{N'^{\superscriptO} }_{{\mathrm{2}}} 
    }{%
        \possiblyWithSub\stageOmetaColor{N^{\superscriptO} }_{{\mathrm{1}}} \  \possiblyWithSub\stageOmetaColor{N^{\superscriptO} }_{{\mathrm{2}}}   \Longrightarrow^{0}   \possiblyWithSub\stageOmetaColor{N'^{\superscriptO} }_{{\mathrm{1}}} \  \possiblyWithSub\stageOmetaColor{N'^{\superscriptO} }_{{\mathrm{2}}}  
    }
  \\[0.7em]
    \derive[P0-Beta]{%
       \possiblyWithSub\stageOmetaColor{N^{\superscriptO} }_{{\mathrm{1}}}  \Longrightarrow^{0}  \possiblyWithSub\stageOmetaColor{N'^{\superscriptO} }_{{\mathrm{1}}} 
    \andalso
       \possiblyWithSub\stageOmetaColor{N^{\superscriptO} }_{{\mathrm{2}}}  \Longrightarrow^{0}  \possiblyWithSub\stageOmetaColor{N'^{\superscriptO} }_{{\mathrm{2}}} 
    }{%
         \openO{(}  \ordO{\lambda} \possiblyWithSub\stageOmetaColor{x}  \relO{:}  \possiblyWithSub\stageOmetaColor{T^{\superscriptO} } \punctO{.}\  \possiblyWithSub\stageOmetaColor{N^{\superscriptO} }_{{\mathrm{1}}}  \closeO{)}  \  \possiblyWithSub\stageOmetaColor{N^{\superscriptO} }_{{\mathrm{2}}}   \Longrightarrow^{0}    [  \possiblyWithSub\stageOmetaColor{N'^{\superscriptO} }_{{\mathrm{2}}}  /  \possiblyWithSub\stageOmetaColor{x}  ]    \possiblyWithSub\stageOmetaColor{N'^{\superscriptO} }_{{\mathrm{1}}}  
    }
  \qquad
    \derive[P0-Delta]{%
      \delta(  \possiblyWithSub\stageOmetaColor{a}_{{\mathrm{1}}}  \    \possiblyWithSub\stageOmetaColor{c}_{{\mathrm{2}}}   ) = \possiblyWithSub\stageOmetaColor{q}
    }{%
         \possiblyWithSub\stageOmetaColor{a}_{{\mathrm{1}}}  \    \possiblyWithSub\stageOmetaColor{c}_{{\mathrm{2}}}     \Longrightarrow^{0}   \possiblyWithSub\stageOmetaColor{q}  
    }
  \\[0.7em]
    \derive[P0-Ass]{%
       \possiblyWithSub\stageImetaColor{T^{\superscriptI} }_{{\mathrm{1}}}  \Longrightarrow^{1}  \possiblyWithSub\stageImetaColor{T'^{\superscriptI} }_{{\mathrm{1}}} 
    \andalso
       \possiblyWithSub\stageImetaColor{T^{\superscriptI} }_{{\mathrm{2}}}  \Longrightarrow^{1}  \possiblyWithSub\stageImetaColor{T'^{\superscriptI} }_{{\mathrm{2}}} 
    }{%
        \LeftAssertParen\openO{\langle} \possiblyWithSub\stageImetaColor{T^{\superscriptI} }_{{\mathrm{1}}} \closeO{\rangle} \relO{\CastArrow} \openO{\langle} \possiblyWithSub\stageImetaColor{T^{\superscriptI} }_{{\mathrm{2}}} \closeO{\rangle}\RightAssertParen^{ L }   \Longrightarrow^{0}   \LeftAssertParen\openO{\langle} \possiblyWithSub\stageImetaColor{T'^{\superscriptI} }_{{\mathrm{1}}} \closeO{\rangle} \relO{\CastArrow} \openO{\langle} \possiblyWithSub\stageImetaColor{T'^{\superscriptI} }_{{\mathrm{2}}} \closeO{\rangle}\RightAssertParen^{ L }  
    }
  \qquad
    \derive[P0-AssPass]{}{%
        \LeftAssertParen\openO{\langle}  \possiblyWithSub\stageImetaColor{\tau^{\superscriptI} }  \closeO{\rangle} \relO{\CastArrow} \openO{\langle}  \possiblyWithSub\stageImetaColor{\tau^{\superscriptI} }  \closeO{\rangle}\RightAssertParen^{ L }   \Longrightarrow^{0}   \ordO{\lambda} \possiblyWithSub\stageOmetaColor{x}  \relO{:}   \openO{\langle}  \possiblyWithSub\stageImetaColor{\tau^{\superscriptI} }  \closeO{\rangle}  \punctO{.}\   \possiblyWithSub\stageOmetaColor{x}   
    }
  \\[0.7em]
    \derive[P0-Brkt]{%
       \possiblyWithSub\stageImetaColor{N^{\superscriptI} }  \Longrightarrow^{1}  \possiblyWithSub\stageImetaColor{N'^{\superscriptI} } 
    }{%
        \openO{\langle} \possiblyWithSub\stageImetaColor{N^{\superscriptI} } \closeO{\rangle}   \Longrightarrow^{0}   \openO{\langle} \possiblyWithSub\stageImetaColor{N'^{\superscriptI} } \closeO{\rangle}  
    }
  \qquad
    \derive[P0-Rfn]{%
       \possiblyWithSub\stageOmetaColor{N^{\superscriptO} }  \Longrightarrow^{0}  \possiblyWithSub\stageOmetaColor{N'^{\superscriptO} } 
    }{%
        \LeftAssertParen \relO{\CastArrow}   \openO{\{} \possiblyWithSub\stageOmetaColor{\nu}  \relO{:}  \possiblyWithSub\stageOmetaColor{B}  \relO{\mid}  \possiblyWithSub\stageOmetaColor{N^{\superscriptO} } \closeO{\} }   \RightAssertParen^{ L }   \Longrightarrow^{0}   \LeftAssertParen \relO{\CastArrow}   \openO{\{} \possiblyWithSub\stageOmetaColor{\nu}  \relO{:}  \possiblyWithSub\stageOmetaColor{B}  \relO{\mid}  \possiblyWithSub\stageOmetaColor{N'^{\superscriptO} } \closeO{\} }   \RightAssertParen^{ L }  
    }
  \\[0.7em]
    \derive[P0-RfnStart]{%
       \possiblyWithSub\stageOmetaColor{N^{\superscriptO} }  \Longrightarrow^{0}  \possiblyWithSub\stageOmetaColor{N'^{\superscriptO} } 
    }{%
         \LeftAssertParen \relO{\CastArrow}   \openO{\{} \possiblyWithSub\stageOmetaColor{\nu}  \relO{:}  \possiblyWithSub\stageOmetaColor{B}  \relO{\mid}  \possiblyWithSub\stageOmetaColor{N^{\superscriptO} } \closeO{\} }   \RightAssertParen^{ L }  \    \possiblyWithSub\stageOmetaColor{c}_{{\mathrm{2}}}     \Longrightarrow^{0}   \LeftAssertParen   \openO{\{} \possiblyWithSub\stageOmetaColor{\nu}  \relO{:}  \possiblyWithSub\stageOmetaColor{B}  \relO{\mid}  \possiblyWithSub\stageOmetaColor{N'^{\superscriptO} } \closeO{\} }  \punctO{,}    [    \possiblyWithSub\stageOmetaColor{c}_{{\mathrm{2}}}    /  \possiblyWithSub\stageOmetaColor{\nu}  ]    \possiblyWithSub\stageOmetaColor{N'^{\superscriptO} }  \punctO{,}  \possiblyWithSub\stageOmetaColor{c}_{{\mathrm{2}}}  \RightAssertParen^{ L }  
    }
  \\[0.7em]
    \derive[P0-RfnAct]{%
       \possiblyWithSub\stageOmetaColor{N^{\superscriptO} }_{{\mathrm{1}}}  \Longrightarrow^{0}  \possiblyWithSub\stageOmetaColor{N'^{\superscriptO} }_{{\mathrm{1}}} 
    \andalso
       \possiblyWithSub\stageOmetaColor{N^{\superscriptO} }_{{\mathrm{2}}}  \Longrightarrow^{0}  \possiblyWithSub\stageOmetaColor{N'^{\superscriptO} }_{{\mathrm{2}}} 
    }{%
        \LeftAssertParen   \openO{\{} \possiblyWithSub\stageOmetaColor{\nu}  \relO{:}  \possiblyWithSub\stageOmetaColor{B}  \relO{\mid}  \possiblyWithSub\stageOmetaColor{N^{\superscriptO} }_{{\mathrm{1}}} \closeO{\} }  \punctO{,}  \possiblyWithSub\stageOmetaColor{N^{\superscriptO} }_{{\mathrm{2}}} \punctO{,}  \possiblyWithSub\stageOmetaColor{c}  \RightAssertParen^{ L }   \Longrightarrow^{0}   \LeftAssertParen   \openO{\{} \possiblyWithSub\stageOmetaColor{\nu}  \relO{:}  \possiblyWithSub\stageOmetaColor{B}  \relO{\mid}  \possiblyWithSub\stageOmetaColor{N'^{\superscriptO} }_{{\mathrm{1}}} \closeO{\} }  \punctO{,}  \possiblyWithSub\stageOmetaColor{N'^{\superscriptO} }_{{\mathrm{2}}} \punctO{,}  \possiblyWithSub\stageOmetaColor{c}  \RightAssertParen^{ L }  
    }
  \\[0.7em]
    \derive[P0-RfnPass]{}{%
        \LeftAssertParen   \openO{\{} \possiblyWithSub\stageOmetaColor{\nu}  \relO{:}  \possiblyWithSub\stageOmetaColor{B}  \relO{\mid}  \possiblyWithSub\stageOmetaColor{N^{\superscriptO} } \closeO{\} }  \punctO{,}     \ttO{true}    \punctO{,}  \possiblyWithSub\stageOmetaColor{c}  \RightAssertParen^{ L }   \Longrightarrow^{0}    \possiblyWithSub\stageOmetaColor{c}   
    }
  \end{center}
  \vspace{1em}
  \begin{flushleft}
    \fbox{\( \possiblyWithSub\stageImetaColor{N^{\superscriptI} }  \Longrightarrow^{1}  \possiblyWithSub\stageImetaColor{N'^{\superscriptI} } \)}
  \end{flushleft}
  \vspace{-2.25em}
  \begin{center}
    \hspace{6em}%
    \derive[P1-CstP]{%
    }{%
        \possiblyWithSub\stageImetaColor{c}   \Longrightarrow^{1}   \possiblyWithSub\stageImetaColor{c}  
    }
  \qquad
    \derive[P1-Var]{%
    }{%
        \possiblyWithSub\stageImetaColor{x}   \Longrightarrow^{1}   \possiblyWithSub\stageImetaColor{x}  
    }
  \qquad
    \derive[P1-Abs]{%
       \possiblyWithSub\stageImetaColor{T^{\superscriptI} }  \Longrightarrow^{1}  \possiblyWithSub\stageImetaColor{T'^{\superscriptI} } 
    \andalso
       \possiblyWithSub\stageImetaColor{N^{\superscriptI} }  \Longrightarrow^{1}  \possiblyWithSub\stageImetaColor{N'^{\superscriptI} } 
    }{%
        \ordI{\lambda} \possiblyWithSub\stageImetaColor{x}  \relI{:}  \possiblyWithSub\stageImetaColor{T^{\superscriptI} } \punctI{.}\  \possiblyWithSub\stageImetaColor{N^{\superscriptI} }   \Longrightarrow^{1}   \ordI{\lambda} \possiblyWithSub\stageImetaColor{x}  \relI{:}  \possiblyWithSub\stageImetaColor{T'^{\superscriptI} } \punctI{.}\  \possiblyWithSub\stageImetaColor{N'^{\superscriptI} }  
    }
  \\[0.7em]
    \derive[P1-App]{%
       \possiblyWithSub\stageImetaColor{N^{\superscriptI} }_{{\mathrm{1}}}  \Longrightarrow^{1}  \possiblyWithSub\stageImetaColor{N'^{\superscriptI} }_{{\mathrm{1}}} 
    \andalso
       \possiblyWithSub\stageImetaColor{N^{\superscriptI} }_{{\mathrm{2}}}  \Longrightarrow^{1}  \possiblyWithSub\stageImetaColor{N'^{\superscriptI} }_{{\mathrm{2}}} 
    }{%
        \possiblyWithSub\stageImetaColor{N^{\superscriptI} }_{{\mathrm{1}}} \  \possiblyWithSub\stageImetaColor{N^{\superscriptI} }_{{\mathrm{2}}}   \Longrightarrow^{1}   \possiblyWithSub\stageImetaColor{N'^{\superscriptI} }_{{\mathrm{1}}} \  \possiblyWithSub\stageImetaColor{N'^{\superscriptI} }_{{\mathrm{2}}}  
    }
  \qquad
    \derive[P1-Esc]{%
       \possiblyWithSub\stageOmetaColor{N^{\superscriptO} }  \Longrightarrow^{0}  \possiblyWithSub\stageOmetaColor{N'^{\superscriptO} } 
    }{%
        \ordI{\sim} \possiblyWithSub\stageOmetaColor{N^{\superscriptO} }   \Longrightarrow^{1}   \ordI{\sim} \possiblyWithSub\stageOmetaColor{N'^{\superscriptO} }  
    }
  \qquad
    \derive[P1-Cancel]{%
       \possiblyWithSub\stageImetaColor{N^{\superscriptI} }  \Longrightarrow^{1}  \possiblyWithSub\stageImetaColor{N'^{\superscriptI} } 
    }{%
        \ordI{\sim}  \openO{\langle} \possiblyWithSub\stageImetaColor{N^{\superscriptI} } \closeO{\rangle}    \Longrightarrow^{1}  \possiblyWithSub\stageImetaColor{N'^{\superscriptI} } 
    }
  \end{center}
  \vspace{1em}
  \begin{flushleft}
    \fbox{\( \possiblyWithSub\stageOmetaColor{T^{\superscriptO} }  \Longrightarrow^{0}  \possiblyWithSub\stageOmetaColor{T'^{\superscriptO} } \)}
  \end{flushleft}
  \vspace{-2.25em}
  \begin{center}
    \hspace{5em}%
    \derive[PT0-Rfn]{%
       \possiblyWithSub\stageOmetaColor{N^{\superscriptO} }  \Longrightarrow^{0}  \possiblyWithSub\stageOmetaColor{N'^{\superscriptO} } 
    }{%
         \openO{\{} \possiblyWithSub\stageOmetaColor{\nu}  \relO{:}  \possiblyWithSub\stageOmetaColor{B}  \relO{\mid}  \possiblyWithSub\stageOmetaColor{N^{\superscriptO} } \closeO{\} }    \Longrightarrow^{0}    \openO{\{} \possiblyWithSub\stageOmetaColor{\nu}  \relO{:}  \possiblyWithSub\stageOmetaColor{B}  \relO{\mid}  \possiblyWithSub\stageOmetaColor{N'^{\superscriptO} } \closeO{\} }   
    }
  \qquad
    \derive[PT0-Tensor]{}{%
        \ttO{Tensor}\  \possiblyWithSub\stageOmetaColor{s}   \Longrightarrow^{0}   \ttO{Tensor}\  \possiblyWithSub\stageOmetaColor{s}  
    }
  \\[0.7em]
    \derive[PT0-Code]{%
       \possiblyWithSub\stageImetaColor{T^{\superscriptI} }  \Longrightarrow^{1}  \possiblyWithSub\stageImetaColor{T'^{\superscriptI} } 
    }{%
        \openO{\langle} \possiblyWithSub\stageImetaColor{T^{\superscriptI} }_{{\mathrm{1}}} \closeO{\rangle}   \Longrightarrow^{0}   \openO{\langle} \possiblyWithSub\stageImetaColor{T'^{\superscriptI} }_{{\mathrm{1}}} \closeO{\rangle}  
    }
  \qquad
    \derive[PT0-Arr]{%
       \possiblyWithSub\stageOmetaColor{T^{\superscriptO} }_{{\mathrm{1}}}  \Longrightarrow^{0}  \possiblyWithSub\stageOmetaColor{T'^{\superscriptO} }_{{\mathrm{1}}} 
    \andalso
       \possiblyWithSub\stageOmetaColor{T^{\superscriptO} }_{{\mathrm{2}}}  \Longrightarrow^{0}  \possiblyWithSub\stageOmetaColor{T'^{\superscriptO} }_{{\mathrm{2}}} 
    }{%
        \openO{(} \possiblyWithSub\stageOmetaColor{x}  \relO{:}  \possiblyWithSub\stageOmetaColor{T^{\superscriptO} }_{{\mathrm{1}}} \closeO{)} \relO{\to}  \possiblyWithSub\stageOmetaColor{T^{\superscriptO} }_{{\mathrm{2}}}   \Longrightarrow^{0}   \openO{(} \possiblyWithSub\stageOmetaColor{x}  \relO{:}  \possiblyWithSub\stageOmetaColor{T'^{\superscriptO} }_{{\mathrm{1}}} \closeO{)} \relO{\to}  \possiblyWithSub\stageOmetaColor{T'^{\superscriptO} }_{{\mathrm{2}}}  
    }
  \end{center}
  \vspace{1em}
  \begin{flushleft}
    \fbox{\( \possiblyWithSub\stageImetaColor{T^{\superscriptI} }  \Longrightarrow^{1}  \possiblyWithSub\stageImetaColor{T'^{\superscriptI} } \)}
  \end{flushleft}
  \vspace{-2.25em}
  \begin{center}
    \hspace{5em}%
    \derive[PT1-Tensor]{%
       \possiblyWithSub\stageOmetaColor{N^{\superscriptO} }  \Longrightarrow^{0}  \possiblyWithSub\stageOmetaColor{N'^{\superscriptO} } 
    }{%
        \ttI{Tensor}\ \ordI{\%} \possiblyWithSub\stageOmetaColor{N^{\superscriptO} }   \Longrightarrow^{1}   \ttI{Tensor}\ \ordI{\%} \possiblyWithSub\stageOmetaColor{N'^{\superscriptO} }  
    }
  \\[0.7em]
    \derive[PT1-Base]{%
    }{%
        \possiblyWithSub\stageImetaColor{B}   \Longrightarrow^{1}   \possiblyWithSub\stageImetaColor{B}  
    }
  \qquad
    \derive[PT1-Arr]{%
       \possiblyWithSub\stageImetaColor{T^{\superscriptI} }_{{\mathrm{1}}}  \Longrightarrow^{1}  \possiblyWithSub\stageImetaColor{T'^{\superscriptI} }_{{\mathrm{1}}} 
    \andalso
       \possiblyWithSub\stageImetaColor{T^{\superscriptI} }_{{\mathrm{2}}}  \Longrightarrow^{1}  \possiblyWithSub\stageImetaColor{T'^{\superscriptI} }_{{\mathrm{2}}} 
    }{%
        \possiblyWithSub\stageImetaColor{T^{\superscriptI} }_{{\mathrm{1}}}  \relI{\to}  \possiblyWithSub\stageImetaColor{T^{\superscriptI} }_{{\mathrm{2}}}   \Longrightarrow^{1}   \possiblyWithSub\stageImetaColor{T'^{\superscriptI} }_{{\mathrm{1}}}  \relI{\to}  \possiblyWithSub\stageImetaColor{T'^{\superscriptI} }_{{\mathrm{2}}}  
    }
  \end{center}
  \caption{Parallel reductions}
  \label{fig:parallel-reduction}
\end{figure}
  \indent
    We use (1)~\(\beta\)-equivalences~\( \possiblyWithSub\stageOmetaColor{N^{\superscriptO} }  \cong^{0}  \possiblyWithSub\stageOmetaColor{N'^{\superscriptO} } \),
    \( \possiblyWithSub\stageImetaColor{N^{\superscriptI} }  \cong^{1}  \possiblyWithSub\stageImetaColor{N'^{\superscriptI} } \), \( \possiblyWithSub\stageOmetaColor{T^{\superscriptO} }  \cong^{0}  \possiblyWithSub\stageOmetaColor{T'^{\superscriptO} } \), and \( \possiblyWithSub\stageImetaColor{T^{\superscriptI} }  \cong^{1}  \possiblyWithSub\stageImetaColor{T'^{\superscriptI} } \), and
    (2)~their corresponding \dfn{parallel reduction} relations~\( \possiblyWithSub\stageOmetaColor{N^{\superscriptO} }  \Longrightarrow^{0}  \possiblyWithSub\stageOmetaColor{N'^{\superscriptO} } \),
    \( \possiblyWithSub\stageImetaColor{N^{\superscriptI} }  \Longrightarrow^{1}  \possiblyWithSub\stageImetaColor{N'^{\superscriptI} } \), \( \possiblyWithSub\stageOmetaColor{T^{\superscriptO} }  \Longrightarrow^{0}  \possiblyWithSub\stageOmetaColor{T'^{\superscriptO} } \), and \( \possiblyWithSub\stageImetaColor{T^{\superscriptI} }  \Longrightarrow^{1}  \possiblyWithSub\stageImetaColor{T'^{\superscriptI} } \),
    as displayed in Figures~\ref{fig:beta-type-equivalence}, \ref{fig:beta-term-equivalence},
    and \ref{fig:parallel-reduction}.
  \par
  \begin{lemma}[Reduction conforms to \(\beta\)-equivalence]\label{lem:reduction-implies-beta-equiv}
    \noindent
    \begin{enumerate}
      \item \( \possiblyWithSub\stageOmetaColor{N^{\superscriptO} }  \longrightarrow^{0}   \possiblyWithSub\stageOmetaColor{N'^{\superscriptO} }  \) implies \( \possiblyWithSub\stageOmetaColor{N^{\superscriptO} }  \cong^{0}  \possiblyWithSub\stageOmetaColor{N'^{\superscriptO} } \).
      \item \( \possiblyWithSub\stageImetaColor{N^{\superscriptI} }  \longrightarrow^{1}   \possiblyWithSub\stageImetaColor{N'^{\superscriptI} }  \) implies \( \possiblyWithSub\stageImetaColor{N^{\superscriptI} }  \cong^{1}  \possiblyWithSub\stageImetaColor{N'^{\superscriptI} } \).
      \item \( \possiblyWithSub\stageImetaColor{T^{\superscriptI} }  \longrightarrow^{1}   \possiblyWithSub\stageImetaColor{T'^{\superscriptI} }  \) implies \( \possiblyWithSub\stageImetaColor{T^{\superscriptI} }  \cong^{1}  \possiblyWithSub\stageImetaColor{T'^{\superscriptI} } \).
    \end{enumerate}
  \end{lemma}
  \begin{proof}
    By straightforward induction on the derivation.
  \end{proof}
  \begin{lemma}\label{lem:csr-terms-are-equiv}
    \noindent
    \begin{enumerate}
      \item \( \sigma  \longrightarrow  \sigma' \) implies \(  \sigma   \possiblyWithSub\stageOmetaColor{N^{\superscriptO} }   \cong^{0}   \sigma'   \possiblyWithSub\stageOmetaColor{N^{\superscriptO} }  \).
      \item \( \sigma  \longrightarrow  \sigma' \) implies \(  \sigma   \possiblyWithSub\stageImetaColor{N^{\superscriptI} }   \cong^{1}   \sigma'   \possiblyWithSub\stageImetaColor{N^{\superscriptI} }  \).
      \item \( \sigma  \longrightarrow  \sigma' \) implies \(  \sigma   \possiblyWithSub\stageOmetaColor{T^{\superscriptO} }   \cong^{0}   \sigma'   \possiblyWithSub\stageOmetaColor{T^{\superscriptO} }  \).
      \item \( \sigma  \longrightarrow  \sigma' \) implies \(  \sigma   \possiblyWithSub\stageImetaColor{T^{\superscriptI} }   \cong^{1}   \sigma'   \possiblyWithSub\stageImetaColor{T^{\superscriptI} }  \).
    \end{enumerate}
  \end{lemma}
  \begin{proof}
    By mutual induction on the structure of \(\possiblyWithSub\stageOmetaColor{N^{\superscriptO} }\), \(\possiblyWithSub\stageImetaColor{N^{\superscriptI} }\), \(\possiblyWithSub\stageOmetaColor{T^{\superscriptO} }\), and \(\possiblyWithSub\stageImetaColor{T^{\superscriptI} }\).
    \begin{enumerate}
      \item
        \begin{itemize}
          \item Case~\(\possiblyWithSub\stageOmetaColor{N^{\superscriptO} } = \possiblyWithSub\stageOmetaColor{x}\):
            \begin{itemize}
              \item If~\(\possiblyWithSub\stageOmetaColor{x} \in \dom \sigma = \dom \sigma'\):
                By the definition of \( \sigma  \longrightarrow  \sigma' \),
                we have \(  \sigma ( \possiblyWithSub\stageOmetaColor{x} )   \longrightarrow^{0\,\ast}    \sigma' ( \possiblyWithSub\stageOmetaColor{x} )   \).
                Thus, by Lemma~\ref{lem:reduction-implies-beta-equiv},
                we have \(  \sigma ( \possiblyWithSub\stageOmetaColor{x} )   \cong^{0}   \sigma' ( \possiblyWithSub\stageOmetaColor{x} )  \).
              \item If~\(\possiblyWithSub\stageOmetaColor{x} \not\in \dom \sigma = \dom \sigma'\):
                Since \( \sigma   \possiblyWithSub\stageOmetaColor{N^{\superscriptO} }  =  \sigma'   \possiblyWithSub\stageOmetaColor{N^{\superscriptO} }  = \possiblyWithSub\stageOmetaColor{x}\),
                we can immediately finish the proof by
                \derive[Bq0-Refl]{}{%
                    \possiblyWithSub\stageOmetaColor{x}   \cong^{0}   \possiblyWithSub\stageOmetaColor{x}  
                }.
            \end{itemize}
          \item Case~\(\possiblyWithSub\stageOmetaColor{N^{\superscriptO} } =  \openO{(}  \ordO{\lambda} \possiblyWithSub\stageOmetaColor{x}  \relO{:}  \possiblyWithSub\stageOmetaColor{T^{\superscriptO} }_{{\mathrm{1}}} \punctO{.}\  \possiblyWithSub\stageOmetaColor{N^{\superscriptO} }_{{\mathrm{2}}}  \closeO{)} \):
            By the Barendregt convention, we can safely assume
            \(\possiblyWithSub\stageOmetaColor{x} \not\in \dom \sigma = \dom \sigma'\),
            and thereby have
            \( \sigma   \possiblyWithSub\stageOmetaColor{N^{\superscriptO} }  =  \openO{(}  \ordO{\lambda} \possiblyWithSub\stageOmetaColor{x}  \relO{:}   \sigma   \possiblyWithSub\stageOmetaColor{T^{\superscriptO} }_{{\mathrm{1}}}  \punctO{.}\   \sigma   \possiblyWithSub\stageOmetaColor{N^{\superscriptO} }_{{\mathrm{2}}}   \closeO{)} \) and
            \( \sigma'   \possiblyWithSub\stageOmetaColor{N^{\superscriptO} }  =  \openO{(}  \ordO{\lambda} \possiblyWithSub\stageOmetaColor{x}  \relO{:}   \sigma'   \possiblyWithSub\stageOmetaColor{T^{\superscriptO} }_{{\mathrm{1}}}  \punctO{.}\   \sigma'   \possiblyWithSub\stageOmetaColor{N^{\superscriptO} }_{{\mathrm{2}}}   \closeO{)} \).
            By IH, we have
            \(  \sigma   \possiblyWithSub\stageOmetaColor{T^{\superscriptO} }_{{\mathrm{1}}}   \cong^{0}   \sigma'   \possiblyWithSub\stageOmetaColor{T^{\superscriptO} }_{{\mathrm{1}}}  \) and
            \(  \sigma   \possiblyWithSub\stageOmetaColor{N^{\superscriptO} }_{{\mathrm{2}}}   \cong^{0}   \sigma'   \possiblyWithSub\stageOmetaColor{N^{\superscriptO} }_{{\mathrm{2}}}  \).
            Therefore, we can derive
            \begin{center}
              \derive[Bq0-Abs]{%
                  \sigma   \possiblyWithSub\stageOmetaColor{T^{\superscriptO} }_{{\mathrm{1}}}   \cong^{0}   \sigma'   \possiblyWithSub\stageOmetaColor{T^{\superscriptO} }_{{\mathrm{1}}}  
              \andalso
                  \sigma   \possiblyWithSub\stageOmetaColor{N^{\superscriptO} }_{{\mathrm{2}}}   \cong^{0}   \sigma'   \possiblyWithSub\stageOmetaColor{N^{\superscriptO} }_{{\mathrm{2}}}  
              }{%
                  \ordO{\lambda} \possiblyWithSub\stageOmetaColor{x}  \relO{:}   \sigma   \possiblyWithSub\stageOmetaColor{T^{\superscriptO} }_{{\mathrm{1}}}  \punctO{.}\   \sigma   \possiblyWithSub\stageOmetaColor{N^{\superscriptO} }_{{\mathrm{2}}}    \cong^{0}   \ordO{\lambda} \possiblyWithSub\stageOmetaColor{x}  \relO{:}   \sigma'   \possiblyWithSub\stageOmetaColor{T^{\superscriptO} }_{{\mathrm{1}}}  \punctO{.}\   \sigma'   \possiblyWithSub\stageOmetaColor{N^{\superscriptO} }_{{\mathrm{2}}}   
              }.
            \end{center}
          \item Case~\(\possiblyWithSub\stageOmetaColor{N^{\superscriptO} } =  \LeftAssertParen \relO{\CastArrow}   \openO{\{} \possiblyWithSub\stageOmetaColor{\nu}  \relO{:}  \possiblyWithSub\stageOmetaColor{B}  \relO{\mid}  \possiblyWithSub\stageOmetaColor{N^{\superscriptO} }_{{\mathrm{1}}} \closeO{\} }   \RightAssertParen^{ L } \):
            Again, by the Barendregt convention, we can safely assume
            \(\possiblyWithSub\stageOmetaColor{\nu} \not\in \dom \sigma = \dom \sigma'\),
            and thereby have
            \( \sigma   \possiblyWithSub\stageOmetaColor{N^{\superscriptO} }  =  \LeftAssertParen \relO{\CastArrow}   \openO{\{} \possiblyWithSub\stageOmetaColor{\nu}  \relO{:}  \possiblyWithSub\stageOmetaColor{B}  \relO{\mid}   \sigma   \possiblyWithSub\stageOmetaColor{N^{\superscriptO} }_{{\mathrm{1}}}  \closeO{\} }   \RightAssertParen^{ L } \) and
            \( \sigma'   \possiblyWithSub\stageOmetaColor{N^{\superscriptO} }  =  \LeftAssertParen \relO{\CastArrow}   \openO{\{} \possiblyWithSub\stageOmetaColor{\nu}  \relO{:}  \possiblyWithSub\stageOmetaColor{B}  \relO{\mid}   \sigma'   \possiblyWithSub\stageOmetaColor{N^{\superscriptO} }_{{\mathrm{1}}}  \closeO{\} }   \RightAssertParen^{ L } \).
            Since we have \(  \sigma   \possiblyWithSub\stageOmetaColor{N^{\superscriptO} }_{{\mathrm{1}}}   \cong^{0}   \sigma'   \possiblyWithSub\stageOmetaColor{N^{\superscriptO} }_{{\mathrm{1}}}  \),
            we can derive
            \begin{center}
              \derive[Bq0-Rfn]{%
                  \sigma   \possiblyWithSub\stageOmetaColor{N^{\superscriptO} }_{{\mathrm{1}}}   \cong^{0}   \sigma'   \possiblyWithSub\stageOmetaColor{N^{\superscriptO} }_{{\mathrm{1}}}  
              }{%
                  \LeftAssertParen \relO{\CastArrow}   \openO{\{} \possiblyWithSub\stageOmetaColor{\nu}  \relO{:}  \possiblyWithSub\stageOmetaColor{B}  \relO{\mid}   \sigma   \possiblyWithSub\stageOmetaColor{N^{\superscriptO} }_{{\mathrm{1}}}  \closeO{\} }   \RightAssertParen^{ L }   \cong^{0}   \LeftAssertParen \relO{\CastArrow}   \openO{\{} \possiblyWithSub\stageOmetaColor{\nu}  \relO{:}  \possiblyWithSub\stageOmetaColor{B}  \relO{\mid}   \sigma'   \possiblyWithSub\stageOmetaColor{N^{\superscriptO} }_{{\mathrm{1}}}  \closeO{\} }   \RightAssertParen^{ L }  
              }.
            \end{center}
          \item Case~\(\possiblyWithSub\stageOmetaColor{N^{\superscriptO} } =  \LeftAssertParen   \openO{\{} \possiblyWithSub\stageOmetaColor{\nu}  \relO{:}  \possiblyWithSub\stageOmetaColor{B}  \relO{\mid}  \possiblyWithSub\stageOmetaColor{N^{\superscriptO} }_{{\mathrm{1}}} \closeO{\} }  \punctO{,}  \possiblyWithSub\stageOmetaColor{N^{\superscriptO} }_{{\mathrm{2}}} \punctO{,}  \possiblyWithSub\stageOmetaColor{c}  \RightAssertParen^{ L } \):
            This can be proved in a manner similar to the previous one.
          \item The other cases are all straightforward.
        \end{itemize}
      \item
        All the cases are straightforward.
      \item
        \begin{itemize}
          \item Case~\(\possiblyWithSub\stageOmetaColor{T^{\superscriptO} } =  \openO{(} \possiblyWithSub\stageOmetaColor{x}  \relO{:}  \possiblyWithSub\stageOmetaColor{T^{\superscriptO} }_{{\mathrm{1}}} \closeO{)} \relO{\to}  \possiblyWithSub\stageOmetaColor{T^{\superscriptO} }_{{\mathrm{2}}} \):
            Straightforward by IH since we can safely assume
            \(\possiblyWithSub\stageOmetaColor{x} \not\in \dom \sigma = \dom \sigma'\) by the Barendregt convention.
          \item Case~\(\possiblyWithSub\stageOmetaColor{T^{\superscriptO} } =  \openO{\{} \possiblyWithSub\stageOmetaColor{\nu}  \relO{:}  \possiblyWithSub\stageOmetaColor{B}  \relO{\mid}  \possiblyWithSub\stageOmetaColor{N^{\superscriptO} } \closeO{\} } \):
            This is also straightforward by the Barendregt convention and IH.
          \item The other cases are all straightforward.
        \end{itemize}
      \item
        \begin{itemize}
          \item Case~\(\possiblyWithSub\stageImetaColor{T^{\superscriptI} } =  \ttI{Tensor}\ \ordI{\%} \possiblyWithSub\stageOmetaColor{N^{\superscriptO} } \):
            By IH, we have \(  \sigma   \possiblyWithSub\stageOmetaColor{N^{\superscriptO} }   \cong^{0}   \sigma'   \possiblyWithSub\stageOmetaColor{N^{\superscriptO} }  \), and thereby we can derive
            \begin{center}
              \derive[BqT1-Tensor]{%
                  \sigma   \possiblyWithSub\stageOmetaColor{N^{\superscriptO} }   \cong^{0}   \sigma'   \possiblyWithSub\stageOmetaColor{N^{\superscriptO} }  
              }{%
                  \ttI{Tensor}\ \ordI{\%}  \openO{(}  \sigma   \possiblyWithSub\stageOmetaColor{N^{\superscriptO} }  \closeO{)}    \cong^{1}   \ttI{Tensor}\ \ordI{\%}  \openO{(}  \sigma'   \possiblyWithSub\stageOmetaColor{N^{\superscriptO} }  \closeO{)}   
              }.
            \end{center}
          \item
            The other cases are also straightforward.
        \end{itemize}
    \end{enumerate}
  \end{proof}
  \begin{lemma}[Stage-1 CSR equivalence implies \(\beta\)-equivalence]\label{lem:csr-equiv-implies-beta-equiv-1}
    \( \possiblyWithSub\stageImetaColor{T^{\superscriptI} }  \equiv^{1}  \possiblyWithSub\stageImetaColor{T'^{\superscriptI} } \) implies \( \possiblyWithSub\stageImetaColor{T^{\superscriptI} }  \cong^{1}  \possiblyWithSub\stageImetaColor{T'^{\superscriptI} } \).
  \end{lemma}
  \begin{proof}
    By induction on the derivation of \( \possiblyWithSub\stageImetaColor{T^{\superscriptI} }  \equiv^{1}  \possiblyWithSub\stageImetaColor{T'^{\superscriptI} } \).
    \begin{itemize}
      \item Case \derive[CqT1-Tensor]{%
         \sigma  \longrightarrow  \sigma' 
      }{%
          \ttI{Tensor}\ \ordI{\%}  \openO{(}  \sigma   \possiblyWithSub\stageOmetaColor{N^{\superscriptO} }  \closeO{)}    \equiv^{1}   \ttI{Tensor}\ \ordI{\%}  \openO{(}  \sigma'   \possiblyWithSub\stageOmetaColor{N^{\superscriptO} }  \closeO{)}   
      }:
        By Lemma~\ref{lem:csr-terms-are-equiv}, we have \(  \sigma   \possiblyWithSub\stageOmetaColor{N^{\superscriptO} }   \cong^{0}   \sigma'   \possiblyWithSub\stageOmetaColor{N^{\superscriptO} }  \).
        Thus, we can derive
        \begin{center}
          \derive[BqT1-Tensor]{%
              \sigma   \possiblyWithSub\stageOmetaColor{N^{\superscriptO} }   \cong^{0}   \sigma'   \possiblyWithSub\stageOmetaColor{N^{\superscriptO} }  
          }{%
              \ttI{Tensor}\ \ordI{\%}  \openO{(}  \sigma   \possiblyWithSub\stageOmetaColor{N^{\superscriptO} }  \closeO{)}    \cong^{1}   \ttI{Tensor}\ \ordI{\%}  \openO{(}  \sigma'   \possiblyWithSub\stageOmetaColor{N^{\superscriptO} }  \closeO{)}   
          }.
        \end{center}
      \item The other cases (i.e.,~\rulename{CqT1-Refl}, \rulename{CqT1-Sym},
      \rulename{CqT1-Trans}, and \rulename{CqT1-Arr}) are straightforward.
    \end{itemize}
  \end{proof}
  \begin{lemma}[Stage-0 CSR equivalence implies \(\beta\)-equivalence]\label{lem:csr-equiv-implies-beta-equiv-0}
    \( \possiblyWithSub\stageOmetaColor{T^{\superscriptO} }  \equiv^{0}  \possiblyWithSub\stageOmetaColor{T'^{\superscriptO} } \) implies \( \possiblyWithSub\stageOmetaColor{T^{\superscriptO} }  \cong^{0}  \possiblyWithSub\stageOmetaColor{T'^{\superscriptO} } \).
  \end{lemma}
  \begin{proof}
    By induction on the derivation of \( \possiblyWithSub\stageOmetaColor{T^{\superscriptO} }  \equiv^{0}  \possiblyWithSub\stageOmetaColor{T'^{\superscriptO} } \).
    \begin{itemize}
      \item Case \derive[CqT0-Code]{%
         \possiblyWithSub\stageImetaColor{T^{\superscriptI} }_{{\mathrm{1}}}  \equiv^{1}  \possiblyWithSub\stageImetaColor{T^{\superscriptI} }_{{\mathrm{2}}} 
      }{%
          \openO{\langle} \possiblyWithSub\stageImetaColor{T^{\superscriptI} }_{{\mathrm{1}}} \closeO{\rangle}   \equiv^{0}   \openO{\langle} \possiblyWithSub\stageImetaColor{T^{\superscriptI} }_{{\mathrm{2}}} \closeO{\rangle}  
      }:
        Straightforward by Lemma~\ref{lem:csr-equiv-implies-beta-equiv-1}.
      \item Case \derive[CqT0-Rfn]{%
         \sigma  \longrightarrow  \sigma' 
      }{%
           \openO{\{} \possiblyWithSub\stageOmetaColor{\nu}  \relO{:}  \possiblyWithSub\stageOmetaColor{B}  \relO{\mid}   \sigma   \possiblyWithSub\stageOmetaColor{N^{\superscriptO} }  \closeO{\} }    \equiv^{0}    \openO{\{} \possiblyWithSub\stageOmetaColor{\nu}  \relO{:}  \possiblyWithSub\stageOmetaColor{B}  \relO{\mid}   \sigma'   \possiblyWithSub\stageOmetaColor{N^{\superscriptO} }  \closeO{\} }   
      }:
        By Lemma~\ref{lem:csr-terms-are-equiv}, we have \(  \sigma   \possiblyWithSub\stageOmetaColor{N^{\superscriptO} }   \cong^{0}   \sigma'   \possiblyWithSub\stageOmetaColor{N^{\superscriptO} }  \).
        Thus, we can derive
        \begin{center}
          \derive[BqT0-Rfn]{%
              \sigma   \possiblyWithSub\stageOmetaColor{N^{\superscriptO} }   \cong^{0}   \sigma'   \possiblyWithSub\stageOmetaColor{N^{\superscriptO} }  
          }{%
               \openO{\{} \possiblyWithSub\stageOmetaColor{\nu}  \relO{:}  \possiblyWithSub\stageOmetaColor{B}  \relO{\mid}   \sigma   \possiblyWithSub\stageOmetaColor{N^{\superscriptO} }  \closeO{\} }    \cong^{0}    \openO{\{} \possiblyWithSub\stageOmetaColor{\nu}  \relO{:}  \possiblyWithSub\stageOmetaColor{B}  \relO{\mid}   \sigma'   \possiblyWithSub\stageOmetaColor{N^{\superscriptO} }  \closeO{\} }   
          }.
        \end{center}
      \item The other cases (i.e.,~\rulename{CqT0-Refl}, \rulename{CqT0-Sym},
      \rulename{CqT0-Trans}, and \rulename{CqT0-Arr}) are straightforward.
    \end{itemize}
  \end{proof}
  \begin{lemma}[Parallel reduction includes equality]\label{lem:eq-implies-par}
    \noindent
    \begin{enumerate}
      \item \( \possiblyWithSub\stageOmetaColor{N^{\superscriptO} }  \Longrightarrow^{0}  \possiblyWithSub\stageOmetaColor{N^{\superscriptO} } \) for any \(\possiblyWithSub\stageOmetaColor{N^{\superscriptO} }\).
      \item \( \possiblyWithSub\stageImetaColor{N^{\superscriptI} }  \Longrightarrow^{1}  \possiblyWithSub\stageImetaColor{N^{\superscriptI} } \) for any \(\possiblyWithSub\stageImetaColor{N^{\superscriptI} }\).
      \item \( \possiblyWithSub\stageOmetaColor{T^{\superscriptO} }  \Longrightarrow^{0}  \possiblyWithSub\stageOmetaColor{T^{\superscriptO} } \) for any \(\possiblyWithSub\stageOmetaColor{T^{\superscriptO} }\).
      \item \( \possiblyWithSub\stageImetaColor{T^{\superscriptI} }  \Longrightarrow^{1}  \possiblyWithSub\stageImetaColor{T^{\superscriptI} } \) for any \(\possiblyWithSub\stageImetaColor{T^{\superscriptI} }\).
    \end{enumerate}
  \end{lemma}
  \begin{proof}
    By straightforward mutual induction on the structure of
    \(\possiblyWithSub\stageOmetaColor{N^{\superscriptO} }\), \(\possiblyWithSub\stageImetaColor{N^{\superscriptI} }\), \(\possiblyWithSub\stageOmetaColor{T^{\superscriptO} }\), and \(\possiblyWithSub\stageImetaColor{T^{\superscriptI} }\).
  \end{proof}
  \begin{lemma}[Parallel reduction conforms to \(\beta\)-equivalence]\label{lem:par-implies-equiv}
    \noindent
    \begin{enumerate}
      \item \( \possiblyWithSub\stageOmetaColor{T^{\superscriptO} }  \Longrightarrow^{0}  \possiblyWithSub\stageOmetaColor{T'^{\superscriptO} } \) implies \( \possiblyWithSub\stageOmetaColor{T^{\superscriptO} }  \cong^{0}  \possiblyWithSub\stageOmetaColor{T'^{\superscriptO} } \).
      \item \( \possiblyWithSub\stageImetaColor{T^{\superscriptI} }  \Longrightarrow^{1}  \possiblyWithSub\stageImetaColor{T'^{\superscriptI} } \) implies \( \possiblyWithSub\stageImetaColor{T^{\superscriptI} }  \cong^{1}  \possiblyWithSub\stageImetaColor{T'^{\superscriptI} } \).
    \end{enumerate}
  \end{lemma}
  \begin{proof}
    By straightforward mutual induction on the derivations.
  \end{proof}
  \begin{lemma}\label{lem:refl-trans-symm-par-propagates-type-structure}
    \noindent
    \begin{enumerate}
      \item
        If \( \possiblyWithSub\stageImetaColor{T^{\superscriptI} }_{{\mathrm{1}}}  \Longleftrightarrow^{1\,\ast}  \possiblyWithSub\stageImetaColor{T^{\superscriptI} }_{{\mathrm{2}}} \), then \(  \openO{\langle} \possiblyWithSub\stageImetaColor{T^{\superscriptI} }_{{\mathrm{1}}} \closeO{\rangle}   \Longleftrightarrow^{0\,\ast}   \openO{\langle} \possiblyWithSub\stageImetaColor{T^{\superscriptI} }_{{\mathrm{2}}} \closeO{\rangle}  \).
      \item
        If \( \possiblyWithSub\stageOmetaColor{T^{\superscriptO} }_{{\mathrm{1}}}  \Longleftrightarrow^{0\,\ast}  \possiblyWithSub\stageOmetaColor{T^{\superscriptO} }_{{\mathrm{2}}} \), then \(  \openO{(} \possiblyWithSub\stageOmetaColor{x}  \relO{:}  \possiblyWithSub\stageOmetaColor{T^{\superscriptO} }_{{\mathrm{1}}} \closeO{)} \relO{\to}  \possiblyWithSub\stageOmetaColor{T^{\superscriptO} }   \Longleftrightarrow^{0\,\ast}   \openO{(} \possiblyWithSub\stageOmetaColor{x}  \relO{:}  \possiblyWithSub\stageOmetaColor{T^{\superscriptO} }_{{\mathrm{2}}} \closeO{)} \relO{\to}  \possiblyWithSub\stageOmetaColor{T^{\superscriptO} }  \).
      \item
        If \( \possiblyWithSub\stageOmetaColor{T^{\superscriptO} }_{{\mathrm{1}}}  \Longleftrightarrow^{0\,\ast}  \possiblyWithSub\stageOmetaColor{T^{\superscriptO} }_{{\mathrm{2}}} \), then \(  \openO{(} \possiblyWithSub\stageOmetaColor{x}  \relO{:}  \possiblyWithSub\stageOmetaColor{T^{\superscriptO} } \closeO{)} \relO{\to}  \possiblyWithSub\stageOmetaColor{T^{\superscriptO} }_{{\mathrm{1}}}   \Longleftrightarrow^{0\,\ast}   \openO{(} \possiblyWithSub\stageOmetaColor{x}  \relO{:}  \possiblyWithSub\stageOmetaColor{T^{\superscriptO} } \closeO{)} \relO{\to}  \possiblyWithSub\stageOmetaColor{T^{\superscriptO} }_{{\mathrm{2}}}  \).
      \item
        If \( \possiblyWithSub\stageOmetaColor{N^{\superscriptO} }_{{\mathrm{1}}}  \Longleftrightarrow^{0\,\ast}  \possiblyWithSub\stageOmetaColor{N^{\superscriptO} }_{{\mathrm{2}}} \), then \(   \openO{\{} \possiblyWithSub\stageOmetaColor{\nu}  \relO{:}  \possiblyWithSub\stageOmetaColor{B}  \relO{\mid}  \possiblyWithSub\stageOmetaColor{N^{\superscriptO} }_{{\mathrm{1}}} \closeO{\} }    \Longleftrightarrow^{0\,\ast}    \openO{\{} \possiblyWithSub\stageOmetaColor{\nu}  \relO{:}  \possiblyWithSub\stageOmetaColor{B}  \relO{\mid}  \possiblyWithSub\stageOmetaColor{N^{\superscriptO} }_{{\mathrm{2}}} \closeO{\} }   \).
      \item
        If \( \possiblyWithSub\stageImetaColor{T^{\superscriptI} }_{{\mathrm{1}}}  \Longleftrightarrow^{1\,\ast}  \possiblyWithSub\stageImetaColor{T^{\superscriptI} }_{{\mathrm{2}}} \), then \(  \possiblyWithSub\stageImetaColor{T^{\superscriptI} }_{{\mathrm{1}}}  \relI{\to}  \possiblyWithSub\stageImetaColor{T^{\superscriptI} }   \Longleftrightarrow^{1\,\ast}   \possiblyWithSub\stageImetaColor{T^{\superscriptI} }_{{\mathrm{2}}}  \relI{\to}  \possiblyWithSub\stageImetaColor{T^{\superscriptI} }  \).
      \item
        If \( \possiblyWithSub\stageImetaColor{T^{\superscriptI} }_{{\mathrm{1}}}  \Longleftrightarrow^{1\,\ast}  \possiblyWithSub\stageImetaColor{T^{\superscriptI} }_{{\mathrm{2}}} \), then \(  \possiblyWithSub\stageImetaColor{T^{\superscriptI} }  \relI{\to}  \possiblyWithSub\stageImetaColor{T^{\superscriptI} }_{{\mathrm{1}}}   \Longleftrightarrow^{1\,\ast}   \possiblyWithSub\stageImetaColor{T^{\superscriptI} }  \relI{\to}  \possiblyWithSub\stageImetaColor{T^{\superscriptI} }_{{\mathrm{2}}}  \).
      \item
        If \( \possiblyWithSub\stageOmetaColor{N^{\superscriptO} }_{{\mathrm{1}}}  \Longleftrightarrow^{0\,\ast}  \possiblyWithSub\stageOmetaColor{N^{\superscriptO} }_{{\mathrm{2}}} \), then \(  \ttI{Tensor}\ \ordI{\%} \possiblyWithSub\stageOmetaColor{N^{\superscriptO} }_{{\mathrm{1}}}   \Longleftrightarrow^{1\,\ast}   \ttI{Tensor}\ \ordI{\%} \possiblyWithSub\stageOmetaColor{N^{\superscriptO} }_{{\mathrm{2}}}  \).
    \end{enumerate}
  \end{lemma}
  \begin{proof}
    \begin{enumerate}
      \item
        By induction on the number of steps of \( \possiblyWithSub\stageImetaColor{T^{\superscriptI} }_{{\mathrm{1}}}  \Longleftrightarrow^{1\,\ast}  \possiblyWithSub\stageImetaColor{T^{\superscriptI} }_{{\mathrm{2}}} \).
        \begin{itemize}
          \item Case \(\possiblyWithSub\stageImetaColor{T^{\superscriptI} }_{{\mathrm{1}}} = \possiblyWithSub\stageImetaColor{T^{\superscriptI} }_{{\mathrm{2}}}\) is immediate.
          \item Case \( \possiblyWithSub\stageImetaColor{T'^{\superscriptI} }_{{\mathrm{1}}}  \Longrightarrow^{1}  \possiblyWithSub\stageImetaColor{T^{\superscriptI} }_{{\mathrm{1}}} \) and \( \possiblyWithSub\stageImetaColor{T'^{\superscriptI} }_{{\mathrm{1}}}  \Longleftrightarrow^{1\,\ast}  \possiblyWithSub\stageImetaColor{T^{\superscriptI} }_{{\mathrm{2}}} \):
            First, we can derive
            \begin{center}
              \derive[PT0-Code]{%
                 \possiblyWithSub\stageImetaColor{T'^{\superscriptI} }_{{\mathrm{1}}}  \Longrightarrow^{1}  \possiblyWithSub\stageImetaColor{T^{\superscriptI} }_{{\mathrm{1}}} 
              }{%
                  \openO{\langle} \possiblyWithSub\stageImetaColor{T'^{\superscriptI} }_{{\mathrm{1}}} \closeO{\rangle}   \Longrightarrow^{0}   \openO{\langle} \possiblyWithSub\stageImetaColor{T^{\superscriptI} }_{{\mathrm{1}}} \closeO{\rangle}  
              }.
            \end{center}
            Also, by IH, we have \(  \openO{\langle} \possiblyWithSub\stageImetaColor{T'^{\superscriptI} }_{{\mathrm{1}}} \closeO{\rangle}   \Longleftrightarrow^{0\,\ast}   \openO{\langle} \possiblyWithSub\stageImetaColor{T^{\superscriptI} }_{{\mathrm{2}}} \closeO{\rangle}  \).
            Therefore, we have \(  \openO{\langle} \possiblyWithSub\stageImetaColor{T^{\superscriptI} }_{{\mathrm{1}}} \closeO{\rangle}   \Longleftrightarrow^{0\,\ast}   \openO{\langle} \possiblyWithSub\stageImetaColor{T^{\superscriptI} }_{{\mathrm{2}}} \closeO{\rangle}  \).
          \item Case \( \possiblyWithSub\stageImetaColor{T^{\superscriptI} }_{{\mathrm{1}}}  \Longrightarrow^{1}  \possiblyWithSub\stageImetaColor{T'^{\superscriptI} }_{{\mathrm{1}}} \) and \( \possiblyWithSub\stageImetaColor{T'^{\superscriptI} }_{{\mathrm{1}}}  \Longleftrightarrow^{1\,\ast}  \possiblyWithSub\stageImetaColor{T^{\superscriptI} }_{{\mathrm{2}}} \):
            Can be proved in the same manner as the previous one.
        \end{itemize}
      \item
        By induction on the number of steps of \( \possiblyWithSub\stageOmetaColor{T^{\superscriptO} }_{{\mathrm{1}}}  \Longleftrightarrow^{0\,\ast}  \possiblyWithSub\stageOmetaColor{T^{\superscriptO} }_{{\mathrm{2}}} \).
        \begin{itemize}
          \item Case \(\possiblyWithSub\stageOmetaColor{T^{\superscriptO} }_{{\mathrm{1}}} = \possiblyWithSub\stageOmetaColor{T^{\superscriptO} }_{{\mathrm{2}}}\) is immediate.
          \item Case \( \possiblyWithSub\stageOmetaColor{T'^{\superscriptO} }_{{\mathrm{1}}}  \Longrightarrow^{0}  \possiblyWithSub\stageOmetaColor{T^{\superscriptO} }_{{\mathrm{1}}} \) and \( \possiblyWithSub\stageOmetaColor{T'^{\superscriptO} }_{{\mathrm{1}}}  \Longleftrightarrow^{0\,\ast}  \possiblyWithSub\stageOmetaColor{T^{\superscriptO} }_{{\mathrm{2}}} \):
            First, since \( \possiblyWithSub\stageOmetaColor{T^{\superscriptO} }  \Longrightarrow^{0}  \possiblyWithSub\stageOmetaColor{T^{\superscriptO} } \) holds by Lemma~\ref{lem:eq-implies-par},
            we can derive
            \begin{center}
              \derive[PT0-Arr]{%
                 \possiblyWithSub\stageOmetaColor{T'^{\superscriptO} }_{{\mathrm{1}}}  \Longrightarrow^{0}  \possiblyWithSub\stageOmetaColor{T^{\superscriptO} }_{{\mathrm{1}}} 
              \andalso
                 \possiblyWithSub\stageOmetaColor{T^{\superscriptO} }  \Longrightarrow^{0}  \possiblyWithSub\stageOmetaColor{T^{\superscriptO} } 
              }{%
                  \openO{(} \possiblyWithSub\stageOmetaColor{x}  \relO{:}  \possiblyWithSub\stageOmetaColor{T'^{\superscriptO} }_{{\mathrm{1}}} \closeO{)} \relO{\to}  \possiblyWithSub\stageOmetaColor{T^{\superscriptO} }   \Longrightarrow^{0}   \openO{(} \possiblyWithSub\stageOmetaColor{x}  \relO{:}  \possiblyWithSub\stageOmetaColor{T^{\superscriptO} }_{{\mathrm{1}}} \closeO{)} \relO{\to}  \possiblyWithSub\stageOmetaColor{T^{\superscriptO} }  
              }.
            \end{center}
            Also, by IH, we have \(  \openO{(} \possiblyWithSub\stageOmetaColor{x}  \relO{:}  \possiblyWithSub\stageOmetaColor{T'^{\superscriptO} }_{{\mathrm{1}}} \closeO{)} \relO{\to}  \possiblyWithSub\stageOmetaColor{T^{\superscriptO} }   \Longleftrightarrow^{0\,\ast}   \openO{(} \possiblyWithSub\stageOmetaColor{x}  \relO{:}  \possiblyWithSub\stageOmetaColor{T^{\superscriptO} }_{{\mathrm{2}}} \closeO{)} \relO{\to}  \possiblyWithSub\stageOmetaColor{T^{\superscriptO} }  \).
            Therefore, we have \(  \openO{(} \possiblyWithSub\stageOmetaColor{x}  \relO{:}  \possiblyWithSub\stageOmetaColor{T^{\superscriptO} }_{{\mathrm{1}}} \closeO{)} \relO{\to}  \possiblyWithSub\stageOmetaColor{T^{\superscriptO} }   \Longleftrightarrow^{0\,\ast}   \openO{(} \possiblyWithSub\stageOmetaColor{x}  \relO{:}  \possiblyWithSub\stageOmetaColor{T^{\superscriptO} }_{{\mathrm{2}}} \closeO{)} \relO{\to}  \possiblyWithSub\stageOmetaColor{T^{\superscriptO} }  \).
          \item Case \( \possiblyWithSub\stageOmetaColor{T^{\superscriptO} }_{{\mathrm{1}}}  \Longrightarrow^{0}  \possiblyWithSub\stageOmetaColor{T'^{\superscriptO} }_{{\mathrm{1}}} \) and \( \possiblyWithSub\stageOmetaColor{T'^{\superscriptO} }_{{\mathrm{1}}}  \Longleftrightarrow^{0\,\ast}  \possiblyWithSub\stageOmetaColor{T^{\superscriptO} }_{{\mathrm{2}}} \):
            Can be proved in the same manner as the previous one.
        \end{itemize}
    \end{enumerate}
    (3)--(7) can be proved in similar ways.
  \end{proof}
  \begin{lemma}\label{lem:refl-trans-symm-par-propagates-term-0-structure}
    \noindent
    \begin{enumerate}
      \item
        If \( \possiblyWithSub\stageImetaColor{N^{\superscriptI} }_{{\mathrm{1}}}  \Longleftrightarrow^{1\,\ast}  \possiblyWithSub\stageImetaColor{N^{\superscriptI} }_{{\mathrm{2}}} \), then \(  \openO{\langle} \possiblyWithSub\stageImetaColor{N^{\superscriptI} }_{{\mathrm{1}}} \closeO{\rangle}   \Longleftrightarrow^{0\,\ast}   \openO{\langle} \possiblyWithSub\stageImetaColor{N^{\superscriptI} }_{{\mathrm{2}}} \closeO{\rangle}  \).
      \item
        If \( \possiblyWithSub\stageImetaColor{T^{\superscriptI} }_{{\mathrm{1}}}  \Longleftrightarrow^{1\,\ast}  \possiblyWithSub\stageImetaColor{T^{\superscriptI} }_{{\mathrm{2}}} \), then \(  \LeftAssertParen\openO{\langle} \possiblyWithSub\stageImetaColor{T^{\superscriptI} }_{{\mathrm{1}}} \closeO{\rangle} \relO{\CastArrow} \openO{\langle} \possiblyWithSub\stageImetaColor{T^{\superscriptI} } \closeO{\rangle}\RightAssertParen^{ L }   \Longleftrightarrow^{0\,\ast}   \LeftAssertParen\openO{\langle} \possiblyWithSub\stageImetaColor{T^{\superscriptI} }_{{\mathrm{2}}} \closeO{\rangle} \relO{\CastArrow} \openO{\langle} \possiblyWithSub\stageImetaColor{T^{\superscriptI} } \closeO{\rangle}\RightAssertParen^{ L }  \).
      \item
        If \( \possiblyWithSub\stageImetaColor{T^{\superscriptI} }_{{\mathrm{1}}}  \Longleftrightarrow^{1\,\ast}  \possiblyWithSub\stageImetaColor{T^{\superscriptI} }_{{\mathrm{2}}} \), then \(  \LeftAssertParen\openO{\langle} \possiblyWithSub\stageImetaColor{T^{\superscriptI} } \closeO{\rangle} \relO{\CastArrow} \openO{\langle} \possiblyWithSub\stageImetaColor{T^{\superscriptI} }_{{\mathrm{1}}} \closeO{\rangle}\RightAssertParen^{ L }   \Longleftrightarrow^{0\,\ast}   \LeftAssertParen\openO{\langle} \possiblyWithSub\stageImetaColor{T^{\superscriptI} } \closeO{\rangle} \relO{\CastArrow} \openO{\langle} \possiblyWithSub\stageImetaColor{T^{\superscriptI} }_{{\mathrm{2}}} \closeO{\rangle}\RightAssertParen^{ L }  \).
      \item
        If \( \possiblyWithSub\stageOmetaColor{T^{\superscriptO} }_{{\mathrm{1}}}  \Longleftrightarrow^{0\,\ast}  \possiblyWithSub\stageOmetaColor{T^{\superscriptO} }_{{\mathrm{2}}} \), then \(  \ordO{\lambda} \possiblyWithSub\stageOmetaColor{x}  \relO{:}  \possiblyWithSub\stageOmetaColor{T^{\superscriptO} }_{{\mathrm{1}}} \punctO{.}\  \possiblyWithSub\stageOmetaColor{N^{\superscriptO} }   \Longleftrightarrow^{0\,\ast}   \ordO{\lambda} \possiblyWithSub\stageOmetaColor{x}  \relO{:}  \possiblyWithSub\stageOmetaColor{T^{\superscriptO} }_{{\mathrm{2}}} \punctO{.}\  \possiblyWithSub\stageOmetaColor{N^{\superscriptO} }  \).
      \item
        If \( \possiblyWithSub\stageOmetaColor{N^{\superscriptO} }_{{\mathrm{1}}}  \Longleftrightarrow^{0\,\ast}  \possiblyWithSub\stageOmetaColor{N^{\superscriptO} }_{{\mathrm{2}}} \), then \(  \ordO{\lambda} \possiblyWithSub\stageOmetaColor{x}  \relO{:}  \possiblyWithSub\stageOmetaColor{T^{\superscriptO} } \punctO{.}\  \possiblyWithSub\stageOmetaColor{N^{\superscriptO} }_{{\mathrm{1}}}   \Longleftrightarrow^{0\,\ast}   \ordO{\lambda} \possiblyWithSub\stageOmetaColor{x}  \relO{:}  \possiblyWithSub\stageOmetaColor{T^{\superscriptO} } \punctO{.}\  \possiblyWithSub\stageOmetaColor{N^{\superscriptO} }_{{\mathrm{2}}}  \).
      \item
        If \( \possiblyWithSub\stageOmetaColor{N^{\superscriptO} }_{{\mathrm{1}}}  \Longleftrightarrow^{0\,\ast}  \possiblyWithSub\stageOmetaColor{N^{\superscriptO} }_{{\mathrm{2}}} \), then \(  \possiblyWithSub\stageOmetaColor{N^{\superscriptO} }_{{\mathrm{1}}} \  \possiblyWithSub\stageOmetaColor{N^{\superscriptO} }   \Longleftrightarrow^{0\,\ast}   \possiblyWithSub\stageOmetaColor{N^{\superscriptO} }_{{\mathrm{2}}} \  \possiblyWithSub\stageOmetaColor{N^{\superscriptO} }  \).
      \item
        If \( \possiblyWithSub\stageOmetaColor{N^{\superscriptO} }_{{\mathrm{1}}}  \Longleftrightarrow^{0\,\ast}  \possiblyWithSub\stageOmetaColor{N^{\superscriptO} }_{{\mathrm{2}}} \), then \(  \possiblyWithSub\stageOmetaColor{N^{\superscriptO} } \  \possiblyWithSub\stageOmetaColor{N^{\superscriptO} }_{{\mathrm{1}}}   \Longleftrightarrow^{0\,\ast}   \possiblyWithSub\stageOmetaColor{N^{\superscriptO} } \  \possiblyWithSub\stageOmetaColor{N^{\superscriptO} }_{{\mathrm{2}}}  \).
      \item
        If \( \possiblyWithSub\stageOmetaColor{N^{\superscriptO} }_{{\mathrm{1}}}  \Longleftrightarrow^{0\,\ast}  \possiblyWithSub\stageOmetaColor{N^{\superscriptO} }_{{\mathrm{2}}} \), then
        \(  \LeftAssertParen \relO{\CastArrow}   \openO{\{} \possiblyWithSub\stageOmetaColor{\nu}  \relO{:}  \possiblyWithSub\stageOmetaColor{B}  \relO{\mid}  \possiblyWithSub\stageOmetaColor{N^{\superscriptO} }_{{\mathrm{1}}} \closeO{\} }   \RightAssertParen^{ L }   \Longleftrightarrow^{0\,\ast}   \LeftAssertParen \relO{\CastArrow}   \openO{\{} \possiblyWithSub\stageOmetaColor{\nu}  \relO{:}  \possiblyWithSub\stageOmetaColor{B}  \relO{\mid}  \possiblyWithSub\stageOmetaColor{N^{\superscriptO} }_{{\mathrm{2}}} \closeO{\} }   \RightAssertParen^{ L }  \).
      \item
        If \( \possiblyWithSub\stageOmetaColor{N^{\superscriptO} }_{{\mathrm{1}}}  \Longleftrightarrow^{0\,\ast}  \possiblyWithSub\stageOmetaColor{N^{\superscriptO} }_{{\mathrm{2}}} \), then
        \(  \LeftAssertParen   \openO{\{} \possiblyWithSub\stageOmetaColor{\nu}  \relO{:}  \possiblyWithSub\stageOmetaColor{B}  \relO{\mid}  \possiblyWithSub\stageOmetaColor{N^{\superscriptO} }_{{\mathrm{1}}} \closeO{\} }  \punctO{,}  \possiblyWithSub\stageOmetaColor{N^{\superscriptO} } \punctO{,}  \possiblyWithSub\stageOmetaColor{c}  \RightAssertParen^{ L }   \Longleftrightarrow^{0\,\ast}   \LeftAssertParen   \openO{\{} \possiblyWithSub\stageOmetaColor{\nu}  \relO{:}  \possiblyWithSub\stageOmetaColor{B}  \relO{\mid}  \possiblyWithSub\stageOmetaColor{N^{\superscriptO} }_{{\mathrm{2}}} \closeO{\} }  \punctO{,}  \possiblyWithSub\stageOmetaColor{N^{\superscriptO} } \punctO{,}  \possiblyWithSub\stageOmetaColor{c}  \RightAssertParen^{ L }  \).
      \item
        If \( \possiblyWithSub\stageOmetaColor{N^{\superscriptO} }_{{\mathrm{1}}}  \Longleftrightarrow^{0\,\ast}  \possiblyWithSub\stageOmetaColor{N^{\superscriptO} }_{{\mathrm{2}}} \), then
        \(  \LeftAssertParen   \openO{\{} \possiblyWithSub\stageOmetaColor{\nu}  \relO{:}  \possiblyWithSub\stageOmetaColor{B}  \relO{\mid}  \possiblyWithSub\stageOmetaColor{N^{\superscriptO} } \closeO{\} }  \punctO{,}  \possiblyWithSub\stageOmetaColor{N^{\superscriptO} }_{{\mathrm{1}}} \punctO{,}  \possiblyWithSub\stageOmetaColor{c}  \RightAssertParen^{ L }   \Longleftrightarrow^{0\,\ast}   \LeftAssertParen   \openO{\{} \possiblyWithSub\stageOmetaColor{\nu}  \relO{:}  \possiblyWithSub\stageOmetaColor{B}  \relO{\mid}  \possiblyWithSub\stageOmetaColor{N^{\superscriptO} } \closeO{\} }  \punctO{,}  \possiblyWithSub\stageOmetaColor{N^{\superscriptO} }_{{\mathrm{2}}} \punctO{,}  \possiblyWithSub\stageOmetaColor{c}  \RightAssertParen^{ L }  \).
      \item
        If \( \possiblyWithSub\stageOmetaColor{N^{\superscriptO} }_{{\mathrm{1}}}  \Longleftrightarrow^{0\,\ast}  \possiblyWithSub\stageOmetaColor{N^{\superscriptO} }_{{\mathrm{2}}} \), then \(  \ordI{\sim} \possiblyWithSub\stageOmetaColor{N^{\superscriptO} }_{{\mathrm{1}}}   \Longleftrightarrow^{1\,\ast}   \ordI{\sim} \possiblyWithSub\stageOmetaColor{N^{\superscriptO} }_{{\mathrm{2}}}  \).
      \item
        If \( \possiblyWithSub\stageImetaColor{N^{\superscriptI} }_{{\mathrm{1}}}  \Longleftrightarrow^{1\,\ast}  \possiblyWithSub\stageImetaColor{N^{\superscriptI} }_{{\mathrm{2}}} \), then \(  \possiblyWithSub\stageImetaColor{N^{\superscriptI} }_{{\mathrm{1}}} \  \possiblyWithSub\stageImetaColor{N^{\superscriptI} }   \Longleftrightarrow^{1\,\ast}   \possiblyWithSub\stageImetaColor{N^{\superscriptI} }_{{\mathrm{2}}} \  \possiblyWithSub\stageImetaColor{N^{\superscriptI} }  \).
      \item
        If \( \possiblyWithSub\stageImetaColor{N^{\superscriptI} }_{{\mathrm{1}}}  \Longleftrightarrow^{1\,\ast}  \possiblyWithSub\stageImetaColor{N^{\superscriptI} }_{{\mathrm{2}}} \), then \(  \possiblyWithSub\stageImetaColor{N^{\superscriptI} } \  \possiblyWithSub\stageImetaColor{N^{\superscriptI} }_{{\mathrm{1}}}   \Longleftrightarrow^{1\,\ast}   \possiblyWithSub\stageImetaColor{N^{\superscriptI} } \  \possiblyWithSub\stageImetaColor{N^{\superscriptI} }_{{\mathrm{2}}}  \).
      \item
        If \( \possiblyWithSub\stageImetaColor{T^{\superscriptI} }_{{\mathrm{1}}}  \Longleftrightarrow^{1\,\ast}  \possiblyWithSub\stageImetaColor{T^{\superscriptI} }_{{\mathrm{2}}} \), then \(  \ordI{\lambda} \possiblyWithSub\stageImetaColor{x}  \relI{:}  \possiblyWithSub\stageImetaColor{T^{\superscriptI} }_{{\mathrm{1}}} \punctI{.}\  \possiblyWithSub\stageImetaColor{N^{\superscriptI} }   \Longleftrightarrow^{1\,\ast}   \ordI{\lambda} \possiblyWithSub\stageImetaColor{x}  \relI{:}  \possiblyWithSub\stageImetaColor{T^{\superscriptI} }_{{\mathrm{2}}} \punctI{.}\  \possiblyWithSub\stageImetaColor{N^{\superscriptI} }  \).
      \item
        If \( \possiblyWithSub\stageImetaColor{N^{\superscriptI} }_{{\mathrm{1}}}  \Longleftrightarrow^{1\,\ast}  \possiblyWithSub\stageImetaColor{N^{\superscriptI} }_{{\mathrm{2}}} \), then \(  \ordI{\lambda} \possiblyWithSub\stageImetaColor{x}  \relI{:}  \possiblyWithSub\stageImetaColor{T^{\superscriptI} } \punctI{.}\  \possiblyWithSub\stageImetaColor{N^{\superscriptI} }_{{\mathrm{1}}}   \Longleftrightarrow^{1\,\ast}   \ordI{\lambda} \possiblyWithSub\stageImetaColor{x}  \relI{:}  \possiblyWithSub\stageImetaColor{T^{\superscriptI} } \punctI{.}\  \possiblyWithSub\stageImetaColor{N^{\superscriptI} }_{{\mathrm{2}}}  \).
    \end{enumerate}
  \end{lemma}
  \begin{proof}
    All the items are by straightforward induction on the number of reduction steps,
    similarly to Lemma~\ref{lem:refl-trans-symm-par-propagates-type-structure}.
  \end{proof}
  \begin{lemma}[\(\beta\)-Equiv. is spanned by refl. trans. symm. closure of parallel reduction]\label{lem:equiv-implies-refl-trans-symm-par}
    \noindent
    \begin{enumerate}
      \item \( \possiblyWithSub\stageOmetaColor{N^{\superscriptO} }_{{\mathrm{1}}}  \cong^{0}  \possiblyWithSub\stageOmetaColor{N^{\superscriptO} }_{{\mathrm{2}}} \) implies \( \possiblyWithSub\stageOmetaColor{N^{\superscriptO} }_{{\mathrm{1}}}  \Longleftrightarrow^{0\,\ast}  \possiblyWithSub\stageOmetaColor{N^{\superscriptO} }_{{\mathrm{2}}} \).
      \item \( \possiblyWithSub\stageImetaColor{N^{\superscriptI} }_{{\mathrm{1}}}  \cong^{1}  \possiblyWithSub\stageImetaColor{N^{\superscriptI} }_{{\mathrm{2}}} \) implies \( \possiblyWithSub\stageImetaColor{N^{\superscriptI} }_{{\mathrm{1}}}  \Longleftrightarrow^{1\,\ast}  \possiblyWithSub\stageImetaColor{N^{\superscriptI} }_{{\mathrm{2}}} \).
      \item \( \possiblyWithSub\stageOmetaColor{T^{\superscriptO} }_{{\mathrm{1}}}  \cong^{0}  \possiblyWithSub\stageOmetaColor{T^{\superscriptO} }_{{\mathrm{2}}} \) implies \( \possiblyWithSub\stageOmetaColor{T^{\superscriptO} }_{{\mathrm{1}}}  \Longleftrightarrow^{0\,\ast}  \possiblyWithSub\stageOmetaColor{T^{\superscriptO} }_{{\mathrm{2}}} \).
      \item \( \possiblyWithSub\stageImetaColor{T^{\superscriptI} }_{{\mathrm{1}}}  \cong^{1}  \possiblyWithSub\stageImetaColor{T^{\superscriptI} }_{{\mathrm{2}}} \) implies \( \possiblyWithSub\stageImetaColor{T^{\superscriptI} }_{{\mathrm{1}}}  \Longleftrightarrow^{1\,\ast}  \possiblyWithSub\stageImetaColor{T^{\superscriptI} }_{{\mathrm{2}}} \).
    \end{enumerate}
  \end{lemma}
  \begin{proof}
    By mutual induction on the derivations.
    \begin{enumerate}
      \item
        \begin{itemize}
          \item Case~\rulename{Bq0-Refl} is immediate.
          \item Case~\rulename{Bq0-Sym} and \rulename{Bq0-Trans} are straightforward by IH.
          \item Case~\derive[Bq0-AssPass]{}{%
              \LeftAssertParen\openO{\langle}  \possiblyWithSub\stageImetaColor{\tau^{\superscriptI} }  \closeO{\rangle} \relO{\CastArrow} \openO{\langle}  \possiblyWithSub\stageImetaColor{\tau^{\superscriptI} }  \closeO{\rangle}\RightAssertParen^{ L }   \cong^{0}   \ordO{\lambda} \possiblyWithSub\stageOmetaColor{x}  \relO{:}   \openO{\langle}  \possiblyWithSub\stageImetaColor{\tau^{\superscriptI} }  \closeO{\rangle}  \punctO{.}\   \possiblyWithSub\stageOmetaColor{x}   
          }:
            Immediate by \rulename{P0-AssPass}.
          \item Case~\derive[Bq0-Beta]{}{%
               \openO{(}  \ordO{\lambda} \possiblyWithSub\stageOmetaColor{x}  \relO{:}  \possiblyWithSub\stageOmetaColor{T^{\superscriptO} } \punctO{.}\  \possiblyWithSub\stageOmetaColor{N^{\superscriptO} }_{{\mathrm{1}}}  \closeO{)}  \  \possiblyWithSub\stageOmetaColor{N^{\superscriptO} }_{{\mathrm{2}}}   \cong^{0}    [  \possiblyWithSub\stageOmetaColor{N^{\superscriptO} }_{{\mathrm{2}}}  /  \possiblyWithSub\stageOmetaColor{x}  ]    \possiblyWithSub\stageOmetaColor{N^{\superscriptO} }_{{\mathrm{1}}}  
          }:
            Since \( \possiblyWithSub\stageOmetaColor{N^{\superscriptO} }_{{\mathrm{1}}}  \Longrightarrow^{0}  \possiblyWithSub\stageOmetaColor{N^{\superscriptO} }_{{\mathrm{1}}} \) and \( \possiblyWithSub\stageOmetaColor{N^{\superscriptO} }_{{\mathrm{2}}}  \Longrightarrow^{0}  \possiblyWithSub\stageOmetaColor{N^{\superscriptO} }_{{\mathrm{2}}} \)
            hold by Lemma~\ref{lem:eq-implies-par},
            we can derive
            \begin{center}
              \derive[P0-Beta]{%
                 \possiblyWithSub\stageOmetaColor{N^{\superscriptO} }_{{\mathrm{1}}}  \Longrightarrow^{0}  \possiblyWithSub\stageOmetaColor{N^{\superscriptO} }_{{\mathrm{1}}} 
              \andalso
                 \possiblyWithSub\stageOmetaColor{N^{\superscriptO} }_{{\mathrm{2}}}  \Longrightarrow^{0}  \possiblyWithSub\stageOmetaColor{N^{\superscriptO} }_{{\mathrm{2}}} 
              }{%
                   \openO{(}  \ordO{\lambda} \possiblyWithSub\stageOmetaColor{x}  \relO{:}  \possiblyWithSub\stageOmetaColor{T^{\superscriptO} } \punctO{.}\  \possiblyWithSub\stageOmetaColor{N^{\superscriptO} }_{{\mathrm{1}}}  \closeO{)}  \  \possiblyWithSub\stageOmetaColor{N^{\superscriptO} }_{{\mathrm{2}}}   \Longrightarrow^{0}    [  \possiblyWithSub\stageOmetaColor{N^{\superscriptO} }_{{\mathrm{2}}}  /  \possiblyWithSub\stageOmetaColor{x}  ]    \possiblyWithSub\stageOmetaColor{N^{\superscriptO} }_{{\mathrm{1}}}  
              }.
            \end{center}
          \item Case~\derive[Bq0-Delta]{%
            \delta(  \possiblyWithSub\stageOmetaColor{a}_{{\mathrm{1}}}  \    \possiblyWithSub\stageOmetaColor{c}_{{\mathrm{2}}}   ) = \possiblyWithSub\stageOmetaColor{q}
          }{%
               \possiblyWithSub\stageOmetaColor{a}_{{\mathrm{1}}}  \    \possiblyWithSub\stageOmetaColor{c}_{{\mathrm{2}}}     \cong^{0}   \possiblyWithSub\stageOmetaColor{q}  
          }:
            We can immediately derive
            \begin{center}
              \derive[P0-Delta]{%
                \delta(  \possiblyWithSub\stageOmetaColor{a}_{{\mathrm{1}}}  \    \possiblyWithSub\stageOmetaColor{c}_{{\mathrm{2}}}   ) = \possiblyWithSub\stageOmetaColor{q}
              }{%
                   \possiblyWithSub\stageOmetaColor{a}_{{\mathrm{1}}}  \    \possiblyWithSub\stageOmetaColor{c}_{{\mathrm{2}}}     \Longrightarrow^{0}   \possiblyWithSub\stageOmetaColor{q}  
              }.
            \end{center}
          \item Case~\derive[Bq0-RfnStart]{}{%
               \LeftAssertParen \relO{\CastArrow}   \openO{\{} \possiblyWithSub\stageOmetaColor{\nu}  \relO{:}  \possiblyWithSub\stageOmetaColor{B}  \relO{\mid}  \possiblyWithSub\stageOmetaColor{N^{\superscriptO} } \closeO{\} }   \RightAssertParen^{ L }  \    \possiblyWithSub\stageOmetaColor{c}     \cong^{0}   \LeftAssertParen   \openO{\{} \possiblyWithSub\stageOmetaColor{\nu}  \relO{:}  \possiblyWithSub\stageOmetaColor{B}  \relO{\mid}  \possiblyWithSub\stageOmetaColor{N^{\superscriptO} } \closeO{\} }  \punctO{,}    [    \possiblyWithSub\stageOmetaColor{c}    /  \possiblyWithSub\stageOmetaColor{\nu}  ]    \possiblyWithSub\stageOmetaColor{N^{\superscriptO} }  \punctO{,}  \possiblyWithSub\stageOmetaColor{c}  \RightAssertParen^{ L }  
          }:
            Again, since \( \possiblyWithSub\stageOmetaColor{N^{\superscriptO} }  \Longrightarrow^{0}  \possiblyWithSub\stageOmetaColor{N^{\superscriptO} } \) holds by Lemma~\ref{lem:eq-implies-par},
            we can derive
            \begin{center}
              \derive[P0-RfnStart]{%
                 \possiblyWithSub\stageOmetaColor{N^{\superscriptO} }  \Longrightarrow^{0}  \possiblyWithSub\stageOmetaColor{N^{\superscriptO} } 
              }{%
                   \LeftAssertParen \relO{\CastArrow}   \openO{\{} \possiblyWithSub\stageOmetaColor{\nu}  \relO{:}  \possiblyWithSub\stageOmetaColor{B}  \relO{\mid}  \possiblyWithSub\stageOmetaColor{N^{\superscriptO} } \closeO{\} }   \RightAssertParen^{ L }  \    \possiblyWithSub\stageOmetaColor{c}_{{\mathrm{2}}}     \Longrightarrow^{0}   \LeftAssertParen   \openO{\{} \possiblyWithSub\stageOmetaColor{\nu}  \relO{:}  \possiblyWithSub\stageOmetaColor{B}  \relO{\mid}  \possiblyWithSub\stageOmetaColor{N^{\superscriptO} } \closeO{\} }  \punctO{,}    [    \possiblyWithSub\stageOmetaColor{c}_{{\mathrm{2}}}    /  \possiblyWithSub\stageOmetaColor{\nu}  ]    \possiblyWithSub\stageOmetaColor{N^{\superscriptO} }  \punctO{,}  \possiblyWithSub\stageOmetaColor{c}_{{\mathrm{2}}}  \RightAssertParen^{ L }  
              }.
            \end{center}
          \item Case~\derive[Bq0-Brkt]{%
             \possiblyWithSub\stageImetaColor{N^{\superscriptI} }_{{\mathrm{1}}}  \cong^{1}  \possiblyWithSub\stageImetaColor{N^{\superscriptI} }_{{\mathrm{2}}} 
          }{%
              \openO{\langle} \possiblyWithSub\stageImetaColor{N^{\superscriptI} }_{{\mathrm{1}}} \closeO{\rangle}   \cong^{0}   \openO{\langle} \possiblyWithSub\stageImetaColor{N^{\superscriptI} }_{{\mathrm{2}}} \closeO{\rangle}  
          }:
            Immediate by IH and Lemma~\ref{lem:refl-trans-symm-par-propagates-term-0-structure}~(1).
          \item Case~\derive[Bq0-Ass]{%
             \possiblyWithSub\stageImetaColor{T^{\superscriptI} }_{{\mathrm{11}}}  \cong^{1}  \possiblyWithSub\stageImetaColor{T^{\superscriptI} }_{{\mathrm{21}}} 
          \andalso
             \possiblyWithSub\stageImetaColor{T^{\superscriptI} }_{{\mathrm{12}}}  \cong^{1}  \possiblyWithSub\stageImetaColor{T^{\superscriptI} }_{{\mathrm{22}}} 
          }{%
              \LeftAssertParen\openO{\langle} \possiblyWithSub\stageImetaColor{T^{\superscriptI} }_{{\mathrm{11}}} \closeO{\rangle} \relO{\CastArrow} \openO{\langle} \possiblyWithSub\stageImetaColor{T^{\superscriptI} }_{{\mathrm{12}}} \closeO{\rangle}\RightAssertParen^{ L }   \cong^{0}   \LeftAssertParen\openO{\langle} \possiblyWithSub\stageImetaColor{T^{\superscriptI} }_{{\mathrm{21}}} \closeO{\rangle} \relO{\CastArrow} \openO{\langle} \possiblyWithSub\stageImetaColor{T^{\superscriptI} }_{{\mathrm{22}}} \closeO{\rangle}\RightAssertParen^{ L }  
          }:
            By IH and Lemmata~\ref{lem:eq-implies-par}
            and \ref{lem:refl-trans-symm-par-propagates-term-0-structure}~(2)--(3).
          \item Case~\derive[Bq0-Abs]{%
             \possiblyWithSub\stageOmetaColor{T^{\superscriptO} }_{{\mathrm{1}}}  \cong^{0}  \possiblyWithSub\stageOmetaColor{T^{\superscriptO} }_{{\mathrm{2}}} 
          \andalso
             \possiblyWithSub\stageOmetaColor{N^{\superscriptO} }_{{\mathrm{1}}}  \cong^{0}  \possiblyWithSub\stageOmetaColor{N^{\superscriptO} }_{{\mathrm{2}}} 
          }{%
              \ordO{\lambda} \possiblyWithSub\stageOmetaColor{x}  \relO{:}  \possiblyWithSub\stageOmetaColor{T^{\superscriptO} }_{{\mathrm{1}}} \punctO{.}\  \possiblyWithSub\stageOmetaColor{N^{\superscriptO} }_{{\mathrm{1}}}   \cong^{0}   \ordO{\lambda} \possiblyWithSub\stageOmetaColor{x}  \relO{:}  \possiblyWithSub\stageOmetaColor{T^{\superscriptO} }_{{\mathrm{2}}} \punctO{.}\  \possiblyWithSub\stageOmetaColor{N^{\superscriptO} }_{{\mathrm{2}}}  
          }:
            By IH and Lemmata~\ref{lem:eq-implies-par}
            and \ref{lem:refl-trans-symm-par-propagates-term-0-structure}~(4)--(5).
          \item Case~\derive[Bq0-App]{%
             \possiblyWithSub\stageOmetaColor{N^{\superscriptO} }_{{\mathrm{11}}}  \cong^{0}  \possiblyWithSub\stageOmetaColor{N^{\superscriptO} }_{{\mathrm{21}}} 
          \andalso
             \possiblyWithSub\stageOmetaColor{N^{\superscriptO} }_{{\mathrm{12}}}  \cong^{0}  \possiblyWithSub\stageOmetaColor{N^{\superscriptO} }_{{\mathrm{22}}} 
          }{%
              \possiblyWithSub\stageOmetaColor{N^{\superscriptO} }_{{\mathrm{11}}} \  \possiblyWithSub\stageOmetaColor{N^{\superscriptO} }_{{\mathrm{12}}}   \cong^{0}   \possiblyWithSub\stageOmetaColor{N^{\superscriptO} }_{{\mathrm{21}}} \  \possiblyWithSub\stageOmetaColor{N^{\superscriptO} }_{{\mathrm{22}}}  
          }:
            By IH and Lemmata~\ref{lem:eq-implies-par}
            and \ref{lem:refl-trans-symm-par-propagates-term-0-structure}~(6)--(7).
          \item Case~\derive[Bq0-Rfn]{%
             \possiblyWithSub\stageOmetaColor{N^{\superscriptO} }_{{\mathrm{1}}}  \cong^{0}  \possiblyWithSub\stageOmetaColor{N^{\superscriptO} }_{{\mathrm{2}}} 
          }{%
              \LeftAssertParen \relO{\CastArrow}   \openO{\{} \possiblyWithSub\stageOmetaColor{\nu}  \relO{:}  \possiblyWithSub\stageOmetaColor{B}  \relO{\mid}  \possiblyWithSub\stageOmetaColor{N^{\superscriptO} }_{{\mathrm{1}}} \closeO{\} }   \RightAssertParen^{ L }   \cong^{0}   \LeftAssertParen \relO{\CastArrow}   \openO{\{} \possiblyWithSub\stageOmetaColor{\nu}  \relO{:}  \possiblyWithSub\stageOmetaColor{B}  \relO{\mid}  \possiblyWithSub\stageOmetaColor{N^{\superscriptO} }_{{\mathrm{2}}} \closeO{\} }   \RightAssertParen^{ L }  
          }:
            Immediate by IH and Lemma~\ref{lem:refl-trans-symm-par-propagates-term-0-structure}~(8).
          \item Case~\derive[Bq0-RfnAct]{%
             \possiblyWithSub\stageOmetaColor{N^{\superscriptO} }_{{\mathrm{11}}}  \cong^{0}  \possiblyWithSub\stageOmetaColor{N^{\superscriptO} }_{{\mathrm{21}}} 
          \andalso
             \possiblyWithSub\stageOmetaColor{N^{\superscriptO} }_{{\mathrm{12}}}  \cong^{0}  \possiblyWithSub\stageOmetaColor{N^{\superscriptO} }_{{\mathrm{22}}} 
          }{%
              \LeftAssertParen   \openO{\{} \possiblyWithSub\stageOmetaColor{\nu}  \relO{:}  \possiblyWithSub\stageOmetaColor{B}  \relO{\mid}  \possiblyWithSub\stageOmetaColor{N^{\superscriptO} }_{{\mathrm{11}}} \closeO{\} }  \punctO{,}  \possiblyWithSub\stageOmetaColor{N^{\superscriptO} }_{{\mathrm{12}}} \punctO{,}  \possiblyWithSub\stageOmetaColor{c}  \RightAssertParen^{ L }   \cong^{0}   \LeftAssertParen   \openO{\{} \possiblyWithSub\stageOmetaColor{\nu}  \relO{:}  \possiblyWithSub\stageOmetaColor{B}  \relO{\mid}  \possiblyWithSub\stageOmetaColor{N^{\superscriptO} }_{{\mathrm{21}}} \closeO{\} }  \punctO{,}  \possiblyWithSub\stageOmetaColor{N^{\superscriptO} }_{{\mathrm{22}}} \punctO{,}  \possiblyWithSub\stageOmetaColor{c}  \RightAssertParen^{ L }  
          }:
            By IH and Lemmata~\ref{lem:eq-implies-par}
            and \ref{lem:refl-trans-symm-par-propagates-term-0-structure}~(9)--(10).
        \end{itemize}
      \item
        \begin{itemize}
          \item Case~\rulename{Bq1-Refl} is immediate.
          \item Cases~\rulename{Bq1-Sym} and \rulename{Bq1-Trans} are straightforward by IH.
          \item Case~\derive[Bq1-Cancel]{}{%
              \ordI{\sim}  \openO{\langle} \possiblyWithSub\stageImetaColor{N^{\superscriptI} } \closeO{\rangle}    \cong^{1}  \possiblyWithSub\stageImetaColor{N^{\superscriptI} } 
          }:
            Since \( \possiblyWithSub\stageImetaColor{N^{\superscriptI} }  \Longrightarrow^{1}  \possiblyWithSub\stageImetaColor{N^{\superscriptI} } \) holds by Lemma~\ref{lem:eq-implies-par},
            we can derive
            \begin{center}
              \derive[P1-Cancel]{%
                 \possiblyWithSub\stageImetaColor{N^{\superscriptI} }  \Longrightarrow^{1}  \possiblyWithSub\stageImetaColor{N^{\superscriptI} } 
              }{%
                  \ordI{\sim}  \openO{\langle} \possiblyWithSub\stageImetaColor{N^{\superscriptI} } \closeO{\rangle}    \Longrightarrow^{1}  \possiblyWithSub\stageImetaColor{N^{\superscriptI} } 
              }.
            \end{center}
          \item Case~\derive[Bq1-Esc]{%
             \possiblyWithSub\stageOmetaColor{N^{\superscriptO} }_{{\mathrm{1}}}  \cong^{0}  \possiblyWithSub\stageOmetaColor{N^{\superscriptO} }_{{\mathrm{2}}} 
          }{%
              \ordI{\sim} \possiblyWithSub\stageOmetaColor{N^{\superscriptO} }_{{\mathrm{1}}}   \cong^{1}   \ordI{\sim} \possiblyWithSub\stageOmetaColor{N^{\superscriptO} }_{{\mathrm{2}}}  
          }:
            Immediate by IH and Lemma~\ref{lem:refl-trans-symm-par-propagates-term-0-structure}~(11).
          \item Case~\derive[Bq1-App]{%
             \possiblyWithSub\stageImetaColor{N^{\superscriptI} }_{{\mathrm{11}}}  \cong^{1}  \possiblyWithSub\stageImetaColor{N^{\superscriptI} }_{{\mathrm{21}}} 
          \andalso
             \possiblyWithSub\stageImetaColor{N^{\superscriptI} }_{{\mathrm{12}}}  \cong^{1}  \possiblyWithSub\stageImetaColor{N^{\superscriptI} }_{{\mathrm{22}}} 
          }{%
              \possiblyWithSub\stageImetaColor{N^{\superscriptI} }_{{\mathrm{11}}} \  \possiblyWithSub\stageImetaColor{N^{\superscriptI} }_{{\mathrm{12}}}   \cong^{1}   \possiblyWithSub\stageImetaColor{N^{\superscriptI} }_{{\mathrm{21}}} \  \possiblyWithSub\stageImetaColor{N^{\superscriptI} }_{{\mathrm{22}}}  
          }:
            By IH and Lemmata~\ref{lem:eq-implies-par}
            and \ref{lem:refl-trans-symm-par-propagates-term-0-structure}~(12)--(13).
          \item Case~\derive[Bq1-Abs]{%
             \possiblyWithSub\stageImetaColor{T^{\superscriptI} }_{{\mathrm{11}}}  \cong^{1}  \possiblyWithSub\stageImetaColor{T^{\superscriptI} }_{{\mathrm{21}}} 
          \andalso
             \possiblyWithSub\stageImetaColor{N^{\superscriptI} }_{{\mathrm{12}}}  \cong^{1}  \possiblyWithSub\stageImetaColor{N^{\superscriptI} }_{{\mathrm{22}}} 
          }{%
              \ordI{\lambda} \possiblyWithSub\stageImetaColor{x}  \relI{:}  \possiblyWithSub\stageImetaColor{T^{\superscriptI} }_{{\mathrm{11}}} \punctI{.}\  \possiblyWithSub\stageImetaColor{N^{\superscriptI} }_{{\mathrm{12}}}   \cong^{1}   \ordI{\lambda} \possiblyWithSub\stageImetaColor{x}  \relI{:}  \possiblyWithSub\stageImetaColor{T^{\superscriptI} }_{{\mathrm{21}}} \punctI{.}\  \possiblyWithSub\stageImetaColor{N^{\superscriptI} }_{{\mathrm{22}}}  
          }:
            By IH and Lemmata~\ref{lem:eq-implies-par}
            and \ref{lem:refl-trans-symm-par-propagates-term-0-structure}~(14)--(15).
        \end{itemize}
      \item
        \begin{itemize}
          \item Case~\rulename{BqT0-Refl} is immediate.
          \item Cases~\rulename{BqT0-Sym} and \rulename{BqT0-Trans} are straightforward by IH.
          \item Case \derive[BqT0-Code]{%
             \possiblyWithSub\stageImetaColor{T^{\superscriptI} }_{{\mathrm{1}}}  \cong^{1}  \possiblyWithSub\stageImetaColor{T^{\superscriptI} }_{{\mathrm{2}}} 
          }{%
              \openO{\langle} \possiblyWithSub\stageImetaColor{T^{\superscriptI} }_{{\mathrm{1}}} \closeO{\rangle}   \cong^{0}   \openO{\langle} \possiblyWithSub\stageImetaColor{T^{\superscriptI} }_{{\mathrm{2}}} \closeO{\rangle}  
          }:
            By IH, we have \( \possiblyWithSub\stageImetaColor{T^{\superscriptI} }_{{\mathrm{1}}}  \Longleftrightarrow^{1\,\ast}  \possiblyWithSub\stageImetaColor{T^{\superscriptI} }_{{\mathrm{2}}} \).
            Therefore, by Lemma~\ref{lem:refl-trans-symm-par-propagates-type-structure}~(1),
            we have \(  \openO{\langle} \possiblyWithSub\stageImetaColor{T^{\superscriptI} }_{{\mathrm{1}}} \closeO{\rangle}   \Longleftrightarrow^{0\,\ast}   \openO{\langle} \possiblyWithSub\stageImetaColor{T^{\superscriptI} }_{{\mathrm{2}}} \closeO{\rangle}  \).
          \item Case \derive[BqT0-Arr]{%
             \possiblyWithSub\stageOmetaColor{T^{\superscriptO} }_{{\mathrm{11}}}  \cong^{0}  \possiblyWithSub\stageOmetaColor{T^{\superscriptO} }_{{\mathrm{21}}} 
          \andalso
             \possiblyWithSub\stageOmetaColor{T^{\superscriptO} }_{{\mathrm{12}}}  \cong^{0}  \possiblyWithSub\stageOmetaColor{T^{\superscriptO} }_{{\mathrm{22}}} 
          }{%
              \openO{(} \possiblyWithSub\stageOmetaColor{x}  \relO{:}  \possiblyWithSub\stageOmetaColor{T^{\superscriptO} }_{{\mathrm{11}}} \closeO{)} \relO{\to}  \possiblyWithSub\stageOmetaColor{T^{\superscriptO} }_{{\mathrm{12}}}   \cong^{0}   \openO{(} \possiblyWithSub\stageOmetaColor{x}  \relO{:}  \possiblyWithSub\stageOmetaColor{T^{\superscriptO} }_{{\mathrm{21}}} \closeO{)} \relO{\to}  \possiblyWithSub\stageOmetaColor{T^{\superscriptO} }_{{\mathrm{22}}}  
          }:
            By IH, we have \( \possiblyWithSub\stageOmetaColor{T^{\superscriptO} }_{{\mathrm{11}}}  \Longleftrightarrow^{0\,\ast}  \possiblyWithSub\stageOmetaColor{T^{\superscriptO} }_{{\mathrm{21}}} \) and \( \possiblyWithSub\stageOmetaColor{T^{\superscriptO} }_{{\mathrm{12}}}  \Longleftrightarrow^{0\,\ast}  \possiblyWithSub\stageOmetaColor{T^{\superscriptO} }_{{\mathrm{22}}} \).
            Thus, by Lemmata~\ref{lem:eq-implies-par}
            and~\ref{lem:refl-trans-symm-par-propagates-type-structure}~(2)--(3),
            we have
            \(  \openO{(} \possiblyWithSub\stageOmetaColor{x}  \relO{:}  \possiblyWithSub\stageOmetaColor{T^{\superscriptO} }_{{\mathrm{11}}} \closeO{)} \relO{\to}  \possiblyWithSub\stageOmetaColor{T^{\superscriptO} }_{{\mathrm{12}}}   \Longleftrightarrow^{0\,\ast}   \openO{(} \possiblyWithSub\stageOmetaColor{x}  \relO{:}  \possiblyWithSub\stageOmetaColor{T^{\superscriptO} }_{{\mathrm{21}}} \closeO{)} \relO{\to}  \possiblyWithSub\stageOmetaColor{T^{\superscriptO} }_{{\mathrm{12}}}  
              \Longleftrightarrow^{\ast\,0}  \openO{(} \possiblyWithSub\stageOmetaColor{x}  \relO{:}  \possiblyWithSub\stageOmetaColor{T^{\superscriptO} }_{{\mathrm{21}}} \closeO{)} \relO{\to}  \possiblyWithSub\stageOmetaColor{T^{\superscriptO} }_{{\mathrm{22}}} \).
          \item Case \derive[BqT0-Rfn]{%
             \possiblyWithSub\stageOmetaColor{N^{\superscriptO} }_{{\mathrm{1}}}  \cong^{0}  \possiblyWithSub\stageOmetaColor{N^{\superscriptO} }_{{\mathrm{2}}} 
          }{%
               \openO{\{} \possiblyWithSub\stageOmetaColor{\nu}  \relO{:}  \possiblyWithSub\stageOmetaColor{B}  \relO{\mid}  \possiblyWithSub\stageOmetaColor{N^{\superscriptO} }_{{\mathrm{1}}} \closeO{\} }    \cong^{0}    \openO{\{} \possiblyWithSub\stageOmetaColor{\nu}  \relO{:}  \possiblyWithSub\stageOmetaColor{B}  \relO{\mid}  \possiblyWithSub\stageOmetaColor{N^{\superscriptO} }_{{\mathrm{2}}} \closeO{\} }   
          }:
            By IH, we have \( \possiblyWithSub\stageOmetaColor{N^{\superscriptO} }_{{\mathrm{1}}}  \Longleftrightarrow^{0\,\ast}  \possiblyWithSub\stageOmetaColor{N^{\superscriptO} }_{{\mathrm{2}}} \).
            Thus, by Lemma~\ref{lem:refl-trans-symm-par-propagates-type-structure}~(4), we have
            \(   \openO{\{} \possiblyWithSub\stageOmetaColor{\nu}  \relO{:}  \possiblyWithSub\stageOmetaColor{B}  \relO{\mid}  \possiblyWithSub\stageOmetaColor{N^{\superscriptO} }_{{\mathrm{1}}} \closeO{\} }    \Longleftrightarrow^{0\,\ast}    \openO{\{} \possiblyWithSub\stageOmetaColor{\nu}  \relO{:}  \possiblyWithSub\stageOmetaColor{B}  \relO{\mid}  \possiblyWithSub\stageOmetaColor{N^{\superscriptO} }_{{\mathrm{2}}} \closeO{\} }   \).
        \end{itemize}
      \item
        \begin{itemize}
          \item Case~\rulename{BqT1-Refl} is immediate.
          \item Cases~\rulename{BqT1-Sym} and \rulename{BqT1-Trans} are straightforward by IH.
          \item Case \derive[BqT1-Arr]{%
             \possiblyWithSub\stageImetaColor{T^{\superscriptI} }_{{\mathrm{11}}}  \cong^{1}  \possiblyWithSub\stageImetaColor{T^{\superscriptI} }_{{\mathrm{21}}} 
          \andalso
             \possiblyWithSub\stageImetaColor{T^{\superscriptI} }_{{\mathrm{12}}}  \cong^{1}  \possiblyWithSub\stageImetaColor{T^{\superscriptI} }_{{\mathrm{22}}} 
          }{%
              \possiblyWithSub\stageImetaColor{T^{\superscriptI} }_{{\mathrm{11}}}  \relI{\to}  \possiblyWithSub\stageImetaColor{T^{\superscriptI} }_{{\mathrm{12}}}   \cong^{1}   \possiblyWithSub\stageImetaColor{T^{\superscriptI} }_{{\mathrm{21}}}  \relI{\to}  \possiblyWithSub\stageImetaColor{T^{\superscriptI} }_{{\mathrm{22}}}  
          }:
            By IH and Lemmata~\ref{lem:eq-implies-par}
            and \ref{lem:refl-trans-symm-par-propagates-type-structure}~(5)--(6).
          \item Case \derive[BqT1-Tensor]{%
             \possiblyWithSub\stageOmetaColor{N^{\superscriptO} }_{{\mathrm{1}}}  \cong^{0}  \possiblyWithSub\stageOmetaColor{N^{\superscriptO} }_{{\mathrm{2}}} 
          }{%
              \ttI{Tensor}\ \ordI{\%} \possiblyWithSub\stageOmetaColor{N^{\superscriptO} }_{{\mathrm{1}}}   \cong^{1}   \ttI{Tensor}\ \ordI{\%} \possiblyWithSub\stageOmetaColor{N^{\superscriptO} }_{{\mathrm{2}}}  
          }:
            Immediate by IH and Lemma~\ref{lem:refl-trans-symm-par-propagates-type-structure}~(7).
        \end{itemize}
    \end{enumerate}
  \end{proof}
  \begin{corollary}\label{cor:equivalence-of-equiv-and-refl-trans-symm-par}
    \( \possiblyWithSub\stageImetaColor{T^{\superscriptI} }_{{\mathrm{1}}}  \cong^{1}  \possiblyWithSub\stageImetaColor{T^{\superscriptI} }_{{\mathrm{2}}} \) if and only if \( \possiblyWithSub\stageImetaColor{T^{\superscriptI} }_{{\mathrm{1}}}  \Longleftrightarrow^{1\,\ast}  \possiblyWithSub\stageImetaColor{T^{\superscriptI} }_{{\mathrm{2}}} \).
  \end{corollary}
  \begin{proof}
    Straightforward from
    Lemmata~\ref{lem:par-implies-equiv} and \ref{lem:equiv-implies-refl-trans-symm-par}.
  \end{proof}
\begin{figure}[tbp]
  \begin{gather*}
     (  \possiblyWithSub\stageImetaColor{B}  )^{\ast}  := \possiblyWithSub\stageImetaColor{B}
  \qquad
     (  \ttI{Tensor}\ \ordI{\%} \possiblyWithSub\stageOmetaColor{N^{\superscriptO} }  )^{\ast}  :=  \ttI{Tensor}\ \ordI{\%}  ( \possiblyWithSub\stageOmetaColor{N^{\superscriptO} } )^{\ast}  
  \qquad
     (  \possiblyWithSub\stageImetaColor{T^{\superscriptI} }_{{\mathrm{1}}}  \relI{\to}  \possiblyWithSub\stageImetaColor{T^{\superscriptI} }_{{\mathrm{2}}}  )^{\ast}  :=   ( \possiblyWithSub\stageImetaColor{T^{\superscriptI} }_{{\mathrm{1}}} )^{\ast}   \relI{\to}   ( \possiblyWithSub\stageImetaColor{T^{\superscriptI} }_{{\mathrm{2}}} )^{\ast}  
  \\
     (   \openO{\{} \possiblyWithSub\stageOmetaColor{\nu}  \relO{:}  \possiblyWithSub\stageOmetaColor{B}  \relO{\mid}  \possiblyWithSub\stageOmetaColor{N^{\superscriptO} } \closeO{\} }   )^{\ast}  :=  \openO{\{} \possiblyWithSub\stageOmetaColor{\nu}  \relO{:}  \possiblyWithSub\stageOmetaColor{B}  \relO{\mid}   ( \possiblyWithSub\stageOmetaColor{N^{\superscriptO} } )^{\ast}  \closeO{\} } 
  \qquad
     (  \ttO{Tensor}\  \possiblyWithSub\stageOmetaColor{s}  )^{\ast}  :=  \ttO{Tensor}\  \possiblyWithSub\stageOmetaColor{s} 
  \\
     (  \openO{(} \possiblyWithSub\stageOmetaColor{x}  \relO{:}  \possiblyWithSub\stageOmetaColor{T^{\superscriptO} }_{{\mathrm{1}}} \closeO{)} \relO{\to}  \possiblyWithSub\stageOmetaColor{T^{\superscriptO} }_{{\mathrm{2}}}  )^{\ast}  :=  \openO{(} \possiblyWithSub\stageOmetaColor{x}  \relO{:}   ( \possiblyWithSub\stageOmetaColor{T^{\superscriptO} }_{{\mathrm{1}}} )^{\ast}  \closeO{)} \relO{\to}   ( \possiblyWithSub\stageOmetaColor{T^{\superscriptO} }_{{\mathrm{2}}} )^{\ast}  
  \qquad
     (  \openO{\langle} \possiblyWithSub\stageImetaColor{T^{\superscriptI} } \closeO{\rangle}  )^{\ast}  :=  \openO{\langle}  ( \possiblyWithSub\stageImetaColor{T^{\superscriptI} } )^{\ast}  \closeO{\rangle} 
  \\
     (   \possiblyWithSub\stageOmetaColor{p}   )^{\ast}  := \possiblyWithSub\stageOmetaColor{p}
  \qquad
     (   \possiblyWithSub\stageOmetaColor{c}   )^{\ast}  := \possiblyWithSub\stageOmetaColor{c}
  \qquad
     (  \possiblyWithSub\stageOmetaColor{x}  )^{\ast}  := \possiblyWithSub\stageOmetaColor{x}
  \qquad
     (  \ordO{\lambda} \possiblyWithSub\stageOmetaColor{x}  \relO{:}  \possiblyWithSub\stageOmetaColor{T^{\superscriptO} } \punctO{.}\  \possiblyWithSub\stageOmetaColor{N^{\superscriptO} }  )^{\ast}  :=  \ordO{\lambda} \possiblyWithSub\stageOmetaColor{x}  \relO{:}   ( \possiblyWithSub\stageOmetaColor{T^{\superscriptO} } )^{\ast}  \punctO{.}\   ( \possiblyWithSub\stageOmetaColor{N^{\superscriptO} } )^{\ast}  
  \\
     (  \possiblyWithSub\stageOmetaColor{N^{\superscriptO} }_{{\mathrm{1}}} \  \possiblyWithSub\stageOmetaColor{N^{\superscriptO} }_{{\mathrm{2}}}  )^{\ast}  :=
      \begin{cases}
          [   ( \possiblyWithSub\stageOmetaColor{N^{\superscriptO} }_{{\mathrm{2}}} )^{\ast}   /  \possiblyWithSub\stageOmetaColor{x}  ]     ( \possiblyWithSub\stageOmetaColor{N^{\superscriptO} } )^{\ast}  
          &\text{(if \(\possiblyWithSub\stageOmetaColor{N^{\superscriptO} }_{{\mathrm{1}}} =  \ordO{\lambda} \possiblyWithSub\stageOmetaColor{x}  \relO{:}  \possiblyWithSub\stageOmetaColor{T^{\superscriptO} } \punctO{.}\  \possiblyWithSub\stageOmetaColor{N^{\superscriptO} } \))}
      \\
         \LeftAssertParen   \openO{\{} \possiblyWithSub\stageOmetaColor{\nu}  \relO{:}  \possiblyWithSub\stageOmetaColor{B}  \relO{\mid}   ( \possiblyWithSub\stageOmetaColor{N^{\superscriptO} }_{{\mathrm{11}}} )^{\ast}  \closeO{\} }  \punctO{,}    [    \possiblyWithSub\stageOmetaColor{c}_{{\mathrm{2}}}    /  \possiblyWithSub\stageOmetaColor{\nu}  ]     ( \possiblyWithSub\stageOmetaColor{N^{\superscriptO} }_{{\mathrm{11}}} )^{\ast}   \punctO{,}  \possiblyWithSub\stageOmetaColor{c}_{{\mathrm{2}}}  \RightAssertParen^{ L } 
          &\text{(if \(\possiblyWithSub\stageOmetaColor{N^{\superscriptO} }_{{\mathrm{1}}} =  \LeftAssertParen \relO{\CastArrow}   \openO{\{} \possiblyWithSub\stageOmetaColor{\nu}  \relO{:}  \possiblyWithSub\stageOmetaColor{B}  \relO{\mid}  \possiblyWithSub\stageOmetaColor{N^{\superscriptO} }_{{\mathrm{11}}} \closeO{\} }   \RightAssertParen^{ L } \) and \(\possiblyWithSub\stageOmetaColor{N^{\superscriptO} }_{{\mathrm{2}}} = \possiblyWithSub\stageOmetaColor{c}_{{\mathrm{2}}}\))}
      \\
        \possiblyWithSub\stageOmetaColor{q}
          &\text{(if \(\possiblyWithSub\stageOmetaColor{N^{\superscriptO} }_{{\mathrm{1}}} = \possiblyWithSub\stageOmetaColor{a}_{{\mathrm{1}}}\), \(\possiblyWithSub\stageOmetaColor{N^{\superscriptO} }_{{\mathrm{2}}} = \possiblyWithSub\stageOmetaColor{c}_{{\mathrm{2}}}\), and \(\delta(  \possiblyWithSub\stageOmetaColor{a}_{{\mathrm{1}}}  \    \possiblyWithSub\stageOmetaColor{c}_{{\mathrm{2}}}   ) = \possiblyWithSub\stageOmetaColor{q}\))}
      \\
          ( \possiblyWithSub\stageOmetaColor{N^{\superscriptO} }_{{\mathrm{1}}} )^{\ast}  \   ( \possiblyWithSub\stageOmetaColor{N^{\superscriptO} }_{{\mathrm{2}}} )^{\ast}  
          &\text{(otherwise)}
      \end{cases}
  \\
     (  \LeftAssertParen\openO{\langle} \possiblyWithSub\stageImetaColor{T^{\superscriptI} }_{{\mathrm{1}}} \closeO{\rangle} \relO{\CastArrow} \openO{\langle} \possiblyWithSub\stageImetaColor{T^{\superscriptI} }_{{\mathrm{2}}} \closeO{\rangle}\RightAssertParen^{ L }  )^{\ast}  :=
      \begin{cases}
         \ordO{\lambda} \possiblyWithSub\stageOmetaColor{x}  \relO{:}   \openO{\langle}  \possiblyWithSub\stageImetaColor{\tau^{\superscriptI} }  \closeO{\rangle}  \punctO{.}\   \possiblyWithSub\stageOmetaColor{x}  
          &\text{(if \(\possiblyWithSub\stageImetaColor{T^{\superscriptI} }_{{\mathrm{1}}}\) and \(\possiblyWithSub\stageImetaColor{T^{\superscriptI} }_{{\mathrm{2}}}\) are a common value~\(\possiblyWithSub\stageImetaColor{\tau^{\superscriptI} }\))}
      \\
         \LeftAssertParen\openO{\langle}  ( \possiblyWithSub\stageImetaColor{T^{\superscriptI} }_{{\mathrm{1}}} )^{\ast}  \closeO{\rangle} \relO{\CastArrow} \openO{\langle}  ( \possiblyWithSub\stageImetaColor{T^{\superscriptI} }_{{\mathrm{2}}} )^{\ast}  \closeO{\rangle}\RightAssertParen^{ L } 
          &\text{(otherwise)}
      \end{cases}
  \\
     (  \openO{\langle} \possiblyWithSub\stageImetaColor{N^{\superscriptI} } \closeO{\rangle}  )^{\ast}  :=  \openO{\langle}  ( \possiblyWithSub\stageImetaColor{N^{\superscriptI} } )^{\ast}  \closeO{\rangle} 
  \qquad
     (  \LeftAssertParen \relO{\CastArrow}   \openO{\{} \possiblyWithSub\stageOmetaColor{\nu}  \relO{:}  \possiblyWithSub\stageOmetaColor{B}  \relO{\mid}  \possiblyWithSub\stageOmetaColor{N^{\superscriptO} } \closeO{\} }   \RightAssertParen^{ L }  )^{\ast}  :=
       \LeftAssertParen \relO{\CastArrow}   \openO{\{} \possiblyWithSub\stageOmetaColor{\nu}  \relO{:}  \possiblyWithSub\stageOmetaColor{B}  \relO{\mid}   ( \possiblyWithSub\stageOmetaColor{N^{\superscriptO} } )^{\ast}  \closeO{\} }   \RightAssertParen^{ L } 
  \\
     (  \LeftAssertParen   \openO{\{} \possiblyWithSub\stageOmetaColor{\nu}  \relO{:}  \possiblyWithSub\stageOmetaColor{B}  \relO{\mid}  \possiblyWithSub\stageOmetaColor{N^{\superscriptO} } \closeO{\} }  \punctO{,}  \possiblyWithSub\stageOmetaColor{N'^{\superscriptO} } \punctO{,}  \possiblyWithSub\stageOmetaColor{c}  \RightAssertParen^{ L }  )^{\ast}  :=
      \begin{cases}
        \possiblyWithSub\stageOmetaColor{c}
          &\text{(if \(\possiblyWithSub\stageOmetaColor{N'^{\superscriptO} } =   \ttO{true}  \))}
      \\
         \LeftAssertParen   \openO{\{} \possiblyWithSub\stageOmetaColor{\nu}  \relO{:}  \possiblyWithSub\stageOmetaColor{B}  \relO{\mid}   ( \possiblyWithSub\stageOmetaColor{N^{\superscriptO} } )^{\ast}  \closeO{\} }  \punctO{,}   ( \possiblyWithSub\stageOmetaColor{N'^{\superscriptO} } )^{\ast}  \punctO{,}  \possiblyWithSub\stageOmetaColor{c}  \RightAssertParen^{ L } 
          &\text{(otherwise)}
      \end{cases}
  \\
     (  \possiblyWithSub\stageImetaColor{c}  )^{\ast}  := \possiblyWithSub\stageImetaColor{c}
  \qquad
     (  \possiblyWithSub\stageImetaColor{x}  )^{\ast}  := \possiblyWithSub\stageImetaColor{x}
  \qquad
     (  \ordI{\lambda} \possiblyWithSub\stageImetaColor{x}  \relI{:}  \possiblyWithSub\stageImetaColor{T^{\superscriptI} } \punctI{.}\  \possiblyWithSub\stageImetaColor{N^{\superscriptI} }  )^{\ast}  :=  \ordI{\lambda} \possiblyWithSub\stageImetaColor{x}  \relI{:}   ( \possiblyWithSub\stageImetaColor{T^{\superscriptI} } )^{\ast}  \punctI{.}\   ( \possiblyWithSub\stageImetaColor{N^{\superscriptI} } )^{\ast}  
  \\
     (  \possiblyWithSub\stageImetaColor{N^{\superscriptI} }_{{\mathrm{1}}} \  \possiblyWithSub\stageImetaColor{N^{\superscriptI} }_{{\mathrm{2}}}  )^{\ast}  :=   ( \possiblyWithSub\stageImetaColor{N^{\superscriptI} }_{{\mathrm{1}}} )^{\ast}  \   ( \possiblyWithSub\stageImetaColor{N^{\superscriptI} }_{{\mathrm{2}}} )^{\ast}  
  \qquad
     (  \ordI{\sim} \possiblyWithSub\stageOmetaColor{N^{\superscriptO} }  )^{\ast}  :=
      \begin{cases}
         ( \possiblyWithSub\stageImetaColor{N^{\superscriptI} } )^{\ast} 
          &\text{(if \(\possiblyWithSub\stageOmetaColor{N^{\superscriptO} } =  \openO{\langle} \possiblyWithSub\stageImetaColor{N^{\superscriptI} } \closeO{\rangle} \))}
      \\
         \ordI{\sim}  ( \possiblyWithSub\stageOmetaColor{N^{\superscriptO} } )^{\ast}  
          &\text{(otherwise)}
      \end{cases}
  \end{gather*}
  \caption{Conversion \`{a} la Takahashi~\cite{Takahashi1995}}
  \label{fig:star-conversion}
\end{figure}
  \begin{lemma}\label{lem:par-const-normal-form}
    \(   \possiblyWithSub\stageOmetaColor{c}    \Longrightarrow^{0}  \possiblyWithSub\stageOmetaColor{N'^{\superscriptO} } \) implies \(\possiblyWithSub\stageOmetaColor{N'^{\superscriptO} } = \possiblyWithSub\stageOmetaColor{c}\).
  \end{lemma}
  \begin{proof}
    Straightforward from the definition of \(\Longrightarrow^0\).
  \end{proof}
  \begin{lemma}\label{lem:par-prim-partial-app-normal-form}
    For any value of the form~\(\possiblyWithSub\stageOmetaColor{a}\),
    \(  \possiblyWithSub\stageOmetaColor{a}   \Longrightarrow^{0}  \possiblyWithSub\stageOmetaColor{N'^{\superscriptO} } \) implies \(\possiblyWithSub\stageOmetaColor{N'^{\superscriptO} } = \possiblyWithSub\stageOmetaColor{a}\).
  \end{lemma}
  \begin{proof}
    By straightforward induction on the number of arguments of \(\possiblyWithSub\stageOmetaColor{a}\).
  \end{proof}
  \begin{lemma}\label{lem:par-type-normal-form}
    \(  \possiblyWithSub\stageImetaColor{\tau^{\superscriptI} }   \Longrightarrow^{1}  \possiblyWithSub\stageImetaColor{T^{\superscriptI} } \) implies \(\possiblyWithSub\stageImetaColor{T^{\superscriptI} } = \possiblyWithSub\stageImetaColor{\tau^{\superscriptI} }\).
  \end{lemma}
  \begin{proof}
    By straightforward induction using Lemma~\ref{lem:par-const-normal-form}.
  \end{proof}
  \begin{lemma}\label{lem:ass-pass-par}
    If \(  \LeftAssertParen\openO{\langle}  \possiblyWithSub\stageImetaColor{\tau^{\superscriptI} }  \closeO{\rangle} \relO{\CastArrow} \openO{\langle}  \possiblyWithSub\stageImetaColor{\tau^{\superscriptI} }  \closeO{\rangle}\RightAssertParen^{ L }   \Longrightarrow^{0}  \possiblyWithSub\stageOmetaColor{N'^{\superscriptO} } \), then we have
    one of the following:
    \begin{enumerate}
      \item \(\possiblyWithSub\stageOmetaColor{N'^{\superscriptO} } =  \LeftAssertParen\openO{\langle}  \possiblyWithSub\stageImetaColor{\tau^{\superscriptI} }  \closeO{\rangle} \relO{\CastArrow} \openO{\langle}  \possiblyWithSub\stageImetaColor{\tau^{\superscriptI} }  \closeO{\rangle}\RightAssertParen^{ L } \), or
      \item \(\possiblyWithSub\stageOmetaColor{N'^{\superscriptO} } =  \ordO{\lambda} \possiblyWithSub\stageOmetaColor{x}  \relO{:}   \openO{\langle}  \possiblyWithSub\stageImetaColor{\tau^{\superscriptI} }  \closeO{\rangle}  \punctO{.}\   \possiblyWithSub\stageOmetaColor{x}  \) for some \(\possiblyWithSub\stageOmetaColor{x}\).
    \end{enumerate}
  \end{lemma}
  \begin{proof}
    By tracing back the derivation of \(  \LeftAssertParen\openO{\langle}  \possiblyWithSub\stageImetaColor{\tau^{\superscriptI} }  \closeO{\rangle} \relO{\CastArrow} \openO{\langle}  \possiblyWithSub\stageImetaColor{\tau^{\superscriptI} }  \closeO{\rangle}\RightAssertParen^{ L }   \Longrightarrow^{0}  \possiblyWithSub\stageOmetaColor{N'^{\superscriptO} } \),
    we have the following two cases:
    \begin{itemize}
      \item Case \derive[P0-AssPass]{%
      }{%
          \LeftAssertParen\openO{\langle}  \possiblyWithSub\stageImetaColor{\tau^{\superscriptI} }  \closeO{\rangle} \relO{\CastArrow} \openO{\langle}  \possiblyWithSub\stageImetaColor{\tau^{\superscriptI} }  \closeO{\rangle}\RightAssertParen^{ L }   \Longrightarrow^{0}   \ordO{\lambda} \possiblyWithSub\stageOmetaColor{x}  \relO{:}   \openO{\langle}  \possiblyWithSub\stageImetaColor{\tau^{\superscriptI} }  \closeO{\rangle}  \punctO{.}\   \possiblyWithSub\stageOmetaColor{x}   
      }:
        This clearly satisfies (2).
      \item Case \derive[P0-Ass]{%
          \possiblyWithSub\stageImetaColor{\tau^{\superscriptI} }_{{\mathrm{1}}}   \Longrightarrow^{1}  \possiblyWithSub\stageImetaColor{T'^{\superscriptI} }_{{\mathrm{1}}} 
      \andalso
          \possiblyWithSub\stageImetaColor{\tau^{\superscriptI} }_{{\mathrm{2}}}   \Longrightarrow^{1}  \possiblyWithSub\stageImetaColor{T'^{\superscriptI} }_{{\mathrm{2}}} 
      }{%
          \LeftAssertParen\openO{\langle}  \possiblyWithSub\stageImetaColor{\tau^{\superscriptI} }_{{\mathrm{1}}}  \closeO{\rangle} \relO{\CastArrow} \openO{\langle}  \possiblyWithSub\stageImetaColor{\tau^{\superscriptI} }_{{\mathrm{2}}}  \closeO{\rangle}\RightAssertParen^{ L }   \Longrightarrow^{0}   \LeftAssertParen\openO{\langle} \possiblyWithSub\stageImetaColor{T'^{\superscriptI} }_{{\mathrm{1}}} \closeO{\rangle} \relO{\CastArrow} \openO{\langle} \possiblyWithSub\stageImetaColor{T'^{\superscriptI} }_{{\mathrm{2}}} \closeO{\rangle}\RightAssertParen^{ L }  
      }:
        By Lemma~\ref{lem:par-type-normal-form},
        from \(  \possiblyWithSub\stageImetaColor{\tau^{\superscriptI} }_{{\mathrm{1}}}   \Longrightarrow^{1}  \possiblyWithSub\stageImetaColor{T'^{\superscriptI} }_{{\mathrm{1}}} \) and \(  \possiblyWithSub\stageImetaColor{\tau^{\superscriptI} }_{{\mathrm{2}}}   \Longrightarrow^{1}  \possiblyWithSub\stageImetaColor{T'^{\superscriptI} }_{{\mathrm{2}}} \),
        we have \(\possiblyWithSub\stageImetaColor{T'^{\superscriptI} }_{{\mathrm{1}}} = \possiblyWithSub\stageImetaColor{\tau^{\superscriptI} }_{{\mathrm{1}}}\) and \(\possiblyWithSub\stageImetaColor{T'^{\superscriptI} }_{{\mathrm{2}}} = \possiblyWithSub\stageImetaColor{\tau^{\superscriptI} }_{{\mathrm{2}}}\).
        Therefore, we have
        \(\possiblyWithSub\stageOmetaColor{N'^{\superscriptO} } =  \LeftAssertParen\openO{\langle} \possiblyWithSub\stageImetaColor{T'^{\superscriptI} }_{{\mathrm{1}}} \closeO{\rangle} \relO{\CastArrow} \openO{\langle} \possiblyWithSub\stageImetaColor{T'^{\superscriptI} }_{{\mathrm{2}}} \closeO{\rangle}\RightAssertParen^{ L }  =  \LeftAssertParen\openO{\langle}  \possiblyWithSub\stageImetaColor{\tau^{\superscriptI} }_{{\mathrm{1}}}  \closeO{\rangle} \relO{\CastArrow} \openO{\langle}  \possiblyWithSub\stageImetaColor{\tau^{\superscriptI} }_{{\mathrm{2}}}  \closeO{\rangle}\RightAssertParen^{ L } \).
    \end{itemize}
  \end{proof}
  \begin{lemma}[Substitution preserves parallel reduction]\label{lem:par-subst}
    Suppose \( \possiblyWithSub\stageOmetaColor{N^{\superscriptO} }_{{\mathrm{0}}}  \Longrightarrow^{0}  \possiblyWithSub\stageOmetaColor{N'^{\superscriptO} }_{{\mathrm{0}}} \).
    \begin{enumerate}
      \item
        If \( \possiblyWithSub\stageOmetaColor{N^{\superscriptO} }  \Longrightarrow^{0}  \possiblyWithSub\stageOmetaColor{N'^{\superscriptO} } \), then \(   [  \possiblyWithSub\stageOmetaColor{N^{\superscriptO} }_{{\mathrm{0}}}  /  \possiblyWithSub\stageOmetaColor{x}  ]    \possiblyWithSub\stageOmetaColor{N^{\superscriptO} }   \Longrightarrow^{0}    [  \possiblyWithSub\stageOmetaColor{N'^{\superscriptO} }_{{\mathrm{0}}}  /  \possiblyWithSub\stageOmetaColor{x}  ]    \possiblyWithSub\stageOmetaColor{N'^{\superscriptO} }  \).
      \item
        If \( \possiblyWithSub\stageImetaColor{N^{\superscriptI} }  \Longrightarrow^{1}  \possiblyWithSub\stageImetaColor{N'^{\superscriptI} } \), then \(   [  \possiblyWithSub\stageOmetaColor{N^{\superscriptO} }_{{\mathrm{0}}}  /  \possiblyWithSub\stageOmetaColor{x}  ]    \possiblyWithSub\stageImetaColor{N^{\superscriptI} }   \Longrightarrow^{1}    [  \possiblyWithSub\stageOmetaColor{N'^{\superscriptO} }_{{\mathrm{0}}}  /  \possiblyWithSub\stageOmetaColor{x}  ]    \possiblyWithSub\stageImetaColor{N'^{\superscriptI} }  \).
      \item
        If \( \possiblyWithSub\stageOmetaColor{T^{\superscriptO} }  \Longrightarrow^{0}  \possiblyWithSub\stageOmetaColor{T'^{\superscriptO} } \), then \(   [  \possiblyWithSub\stageOmetaColor{N^{\superscriptO} }_{{\mathrm{0}}}  /  \possiblyWithSub\stageOmetaColor{x}  ]    \possiblyWithSub\stageOmetaColor{T^{\superscriptO} }   \Longrightarrow^{0}    [  \possiblyWithSub\stageOmetaColor{N'^{\superscriptO} }_{{\mathrm{0}}}  /  \possiblyWithSub\stageOmetaColor{x}  ]    \possiblyWithSub\stageOmetaColor{T'^{\superscriptO} }  \).
      \item
        If \( \possiblyWithSub\stageImetaColor{T^{\superscriptI} }  \Longrightarrow^{1}  \possiblyWithSub\stageImetaColor{T'^{\superscriptI} } \), then \(   [  \possiblyWithSub\stageOmetaColor{N^{\superscriptO} }_{{\mathrm{0}}}  /  \possiblyWithSub\stageOmetaColor{x}  ]    \possiblyWithSub\stageImetaColor{T^{\superscriptI} }   \Longrightarrow^{1}    [  \possiblyWithSub\stageOmetaColor{N'^{\superscriptO} }_{{\mathrm{0}}}  /  \possiblyWithSub\stageOmetaColor{x}  ]    \possiblyWithSub\stageImetaColor{T'^{\superscriptI} }  \).
    \end{enumerate}
  \end{lemma}
  \begin{proof}
    By mutual induction on the derivation of
    \( \possiblyWithSub\stageOmetaColor{N^{\superscriptO} }  \Longrightarrow^{0}  \possiblyWithSub\stageOmetaColor{N'^{\superscriptO} } \), \( \possiblyWithSub\stageImetaColor{N^{\superscriptI} }  \Longrightarrow^{1}  \possiblyWithSub\stageImetaColor{N'^{\superscriptI} } \),
    \( \possiblyWithSub\stageOmetaColor{T^{\superscriptO} }  \Longrightarrow^{0}  \possiblyWithSub\stageOmetaColor{T'^{\superscriptO} } \), and \( \possiblyWithSub\stageImetaColor{T^{\superscriptI} }  \Longrightarrow^{1}  \possiblyWithSub\stageImetaColor{T'^{\superscriptI} } \).
    \begin{enumerate}
      \item
        \begin{itemize}
          \item Cases~\rulename{P0-Cst0}, \rulename{P0-CstP},
          \rulename{P0-Delta}, \rulename{P0-AssPass}, and \rulename{P0-RfnPass}
          are trivial.
          \item Case~\derive[P0-Var]{}{%
              \possiblyWithSub\stageOmetaColor{x'}   \Longrightarrow^{0}   \possiblyWithSub\stageOmetaColor{x'}  
          }:
            We further do the following case analysis:
            \begin{itemize}
              \item Case~\(\possiblyWithSub\stageOmetaColor{x'} \neq \possiblyWithSub\stageOmetaColor{x}\):
                This is immediate since we have
                \(  [  \possiblyWithSub\stageOmetaColor{N^{\superscriptO} }_{{\mathrm{0}}}  /  \possiblyWithSub\stageOmetaColor{x}  ]    \possiblyWithSub\stageOmetaColor{N^{\superscriptO} }  =   [  \possiblyWithSub\stageOmetaColor{N'^{\superscriptO} }_{{\mathrm{0}}}  /  \possiblyWithSub\stageOmetaColor{x}  ]    \possiblyWithSub\stageOmetaColor{N'^{\superscriptO} }  = \possiblyWithSub\stageOmetaColor{x'}\).
              \item Case~\(\possiblyWithSub\stageOmetaColor{x'} = \possiblyWithSub\stageOmetaColor{x}\):
                Since \(  [  \possiblyWithSub\stageOmetaColor{N^{\superscriptO} }_{{\mathrm{0}}}  /  \possiblyWithSub\stageOmetaColor{x}  ]    \possiblyWithSub\stageOmetaColor{N^{\superscriptO} }  = \possiblyWithSub\stageOmetaColor{N^{\superscriptO} }_{{\mathrm{0}}}\)
                and \(  [  \possiblyWithSub\stageOmetaColor{N'^{\superscriptO} }_{{\mathrm{0}}}  /  \possiblyWithSub\stageOmetaColor{x}  ]    \possiblyWithSub\stageOmetaColor{N'^{\superscriptO} }  = \possiblyWithSub\stageOmetaColor{N'^{\superscriptO} }_{{\mathrm{0}}}\),
                we can finish the proof by the assumption~\( \possiblyWithSub\stageOmetaColor{N^{\superscriptO} }_{{\mathrm{0}}}  \Longrightarrow^{0}  \possiblyWithSub\stageOmetaColor{N'^{\superscriptO} }_{{\mathrm{0}}} \).
            \end{itemize}
          \item Case~\derive[P0-Abs]{%
             \possiblyWithSub\stageOmetaColor{T^{\superscriptO} }_{{\mathrm{1}}}  \Longrightarrow^{0}  \possiblyWithSub\stageOmetaColor{T'^{\superscriptO} }_{{\mathrm{1}}} 
          \andalso
             \possiblyWithSub\stageOmetaColor{N^{\superscriptO} }_{{\mathrm{2}}}  \Longrightarrow^{0}  \possiblyWithSub\stageOmetaColor{N'^{\superscriptO} }_{{\mathrm{2}}} 
          }{%
              \ordO{\lambda} \possiblyWithSub\stageOmetaColor{x'}  \relO{:}  \possiblyWithSub\stageOmetaColor{T^{\superscriptO} }_{{\mathrm{1}}} \punctO{.}\  \possiblyWithSub\stageOmetaColor{N^{\superscriptO} }_{{\mathrm{2}}}   \Longrightarrow^{0}   \ordO{\lambda} \possiblyWithSub\stageOmetaColor{x'}  \relO{:}  \possiblyWithSub\stageOmetaColor{T'^{\superscriptO} }_{{\mathrm{1}}} \punctO{.}\  \possiblyWithSub\stageOmetaColor{N'^{\superscriptO} }_{{\mathrm{2}}}  
          }:
            W.l.o.g., we can assume \(\possiblyWithSub\stageOmetaColor{x'} \neq \possiblyWithSub\stageOmetaColor{x}\).
            By IH, we have \(   [  \possiblyWithSub\stageOmetaColor{N^{\superscriptO} }_{{\mathrm{0}}}  /  \possiblyWithSub\stageOmetaColor{x}  ]    \possiblyWithSub\stageOmetaColor{T^{\superscriptO} }_{{\mathrm{1}}}   \Longrightarrow^{0}    [  \possiblyWithSub\stageOmetaColor{N'^{\superscriptO} }_{{\mathrm{0}}}  /  \possiblyWithSub\stageOmetaColor{x}  ]    \possiblyWithSub\stageOmetaColor{T'^{\superscriptO} }_{{\mathrm{1}}}  \)
            and \(   [  \possiblyWithSub\stageOmetaColor{N^{\superscriptO} }_{{\mathrm{0}}}  /  \possiblyWithSub\stageOmetaColor{x}  ]    \possiblyWithSub\stageOmetaColor{N^{\superscriptO} }_{{\mathrm{2}}}   \Longrightarrow^{0}    [  \possiblyWithSub\stageOmetaColor{N'^{\superscriptO} }_{{\mathrm{0}}}  /  \possiblyWithSub\stageOmetaColor{x}  ]    \possiblyWithSub\stageOmetaColor{N'^{\superscriptO} }_{{\mathrm{2}}}  \).
            Thus, we can derive
            \begin{center}
              \derive[P0-Abs]{%
                   [  \possiblyWithSub\stageOmetaColor{N^{\superscriptO} }_{{\mathrm{0}}}  /  \possiblyWithSub\stageOmetaColor{x}  ]    \possiblyWithSub\stageOmetaColor{T^{\superscriptO} }_{{\mathrm{1}}}   \Longrightarrow^{0}    [  \possiblyWithSub\stageOmetaColor{N'^{\superscriptO} }_{{\mathrm{0}}}  /  \possiblyWithSub\stageOmetaColor{x}  ]    \possiblyWithSub\stageOmetaColor{T'^{\superscriptO} }_{{\mathrm{1}}}  
              \andalso
                   [  \possiblyWithSub\stageOmetaColor{N^{\superscriptO} }_{{\mathrm{0}}}  /  \possiblyWithSub\stageOmetaColor{x}  ]    \possiblyWithSub\stageOmetaColor{N^{\superscriptO} }_{{\mathrm{2}}}   \Longrightarrow^{0}    [  \possiblyWithSub\stageOmetaColor{N'^{\superscriptO} }_{{\mathrm{0}}}  /  \possiblyWithSub\stageOmetaColor{x}  ]    \possiblyWithSub\stageOmetaColor{N'^{\superscriptO} }_{{\mathrm{2}}}  
              }{%
                   [  \possiblyWithSub\stageOmetaColor{N^{\superscriptO} }_{{\mathrm{0}}}  /  \possiblyWithSub\stageOmetaColor{x}  ]     \openO{(}  \ordO{\lambda} \possiblyWithSub\stageOmetaColor{x'}  \relO{:}  \possiblyWithSub\stageOmetaColor{T^{\superscriptO} }_{{\mathrm{1}}} \punctO{.}\  \possiblyWithSub\stageOmetaColor{N^{\superscriptO} }_{{\mathrm{2}}}  \closeO{)}    \Longrightarrow^{0}    [  \possiblyWithSub\stageOmetaColor{N'^{\superscriptO} }_{{\mathrm{0}}}  /  \possiblyWithSub\stageOmetaColor{x}  ]     \openO{(}  \ordO{\lambda} \possiblyWithSub\stageOmetaColor{x'}  \relO{:}  \possiblyWithSub\stageOmetaColor{T'^{\superscriptO} }_{{\mathrm{1}}} \punctO{.}\  \possiblyWithSub\stageOmetaColor{N'^{\superscriptO} }_{{\mathrm{2}}}  \closeO{)}   
              }.
            \end{center}
          \item Case~\derive[P0-Beta]{%
             \possiblyWithSub\stageOmetaColor{N^{\superscriptO} }_{{\mathrm{1}}}  \Longrightarrow^{0}  \possiblyWithSub\stageOmetaColor{N'^{\superscriptO} }_{{\mathrm{1}}} 
          \andalso
             \possiblyWithSub\stageOmetaColor{N^{\superscriptO} }_{{\mathrm{2}}}  \Longrightarrow^{0}  \possiblyWithSub\stageOmetaColor{N'^{\superscriptO} }_{{\mathrm{2}}} 
          }{%
               \openO{(}  \ordO{\lambda} \possiblyWithSub\stageOmetaColor{x'}  \relO{:}  \possiblyWithSub\stageOmetaColor{T^{\superscriptO} } \punctO{.}\  \possiblyWithSub\stageOmetaColor{N^{\superscriptO} }_{{\mathrm{1}}}  \closeO{)}  \  \possiblyWithSub\stageOmetaColor{N^{\superscriptO} }_{{\mathrm{2}}}   \Longrightarrow^{0}    [  \possiblyWithSub\stageOmetaColor{N'^{\superscriptO} }_{{\mathrm{2}}}  /  \possiblyWithSub\stageOmetaColor{x'}  ]    \possiblyWithSub\stageOmetaColor{N'^{\superscriptO} }_{{\mathrm{1}}}  
          }:
            By IH, we have \(   [  \possiblyWithSub\stageOmetaColor{N^{\superscriptO} }_{{\mathrm{0}}}  /  \possiblyWithSub\stageOmetaColor{x}  ]    \possiblyWithSub\stageOmetaColor{N^{\superscriptO} }_{{\mathrm{1}}}   \Longrightarrow^{0}    [  \possiblyWithSub\stageOmetaColor{N'^{\superscriptO} }_{{\mathrm{0}}}  /  \possiblyWithSub\stageOmetaColor{x}  ]    \possiblyWithSub\stageOmetaColor{N'^{\superscriptO} }_{{\mathrm{1}}}  \)
            and \(   [  \possiblyWithSub\stageOmetaColor{N^{\superscriptO} }_{{\mathrm{0}}}  /  \possiblyWithSub\stageOmetaColor{x}  ]    \possiblyWithSub\stageOmetaColor{N^{\superscriptO} }_{{\mathrm{2}}}   \Longrightarrow^{0}    [  \possiblyWithSub\stageOmetaColor{N'^{\superscriptO} }_{{\mathrm{0}}}  /  \possiblyWithSub\stageOmetaColor{x}  ]    \possiblyWithSub\stageOmetaColor{N'^{\superscriptO} }_{{\mathrm{2}}}  \).
            W.l.o.g., we can assume \(\possiblyWithSub\stageOmetaColor{x'} \neq \possiblyWithSub\stageOmetaColor{x}\),
            and thereby we can derive
            \(   [  \possiblyWithSub\stageOmetaColor{N^{\superscriptO} }_{{\mathrm{0}}}  /  \possiblyWithSub\stageOmetaColor{x}  ]     \openO{(}   \openO{(}  \ordO{\lambda} \possiblyWithSub\stageOmetaColor{x'}  \relO{:}  \possiblyWithSub\stageOmetaColor{T^{\superscriptO} } \punctO{.}\  \possiblyWithSub\stageOmetaColor{N^{\superscriptO} }_{{\mathrm{1}}}  \closeO{)}  \  \possiblyWithSub\stageOmetaColor{N^{\superscriptO} }_{{\mathrm{2}}}  \closeO{)}    \Longrightarrow^{0}    [  \possiblyWithSub\stageOmetaColor{N'^{\superscriptO} }_{{\mathrm{0}}}  /  \possiblyWithSub\stageOmetaColor{x}  ]      [  \possiblyWithSub\stageOmetaColor{N'^{\superscriptO} }_{{\mathrm{2}}}  /  \possiblyWithSub\stageOmetaColor{x'}  ]    \possiblyWithSub\stageOmetaColor{N'^{\superscriptO} }_{{\mathrm{1}}}   \)
            as follows:
            \begin{center}
              \derive[P0-Beta]{%
                   [  \possiblyWithSub\stageOmetaColor{N^{\superscriptO} }_{{\mathrm{0}}}  /  \possiblyWithSub\stageOmetaColor{x}  ]    \possiblyWithSub\stageOmetaColor{N^{\superscriptO} }_{{\mathrm{1}}}   \Longrightarrow^{0}    [  \possiblyWithSub\stageOmetaColor{N'^{\superscriptO} }_{{\mathrm{0}}}  /  \possiblyWithSub\stageOmetaColor{x}  ]    \possiblyWithSub\stageOmetaColor{N'^{\superscriptO} }_{{\mathrm{1}}}  
              \andalso
                   [  \possiblyWithSub\stageOmetaColor{N^{\superscriptO} }_{{\mathrm{0}}}  /  \possiblyWithSub\stageOmetaColor{x}  ]    \possiblyWithSub\stageOmetaColor{N^{\superscriptO} }_{{\mathrm{2}}}   \Longrightarrow^{0}    [  \possiblyWithSub\stageOmetaColor{N'^{\superscriptO} }_{{\mathrm{0}}}  /  \possiblyWithSub\stageOmetaColor{x}  ]    \possiblyWithSub\stageOmetaColor{N'^{\superscriptO} }_{{\mathrm{2}}}  
              }{%
                   \openO{(}  \ordO{\lambda} \possiblyWithSub\stageOmetaColor{x'}  \relO{:}  \possiblyWithSub\stageOmetaColor{T^{\superscriptO} } \punctO{.}\    [  \possiblyWithSub\stageOmetaColor{N^{\superscriptO} }_{{\mathrm{0}}}  /  \possiblyWithSub\stageOmetaColor{x}  ]    \possiblyWithSub\stageOmetaColor{N^{\superscriptO} }_{{\mathrm{1}}}   \closeO{)}  \    [  \possiblyWithSub\stageOmetaColor{N^{\superscriptO} }_{{\mathrm{0}}}  /  \possiblyWithSub\stageOmetaColor{x}  ]    \possiblyWithSub\stageOmetaColor{N^{\superscriptO} }_{{\mathrm{2}}}    \Longrightarrow^{0}    [    [  \possiblyWithSub\stageOmetaColor{N'^{\superscriptO} }_{{\mathrm{0}}}  /  \possiblyWithSub\stageOmetaColor{x}  ]    \possiblyWithSub\stageOmetaColor{N'^{\superscriptO} }_{{\mathrm{2}}}   /  \possiblyWithSub\stageOmetaColor{x'}  ]      [  \possiblyWithSub\stageOmetaColor{N'^{\superscriptO} }_{{\mathrm{0}}}  /  \possiblyWithSub\stageOmetaColor{x}  ]    \possiblyWithSub\stageOmetaColor{N'^{\superscriptO} }_{{\mathrm{1}}}   
              }
            \end{center}
          \item Cases~\rulename{P0-App}, \rulename{P0-Ass}, \rulename{P0-Brkt},
          \rulename{P0-Rfn}, \rulename{P0-RfnStart}, and \rulename{P0-RfnAct}
          are also straightforward by IH (possibly using the Barendregt convention),
          similarly to the case of \rulename{P0-Abs}.
        \end{itemize}
      \item
        All the cases are immediate or straightforward by IH.
      \item
        \begin{itemize}
          \item Case~\rulename{PT0-Tensor}:
            Trivial.
          \item Case~\rulename{PT0-Code}:
            Straightforward by IH.
          \item Cases~\rulename{PT0-Rfn} and \rulename{PT0-Arr}:
            Straightforward by IH, using the Barendregt convention.
        \end{itemize}
      \item
        All the cases (including \rulename{PT1-Tensor}) are immediate or straightforward by IH.
    \end{enumerate}
  \end{proof}
  \indent
    Following the approach taken by Takahashi~\cite{Takahashi1995},
    we use conversions~\( ( \possiblyWithSub\stageImetaColor{T^{\superscriptI} } )^{\ast}  = \possiblyWithSub\stageImetaColor{T'^{\superscriptI} }\), \( ( \possiblyWithSub\stageOmetaColor{T^{\superscriptO} } )^{\ast}  = \possiblyWithSub\stageOmetaColor{T'^{\superscriptO} }\),
    \( ( \possiblyWithSub\stageOmetaColor{N^{\superscriptO} } )^{\ast}  = \possiblyWithSub\stageOmetaColor{N'^{\superscriptO} }\), and \( ( \possiblyWithSub\stageImetaColor{N^{\superscriptI} } )^{\ast}  = \possiblyWithSub\stageImetaColor{N'^{\superscriptI} }\)
    defined in Figure~\ref{fig:star-conversion} henceforth.
  \par
  \begin{lemma}\label{lem:star-identical-type-value}
    \( (  \possiblyWithSub\stageImetaColor{\tau^{\superscriptI} }  )^{\ast}  = \possiblyWithSub\stageImetaColor{\tau^{\superscriptI} }\) for any type value~\(\possiblyWithSub\stageImetaColor{\tau^{\superscriptI} }\).
  \end{lemma}
  \begin{proof}
    By straightforward induction on the structure of \(\possiblyWithSub\stageImetaColor{\tau^{\superscriptI} }\).
  \end{proof}
  \begin{lemma}[Confluence by conversion \`{a} la Takahashi]\label{lem:confluence-by-star}
    \noindent
    \begin{enumerate}
      \item \( \possiblyWithSub\stageOmetaColor{N^{\superscriptO} }  \Longrightarrow^{0}  \possiblyWithSub\stageOmetaColor{N'^{\superscriptO} } \) implies \( \possiblyWithSub\stageOmetaColor{N'^{\superscriptO} }  \Longrightarrow^{0}   ( \possiblyWithSub\stageOmetaColor{N^{\superscriptO} } )^{\ast}  \).
      \item \( \possiblyWithSub\stageImetaColor{N^{\superscriptI} }  \Longrightarrow^{1}  \possiblyWithSub\stageImetaColor{N'^{\superscriptI} } \) implies \( \possiblyWithSub\stageImetaColor{N'^{\superscriptI} }  \Longrightarrow^{1}   ( \possiblyWithSub\stageImetaColor{N^{\superscriptI} } )^{\ast}  \).
      \item \( \possiblyWithSub\stageOmetaColor{T^{\superscriptO} }  \Longrightarrow^{0}  \possiblyWithSub\stageOmetaColor{T'^{\superscriptO} } \) implies \( \possiblyWithSub\stageOmetaColor{T'^{\superscriptO} }  \Longrightarrow^{0}   ( \possiblyWithSub\stageOmetaColor{T^{\superscriptO} } )^{\ast}  \).
      \item \( \possiblyWithSub\stageImetaColor{T^{\superscriptI} }  \Longrightarrow^{1}  \possiblyWithSub\stageImetaColor{T'^{\superscriptI} } \) implies \( \possiblyWithSub\stageImetaColor{T'^{\superscriptI} }  \Longrightarrow^{1}   ( \possiblyWithSub\stageImetaColor{T^{\superscriptI} } )^{\ast}  \).
    \end{enumerate}
  \end{lemma}
  \begin{proof}
    By mutual induction on the structure of
    \(\possiblyWithSub\stageOmetaColor{N^{\superscriptO} }\), \(\possiblyWithSub\stageImetaColor{N^{\superscriptI} }\), \(\possiblyWithSub\stageOmetaColor{T^{\superscriptO} }\), and \(\possiblyWithSub\stageImetaColor{T^{\superscriptI} }\).
    \begin{enumerate}
      \item
        \begin{itemize}
          \item Case \(\possiblyWithSub\stageOmetaColor{N^{\superscriptO} } =  \LeftAssertParen\openO{\langle} \possiblyWithSub\stageImetaColor{T^{\superscriptI} }_{{\mathrm{1}}} \closeO{\rangle} \relO{\CastArrow} \openO{\langle} \possiblyWithSub\stageImetaColor{T^{\superscriptI} }_{{\mathrm{2}}} \closeO{\rangle}\RightAssertParen^{ L } \) where
          either \(\possiblyWithSub\stageImetaColor{T^{\superscriptI} }_{{\mathrm{1}}}\) or \(\possiblyWithSub\stageImetaColor{T^{\superscriptI} }_{{\mathrm{2}}}\) is not a type value:
            By the definition of \(\Longrightarrow^{0}\),
            \(\possiblyWithSub\stageOmetaColor{N'^{\superscriptO} }\) is of the form~\( \LeftAssertParen\openO{\langle} \possiblyWithSub\stageImetaColor{T'^{\superscriptI} }_{{\mathrm{1}}} \closeO{\rangle} \relO{\CastArrow} \openO{\langle} \possiblyWithSub\stageImetaColor{T'^{\superscriptI} }_{{\mathrm{2}}} \closeO{\rangle}\RightAssertParen^{ L } \), and
            we have \( \possiblyWithSub\stageImetaColor{T^{\superscriptI} }_{{\mathrm{1}}}  \Longrightarrow^{1}  \possiblyWithSub\stageImetaColor{T'^{\superscriptI} }_{{\mathrm{1}}} \) and \( \possiblyWithSub\stageImetaColor{T^{\superscriptI} }_{{\mathrm{2}}}  \Longrightarrow^{1}  \possiblyWithSub\stageImetaColor{T'^{\superscriptI} }_{{\mathrm{2}}} \).
            Then, by IH, we have \( \possiblyWithSub\stageImetaColor{T'^{\superscriptI} }_{{\mathrm{1}}}  \Longrightarrow^{1}   ( \possiblyWithSub\stageImetaColor{T^{\superscriptI} }_{{\mathrm{1}}} )^{\ast}  \) and \( \possiblyWithSub\stageImetaColor{T'^{\superscriptI} }_{{\mathrm{2}}}  \Longrightarrow^{1}   ( \possiblyWithSub\stageImetaColor{T^{\superscriptI} }_{{\mathrm{2}}} )^{\ast}  \),
            and we can thus derive
            \begin{center}
              \derive[P0-Ass]{%
                 \possiblyWithSub\stageImetaColor{T'^{\superscriptI} }_{{\mathrm{1}}}  \Longrightarrow^{1}   ( \possiblyWithSub\stageImetaColor{T^{\superscriptI} }_{{\mathrm{1}}} )^{\ast}  
              \andalso
                 \possiblyWithSub\stageImetaColor{T'^{\superscriptI} }_{{\mathrm{2}}}  \Longrightarrow^{1}   ( \possiblyWithSub\stageImetaColor{T^{\superscriptI} }_{{\mathrm{2}}} )^{\ast}  
              }{%
                  \LeftAssertParen\openO{\langle} \possiblyWithSub\stageImetaColor{T'^{\superscriptI} }_{{\mathrm{1}}} \closeO{\rangle} \relO{\CastArrow} \openO{\langle} \possiblyWithSub\stageImetaColor{T'^{\superscriptI} }_{{\mathrm{2}}} \closeO{\rangle}\RightAssertParen^{ L }   \Longrightarrow^{0}   \LeftAssertParen\openO{\langle}  ( \possiblyWithSub\stageImetaColor{T^{\superscriptI} }_{{\mathrm{1}}} )^{\ast}  \closeO{\rangle} \relO{\CastArrow} \openO{\langle}  ( \possiblyWithSub\stageImetaColor{T^{\superscriptI} }_{{\mathrm{2}}} )^{\ast}  \closeO{\rangle}\RightAssertParen^{ L }  
              }.
            \end{center}
          \item Case \(\possiblyWithSub\stageOmetaColor{N^{\superscriptO} } =  \LeftAssertParen\openO{\langle}  \possiblyWithSub\stageImetaColor{\tau^{\superscriptI} }  \closeO{\rangle} \relO{\CastArrow} \openO{\langle}  \possiblyWithSub\stageImetaColor{\tau^{\superscriptI} }  \closeO{\rangle}\RightAssertParen^{ L } \):
            First, \( ( \possiblyWithSub\stageOmetaColor{N^{\superscriptO} } )^{\ast}  =  \openO{(}  \ordO{\lambda} \possiblyWithSub\stageOmetaColor{x}  \relO{:}   \openO{\langle}  \possiblyWithSub\stageImetaColor{\tau^{\superscriptI} }  \closeO{\rangle}  \punctO{.}\   \possiblyWithSub\stageOmetaColor{x}   \closeO{)} \) holds.
            By Lemma~\ref{lem:ass-pass-par}, we can do the following case analysis:
            \begin{itemize}
              \item Case \(\possiblyWithSub\stageOmetaColor{N'^{\superscriptO} } =  \LeftAssertParen\openO{\langle}  \possiblyWithSub\stageImetaColor{\tau^{\superscriptI} }  \closeO{\rangle} \relO{\CastArrow} \openO{\langle}  \possiblyWithSub\stageImetaColor{\tau^{\superscriptI} }  \closeO{\rangle}\RightAssertParen^{ L } \):
                we immediately have
                \begin{center}
                  \derive[P0-AssPass]{}{%
                     \possiblyWithSub\stageOmetaColor{N'^{\superscriptO} }  \Longrightarrow^{0}   \ordO{\lambda} \possiblyWithSub\stageOmetaColor{x}  \relO{:}   \openO{\langle}  \possiblyWithSub\stageImetaColor{\tau^{\superscriptI} }  \closeO{\rangle}  \punctO{.}\   \possiblyWithSub\stageOmetaColor{x}   
                  }.
                \end{center}
              \item Case \(\possiblyWithSub\stageOmetaColor{N'^{\superscriptO} } =  \openO{(}  \ordO{\lambda} \possiblyWithSub\stageOmetaColor{x}  \relO{:}   \openO{\langle}  \possiblyWithSub\stageImetaColor{\tau^{\superscriptI} }  \closeO{\rangle}  \punctO{.}\   \possiblyWithSub\stageOmetaColor{x}   \closeO{)} \) for some \(\possiblyWithSub\stageOmetaColor{x}\):
                Since \(  \possiblyWithSub\stageImetaColor{\tau^{\superscriptI} }   \Longrightarrow^{1}   \possiblyWithSub\stageImetaColor{\tau^{\superscriptI} }  \) holds
                by Lemma~\ref{lem:eq-implies-par},
                we can derive
                \begin{center}
                  \infer[P0-Lam]{%
                      \possiblyWithSub\stageImetaColor{\tau^{\superscriptI} }   \Longrightarrow^{1}   \possiblyWithSub\stageImetaColor{\tau^{\superscriptI} }  
                  \andalso
                    \infer[P0-Var]{}{%
                        \possiblyWithSub\stageOmetaColor{x}   \Longrightarrow^{0}   \possiblyWithSub\stageOmetaColor{x}  
                    }
                  }{%
                      \ordO{\lambda} \possiblyWithSub\stageOmetaColor{x}  \relO{:}   \openO{\langle}  \possiblyWithSub\stageImetaColor{\tau^{\superscriptI} }  \closeO{\rangle}  \punctO{.}\   \possiblyWithSub\stageOmetaColor{x}    \Longrightarrow^{0}   \ordO{\lambda} \possiblyWithSub\stageOmetaColor{x}  \relO{:}   \openO{\langle}  \possiblyWithSub\stageImetaColor{\tau^{\superscriptI} }  \closeO{\rangle}  \punctO{.}\   \possiblyWithSub\stageOmetaColor{x}   
                  }.
                \end{center}
            \end{itemize}
          \item Case \(\possiblyWithSub\stageOmetaColor{N^{\superscriptO} } =  \LeftAssertParen\openO{\langle}  \possiblyWithSub\stageImetaColor{\tau^{\superscriptI} }_{{\mathrm{1}}}  \closeO{\rangle} \relO{\CastArrow} \openO{\langle}  \possiblyWithSub\stageImetaColor{\tau^{\superscriptI} }_{{\mathrm{2}}}  \closeO{\rangle}\RightAssertParen^{ L } \) where \(\possiblyWithSub\stageImetaColor{\tau^{\superscriptI} }_{{\mathrm{1}}} \neq \possiblyWithSub\stageImetaColor{\tau^{\superscriptI} }_{{\mathrm{2}}}\):
            By tracing back the derivation of \( \possiblyWithSub\stageOmetaColor{N^{\superscriptO} }  \Longrightarrow^{0}  \possiblyWithSub\stageOmetaColor{N'^{\superscriptO} } \),
            we only have the case where \(\possiblyWithSub\stageOmetaColor{N'^{\superscriptO} } =  \LeftAssertParen\openO{\langle} \possiblyWithSub\stageImetaColor{T'^{\superscriptI} }_{{\mathrm{1}}} \closeO{\rangle} \relO{\CastArrow} \openO{\langle} \possiblyWithSub\stageImetaColor{T'^{\superscriptI} }_{{\mathrm{2}}} \closeO{\rangle}\RightAssertParen^{ L } \)
            for some \(\possiblyWithSub\stageImetaColor{T'^{\superscriptI} }_{{\mathrm{1}}}\) and \(\possiblyWithSub\stageImetaColor{T'^{\superscriptI} }_{{\mathrm{2}}}\) by the following:
            \begin{center}
              \derive[P0-Ass]{%
                  \possiblyWithSub\stageImetaColor{\tau^{\superscriptI} }_{{\mathrm{1}}}   \Longrightarrow^{1}  \possiblyWithSub\stageImetaColor{T'^{\superscriptI} }_{{\mathrm{1}}} 
              \andalso
                  \possiblyWithSub\stageImetaColor{\tau^{\superscriptI} }_{{\mathrm{2}}}   \Longrightarrow^{1}  \possiblyWithSub\stageImetaColor{T'^{\superscriptI} }_{{\mathrm{2}}} 
              }{%
                  \LeftAssertParen\openO{\langle}  \possiblyWithSub\stageImetaColor{\tau^{\superscriptI} }_{{\mathrm{1}}}  \closeO{\rangle} \relO{\CastArrow} \openO{\langle}  \possiblyWithSub\stageImetaColor{\tau^{\superscriptI} }_{{\mathrm{2}}}  \closeO{\rangle}\RightAssertParen^{ L }   \Longrightarrow^{0}   \LeftAssertParen\openO{\langle} \possiblyWithSub\stageImetaColor{T'^{\superscriptI} }_{{\mathrm{1}}} \closeO{\rangle} \relO{\CastArrow} \openO{\langle} \possiblyWithSub\stageImetaColor{T'^{\superscriptI} }_{{\mathrm{2}}} \closeO{\rangle}\RightAssertParen^{ L }  
              }.
            \end{center}
            Then, by Lemma~\ref{lem:par-type-normal-form},
            we have \(\possiblyWithSub\stageImetaColor{T'^{\superscriptI} }_{{\mathrm{1}}} = \possiblyWithSub\stageImetaColor{\tau^{\superscriptI} }_{{\mathrm{1}}}\) and \(\possiblyWithSub\stageImetaColor{T'^{\superscriptI} }_{{\mathrm{2}}} = \possiblyWithSub\stageImetaColor{\tau^{\superscriptI} }_{{\mathrm{2}}}\),
            and thus \(\possiblyWithSub\stageOmetaColor{N'^{\superscriptO} } =  \LeftAssertParen\openO{\langle}  \possiblyWithSub\stageImetaColor{\tau^{\superscriptI} }_{{\mathrm{1}}}  \closeO{\rangle} \relO{\CastArrow} \openO{\langle}  \possiblyWithSub\stageImetaColor{\tau^{\superscriptI} }_{{\mathrm{2}}}  \closeO{\rangle}\RightAssertParen^{ L }  = \possiblyWithSub\stageOmetaColor{N^{\superscriptO} }\) holds.
            Therefore, by Lemmata~\ref{lem:eq-implies-par} and \ref{lem:star-identical-type-value},
            we have \( \possiblyWithSub\stageOmetaColor{N'^{\superscriptO} }  \Longrightarrow^{0}   ( \possiblyWithSub\stageOmetaColor{N^{\superscriptO} } )^{\ast}  \).
          \item Case \(\possiblyWithSub\stageOmetaColor{N^{\superscriptO} } =  \possiblyWithSub\stageOmetaColor{N^{\superscriptO} }_{{\mathrm{1}}} \  \possiblyWithSub\stageOmetaColor{N^{\superscriptO} }_{{\mathrm{2}}} \):
            We do a further case analysis as follows:
            \begin{itemize}
              \item Case \(\possiblyWithSub\stageOmetaColor{N^{\superscriptO} }_{{\mathrm{1}}} =  \openO{(}  \ordO{\lambda} \possiblyWithSub\stageOmetaColor{x}  \relO{:}  \possiblyWithSub\stageOmetaColor{T^{\superscriptO} }_{{\mathrm{11}}} \punctO{.}\  \possiblyWithSub\stageOmetaColor{N^{\superscriptO} }_{{\mathrm{12}}}  \closeO{)} \):
                We first have \( ( \possiblyWithSub\stageOmetaColor{N^{\superscriptO} } )^{\ast}  =   [   ( \possiblyWithSub\stageOmetaColor{N^{\superscriptO} }_{{\mathrm{2}}} )^{\ast}   /  \possiblyWithSub\stageOmetaColor{x}  ]     ( \possiblyWithSub\stageOmetaColor{N^{\superscriptO} }_{{\mathrm{12}}} )^{\ast}  \).
                By tracing back the derivation of \( \possiblyWithSub\stageOmetaColor{N^{\superscriptO} }  \Longrightarrow^{0}  \possiblyWithSub\stageOmetaColor{N'^{\superscriptO} } \),
                we have the following two cases:
                \begin{itemize}
                  \item Case \derive[P0-App]{%
                      \ordO{\lambda} \possiblyWithSub\stageOmetaColor{x}  \relO{:}  \possiblyWithSub\stageOmetaColor{T^{\superscriptO} }_{{\mathrm{11}}} \punctO{.}\  \possiblyWithSub\stageOmetaColor{N^{\superscriptO} }_{{\mathrm{12}}}   \Longrightarrow^{0}  \possiblyWithSub\stageOmetaColor{N'^{\superscriptO} }_{{\mathrm{1}}} 
                  \andalso
                     \possiblyWithSub\stageOmetaColor{N^{\superscriptO} }_{{\mathrm{2}}}  \Longrightarrow^{0}  \possiblyWithSub\stageOmetaColor{N'^{\superscriptO} }_{{\mathrm{2}}} 
                  }{%
                       \openO{(}  \ordO{\lambda} \possiblyWithSub\stageOmetaColor{x}  \relO{:}  \possiblyWithSub\stageOmetaColor{T^{\superscriptO} }_{{\mathrm{11}}} \punctO{.}\  \possiblyWithSub\stageOmetaColor{N^{\superscriptO} }_{{\mathrm{12}}}  \closeO{)}  \  \possiblyWithSub\stageOmetaColor{N^{\superscriptO} }_{{\mathrm{2}}}   \Longrightarrow^{0}   \possiblyWithSub\stageOmetaColor{N'^{\superscriptO} }_{{\mathrm{1}}} \  \possiblyWithSub\stageOmetaColor{N'^{\superscriptO} }_{{\mathrm{2}}}  
                  }:
                    We can trace back the derivation of \(  \ordO{\lambda} \possiblyWithSub\stageOmetaColor{x}  \relO{:}  \possiblyWithSub\stageOmetaColor{T^{\superscriptO} }_{{\mathrm{11}}} \punctO{.}\  \possiblyWithSub\stageOmetaColor{N^{\superscriptO} }_{{\mathrm{12}}}   \Longrightarrow^{0}  \possiblyWithSub\stageOmetaColor{N'^{\superscriptO} }_{{\mathrm{1}}} \)
                    as follows:
                    \begin{center}
                      \derive[P0-Lam]{%
                         \possiblyWithSub\stageOmetaColor{T^{\superscriptO} }_{{\mathrm{11}}}  \Longrightarrow^{0}  \possiblyWithSub\stageOmetaColor{T'^{\superscriptO} }_{{\mathrm{11}}} 
                      \andalso
                         \possiblyWithSub\stageOmetaColor{N^{\superscriptO} }_{{\mathrm{12}}}  \Longrightarrow^{0}  \possiblyWithSub\stageOmetaColor{N'^{\superscriptO} }_{{\mathrm{12}}} 
                      }{%
                          \ordO{\lambda} \possiblyWithSub\stageOmetaColor{x}  \relO{:}  \possiblyWithSub\stageOmetaColor{T^{\superscriptO} }_{{\mathrm{11}}} \punctO{.}\  \possiblyWithSub\stageOmetaColor{N^{\superscriptO} }_{{\mathrm{12}}}   \Longrightarrow^{0}   \ordO{\lambda} \possiblyWithSub\stageOmetaColor{x}  \relO{:}  \possiblyWithSub\stageOmetaColor{T'^{\superscriptO} }_{{\mathrm{11}}} \punctO{.}\  \possiblyWithSub\stageOmetaColor{N'^{\superscriptO} }_{{\mathrm{12}}}  
                      }
                    \end{center}
                    By IH, we have \( \possiblyWithSub\stageOmetaColor{N'^{\superscriptO} }_{{\mathrm{12}}}  \Longrightarrow^{0}   ( \possiblyWithSub\stageOmetaColor{N^{\superscriptO} }_{{\mathrm{12}}} )^{\ast}  \) and \( \possiblyWithSub\stageOmetaColor{N'^{\superscriptO} }_{{\mathrm{2}}}  \Longrightarrow^{0}   ( \possiblyWithSub\stageOmetaColor{N^{\superscriptO} }_{{\mathrm{2}}} )^{\ast}  \).
                    This enables us to derive
                    \begin{center}
                      \derive[P0-Beta]{%
                         \possiblyWithSub\stageOmetaColor{N'^{\superscriptO} }_{{\mathrm{12}}}  \Longrightarrow^{0}   ( \possiblyWithSub\stageOmetaColor{N^{\superscriptO} }_{{\mathrm{12}}} )^{\ast}  
                      \andalso
                         \possiblyWithSub\stageOmetaColor{N'^{\superscriptO} }_{{\mathrm{2}}}  \Longrightarrow^{0}   ( \possiblyWithSub\stageOmetaColor{N^{\superscriptO} }_{{\mathrm{2}}} )^{\ast}  
                      }{%
                           \openO{(}  \ordO{\lambda} \possiblyWithSub\stageOmetaColor{x}  \relO{:}  \possiblyWithSub\stageOmetaColor{T'^{\superscriptO} }_{{\mathrm{11}}} \punctO{.}\  \possiblyWithSub\stageOmetaColor{N'^{\superscriptO} }_{{\mathrm{12}}}  \closeO{)}  \  \possiblyWithSub\stageOmetaColor{N'^{\superscriptO} }_{{\mathrm{2}}}   \Longrightarrow^{0}    [   ( \possiblyWithSub\stageOmetaColor{N^{\superscriptO} }_{{\mathrm{2}}} )^{\ast}   /  \possiblyWithSub\stageOmetaColor{x}  ]     ( \possiblyWithSub\stageOmetaColor{N^{\superscriptO} }_{{\mathrm{12}}} )^{\ast}   
                      }.
                    \end{center}
                  \item Case \derive[P0-Beta]{%
                     \possiblyWithSub\stageOmetaColor{N^{\superscriptO} }_{{\mathrm{12}}}  \Longrightarrow^{0}  \possiblyWithSub\stageOmetaColor{N'^{\superscriptO} }_{{\mathrm{12}}} 
                  \andalso
                     \possiblyWithSub\stageOmetaColor{N^{\superscriptO} }_{{\mathrm{2}}}  \Longrightarrow^{0}  \possiblyWithSub\stageOmetaColor{N'^{\superscriptO} }_{{\mathrm{2}}} 
                  }{%
                       \openO{(}  \ordO{\lambda} \possiblyWithSub\stageOmetaColor{x}  \relO{:}  \possiblyWithSub\stageOmetaColor{T^{\superscriptO} }_{{\mathrm{11}}} \punctO{.}\  \possiblyWithSub\stageOmetaColor{N^{\superscriptO} }_{{\mathrm{12}}}  \closeO{)}  \  \possiblyWithSub\stageOmetaColor{N^{\superscriptO} }_{{\mathrm{2}}}   \Longrightarrow^{0}    [  \possiblyWithSub\stageOmetaColor{N'^{\superscriptO} }_{{\mathrm{2}}}  /  \possiblyWithSub\stageOmetaColor{x}  ]    \possiblyWithSub\stageOmetaColor{N'^{\superscriptO} }_{{\mathrm{12}}}  
                  }:
                    By IH, we have \( \possiblyWithSub\stageOmetaColor{N'^{\superscriptO} }_{{\mathrm{12}}}  \Longrightarrow^{0}   ( \possiblyWithSub\stageOmetaColor{N^{\superscriptO} }_{{\mathrm{12}}} )^{\ast}  \) and \( \possiblyWithSub\stageOmetaColor{N'^{\superscriptO} }_{{\mathrm{2}}}  \Longrightarrow^{0}   ( \possiblyWithSub\stageOmetaColor{N^{\superscriptO} }_{{\mathrm{2}}} )^{\ast}  \).
                    Therefore, by Lemma~\ref{lem:par-subst},
                    we have \(   [  \possiblyWithSub\stageOmetaColor{N'^{\superscriptO} }_{{\mathrm{2}}}  /  \possiblyWithSub\stageOmetaColor{x}  ]    \possiblyWithSub\stageOmetaColor{N'^{\superscriptO} }_{{\mathrm{12}}}   \Longrightarrow^{0}    [   ( \possiblyWithSub\stageOmetaColor{N^{\superscriptO} }_{{\mathrm{2}}} )^{\ast}   /  \possiblyWithSub\stageOmetaColor{x}  ]     ( \possiblyWithSub\stageOmetaColor{N^{\superscriptO} }_{{\mathrm{12}}} )^{\ast}   \).
                \end{itemize}
              \item Case where \(\possiblyWithSub\stageOmetaColor{N^{\superscriptO} }_{{\mathrm{1}}} = \possiblyWithSub\stageOmetaColor{a}_{{\mathrm{1}}}\), \(\possiblyWithSub\stageOmetaColor{N^{\superscriptO} }_{{\mathrm{2}}} = \possiblyWithSub\stageOmetaColor{c}_{{\mathrm{2}}}\),
              and \(\delta(  \possiblyWithSub\stageOmetaColor{a}_{{\mathrm{1}}}  \    \possiblyWithSub\stageOmetaColor{c}_{{\mathrm{2}}}   ) = \possiblyWithSub\stageOmetaColor{q}\):
                We first have \( ( \possiblyWithSub\stageOmetaColor{N^{\superscriptO} } )^{\ast}  = \possiblyWithSub\stageOmetaColor{q}\).
                By tracing back the derivation of \( \possiblyWithSub\stageOmetaColor{N^{\superscriptO} }  \Longrightarrow^{0}  \possiblyWithSub\stageOmetaColor{N'^{\superscriptO} } \),
                we have the following two cases:
                \begin{itemize}
                  \item Case \derive[P0-App]{%
                      \possiblyWithSub\stageOmetaColor{a}_{{\mathrm{1}}}   \Longrightarrow^{0}  \possiblyWithSub\stageOmetaColor{N'^{\superscriptO} }_{{\mathrm{1}}} 
                  \andalso
                       \possiblyWithSub\stageOmetaColor{c}_{{\mathrm{2}}}    \Longrightarrow^{0}  \possiblyWithSub\stageOmetaColor{N'^{\superscriptO} }_{{\mathrm{2}}} 
                  }{%
                       \possiblyWithSub\stageOmetaColor{a}_{{\mathrm{1}}}  \    \possiblyWithSub\stageOmetaColor{c}_{{\mathrm{2}}}     \Longrightarrow^{0}   \possiblyWithSub\stageOmetaColor{N'^{\superscriptO} }_{{\mathrm{1}}} \  \possiblyWithSub\stageOmetaColor{N'^{\superscriptO} }_{{\mathrm{2}}}  
                  }:
                    By Lemmata~\ref{lem:par-prim-partial-app-normal-form}
                    and \ref{lem:par-const-normal-form},
                    we have \(\possiblyWithSub\stageOmetaColor{N'^{\superscriptO} }_{{\mathrm{1}}} = \possiblyWithSub\stageOmetaColor{a}_{{\mathrm{1}}}\) and \(\possiblyWithSub\stageOmetaColor{N'^{\superscriptO} }_{{\mathrm{2}}} = \possiblyWithSub\stageOmetaColor{c}_{{\mathrm{2}}}\), respectively.
                    Thus, we can derive \( \possiblyWithSub\stageOmetaColor{N'^{\superscriptO} }  \Longrightarrow^{0}   ( \possiblyWithSub\stageOmetaColor{N^{\superscriptO} } )^{\ast}  \) simply as follows:
                    \begin{center}
                      \derive[P0-Delta]{%
                        \delta(  \possiblyWithSub\stageOmetaColor{a}_{{\mathrm{1}}}  \    \possiblyWithSub\stageOmetaColor{c}_{{\mathrm{2}}}   ) = \possiblyWithSub\stageOmetaColor{q}
                      }{%
                           \possiblyWithSub\stageOmetaColor{a}_{{\mathrm{1}}}  \    \possiblyWithSub\stageOmetaColor{c}_{{\mathrm{2}}}     \Longrightarrow^{0}   \possiblyWithSub\stageOmetaColor{q}  
                      }
                    \end{center}
                  \item Case \derive[P0-Delta]{%
                    \delta(  \possiblyWithSub\stageOmetaColor{a}_{{\mathrm{1}}}  \    \possiblyWithSub\stageOmetaColor{c}_{{\mathrm{2}}}   ) = \possiblyWithSub\stageOmetaColor{q}
                  }{%
                       \possiblyWithSub\stageOmetaColor{a}_{{\mathrm{1}}}  \    \possiblyWithSub\stageOmetaColor{c}_{{\mathrm{2}}}     \Longrightarrow^{0}   \possiblyWithSub\stageOmetaColor{q}  
                  }:
                    Since \(\possiblyWithSub\stageOmetaColor{N'^{\superscriptO} } = \possiblyWithSub\stageOmetaColor{q} = \delta(  \possiblyWithSub\stageOmetaColor{a}_{{\mathrm{1}}}  \    \possiblyWithSub\stageOmetaColor{c}_{{\mathrm{2}}}   ) =  ( \possiblyWithSub\stageOmetaColor{N^{\superscriptO} } )^{\ast} \) holds,
                    we clearly have \( \possiblyWithSub\stageOmetaColor{N'^{\superscriptO} }  \Longrightarrow^{0}   ( \possiblyWithSub\stageOmetaColor{N^{\superscriptO} } )^{\ast}  \)
                    by Lemma~\ref{lem:eq-implies-par}.
                \end{itemize}
              \item Case where \(\possiblyWithSub\stageOmetaColor{N^{\superscriptO} }_{{\mathrm{1}}} =  \LeftAssertParen \relO{\CastArrow}   \openO{\{} \possiblyWithSub\stageOmetaColor{\nu}  \relO{:}  \possiblyWithSub\stageOmetaColor{B}  \relO{\mid}  \possiblyWithSub\stageOmetaColor{N^{\superscriptO} }_{{\mathrm{11}}} \closeO{\} }   \RightAssertParen^{ L } \) and \(\possiblyWithSub\stageOmetaColor{N^{\superscriptO} }_{{\mathrm{2}}} = \possiblyWithSub\stageOmetaColor{c}_{{\mathrm{2}}}\):
                We first have
                \( ( \possiblyWithSub\stageOmetaColor{N^{\superscriptO} } )^{\ast}  =  \LeftAssertParen   \openO{\{} \possiblyWithSub\stageOmetaColor{\nu}  \relO{:}  \possiblyWithSub\stageOmetaColor{B}  \relO{\mid}   ( \possiblyWithSub\stageOmetaColor{N^{\superscriptO} }_{{\mathrm{11}}} )^{\ast}  \closeO{\} }  \punctO{,}    [    \possiblyWithSub\stageOmetaColor{c}_{{\mathrm{2}}}    /  \possiblyWithSub\stageOmetaColor{\nu}  ]     ( \possiblyWithSub\stageOmetaColor{N^{\superscriptO} }_{{\mathrm{11}}} )^{\ast}   \punctO{,}  \possiblyWithSub\stageOmetaColor{c}_{{\mathrm{2}}}  \RightAssertParen^{ L } \).
                By tracing back the derivation of \( \possiblyWithSub\stageOmetaColor{N^{\superscriptO} }  \Longrightarrow^{0}  \possiblyWithSub\stageOmetaColor{N'^{\superscriptO} } \),
                we have the following two cases:
                \begin{itemize}
                  \item Case \derive[P0-App]{%
                      \LeftAssertParen \relO{\CastArrow}   \openO{\{} \possiblyWithSub\stageOmetaColor{\nu}  \relO{:}  \possiblyWithSub\stageOmetaColor{B}  \relO{\mid}  \possiblyWithSub\stageOmetaColor{N^{\superscriptO} }_{{\mathrm{11}}} \closeO{\} }   \RightAssertParen^{ L }   \Longrightarrow^{0}  \possiblyWithSub\stageOmetaColor{N'^{\superscriptO} }_{{\mathrm{1}}} 
                  \andalso
                       \possiblyWithSub\stageOmetaColor{c}_{{\mathrm{2}}}    \Longrightarrow^{0}  \possiblyWithSub\stageOmetaColor{N'^{\superscriptO} }_{{\mathrm{2}}} 
                  }{%
                       \LeftAssertParen \relO{\CastArrow}   \openO{\{} \possiblyWithSub\stageOmetaColor{\nu}  \relO{:}  \possiblyWithSub\stageOmetaColor{B}  \relO{\mid}  \possiblyWithSub\stageOmetaColor{N^{\superscriptO} }_{{\mathrm{11}}} \closeO{\} }   \RightAssertParen^{ L }  \    \possiblyWithSub\stageOmetaColor{c}_{{\mathrm{2}}}     \Longrightarrow^{0}   \possiblyWithSub\stageOmetaColor{N'^{\superscriptO} }_{{\mathrm{1}}} \  \possiblyWithSub\stageOmetaColor{N'^{\superscriptO} }_{{\mathrm{2}}}  
                  }:
                    By IH on \(  \LeftAssertParen \relO{\CastArrow}   \openO{\{} \possiblyWithSub\stageOmetaColor{\nu}  \relO{:}  \possiblyWithSub\stageOmetaColor{B}  \relO{\mid}  \possiblyWithSub\stageOmetaColor{N^{\superscriptO} }_{{\mathrm{11}}} \closeO{\} }   \RightAssertParen^{ L }   \Longrightarrow^{0}  \possiblyWithSub\stageOmetaColor{N'^{\superscriptO} }_{{\mathrm{1}}} \),
                    we have \( \possiblyWithSub\stageOmetaColor{N'^{\superscriptO} }_{{\mathrm{1}}}  \Longrightarrow^{0}   \LeftAssertParen \relO{\CastArrow}   \openO{\{} \possiblyWithSub\stageOmetaColor{\nu}  \relO{:}  \possiblyWithSub\stageOmetaColor{B}  \relO{\mid}   ( \possiblyWithSub\stageOmetaColor{N^{\superscriptO} }_{{\mathrm{11}}} )^{\ast}  \closeO{\} }   \RightAssertParen^{ L }  \).
                    Also, by Lemma~\ref{lem:par-const-normal-form},
                    we have \(\possiblyWithSub\stageOmetaColor{N'^{\superscriptO} }_{{\mathrm{2}}} = \possiblyWithSub\stageOmetaColor{c}_{{\mathrm{2}}}\).
                    Thus, we can derive
                    \begin{center}
                      \derive[P0-RfnStart]{%
                         \possiblyWithSub\stageOmetaColor{N'^{\superscriptO} }_{{\mathrm{1}}}  \Longrightarrow^{0}   \LeftAssertParen \relO{\CastArrow}   \openO{\{} \possiblyWithSub\stageOmetaColor{\nu}  \relO{:}  \possiblyWithSub\stageOmetaColor{B}  \relO{\mid}   ( \possiblyWithSub\stageOmetaColor{N^{\superscriptO} }_{{\mathrm{11}}} )^{\ast}  \closeO{\} }   \RightAssertParen^{ L }  
                      }{%
                          \possiblyWithSub\stageOmetaColor{N'^{\superscriptO} }_{{\mathrm{1}}} \    \possiblyWithSub\stageOmetaColor{c}_{{\mathrm{2}}}     \Longrightarrow^{0}   \LeftAssertParen   \openO{\{} \possiblyWithSub\stageOmetaColor{\nu}  \relO{:}  \possiblyWithSub\stageOmetaColor{B}  \relO{\mid}   ( \possiblyWithSub\stageOmetaColor{N^{\superscriptO} }_{{\mathrm{11}}} )^{\ast}  \closeO{\} }  \punctO{,}    [    \possiblyWithSub\stageOmetaColor{c}_{{\mathrm{2}}}    /  \possiblyWithSub\stageOmetaColor{\nu}  ]     ( \possiblyWithSub\stageOmetaColor{N^{\superscriptO} }_{{\mathrm{11}}} )^{\ast}   \punctO{,}  \possiblyWithSub\stageOmetaColor{c}_{{\mathrm{2}}}  \RightAssertParen^{ L }  
                      }.
                    \end{center}
                  \item Case \derive[P0-RfnStart]{%
                     \possiblyWithSub\stageOmetaColor{N^{\superscriptO} }_{{\mathrm{11}}}  \Longrightarrow^{0}  \possiblyWithSub\stageOmetaColor{N'^{\superscriptO} }_{{\mathrm{11}}} 
                  }{%
                       \LeftAssertParen \relO{\CastArrow}   \openO{\{} \possiblyWithSub\stageOmetaColor{\nu}  \relO{:}  \possiblyWithSub\stageOmetaColor{B}  \relO{\mid}  \possiblyWithSub\stageOmetaColor{N^{\superscriptO} }_{{\mathrm{11}}} \closeO{\} }   \RightAssertParen^{ L }  \    \possiblyWithSub\stageOmetaColor{c}_{{\mathrm{2}}}     \Longrightarrow^{0}   \LeftAssertParen   \openO{\{} \possiblyWithSub\stageOmetaColor{\nu}  \relO{:}  \possiblyWithSub\stageOmetaColor{B}  \relO{\mid}  \possiblyWithSub\stageOmetaColor{N'^{\superscriptO} }_{{\mathrm{11}}} \closeO{\} }  \punctO{,}    [    \possiblyWithSub\stageOmetaColor{c}_{{\mathrm{2}}}    /  \possiblyWithSub\stageOmetaColor{\nu}  ]    \possiblyWithSub\stageOmetaColor{N'^{\superscriptO} }_{{\mathrm{11}}}  \punctO{,}  \possiblyWithSub\stageOmetaColor{c}_{{\mathrm{2}}}  \RightAssertParen^{ L }  
                  }:
                    By IH on \( \possiblyWithSub\stageOmetaColor{N^{\superscriptO} }_{{\mathrm{11}}}  \Longrightarrow^{0}  \possiblyWithSub\stageOmetaColor{N'^{\superscriptO} }_{{\mathrm{11}}} \), we have \( \possiblyWithSub\stageOmetaColor{N'^{\superscriptO} }_{{\mathrm{11}}}  \Longrightarrow^{0}   ( \possiblyWithSub\stageOmetaColor{N^{\superscriptO} }_{{\mathrm{11}}} )^{\ast}  \).
                    Then, by Lemmata~\ref{lem:eq-implies-par} and \ref{lem:par-subst},
                    we have \(   [    \possiblyWithSub\stageOmetaColor{c}_{{\mathrm{2}}}    /  \possiblyWithSub\stageOmetaColor{\nu}  ]    \possiblyWithSub\stageOmetaColor{N'^{\superscriptO} }_{{\mathrm{11}}}   \Longrightarrow^{0}    [    \possiblyWithSub\stageOmetaColor{c}_{{\mathrm{2}}}    /  \possiblyWithSub\stageOmetaColor{\nu}  ]     ( \possiblyWithSub\stageOmetaColor{N^{\superscriptO} }_{{\mathrm{11}}} )^{\ast}   \).
                    Therefore, we have
                    \begin{center}
                      \derive[P0-RfnAct]{%
                         \possiblyWithSub\stageOmetaColor{N'^{\superscriptO} }_{{\mathrm{11}}}  \Longrightarrow^{0}   ( \possiblyWithSub\stageOmetaColor{N^{\superscriptO} }_{{\mathrm{11}}} )^{\ast}  
                      \andalso
                           [    \possiblyWithSub\stageOmetaColor{c}_{{\mathrm{2}}}    /  \possiblyWithSub\stageOmetaColor{\nu}  ]    \possiblyWithSub\stageOmetaColor{N'^{\superscriptO} }_{{\mathrm{11}}}   \Longrightarrow^{0}    [    \possiblyWithSub\stageOmetaColor{c}_{{\mathrm{2}}}    /  \possiblyWithSub\stageOmetaColor{\nu}  ]     ( \possiblyWithSub\stageOmetaColor{N^{\superscriptO} }_{{\mathrm{11}}} )^{\ast}   
                      }{%
                          \LeftAssertParen   \openO{\{} \possiblyWithSub\stageOmetaColor{\nu}  \relO{:}  \possiblyWithSub\stageOmetaColor{B}  \relO{\mid}  \possiblyWithSub\stageOmetaColor{N'^{\superscriptO} }_{{\mathrm{11}}} \closeO{\} }  \punctO{,}    [    \possiblyWithSub\stageOmetaColor{c}_{{\mathrm{2}}}    /  \possiblyWithSub\stageOmetaColor{\nu}  ]    \possiblyWithSub\stageOmetaColor{N'^{\superscriptO} }_{{\mathrm{11}}}  \punctO{,}  \possiblyWithSub\stageOmetaColor{c}_{{\mathrm{2}}}  \RightAssertParen^{ L }   \Longrightarrow^{0}   \LeftAssertParen   \openO{\{} \possiblyWithSub\stageOmetaColor{\nu}  \relO{:}  \possiblyWithSub\stageOmetaColor{B}  \relO{\mid}   ( \possiblyWithSub\stageOmetaColor{N^{\superscriptO} }_{{\mathrm{11}}} )^{\ast}  \closeO{\} }  \punctO{,}    [    \possiblyWithSub\stageOmetaColor{c}_{{\mathrm{2}}}    /  \possiblyWithSub\stageOmetaColor{\nu}  ]     ( \possiblyWithSub\stageOmetaColor{N^{\superscriptO} }_{{\mathrm{11}}} )^{\ast}   \punctO{,}  \possiblyWithSub\stageOmetaColor{c}_{{\mathrm{2}}}  \RightAssertParen^{ L }  
                      }.
                    \end{center}
                \end{itemize}
              \item Otherwise:
                We can trace back the derivation of \( \possiblyWithSub\stageOmetaColor{N^{\superscriptO} }  \Longrightarrow^{0}  \possiblyWithSub\stageOmetaColor{N'^{\superscriptO} } \) as follows:
                \begin{center}
                  \derive[P0-App]{%
                     \possiblyWithSub\stageOmetaColor{N^{\superscriptO} }_{{\mathrm{1}}}  \Longrightarrow^{0}  \possiblyWithSub\stageOmetaColor{N'^{\superscriptO} }_{{\mathrm{1}}} 
                  \andalso
                     \possiblyWithSub\stageOmetaColor{N^{\superscriptO} }_{{\mathrm{2}}}  \Longrightarrow^{0}  \possiblyWithSub\stageOmetaColor{N'^{\superscriptO} }_{{\mathrm{2}}} 
                  }{%
                      \possiblyWithSub\stageOmetaColor{N^{\superscriptO} }_{{\mathrm{1}}} \  \possiblyWithSub\stageOmetaColor{N^{\superscriptO} }_{{\mathrm{2}}}   \Longrightarrow^{0}   \possiblyWithSub\stageOmetaColor{N'^{\superscriptO} }_{{\mathrm{1}}} \  \possiblyWithSub\stageOmetaColor{N'^{\superscriptO} }_{{\mathrm{2}}}  
                  }
                \end{center}
                By IH, we have \( \possiblyWithSub\stageOmetaColor{N'^{\superscriptO} }_{{\mathrm{12}}}  \Longrightarrow^{0}   ( \possiblyWithSub\stageOmetaColor{N^{\superscriptO} }_{{\mathrm{12}}} )^{\ast}  \) and \( \possiblyWithSub\stageOmetaColor{N'^{\superscriptO} }_{{\mathrm{2}}}  \Longrightarrow^{0}   ( \possiblyWithSub\stageOmetaColor{N^{\superscriptO} }_{{\mathrm{2}}} )^{\ast}  \).
                This enables us to derive
                \begin{center}
                  \derive[P0-App]{%
                     \possiblyWithSub\stageOmetaColor{N'^{\superscriptO} }_{{\mathrm{1}}}  \Longrightarrow^{0}   ( \possiblyWithSub\stageOmetaColor{N^{\superscriptO} }_{{\mathrm{1}}} )^{\ast}  
                  \andalso
                     \possiblyWithSub\stageOmetaColor{N'^{\superscriptO} }_{{\mathrm{2}}}  \Longrightarrow^{0}   ( \possiblyWithSub\stageOmetaColor{N^{\superscriptO} }_{{\mathrm{2}}} )^{\ast}  
                  }{%
                      \possiblyWithSub\stageOmetaColor{N'^{\superscriptO} }_{{\mathrm{1}}} \  \possiblyWithSub\stageOmetaColor{N'^{\superscriptO} }_{{\mathrm{2}}}   \Longrightarrow^{0}    ( \possiblyWithSub\stageOmetaColor{N^{\superscriptO} }_{{\mathrm{1}}} )^{\ast}  \   ( \possiblyWithSub\stageOmetaColor{N^{\superscriptO} }_{{\mathrm{2}}} )^{\ast}   
                  }.
                \end{center}
            \end{itemize}
          \item Case \(\possiblyWithSub\stageOmetaColor{N^{\superscriptO} } =  \LeftAssertParen   \openO{\{} \possiblyWithSub\stageOmetaColor{\nu}  \relO{:}  \possiblyWithSub\stageOmetaColor{B}  \relO{\mid}  \possiblyWithSub\stageOmetaColor{N^{\superscriptO} }_{{\mathrm{1}}} \closeO{\} }  \punctO{,}  \possiblyWithSub\stageOmetaColor{N^{\superscriptO} }_{{\mathrm{2}}} \punctO{,}  \possiblyWithSub\stageOmetaColor{c}  \RightAssertParen^{ L } \):
            We do the following case analysis:
            \begin{itemize}
              \item Case \(\possiblyWithSub\stageOmetaColor{N^{\superscriptO} }_{{\mathrm{2}}} =   \ttO{true}  \):
                First, we have \( ( \possiblyWithSub\stageOmetaColor{N^{\superscriptO} } )^{\ast}  = \possiblyWithSub\stageOmetaColor{c}\).
                By tracing back the derivation of \( \possiblyWithSub\stageOmetaColor{N^{\superscriptO} }  \Longrightarrow^{0}  \possiblyWithSub\stageOmetaColor{N'^{\superscriptO} } \),
                we have the following two cases:
                \begin{itemize}
                  \item Case \derive[P0-RfnAct]{%
                     \possiblyWithSub\stageOmetaColor{N^{\superscriptO} }_{{\mathrm{1}}}  \Longrightarrow^{0}  \possiblyWithSub\stageOmetaColor{N'^{\superscriptO} }_{{\mathrm{1}}} 
                  \andalso
                        \ttO{true}     \Longrightarrow^{0}  \possiblyWithSub\stageOmetaColor{N'^{\superscriptO} }_{{\mathrm{2}}} 
                  }{%
                      \LeftAssertParen   \openO{\{} \possiblyWithSub\stageOmetaColor{\nu}  \relO{:}  \possiblyWithSub\stageOmetaColor{B}  \relO{\mid}  \possiblyWithSub\stageOmetaColor{N^{\superscriptO} }_{{\mathrm{1}}} \closeO{\} }  \punctO{,}     \ttO{true}    \punctO{,}  \possiblyWithSub\stageOmetaColor{c}  \RightAssertParen^{ L }   \Longrightarrow^{0}   \LeftAssertParen   \openO{\{} \possiblyWithSub\stageOmetaColor{\nu}  \relO{:}  \possiblyWithSub\stageOmetaColor{B}  \relO{\mid}  \possiblyWithSub\stageOmetaColor{N'^{\superscriptO} }_{{\mathrm{1}}} \closeO{\} }  \punctO{,}  \possiblyWithSub\stageOmetaColor{N'^{\superscriptO} }_{{\mathrm{2}}} \punctO{,}  \possiblyWithSub\stageOmetaColor{c}  \RightAssertParen^{ L }  
                  }:
                    By Lemma~\ref{lem:par-const-normal-form}, \(\possiblyWithSub\stageOmetaColor{N'^{\superscriptO} }_{{\mathrm{2}}} =   \ttO{true}  \) holds.
                    This enables us to derive
                    \begin{center}
                      \derive[P0-RfnPass]{}{%
                          \LeftAssertParen   \openO{\{} \possiblyWithSub\stageOmetaColor{\nu}  \relO{:}  \possiblyWithSub\stageOmetaColor{B}  \relO{\mid}  \possiblyWithSub\stageOmetaColor{N'^{\superscriptO} }_{{\mathrm{1}}} \closeO{\} }  \punctO{,}     \ttO{true}    \punctO{,}  \possiblyWithSub\stageOmetaColor{c}  \RightAssertParen^{ L }   \Longrightarrow^{0}    \possiblyWithSub\stageOmetaColor{c}   
                      }.
                    \end{center}
                  \item Case \derive[P0-RfnPass]{}{%
                      \LeftAssertParen   \openO{\{} \possiblyWithSub\stageOmetaColor{\nu}  \relO{:}  \possiblyWithSub\stageOmetaColor{B}  \relO{\mid}  \possiblyWithSub\stageOmetaColor{N^{\superscriptO} }_{{\mathrm{1}}} \closeO{\} }  \punctO{,}     \ttO{true}    \punctO{,}  \possiblyWithSub\stageOmetaColor{c}  \RightAssertParen^{ L }   \Longrightarrow^{0}    \possiblyWithSub\stageOmetaColor{c}   
                  }:
                    This case is immediate by Lemma~\ref{lem:eq-implies-par}.
                \end{itemize}
              \item Otherwise:
                We have \( ( \possiblyWithSub\stageOmetaColor{N^{\superscriptO} } )^{\ast}  =  \LeftAssertParen   \openO{\{} \possiblyWithSub\stageOmetaColor{\nu}  \relO{:}  \possiblyWithSub\stageOmetaColor{B}  \relO{\mid}   ( \possiblyWithSub\stageOmetaColor{N^{\superscriptO} }_{{\mathrm{1}}} )^{\ast}  \closeO{\} }  \punctO{,}   ( \possiblyWithSub\stageOmetaColor{N^{\superscriptO} }_{{\mathrm{2}}} )^{\ast}  \punctO{,}  \possiblyWithSub\stageOmetaColor{c}  \RightAssertParen^{ L } \).
                The derivation of \( \possiblyWithSub\stageOmetaColor{N^{\superscriptO} }  \Longrightarrow^{0}  \possiblyWithSub\stageOmetaColor{N'^{\superscriptO} } \) can only be the following:
                \begin{center}
                  \derive[P0-RfnAct]{%
                     \possiblyWithSub\stageOmetaColor{N^{\superscriptO} }_{{\mathrm{1}}}  \Longrightarrow^{0}  \possiblyWithSub\stageOmetaColor{N'^{\superscriptO} }_{{\mathrm{1}}} 
                  \andalso
                     \possiblyWithSub\stageOmetaColor{N^{\superscriptO} }_{{\mathrm{2}}}  \Longrightarrow^{0}  \possiblyWithSub\stageOmetaColor{N'^{\superscriptO} }_{{\mathrm{2}}} 
                  }{%
                      \LeftAssertParen   \openO{\{} \possiblyWithSub\stageOmetaColor{\nu}  \relO{:}  \possiblyWithSub\stageOmetaColor{B}  \relO{\mid}  \possiblyWithSub\stageOmetaColor{N^{\superscriptO} }_{{\mathrm{1}}} \closeO{\} }  \punctO{,}  \possiblyWithSub\stageOmetaColor{N^{\superscriptO} }_{{\mathrm{2}}} \punctO{,}  \possiblyWithSub\stageOmetaColor{c}  \RightAssertParen^{ L }   \Longrightarrow^{0}   \LeftAssertParen   \openO{\{} \possiblyWithSub\stageOmetaColor{\nu}  \relO{:}  \possiblyWithSub\stageOmetaColor{B}  \relO{\mid}  \possiblyWithSub\stageOmetaColor{N'^{\superscriptO} }_{{\mathrm{1}}} \closeO{\} }  \punctO{,}  \possiblyWithSub\stageOmetaColor{N'^{\superscriptO} }_{{\mathrm{2}}} \punctO{,}  \possiblyWithSub\stageOmetaColor{c}  \RightAssertParen^{ L }  
                  }:
                    By IH, we have \( \possiblyWithSub\stageOmetaColor{N'^{\superscriptO} }_{{\mathrm{1}}}  \Longrightarrow^{0}   ( \possiblyWithSub\stageOmetaColor{N^{\superscriptO} }_{{\mathrm{1}}} )^{\ast}  \) and \( \possiblyWithSub\stageOmetaColor{N'^{\superscriptO} }_{{\mathrm{2}}}  \Longrightarrow^{0}   ( \possiblyWithSub\stageOmetaColor{N^{\superscriptO} }_{{\mathrm{2}}} )^{\ast}  \),
                    and thus we can derive
                    \begin{center}
                      \derive[P0-RfnAct]{%
                         \possiblyWithSub\stageOmetaColor{N'^{\superscriptO} }_{{\mathrm{1}}}  \Longrightarrow^{0}   ( \possiblyWithSub\stageOmetaColor{N^{\superscriptO} }_{{\mathrm{1}}} )^{\ast}  
                      \andalso
                         \possiblyWithSub\stageOmetaColor{N'^{\superscriptO} }_{{\mathrm{2}}}  \Longrightarrow^{0}   ( \possiblyWithSub\stageOmetaColor{N^{\superscriptO} }_{{\mathrm{2}}} )^{\ast}  
                      }{%
                          \LeftAssertParen   \openO{\{} \possiblyWithSub\stageOmetaColor{\nu}  \relO{:}  \possiblyWithSub\stageOmetaColor{B}  \relO{\mid}  \possiblyWithSub\stageOmetaColor{N'^{\superscriptO} }_{{\mathrm{1}}} \closeO{\} }  \punctO{,}  \possiblyWithSub\stageOmetaColor{N'^{\superscriptO} }_{{\mathrm{2}}} \punctO{,}  \possiblyWithSub\stageOmetaColor{c}  \RightAssertParen^{ L }   \Longrightarrow^{0}   \LeftAssertParen   \openO{\{} \possiblyWithSub\stageOmetaColor{\nu}  \relO{:}  \possiblyWithSub\stageOmetaColor{B}  \relO{\mid}   ( \possiblyWithSub\stageOmetaColor{N^{\superscriptO} }_{{\mathrm{1}}} )^{\ast}  \closeO{\} }  \punctO{,}   ( \possiblyWithSub\stageOmetaColor{N^{\superscriptO} }_{{\mathrm{2}}} )^{\ast}  \punctO{,}  \possiblyWithSub\stageOmetaColor{c}  \RightAssertParen^{ L }  
                      }.
                    \end{center}
                \end{center}
            \end{itemize}
          \item The other cases are immediate or straightforward by IH.
        \end{itemize}
      \item
        \begin{itemize}
          \item Case \(\possiblyWithSub\stageImetaColor{N^{\superscriptI} } =  \ordI{\sim} \possiblyWithSub\stageOmetaColor{N^{\superscriptO} } \):
            We do a further case analysis on \(\possiblyWithSub\stageOmetaColor{N^{\superscriptO} }\):
            \begin{itemize}
              \item Case \(\possiblyWithSub\stageOmetaColor{N^{\superscriptO} } =  \openO{\langle} \possiblyWithSub\stageImetaColor{N^{\superscriptI} }_{{\mathrm{0}}} \closeO{\rangle} \):
                We first have \( ( \possiblyWithSub\stageImetaColor{N^{\superscriptI} } )^{\ast}  =  (  \ordI{\sim}  \openO{\langle} \possiblyWithSub\stageImetaColor{N^{\superscriptI} }_{{\mathrm{0}}} \closeO{\rangle}   )^{\ast}  =  ( \possiblyWithSub\stageImetaColor{N^{\superscriptI} }_{{\mathrm{0}}} )^{\ast} \).
                Considering the derivation of \( \possiblyWithSub\stageImetaColor{N^{\superscriptI} }  \Longrightarrow^{1}  \possiblyWithSub\stageImetaColor{N'^{\superscriptI} } \),
                we have the following two cases:
                \begin{itemize}
                  \item Case \derive[P1-Esc]{%
                      \openO{\langle} \possiblyWithSub\stageImetaColor{N^{\superscriptI} }_{{\mathrm{0}}} \closeO{\rangle}   \Longrightarrow^{0}  \possiblyWithSub\stageOmetaColor{N'^{\superscriptO} } 
                  }{%
                      \ordI{\sim}  \openO{\langle} \possiblyWithSub\stageImetaColor{N^{\superscriptI} }_{{\mathrm{0}}} \closeO{\rangle}    \Longrightarrow^{1}   \ordI{\sim} \possiblyWithSub\stageOmetaColor{N'^{\superscriptO} }  
                  }:
                    We can further trace back the derivation of \(  \openO{\langle} \possiblyWithSub\stageImetaColor{N^{\superscriptI} }_{{\mathrm{0}}} \closeO{\rangle}   \Longrightarrow^{0}  \possiblyWithSub\stageOmetaColor{N'^{\superscriptO} } \) as follows:
                    \begin{center}
                      \derive[P0-Brkt]{%
                         \possiblyWithSub\stageImetaColor{N^{\superscriptI} }_{{\mathrm{0}}}  \Longrightarrow^{1}  \possiblyWithSub\stageImetaColor{N'^{\superscriptI} }_{{\mathrm{0}}} 
                      }{%
                          \openO{\langle} \possiblyWithSub\stageImetaColor{N^{\superscriptI} }_{{\mathrm{0}}} \closeO{\rangle}   \Longrightarrow^{0}   \openO{\langle} \possiblyWithSub\stageImetaColor{N'^{\superscriptI} }_{{\mathrm{0}}} \closeO{\rangle}  
                      }
                    \end{center}
                    Then, by IH, from \( \possiblyWithSub\stageImetaColor{N^{\superscriptI} }_{{\mathrm{0}}}  \Longrightarrow^{1}  \possiblyWithSub\stageImetaColor{N'^{\superscriptI} }_{{\mathrm{0}}} \),
                    we have \( \possiblyWithSub\stageImetaColor{N'^{\superscriptI} }_{{\mathrm{0}}}  \Longrightarrow^{1}   ( \possiblyWithSub\stageImetaColor{N^{\superscriptI} }_{{\mathrm{0}}} )^{\ast}  \).
                    Since \(\possiblyWithSub\stageImetaColor{N'^{\superscriptI} } =  \ordI{\sim} \possiblyWithSub\stageOmetaColor{N'^{\superscriptO} }  =  \ordI{\sim}  \openO{\langle} \possiblyWithSub\stageImetaColor{N'^{\superscriptI} }_{{\mathrm{0}}} \closeO{\rangle}  \),
                    we can derive \( \possiblyWithSub\stageImetaColor{N'^{\superscriptI} }  \Longrightarrow^{1}   ( \possiblyWithSub\stageImetaColor{N^{\superscriptI} } )^{\ast}  \) as follows:
                    \begin{center}
                      \derive[P1-Cancel]{%
                         \possiblyWithSub\stageImetaColor{N'^{\superscriptI} }_{{\mathrm{0}}}  \Longrightarrow^{1}   ( \possiblyWithSub\stageImetaColor{N^{\superscriptI} }_{{\mathrm{0}}} )^{\ast}  
                      }{%
                          \ordI{\sim}  \openO{\langle} \possiblyWithSub\stageImetaColor{N'^{\superscriptI} }_{{\mathrm{0}}} \closeO{\rangle}    \Longrightarrow^{1}   ( \possiblyWithSub\stageImetaColor{N^{\superscriptI} }_{{\mathrm{0}}} )^{\ast}  
                      }
                    \end{center}
                  \item Case \derive[P1-Cancel]{%
                     \possiblyWithSub\stageImetaColor{N^{\superscriptI} }_{{\mathrm{0}}}  \Longrightarrow^{1}  \possiblyWithSub\stageImetaColor{N'^{\superscriptI} } 
                  }{%
                      \ordI{\sim}  \openO{\langle} \possiblyWithSub\stageImetaColor{N^{\superscriptI} }_{{\mathrm{0}}} \closeO{\rangle}    \Longrightarrow^{1}  \possiblyWithSub\stageImetaColor{N'^{\superscriptI} } 
                  }:
                    By IH, from \( \possiblyWithSub\stageImetaColor{N^{\superscriptI} }_{{\mathrm{0}}}  \Longrightarrow^{1}  \possiblyWithSub\stageImetaColor{N'^{\superscriptI} } \), we have \( \possiblyWithSub\stageImetaColor{N'^{\superscriptI} }  \Longrightarrow^{1}   ( \possiblyWithSub\stageImetaColor{N^{\superscriptI} }_{{\mathrm{0}}} )^{\ast}  \).
                    Since \( ( \possiblyWithSub\stageImetaColor{N^{\superscriptI} } )^{\ast}  =  ( \possiblyWithSub\stageImetaColor{N^{\superscriptI} }_{{\mathrm{0}}} )^{\ast} \), we have \( \possiblyWithSub\stageImetaColor{N'^{\superscriptI} }  \Longrightarrow^{1}   ( \possiblyWithSub\stageImetaColor{N^{\superscriptI} } )^{\ast}  \).
                \end{itemize}
              \item Otherwise:
                We first have \( ( \possiblyWithSub\stageImetaColor{N^{\superscriptI} } )^{\ast}  =  \ordI{\sim}  ( \possiblyWithSub\stageOmetaColor{N^{\superscriptO} } )^{\ast}  \).
                We can trace back the derivation of \( \possiblyWithSub\stageImetaColor{N^{\superscriptI} }  \Longrightarrow^{1}  \possiblyWithSub\stageImetaColor{N'^{\superscriptI} } \) as follows:
                \begin{center}
                  \derive[P1-Esc]{%
                     \possiblyWithSub\stageOmetaColor{N^{\superscriptO} }  \Longrightarrow^{0}  \possiblyWithSub\stageOmetaColor{N'^{\superscriptO} } 
                  }{%
                      \ordI{\sim} \possiblyWithSub\stageOmetaColor{N^{\superscriptO} }   \Longrightarrow^{1}   \ordI{\sim} \possiblyWithSub\stageOmetaColor{N'^{\superscriptO} }  
                  }
                \end{center}
                By IH, from \( \possiblyWithSub\stageOmetaColor{N^{\superscriptO} }  \Longrightarrow^{0}  \possiblyWithSub\stageOmetaColor{N'^{\superscriptO} } \), we have \( \possiblyWithSub\stageOmetaColor{N'^{\superscriptO} }  \Longrightarrow^{0}   ( \possiblyWithSub\stageOmetaColor{N^{\superscriptO} } )^{\ast}  \).
                Therefore, we can derive
                \begin{center}
                  \derive[P1-Esc]{%
                     \possiblyWithSub\stageOmetaColor{N'^{\superscriptO} }  \Longrightarrow^{0}   ( \possiblyWithSub\stageOmetaColor{N^{\superscriptO} } )^{\ast}  
                  }{%
                      \ordI{\sim} \possiblyWithSub\stageOmetaColor{N'^{\superscriptO} }   \Longrightarrow^{1}   \ordI{\sim}  ( \possiblyWithSub\stageOmetaColor{N^{\superscriptO} } )^{\ast}   
                  }.
                \end{center}
            \end{itemize}
          \item The other cases are immediate or straightforward by IH.
        \end{itemize}
      \item All the cases are immediate or straightforward by IH.
      \item Again, all the cases are immediate or straightforward by IH.
    \end{enumerate}
  \end{proof}
  \begin{corollary}[Confluence]\label{cor:confluence}
    If \( \possiblyWithSub\stageImetaColor{T^{\superscriptI} }  \Longrightarrow^{1}  \possiblyWithSub\stageImetaColor{T^{\superscriptI} }_{{\mathrm{1}}} \) and \( \possiblyWithSub\stageImetaColor{T^{\superscriptI} }  \Longrightarrow^{1}  \possiblyWithSub\stageImetaColor{T^{\superscriptI} }_{{\mathrm{2}}} \),
    then there exists \(\possiblyWithSub\stageImetaColor{T'^{\superscriptI} }\) such that
    \( \possiblyWithSub\stageImetaColor{T^{\superscriptI} }_{{\mathrm{1}}}  \Longrightarrow^{1}  \possiblyWithSub\stageImetaColor{T'^{\superscriptI} } \) and \( \possiblyWithSub\stageImetaColor{T^{\superscriptI} }_{{\mathrm{2}}}  \Longrightarrow^{1}  \possiblyWithSub\stageImetaColor{T'^{\superscriptI} } \).
  \end{corollary}
  \begin{proof}
    Immediate from Lemma~\ref{lem:confluence-by-star}.
  \end{proof}
  \begin{lemma}[Church---Rosser]\label{lem:church-rosser}
    If \( \possiblyWithSub\stageImetaColor{T^{\superscriptI} }  \Longrightarrow^{1\,\ast}  \possiblyWithSub\stageImetaColor{T^{\superscriptI} }_{{\mathrm{1}}} \) and \( \possiblyWithSub\stageImetaColor{T^{\superscriptI} }  \Longrightarrow^{1\,\ast}  \possiblyWithSub\stageImetaColor{T^{\superscriptI} }_{{\mathrm{2}}} \),
    then there exists \(\possiblyWithSub\stageImetaColor{T'^{\superscriptI} }\) such that
    \( \possiblyWithSub\stageImetaColor{T^{\superscriptI} }_{{\mathrm{1}}}  \Longrightarrow^{1\,\ast}  \possiblyWithSub\stageImetaColor{T'^{\superscriptI} } \) and \( \possiblyWithSub\stageImetaColor{T^{\superscriptI} }_{{\mathrm{2}}}  \Longrightarrow^{1\,\ast}  \possiblyWithSub\stageImetaColor{T'^{\superscriptI} } \).
  \end{lemma}
  \begin{proof}
    By repeated use of Corollary~\ref{cor:confluence}.
  \end{proof}
  \begin{lemma}\label{lem:refl-trans-symm-par-implies-confluence}
    If \( \possiblyWithSub\stageImetaColor{T^{\superscriptI} }_{{\mathrm{1}}}  \Longleftrightarrow^{1\,\ast}  \possiblyWithSub\stageImetaColor{T^{\superscriptI} }_{{\mathrm{2}}} \), then there exists \(\possiblyWithSub\stageImetaColor{T'^{\superscriptI} }\) such that
    \( \possiblyWithSub\stageImetaColor{T^{\superscriptI} }_{{\mathrm{1}}}  \Longrightarrow^{1\,\ast}  \possiblyWithSub\stageImetaColor{T'^{\superscriptI} } \) and \( \possiblyWithSub\stageImetaColor{T^{\superscriptI} }_{{\mathrm{2}}}  \Longrightarrow^{1\,\ast}  \possiblyWithSub\stageImetaColor{T'^{\superscriptI} } \).
  \end{lemma}
  \begin{proof}
    By induction on the ``length'' of \( \possiblyWithSub\stageImetaColor{T^{\superscriptI} }_{{\mathrm{1}}}  \Longleftrightarrow^{1\,\ast}  \possiblyWithSub\stageImetaColor{T^{\superscriptI} }_{{\mathrm{2}}} \).
    \begin{itemize}
      \item Case \(\possiblyWithSub\stageImetaColor{T^{\superscriptI} }_{{\mathrm{1}}} = \possiblyWithSub\stageImetaColor{T^{\superscriptI} }_{{\mathrm{2}}}\):
        Immediate by Lemma~\ref{lem:eq-implies-par}.
      \item Case where there exists \(\possiblyWithSub\stageImetaColor{T^{\superscriptI} }_{{\mathrm{0}}}\) such that
      \( \possiblyWithSub\stageImetaColor{T^{\superscriptI} }_{{\mathrm{1}}}  \Longrightarrow^{1}  \possiblyWithSub\stageImetaColor{T^{\superscriptI} }_{{\mathrm{0}}} \) and \( \possiblyWithSub\stageImetaColor{T^{\superscriptI} }_{{\mathrm{0}}}  \Longleftrightarrow^{1\,\ast}  \possiblyWithSub\stageImetaColor{T^{\superscriptI} }_{{\mathrm{2}}} \):
        By IH on \( \possiblyWithSub\stageImetaColor{T^{\superscriptI} }_{{\mathrm{0}}}  \Longleftrightarrow^{1\,\ast}  \possiblyWithSub\stageImetaColor{T^{\superscriptI} }_{{\mathrm{2}}} \), which is a ``strictly shorter'' sequence,
        there exists \(\possiblyWithSub\stageImetaColor{T'^{\superscriptI} }\) such that
        \( \possiblyWithSub\stageImetaColor{T^{\superscriptI} }_{{\mathrm{0}}}  \Longrightarrow^{1\,\ast}  \possiblyWithSub\stageImetaColor{T'^{\superscriptI} } \) and \( \possiblyWithSub\stageImetaColor{T^{\superscriptI} }_{{\mathrm{2}}}  \Longrightarrow^{1\,\ast}  \possiblyWithSub\stageImetaColor{T'^{\superscriptI} } \).
        By transitivity, from \( \possiblyWithSub\stageImetaColor{T^{\superscriptI} }_{{\mathrm{1}}}  \Longrightarrow^{1}  \possiblyWithSub\stageImetaColor{T^{\superscriptI} }_{{\mathrm{0}}} \) and \( \possiblyWithSub\stageImetaColor{T^{\superscriptI} }_{{\mathrm{0}}}  \Longrightarrow^{1\,\ast}  \possiblyWithSub\stageImetaColor{T'^{\superscriptI} } \),
        we also have \( \possiblyWithSub\stageImetaColor{T^{\superscriptI} }_{{\mathrm{1}}}  \Longrightarrow^{1\,\ast}  \possiblyWithSub\stageImetaColor{T'^{\superscriptI} } \).
      \item Case where there exists \(\possiblyWithSub\stageImetaColor{T^{\superscriptI} }_{{\mathrm{0}}}\) such that
      \( \possiblyWithSub\stageImetaColor{T^{\superscriptI} }_{{\mathrm{0}}}  \Longrightarrow^{1}  \possiblyWithSub\stageImetaColor{T^{\superscriptI} }_{{\mathrm{1}}} \) and \( \possiblyWithSub\stageImetaColor{T^{\superscriptI} }_{{\mathrm{0}}}  \Longleftrightarrow^{1\,\ast}  \possiblyWithSub\stageImetaColor{T^{\superscriptI} }_{{\mathrm{2}}} \):
        Similarly to the previous case, by IH, there exists \(\possiblyWithSub\stageImetaColor{T'^{\superscriptI} }_{{\mathrm{0}}}\) such that
        \( \possiblyWithSub\stageImetaColor{T^{\superscriptI} }_{{\mathrm{0}}}  \Longrightarrow^{1\,\ast}  \possiblyWithSub\stageImetaColor{T'^{\superscriptI} }_{{\mathrm{0}}} \) and \( \possiblyWithSub\stageImetaColor{T^{\superscriptI} }_{{\mathrm{2}}}  \Longrightarrow^{1\,\ast}  \possiblyWithSub\stageImetaColor{T'^{\superscriptI} }_{{\mathrm{0}}} \).
        Then, by Lemma~\ref{lem:church-rosser},
        from \( \possiblyWithSub\stageImetaColor{T^{\superscriptI} }_{{\mathrm{0}}}  \Longrightarrow^{1}  \possiblyWithSub\stageImetaColor{T^{\superscriptI} }_{{\mathrm{1}}} \) and \( \possiblyWithSub\stageImetaColor{T^{\superscriptI} }_{{\mathrm{0}}}  \Longrightarrow^{1\,\ast}  \possiblyWithSub\stageImetaColor{T'^{\superscriptI} }_{{\mathrm{0}}} \),
        there exists \(\possiblyWithSub\stageImetaColor{T'^{\superscriptI} }\) such that
        \( \possiblyWithSub\stageImetaColor{T^{\superscriptI} }_{{\mathrm{1}}}  \Longrightarrow^{1\,\ast}  \possiblyWithSub\stageImetaColor{T'^{\superscriptI} } \) and \( \possiblyWithSub\stageImetaColor{T'^{\superscriptI} }_{{\mathrm{0}}}  \Longrightarrow^{1\,\ast}  \possiblyWithSub\stageImetaColor{T'^{\superscriptI} } \).
        By transitivity, from \( \possiblyWithSub\stageImetaColor{T^{\superscriptI} }_{{\mathrm{2}}}  \Longrightarrow^{1\,\ast}  \possiblyWithSub\stageImetaColor{T'^{\superscriptI} }_{{\mathrm{0}}} \) and \( \possiblyWithSub\stageImetaColor{T'^{\superscriptI} }_{{\mathrm{0}}}  \Longrightarrow^{1\,\ast}  \possiblyWithSub\stageImetaColor{T'^{\superscriptI} } \),
        we also have \( \possiblyWithSub\stageImetaColor{T^{\superscriptI} }_{{\mathrm{2}}}  \Longrightarrow^{1\,\ast}  \possiblyWithSub\stageImetaColor{T'^{\superscriptI} } \).
    \end{itemize}
  \end{proof}
  \begin{corollary}\label{cor:beta-equiv-implies-confluence}
    If \( \possiblyWithSub\stageImetaColor{T^{\superscriptI} }_{{\mathrm{1}}}  \cong^{1}  \possiblyWithSub\stageImetaColor{T^{\superscriptI} }_{{\mathrm{2}}} \), then there exists \(\possiblyWithSub\stageImetaColor{T'^{\superscriptI} }\) such that
    \( \possiblyWithSub\stageImetaColor{T^{\superscriptI} }_{{\mathrm{1}}}  \Longrightarrow^{1\,\ast}  \possiblyWithSub\stageImetaColor{T'^{\superscriptI} } \) and \( \possiblyWithSub\stageImetaColor{T^{\superscriptI} }_{{\mathrm{2}}}  \Longrightarrow^{1\,\ast}  \possiblyWithSub\stageImetaColor{T'^{\superscriptI} } \).
  \end{corollary}
  \begin{proof}
    Immediate from Corollary~\ref{cor:equivalence-of-equiv-and-refl-trans-symm-par}
    and Lemma~\ref{lem:refl-trans-symm-par-implies-confluence}.
  \end{proof}
  \begin{lemma}\label{lem:type-values-are-normal}
    \(  \possiblyWithSub\stageImetaColor{\tau^{\superscriptI} }   \Longrightarrow^{1\,\ast}   \possiblyWithSub\stageImetaColor{\tau'^{\superscriptI} }  \) implies \(\possiblyWithSub\stageImetaColor{\tau'^{\superscriptI} } = \possiblyWithSub\stageImetaColor{\tau^{\superscriptI} }\).
  \end{lemma}
  \begin{proof}
    By induction on the ``length'' of \(  \possiblyWithSub\stageImetaColor{\tau^{\superscriptI} }   \Longrightarrow^{1\,\ast}   \possiblyWithSub\stageImetaColor{\tau'^{\superscriptI} }  \).
    \begin{itemize}
      \item Case \(\possiblyWithSub\stageImetaColor{\tau^{\superscriptI} } = \possiblyWithSub\stageImetaColor{\tau'^{\superscriptI} }\) is immediate.
      \item Case where \(  \possiblyWithSub\stageImetaColor{\tau^{\superscriptI} }   \Longrightarrow^{1}  \possiblyWithSub\stageImetaColor{T^{\superscriptI} } \) and \( \possiblyWithSub\stageImetaColor{T^{\superscriptI} }  \Longrightarrow^{1\,\ast}   \possiblyWithSub\stageImetaColor{\tau'^{\superscriptI} }  \) for some \(\possiblyWithSub\stageImetaColor{T^{\superscriptI} }\):
        By \(  \possiblyWithSub\stageImetaColor{\tau^{\superscriptI} }   \Longrightarrow^{1}  \possiblyWithSub\stageImetaColor{T^{\superscriptI} } \) and Lemma~\ref{lem:par-type-normal-form},
        we have \(\possiblyWithSub\stageImetaColor{T^{\superscriptI} } = \possiblyWithSub\stageImetaColor{\tau^{\superscriptI} }\). Therefore, by IH on \( \possiblyWithSub\stageImetaColor{T^{\superscriptI} }  \Longrightarrow^{1\,\ast}   \possiblyWithSub\stageImetaColor{\tau'^{\superscriptI} }  \),
        which is a ``strictly shorter'' reduction sequence,
        we have \(\possiblyWithSub\stageImetaColor{\tau'^{\superscriptI} } = \possiblyWithSub\stageImetaColor{\tau^{\superscriptI} }\).
    \end{itemize}
  \end{proof}
  \begin{lemma}\label{lem:value-type-beta-equiv-implies-equal}
    \(  \possiblyWithSub\stageImetaColor{\tau^{\superscriptI} }_{{\mathrm{1}}}   \cong^{1}   \possiblyWithSub\stageImetaColor{\tau^{\superscriptI} }_{{\mathrm{2}}}  \) implies \(\possiblyWithSub\stageImetaColor{\tau^{\superscriptI} }_{{\mathrm{1}}} = \possiblyWithSub\stageImetaColor{\tau^{\superscriptI} }_{{\mathrm{2}}}\).
  \end{lemma}
  \begin{proof}
    By \(  \possiblyWithSub\stageImetaColor{\tau^{\superscriptI} }_{{\mathrm{1}}}   \cong^{1}   \possiblyWithSub\stageImetaColor{\tau^{\superscriptI} }_{{\mathrm{2}}}  \) and Corollary~\ref{cor:beta-equiv-implies-confluence},
    there exists \(\possiblyWithSub\stageImetaColor{T'^{\superscriptI} }\) such that
    \(  \possiblyWithSub\stageImetaColor{\tau^{\superscriptI} }_{{\mathrm{1}}}   \Longrightarrow^{1\,\ast}  \possiblyWithSub\stageImetaColor{T'^{\superscriptI} } \) and \(  \possiblyWithSub\stageImetaColor{\tau^{\superscriptI} }_{{\mathrm{2}}}   \Longrightarrow^{1\,\ast}  \possiblyWithSub\stageImetaColor{T'^{\superscriptI} } \).
    Therefore, by Lemma~\ref{lem:type-values-are-normal},
    we have \(\possiblyWithSub\stageImetaColor{\tau^{\superscriptI} }_{{\mathrm{1}}} = \possiblyWithSub\stageImetaColor{T'^{\superscriptI} } = \possiblyWithSub\stageImetaColor{\tau^{\superscriptI} }_{{\mathrm{2}}}\).
  \end{proof}
  \recalllemma{lem:value-type-csr-equiv-implies-equal}{%
    \(  \possiblyWithSub\stageImetaColor{\tau^{\superscriptI} }_{{\mathrm{1}}}   \equiv^{1}   \possiblyWithSub\stageImetaColor{\tau^{\superscriptI} }_{{\mathrm{2}}}  \) implies \(\possiblyWithSub\stageImetaColor{\tau^{\superscriptI} }_{{\mathrm{1}}} = \possiblyWithSub\stageImetaColor{\tau^{\superscriptI} }_{{\mathrm{2}}}\).
  }
  \begin{proof}
    Immediate from Lemmata~\ref{lem:csr-equiv-implies-beta-equiv-1}
    and \ref{lem:value-type-beta-equiv-implies-equal}.
  \end{proof}
  \begin{lemma}[CSR equivalence preserves the form of stage-\(1\) function types]\label{lem:stage-1-arrow-csr-type-equiv-form}
    If \(  \possiblyWithSub\stageImetaColor{T^{\superscriptI} }_{{\mathrm{11}}}  \relI{\to}  \possiblyWithSub\stageImetaColor{T^{\superscriptI} }_{{\mathrm{12}}}   \equiv^{1}  \possiblyWithSub\stageImetaColor{T^{\superscriptI} } \), then \(\possiblyWithSub\stageImetaColor{T^{\superscriptI} }\) is of the form~\( \possiblyWithSub\stageImetaColor{T^{\superscriptI} }_{{\mathrm{21}}}  \relI{\to}  \possiblyWithSub\stageImetaColor{T^{\superscriptI} }_{{\mathrm{22}}} \),
    and we have \( \possiblyWithSub\stageImetaColor{T^{\superscriptI} }_{{\mathrm{11}}}  \equiv^{1}  \possiblyWithSub\stageImetaColor{T^{\superscriptI} }_{{\mathrm{21}}} \) and \( \possiblyWithSub\stageImetaColor{T^{\superscriptI} }_{{\mathrm{12}}}  \equiv^{1}  \possiblyWithSub\stageImetaColor{T^{\superscriptI} }_{{\mathrm{22}}} \).
  \end{lemma}
  \begin{proof}
    By straightforward induction similar to Lemma~\ref{lem:arrow-type-csr-equiv-form}.
  \end{proof}
  \recalllemma{lem:final-target-typing}{%
    If \( \vdash^{1}  \mathit{\Gamma} \), \( \mathit{\Gamma}  \vdash^{1}   \possiblyWithSub\stageImetaColor{v^{\superscriptI} }   :  \possiblyWithSub\stageImetaColor{T^{\superscriptI} } \), and \( \mathit{\Gamma}  \equiv  \gamma \),
    then there exists \(\possiblyWithSub\stageImetaColor{\tau^{\superscriptI} }\) such that \( \gamma  \PositionZeroTurnstile  \possiblyWithSub\stageImetaColor{v^{\superscriptI} }  :  \possiblyWithSub\stageImetaColor{\tau^{\superscriptI} } \) and \( \possiblyWithSub\stageImetaColor{T^{\superscriptI} }  \equiv^{1}   \possiblyWithSub\stageImetaColor{\tau^{\superscriptI} }  \).
  }
  \begin{proof}
    By induction on the derivation of \( \mathit{\Gamma}  \vdash^{1}   \possiblyWithSub\stageImetaColor{v^{\superscriptI} }   :  \possiblyWithSub\stageImetaColor{T^{\superscriptI} } \).
    \begin{itemize}
      \item Case \derive[T1-TyEquiv]{%
         \mathit{\Gamma}  \vdash^{1}   \possiblyWithSub\stageImetaColor{v^{\superscriptI} }   :  \possiblyWithSub\stageImetaColor{T'^{\superscriptI} } 
      \andalso
         \possiblyWithSub\stageImetaColor{T'^{\superscriptI} }  \equiv^{1}  \possiblyWithSub\stageImetaColor{T^{\superscriptI} } 
      \andalso
         \mathit{\Gamma}  \vdash^{1}  \possiblyWithSub\stageImetaColor{T^{\superscriptI} } 
      }{%
         \mathit{\Gamma}  \vdash^{1}   \possiblyWithSub\stageImetaColor{v^{\superscriptI} }   :  \possiblyWithSub\stageImetaColor{T^{\superscriptI} } 
      }:
        By IH, there exists \(\possiblyWithSub\stageImetaColor{\tau^{\superscriptI} }\) such that \( \gamma  \PositionZeroTurnstile  \possiblyWithSub\stageImetaColor{v^{\superscriptI} }  :  \possiblyWithSub\stageImetaColor{\tau^{\superscriptI} } \) and \( \possiblyWithSub\stageImetaColor{T'^{\superscriptI} }  \equiv^{1}   \possiblyWithSub\stageImetaColor{\tau^{\superscriptI} }  \).
        This enables us to derive
        \begin{center}
          \infer[CqT1-Trans]{%
            \infer[CqT1-Sym]{%
               \possiblyWithSub\stageImetaColor{T'^{\superscriptI} }  \equiv^{1}  \possiblyWithSub\stageImetaColor{T^{\superscriptI} } 
            }{%
               \possiblyWithSub\stageImetaColor{T^{\superscriptI} }  \equiv^{1}  \possiblyWithSub\stageImetaColor{T'^{\superscriptI} } 
            }
          \andalso
             \possiblyWithSub\stageImetaColor{T'^{\superscriptI} }  \equiv^{1}   \possiblyWithSub\stageImetaColor{\tau^{\superscriptI} }  
          }{%
             \possiblyWithSub\stageImetaColor{T^{\superscriptI} }  \equiv^{1}   \possiblyWithSub\stageImetaColor{\tau^{\superscriptI} }  
          }.
        \end{center}
      \item Case \derive[T1-Var]{%
         \vdash  \mathit{\Gamma} 
      \andalso
        \mathit{\Gamma}(\possiblyWithSub\stageImetaColor{x}) = (\possiblyWithSub\stageImetaColor{T^{\superscriptI} })^{1}
      }{%
         \mathit{\Gamma}  \vdash^{1}   \possiblyWithSub\stageImetaColor{x}   :  \possiblyWithSub\stageImetaColor{T^{\superscriptI} } 
      }:
        Since \( \mathit{\Gamma}  \equiv  \gamma \) and \(\mathit{\Gamma}(\possiblyWithSub\stageImetaColor{x}) = (\possiblyWithSub\stageImetaColor{T^{\superscriptI} })^{1}\),
        by letting \(\possiblyWithSub\stageImetaColor{\tau^{\superscriptI} } := \gamma(\possiblyWithSub\stageImetaColor{x})\), we clearly have \( \possiblyWithSub\stageImetaColor{T^{\superscriptI} }  \equiv^{1}   \possiblyWithSub\stageImetaColor{\tau^{\superscriptI} }  \)
        and
        \derive[G-Var]{%
          \gamma(\possiblyWithSub\stageImetaColor{x}) = \possiblyWithSub\stageImetaColor{\tau^{\superscriptI} }
        }{%
           \gamma  \PositionZeroTurnstile   \possiblyWithSub\stageImetaColor{x}   :  \possiblyWithSub\stageImetaColor{\tau^{\superscriptI} } 
        }.
      \item Case \derive[T1-Abs]{%
         \mathit{\Gamma}  \vdash^{1}   \possiblyWithSub\stageImetaColor{\tau^{\superscriptI} }_{{\mathrm{1}}}  
      \andalso
          \mathit{\Gamma} ,  \possiblyWithSub\stageImetaColor{x}  : (  \possiblyWithSub\stageImetaColor{\tau^{\superscriptI} }_{{\mathrm{1}}}  )^{1}   \vdash^{1}   \possiblyWithSub\stageImetaColor{v^{\superscriptI} }_{{\mathrm{2}}}   :  \possiblyWithSub\stageImetaColor{T^{\superscriptI} }_{{\mathrm{2}}} 
      \andalso
        \possiblyWithSub\stageImetaColor{x} \not\in \fv(\possiblyWithSub\stageImetaColor{\tau^{\superscriptI} }_{{\mathrm{2}}})
      }{%
         \mathit{\Gamma}  \vdash^{1}   \openI{(}  \ordI{\lambda} \possiblyWithSub\stageImetaColor{x}  \relI{:}   \possiblyWithSub\stageImetaColor{\tau^{\superscriptI} }_{{\mathrm{1}}}  \punctI{.}\   \possiblyWithSub\stageImetaColor{v^{\superscriptI} }_{{\mathrm{2}}}   \closeI{)}   :    \possiblyWithSub\stageImetaColor{\tau^{\superscriptI} }_{{\mathrm{1}}}   \relI{\to}  \possiblyWithSub\stageImetaColor{T^{\superscriptI} }_{{\mathrm{2}}}  
      }:
        Since \( \vdash   \mathit{\Gamma} ,  \possiblyWithSub\stageImetaColor{x}  : (  \possiblyWithSub\stageImetaColor{\tau^{\superscriptI} }_{{\mathrm{1}}}  )^{1}  \) and \(  \mathit{\Gamma} ,  \possiblyWithSub\stageImetaColor{x}  : (  \possiblyWithSub\stageImetaColor{\tau^{\superscriptI} }_{{\mathrm{1}}}  )^{1}   \equiv   \gamma ,  \possiblyWithSub\stageImetaColor{x}  :  \possiblyWithSub\stageImetaColor{\tau^{\superscriptI} }_{{\mathrm{1}}}  \) clearly hold,
        by IH, there exists \(\possiblyWithSub\stageImetaColor{\tau^{\superscriptI} }_{{\mathrm{2}}}\) such that
        \(  \gamma ,  \possiblyWithSub\stageImetaColor{x}  :  \possiblyWithSub\stageImetaColor{\tau^{\superscriptI} }_{{\mathrm{1}}}   \PositionZeroTurnstile  \possiblyWithSub\stageImetaColor{v^{\superscriptI} }_{{\mathrm{2}}}  :  \possiblyWithSub\stageImetaColor{\tau^{\superscriptI} }_{{\mathrm{2}}} \) and \( \possiblyWithSub\stageImetaColor{T^{\superscriptI} }_{{\mathrm{2}}}  \equiv^{1}   \possiblyWithSub\stageImetaColor{\tau^{\superscriptI} }_{{\mathrm{2}}}  \).
        These enable us to derive the following two:
        \begin{center}
          \infer{%
              \gamma ,  \possiblyWithSub\stageImetaColor{x}  :  \possiblyWithSub\stageImetaColor{\tau^{\superscriptI} }_{{\mathrm{1}}}   \PositionZeroTurnstile  \possiblyWithSub\stageImetaColor{v^{\superscriptI} }_{{\mathrm{2}}}  :  \possiblyWithSub\stageImetaColor{\tau^{\superscriptI} }_{{\mathrm{2}}} 
          }{%
             \gamma  \PositionZeroTurnstile   \openI{(}  \ordI{\lambda} \possiblyWithSub\stageImetaColor{x}  \relI{:}  \possiblyWithSub\stageImetaColor{\tau^{\superscriptI} }_{{\mathrm{1}}} \punctI{.}\  \possiblyWithSub\stageImetaColor{v^{\superscriptI} }_{{\mathrm{2}}}  \closeI{)}   :   \possiblyWithSub\stageImetaColor{\tau^{\superscriptI} }_{{\mathrm{1}}}  \relI{\to}  \possiblyWithSub\stageImetaColor{\tau^{\superscriptI} }_{{\mathrm{2}}}  
          }
        \qquad
          \infer{%
            \infer{}{%
                \possiblyWithSub\stageImetaColor{\tau^{\superscriptI} }_{{\mathrm{1}}}   \equiv^{1}   \possiblyWithSub\stageImetaColor{\tau^{\superscriptI} }_{{\mathrm{1}}}  
            }
          \andalso
             \possiblyWithSub\stageImetaColor{T^{\superscriptI} }_{{\mathrm{2}}}  \equiv^{1}   \possiblyWithSub\stageImetaColor{\tau^{\superscriptI} }_{{\mathrm{2}}}  
          }{%
               \possiblyWithSub\stageImetaColor{\tau^{\superscriptI} }_{{\mathrm{1}}}   \relI{\to}   \possiblyWithSub\stageImetaColor{\tau^{\superscriptI} }_{{\mathrm{2}}}    \equiv^{1}    \possiblyWithSub\stageImetaColor{\tau^{\superscriptI} }_{{\mathrm{1}}}   \relI{\to}  \possiblyWithSub\stageImetaColor{T^{\superscriptI} }_{{\mathrm{2}}}  
          }
        \end{center}
        i.e.,~\(\possiblyWithSub\stageImetaColor{\tau^{\superscriptI} } :=  \possiblyWithSub\stageImetaColor{\tau^{\superscriptI} }_{{\mathrm{1}}}  \relI{\to}  \possiblyWithSub\stageImetaColor{\tau^{\superscriptI} }_{{\mathrm{2}}} \) satisfies the desired properties.
      \item Case \derive[T1-App]{%
         \mathit{\Gamma}  \vdash^{1}   \possiblyWithSub\stageImetaColor{v^{\superscriptI} }_{{\mathrm{1}}}   :   \possiblyWithSub\stageImetaColor{T^{\superscriptI} }_{{\mathrm{2}}}  \relI{\to}  \possiblyWithSub\stageImetaColor{T^{\superscriptI} }  
      \andalso
         \mathit{\Gamma}  \vdash^{1}   \possiblyWithSub\stageImetaColor{v^{\superscriptI} }_{{\mathrm{2}}}   :  \possiblyWithSub\stageImetaColor{T^{\superscriptI} }_{{\mathrm{2}}} 
      }{%
         \mathit{\Gamma}  \vdash^{1}    \possiblyWithSub\stageImetaColor{v^{\superscriptI} }_{{\mathrm{1}}}  \   \possiblyWithSub\stageImetaColor{v^{\superscriptI} }_{{\mathrm{2}}}    :  \possiblyWithSub\stageImetaColor{T^{\superscriptI} } 
      }:
        By IH, there exist \(\possiblyWithSub\stageImetaColor{\tau^{\superscriptI} }_{{\mathrm{1}}}\) and \(\possiblyWithSub\stageImetaColor{\tau^{\superscriptI} }_{{\mathrm{2}}}\) such that
        \( \gamma  \PositionZeroTurnstile  \possiblyWithSub\stageImetaColor{v^{\superscriptI} }_{{\mathrm{1}}}  :  \possiblyWithSub\stageImetaColor{\tau^{\superscriptI} }_{{\mathrm{1}}} \), \(  \possiblyWithSub\stageImetaColor{T^{\superscriptI} }_{{\mathrm{2}}}  \relI{\to}  \possiblyWithSub\stageImetaColor{T^{\superscriptI} }   \equiv^{1}   \possiblyWithSub\stageImetaColor{\tau^{\superscriptI} }_{{\mathrm{1}}}  \),
        \( \gamma  \PositionZeroTurnstile  \possiblyWithSub\stageImetaColor{v^{\superscriptI} }_{{\mathrm{2}}}  :  \possiblyWithSub\stageImetaColor{\tau^{\superscriptI} }_{{\mathrm{2}}} \), and \( \possiblyWithSub\stageImetaColor{T^{\superscriptI} }_{{\mathrm{2}}}  \equiv^{1}   \possiblyWithSub\stageImetaColor{\tau^{\superscriptI} }_{{\mathrm{2}}}  \).
        Then, by Lemma~\ref{lem:stage-1-arrow-csr-type-equiv-form},
        from \(  \possiblyWithSub\stageImetaColor{T^{\superscriptI} }_{{\mathrm{2}}}  \relI{\to}  \possiblyWithSub\stageImetaColor{T^{\superscriptI} }   \equiv^{1}   \possiblyWithSub\stageImetaColor{\tau^{\superscriptI} }_{{\mathrm{1}}}  \),
        there exist \(\possiblyWithSub\stageImetaColor{\tau^{\superscriptI} }_{{\mathrm{11}}}\) and \(\possiblyWithSub\stageImetaColor{\tau^{\superscriptI} }_{{\mathrm{12}}}\) such that
        \(\possiblyWithSub\stageImetaColor{\tau^{\superscriptI} }_{{\mathrm{1}}} =  \possiblyWithSub\stageImetaColor{\tau^{\superscriptI} }_{{\mathrm{11}}}  \relI{\to}  \possiblyWithSub\stageImetaColor{\tau^{\superscriptI} }_{{\mathrm{12}}} \), \(  \possiblyWithSub\stageImetaColor{\tau^{\superscriptI} }_{{\mathrm{11}}}   \equiv^{1}  \possiblyWithSub\stageImetaColor{T^{\superscriptI} }_{{\mathrm{2}}} \), and \(  \possiblyWithSub\stageImetaColor{\tau^{\superscriptI} }_{{\mathrm{12}}}   \equiv^{1}  \possiblyWithSub\stageImetaColor{T^{\superscriptI} } \),
        and thus we can derive
        \begin{center}
          \derive[CqT0-Trans]{%
              \possiblyWithSub\stageImetaColor{\tau^{\superscriptI} }_{{\mathrm{11}}}   \equiv^{1}  \possiblyWithSub\stageImetaColor{T^{\superscriptI} }_{{\mathrm{2}}} 
          \andalso
             \possiblyWithSub\stageImetaColor{T^{\superscriptI} }_{{\mathrm{2}}}  \equiv^{1}   \possiblyWithSub\stageImetaColor{\tau^{\superscriptI} }_{{\mathrm{2}}}  
          }{%
              \possiblyWithSub\stageImetaColor{\tau^{\superscriptI} }_{{\mathrm{11}}}   \equiv^{1}   \possiblyWithSub\stageImetaColor{\tau^{\superscriptI} }_{{\mathrm{2}}}  
          }.
        \end{center}
        Here, by Lemma~\ref{lem:value-type-csr-equiv-implies-equal},
        we have \(\possiblyWithSub\stageImetaColor{\tau^{\superscriptI} }_{{\mathrm{11}}} = \possiblyWithSub\stageImetaColor{\tau^{\superscriptI} }_{{\mathrm{2}}}\) and thereby \(\possiblyWithSub\stageImetaColor{\tau^{\superscriptI} }_{{\mathrm{1}}} =  \possiblyWithSub\stageImetaColor{\tau^{\superscriptI} }_{{\mathrm{2}}}  \relI{\to}  \possiblyWithSub\stageImetaColor{\tau^{\superscriptI} }_{{\mathrm{12}}} \).
        This enables us to derive
        \begin{center}
          \derive[G-App]{%
             \gamma  \PositionZeroTurnstile  \possiblyWithSub\stageImetaColor{v^{\superscriptI} }_{{\mathrm{1}}}  :   \possiblyWithSub\stageImetaColor{\tau^{\superscriptI} }_{{\mathrm{2}}}  \relI{\to}  \possiblyWithSub\stageImetaColor{\tau^{\superscriptI} }_{{\mathrm{12}}}  
          \andalso
             \gamma  \PositionZeroTurnstile  \possiblyWithSub\stageImetaColor{v^{\superscriptI} }_{{\mathrm{2}}}  :  \possiblyWithSub\stageImetaColor{\tau^{\superscriptI} }_{{\mathrm{2}}} 
          }{%
             \gamma  \PositionZeroTurnstile   \possiblyWithSub\stageImetaColor{v^{\superscriptI} }_{{\mathrm{1}}} \  \possiblyWithSub\stageImetaColor{v^{\superscriptI} }_{{\mathrm{2}}}   :  \possiblyWithSub\stageImetaColor{\tau^{\superscriptI} }_{{\mathrm{12}}} 
          }.
        \end{center}
        Since \(  \possiblyWithSub\stageImetaColor{\tau^{\superscriptI} }_{{\mathrm{12}}}   \equiv^{1}  \possiblyWithSub\stageImetaColor{T^{\superscriptI} } \) already holds,
        \(\possiblyWithSub\stageImetaColor{\tau^{\superscriptI} }_{{\mathrm{12}}}\) satisfies the desired properties.
      \item Case \derive[T1-CstP]{%
         \vdash  \mathit{\Gamma} 
      \andalso
        \ConstEnvPers(c) = \possiblyWithSub\stageImetaColor{\tau^{\superscriptI} }
      }{%
         \mathit{\Gamma}  \vdash^{1}   \possiblyWithSub\stageImetaColor{c}   :   \possiblyWithSub\stageImetaColor{\tau^{\superscriptI} }  
      }:
        We immediately have
        \derive[CqT1-Refl]{}{%
            \possiblyWithSub\stageImetaColor{\tau^{\superscriptI} }   \equiv^{1}   \possiblyWithSub\stageImetaColor{\tau^{\superscriptI} }  
        }
        and can derive
        \derive[G-Cst]{%
          \ConstEnvPers(c) = \possiblyWithSub\stageImetaColor{\tau^{\superscriptI} }
        }{%
           \gamma  \PositionZeroTurnstile   \possiblyWithSub\stageImetaColor{c}   :  \possiblyWithSub\stageImetaColor{\tau^{\superscriptI} } 
        }.
      \item Case \rulename{T1-Esc} contradicts the form of \(\possiblyWithSub\stageImetaColor{v^{\superscriptI} }\).
    \end{itemize}
  \end{proof}

\end{document}